%% file: main.tex
\documentclass[
twoside,openright,titlepage,numbers=noenddot,headinclude,
footinclude=true,cleardoublepage=empty,abstractoff, 
BCOR=5mm,paper=a4,fontsize=11pt,
ngerman,american,%
]{scrreprt}

 \newif\ifdraft 
 \newif\ifsembunyi 

 \newif\ifhideintro 
 \newif\ifhideconclusion 
 \newif\ifhidecontent 
 \newif\ifhidefrontmatter 
 \newif\ifhideshade 

 \newif\iffinal 

 \drafttrue
 \sembunyitrue
 \sembunyifalse
 \hideintrotrue
 \hideintrofalse
 \hideconclusiontrue
 \hideconclusionfalse
 \hidecontenttrue
 \hidecontentfalse
 \hidefrontmattertrue
 \hidefrontmatterfalse
 \hideshadetrue
 \hideshadefalse

\finaltrue

\iffinal

\draftfalse
\sembunyifalse
\hideintrofalse
\hideconclusionfalse
\hidecontentfalse
\hidefrontmatterfalse
\hideshadefalse

\else

\fi

\usepackage[a-1b]{pdfx}
\input{0.setting/main-config}

\input{0.setting/package-import}

\input{0.setting/my-config}


\input{macros/macros-writing}

\input{macros/macros-intersection}

\input{macros/macros-obda}

\input{macros/macros-dcds}

\input{macros/macros-kab}

\input{macros/macros-gkab}

\input{macros/macros-context}
\input{macros/macros-iacsgkab}

\input{macros/macros-agkab}

\input{macros/macros-sedap}

\input{macros/macros-example}

\begin{document}
\frenchspacing
\raggedbottom
\selectlanguage{american} 
\pagenumbering{roman}
\pagestyle{plain}

\ifhidefrontmatter
\else
\include{1.front-back-matter/cover}
\cleardoublepage\include{1.front-back-matter/dirty-title-page}
\cleardoublepage\include{1.front-back-matter/title-page}
\include{1.front-back-matter/title-back}
\cleardoublepage\include{1.front-back-matter/dedication}
\cleardoublepage\include{1.front-back-matter/abstract}
\cleardoublepage\include{1.front-back-matter/acknowledgments}

\fi

\pagestyle{scrheadings}

\ifhidefrontmatter
\else
\cleardoublepage\include{1.front-back-matter/contents}

\fi


\pagenumbering{arabic}
\cleardoublepage

\clearpage
\input{2.chapters/1-intro-part1}

\input{2.chapters/1-intro-part2}

\cleardoublepage

\include{2.chapters/2-preliminaries}

\cleardoublepage

\include{2.chapters/3-kab}
\cleardoublepage

\include{2.chapters/4-gkab}

\cleardoublepage


\include{2.chapters/5-ia-gkab}

\cleardoublepage


\include{2.chapters/6-cs-gkab}

\cleardoublepage


\include{2.chapters/7-cs-ia-gkab}

\cleardoublepage


\include{2.chapters/8-a-gkab}

\cleardoublepage


\include{2.chapters/9-sedap}

\cleardoublepage


\include{2.chapters/10-conclusion}

\cleardoublepage




\cleardoublepage\include{3.bibliography/bibliography}

\ifsembunyi
\else
\cleardoublepage\include{1.front-back-matter/closing}

\fi


\end{document} 


%% file: 0.setting/main-config.tex
\ifdraft 

\PassOptionsToPackage{ eulerchapternumbers, listings,
  drafting, pdfspacing,
  linedheaders, subfig, beramono,
  parts} {classicthesis} 

\else 

\PassOptionsToPackage{ eulerchapternumbers, listings, 
  pdfspacing,
  linedheaders, subfig, beramono,
  dottedtoc,
 parts} {classicthesis}

\fi

\usepackage{ifthen}
\newboolean{enable-backrefs} 
\setboolean{enable-backrefs}{true} 

\newcommand{\myTitle}{Verification of Data-aware Business Processes in the
Presence of Ontologies\xspace}
\newcommand{\myTitleBU}{VERIFICATION OF DATA-AWARE BUSINESS PROCESSES \\ IN THE
  PRESENCE OF ONTOLOGIES\xspace}

\newcommand{\myName}{Ario Santoso\xspace}

\newcommand{\mySupervisor}{Diego Calvanese}
\newcommand{\mySupervisorWebsite}{\url{http://www.inf.unibz.it/~calvanese}}
\newcommand{\mySupervisorInstitution}{Libera Universit{\`a} di Bolzano}

\newcommand{\myCoSupervisor}{Marco Montali}
\newcommand{\myCoSupervisorWebsite}{\url{http://www.inf.unibz.it/~montali}}
\newcommand{\myCoSupervisorInstitution}{Libera Universit{\`a} di Bolzano}

\newcommand{\myExternalSupervisor}{Franz Baader}
\newcommand{\myExternalSupervisorWebsite}{\url{http://lat.inf.tu-dresden.de/~baader}}
\newcommand{\myExternalSupervisorInstitution}{Technische Universit\"at
  Dresden}

\newcommand{\myFaculty}{Faculty of Computer Science\xspace}
\newcommand{\myUni}{Free University of Bozen-Bolzano\xspace}

\newcommand{\myTime}{April 2016\xspace}
\newcommand{\myWebsite}{\url{http://www.inf.unibz.it/~asantoso}\xspace}

\newcounter{dummy} 
\providecommand{\mLyX}{L\kern-.1667em\lower.25em\hbox{Y}\kern-.125emX\@}

\PassOptionsToPackage{latin9}{inputenc}	
 \usepackage{inputenc}				

\PassOptionsToPackage{british,UKenglish,USenglish,english,american}{babel}   
 \usepackage{babel}

\PassOptionsToPackage{square,numbers}{natbib}
 \usepackage{natbib}				

 \usepackage{amsmath}

\PassOptionsToPackage{T1}{fontenc} 
	\usepackage{fontenc}     
\usepackage{textcomp} 
\usepackage{scrhack} 
\usepackage{xspace} 
\usepackage{mparhack} 
\usepackage{fixltx2e} 
\PassOptionsToPackage{printonlyused,smaller}{acronym}
	\usepackage{acronym} 

\usepackage{tabularx} 
\usepackage{caption}
\usepackage{subfig}  

\usepackage{listings} 
	\usepackage{hyperref}  
\PassOptionsToPackage{pdftex}{graphicx}
	\usepackage{graphicx} 

\newcommand{\backrefnotcitedstring}{\relax}
\newcommand{\backrefcitedsinglestring}[1]{(Cited on page~#1.)}
\newcommand{\backrefcitedmultistring}[1]{(Cited on pages~#1.)}
\ifthenelse{\boolean{enable-backrefs}}%
{%
		\PassOptionsToPackage{hyperpageref}{backref}
		\usepackage{backref} 
		   \renewcommand*{\backref}[1]{}  
		   \renewcommand*{\backrefalt}[4]{
		      \ifcase #1 %
		         \backrefnotcitedstring%
		      \or%
		         \backrefcitedsinglestring{#2}%
		      \else%
		         \backrefcitedmultistring{#2}%
		      \fi}%
}{\relax}    

\hypersetup{%
    colorlinks=true, linktocpage=true, pdfstartpage=3, pdfstartview=FitV,%
    breaklinks=true, pdfpagemode=UseNone, pageanchor=true, pdfpagemode=UseOutlines,%
    plainpages=false, bookmarksnumbered, bookmarksopen=true, bookmarksopenlevel=1,%
    hypertexnames=true, pdfhighlight=/O,
    urlcolor=webbrown, linkcolor=RoyalBlue, citecolor=webgreen, 
    pdftitle={\myTitle},%
    pdfauthor={\textcopyright\ \myName, \myUni, \myFaculty},%
    pdfsubject={},%
    pdfkeywords={},%
    pdfcreator={pdfLaTeX},%
    pdfproducer={LaTeX with hyperref and classicthesis},%
    pdfencoding=unicode 
}   
\inputencoding{utf8}
\makeatother 

\makeatletter
\@ifpackageloaded{babel}%
    {%
       \addto\extrasamerican{%
				}%
       \addto\extrasngerman{%
				}%
    }{\relax}
\makeatother

\usepackage{classicthesis} 

\areaset[current]{405pt}{753pt} 

\usepackage{lmodern} 

%% file: 0.setting/package-import.tex
\usepackage{microtype} 
\usepackage{latexsym}
\usepackage{amssymb}
\usepackage{amsmath}
\usepackage{amsthm}

\usepackage{xspace}
\xspaceaddexceptions{]\}}

\usepackage{url}
\usepackage{comment}
\usepackage{enumerate}
\usepackage[defblank]{paralist}

\usepackage{lipsum}

\usepackage{mathtools}

\usepackage[english]{cleveref}

\usepackage{anyfontsize}

\usepackage{framed}
\definecolor{shadecolor}{gray}{0.90}

\usepackage[titles]{tocloft}

\usepackage{eso-pic}

%% file: 0.setting/my-config.tex
\makeatletter  
\titleformat{\section}{\bf\large}{\large\thesection}{1em}{}
\titleformat{\subsection}{\bf\normalsize}{\large\thesubsection}{1em}{}
\titleformat{\subsubsection}{\bf\normalsize}{\large\thesubsubsection}{1em}{}
\makeatother

\makeatletter
\ifthenelse{\boolean{@linedheaders}}%
    {
    \titleformat{\chapter}[display]%
        {\relax}
       {\mbox{}\oldmarginpar{\vspace*{-3.0\baselineskip}\color{halfgray}\chapterNumber\thechapter}}
        {0pt}%
        {\color{black!70!darkgray}\LARGE\raggedright\spacedallcaps}
        [\normalcolor\normalsize\vspace*{.3\baselineskip}\titlerule\vspace*{.3\baselineskip}]%
    }
    {
    \titleformat{\chapter}[display]%
        {\relax}
        {\mbox{}\oldmarginpar{\vspace*{-2\baselineskip}\color{halfgray}\chapterNumber\thechapter}}
        {0pt}%
        {\color{black!70!darkgray}\Large\raggedright\spacedallcaps}
        [\normalcolor\normalsize\vspace*{.3\baselineskip}\titlerule\vspace*{.3\baselineskip}]%
    }
\makeatother  

\AtBeginDocument{\renewcommand{\thepart}{\Roman{part}}}

\renewcommand{\spacedlowsmallcaps}[1]{\textsc{#1}}%
\renewcommand{\spacedallcaps}[1]{\textls[170]{\MakeTextUppercase{#1}}}%

\allowdisplaybreaks

%% file: macros/macros-writing.tex
\newcommand{\qedw}{\hfill\ensuremath{\square}}

\newcommand{\qedboxfull}{\vrule height 5pt width 5pt depth 0pt}
\newcommand{\qedfull}{\hfill{\qedboxfull}}

\newtheorem{counter}{Counter}[chapter] 

\theoremstyle{definition}
\newtheorem{mydefinition}[counter]{Definition}
\newenvironment{definition}{\begin{mydefinition}}{\null\hfill\qedfull\smallskip\end{mydefinition}}
\crefname{mydefinition}{definition}{definitions}

\newtheorem{myexample}[counter]{Example}

\ifhideshade
\newenvironment{example}{\begin{myexample}}{\end{myexample}}
\else
\newenvironment{example}{\begin{myexample}\begin{shaded}}{\end{shaded}\end{myexample}}
\fi

\crefname{myexample}{example}{examples}

\theoremstyle{remark}

\crefname{remark}{remark}{remarks}

\theoremstyle{plain}

\crefname{conjecture}{conjecture}{conjectures}

\crefname{claim}{claim}{claims}

\crefname{proposition}{proposition}{propositions}
\newtheorem{lemma}[counter]{Lemma}
\crefname{lemma}{lemma}{lemmas}
\newtheorem{theorem}[counter]{Theorem}
\crefname{theorem}{theorem}{theorems}

\crefname{corollary}{corollary}{corollaries}

\newcommand{\A}{\mathcal{\uppercase{A}}}
\newcommand{\B}{\mathcal{\uppercase{B}}}
\newcommand{\Ccal}{\mathcal{\uppercase{C}}}

\newcommand{\E}{\mathcal{\uppercase{E}}}
\newcommand{\F}{\mathcal{\uppercase{F}}}
\newcommand{\Gcal}{\mathcal{\uppercase{G}}}

\newcommand{\I}{\mathcal{\uppercase{I}}}

\newcommand{\K}{\mathcal{\uppercase{K}}}

\renewcommand{\L}{ \mathcal{\uppercase{L}} }

\newcommand{\M}{\mathcal{\uppercase{M}}}

\renewcommand{\O}{\mathcal{\uppercase{O}}}

\newcommand{\Q}{\mathcal{\uppercase{Q}}}
\newcommand{\R}{\mathcal{\uppercase{R}}}
\renewcommand{\S}{\mathcal{\uppercase{S}}}
\newcommand{\T}{\mathcal{\uppercase{T}}}

\newcommand{\V}{\mathcal{\uppercase{V}}}

\newcommand{\true}{\mathsf{true}}
\newcommand{\false}{\mathsf{false}}

\renewcommand{\bf}{\normalfont \bfseries }
\renewcommand{\it}{\normalfont \itshape }
\renewcommand{\tt}{\ttfamily }

\newcommand{\ra}{\rightarrow}

\newcommand{\lra}{\leftrightarrow}

\newcommand{\Lora}{\Longrightarrow}

\newcommand{\Lola}{\Longleftarrow}

\newcommand{\set}[1]{\{#1\}} 

\newcommand{\card}[1]{|{#1}|} 

\newcommand{\tup}[1]{\langle #1\rangle} 

\newcommand{\tap}[1]{[#1]} 

\newcommand{\sidetext}[1]{\graffito{#1}} 

\newcommand{\sidetextb}[1]{\graffito{#1}} 

%% file: macros/macros-intersection.tex
\newcommand{\ANS}{\textsc{ans}}
\newcommand{\Ans}{\textsc{cert}}
\newcommand{\ans}{\mathit{cert}}

\newcommand{\rew}{\mathit{rew}}
\newcommand{\conj}{\mathit{conj}}

\newcommand{\adom}[1]{\textsc{adom}(#1)} 

\newcommand{\difol}{\text{DI-FOL}\xspace} 
\newcommand{\diecq}{\text{DI-ECQ}\xspace} 

\newcommand{\const}{\Delta} 
\newcommand{\iconst}{\Delta_0} 

\newcommand{\oset}{\Delta_O} 
\newcommand{\vset}{\Delta_V} 

\newcommand{\domain}[1]{\ensuremath{\textsc{dom}(#1)}\xspace}

\newcommand{\kdbsim}{\sim_{\textsc{kd}}}

\newcommand{\jbsim}{\sim_{\textsc{j}}}

\newcommand{\sbsim}{\sim_{\textsc{so}}}

\newcommand{\ebsim}{\sim_{\textsc{e}}}

\newcommand{\stbsim}{\sim_{\textsc{st}}}

\newcommand{\arset}[1]{\textsc{b-rep}(#1)}

\newcommand{\iarset}[1]{\textsc{c-rep}(#1)}

\newcommand{\evol}{\textsc{evol}}

\newcommand{\qunsatf}{q^f_{\textnormal{unsat}}}
\newcommand{\qunsatn}{q^n_{\textnormal{unsat}}}
\newcommand{\qunsatfol}[1]{Q^{#1}_{\textnormal{unsatFOL}}}
\newcommand{\qunsatecq}[1]{Q^{#1}_{\textnormal{unsatECQ}}}
\newcommand{\csqunsatecq}[1]{Q^{#1}_{\textnormal{unsatECQ}}}

%% file: macros/macros-obda.tex
\newcommand{\dllite}{\textit{\uppercase{DL}-\uppercase{L}ite}\xspace}
\newcommand{\dlliter}{\textit{\uppercase{DL}-\uppercase{L}ite}\ensuremath{_{\R}}\xspace}
\newcommand{\dllitea}{\textit{\uppercase{DL}-\uppercase{L}ite}\ensuremath{_{\!\A}}\xspace}

\newcommand{\dlalc}{\A\L\Ccal}

\newcommand{\INV}[1]{#1^{-}}

\newcommand{\SOMET}[1]{\exists #1}

\newcommand{\DOMAIN}[1]{\delta(#1)}

\newcommand{\funct}[1]{(\mathsf{funct}~#1)}

\newcommand{\ISA}{\sqsubseteq}

\newcommand{\voc}{\textsc{voc}}

\newcommand{\dom}[1][\I]{\Delta^{#1}}  
\newcommand{\Int}[2][\I]{#2^{#1}}      
\newcommand{\INT}[2][\I]{(#2)^{#1}}   
\newcommand{\inter}[1][\I]{(\dom[#1],\Int[#1]{\cdot})}   

%% file: macros/macros-dcds.tex
\newcommand{\dcdssym}{\S} 

\newcommand{\dcomp}{D} 
\newcommand{\ec}{\ensuremath{E}}
\newcommand{\ecset}{\ensuremath{\E}}
\newcommand{\idb}{\dbinst_{0}}

\newcommand{\pcomp}{P} 
\newcommand{\dservcall}{\F} 
\newcommand{\dscall}{\mbox{f}} 
\newcommand{\deff}[1]{\textsc{Eff}(#1)} 
\newcommand{\dact}{\alpha} \newcommand{\dactset}{\A}
\newcommand{\dprocset}{\ensuremath{\varrho}}

\newcommand{\dstateset}{\Sigma} 
\newcommand{\dtrans}{\Rightarrow} 
\newcommand{\db}{\mathit{db}} 

\newcommand{\dscmap}{\ensuremath{m}\xspace} 
\newcommand{\dscset}{\ensuremath{\mathbb{SC}}\xspace} 

 \newcommand{\ddoo}[1]{\textsc{do}_{\textsc{dcds}}(#1)} 
\newcommand{\dcalls}[1]{{\textsc{calls}({#1})}} 
\newcommand{\deval}[1]{{\textsc{eval}(#1)}} 
\newcommand{\dexect}[1]{\xrightarrow[]{#1}} 

\newcommand{\dbschema}{\R} 
\newcommand{\dbinst}{\textbf{I}} 

%% file: macros/macros-kab.tex
\newcommand{\initabox}{A_0} 

\newcommand{\act}{\alpha} 
\newcommand{\actset}{\Gamma} 
\newcommand{\add}{\textbf{add \xspace}}
\newcommand{\del}{\textbf{del \xspace}}
\newcommand{\eff}[1]{\textsc{Eff}(#1)}
\newcommand{\facta}{F^+}
\newcommand{\factd}{F^-}
\newcommand{\effd}[1]{\textsc{Eff}_{d}(#1)}

\newcommand{\procset}{\Pi} 

\newcommand{\kabsym}{\K}

\newcommand{\servcall}{\F} 
\newcommand{\scall}{\mbox{f}} 
\newcommand{\map}[2]{#1 \rightsquigarrow #2} 
\newcommand{\carule}[2]{#1 \mapsto #2} 

\newcommand{\stateset}{\Sigma} 
\newcommand{\trans}{\Rightarrow} 
\newcommand{\abox}{\mathit{abox}} 

\newcommand{\ts}[1]{\varUpsilon_{#1}} 

\newcommand{\scmap}{\ensuremath{m}\xspace} 

\newcommand{\doo}[1]{\textsc{do}(#1)} 
\newcommand{\calls}[1]{{\textsc{calls}({#1})}} 
\newcommand{\eval}[1]{{\textsc{eval}(#1)}} 
\newcommand{\exec}[1]{\textsc{exec}_{#1}\xspace} 
\newcommand{\exect}[1]{\xrightarrow[]{#1}} 

\newcommand{\addfactssym}{\textsc{add}}
\newcommand{\delfactssym}{\textsc{del}}

\newcommand{\addfacts}[1]{\addfactssym(#1)}
\newcommand{\delfacts}[1]{\delfactssym(#1)}

\newcommand{\muL}{\mu\mathcal{\uppercase{L}}}

\newcommand{\muladom}{\ensuremath{\mu\mathcal{\uppercase{L}}_{\uppercase{A}}^{{\textnormal{\uppercase{EQL}}}}}\xspace}

\newcommand{\ladom}{\ensuremath{\mathcal{\uppercase{L}}_{\uppercase{A}}^{{\textnormal{\uppercase{EQL}}}}}\xspace}

\newcommand{\mula}{\ensuremath{\mu\mathcal{\uppercase{L}}_{\uppercase{A}}}\xspace}

\newcommand{\BOX}[1]{ [\!-\!] #1}
\newcommand{\DIAM}[1]{\langle \!-\! \rangle #1}

\newcommand{\vfo}{\ensuremath{v}} 
\newcommand{\vso}{\ensuremath{V}} 

\newcommand{\MOD}[1]{(#1)^{\ts{}}}
\newcommand{\MODA}[1]{(#1)_{\vfo,\vso}^{\ts{}}}
\newcommand{\MODAX}[2]{(#1)_{\vfo #2,\vso}^{\ts{}}}

\newcommand{\tdcds}{\tau_{dcds}}

%% file: macros/macros-gkab.tex
\newcommand{\gkabsym}{\Gcal}
\newcommand{\ginitprog}{\delta}

\newcommand{\gemptyprog}{\varepsilon}

\newcommand{\gPICK}{\textbf{pick}}
\newcommand{\gWHILE}{\textbf{while}}
\newcommand{\gDO}{\textbf{do}}

\newcommand{\gIF}{\textbf{if}}
\newcommand{\gTHEN}{\textbf{then}}
\newcommand{\gELSE}{\textbf{else}}

\newcommand{\gif}[3]{\gIF~#1~\gTHEN~#2~\gELSE~#3}
\newcommand{\gact}[2]{\gPICK~#1.#2}
\newcommand{\gwhile}[2]{\gWHILE~#1~\gDO~#2}

\newcommand{\gprogtrans}[1]{\xrightarrow{#1}}

\newcommand{\gexectrans}{\ra}

\newcommand{\pid}{pid}
\newcommand{\tpid}{\tau_{id}}

\newcommand{\gfin}{\mathbb{F}}

\newcommand{\final}[1]{#1\in\gfin}

\newcommand{\ask}{\textsc{ask}\xspace}
\newcommand{\tell}{\textsc{tell}\xspace}
\newcommand{\filter}{f}
\newcommand{\filterb}{\mathbf{f}}

\newcommand{\progres}{\textsc{res}} 

\newcommand{\tkabs}{\tau_S} 

\newcommand{\ppre}{pre} 
\newcommand{\ppost}{post}

\newcommand{\mimic}{\cong}

\newcommand{\tgkab}{\tau_{\gkabsym}} 
\newcommand{\tgprog}{t_{\gkabsym}} 

\newcommand{\eqm}{\simeq}

\newcommand{\nnf}{\textsc{nnf}}

\newcommand{\setinvocation}{\Lambda}

\newcommand{\tmpconceptname}{\mathsf{State}}
\newcommand{\tmpconst}{\mathit{temp}}
\newcommand{\tmp}{\tmpconceptname(\tmpconst)}

\newcommand{\pre}{\mathit{st}} 
\newcommand{\post}{\mathit{ed}} 
\newcommand{\flagconceptname}{\mathsf{Flag}} 
\newcommand{\flagconcept}[1]{\flagconceptname(#1)} 

\newcommand{\noopconceptname}{\mathsf{Noop}} 
\newcommand{\noopconcept}[1]{\noopconceptname(#1)} 

\newcommand{\tforb}{t_B} 
\newcommand{\tforj}{t_{j}} 
\newcommand{\tford}{t_{dup}} 

\newcommand{\inc}{\textsc{inc}}

\newcommand{\tgkabb}{\tau_B} 
\newcommand{\tgkabc}{\tau_C} 
\newcommand{\tgkabe}{\tau_E} 

\newcommand{\tgprogb}{\kappa_B} 
\newcommand{\tgprogc}{\kappa_C} 
\newcommand{\tgproge}{\kappa_E} 

\newcommand{\actsettmpa}{\actset_{\varepsilon}^+}
\newcommand{\actsettmpd}{\actset_{\varepsilon}^-}

\newcommand{\tpick}{\tau_\pi}

 \newcommand{\tforsb}{t_{sb}} 

 \newcommand{\tgkabsb}{\tau_{sb}} 

 \newcommand{\tgprogsb}{\kappa_{sb}} 

%% file: macros/macros-context.tex
\newcommand{\cstell}{\textsc{cs-tell}\xspace}
\newcommand{\csfilter}{f^{\ctxb}}

\newcommand{\csgkab}{CSGKAB\xspace}
\newcommand{\csgkabs}{CSGKABs\xspace}
\newcommand{\csgkabsym}{\gkabsym_{\ctxb}}

\newcommand{\scsgkab}{S-CSGKAB\xspace}
\newcommand{\scsgkabs}{S-CSGKABs\xspace}

\newcommand{\gactc}[3]{\gPICK~\tup{#1, #2}.#3}

\newcommand{\ctxchg}{\textsc{ctx-chg}} 

\newcommand{\ctxprocset}{\Pi_\ctx} %

\newcommand{\initctx}{\ctx_0}

\newcommand{\carulex}[3]{\tup{#1,#2} \mapsto \set{#3}} 

\newcommand{\cntx}{\mathit{ctx}} 

\newcommand{\cdimset}{\mathbb{D}\xspace}

\newcommand{\cdom}[1][d]{\ensuremath{\mathit{Dom}(#1)}\xspace} 
\newcommand{\topv}[1][d]{\top_{#1}}
\newcommand{\cover}[1][d]{\prec_{#1}} 

\newcommand{\cval}[2]{[#1\leadsto #2]} 

\newcommand{\ctxall}[1]{\textsc{ctx}(#1)}

\newcommand{\ctxconst}{\textbf{c}}

\newcommand{\ctxlang}{\L_{\ctxb}} 
\newcommand{\ctxe}{\varphi_{\ctx}} 

\newcommand{\ctx}{C}
\newcommand{\ctxb}{\mathit{cx}}

\newcommand{\ctxth}{\Phi_{\cdimset}}

\newcommand{\ctbox}{T_\ctxb}

\newcommand{\ctxmul}{\textsc{ctx}}
\newcommand{\mulcs}{\ensuremath{\muL_{\ctxmul}}\xspace}

\newcommand{\tgkabcs}{\tau_{\ctxb}} 

\newcommand{\tgprogcs}{\kappa_\ctxb} 

\newcommand{\tcsmula}{t_{\ctxb}} 

\newcommand{\ctxproc}{\Lambda_\ctx}

\newcommand{\inccon}{Inc}

\newcommand{\tfort}{t_{trip}} 

\newcommand{\eqc}{=_\ctxb}

\newcommand{\cdcq}{\textbf{D}} 

\newcommand{\cdcc}{D}

\newcommand{\tgkabscs}{\tau_{sc}} 

\newcommand{\tgprogscs}{\kappa_{sc}} 

%% file: macros/macros-iacsgkab.tex
\newcommand{\tgkabsic}{\tau_{sic}} 

\newcommand{\tgprogsic}{\kappa_{sic}} 

\newcommand{\icgkabs}{I-CSGKABs\xspace}

\newcommand{\bicgkab}{B-CSGKAB\xspace}
\newcommand{\bicgkabs}{B-CSGKABs\xspace}

\newcommand{\cicgkab}{C-CSGKAB\xspace}
\newcommand{\cicgkabs}{C-CSGKABs\xspace}

\newcommand{\eicgkab}{E-CSGKAB\xspace}
\newcommand{\eicgkabs}{E-CSGKABs\xspace}

\newcommand{\tgkabbcs}{\tau^{\ctxb}_B} 
\newcommand{\tgkabccs}{\tau^{\ctxb}_C} 
\newcommand{\tgkabecs}{\tau^{\ctxb}_E} 

\newcommand{\tgprogbcs}{\kappa^{\ctxb}_B} 
\newcommand{\tgprogccs}{\kappa^{\ctxb}_C} 
\newcommand{\tgprogecs}{\kappa^{\ctxb}_E} 

\newcommand{\tforjcs}{t^{\ctxb}_{j}} 

\newcommand{\cjbsimabr}{CJ}
\newcommand{\cjbsim}{\sim_{\textsc{cj}}}

%% file: macros/macros-agkab.tex
\newcommand{\agkab}{AGKAB\xspace}
\newcommand{\agkabs}{AGKABs\xspace}
\newcommand{\agkabsym}{\gkabsym_{\A}}

\newcommand{\bagkab}{B-AGKAB\xspace}
\newcommand{\bagkabs}{B-AGKABs\xspace}

\newcommand{\cagkab}{C-AGKAB\xspace}
\newcommand{\cagkabs}{C-AGKABs\xspace}

\newcommand{\eagkab}{E-AGKAB\xspace}
\newcommand{\eagkabs}{E-AGKABs\xspace}

\newcommand{\actsrc}{\mathit{actsrc}}

\newcommand{\fadd}{\mathit{fa}}

\newcommand{\fdel}{\mathit{fd}}

\newcommand{\actpar}{\mathit{actpar}}

\newcommand{\mulcsa}{\ensuremath{\muL^{Alt}_{\ctxmul}}\xspace}

\newcommand{\ajbsimabr}{AJ} 
\newcommand{\ajbsim}{\sim_{\textsc{aj}}} 

\newcommand{\asbsimabr}{AS} 
\newcommand{\asbsim}{\sim_{\textsc{AS}}} 

\newcommand{\eqmc}{\simeq_\ctxb}

\newcommand{\tgkabba}{\tau^{\A}_B} 
\newcommand{\tgkabca}{\tau^{\A}_C} 
\newcommand{\tgkabea}{\tau^{\A}_E} 

\newcommand{\tgprogba}{\kappa^{\A}_B} 
\newcommand{\tgprogca}{\kappa^{\A}_C} 
\newcommand{\tgprogea}{\kappa^{\A}_E} 

\newcommand{\tforja}{t_{j}^{\A}} 
\newcommand{\tforsa}{t_{s}^{\A}} 

\newcommand{\tgkabsba}{\tau_{sba}} 

\newcommand{\ssevbsimabr}{S7} 
\newcommand{\ssevbsim}{\sim_{\textsc{S7}}} 

\newcommand{\tforsba}{t_{sba}} 

\newcommand{\ststateset}{\Sigma_{st}} 
\newcommand{\imstateset}{\Sigma_{im}} 

\newcommand{\sccname}{\mathsf{St}}
\newcommand{\scconst}{\mathit{servCl}}
\newcommand{\sctmp}{\sccname(\scconst)}
\newcommand{\scstateset}{\Sigma_{sc}}

\newcommand{\cxcname}{\mathsf{St}}
\newcommand{\cxconst}{\mathit{ctxChg}}
\newcommand{\cxtmp}{\cxcname(\cxconst)}
\newcommand{\cxstateset}{\Sigma_{cx}}

\newcommand{\ftcname}{\mathsf{St}}
\newcommand{\ftconst}{\mathit{filt}}
\newcommand{\fttmp}{\ftcname(\ftconst)}
\newcommand{\ftstateset}{\Sigma_{ft}}

\newcommand{\fticname}{\mathsf{St}}
\newcommand{\fticonst}{\mathit{int}}
\newcommand{\ftitmp}{\fticname(\fticonst)}

\newcommand{\parconceptname}[1]{{Par}_{#1}}
\newcommand{\actsrcconceptname}{Actsrc}

\newcommand{\addass}[1]{\textsc{add}(#1)}
\newcommand{\delass}[1]{\textsc{del}(#1)}
\newcommand{\parass}[1]{\textsc{par}(#1)}
\newcommand{\ctxass}[1]{\textsc{ctx}(#1)}

\newcommand{\lab}{\textsc{label}}
\newcommand{\viol}{\textsc{viol}}
\newcommand{\violcon}{\textsf{Viol}}

%% file: macros/macros-sedap.tex
\newcommand{\sgds}{SEDAP\xspace}
\newcommand{\sgdss}{SEDAP\lowercase{s}\xspace}
\newcommand{\sgdssym}{\S}

\newcommand{\valset}{\V}
\newcommand{\funcsym}{\Lambda}

\newcommand{\val}{\textsl{val}}

\newcommand{\unfold}[2]{{\textsc{unfold}(#1,#2)}}

\newcommand{\sactset}{\dactset}

\newcommand{\sprocset}{\dprocset}

\newcommand{\rts}[1]{\varUpsilon^R_{#1}}

\newcommand{\sts}[1]{\varUpsilon^S_{#1}}

\newcommand{\fsset}[1]{\textsc{fs}(#1)}

\newcommand{\obdasys}{\O}
\newcommand{\obdamap}{\M}

\newcommand{\todcds}{\varsigma}

\newcommand{\obgsm}{OBGSM\xspace}

\newcommand{\ctla}{ \textnormal{CTL}\ensuremath{_A}\xspace}
\newcommand{\ctladl}{\ensuremath{\ctla^{{\textnormal{EQL}}}}\xspace}

\newcommand{\obgsmmap}{\obdamap}

\newcommand{\AG}{\textsc{AG}}
\newcommand{\EG}{\textsc{EG}}
\newcommand{\AF}{\textsc{AF}}
\newcommand{\EF}{\textsc{EF}}
\newcommand{\AX}{\textsc{AX}}
\newcommand{\EX}{\textsc{EX}}

\newcommand{\Auntil}[2]{\textsc{A}~(#1~\textsc{until}~#2)}
\newcommand{\Euntil}[2]{\textsc{E}~(#1~\textsc{until}~#2)}

\newcommand{\andTemp}[2]{#1~\textsc{and}~#2}
\newcommand{\orTemp}[2]{#1~\textsc{or}~#2}
\newcommand{\orTempB}{~\textsc{or}~}
\newcommand{\notTemp}{\textsc{!}}
\newcommand{\implTemp}[2]{#1~\textsc{->}~#2}
\newcommand{\implTempB}{~\textsc{->}~}

\newcommand{\existsTemp}{\textsc{exists}\xspace}
\newcommand{\forallTemp}{\textsc{forall}\xspace}

\newcommand{\existsQuantification}{{\texttt{existsQuantification}}\xspace}
\newcommand{\forallQuantification}{{\texttt{forallQuantification}}\xspace}

\newcommand{\formula}{{\texttt{formula}}\xspace}
\newcommand{\query}{{\texttt{query}}\xspace}

\newcommand{\variables}{{\textbf{Var}}\xspace}

\newcommand{\triples}{{\mathbf{Triples}}\xspace}
\newcommand{\prefixDeclaration}{{\mathbf{PrefixDeclarations}}\xspace}

\newcommand{\sparqlSELECT}{\textsc{Select}\xspace}

\newcommand{\sparqlWHERE}{\textsc{Where}\xspace}

\newcommand{\sparqlFILTER}{{\texttt{filter}}\xspace}
\newcommand{\sparqlFILTEREXPRESSION}{{\texttt{filterExpression}}\xspace}
\newcommand{\sparqlFILTERAND}{{\texttt{\&\&}}\xspace}

\newcommand{\sparqlLT}{{\texttt{<}}\xspace}
\newcommand{\sparqlLTE}{{\texttt{<=}}\xspace}
\newcommand{\sparqlGT}{{\texttt{>}}\xspace}
\newcommand{\sparqlGTE}{{\texttt{>=}}\xspace}
\newcommand{\sparqlEQ}{{\texttt{=}}\xspace}
\newcommand{\sparqlNEQ}{{\texttt{!=}}\xspace}

\newcommand{\sparqlVarConst}{{\texttt{var\_const}}\xspace}

\newcommand{\sparqlString}{{\texttt{string}}\xspace}
\newcommand{\sparqlInteger}{{\texttt{integer}}\xspace}

\newcommand{\doubleCarret}{\mbox{\^{}\^{}}}

\newcommand{\prefixDeclarationMapping}{{\texttt{[PrefixDeclaration]}}\xspace}
\newcommand{\classDeclarationMapping}{{\texttt{[ClassDeclaration]}}\xspace}
\newcommand{\objectPropertyDeclarationMapping}{{\texttt{[ObjectPropertyDeclaration]}}\xspace}
\newcommand{\dataPropertyDeclarationMapping}{{\texttt{[DataPropertyDeclaration]}}\xspace}
\newcommand{\mappingDeclaration}{{\texttt{[MappingDeclaration]}}\xspace}

\newcommand{\expression}{{\texttt{expression}}}

\newcommand{\notGSM}{{\texttt{!}}}
\newcommand{\andGSM}{\textsc{and}}
\newcommand{\orGSM}{\textsc{or}}

\newcommand{\getGSM}{\textbf{get}}
\newcommand{\forallGSM}{\textbf{forall}}
\newcommand{\existsGSM}{\textbf{exists}}

\newcommand{\variableGSM}{variable}
\newcommand{\constantGSM}{{\texttt{constant}}}

\newcommand{\aopGSM}{\textbf{aop}}
\newcommand{\lopGSM}{\textbf{lop}}

\newcommand{\ontop}{\textsc{-ontop-}\xspace}

\newcommand{\sparql}{SPARQL 1.1\xspace}

\newcommand{\Jena}{Apache Jena$^{\textsc{TM}}$\xspace}

%% file: macros/macros-example.tex
\newcommand{\exc}[1]{\textsc{#1}} 

\newcommand{\exv}[1]{\texttt{#1}}

\newcommand{\exo}[1]{\ensuremath{\mathsf{#1}}\xspace} 

\newcommand{\exa}[1]{\ensuremath{\mathsf{#1}}\xspace} 

\newcommand{\exvar}[1]{\ensuremath{\textit{$#1$}}\xspace}

\newcommand{\excon}[1]{\ensuremath{\text{#1}}\xspace}

\newcommand{\exs}[1]{\ensuremath{\textsc{#1}}\xspace}

\newcommand{\exr}[1]{\ensuremath{\uppercase{\textsf{#1}}}\xspace}
\newcommand{\exra}[1]{\ensuremath{\lowercase{\textsf{#1}}}\xspace}
\newcommand{\exf}[1]{\ensuremath{\lowercase{\textit{#1}}}}

%% file: 1.front-back-matter/cover.tex
 \pdfbookmark[1]{Cover}{cover}
 \thispagestyle{empty}













\newcommand\BackgroundPic{%
\put(0,0){%
\parbox[b][\paperheight]{\paperwidth}{%
\vfill
\centering
\includegraphics[width=1.01\paperwidth,height=\paperheight,%
keepaspectratio]{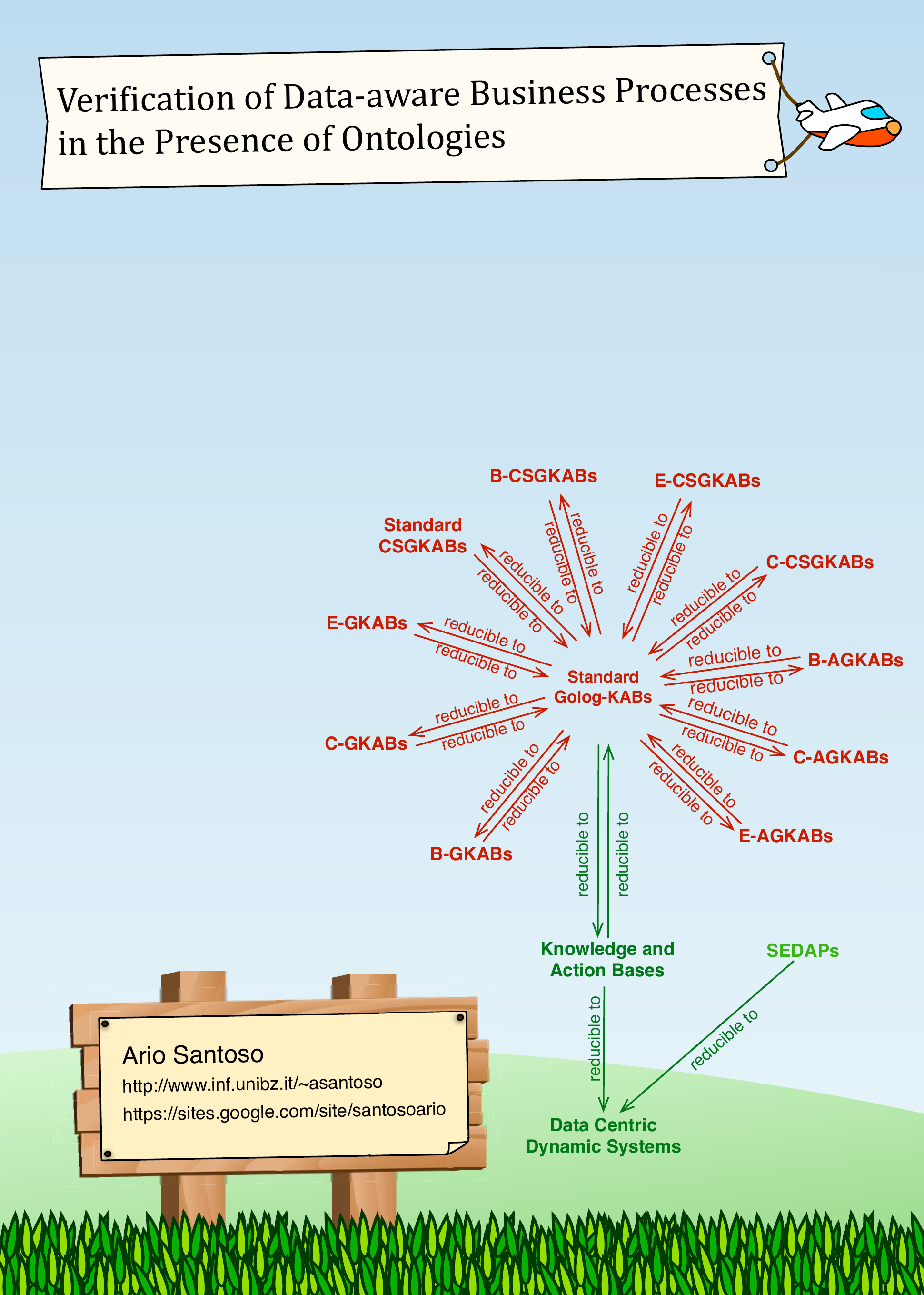}%
\vfill
}}}

\AddToShipoutPicture*{\BackgroundPic}
\ClearShipoutPicture

\ \ 
\cleardoublepage


%% file: 1.front-back-matter/dirty-title-page.tex
\pdfbookmark[1]{Title}{title}
\thispagestyle{empty}
\begin{center}

\vspace*{1cm}
\textcolor{Maroon}{\textbf{\large\textls[30]{\myTitleBU}}} 

\vspace*{7cm}

\includegraphics[width=6cm]{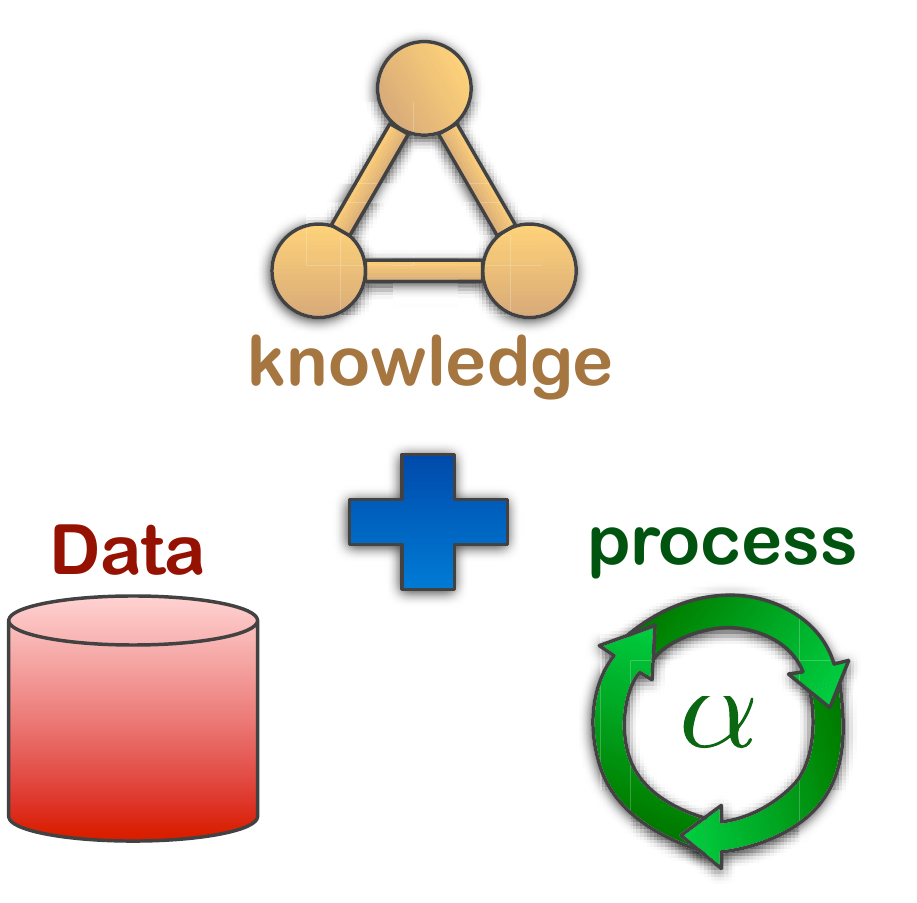} 

\vspace*{5cm}

\spacedlowsmallcaps{\myName}\\
\url{santoso@inf.unibz.it}\\
\url{santoso.ario@gmail.com}\\
\myWebsite\\
\url{https://sites.google.com/site/santosoario}

\end{center}


%% file: 1.front-back-matter/title-page.tex

\thispagestyle{empty}

    \begin{center}
        \large  

        \hfill

        \vfill

        \begingroup
            \color{Maroon}{\textbf{\large\textls[30]{\myTitleBU}}} \\
        \endgroup

        
        \bigskip

        \spacedlowsmallcaps{\myName}

        \vfill

        \vspace*{6.1cm} 
        \includegraphics[height=1.5cm]{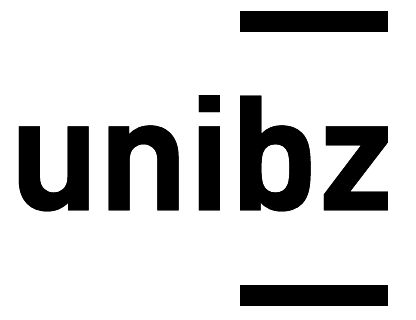} \hspace*{2.1cm}
        \includegraphics[height=1.3cm]{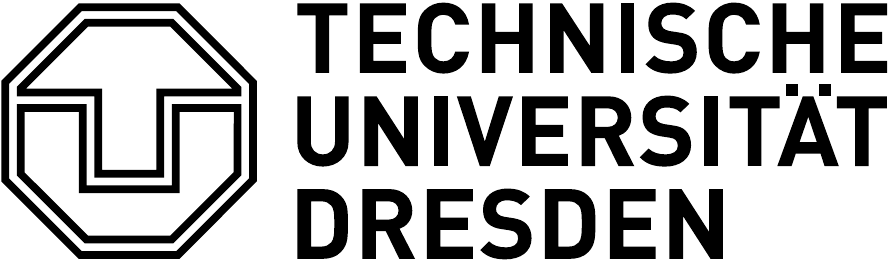}
        \vspace*{3.9cm}

        
        \vfill
        European PhD Program in Computational Logic (EPCL)\\ \ \\
        
        \spacedlowsmallcaps{First institution:}\\
        Facolt{\`a} di Scienze e Tecnologie Informatiche \\
        Libera Universit{\`a} di Bolzano\\ Italy \\ \ \\
        
        \spacedlowsmallcaps{Second institution:}\\
        Fakult{\"a}t Informatik \\
        Technische Universit{\"a}t Dresden\\ Germany \\

        






        \medskip

        \bigskip
        \bigskip
        \bigskip
 
        \bigskip

        \myTime

        \vfill                      

    \end{center}  


%% file: 1.front-back-matter/title-back.tex
\thispagestyle{empty}

\hfill

\vfill

\noindent\myName \\\textit{\myTitle} \\PhD Dissertation, 
\textcopyright\ \myTime

\bigskip

\noindent\spacedlowsmallcaps{Supervisor}: \\
\mySupervisor \ -
\mySupervisorWebsite\\
\mySupervisorInstitution\ (Bolzano - Italy) \\

\noindent\spacedlowsmallcaps{Co-Supervisor}: \\
\myCoSupervisor \ - \myCoSupervisorWebsite\\
\myCoSupervisorInstitution\ (Bolzano - Italy)\\ 

\noindent\spacedlowsmallcaps{External Supervisor}: \\
\myExternalSupervisor \ - \myExternalSupervisorWebsite\\
\myExternalSupervisorInstitution\ (Dresden - Germany) \\

\noindent\spacedlowsmallcaps{Reviewers}: \\
Yves Lesp{\'e}rance \ - \url{http://www.cse.yorku.ca/~lesperan}\\
York University (Toronto - Canada) \vspace*{1mm}\\
Sebastian Sardi{\~n}a \ - \url{http://www1.rmit.edu.au/staff/sebastian-sardina}\\
Royal Melbourne Institute of Technology - RMIT University (Melbourne - Australia) \\

\noindent\spacedlowsmallcaps{PhD Defense Committee}: \\
Sven Helmer \ - \url{http://www.inf.unibz.it/~shelmer}\\
Libera Universit{\`a} di Bolzano (Bolzano - Italy) \vspace*{1mm}\\
Steffen H{\"o}lldobler \ - \url{http://www.computational-logic.org/~sh}\\
Technische Universit\"at Dresden (Dresden - Germany) \vspace*{1mm}\\
Gerhard Lakemeyer \ - \url{http://www-i5.informatik.rwth-aachen.de/~gerhard}\\
RWTH Aachen University (Aachen - Germany) \\


\medskip





%% file: 1.front-back-matter/dedication.tex
\thispagestyle{empty}
\refstepcounter{dummy}





\hfill
\vfill
\begin{flushright}
To You,\vspace*{-1mm}\\
who illuminates my paths.\vspace*{2mm}\\

To all of my teachers, lecturers, and professors,\vspace*{-1mm}\\
who have taught and guided me until I can reach this far.\vspace*{2mm}\\

To everyone who has influenced my journey  up to this moment,\vspace*{-1mm}\\
good or bad, it shapes me to what I am now.\vspace*{2mm}\\

To my living caffeine,\vspace*{-1mm}\\
who always be there and supports me.

\end{flushright}


%% file: 1.front-back-matter/abstract.tex
\pdfbookmark[1]{Abstract}{Abstract}

\chapter*{Abstract}
{\small
The meet up between data, processes and structural knowledge in modeling complex enterprise systems is a challenging task that has led to the study of combining formalisms from knowledge representation, database theory, and process management. Moreover, to ensure system correctness, formal verification also comes into play as a promising approach that offers well-established techniques. In line with this, significant results have been obtained within the research on data-aware business processes, which studies the marriage between static and dynamic aspects of a system within a unified framework. However, several limitations are still present. Various formalisms for data-aware processes that have been studied typically use a simple mechanism  for specifying the system dynamics. The majority of works also assume a rather simple treatment of inconsistency (i.e., reject inconsistent system states). Many researches in this area that consider structural domain knowledge typically also assume that such knowledge remains fixed along the system evolution (context-independent), and this might be too restrictive. Moreover, the information model of data-aware processes sometimes relies on relatively simple structures. This situation might cause an abstraction gap between the high-level conceptual view that business stakeholders have, and the low-level representation of information. When it comes to verification, taking into account all of the aspects above makes the problem more challenging.
}

{\small
In this thesis, we investigate the verification of data-aware processes in the presence of ontologies while at the same time addressing all limitations above. Specifically, we provide the following contributions: (1) We propose a formal framework called Golog-KABs (GKABs), by leveraging on the state of the art formalisms for data-aware processes equipped with ontologies. GKABs enable us to specify semantically-rich data-aware business processes, where the system dynamics are specified using a high-level action language inspired by the Golog programming language. (2) We propose a parametric execution semantics for GKABs that is able to elegantly accommodate a plethora of inconsistency-aware semantics based on the well-known notion of repair, and this leads us to consider several variants of inconsistency-aware GKABs. (3) We enhance GKABs towards context-sensitive GKABs that take into account the contextual information during the system evolution. (4) We marry these two settings and introduce inconsistency-aware context-sensitive GKABs. (5) We introduce the so-called Alternating-GKABs that allow for a more fine-grained analysis over the evolution of inconsistency-aware context-sensitive systems. (6) In addition to GKABs, we introduce a novel framework called Semantically-Enhanced Data-Aware Processes (SEDAPs) that, by utilizing ontologies, enable us to have a high-level conceptual view over the evolution of the underlying system. We provide not only theoretical results, but have also implemented this concept of SEDAPs.
}

{\small
We also provide numerous reductions for the verification of sophisticated first-order temporal properties over all of the settings above, and show that verification can be addressed using existing techniques developed for Data-Centric Dynamic Systems (which is a well-established data-aware processes framework), under suitable boundedness assumptions for the number of objects freshly introduced in the system while it evolves. Notably, all proposed GKAB extensions have no negative impact on computational complexity.
}




%% file: 1.front-back-matter/acknowledgments.tex
\pdfbookmark[1]{Acknowledgments}{acknowledgments}


\bigskip

\begingroup
\let\clearpage\relax
\let\cleardoublepage\relax
\let\cleardoublepage\relax
\chapter*{Acknowledgments}




\noindent
I would like to express my gratitude to Prof. Diego Calvanese, who has
been my great supervisor even since I did my master thesis. Also thanks
to Marco Montali who has become my great co-supervisor. Thanks to both
of them for their unbounded help, guidance and kindness. Their
greatness might even be inexpressible in words.

Sincere thanks to the Free University of Bozen-Bolzano (FUB) and the
KRDB Research Centre for Knowledge and Data, which have provided me
with plenty of support that I can not mention one by one. Thanks to
Viviana Foscarin, the student secretariat personnel in the Faculty of
Computer Science at FUB, for all of the support related to any
administrative issues. Also thanks to Technische Universit\"at Dresden
(TUD) and European PhD Program in Computational Logic (EPCL) for all
of the support related to my study within the EPCL program as well as
during my research visit in TUD.

I would also like to express my gratitude to Prof. Franz Baader who
has been my supervisor within the EPCL Program and also for hosting me
in his group at Lehrstuhl f\"ur Automatentheorie of TUD.  I would also
like to thank all of the colleagues in Lehrstuhl f\"ur
Automatentheorie of TUD for 
the hospitality during my stay. 
Especially, thanks to Anni Yasmin Turhan for all the fascinating
discussions 
and help.  
Thanks to Stephan B\"ohme for all the interesting discussions
regarding the R\"osi project. Thanks to {\.I}smail {\.I}lkan Ceylan
who has introduced me into the works on context and also thanks for
our fruitful discussions and collaborations that have led to nice
outcomes related to the work on context. Thanks to Benjamin
Zarrie{\ss} for all the nice discussions, especially related to Golog,
which has become an important part in this thesis. Thanks to
Mrs. Achtruth who gave me enormous helps on administrative
matters during my stay in TUD. Last but not least, thanks to all of my
colleagues in TUD who gave me great experiences during my stay
there. 

I would also like to express my gratitude to Yves Lesp{\'e}rance and
Sebastian Sardi{\~n}a who have became great reviewers for this
thesis. Thank you so much for the time that you have spend in reading
this thesis and of course for all of your fruitful comments, feedback
and suggestions that really improve the thesis.

I would also like to thank Prof. Steffen H\"olldobler and Christoph
Wernhard who have coordinated the EPCL program. Also thanks to Julia
Kopenhagen and Sylvia W{\"u}nsch for all of the support related to
administrative aspects within EPCL.

Also, I would like to express my gratitude to Babak Bagheri Hariri,
Giuseppe De Giacomo, Evgeny Kharlamov, Domenico Lembo, Dmitry
Solomakhin, and Dmitriy Zheleznyakov for all of the collaborations
that have led to some of the results in this thesis.

I would also like to say thanks to all of my colleagues in KRDB whom I
cannot mention one by one. I would also like to express my thanks to
all of the people who gave me wonderful times during my stay in
Dresden, especially the people in Formid e.V., and also all of my
friends there whom I can not elaborate one by one.
%
%
Also thanks to all of my friends in Bolzano and Trento for giving me
a great time during my stay in Bolzano.

Also deep thanks to Eliza Margaretha and Annisa Ihsani for some
writing enhancements and writing discussions concerning this
thesis. Additionally, thanks to Arya and Alida Widianti for some
discussions regarding business scenarios for the running examples on
this thesis.

Last but not least, special thanks to my living caffeine for several
writing enhancements, and most prominently for her endless support
that always strengthen me in various difficult moments. In the end, I
would like to say thanks to everyone who has supported me in this
journey, though their names might not be in this page, they are always
in my mind and especially in my heart. 

\bigskip
\bigskip
\bigskip
\begin{flushright}
Bolzano - Italy, \myTime \\
Ario Santoso
\end{flushright}

\endgroup

%% file: 1.front-back-matter/contents.tex
\refstepcounter{dummy}
\pdfbookmark[1]{\contentsname}{tableofcontents}
\setcounter{tocdepth}{1} 
\setcounter{secnumdepth}{3} 
\manualmark
\markboth{\spacedlowsmallcaps{\contentsname}}{\spacedlowsmallcaps{\contentsname}}
\tableofcontents 
\automark[section]{chapter}
\renewcommand{\chaptermark}[1]{\markboth{\spacedlowsmallcaps{#1}}{\spacedlowsmallcaps{#1}}}
\renewcommand{\sectionmark}[1]{\markright{\thesection\enspace\spacedlowsmallcaps{#1}}}
\clearpage


%% file: 2.chapters/1-intro-part1.tex
\chapter{Introduction}\label{ch:intro}

\ifhideintro
 
\fi

Data 
%
%
and processes are golden ingredients 
for any information system.
%
%
%
%
%
As usual, data are simply facts that might be used for a specific
purpose, while a (business) process
is 
a sequence of actions/activities that are performed in order to
achieve a certain (business) goal, 
and that might also manipulate data during its 
execution. Within an information system,
data 
are also considered as the elements that characterize the
\emph{static} aspect of the system, while
processes 
characterize the \emph{dynamic} aspect of the system.
Due to the importance of data, they are even often considered as the
driver of an organization. In fact, typically many prominent and
critical (business-related) decisions within an organization are made
based on the data. On the other hand, processes are also vital for any
competitive business. They differentiate between good and outstanding
business performance. Hence, it is inevitable that data and processes
are notable aspects within information systems that influence the
performance of organizations.
%
%
%
%






Although data and processes are fundamentally two different entities,
they are tightly connected.  However, traditional system modeling
approaches model data and processes separately.
%
%
When it comes to process modeling, people often abstract away the
data, and when modeling the data, people often think about the
processes only afterwards~\cite{Ric10,Reic12,Dum12}. This situation
might be unsatisfactory.
As witnessed by~\cite{Reic12,Dum12,Ric10,MSW11,Hull08,CoHu09}, there
is evidence
%
%
of the need to treat 
both data and processes as first class
citizens when building a 
system.
They may even be considered as ``\emph{two sides of the same
  coin}''~\cite{Reic12}.
Thus, focusing on data and processes separately while designing the
system might be insufficient.
In fact, considering both data and processes together while designing
the system could promote us into a better unified holistic view of the
system. Furthermore, it could help us in avoiding various problems of
the traditional system modeling approaches that consider these two
aspects independently (e.g., the system is inadequately covering some
process scenarios~\cite{Reic12}).

Along with the need of focusing on both data and processes
simultaneously,
the \emph{artifact-centric business process}
paradigm~\cite{NiCa03,Hull08,CoHu09} emerges as a promising approach
that combines both static and dynamic aspects while designing a
system. It provides a rich and robust model for devising business
processes in which data and processes are first class citizens. This
initiative was initially pioneered at IBM
research\footnote{International Business Machine (IBM) Corp.\ -
  \url{https://www.ibm.com/}}~\cite{NiCa03} .  Since then, extensive
studies have been accomplished in this area and numerous fruitful
outcomes have been achieved
(e.g.,~\cite{BGHLS07,ABGM09b,BeLP12,BeLP12b,CDDR10,GeBS07,GeSu07}). Moreover,
the artifact-centric paradigm has been successfully applied in various
settings (cf.~\cite{BCKNW07,BGLH*05,CCFHLNNPVW09}). This line of
research is often also called \emph{data-aware (business) processes}.
%
%
%



Orthogonal to processes and data, \emph{ontologies} allow us to have a
formal conceptualization of the structural/intensional knowledge about
the 
domain of interest.
In particular, what do we mean by knowledge is the universal
statements about data. Such statements describe the structure of the
domain as well as 
enable us to infer/derive some implicit information from the explicit
one. 
%
%
Typically, ontologies are formalized in logic-based
languages (e.g., First Order Logic (FOL), or Description Logic
(DL)). As an example, consider a \emph{customer order processing}
scenario within a company. In FOL-based ontologies, we can encode
domain knowledge saying that ``\emph{each assembled order is an
  order}'' as a first order sentence/axiom as follows:
$\forall x. \mbox{\exo{AssembledOrder}}(x) \ra \mbox{\exo{Order}}(x)$.
Besides enabling us to conveniently structure the domain knowledge, a
crucial advantage of ontologies is that they allow us to reason about
the domain. For instance, in our example, whenever we know that
something is an assembled order, we can infer that it is also an
order. Since fundamentally ontology captures the structural knowledge
of the domain of interest, we often also consider it as the structural
knowledge component of a system.







Looking at ontologies and the artifact-centric approach,
there are some researches on data-aware processes formalisms that take
into account ontologies (e.g.,~\cite{BCDD*12,BCMD*13,BCDD11}). Besides
allowing us to focus simultaneously on data and
processes, 
the proposed framework enables us to incorporate the domain knowledge
inside the designed system and leads us to a semantically-rich
system.
%
%

When it comes to the need of ensuring the correctness of the developed
system, 
%
%
there are various techniques that are usually applied such as
(software/system) testing, peer review, simulation and formal
verification. The choice of the method is typically based on the
complexity of the system as well as the required degree of
safety. Each of those techniques has its own advantages and
disadvantages. For instance, testing might be easier to do than formal
verification, but is in general less reliable. 
As stated by the famous computer scientist E.\ Dijkstra,
\emph{``Testing can only show the presence of errors, but not their
  absence''}.
%
%
%
In fact, as reported in the survey of artifact-centric business
processes models~\cite{Hull08}, formal verification for
artifact-centric systems is an important research direction aimed at
establishing
sophisticated techniques to analyze the correctness 
of data-aware business processes.
Model checking~\cite{BaKa08} is a 
widely studied and successful formal verification technique, see,
e.g., \cite{CW96} for notable success stories.
%
%
However, the interactions between data and processes typically makes
the problem more difficult since it makes the system in general
become infinite states. Thereby typical model checking techniques for
finite state systems are inapplicable.

In this thesis, motivated by various works on data, processes and
ontologies, 
we focus on the formal verification 
of several variants of
data-aware business processes that are enriched with
ontologies.
It is noteworthy to remark that this line of research
opens up various fascinating connections among diverse research areas
such as Databases, Formal Verification, Model Checking, Business
Process Management, Knowledge Representation, and specifically
Description Logics, and Reasoning About Actions.

The rest of the chapter is organized as follows: In
\Cref{sec:metup-dpk} we briefly overview some studies related to data,
processes and knowledge as well as their blending.  We then continue
by elaborating our research challenges in \Cref{sec:res-challenge},
and exhibit the core results within this thesis in
\Cref{sec:contribution}. We conclude this chapter by listing the
publications of the results from this thesis in
\Cref{sec:publications} as well as providing a concise outlook to the
thesis structure in \Cref{sec:thesis-structure}.

\section{The Meet Up Between Data, Processes and Knowledge}\label{sec:metup-dpk}

Over the years, there has been plenty of effort in providing means to
model the structure of data. This brought us a plethora of data
modeling languages 
such as UML~\cite{UML05,FoSc97}, ER diagrams~\cite{ElNa07}, and
ORM~\cite{TH10,TH15}. Several tools also have been developed in order
to ease data modeling (e.g., Rational Rose, Enterprise
Architect). Moreover, various researches have been conducted within
this area such as
\begin{inparaenum}[\it (i)]
\item performing a comparative analysis among different modeling
  languages~\cite{HaBl99},
\item studying the correspondence between a certain logic and a
  particular data modeling language~\cite{FMS12},
\item establishing an automated reasoning technique to reason about a
  specific modeling language~\cite{BeCD05}, 
  etc.
\end{inparaenum}
%
%
%

On the other hand, numerous works are concerned with the problem of
establishing mechanisms to specify (business) processes. 
%
%
Various approaches for modeling processes have been studied/proposed
such as Petri Nets, BPMN, Workflow Pattern, YAWL, and BPEL,
(cf.~\cite{BPMN,YAWL,WSBPEL}).  Some studies on critically comparing
or surveying various approaches for business processes modeling
can be found in~\cite{B12,GKP98,LS07,RRIG09}.

Ontologies have become a substantial research direction within the
area of knowledge representation, which is traditionally concerned
with the problem of representing possibly complex knowledge about a
domain of interest. Various approaches have been proposed, such as 
semantic networks~\cite{Wood91,Lehm92,Sowa91}, frame based
systems~\cite{FiKe85}, and logic based approaches~\cite{Bad99,BCMNP07}.
In computer science, knowledge representation focuses on how to
represent knowledge such that it is effectively and efficiently
machine processable. This leads to the important task of reasoning
over the known facts in order to infer unknown/implicit facts
from the existing knowledge.
%
Various languages for expressing ontologies have been proposed,
notably Description Logics (DLs)~\cite{Bad99}, and 
Datalog~\cite{CaLP96,CGLMP10}. The researches on ontologies
typically deal with the trade off between expressivity and the
computational complexity of inference. Trivially, more expressive
languages make reasoning more difficult and vice versa.

In the remainder of this section, we briefly overview various researches
that take into account the combination among data, processes or
knowledge.

\subsection{The Marriage Between Processes and Knowledge}

A combination between processes and knowledge has been carried out in
the context of semantic web services~\cite{MSZ01,Swa02}. The idea of
semantic web services is to provide a semantic markup on web services
to make them understandable and processable by machines. These
semantic markups give more information about the services in a machine
processable format. One of the advantages from this proposal is
enabling automated web service discovery, execution, composition and
interoperation. In this context, the web services are the processes
and the semantic markups are the knowledge, and the area can be
considered as a proposal to get benefits from their combination.  
One interesting line of research related to semantic web service is
that of composition (cf.~\cite{LL06,MBE03}), which is concerned with
the problem of how to compose available services in order to realize a
requested, but still unavailable, service.

Another line of research that marries processes and knowledge is that
of Semantic Business Processes (SBP), whose basic idea is to adopt 
semantic technologies for Business Process Management
(BPM)~\cite{Wesk07,aalst13}. The motivation comes from the need and
the lack of machines accessible semantics in the current business
process
representation~\cite{HLDWF05}. 
Adopting semantic technologies might help the automation of many tasks
related to BPM~\cite{WMFKBLLC07}. This leads to a research area called
Semantic Business Process Management (SBPM) (see~\cite{HLDWF05,
  WMFKBLLC07}).
In \cite{HLDWF05}, the authors argue that the lack of machine-readable
semantics in current business process representation is the major
obstacle in the automation of business process management and they
point out that the semantic web and semantic web service technology
might provide the necessary tools. Hence, they propose to combine the
techniques in SWS and BPM. Continuing the vision of SBPM in
\cite{HLDWF05}, the work in \cite{WMFKBLLC07} describes how ontologies
and semantic web service technologies can be used in the BPM lifecycle
(process modeling, implementation, execution, and analysis). It also
identifies functional requirements of SBPM as well as their benefits,
which are mostly about the support of process automation. Some other
work on SBPM can be found in \cite{DPVDSRNC07} which is about mining
and monitoring of the process, which 
are important parts of analysis in the BPM
lifecycle. 
We mention also work about measuring the similarity between SBP
\cite{EKO07} and a proposal on a framework for compliance management
of SBP \cite{KSMP08}.

\subsection{The Marriage Between Data and Knowledge}

An extensive study on the marriage between data and knowledge is
attested by the research on Ontology Based Data Access
(OBDA)~\cite{LuTW09,HMAM*08,CDLL*09,PLCD*08,RoCa11b,RoLC08,
  CDLL*07b,PoRR08,SKZJ*14,SeAM14,BiRo15,CGHH*13,CGHH*13b,
  CLRSX14,HHHH*13,LRXC15,RoCa08,RoCa11}. The idea is to provide a
conceptual view over (existing) data repositories through ontologies
that abstract away from how such data are maintained.
Technically speaking, such approach adds an ontology over the data repositories,
which captures the domain of interest, and then we can query the data
repositories through the ontology. Within this setting, we obtain:
\begin{compactitem}


\item More sophisticated access to data repositories.

\item Sophisticated query answering ability, which enables us to deal
  with incomplete information and to infer some facts that are not
  stated explicitly in the data. 

\item An ability to impose constraints on the data over the
  conceptual level.

\item A high level abstraction that hides the low level details on how
  the data are stored.

\item A unified view on multiple data sources through the ontology.


\end{compactitem}
%
%
%
However, these advantages do not come for free, and
various efforts have been made to overcome all challenges such as
\begin{compactitem}
\item finding the right formalism for the ontology, 
\item dealing with performance,
\item tackling the impedance mismatch problem. I.e, the problem that
  arises because of the mismatch between what is stored in the data
  repositories (i.e., values) and in the ontologies (i.e., abstract
  objects). Such situation demands a mechanism to establish the links
  between the values in the data repositories and the objects in the
  ontologies. 
\end{compactitem}
Not only theoretical results have been achieved, but also intensive
efforts have been put in realizing these concepts into implemented
systems
(cf.~\cite{RoLC08,SeAM14,RoCa12,CCKEKLRRX15,CDLLPRRRS11}).

\subsection{The Marriage Between Data and Processes}

The marriage between data and processes has been exhibited by various
studies in the area of \emph{data-aware business processes}.
%
%
A survey on this area can be found in~\cite{CDM13}.  
Moreover, some results as well as research directions and challenges
specific on the artifact-centric approach can be found
in~\cite{Hull08,CoHu09}.
%
%
%

In artifact-centric approach, the key business-relevant entities are
modeled as \emph{(business) artifacts}, and an artifact itself is
constituted by an \emph{information model} and a \emph{lifecycle}. The
former captures the artifact's relevant data
while the latter characterizes the evolution of an artifact (i.e.,
specifies the permitted ways to progress the information model). One
can also say that the lifecycle of an artifact captures the possible
``business-relevant stages'' as well as their possible changes (from
one stage to another) during the evolution of an artifact. During
their evolution, the data within an artifact are also manipulated.  An
artifact-centric system is then constituted by a set of artifacts that
might interact with each other and evolve over time.
As a simple example, consider an artifact \textsc{CustomerOrder} that
represents a business entity that captures the order of a
customer. The information model of \textsc{CustomerOrder} might
contain the data about the corresponding order (e.g., a list of ordered
products). On the other hand, the lifecycle of \textsc{CustomerOrder}
characterizes how a \textsc{CustomerOrder} might evolve from one stage
to another one, 
such as from the stage of an order being received to the stage of an
order being processed.

Many variants of artifact-centric systems have been studied.  Some of
the early works on this paradigm are presented
in~\cite{GeSu07,GeBS07}. In those works, the authors investigate
artifact-based systems where the information model is constituted by a
tuple of typed-attributes. 
To characterize the lifecycle of artifacts, the
framework 
uses 
finite state machines. These works mainly focus on investigating 
static analysis techniques for the artifact-centric
framework. 
%
Still among the early studies on the artifact-centric paradigm, the
work in~\cite{DHPV09} investigates artifact systems that are equipped
with a static relational database (i.e., it stays fixed during the
system evolution). The evolution of the system is then characterized
by the services that manipulate updatable data inside the existing
artifacts.  This work investigates the decidability boundaries for the
verification of First Order LTL formulas over this setting.
%
%

The works in~\cite{ABGM09b,AbSV08,AbSV09b} consider artifact-centric
systems that are constituted by a set of (interacting) XML-based
documents called Active XML (AXML) documents~\cite{ABM04,ABM08}.
%
%
An AXML document is an XML-based document that may contain embedded
function, and that evolves over time based on the result of such
function calls. The function calls are differentiated into internal
and external function calls. The former do local computations while
the latter interact with users or other services. 
%
The interesting task in AXML-based artifact-centric systems is to
analyze their behavior, which 
is characterized by the evolution of the documents. In particular,
they intend to verify temporal properties over the runs of the
system. The temporal properties 
are specified by the temporal logic Tree-LTL, in which the atomic
properties are tree-like patterns that can be checked over the state
of the system and the temporal parts are 
as in the usual Linear Temporal Logic (LTL)~\cite{BaKa08}.

The research in~\cite{BCDDF11} studies artifact-centric systems where
data are modeled by relational databases~\cite{ElNa07}. The authors
of~\cite{BCDDF11} primarily focus on investigating the problem of
verifying temporal properties over the evolution of the system
that 
are expressed in a first-order variant of
$\mu$-calculus~\cite{BrSt07}.

Still within the spirit of the artifact-centric paradigm,
the works in~\cite{HDDF*11,DaHV13} propose the so-called
Guard-Stage-Milestone (GSM)
%
as a framework for modeling/specifying artifact-centric systems.
%
%
GSM is equipped with a formal execution semantics~\cite{DaHV13}, which
unambiguously characterizes the artifact progression in response to
external events. Notably, several key constructs of the emerging OMG
standard on Case Management and Model
Notation\footnote{\url{http://www.omg.org/spec/CMMN/}} have been
borrowed from GSM.
Another interesting research direction 
is the research on the artifact-centric model in the context of
multi-agent systems (cf.~\cite{BeLP12,BeLP14,GGL14}), leading to the
so-called 
Artifact-Centric Multi-Agent Systems (AC-MAS) framework.
In this setting, each of the agents has some internal data stored
within them.
The system evolution is then characterized based on the actions that
are performed by agents, which involve both agent interactions and
data manipulation.
%
%
The main reasoning task that was tackled is temporal properties
verification over the system. In addition to 
theoretical results, a model checker for AC-MAS also has been
developed (cf.~\cite{GGL14}). 
Notably, the work in~\cite{GGL14} extends GSM towards the setting of
multi-agent system.



Other prominent results on the marriage between data and processes
are provided by~\cite{BCDDM13,BCDDM12} who propose a 
formal framework called Data-Centric Dynamic Systems
(DCDSs). DCDSs basically capture the evolution of a system that is
characterized by the manipulation of a relational database by
processes (actions). The 
processes in DCDSs are declaratively specified in terms of
condition-action rules that tell when and how an action can be
executed. A fascinating result on the decidability
of 
verification of first-order variants of $\mu$-calculus properties over
DCDSs has been obtained.
%
%
Some attempts in implementing DCDS have also
been carried out (cf.~\cite{RMPM13,CMPR15,CMPR15a}).

%% file: 2.chapters/1-intro-part2.tex
\subsection{When Data, Processes and Knowledge Meet Up}

The meet up among data, processes and knowledge (specifically,
ontologies) can be observed in the work on semantic
artifacts~\cite{BCDD11} and also in those works that combine Knowledge
Bases (KBs) and
actions~\cite{BCMD*13,MoCD14,CDMP13,BLMSW05,BZ13,ZC15}.


In~\cite{BCDD11}, the authors propose artifact-centric systems formed
by semantic artifacts, which utilize Description Logic (DL) KBs as the
mechanism to keep artifact relevant information in a semantically-rich
form.
As usual, a DL KB is constituted by an ABox that stores the data and a
TBox that captures the domain knowledge.
%
%
The progression mechanism of a semantic artifact system is provided in
a declarative manner using condition-action rules similar
to~\cite{BCDDF11}. Furthermore,~\cite{BCDD11} studies the problem of
model checking over the semantic artifact systems temporal properties
that are specified using a first order variant of $\mu$-calculus
\cite{BrSt07}. Although in general the problem is
undecidable,~\cite{BCDD11} has identified a syntactic restriction that
guarantees decidability of verification based on the notion of
weak-acyclicity in data exchange~\cite{FKMP03}.

The works~\cite{BLMSW05,BLMSW05b} introduce a DL-based action
formalism.  The semantics of DL-based actions is specified in terms of
manipulation of DL interpretations (which are first-order
interpretations of the unary and binary predicates corresponding to
concepts and roles, respectively).  Concerning
reasoning,~\cite{BLMSW05,BLMSW05b} study the projection problem, i.e.,
the problem of checking whether the execution result of a certain
sequence of actions satisfies a given DL formula (assertion).
%
%
Building on~\cite{BLMSW05,BLMSW05b}, the works~\cite{BZ13,ZC15} study
Golog programs~\cite{LRLLS97} in which the atomic actions are
formalized as DL-based actions.
%
%
Within this setting,~\cite{BZ13,ZC15} tackle the problem of verifying
temporal properties over the execution of Golog program with respect
to the given DL KB.
%
%


In~\cite{BCDD*12,BCMD*13}, 
the authors propose a formal framework, named Knowledge and Action
Bases (KABs), which allows one to capture the manipulation of a DL
Knowledge Base 
over time. The dynamic aspect of KABs is characterized by
condition-action rules that, together with the data manipulated during
the system evolution, determine the possible sequences of actions that
can be executed over the KB.
In contrast to~\cite{BLMSW05,BLMSW05b,BZ13,ZC15}, where actions
directly manipulate DL-interpretations, the work
in~\cite{BCDD*12,BCMD*13,MoCD14} adopts the Levesque functional
approach~\cite{Leve84} in defining the execution semantics of
actions. 
Under this approach, the KB provides two operations, \textsc{ask} to
extract relevant information, and \textsc{tell} to assert new
knowledge, and actions rely on such operations for their execution.
Specifically, in KABs, the \textsc{ask} operation corresponds to the
computation of \emph{certain answers} to queries over the KB, and the
\textsc{tell} operation corresponds to asserting the ABox facts that
should hold in the resulting state.  During action execution, calls to
external services might be issued in order to acquire \emph{new
  values} from outside of the system.  As a consequence, the number of
possible states is not bounded a priori.
Roughly speaking, the calls to external services can be used to model the
interaction with external systems/entities as well as user input that
might inject new values to the system. 
%
The execution semantics of a KAB is provided by a
possibly infinite state transition systems in which each transition
represents an action execution and each state contains a KB.
Regarding reasoning,~\cite{BCDD*12,BCMD*13} study the verification of
temporal properties over the evolution of KABs, where the temporal
properties are specified by a first order variant of
$\mu$-calculus. Although in general verification is undecidable,
decidability has been obtained under suitable restriction based on the
notion of \emph{weak acyclicity} that is borrowed from the work on
data exchange \cite{FKMP03}.

\section{Research Challenges}\label{sec:res-challenge}

Many data-aware processes formalisms that have been studied so far
consider a simple formalism for specifying the progression
mechanism. For instance,~\cite{BCDDM13,BCDDM12,BCDDF11,BCDD11} only
consider condition-action rules to specify when and how an atomic
action can be executed. Although this approach is quite expressive,
one might desire a better control in specifying the desired order of
actions (e.g., to choose one action or another based on the result of
a condition checked over the current state, or to specify that a
certain sequence of actions should be executed as long as a specified
condition holds). Thus, a more sophisticated formalism is required to
specify the system dynamics at a higher-level of
abstraction. \Cref{ex:rc1} illustrates this issue.

\begin{example}\label{ex:rc1}
  Consider an \emph{order processing scenario} in a furniture provider
  enterprise.  Consider the actions/operations $\exa{approveOrder}$
  and $\exa{prepareOrder}$, where an execution of $\exa{approveOrder}$
  approves a single \emph{received order} and an execution of
  $\exa{prepareOrder}$ prepares a singe \emph{approved order}. Suppose
  that we want to strictly enforce that each single execution of
  $\exa{approveOrder}$ must be followed by $\exa{prepareOrder}$. In
  the typical condition-action rules formalism, we can only specify
  the pre-condition of each action/operation. Therefore, in such
  formalism we have to specify that
  \begin{inparaenum}[]
  \item at the end of the execution of $\exa{approveOrder}$, it should
    make the pre-condition of $\exa{prepareOrder}$ become satisfied
    and, 
  \item the pre-condition of $\exa{approveOrder}$ should not be
    satisfied. 
  \end{inparaenum}
  Considering that there could also be other actions/operations, then
  we also need to make sure that the pre-conditions of all other
  actions are not satisfied.  Thus, looking at this situation, it
  might be desirable if we can directly specify such \emph{expected
    sequence of actions/operations}, i.e., by directly saying
  \[
  \exa{approveOrder}; \exa{prepareOrder}
  \]
  which semantically means that every execution of
  $\exa{approveOrder}$ must be followed by $\exa{prepareOrder}$.

  As another natural example, consider the situation where we need to
  specify that a process should be repeated as long as a particular
  condition holds. For instance,
  after delivering a processed order, the whole process should go back
  to the beginning and repeat the whole process until there is no more
  unprocessed order. It might be desirable if we could specify such
  situation in a high level manner by a kind of ``while loop''
  construction. E.g.,
  \[
  \begin{array}{l@{}l}
    &\gwhile{\mbox{"there exists an unprocessed order"} }{ \{ \\
%
    &\hspace*{7mm}\exa{approveOrder};\ \exa{prepareOrders};\ \ldots;\ \exa{deliverOrders}\\
    &\}}
  \end{array}
  \]
\end{example}

Concerning inconsistency management, the majority of approaches
dealing with verification in data-aware processes assume a rather
simple treatment. In particular, they simply reject inconsistent
system states that are produced by the effects of action executions
(see, e.g.,~\cite{DHPV09,BCDDM13,BCMD*13,BCDD11}). In general, this
mechanism is not satisfactory, since the inconsistency may affect just
a small portion of the entire data, and thus should be treated in a
more careful way. This is in line with what is done in numerous
researches that specifically deal with inconsistencies
(cf.~\cite{LeRu07,Bien12,Bert06,CKMZ12}). \Cref{ex:rc2} illustrates
this issue.

\begin{example}\label{ex:rc2}
  Continuing \Cref{ex:rc1}, consider that there is a process of
  designing and assembling an order. Suppose that to enforce
  segregation of duty, we have a \emph{domain knowledge (constraint)}
  in our system saying that ``\textit{a product designer is not a
    product assembler, and vice versa}''. Suppose that we have the
  fact that ``\textit{john is a product designer}'' (i.e.,
  $\exo{Designer}(\excon{john})$). Now, suppose that after an execution of an
  action $\act$, we have a new fact that ``\textit{john is a product
    assembler}'' (i.e., $\exo{Assembler}(\excon{john})$). Hence, we have that
  the constraint above is violated (i.e., at this state the system
  encounters an inconsistency). In this situation, instead of
  disallowing the execution of $\act$ that leads into this
  inconsistent state (i.e., rejecting this inconsistent state), it
  might be desirable to repair the inconsistent state, for instance by
  throwing the older fact (i.e, $\exo{Designer}(\excon{john})$). In fact, it
  could be the case that John has been recently change his role into
  designer, but the system might not yet been updated.
\end{example}

Many works on data-aware processes that incorporate structural
domain knowledge typically assume that such knowledge remains fixed
along the system evolution (e.g.,~\cite{CDMP13,MoCD14,BCMD*13}), i.e.,
that it is independent from the actual system state.  However, this
assumption might be 
too restrictive, since specific knowledge might hold or be applicable
only in specific, \emph{context-dependent} circumstances.  Ideally,
one should be able to form statements that are known to be true in
certain cases, but not necessarily in all. \Cref{ex:rc3} illustrates
this issue.

\begin{example}\label{ex:rc3}
  Continuing \Cref{ex:rc2}, suppose that we want to have such a
  constraint only hold during the \textit{normal season}, but during
  the \textit{peak season}, to enforce efficiency, we want to have
  that ``\textit{each product designer is a product assembler}''.
  Hence, in this situation, it might be desirable to contextualized
  our domain knowledge such that 
  \begin{compactenum}
  \item in the context of \textit{normal season} a designer must not
    be a product assembler (and vice versa),
  \item in the context of \textit{peak season} each designer is also a
    product assembler
  \end{compactenum}
\ \ 
\end{example}

As witnessed by numerous works on data-aware processes (see
e.g.,~\cite{DHPV09,BeLP12,BCDDM13,BCMD*13,MoCD14}), the verification
problem in this setting is in general difficult (more precisely,
undecidable without suitable restrictions) since the number of systems
states is in general infinite. Thus, off the shelf model checking
technique for finite state system cannot be used directly. The
situation becomes even more challenging when we also need to deal with
inconsistencies and/or take into account the presence of contextual
information.

In some formalisms of data-aware processes, the information model
typically relies on relatively simple structures, such as tuples of
typed-attributes (e.g.\ \cite{GeSu07,GeBS07,HDDF*11,DaHV13}).
This situation might cause an abstraction gap between the high-level
conceptual view that business stakeholders have, 
 and the low-level representation of information.
In addition, the data layer within the system might be complicated and
difficult to interact with. In this light, there is a need to have a
high level conceptual view over the system evolution.

In this thesis, we aim at addressing all the issues mentioned above,
by proposing novel extensions of existing models for data-aware
business processes, and by studying how these extensions affect the
problem of formal verification of expressive temporal properties. In
the remaining part of the chapter, we discuss in detail the original
contributions that we have provided along these lines.



\section{Contributions}\label{sec:contribution}

As a first broad contribution of this thesis, we introduce and study
several variants and extensions of the formalism of KABs.
Specifically, the extensions we introduce are the following:
\begin{enumerate}

\item A formal framework, namely Golog-KABs (GKABs), for specifying
  \emph{semantically-rich data-aware business processes} that is
  obtained by leveraging on the current state of the art data-aware
  processes equipped with ontologies.

\item Several variants of \emph{inconsistency-aware Golog-KAB}, which
  extend GKABs by incorporating various inconsistency handling
  mechanisms that had been proposed in the literatures.

\item An extended version of GKABs, namely \emph{Context-Sensitive Golog-KABs}
  (\csgkabs), which takes into account contextual
  information during the evolution.

\item Several variants of \emph{inconsistency-aware context-sensitive
    Golog-KAB}, which are obtained from \csgkabs by incorporating
  various inconsistency management mechanisms.


\item An extension of GKABs, called \emph{Alternating GKABs}, that
  separates the sources of non-determinism within a single step of
  evolution and allows for a more fine-grained analysis on the system
  evolution, while also employing sophisticated inconsistency handling
  mechanism and taking into account contextual information.




\item A novel framework, called \emph{Semantically-Enhanced Data-Aware
    Processes} (\sgdss), which enables us to have a high-level
  conceptual view over the evolution of a data-aware business processes
  by utilizing ontologies.
%

\end{enumerate}

%
We observe that this thesis establishes two different approaches in
devising a semantically-rich data-aware business processes.
One, based on GKABs and their variants, in which we have a KB that
evolves under the effect of actions, requires us to specify the system
from scratch. The other one, namely \sgdss, enables us to enhance
existing data-aware processes systems towards a semantically-rich
system by connecting an ontology via mappings to a traditional
relational data layer that evolves under action execution.

Within all of the settings above, we tackle the problem of
verification of temporal properties over the system
executions.
This task is more challenging than in the basic setting of KABs, on
which we build, since we need to deal with inconsistency in a more
sophisticated manner and consider the contextual information.
%
%
%
In the following sub-sections, we provide more details on each of
these contributions.

\subsection{Golog-KABs (GKABs)}\label{sec:cont-gkab}


Here we devise a formal framework for specifying
\emph{semantically-rich data-aware business processes systems} by
leveraging on the current state of the art data-aware processes system
equipped with
ontologies~\cite{BCMD*13,BCDD*12,MoCD14,CDMP13}. Specifically, we
build on the Knowledge and Action Bases (KABs) framework that was
initially proposed in~\cite{BCMD*13,BCDD*12}. Fundamentally, KABs
provide a semantically rich representation of a domain in the form of
a KB expressed in the lightweight DL \dllite~\cite{CDLLR07}, while
also simultaneously taking into account the dynamic aspects of the
modeled system. As usual, the \dllite KB is constituted by a
\emph{TBox} that captures the intensional knowledge about the domain
and an \emph{ABox} that keeps the data (extensional parts).
The execution semantics of a KAB is given in terms of a (possibly
infinite) transition system, in which each state is labeled by a DL KB
and each transition represents the manipulation of the ABox by an
action.
%
Concerning action specification, rather than following the original
KABs~\cite{BCMD*13,BCDD*12}, in which at each action execution the
state is reconstructed from scratch, we adopt the action formalism
in~\cite{MoCD14}, in which one specifies only the facts to add and
those to delete from the current state. 
Similar to KABs, an action execution might issue external service
calls that might inject fresh values (constants) into the
system. Roughly speaking, the calls to external services can be used
to model the interaction with external systems/entities as well as
user input.
As for the execution semantics w.r.t.\ service calls, instead of
following~\cite{BCMD*13,BCDD*12}, we use the service call evaluation
semantics as in~\cite{BCDDM13,BCDDM12}, which is considered to be less
abstract, more natural, and closer to reality.  
I.e., we evaluate the service calls in the sense that we substitute
each service call with a concrete value when constructing the
transition system.
Since we use KABs that are slightly different from their original
version in~\cite{BCMD*13,BCDD*12}, in \Cref{ch:kab} we show that the
verification of $\muladom$ properties over KABs can be reduced to the
corresponding verification of $\mula$ over DCDSs~\cite{BCDDM13}, where
$\muladom$ and $\mula$ are variants of first order
$\mu$-calculus~\cite{BrSt07} (one of the most powerful temporal
logics, which subsumes LTL, PSL, and CTL*~\cite{ClGP99}).
The different between $\mula$ and $\muladom$ formulas is in the atomic
parts of the formulas. The former consider Domain Independent First
Order Logics queries \cite{AbHV95} as the atomic components of the
formulas while the latter consider Domain Independent EQL-Lite (UCQ)
\cite{CDLLR07b} queries.
The reduction also preserves \emph{run-boundedness}, which is a
restriction that guarantees the decidability of DCDSs
verification. Thus, exploiting the results on verification of
run-bounded DCDSs, it follows that the verification of run-bounded
KABs is decidable and can be reduced to standard finite state model
checking.

In this thesis, we enrich KABs with a high-level, compact action
language inspired by a well-known action programming language in the
area of Artificial Intelligence (AI), namely Golog~\cite{LRLLS97}. 
We call the resulting formalism \emph{Golog-KABs} (GKABs). Thus,
instead of using simple condition-action rules as in KABs, the
progression mechanism in GKABs is specified using Golog
programs. 
This allows modelers to conveniently specify the processes at a
high-level of abstraction and represent the dynamic aspects of the
systems much more compactly (cf. \Cref{ex:con1}).
Roughly speaking, the Golog program characterizes the evolution of a
GKAB by determining the possible orders of action executions that
evolve the KB over time.

\begin{example}\label{ex:con1}
  Recall our \Cref{ex:rc1}, using Golog constructs, we can specify
  such a sequence of actions as well as the repetition of operations in
  a convenient way. Using the Golog construct ``;'' we can specify the
  sequence of actions $\exa{approveOrder}$ and $\exa{prepareOrder}$,
  i.e.,
  \[
  \exa{approveOrder} ; \exa{prepareOrder}
  \]
  Furthermore, we can also specify the repetition of operations using
  the construct 
  \[
  \gwhile{\varphi}{\delta}
  \]
  which we can use to express that $\delta$ will be executed as long
  as $\varphi$ satisfied. See \Cref{ch:gkab} for more details.
\end{example}


To elegantly accommodate various ways of updating the ABox, we
introduce a parametric execution semantics of GKABs. 
%
%
Technically, we adopt Levesque's functional approach, i.e., we assume
that a GKAB provides two operations:
\begin{compactitem}
\item $\textsc{ask}$, to answer queries over the current KB;
\item $\textsc{tell}$, to update the KB (ABox) through an atomic action.
\end{compactitem}
In this work, the $\textsc{ask}$ operator corresponds to the
\emph{certain answers} computation.  The $\textsc{tell}$ operation is
parameterized by \emph{filter relations}, which are used to refine the
way in which an ABox is updated, based on a set of facts to be added
and deleted (that are specified by the action).

In this light, filter relations provide an abstract mechanism to
accommodate in the execution semantics several inconsistency
management approaches based on the well-known notion of
repair~\cite{LLRRS10,LLRRS11,CKNZ10b}.
%
%
Basically, we can obtain various execution semantics for GKABs,
including \emph{inconsistency-aware semantics}, by simply defining
different kinds of filter relation.  For instance, we define
\emph{GKABs with standard execution semantics}, briefly
\emph{S-GKABs}, by defining a filter relation $\filter_S$ that updates
an ABox based on the facts to be added and deleted, and does nothing
w.r.t.\ inconsistency (i.e., updates that lead to an inconsistent
state are simply rejected).

Concerning the verification of \muladom properties over S-GKABs, we
have shown that we can reduce this problem to verification of KABs and
vice versa.
%
%
%
%
%
To encode KABs into S-GKABs, we simulate the standard execution
semantics using a Golog program that runs forever to
non-deterministically pick an executable action with parameters, or
stops if no action is executable.  For the opposite direction, the key
idea is to inductively interpret a Golog program as a structure
consisting of nested processes, suitably composed through the Golog
operators.  We mark the starting and ending point of each Golog
subprogram, and use accessory facts in the ABox to track states
corresponding to subprograms.  Each subprogram is then inductively
translated into a set of actions and condition-action rules, encoding
its entrance and termination conditions.

\subsection{Inconsistency-Aware GKABs}

%
%
%

\begin{figure}[tbp]
  \centering
  \includegraphics[width=1\textwidth]{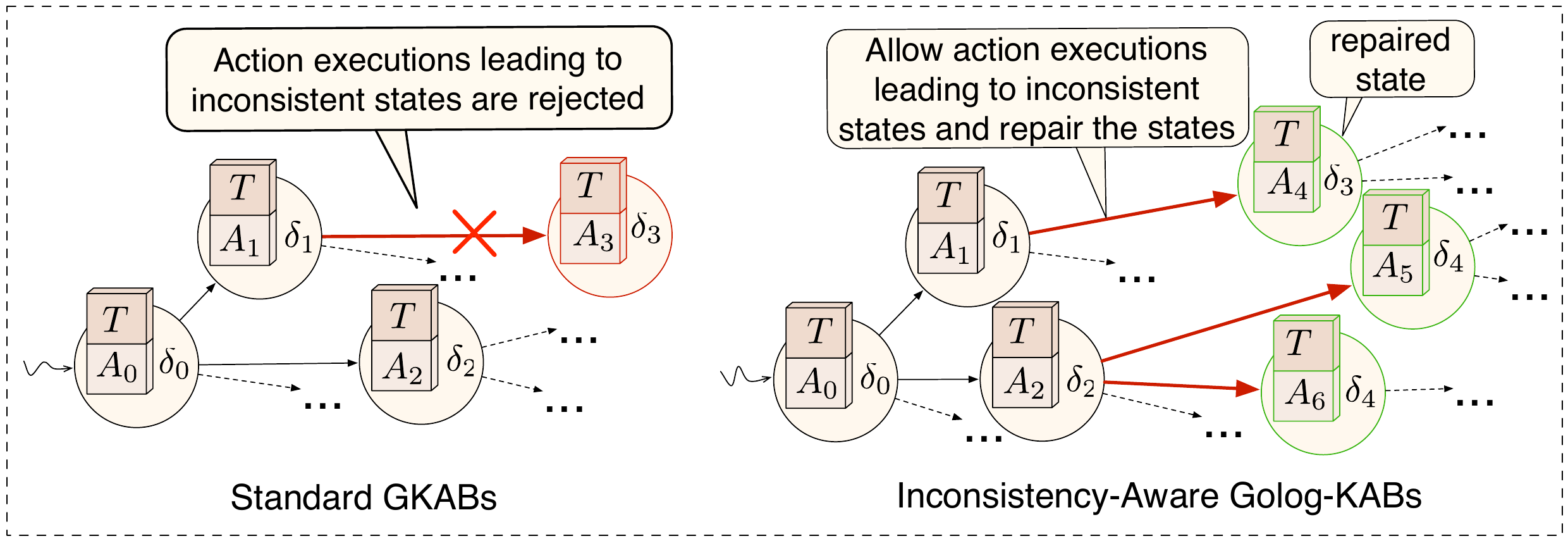}
  \caption{An illustration of Inconsistency-Aware GKABs execution
    compare to Standard GKABs execution (Note: $T$ is a TBox, $A_i$ is
    an ABox, and $\delta_i$ is the remaining program to be executed).}
  \label{fig:igkab-illustration}
\end{figure}

We introduce GKABs with inconsistency-aware semantics by exploiting
the filter relations (i.e., we introduce various kind of filter
relations and plug them in into GKABs). By incorporating
inconsistency-aware semantics, we allow each action that leads to an
inconsistent state and then we repair the
inconsistency. \Cref{fig:igkab-illustration} gives an illustration of
this setting. 
Technically, we introduce filter relations B-filter $\filter_B$,
C-filter $\filter_C$, and B-evol filter $\filter_E$, where
\begin{compactitem}

\item $\filter_B$ incorporates the ABox Repair (AR) semantics
  in~\cite{LLRRS10}. Here we call such approach \emph{bold-repair}
  (\emph{b-repair}), where a b-repair of an ABox $A$ w.r.t.\ TBox $T$
  is a maximal (w.r.t.\ set containment) subset of $A$ that is
  consistent with $T$.


\item $\filter_C$ incorporates the Intersection ABox Repair (IAR)
  semantics in~\cite{LLRRS10}. Here we call such approach
  \emph{certain-repair} (\emph{c-repair}), where a c-repair of an ABox
  $A$ is an ABox that is obtained by intersecting all b-repairs of $A$
  w.r.t.\ $T$.


\item $\filter_E$ updates the ABox using the bold semantics of KB
  evolution~\cite{CKNZ10b}. In this approach, if an inconsistency
  arises due to an update, newly introduced assertions are preferred
  to those already present in the current ABox.
\end{compactitem}
We call the GKABs adopting the execution semantics obtained by
employing those filter relations \emph{B-GKABs}, \emph{C-GKABs}, and
\emph{E-GKABs}, respectively.  We group them under the umbrella of
\emph{inconsistency-aware GKABs} (\emph{I-GKABs}). \Cref{ex:con2}
provide a high level illustration of this setting.

\begin{example}\label{ex:con2}
  Recall \Cref{ex:rc2} where we have an inconsistent state containing
  the facts $\exo{Designer}(\excon{john})$ and
  $\exo{Assembler}(\excon{john})$. Suppose that these two facts are
  the only facts that cause inconsistency, then

  \begin{itemize}

  \item In B-GKABs (i.e., GKABs that employ AR semantics for updating
    the ABox), we repair that state and produce two repair states. One
    repair state containing the fact $\exo{Designer}(\excon{john})$
    while the other one containing the fact
    $\exo{Assembler}(\excon{john})$. I.e., it explores all possible
    repairs.

  \item In C-GKABs (i.e., GKABs that employ IAR semantics for updating
    the ABox), we repair that state and produce a repair state
    containing neither $\exo{Designer}(\excon{john})$ nor
    $\exo{Assembler}(\excon{john})$. I.e., it only keeps the facts
    that do not involve in inconsistency.

  \item In E-GKABs (i.e., GKABs that employ bold semantics of KB
    evolution for updating the ABox), we repair that state and produce
    a repair state containing $\exo{Assembler}(\excon{john})$ (i.e.,
    throw away $\exo{Designer}(\excon{john})$). It reflects the
    situation where new facts are considered to be more correct.

  \end{itemize}
\end{example}

\begin{figure}[hp]
  \centering
  \includegraphics[width=0.5\textwidth]{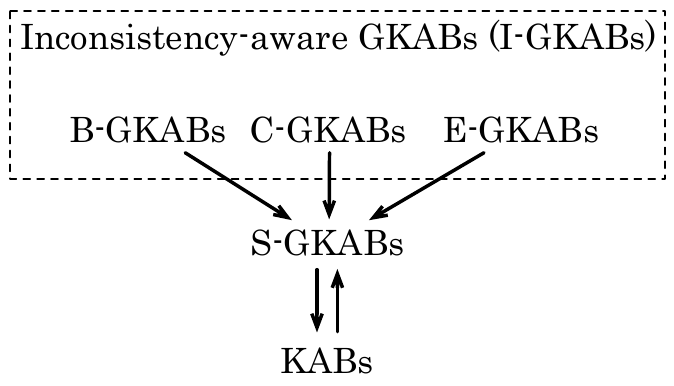}
  \caption{Reductions from I-GKABs (i.e., B-GKABs, C-GKABs, and E-GKABs) to
    KABs}
  \label{fig:igkab-to-kab}
\end{figure}

With respect to verification of \muladom properties over the various types of
GKABs introduced so far, we have proved the results summarized in
Figure~\ref{fig:igkab-to-kab}, where an arrow indicates that we can reduce
verification in (G)KABs in the source to verification in (G)KABs in the
target.
%
%
%
%
%
%
Furthermore, the semantic property of \emph{run-boundedness} (which
guarantees the decidability of KAB verification)~\cite{BCDDM13} is
preserved by all our reductions.
Thus, it follows that verification of \muladom properties over
run-bounded S-GKABs and I-GKABs is decidable, and reducible to
standard $\mu$-calculus finite-state model checking.
For all reductions from I-GKABs to S-GKABs, our general strategy is to
show that S-GKABs are sufficiently expressive to incorporate the
repair-based approaches, so that an action executed under certain
inconsistency-aware semantics can be compiled into a Golog program that
applies the action with the standard semantics, and then explicitly
handles the inconsistency, if needed.
%
%




\subsection{Context-Sensitive GKABs}

\begin{figure}[bp]
  \centering
  \includegraphics[width=1\textwidth]{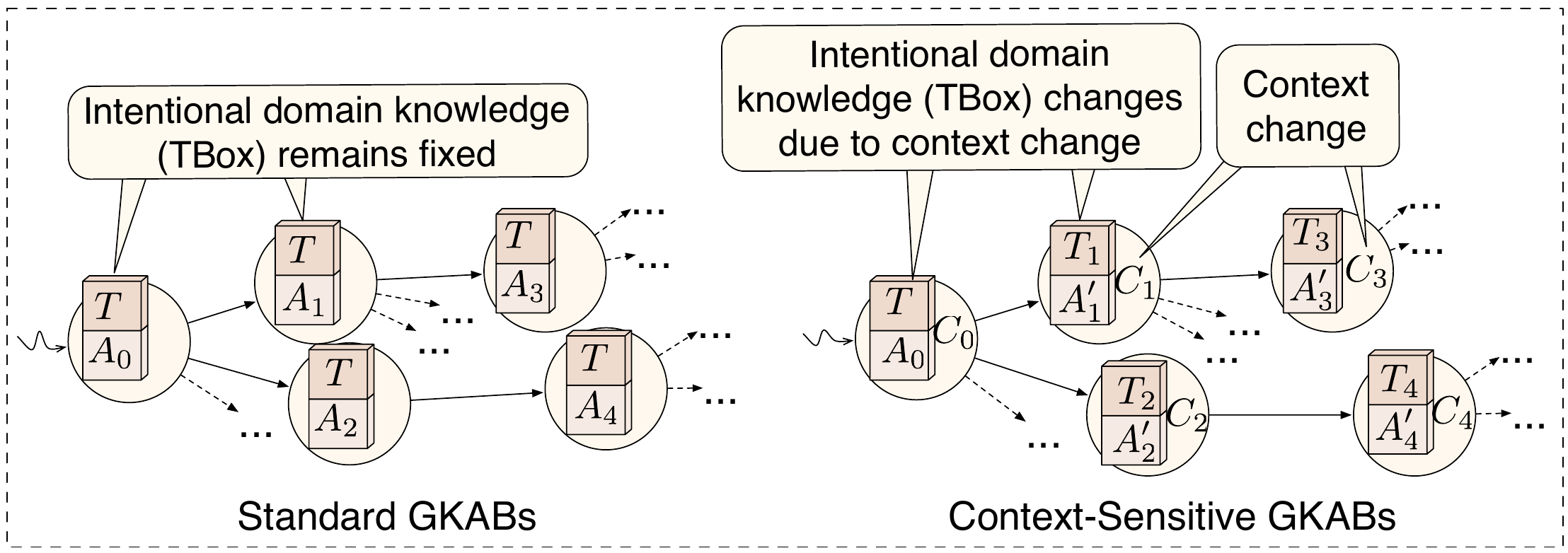}
  \caption{An illustration of Context-Sensitive GKABs execution
    compare to Standard GKABs execution (Note: $T$ is a TBox, $A_i$
    is an ABox, and $C_i$ is a context).}
  \label{fig:csgkab-illustration}
\end{figure}

As the next contributions, we extend GKABs towards Context-Sensitive
GKABs (\csgkabs), which allow us to incorporate contextual information
within the system. The context might change during the system
evolution and influences the system execution in several ways such as:
\begin{compactitem}
\item determining relevant TBox assertions at each state (i.e., TBox
  changes along the system execution depending on the context), and
\item influencing the decision about action executability.
\end{compactitem}
As a consequence of the TBox changes, essentially context also
indirectly affects the results of query answering over the
KB. \Cref{ex:con3} provide a high level intuition of \csgkabs, and
\Cref{fig:csgkab-illustration} illustrates the execution of \csgkabs.

\begin{example}\label{ex:con3}
  Recall \Cref{ex:rc3}, in \csgkabs, we can capture the situation of
  \emph{normal season} and \emph{peak season} as
  ``context''. Moreover, we can encode those context-dependent
  knowledge (i.e., ``\textit{each product designer is a product assembler}''
  and ``\textit{a product designer is not a product assembler (and vice
  versa)}'') such that they only hold on the corresponding desired
  context.  Please see \Cref{ch:cs-gkab} for more details.
\end{example}

Concerning execution semantics, it is worth mentioning that we lift
GKABs into \csgkabs by also retaining their parametric execution
semantics. Therefore, we can easily define various ways of updating
the ABox in \csgkabs by simply ``shaping'' the filter relation, which
is a great basis for integrating various inconsistency management
mechanisms into \csgkabs. 

Regarding verification, to specify the properties to be verified, we
consider a context-sensitive temporal logic $\mulcs$, which extends
$\muladom$ with the possibility of having also ``context expressions''
as an atomic part of the formula. It follows that, using $\mulcs$ we
can also say something about contextual information inside the
properties that we want to verify. In \Cref{ch:cs-gkab}, we study the
verification of \emph{\csgkabs with standard execution semantics},
briefly \emph{\scsgkabs,} that are obtained by using the standard
filter relation.
%
To cope with the problem of verifying \mulcs over \scsgkabs, we reduce
the problem to the corresponding \muladom verification problem over
S-GKABs.




\subsection{Inconsistency-Aware Context-Sensitive GKABs}

We also study the combination of \csgkabs and various inconsistency
management mechanisms (as in I-GKABs), which led us to the
formalization of Inconsistency-aware Context-sensitive GKABs. 
In particular, similar to the way of obtaining I-GKABs, we employ
three filter relations that incorporate the b-repair, c-repair, and
bold-evolution computations. We call \csgkabs adopting the execution
semantics obtained by injecting those filter relations \bicgkabs,
\cicgkabs, and \eicgkabs, respectively.  
We group them under the umbrella of Inconsistency-aware
Context-sensitive GKABs (\icgkabs).

For the verification of \mulcs over \icgkabs, we show that the
verification of \mulcs over \bicgkabs, \cicgkabs, and \eicgkabs can be
reduced to the corresponding verification of \muladom over
S-GKABs. Furthermore, all our reductions also preserve
run-boundedness. It follows that the verification of run-bounded
\scsgkabs, \bicgkabs, \cicgkabs, and \eicgkabs are decidable and
reducible to the standard finite state model checking.



\subsection{Alternating GKABs}

\begin{figure}[tbp]
  \centering
  \includegraphics[width=0.9\textwidth]{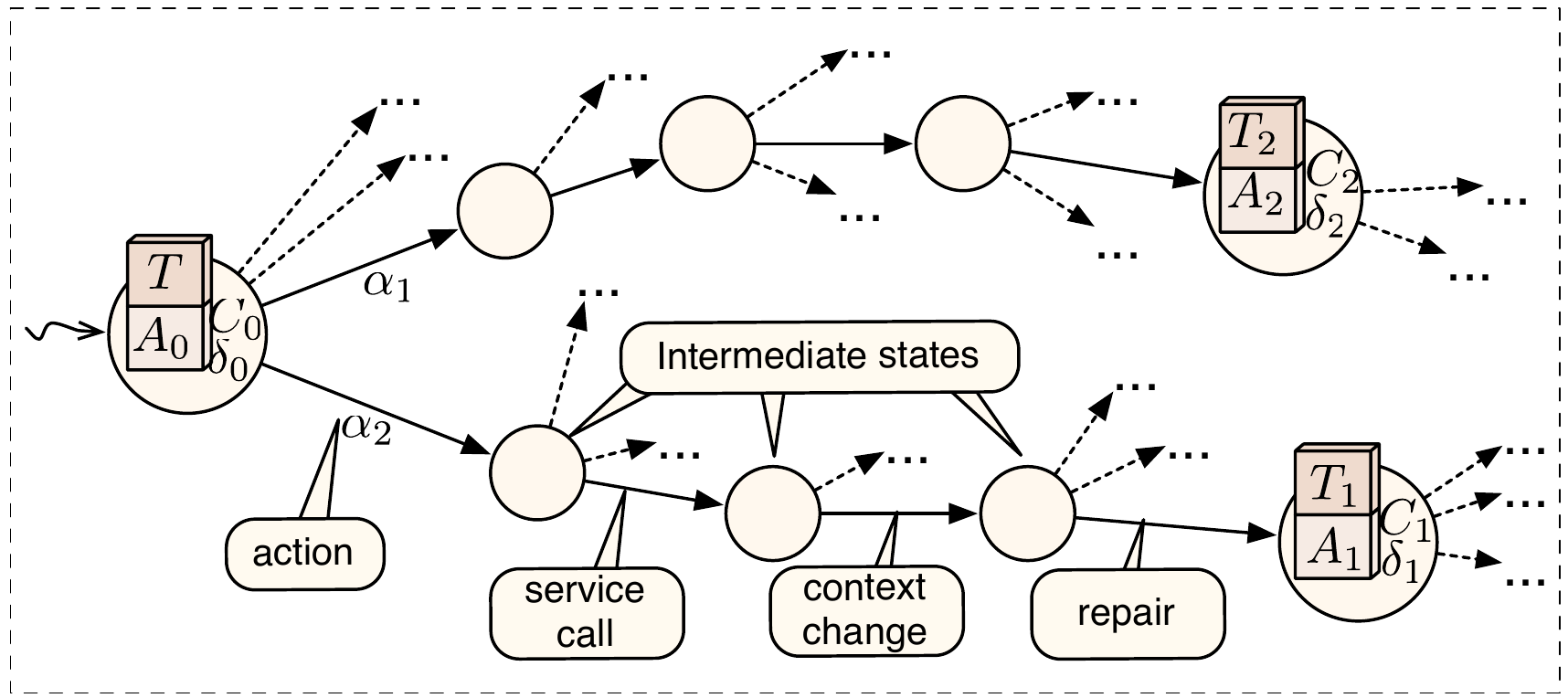}
  \caption{Illustration of Alternating GKABs execution (Note:
    $T$ is a TBox, $A_i$ is an ABox, $C_i$ is a context, and
    $\delta_i$ is the remaining program to be executed).}
  \label{fig:agkab-illustration}
\end{figure}

As a deeper study on GKABs, we introduce \agkabs, which separate
sources of non-determinism during the computation of successor
states. Those sources of non-determinism are:
\begin{compactenum}
\item the choice of grounded actions,
\item the choice of service call results,
\item the choice among all possible new contexts, and
\item the choice of repaired ABoxes when there are several possible
  repairs (which is the case for b-repairs).
\end{compactenum}
In \icgkabs, we encapsulate the computation of all of those sources of
non-determinism in a single transition (i.e., roughly speaking, in a
single transition, non-determinism can be caused by those four
sources). In \agkabs, we separate them such that each state only has
one possible source of non-determinism (one of those four sources).
\Cref{fig:agkab-illustration} gives an illustration of \agkabs
execution, and \Cref{ex:con4} provides an example that gives an
intuition on those sources of non-determinism.

\begin{example}\label{ex:con4}
  Recall our \Cref{ex:rc1,ex:rc2,ex:rc3,ex:con1,ex:con2,ex:con3}.
\begin{compactitem}

\item Concerning non-determinism from the choice of grounded actions,
  now let the action $\exa{approveOrder}$ has a parameter that is the
  order to be approved (i.e., we have $\exa{approveOrder}(\exvar{x})$
  where $\exvar{x}$ is the order to be approved). Suppose that there
  are three received orders (let say $\excon{chair}$, $\excon{table}$,
  and $\excon{cupboard}$), then we can execute the action
  $\exa{approveOrder}/1$ with several possible arguments (i.e., either
  $\exa{approveOrder}(\excon{chair})$,
  $\exa{approveOrder}(\excon{table})$, or
  $\exa{approveOrder}(\excon{cupboard})$). Hence, from that state, we
  have several choice of grounded actions that we can execute, and it
  causes non-determinism in the transition system.

\item About non-determinism from the choice of service call
  results, as it has been mentioned above, within an action execution
  we might issue a service call. 
  Considering the semantics of service calls, since we consider all
  possible return values of a service call, then it is easy to see
  that there are many possibilities of service call results and it
  causes non-determinism in the transition system.

\item Regarding the context change, our model also allow
  non-deterministic changes.

\item As for repairs, we could see in \Cref{ex:con2} that we could
  have several possible repairs when we adopt AR semantics.

\end{compactitem}
\end{example}

Thanks to the separation of the sources of non-determinism, we are
capable to do a more fine-grained analysis over the system
evolution. In particular, we can verify temporal properties that
quantify over each source of non-determinism. For instance, we can
check a property like ``\textit{no matter which action is executed,
  there exists a service call result in which no matter how the
  context is changing, there exists a repair that leads us into a
  certain state that satisfy a certain property}''.

Concerning verification, we introduce \mulcsa, which is a fragment of
\mulcs where we always use the 
modal operators in groups of 4 (e.g.,
$\DIAM{\BOX{\BOX{\DIAM{\Phi}}}}$) in order to quantify separately over
each source of non-determinism. Similar to \icgkabs, we employ three
filter relations that incorporate the b-repair, c-repair, and
bold-evolution computations, obtaining respectively \bagkabs,
\cagkabs, and \eagkabs.

To tackle the problem of \mulcsa verification over \bagkabs, \cagkabs,
and \eagkabs, we prove again that those problems are reducible to the
verification of \muladom over S-GKABs. Also in this case, our
reductions preserve run-boundedness, allowing us again to reduce
verification to standard finite state model checking.



\subsection{Semantically-Enhanced Data-Aware Processes}
\label{sec:cont-sedap}

As a further contribution, we devise a novel framework that enables us
to enhance the existing data-aware business processes system into
a semantically-rich data-aware processes system. In particular we
propose \emph{Semantically-Enhanced Data-Aware Processes} (\sgdss)
which are inspired by the research on Ontology-Based Data Access
(OBDA) \cite{CDLL*09}, where an ontology is used to provide a
conceptual view over (existing) data repositories, to which the
ontology is connected by means of mappings.
Roughly speaking, \sgdss can be considered as an extension of DCDSs
\cite{BCDDM13} where the data layer is constituted by an OBDA system
instead of simply a relational database.
%
%
%
Through the presence of the ontology, a \sgds provides
\begin{inparablank}
\item a unified, high-level conceptual view of the system, reflecting
  the relevant concepts and relations of the domain of interest and
  abstracting away how processes and data are concretely realized and
  stored at the implementation level.
\item This, in turn, is the basis for different important reasoning
  tasks such as verification of conceptual temporal properties,
\item regulating how new processes can be injected into the system,
\item synthesizing new processes starting from high level conceptual
  requirements, and
\item reasoning under implicit and incomplete information.
\end{inparablank}

\begin{figure}[tbp]
\centering
\includegraphics[width=1.0\textwidth]{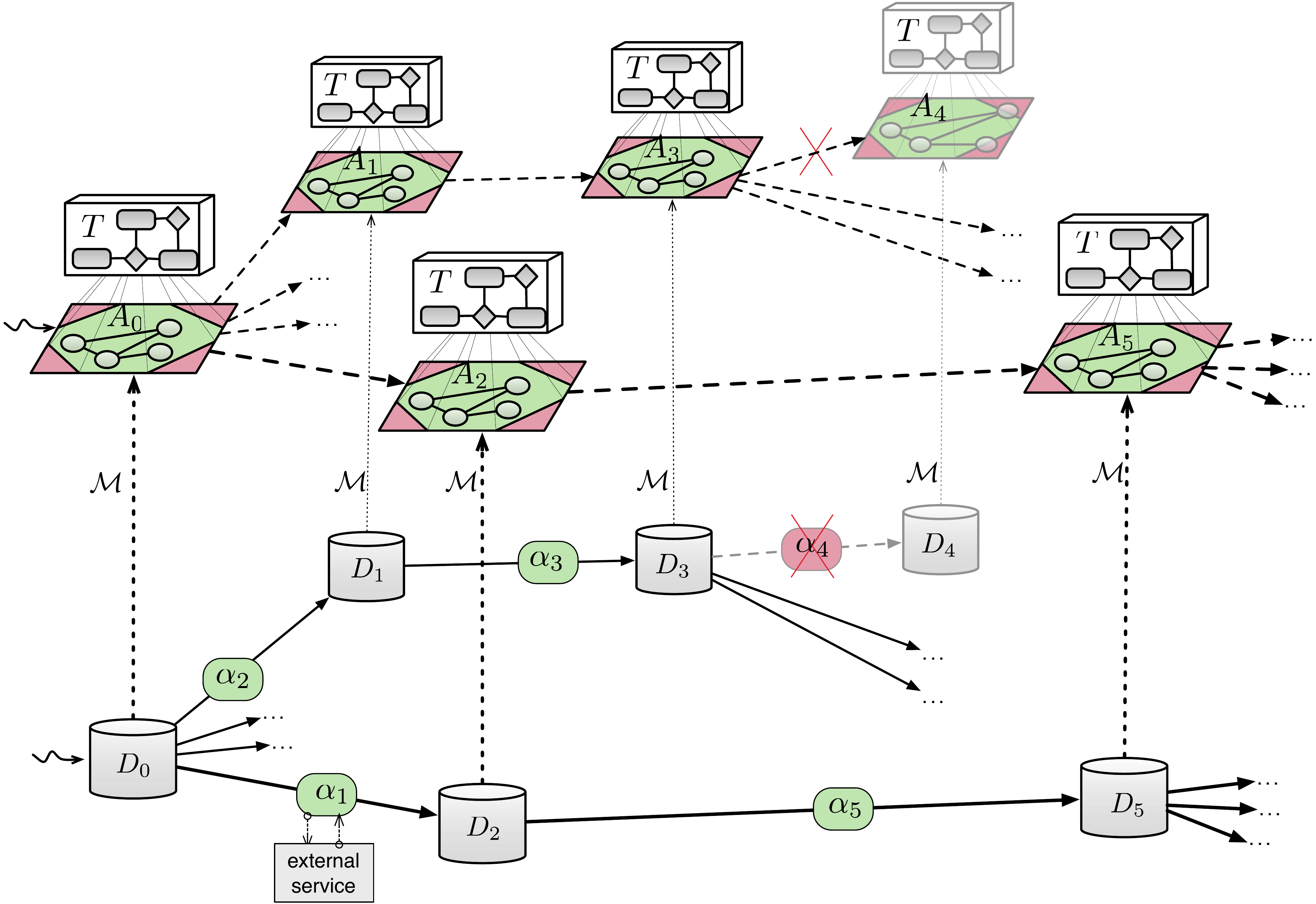}
\caption{Intuition of \sgdss setting \label{sedap-intuition}}
\end{figure}

Basically a \sgds is constituted by three components:
\begin{inparaenum}[\it (i)]
\item an \emph{OBDA system}, which keeps all the data of interest and
  provides a conceptual view over it in terms of a \dllitea
  TBox;
\item a \emph{process component} as in DCDSs, which
  characterizes the evolution (dynamic aspect) of the
  system; and
\item an \emph{initial database instance}.
\end{inparaenum}
Conceptually, a \sgds separates the system into two layers, the
\textit{relational layer} and the \textit{semantic layer}. The
relational layer captures the database evolution (manipulation) done
by the process execution, while the semantic layer exploits the
ontology for providing a conceptual view of the system evolution.
This enables us to
\begin{inparaenum}[\it (i)]
\item \emph{understand} the evolving system through the
  semantic layer, and
\item \emph{govern} the evolution of the system at the semantic layer
  by rejecting those process actions that, currently executed at the
  relational layer, would lead to new system states that violate some
  constraint of the ontology.
\end{inparaenum}
Formally, the semantics of \sgdss is defined in terms of two
transition systems: a \textit{Relational Layer Transition System}
(RTS) and a \textit{Semantic Layer Transition System} (STS).  The RTS
is the same as the transition system of a classical DCDS, which
captures the evolution of the system at the relational layer, tracking
how the database is evolved by the process component. On the other
hand, the STS is a ``virtualization'' of the RTS in the semantic layer
and provides a conceptual view of the system evolution. In particular,
the STS maintains the structure of the RTS unaltered, reflecting that
the process component is executed over the relational layer, but it
associates to each state the set of concept and role assertions
obtained from the application of the mappings starting from the
corresponding database instance. The intuition of the \sgds setting is
depicted in \Cref{sedap-intuition}.

Within \sgds, we address the problem of verifying \emph{conceptual
  temporal properties} that are 
specified at the semantic
layer. 
%
Roughly speaking, to tackle the verification problem, we bring down
the conceptual temporal property from the semantic layer into the
relational layer, by adopting the concept of ``rewriting'' and
``unfolding'' in OBDA, and then exploit the decidability results of
temporal property verification in DCDS. I.e., we show that the
verification of \sgdss can be reduced to the verification of DCDSs.

Going beyond theoretical results only, we have instantiated the
concept of \sgdss into a working tool called \obgsm, in which we use
the standard Guard-Stage-Milestone (GSM) model \cite{HDDF*11,DaHV13}
to represent the system in the relational layer.
%
\obgsm provides a functionality to translate the temporal property
specified at the semantic layer into the temporal property over the
relational layer by applying the ``rewriting'' and ``unfolding''
technique.
It exploits two already existing tools to provide its functionalities:
\begin{compactenum}
\item \ontop\footnote{\url{http://ontop.inf.unibz.it/}}, a JAVA-based
  framework for OBDA, and
\item the \emph{GSMC model checker}, developed within the EU FP7 Project
  ACSI\footnote{``Artifact-Centric Service Interoperation'', see
   \url{http://www.acsi-project.eu/}}, to verify GSM-based artifact-centric
  systems against temporal/dynamic properties \cite{BeLP12b}.
\end{compactenum}
\obgsm also becomes a part of EU FP7 Project ACSI deliverable (see
\cite{ACSI-D2.4.2}), and additionally, we also show how \obgsm can be
used in one of the practical use cases of the EU FP7 Project ACSI.

\subsection{Summary of All Reductions}

In addition to all reductions above, we also show that the
verification of S-GKABs can be reduced to the corresponding
verification of B-GKAB, C-GKAB, E-GKAB, \scsgkabs, \bicgkabs,
\cicgkabs, \eicgkabs, \bagkabs, \cagkabs, and \eagkabs.
Thus, summing it up, we have enriched the state of the art data-aware
business processes equipped with ontologies so that they can
accommodate various prominent scenarios without adding additional
computational complexity.

\begin{figure}[tbp]
\centering
\includegraphics[width=.7\textwidth]{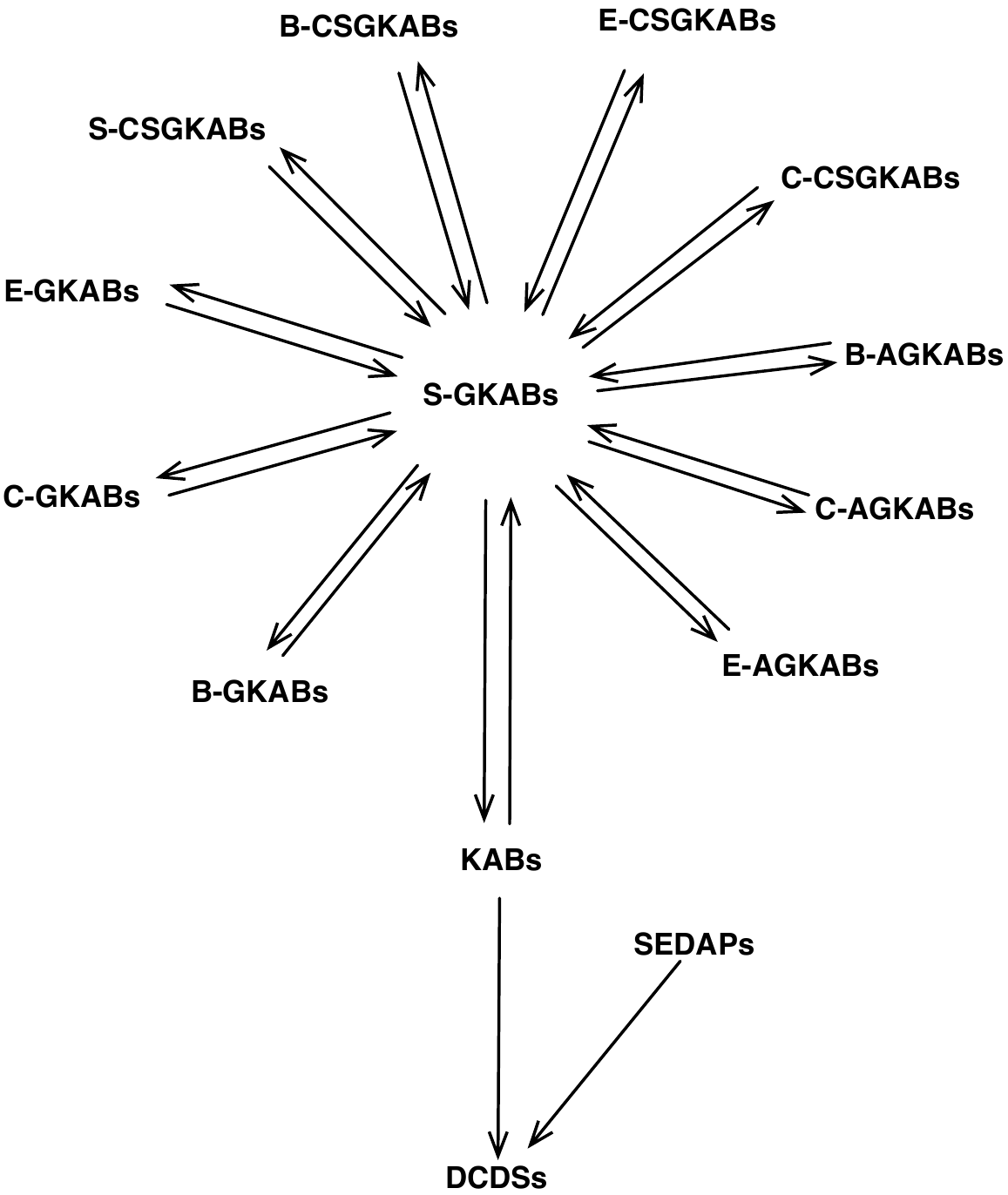}
\caption{Summary of the reductions developed in this thesis. The
  meaning of an arrow from formalisms $A$ to formalism $B$ is that
  verification of $A$ is reducible to verification of
  $B$ \label{fig:flower-all-results}}
\end{figure}

%
All our reductions are visually summarized in
\Cref{fig:flower-all-results}, where an arrow indicates that we can
reduce verification in the formalism at the source of the arrow to
verification in the formalism at the destination of the arrow.

\section{List of Publications}\label{sec:publications}

Some of the core results in this thesis have been published as
detailed below:


\begin{itemize}
\item In conference proceeding:
\begin{enumerate}

\item Diego Calvanese, Marco Montali and Ario
  Santoso. \textit{Verification of generalized inconsistency-aware
    knowledge and action bases}.  In Proceedings of the 24th
  International Joint Conference on Artificial Intelligence (IJCAI),
  pages 2847-2853, AAAI Press, 2015.

\item Diego Calvanese and {\.I}smail {\.I}lkan Ceylan and Marco
  Montali and Ario Santoso. \textit{Verification of context-sensitive
    knowledge and action bases}. In Proceedings of the 14th European
  Conference on Logics in Artificial Intelligence (JELIA), volume 8761
  of Lecture Notes in Computer Science, pages 514-528. Springer, 2014.

\item Babak Bagheri Hariri, Diego Calvanese, Marco Montali, Ario
  Santoso, and Dmitriy Solomakhin. \textit{Verification of
    semantically-enhanced artifact systems}. In Proceedings of the
   11th International Conference on Service Oriented Computing (ICSOC),
  volume 8274 of Lecture Notes in Computer Science, pages
  600-607. Springer, 2013.

\item Diego Calvanese, Evgeny Kharlamov, Marco Montali, Ario Santoso,
  and Dmitriy Zheleznyakov. \textit{Verification of
    inconsistency-aware knowledge and action bases}. In Proceedings
   of the 23rd International Joint Conference on Artificial Intelligence
  (IJCAI), pages 810-816. AAAI Press, 2013.

\item Diego Calvanese, Giuseppe De Giacomo, Domenico Lembo, Marco
  Montali, and Ario Santoso. \textit{Ontology-based governance of
    data-aware processes}.  In Proceedings of the 6th International
  Conference on Web Reasoning and Rule Systems (RR), volume 7497 of
  Lecture Notes in Computer Science, pages 25-41. Springer, 2012.

\item Ario Santoso. \textit{When data, knowledge and processes meet
    together}.  In Proceedings of the 6th International Conference
   on Web Reasoning and Rule Systems (RR), volume 7497 of Lecture Notes in
  Computer Science, pages 291-296. Springer, 2012.
  2012.

\end{enumerate}

\item In workshop proceeding:
\begin{enumerate}

\item Diego Calvanese, Marco Montali and Ario Santoso.
  \textit{Inconsistency management in generalized knowledge and action
    bases}. In Proceedings of the 28th International Workshop on
  Description Logic (DL), volume 1350, 2015.

\item Diego Calvanese, {\.I}smail {\.I}lkan Ceylan, Marco Montali and
  Ario Santoso. \textit{Adding context to knowledge and action
    bases}. In Workshop Notes of the 6th International Workshop on
  Acquisition, Representation and Reasoning about Context with Logic
  (ARCOE-Logic 2014), volume arXiv:1412.7965 of CoRR Technical
  Reports, pages 25-36. arXiv.org e-Print archive, 2014. Available
  at \url{http://arxiv.org/abs/1412.7965}.

\item Diego Calvanese, Evgeny Kharlamov, Marco Montali, Ario Santoso,
  and Dmitriy Zheleznyakov. \textit{Verification of
    inconsistency-aware knowledge and action bases}. In Proceedings
   of the 26th International Workshop on Description Logics (DL), volume
  1014 of CEUR Electronic Workshop Proceedings, \url{http://ceur-ws.org/},
  pages 107-119, 2013. 


\item Diego Calvanese, Giuseppe De Giacomo, Domenico Lembo, Marco
  Montali, and Ario Santoso. \textit{Semantically-governed data-aware
    processes}.  In Proceedings of the 1st International Workshop
   on Knowledge-intensive Business Processes (KiBP), volume 861 of
  CEUR Electronic Workshop Proceedings, \url{http://ceur-ws.org/},
  pages 21-32, 2012.

\end{enumerate}

\item Technical Reports:
\begin{enumerate}

\item Diego Calvanese, Marco Montali and Ario
  Santoso. \textit{Verification of generalized inconsistency-aware
    knowledge and action bases (extended version)}. CoRR Technical
  Report arXiv:1504.08108, arXiv.org e-Print archive, 2015. Available
  at \url{http://arxiv.org/abs/1504.08108}.

\item Babak Bagheri Hariri, Diego Calvanese, Marco Montali, Ario
  Santoso, and Dmitry Solomakhin. \textit{Verification of
    semantically-enhanced artifact systems $($extended
    version$)$}. CoRR Technical Report abs/1308.6292, arXiv.org
  e-Print archive, 2013.  Available at
  \url{http://arxiv.org/abs/1308.6292}.

\item Diego Calvanese, Babak Bagheri Hariri, Riccardo De Masellis,
  Domenico Lembo, Marco Montali, Ario Santoso, Dmitry Solomakhin, and
  Sergio Tessaris. \textit{Techniques and tools for KAB, to manage
    action linkage with the Artifact Layer - Iteration 2}. Deliverable
  ACSI-D2.4.2, ACSI Consortium, May 2013.


\item Diego Calvanese, Evgeny Kharlamov, Marco Montali, Ario Santoso,
  and Dmitriy Zheleznyakov.  \textit{Verification of
    inconsistency-aware knowledge and action bases $($extended
    version$)$}. CoRR Technical Report arXiv:1304.6442, arXiv.org
  e-Print archive, 2013.  Available at \url{http://arxiv.org/abs/1304.6442}.

\item Diego Calvanese, Giuseppe De Giacomo, Domenico Lembo, Marco
  Montali, Marco Ruzzi and Ario Santoso. \textit{Techniques and Tools
    for KAB, to Manage Data Linkage with the Artifact
    Layer}. Deliverable ACSI-D2.3, ACSI Consortium, May 2012.

\end{enumerate}

\end{itemize}

\section{Organization of the Thesis}\label{sec:thesis-structure}

We close this chapter by elaborating on the structure of the
thesis:

\begin{enumerate}
\item In \Cref{ch:intro} provides, we provide a high level
  introduction to the research area, outline the research challenges,
  and summarize the contributions of this thesis.

\item In \Cref{ch:prelim}, we introduce some preliminaries that are
  necessary for the comprehension of the thesis. They include some
  relevant basic notions, agreements on notations, and a quick
  introduction to
  \begin{inparaenum}[\it (i)]
  \item relational databases,
  \item \dllite knowledge bases,
  \item query answering,
  \item history preserving $\mu$-calculus, and
  \item Data Centric Dynamic Systems (DCDSs).
  \end{inparaenum}

\item In \Cref{ch:kab}, we review the basic notion of Knowledge and
  Action Bases (KAB) as proposed by \cite{BCDD*12,BCMD*13}, and
%
%
  we show how to reduce verification of KABs into verification of DCDSs
  \cite{BCDDM13}, which, in the end, opens the door for us to use the
  established results in \cite{BCDDM13}.

\item In \Cref{ch:gkab}, we exhibit our efforts in enriching KABs with
  Golog programs. In particular, we present Golog-KABs (GKABs) and
  define the parametric execution semantics for such framework. We
  also define the standard execution semantics for GKABs and call such
  setting S-GKABs. Last, we show in this chapter that the verification
  of S-GKABs is reducible to that for KABs and vice versa.

\item In \Cref{ch:ia-gkab}, we extend GKABs towards
  Inconsistency-aware GKABs (I-GKABs). The core results presented in
  this chapter are the reductions of verification from various
  Inconsistency-aware GKABs to GKABs with standard execution
  semantics.

\item In \Cref{ch:cs-gkab}, we extend GKABs into Context-sensitive
  GKABs (\csgkabs), which take into account contextual information
  during the evolution.  We show that verification of
  context-sensitive temporal properties over \csgkabs can be reduced
  to the verification of GKABs with standard semantics. Moreover, we
  show that S-GKABs can be easily captured by Context-sensitive GKABs.

\item Building on the results in \Cref{ch:ia-gkab} and
  \Cref{ch:cs-gkab}, in \Cref{ch:cs-ia-gkab}, we present
  Inconsistency-aware Context-Sensitive GKABs, which combine both the
  inconsistency handling mechanism and the contextual
  information. Concerning the core result, here we show that the
  verification of them is reducible to the verification of standard
  GKABS and vice versa

\item In \Cref{ch:a-gkab}, we define Alternating Golog-KABs (\agkabs),
  which separate the sources of non-determinism and enable
  fine-grained analysis over the system evolution. In this chapter, we
  prove that we can reduce the verification of \agkabs with various
  inconsistency handling mechanisms to the
  verification of S-GKABs and vice versa.

\item In \Cref{ch:sedap}, which serves as the last technical chapter
  in this thesis, we focus on our proposed novel framework of
  Semantically-Enhanced Data-Aware Processes (\sgdss) and also show
  the solution for verifying \sgdss.

\item In \Cref{ch:conclusion} we provide conclusions and future works.

\end{enumerate}

%% file: 2.chapters/2-preliminaries.tex
\chapter{Preliminaries}\label{ch:prelim}

\ifhidecontent
 
\fi

This chapter introduces several preliminaries that will be used for
the rest of this thesis.  In particular, we briefly present the notion
of relational databases, \dllite Knowledge Bases (KBs), queries, query
answering over databases as well as over KB, history preserving
$\mu$-calculus (\muladom and \mula), and Data Centric Dynamic Systems
(DCDSs). A convention on some basic notions and notations also will be
presented here.
%
%

We assume some familiarities with the basic notion of Propositional
Logic, and First Order Logic (FOL), for further references, please
consult~\cite{Smul68}. Moreover, in the following we make use of a
countably infinite set $\const$ of constants.



\section{Scenario for the Running Examples}\label{ex:example-scenario}
For the running examples throughout the thesis, we consider a
\emph{furniture provider enterprise} order processing scenario as
follows:
  \begin{compactenum}
  \item First, the company receives some orders.
  \item Second, the company approves the orders (in case they are not yet
    approved).
  \item The company prepares a few things that are needed for further
    order processing steps (such as creating the design and assigning
    the assembling location).
  \item The assembler assembles the orders based on the given design.
  \item The quality controller (QC) checks the assembled orders.
  \item The delivery team delivers the orders to the delivery service.
  \end{compactenum}
  Later on, when necessary, we will give more details on the scenario
  above. In some parts, we also extend this scenario as well as
  develop more stories based on the scenario above.

\section{Some Basic Notions and Notations
  Convention}\label{sec:notation}



Here we briefly sketch some required basic notions and notations as
follows:



Given a set $A$, we write $\card{A}$ to denote the cardinality of set
$A$. I.e., the number of elements in the set $A$.

Let $A$ and $B$ be two arbitrary sets, as usual, a relation
$f: A \times B$ over $A$ and $B$ is a subset of the cartesian
product between $A$ and $B$ (i.e.,
$f \subseteq \set{\tup{a,b} \mid a\in A \mbox{ and } b \in B}$). The
relation $f$ is a function if for each $a \in A$ there exists exactly
one $b \in B$ such that $\tup{a,b} \in f$. 
When 
$f$ is a function, we sometimes write $\tap{a \ra b} \in f$ instead of
$\tup{a, b} \in f$ to denote a tuple in $f$. Moreover, we say
\emph{$f$ maps $a$ to $b$}, written $f(a) = b$, if
$\tap{a \ra b} \in f$. We write $\domain{f}$ to denote the domain of
$f$.



Given a set $V$ of variables, 
a \emph{substitution} $\sigma$ is a function $\sigma: V \ra \const$
which maps each variable in $V$ into a constant in $\const$.
\noindent
Given 
a substitution $\sigma: V \ra \const$, we write $ x/c \in \sigma
$
if $\sigma(x) = c$, i.e., $\sigma$ maps $x$ into $c \in \const$ (or
sometimes we also say $\sigma$ substitutes $x$ with $c \in \const$).
We write $\sigma[x/c]$ to denote a new substitution obtained from
$\sigma$ such that $\sigma[x/c](x) = c$ and
$\sigma[x/c](y) = \sigma(y)$ (for $y \neq x$).

We write
\[
\varphi(x_1, \ldots, x_n)
\]
to denote an open FOL formula $\varphi$ whose free variables are
$x_1, \ldots, x_n$.  Given an FOL formula $\varphi(x_1, \ldots, x_n)$,
an FOL interpretation $\I = \inter$, and a substitution $\sigma$ which
substitutes each free variable $x_i$ with a constant $c \in \dom$,
we write 
\[
\varphi\sigma
\] 
to denote a close FOL formula obtained by applying $\sigma$ to
$\varphi$, i.e., substituting each free variable $x_i$ in $\varphi$
with a constant $c \in \dom$ based on $\sigma$. Moreover,
we write 
\[
\I \models \varphi\sigma
\]
 if $\I$ is a model $\varphi\sigma$.


 For compactness of writing, we often write $\vec{x}$ to denote a
 sequence of variables $x_1, \ldots, x_n$. Moreover, we often say that
 $x$ is a variable in $\vec{x}$ (or we write $x \in \vec{x}$) to say
 that $x$ is the variable $x_i$ in the sequence of variables
 $x_1, \ldots, x_n$ (for an $i \in \set{1, \ldots, n}$). Given two
 sequences of variables $\vec{x}$ and $\vec{y}$, we write
 $\vec{x} \subseteq \vec{y}$, if for each variable $v \in \vec{x}$, we
 have $v \in \vec{y}$. 
%




\section{Relational Databases}\label{sec:databases}

We now proceed to present some preliminaries on relational databases.
For further references, please consult~\cite{AbHV95,ElNa07}.
%
Here we make use the countably infinite set $\const$ of constants as
the domain of the values in a database (we also call the set $\const$
\emph{database domain}).

\begin{definition}[Relation Schema] 
  A \sidetext{Relation Schema} \emph{relation schema} 
  is simply a relation name with some arity $n > 0$.
\end{definition}


\begin{definition}[Database Schema] 
  A \sidetext{Database Schema} \emph{database schema} $\dbschema$ is a
    finite set $\set{R_1, \ldots, R_n}$ of relation schemas.
\end{definition}



\begin{definition}[Database Fact]
  Given \sidetext{Database Fact} a relation schema $R$ with arity $n$,
  a \emph{database fact} (briefly \emph{fact}) over schema $R$ is an
  expression of the form $R(c_1,\ldots,c_n)$ such that $c_i \in \const$
  for $i \in \set{1,\ldots,n}$.
\end{definition}


\begin{definition}[Relation Instance] 
  Given \sidetext{Relation Instance} a relation schema $R$ with arity
  $n$,
  a \emph{relation instance} of $R$ over $\const$ is a finite set of
  facts over $R$.
\end{definition}


\begin{definition}[Database Instance] 
  Given \sidetext{Database Instance} a database schema
  $\dbschema = \set{R_1, \ldots, R_n}$,
  a \emph{database instance which conforms to $\dbschema$} is a finite
  set $\dbinst$ of facts over $R_i$ for $i \in \set{1,\ldots,n}$.
%
%
%
\end{definition}

\begin{example}\label{ex:dbschema-and-dbinst}
  Consider our running example scenario in \Cref{ex:example-scenario},
  as an example of a database schema, we specify a database schema
  $\dbschema$ that contains the following relation schemas\\
  $
  \begin{array}{lr@{ \ }l}
    \bullet&\exr{ORDER}(&\exra{id}, \exra{name}, \exra{processing\_status},
                          \exra{customerID}, \exra{designer}, \\ 
           && \exra{assembler}, \exra{quality\_controller}, \exra{assembling\_loc}, \exra{design} \ ) \\
  \end{array}\\
  \begin{array}{lr@{ \ }l}
    \bullet&\exr{DELIVERED\_ORDER}(&\exra{id}, \exra{delivery\_date}  \ )
  \end{array}
  $\\
  Essentially, the database schema $\dbschema$ contains two relation
  schema, namely $\exr{ORDER}$ and $\exr{DELIVERED\_ORDER}$, where the
  former stores the information about customer orders and the latter
  stores the information about orders that has been delivered.

  An example of a database instance which conforms to $\dbschema$ is
  as follows:
  \[
  \begin{array}{@{}r@{}l@{}}
    \dbinst = \set{&\exr{ORDER}(\excon{123}, \excon{chair}, \excon{received},
                     \excon{456}, \excon{NULL}, \excon{NULL}, \excon{NULL}, \excon{NULL},
                     \excon{ecodesign}), \\
                   &\exr{ORDER}(\excon{321}, \excon{table}, \excon{approved},
                     \excon{654}, \excon{NULL}, \excon{NULL}, \excon{NULL}, \excon{NULL},
                     \excon{NULL})
                     }.
  \end{array}
  \]
\end{example}

\begin{definition}[Active Domain of a Database Instance]\label{def:active-domain-db}
  Given \sidetext{Active Domain of a Database
    Instance} 
  a database instance $\dbinst$, the \emph{active domain} of $\dbinst$
  is the smallest set $\adom{\dbinst} \subseteq \const$ of constants
  such that for each $R(c_1,\ldots, c_n) \in \dbinst$, we have
  $c_i \in \adom{\dbinst}$ for $i \in \set{1, \ldots, n}$.
\end{definition}

\noindent
Intuitively, an active domain of a database instance $\dbinst$ is a
finite set of constants that explicitly present in $\dbinst$.

\begin{definition}[FOL Interpretation Obtained From a Database] \label{def:fol-interpretation-of-db}
  Given \sidetextb{FOL Interpretation Obtained From a
    Database} 
  a database instance $\dbinst$ which conforms to a database schema
  $\dbschema = \set{R_1, \ldots, R_n}$, we define an \emph{FOL
    Interpretation obtained from} $\dbinst$ as a usual FOL
  Interpretation
  $\I_\dbinst = (\dom[\I_\dbinst], \Int[\I_\dbinst]{\cdot})$ such that
\begin{compactenum}
\item $\dom[\I_\dbinst] = \const$,
\item
  $R_i^{\I_\dbinst} = \set{\tup{c_1,\ldots, c_m} \mid R_i(c_1,\ldots,
    c_m) \in \dbinst}$ for $i \in \set{1, \ldots, n}$.
\end{compactenum}
\ \ 
\end{definition}

\noindent
To simplify the notation, when it is clear from the context, we often
just write $\dbinst$ to denote $\I_\dbinst$. For example, given a
close FOL formula $\varphi$, we write $\dbinst \models \varphi$ to
denote $\I_\dbinst \models\varphi$. Informally speaking, in this case
we want to consider a database instance simply as an FOL
interpretation. 



\section{\dllite Knowledge Bases}\label{sec:dllite-KB}

In this thesis we use Description Logic (DL)~\cite{BCMNP07} to express
the Knowledge Bases (KB). In particular, we resort to a specific DL
family, namely \dllite~\cite{CDLLR07,ACKZ09}, which is specifically
tuned to have low complexity of reasoning while still expressive
enough to capture the domain of interest. In this section, we briefly
sketch some members of \dllite family,
namely 
\dllitea~\cite{PLCD*08,CDLL*09}, and \dlliter~\cite{CDLLR07}.
%



\bigskip
\noindent
\textbf{\dllitea. \xspace} \dllitea distinguishes the set of constants
into the set of \emph{objects} 
and the set of \emph{values}. 
%
Hence, for this section only, we assume that the set $\const$ consists
of a set $\oset$ of objects and a set $\vset$ of values such that
$\const = \oset \uplus \vset$.
%
%
%
%
The \dllitea allows for expressing
\begin{inparaenum}[\it (i)]
\item \emph{concepts}, representing sets of objects,
\item \emph{roles}, representing binary relations between objects, and
\item \emph{attributes}, representing binary relations between objects
  and values.
\end{inparaenum}
The syntax of \emph{\dllitea expressions} is given below.

\begin{definition}[The Syntax of \dllitea Expressions]
  The \sidetext{Syntax of \dllitea} syntax of
  \emph{\dllitea expressions} (concept, role and attribute) is defined
  by the following grammar:
  \[
  B ~::=~ N ~\mid~ \SOMET{R} ~\mid~ \DOMAIN{U} \qquad\qquad
   R ~::=~ P ~\mid~\INV{P}
  \]
  where 
\begin{compactitem}
\item $N$, $P$, and $U$ respectively denote a \emph{concept name}, a
  \emph{role name}, and an \emph{attribute name}, 
\item $\INV{P}$ denotes the \emph{inverse of a role}, 
\item $B$ and $R$ respectively denote \emph{concepts} and
  \emph{roles}.
\item The concept $\SOMET{R}$, also called \emph{unqualified
    existential restriction}, denotes the \emph{domain} of a role $R$,
  i.e., the set of objects that $R$ relates to some object.
\item Similarly, the concept $\DOMAIN{U}$ denotes the \emph{attribute
    domain} of $U$, i.e., the set of objects that $U$ relates to some
  value.
\end{compactitem}
%
\end{definition}
\noindent
We assume that the set of concept, role, and attribute names
are disjoint. 
To formally present the definition of \dllitea KBs,
we first explain the notion of \dllitea ABox and TBox as follows:

\begin{definition}[\dllitea ABox]
  A \sidetext{\dllitea ABox} \emph{\dllitea ABox} is a finite set
  $A$ of \emph{ABox assertions} of the form
  \[
    N(o_1), \quad\quad P(o_1,o_2), \quad\quad U(o_1,v),
  \]
  where $o_1, o_2 \in \oset$ denote objects and $v \in \vset$ denotes
  a value. Consecutively, from left to right, $N(o_1)$ is called
  \emph{concept assertion}, $P(o_1,o_2)$ is called \emph{role
    assertion}, and $U(o_1,v)$ is called \emph{attribute assertion}.
\end{definition}

\noindent
\noindent
The notions of \emph{entailment}, \emph{satisfaction}, and
\emph{model} of an ABox is as usual~\cite{CDLLR08b}. 
Informally, an ABox can be considered as a data storage. Notice that
we can also view an ABox as a database instance where the schema is
the set of concept, role, and attribute names. Similar to
\Cref{def:active-domain-db}, we define an \emph{active domain} of ABox
$A$, denoted by $\adom{A}$, as the set of constants from $\const$
(i.e., objects and values) that explicitly present in $A$.

\begin{definition}[Active Domain of an ABox]
  Given \sidetext{Active Domain\\ of an ABox} an ABox $A$, an
  \emph{active domain} of $A$ is a finite set
  $\adom{A} \subseteq \const$ of constants constructed as follows
  \begin{compactenum}
  \item For each concept assertion $N(o) \in A$, we have $o \in \adom{A}$,
  \item For each role assertion $P(o_1, o_2) \in A$, we have $o_1, o_2 \in \adom{A}$,
  \item For each attribute assertion $U(o, v) \in A$, we have
    $o, v \in \adom{A}$.
  \end{compactenum}
 \ \ 
\end{definition}


We now proceed to explain the notion of TBox that intuitively encodes
the domain knowledge as follows.
\begin{definition}[\dllitea TBox]\label{def:dllitea-tbox}
  A \sidetext{\dllitea TBox} \emph{\dllitea TBox} is a finite set
\[
T = T_p \uplus T_n \uplus T_f, 
\]
with
\begin{compactitem}
\item $T_p$ a finite set of \emph{positive inclusion assertions} of
  the form 
  \[
    B_1 \sqsubseteq B_2, \quad\quad R_1 \sqsubseteq R_2, \quad\quad U_1 \sqsubseteq U_2.
  \]
\item $T_n$ a finite set of \emph{negative inclusion assertions} of
  the form
  \[
    B_1 \sqsubseteq \neg B_2, \quad\quad R_1 \sqsubseteq \neg R_2, \quad\quad U_1 \sqsubseteq \neg U_2.
  \]
\item $T_f$ a finite set of \emph{functionality assertions} of the
  form
  \[
  \funct{R}, \quad \quad \quad \funct{U}.
  \]
%
%
\end{compactitem}
where 
\begin{inparaenum}[\it (i)]
\item Each $B_i$ and $R_i$ respectively denote \emph{concepts} and
  \emph{roles}.
\item $U$ denotes an \emph{attribute name}. 
%
\end{inparaenum}
Additionally, as usual in \dllitea TBoxes, we impose that roles and
attributes occurring in functionality assertions
cannot be specialized (i.e., they cannot occur in the right-hand side
of positive inclusions).
\end{definition}

\noindent
The notions of \emph{entailment}, \emph{satisfaction}, and
\emph{model} of a TBox is as usual~\cite{CDLLR08b}.

%
%
%
We \sidetext{TBox Vocabulary} call the set of concept, role, and
attribute names that appear in TBox $T$ a \emph{vocabulary of TBox
  $T$}, denoted by $\voc(T)$.  W.l.o.g. given a TBox $T$, we assume
that $\voc(T)$ contains all possible concept, role, and attribute
names. Notice that we can simply add an assertion $N \sqsubseteq N$
(resp. $P \sqsubseteq P$ and $U \sqsubseteq U$) into the TBox $T$ in
order to add a concept name $N$ (resp. role name $P$, and attribute
name $U$) into $\voc(T)$ such that $\voc(T)$ contains all possible
concept, role, and attribute names, and without changing the expected
set of models of the TBox $T$ (hence, preserving the deductive
closures of $T$).
Moreover, we call an ABox \emph{$A$ is over $\voc(T)$} if it consists
of ABox assertions of the form either $N(o_1)$, $P(o_1, o_2)$, or
$U(o_1, v)$, where $N, P, U \in \voc(T)$.

Having the notion of the \dllitea TBox and ABox in place, we are ready
to introduce a \dllitea Knowledge Bases (KB) as follows:

\begin{definition}[\dllitea Knowledge Base]\label{def:dllitea-kb}
  A \sidetext{\dllitea \\Knowledge Base} \dllitea KB is a pair
  $\tup{T,A}$, where
\begin{inparaenum}[\it (i)]
\item $T$ is a \dllitea TBox,
\item $A$ is a \dllitea ABox over $\voc(T)$
\end{inparaenum}
\end{definition}

For the semantics of \dllitea expressions, as in~\cite{CDLLR08b}, we
adopt the standard FOL semantics of DLs based on FOL
interpretations. 
We also interpret objects as well as values over distinct
domains. Additionally, we adopt the \emph{standard name assumption}, i.e., 
\begin{compactenum}
\item \emph{Unique name assumption} holds (different constants denote
  different objects or values).
\item The interpretation domain 
  contains the set of objects and values and each value (as well as
  object) is interpreted as itself.
\end{compactenum}

\begin{definition}[Semantics of \dllitea Expressions]
  The \sidetext{Semantics of \dllitea} semantics of
  \dllitea expressions is given by an interpretation $\I = \inter$
  where:
  \begin{compactitem}

  \item $\dom = \oset \uplus \vset$ is an \emph{interpretation
      domain}. 

  \item $\Int{\cdot}$ is an \emph{interpretation function} such that 
    \begin{compactitem}
    \item for each object $o \in \oset$, $\Int{o} = o$, 

    \item for each value $v \in \vset$, $\Int{v} = v$,


    \item the conditions in \Cref{tab:dl-semantics} are satisfied.
    \end{compactitem}

\begin{table}

\begin{tabular}{| l | l | l |}
  \hline
  Role name & $P$ & $\Int{P} \subseteq \Int{\oset} \times \Int{\oset}$ \\
  \hline
  Inverse of a role & $P^-$ & $\INT{P^-} = \set{ \tup{o, o'} \mid
                              \tup{o',o} \in \Int{P} }$ \\
  \hline
  \hline
  Attribute name & $U$ & $\Int{U} \subseteq \Int{\oset} \times \Int{\vset}$ \\
  \hline
  \hline
  Concept name & $N$ & $\Int{N} \subseteq \Int{\oset}$ \\
  \hline
  Unqualified existential restriction & $\SOMET{R}$ & $\INT{\SOMET{R}}
                                                      = \set{o \mid
                                                      \exists
                                                      o'. \tup{o,o'}
                                                      \in \Int{R}  }$ \\
  \hline
  Domain of an attribute & $\DOMAIN{U}$ & $\Int{\DOMAIN{U}}  = \set{ o
                                          \mid \exists v. \tup{o,v} \in
                                          \Int{U}}$ \\
  \hline
\end{tabular}

\caption{Semantics \dllitea expressions}\label{tab:dl-semantics}
\end{table}
\end{compactitem}
\ \
\end{definition}

\noindent
The semantics of \dllitea ABox is defined as follows:

\begin{definition}[Semantics of \dllitea ABox]
  An \sidetext{Semantics of \dllitea ABox} interpretation
  $\I = \inter$ satisfies an ABox assertion:
\[
\begin{array}{lcl}
  A(o) & \mbox{ if } & \Int{o} \in \Int{A}; \\
  P(o_1,o_2) & \mbox{ if } & \tup{\Int{o_1}, \Int{o_2}} \in \Int{P}; \\
  U(o,v) & \mbox{ if } & \tup{\Int{o}, \Int{v}} \in \Int{U}.
\end{array}
\]
%
%
\end{definition}

\noindent
The semantics of \dllitea TBox is defined as follows:

\begin{definition}[Semantics of \dllitea TBox]
  An \sidetext{Semantics of \dllitea TBox} interpretation
  $\I = \inter$ satisfies a TBox assertion:
\[
\begin{array}{lcl}
  B_1 \sqsubseteq B_2 & \mbox{ if } & \Int{B_1} \subseteq \Int{B_2}; \\
  R_1 \sqsubseteq R_2 & \mbox{ if } & \Int{R_1} \subseteq \Int{R_2}; \\
  U_1 \sqsubseteq U_2 & \mbox{ if } & \Int{U_1} \subseteq \Int{U_2}; \\
  B_1 \sqsubseteq \neg B_2 & \mbox{ if } & \forall o. o\in \Int{B_1}
                                           \ra o \not\in \Int{B_2}; \\
  R_1 \sqsubseteq \neg R_2 & \mbox{ if } & \forall o, o'. \tup{o,
                                           o'} \in \Int{R_1}
                                           \ra \tup{o,
                                           o'} \not\in \Int{R_2}; \\
  U_1 \sqsubseteq \neg U_2 & \mbox{ if } & \forall o, v. \tup{o,
                                           v} \in \Int{U_1}
                                           \ra \tup{o,
                                           v} \not\in \Int{U_2}; \\
  \funct{R} & \mbox{ if } & \forall o,o',o''. \tup{o, o'} \in
                            \Int{R} \wedge \tup{o, o''} \in \Int{R}
                            \ra o' = o''; \\
  \funct{U} & \mbox{ if } & \forall o,v',v''. \tup{o, v'} \in
                            \Int{R} \wedge \tup{o, v''} \in \Int{R}
                            \ra v' = v''; \\
%
\end{array}
\]
\ \ 
%
\end{definition}

%
%
%
%
The notions of \emph{entailment}, \emph{satisfaction}, and
\emph{model} are as usual~(c.f.\ \cite{CDLLR08b}).
%
%
An interpretation \emph{$\I$ satisfies a TBox $T$}, written
$\I \models T$, if $\I$ satisfies all TBox assertions in
$T$. Similarly, an interpretation \emph{$\I$ satisfies an ABox $A$},
written $\I \models A$, if $\I$ satisfies all ABox assertions in
$A$. Furthermore, an interpretation \emph{$\I$ satisfies a KB
  $\tup{T,A}$}, written $\I \models \tup{T,A}$, if $\I$ satisfies the
TBox $T$ and the ABox $A$.
We say $A$ is \emph{$T$-consistent} if $\tup{T,A}$ is satisfiable,
i.e., admits at least one model, otherwise we say $A$ is
\emph{$T$-inconsistent}.
Additionally, in this thesis we assume that, given a TBox $T$, all
concepts and roles in $\voc(T)$ are satisfiable, i.e., for every
concept $N$ in $T$, there exists at least one model $\I$ of $T$ such
that $\Int{N}$ is non-empty, and similarly for roles.

As an observation, notice that in \dllitea, positive inclusion
assertions alone cannot generate inconsistency. This is an immediate
consequence of Lemma 4.5 in \cite{CDLL*09} which essentially says that
we can always find a model for a \dllitea KB $\tup{T_p,A}$ where $T_p$
only contains positive inclusion assertions. However, positive
inclusion assertions might involve in causing inconsistency by
interacting with negative inclusions.

\begin{example}\label{ex:tbox-and-abox}
  Continuing our running example using the scenario in
  \Cref{ex:example-scenario}. We specify a \dllitea KB $\tup{T,A}$,
  where the TBox $T$ contains the following assertions: \\
$
\begin{array}{rcl}
  \exo{ApprovedOrder} &\sqsubseteq& \exo{Order}\\
  \exo{AssembledOrder} &\sqsubseteq& \exo{Order}\\
  \exo{DeliveredOrder} &\sqsubseteq& \exo{Order}\\
  \exo{ReceivedOrder} &\sqsubseteq& \exo{Order}\\
  \exo{Designer} &\sqsubseteq& \exo{Employee}\\
  \exo{Assembler} &\sqsubseteq& \exo{Employee}\\
  \exo{QualityController} &\sqsubseteq& \exo{Employee}\\
  \exo{Designer} &\sqsubseteq& \neg \exo{Assembler}\\
  \exo{Designer} &\sqsubseteq& \neg \exo{QualityController}\\
  \exo{Assembler} &\sqsubseteq& \neg \exo{QualityController}\\
\end{array}
\qquad
\begin{array}{rcl}
  \SOMET{\exo{assembledBy}^-}&\sqsubseteq& \exo{Employee}\\
  \SOMET{\exo{assembledBy}}&\sqsubseteq& \exo{Order}\\
  \SOMET{\exo{designedBy}^-}&\sqsubseteq& \exo{Employee}\\
  \SOMET{\exo{designedBy}}&\sqsubseteq& \exo{Order}\\
  \SOMET{\exo{checkedBy}^-}&\sqsubseteq& \exo{Employee}\\
  \SOMET{\exo{checkedBy}}&\sqsubseteq& \exo{Order}\\
  \SOMET{\exo{hasAssemblingLoc}^-}&\sqsubseteq& \exo{Location}\\
  \SOMET{\exo{hasAssemblingLoc}}&\sqsubseteq& \exo{Order}\\
  \SOMET{\exo{hasDesign}^-}&\sqsubseteq& \exo{Design}\\
  \SOMET{\exo{hasDesign}}&\sqsubseteq& \exo{Order}\\
\end{array}
$\\
\begin{center}
$
\begin{array}{c}
   \funct{\exo{hasAssemblingLoc}}\\
   \funct{\exo{hasDesign}}\\
\end{array}
$
\end{center}
The intuition of some TBox assertions presented above are as follows:
\begin{compactitem}

\item The assertion $\exo{ApprovedOrder} \sqsubseteq \exo{Order}$ states that every
  approved order is an order.

\item The assertion $\exo{Designer} \sqsubseteq \neg \exo{Assembler}$
  encodes a constraint that a designer is not an assembler.

\item The assertion
  $\exists\exo{hasAssemblingLoc} \sqsubseteq \exo{Order}$ states that
  those that can have an assembling location must be an order.

\item The assertion $\exists\exo{checkedBy-} \sqsubseteq
  \exo{Employee}$ says that something can be only checked by an employee.

\item The assertion $\funct{\exo{hasAssemblingLoc}}$ says that the
  role $\exo{hasAssemblingLoc}$ is functional. Informally, it
  constraints the domain of $\exo{hasAssemblingLoc}$ to have only a
  single range.

\end{compactitem}
Moreover, the ABox $A$ is specified as follows:
\[
A = \set{\exo{ReceivedOrder}(\excon{chair}),
  \exo{ApprovedOrder}(\excon{table})}.
\]
Intuitively, the ABox $A$ states the fact that there are a received
order of chair and an approved order of table.
\end{example}


\smallskip
\noindent
\textbf{\dlliter.  \xspace} \dlliter is a sublanguage of \dllitea
which is obtained by dropping functionality assertions.
%
\dlliter is also the basis of OWL~2~QL (a profile\footnote{In W3C
  terminology, a \emph{profile} is a sublanguage defined by suitable
  syntactic restrictions.} of the Web Ontology Language OWL~2
standardized by the W3C.).  In W3C terminology, a \emph{profile} is a
sublanguage of the full OWL~2, defined by suitable syntactic
restrictions.  OWL~2~QL is specifically designed for building an
ontology layer to wrap possibly very large data sources.  Notably, it
allows for query answering over ontologies with the same data
complexity as plain SQL query evaluation over relational databases.

\section{Query Answering}\label{sec:query-ans}


Roughly speaking, queries are expressions to ask information, and
query answering is a mechanism to extract some information (called
\emph{answer}) from an information source.
In the following, we introduce various kind of queries that will be
used later, and also explain how the answers are computed when we pose
such queries over an information/data source.

As a start, we introduce a very expressive query, namely FOL queries,
as follows:

\begin{definition}[FOL Query]
%
  An \sidetext{FOL Query} \emph{FOL query} $Q$ is an FOL formula
  without function symbols that
  might use some constants in $\const$.
\end{definition}

\noindent
As usual, given an FOL query $Q$, we call $Q$ a \emph{boolean query}
or a \emph{closed query} if $Q$ has no free variables, otherwise we
call it an \emph{open query}.

\subsection{Query Answering Over Databases}\label{subsec:qa-over-db}

\noindent
We use queries to formulate questions to be asked over a
database. Given a database instance $\dbinst$ and a query $q$, query
answering is aiming to obtain the answers to the query $q$ which are
formed by elements of $\const$. 
%
%
%
In the following we introduce some query languages 
that will be used later to query a database and also the notion of
answers to a query. For further references, please
consult~\cite{AbHV95,ElNa07}.
%
%


Given an FOL query $\varphi$, and a database instance $\dbinst$ which
conforms to a database schema $\dbschema$, we say that the FOL query
\emph{$\varphi$ is over $\dbschema$ and $\dbinst$} if the atoms in
$\varphi$ are made from relation schemas in $\dbschema$ and might use
some constants in \adom{$\dbinst$}.
%
%

Now, we present an interesting fragment of FOL queries which will be
used quite often later, namely Union of Conjunctive Query (UCQ).

\begin{definition}[Union of Conjunctive Query Over a Database]
  Given \sidetext{Union of Conjunctive Query (UCQ)} a
  database instance $\dbinst$ which conforms to a database schema
  $\dbschema$, a \emph{Union of Conjunctive Query (UCQ) $q$ over
    $\dbschema$ and $\dbinst$} is an FOL query of the form
  \[
  \exists\vec{y}_1.\conj_1(\vec{x},\vec{y}_1) \lor\cdots\lor
  \exists\vec{y}_n.\conj_n(\vec{x},\vec{y}_n),
  \]
  where each $\conj_i(\vec{x},\vec{y_i})$ in $q$ is a conjunction of
  atoms whose predicates are either 
  \begin{compactitem}
  \item relation schemas in $\dbschema$, or
  \item equality assertions 
  \end{compactitem}
  which involve free variables $\vec{x}$ and existentially quantified
  variables $\vec{y}_1,\ldots,\vec{y}_n$, and/or elements of
  $\adom{\dbinst}$.
\end{definition}

\begin{example}\label{ex:ucq-db}
  Continuing our running example, recall the database schema
  $\dbschema$ and and the database instance $\dbinst$ in
  \Cref{ex:dbschema-and-dbinst}, an example of a UCQ $q$ over
  $\dbschema$ and $\dbinst$ is
  \[
  \exists
  \exvar{x_2}\exvar{x_4}\exvar{x_5}\exvar{x_6}\exvar{x_7}\exvar{x_8}
  \exvar{x_9}.\exr{ORDER}(\exvar{x_1},\exvar{x_2},
  \excon{"received"},\exvar{x_4},\ldots,\exvar{x_9})
  \]
  that retrieves the id of received orders.
\end{example}

\noindent
We sometimes say a UCQ $q$ is over $\dbschema$ to refer to a UCQ whose
atoms are made using relation schemas in $\dbschema$.



Next, we explain the notion of answers to a query $Q$ over a database
instance $\dbinst$ as a result of evaluating $Q$ over $\dbinst$ as
follows.


\begin{definition}[Answers to a Query]\label{def:answer-to-db}
  Given \sidetext{Answers to a Query} a database instance $\dbinst$
  which conforms to a database schema $\dbschema$ and an FOL query $Q$
  over a database schema $\dbschema$.
  The \emph{answers to $Q$ over \dbinst} is the set $\ANS(Q,\dbinst)$
  of substitutions, in which each substitution
  $\sigma \in \ANS(Q,\dbinst)$ substitutes the free variables of $Q$
  with elements of $\const$ 
  such that $\dbinst \models Q\sigma$.
  If $Q$ is a boolean query then its answer is either a singleton set
  of an empty substitution (corresponding to $\true$) or an empty set
  (corresponding to $\false$).
\end{definition}

\noindent
As customary, we can view each substitution $\sigma$ (which is an
answer to a query $Q(x_1, \ldots, x_n)$) simply as a tuple of elements
of $\const$, 
assuming some ordering of the free variables $x_1, \ldots, x_n$ of
$Q$. 
Therefore, given a query $Q(x_1, \ldots, x_n)$ and a substitution
$\sigma \in \ANS(Q(x_1, \ldots, x_n),\dbinst)$ such that
$\sigma(x_i) = c_i$ (for $1 \leq i \leq n$), we sometimes also write
it as either
$\tup{c_1,\ldots,c_n} \in \ANS(Q(x_1, \ldots, x_n),\dbinst)$ or
$\vec{c} \in \ANS(Q(\vec{x}),\dbinst)$.

\begin{example}\label{ex:ans-ucq-db}
  Continuing \Cref{ex:ucq-db}, recall the query $q$ as follows:
  \[
  \exists
  \exvar{x_2}\exvar{x_4}\exvar{x_5}\exvar{x_6}\exvar{x_7}\exvar{x_8}
  \exvar{x_9}.\exr{ORDER}(\exvar{x_1},\exvar{x_2},
  \excon{"received"},\exvar{x_4},\ldots,\exvar{x_9})
  \]
  a substitution $\sigma$ that substitutes $\exvar{x_1}$ to
  ``$\excon{123}$'' is an answer to $q$ over $\dbinst$ (i.e.,
  $\sigma \in \ANS(q,\dbinst)$).
\end{example}

Another interesting fragment of FOL query that will be used later is
\emph{Domain Independent FOL ($\difol$) query}. To formally explain
the notion of $\difol$ query, several preliminaries need to be
introduced as follows.

\begin{definition}[FOL Interpretation Obtained From a Database w.r.t.\
  a Certain Domain] \label{def:fol-interpretation-of-db-spec-dom} Given
  \sidetextb{FOL Interpretation Obtained From a Database w.r.t.\ a
    Certain Domain} 
  a set $\const'$ of constants such that $\const' \subseteq \const$,
  and
  a database instance $\dbinst$ which conforms to a database schema
  $\dbschema = \set{R_1, \ldots, R_n}$, and
  $\adom{\dbinst} \subseteq \const'$, we define an \emph{FOL
    Interpretation obtained from} $\dbinst$ w.r.t.\ $\const'$ as a
  usual FOL Interpretation
  $\I^{\const'}_\dbinst = (\dom[\I^{\const'}_\dbinst],
  \Int[\I^{\const'}_\dbinst]{\cdot})$ such that
\begin{compactitem}
\item $\dom[\I^{\const'}_\dbinst] = \const'$,
\item
  $R_i^{\I^{\const'}_\dbinst} = \set{\tup{c_1,\ldots, c_m} \mid
    R_i(c_1,\ldots, c_m) \in \dbinst}$ for $i \in \set{1, \ldots, n}$.
\end{compactitem}
\ \ 
\end{definition}

\noindent
When it is clear from the context, we often just write
$\dbinst^{\const'}$ to denote $\I^{\const'}_\dbinst$. For example,
given a close FOL query $Q$, we write $\dbinst^{\const'} \models Q$ if
$\I^{\const'}_\dbinst \models Q$.

\begin{definition}[Answers of a Query w.r.t.\ a Certain Domain]
  Given \sidetextb{Answers of a Query w.r.t.\ a Certain Domain} a set
  $\const'$ of constants such that $\const' \subseteq \const$, an FOL
  query $Q(x_1,\ldots x_n)$ over a database schema $\dbschema$, a
  database instance $\dbinst$ which conforms to a database schema
  $\dbschema$, and $\adom{\dbinst} \subseteq \const'$.
%
%
  The \emph{answers to $Q$ over $\dbinst$ w.r.t.\ $\const'$} is the
  set $\ANS_{\const'}(Q, \dbinst)$ of substitutions, in which each
  substitution $\sigma \in \ANS_{\const'}(Q, \dbinst)$ substitutes the
  free variables of $Q$ with elements of $\const'$ 
  such that $\dbinst^{\const'} \models Q\sigma$.
  If $Q$ is a boolean query then its answer is either a singleton set
  of an empty substitution (corresponding to $\true$) or an empty set
  (corresponding to $\false$).
\end{definition}

Having the required preliminaries in hand, we are ready to present the
definition of $\difol$ query as follows.

\begin{definition}[Domain Independent FOL Query] \label{def:domain-independent-query}
  An \sidetext{Domain Independent FOL Query} FOL query $Q$ is a
  \emph{Domain Independent FOL ($\difol$) query} if
  for every database instance $\dbinst$ which conforms to a database
  schema $\dbschema$, and 
  for every database domain $\const'$ and $\const''$ such that 
  $\adom{\dbinst} \subseteq \const' \subseteq \const$ and
  $\adom{\dbinst} \subseteq \const'' \subseteq \const$, we have
  \[
  \ANS_{\const'}(Q, \dbinst) = \ANS_{\const''}(Q, \dbinst)
  \]
\ \ 
\end{definition}

\noindent
The definition above intuitively said that given a database instance
$\dbinst$ and a query $Q$, the answers to $Q$ over $\dbinst$ are the
same no matter whether the domain of database values is the
$\adom{\dbinst}$, the whole set $\const$ of constants, or something
between $\adom{\dbinst}$ and $\const$ (i.e., the answers to the query
is independent from the domain of database values that we use).

\begin{example}
  The query in our previous running example (i.e.,
  \Cref{ex:ucq-db,ex:ans-ucq-db}) 
  \[
  \exists
  \exvar{x_2}\exvar{x_4}\exvar{x_5}\exvar{x_6}\exvar{x_7}\exvar{x_8}
  \exvar{x_9}.\exr{ORDER}(\exvar{x_1},\exvar{x_2},
  \excon{"received"},\exvar{x_4},\ldots,\exvar{x_9})
  \]
  is a domain independent query. On the other hand, the following
  query
  \[
  \neg\exists
  \exvar{x_2}\exvar{x_4}\exvar{x_5}\exvar{x_6}\exvar{x_7}\exvar{x_8}
  \exvar{x_9}.\exr{ORDER}(\exvar{x_1},\exvar{x_2},
  \excon{"received"},\exvar{x_4},\ldots,\exvar{x_9})
  \]
 is an example of a query that is not domain independent.
\end{example}

\subsection{Query Answering Over Knowledge Bases}\label{subsec:qa-over-kb}


We use queries to access KBs and extract values (constants) of
interest, i.e., to ask some questions over a KB and obtain
answers. 
%
%
As in \Cref{sec:dllite-KB}, in the following 
%
%
we assume that the set $\const$ consists of a set $\oset$ of objects
and a set $\vset$ of values such that $\const = \oset \uplus \vset$.
Below, we introduce a particular form of queries that we will use
later to query KBs, namely union of conjunctive queries over KB:


\begin{definition}[Union of Conjunctive Query Over a KB]
  Given \sidetext{Union of Conjunctive Query Over a KB} a KB
  $\tup{T, A}$, a \emph{Union of Conjunctive Query (UCQ) $q$ over
    $\tup{T, A}$} is an FOL query of the form
  \[
  \exists\vec{y}_1.\conj_1(\vec{x},\vec{y}_1) \lor\cdots\lor
  \exists\vec{y}_n.\conj_n(\vec{x},\vec{y}_n),
  \]
  where each $\conj_i(\vec{x},\vec{y_i})$ in $q$ is a conjunction of
  atoms whose predicates are either 
  \begin{compactitem}
  \item concept/role/attribute names in $\voc(T)$, or
  \item equality assertions 
  \end{compactitem}
  which involve free variables $\vec{x}$ and existentially quantified
  variables $\vec{y}_1,\ldots,\vec{y}_n$, and/or elements of
  $\adom{A}$.
\end{definition}


\noindent
We sometimes say a UCQ $q$ is over $T$ to refer to a UCQ whose atoms
are made using concept/role/attribute names in $\voc(T)$.

\begin{example}\label{ex:query-kb}
  Recall the TBox $T$ and the ABox $A$ specified
  in \Cref{ex:tbox-and-abox}, an example of UCQ $q$ over $\tup{T,A}$
  is:
  \[
  \exo{ApprovedOrder}(\exvar{x}) \vee \left(\exo{AssembledOrder}(\exvar{x})
  \wedge \exists y. \exo{checkedBy}(\exvar{x},\exvar{y}) \right)
  \]
  that retrieves either approved orders or assembled orders that has
  been checked by someone.
\end{example}

In the setting of query answering over KBs, we are interested to the
answers which are ``true'' in every model of the corresponding
KB. Such answers are called certain answers and the definition is as
follows:

\begin{definition}[Certain Answer of UCQs]\label{def:certain-answer-ucq}
  Given \sidetext{Certain Answer of UCQs} a KB $\tup{T,A}$ and a UCQ
  $q$ over $\tup{T,A}$, the \emph{(certain) answers} to UCQ $q$ over a
  KB $\tup{T,A}$ is the set $\ans(q,T,A)$ of substitutions $\sigma$ of
  the free variables of $q$ with elements of $\const$ 
  such that $\I \models q\sigma$ for every model $\I$ of $\tup{T,A}$
  (i.e., $q\sigma$ evaluates to true in every model of $\tup{T,A}$).
  If $q$ is a boolean query then its \emph{certain answer} is either a
  singleton set of an empty substitution (corresponding to $\true$) or
  an empty set (corresponding to $\false$).
\end{definition}

\noindent
Similar to answers to a query over a database instance (c.f.\
\Cref{def:answer-to-db}), we can view each substitution
$\sigma \in \ans(q,T,A)$ simply as a tuple of elements of
$\const$, 
assuming some ordering of the free variables of $q$.
I.e., given a query $q(x_1, \ldots, x_n)$ and a substitution
$\sigma \in \ans(q(x_1,\ldots,x_n),T,A)$ such that $\sigma(x_i) = c_i$
(for $1 \leq i \leq n$), we sometimes also write it as either
$\tup{c_1,\ldots,c_n} \in \ans(q(x_1,\ldots,x_n),T,A)$ or
$\vec{c} \in \ans(q(\vec{x}),T,A)$.




\begin{example}\label{ex:ans-ucq-kb}
  Continuing \Cref{ex:query-kb}, recall the query $q$ as follows:
  $ \exo{ApprovedOrder}(\exvar{x}) $.  A substitution $\sigma$ that
  substitutes $\exvar{x}$ to ``$\excon{table}$'' is an answer to $q$ over
  $A$ (i.e., $\sigma \in \ans(q,T,A)$).
\end{example}

In this work, we also consider the extension of UCQs named
\textit{EQL-Lite}(UCQ)~\cite{CDLLR07b} (briefly, ECQs), which is an
FOL query whose atoms are UCQs evaluated according to the certain
answer semantics above (see \Cref{def:certain-answer-ucq}). 

\begin{definition}[EQL-Lite(UCQ)]
%
%
  Formally,\sidetext{EQL-Lite(UCQ)} given a KB $\tup{T, A}$, an \emph{ECQ} over $\tup{T, A}$
  is a (possibly open) formula of the form:
\[
Q ~::=~ [q] ~\mid~ \lnot Q ~\mid~ Q_1\land Q_2 ~\mid~ \exists x. Q
\]
where $q$ is an UCQ over $\tup{T, A}$, and $[q]$ denotes the fact that
$q$ is evaluated under the (minimal) knowledge operator
\cite{CDLLR07b} (We often omit the square brackets for single-atom UCQs). We
call \emph{epistemic atoms} the formula $[q]$ occurring in an ECQ. 
\end{definition}

\begin{example}\label{ex:ecq}
  Recall the TBox $T$ and the ABox $A$ specified in
  \Cref{ex:tbox-and-abox}, an example of ECQ $Q$ over $\tup{T,A}$ is:
  \[
  \exists \exvar{x}.[\exo{Order}(\exvar{x})] \wedge
  \neg[\exo{DeliveredOrder}(\exvar{x})]
  \]
  that checks whether there exists an order that is not yet delivered.
\end{example}

\noindent 
To explain how to compute certain answers to ECQ, we need to introduce
several preliminaries as follows:

\begin{definition}[FOL Query Obtained From an ECQ Query]
  Given \sidetextb{FOL Query of an ECQ Query} an ECQ query $Q(\vec{x})$
  with epistemic atoms $[q_1], \ldots, [q_n]$. We define an \emph{FOL
    query obtained from an ECQ $Q$} (briefly FOL query of $Q$),
  denoted by $Q_{FOL}(\vec{x})$, as query obtained from $Q$ by
  replacing each epistemic atom $[q_i]$ with a new predicate $R_{q_i}$
  where the arity of $R_{q_i}$ is the number of free variables in
  $[q_i]$.
\end{definition}

\begin{definition}[FOL Interpretation for an FOL Query of an ECQ]
  Given \sidetextb{FOL Interpretation for an FOL Query of an ECQ} a KB
  $\tup{T, A}$, and an ECQ $Q(\vec{x})$ with epistemic atoms
  $[q_1], \ldots, [q_n]$. Let $Q_{FOL}(\vec{x})$ be an FOL query
  obtained from $Q(\vec{x})$. We define an \emph{FOL interpretation
    w.r.t.\ $\tup{T, A}$ and $Q(\vec{x})$} as an FOL interpretation
  $\I_{Q, \tup{T,A}} = (\dom[\I_{Q, \tup{T,A}}], \Int[\I_{Q,
    \tup{T,A}}]{\cdot})$ such that
  \begin{compactitem}
  \item $\dom[\I_{Q, \tup{T,A}}] = \const$
  \item for every predicates $R_{q_i}(x_1, \ldots, x_n)$, we have
    \[
    R_{q_i}^{\I_{Q, \tup{T,A}}} = \set{\tup{c_1,\ldots, c_n} \mid
      \sigma \in \ans(q_i(x_1, \ldots, x_n), T, A) \mbox{ and } x_i/c_i \in \sigma}
    \]
  \end{compactitem}
\ \ 
\end{definition}

Having the machinery in hand, we are now ready to explain how to
compute the certain answers of an ECQ. 
Given \sidetext{Certain Answers of an ECQ} a KB $\tup{T, A}$, an ECQ
$Q(\vec{x})$, the \emph{certain answers of $Q(\vec{x})$ over
  $\tup{T, A}$}, denoted by $\Ans(Q(\vec{x}), T, A)$, is a set of
substitutions $\sigma$ of the free variables of $Q(\vec{x})$ with
elements of $\const$ 
such that $\I_{Q, \tup{T,A}} \models Q_{FOL}(\vec{x})\sigma$.
  If $Q$ is a boolean query then its \emph{certain answer} is either a
  singleton set of an empty substitution (corresponding to $\true$) or
  an empty set (corresponding to $\false$).
%
Intuitively, the \emph{certain answers $\Ans(Q,T,A)$ of an ECQ $Q$
  over $\tup{T,A}$} are obtained by computing the certain answers of
the UCQs embedded in $Q$, then composing such answers through the FO
constructs in $Q$. 

Now we introduce an interesting fragment of ECQ namely domain
independent ECQ. 
\begin{definition}[Domain Independent ECQ ($\diecq$)]
  Given \sidetext{Domain Independent ECQ} an ECQ $Q$ with epistemic
  atoms $[q_1], \ldots, [q_n]$,
%
%
  we say $Q$ is a \emph{Domain Independent ECQ ($\diecq$)}, if for
  each FOL interpretation $\I_1 = (\dom[\I_1], \Int[\I_1]{\cdot})$ and
  $\I_2 = (\dom[\I_2], \Int[\I_2]{\cdot})$ for $Q_{FOL}$ such that
  $\dom[\I_1] \subseteq \const$, $\dom[\I_2] \subseteq \const$, and
  $R_{q_i}^{\I_1} = R_{q_i}^{\I_2}$ for all atomic relations
  $R_{q_i}$, we have that
  \begin{center}
    $\I_1 \models Q_{FOL}\sigma$ if and only if
    $\I_2 \models Q_{FOL}\sigma$
  \end{center}
  for any substitution $\sigma$.
\end{definition}

\begin{example}
  The ECQ in \Cref{ex:ecq} is a \diecq.
\end{example}

\subsubsection{FO-Rewritability of \dllite}
Similar to \Cref{def:fol-interpretation-of-db}, considering that an
ABox is also a set of facts, we can define an FOL interpretation
obtained from ABox $A$ as follows:

\begin{definition}[FOL Interpretation Obtained From an
  ABox] \label{def:fol-interpretation-of-abox} Given \sidetextb{FOL
    Interpretation Obtained From an ABox} a KB $\tup{T, A}$,
  we define an \emph{FOL Interpretation obtained from} $A$ as a
  usual FOL Interpretation
  $\I_A = (\dom[\I_A], \Int[\I_A]{\cdot})$ such that 
  \begin{compactitem} 
  \item $\dom[\I_A] = \const$ (Note that $\const$ contains both
    objects and values),
  \item for every concept name $N \in \voc(T)$,
    $N^{\I_A} = \set{o \mid N(o) \in A}$.
  \item for every role name $P \in \voc(T)$,
    $P^{\I_A} = \set{\tup{o_1, o_2} \mid P(o_1, o_2) \in A}$.
  \item for every attribute name $U \in \voc(T)$,
    $U^{\I_A} = \set{\tup{o, v} \mid U(o, v) \in A}$.
  \end{compactitem}
\ \ 
\end{definition}

\noindent
To simplify the notation, when it is clear from the context, we often
just write $A$ to denote $\I_A$. For example, given a close UCQ $q$,
we write $A \models q$ to denote $\I_A \models q$. Informally, we want
to consider an ABox simply as an FOL interpretation.
Furthermore, having \Cref{def:fol-interpretation-of-abox} in
hand, we can also define an evaluation of query $Q$ over an ABox $A$
(similar to \Cref{def:answer-to-db}) as follows.

\begin{definition}[Query Evaluation Over an ABox]\label{def:answer-to-abox}
  Given \sidetextb{Query Evaluation Over an ABox} a KB $\tup{T, A}$ and
  an FOL query $Q$ over $\tup{T, A}$.
  We define \emph{answers to $Q$ over A} as a set $\ANS(Q,A)$ of
  substitutions $\sigma$ of the free variables of $q$ with elements of
  $\const$
  such that $A \models Q\sigma$.
  As before, If $Q$ is a boolean query then its \emph{answer} is
  either a singleton set of an empty substitution (corresponding to
  $\true$) or an empty set (corresponding to $\false$).
\end{definition}


We now recall 
that DL-Lite enjoys the \emph{FO rewritability} property,
which 
means as follows.

\begin{theorem}[FO rewritability of \dllitea~\cite{CDLLR08b}]\label{thm:FO-rewritability}
  Given \sidetext{FO rewritability} a KB $\tup{T, A}$ and a UCQ $q$,
  we have
\[
\ans(q,T,A) = \ANS(\rew(q,T), A),
\]
where $\rew(q,T)$ is a UCQ computed by the query rewriting algorithm
in~\cite{CDLLR08b}. 
\qedw
\end{theorem}

\noindent
\Cref{thm:FO-rewritability} intuitively said that to compute the
certain answers to a UCQ $q$ over a KB $\tup{T,A}$, we can rewrite the
query $q$ to compile away the TBox $T$ as well as incorporate the
domain knowledge encoded in $T$ into $q$, and then we can just
evaluate the rewritten query $\rew(q,T)$ over the ABox $A$.
As in~\cite{CDLL*09}, such query rewriting algorithm is called
\emph{the perfect reformulation algorithm}.
%
%

Furthermore, the perfect reformulation algorithm above can be extended
to ECQs as in~\cite{CDLLR07b}, and that its effect is to ``compile
away'' the TBox. Precisely this statement is stated below.

\begin{theorem}[FO Rewritability of Answering ECQ\cite{CDLLR07b}]\label{thm:FO-rew-ECQ}
Given a KB $\tup{T, A}$, and an ECQ $Q$, we have
\[
\Ans(Q,T,A) = \ANS(\rew(Q,T), A),
\]
Where $\rew(Q,T)$ is an FOL query. Furthermore, if $Q$ is $\diecq$
query, then $\rew(Q, T)$ is $\difol$ query.
\qedw
\end{theorem}

\subsubsection{Consistency Check via Query Answering}

We recall that checking the satisfiability of a \dllitea KB
$\tup{T,A}$ is FO rewritable, i.e., it can be reduced to evaluating a
boolean FOL query
over
$A$ 
\cite{CDLLR07}.
%
For brevity of the notation that we will use later, we introduce
several abbreviations below:
\begin{definition}[Abbreviations For Query Atom]\label{def:abbreviation-query}
  We \sidetextb{Abbreviations For Query Atom} define some notations to
  compactly express various atoms in a query as follows:
  \begin{compactitem}
  \item An atom $B(x)$ denotes
    \begin{compactitem}
    \item $N(x)$ if $B = N$,
    \item $P(x,\_)$ if $B = \exists P$,
    \item $P(\_,x)$ if $B = \exists P^-$,
    \item $U(x,\_)$ if $B = \DOMAIN{U}$,
    \end{compactitem}
    where `$\_$' stands for an anonymous existentially quantified
    variable,
  \item An assertion $R(x, y)$ denotes
    \begin{compactitem}
    \item $P(x, y)$ if $R = P$,
    \item $P(y, x)$ if $R = P^-$,
    \end{compactitem}
  \item An assertion $Z(x, y)$ denotes
    \begin{compactitem}
    \item $U(x, y)$ if $Z = U$,
    \item $P(x, y)$ if $Z = P$,
    \item $P(y, x)$ if $Z = P^-$,
    \end{compactitem}
  \end{compactitem}
\ \ 
\end{definition}

%
%


\begin{definition}[Q-UNSAT-FOL Query Abbreviation]\label{def:qunsat-abbreviation}
  We \sidetextb{Q-UNSAT-FOL Query Abbreviation} define several
  abbreviations for FOL queries, which make use the abbreviations in
  \Cref{def:abbreviation-query}, as follows:
\[
\begin{array}{lll}
  \qunsatf(\funct{Z},x,y,z) &=& Z(x,y) \land Z(x,z) \land y  \neq z;\\
  \qunsatn(B_1 \ISA \neg B_2, x) &=& B_1(x) \land B_2(x); \\
  \qunsatn(R_1 \ISA \neg R_2,x,y) &=& R_1(x,y) \land R_2(x,y); \\
  \qunsatn(U_1 \ISA \neg U_2,x,y) &=& U_1(x,y) \land U_2(x,y)
\end{array}
\]
 \ \ 
\end{definition}

\noindent
Notice that the queries above can also be used to check whether the
corresponding TBox assertion is violated or not. Next, in the
following we define a boolean query $\qunsatfol{T}$ that can be used
to check the consistency of a KB.


\begin{definition}[Q-UNSAT-FOL]\label{def:qunsat-fol}
  Given \sidetext{Q-UNSAT-FOL} a \dllitea TBox $T$, a \emph{query
    $\qunsatfol{T}$} is a boolean FOL query of the following form:
\[
\qunsatfol{T} =
\begin{array}[t]{l}
  \bigvee_{T \models \funct{Z}} \exists x,y,z.\qunsatf(\funct{Z}, x, y, z)   \vee\\
  \bigvee_{T \models B_1 \ISA \neg B_2} \exists x.\qunsatn(B_1 \ISA \neg B_2, x)  \vee\\
  \bigvee_{T \models R_1 \ISA \neg R_2} \exists x,y.\qunsatn(R_1 \ISA \neg R_2, x, y)  \vee\\
  \bigvee_{T \models U_1 \ISA \neg U_2} \exists x,y.\qunsatn(U_1 \ISA \neg U_2, x, y)  
\end{array}
\]
%
 \ \ 
\end{definition}

\noindent
Later on, 
when we do not want to distinguish between values and objects (thus we
drop attributes), we drop the 
last disjunction of the query above.

\begin{theorem}[FO Rewritability of Satisfiability Check in \dllitea~\cite{CDLLR07}]\label{thm:qunsat}
  Given a KB $\tup{T,A}$, we have $\ANS(\qunsatfol{T}, A) = \true$
  (i.e., $A \models \qunsatfol{T}$) if and only if $A$ is
  $T$-inconsistent.
\qedw
\end{theorem}

\section{History Preserving \texorpdfstring{$\mu$}{^^ce^^bc}-Calculus}\label{sec:mula}


We now shift to reasoning about processes (the dynamic aspect). As a
start, we briefly explain a temporal logic named History preserving
$\mu$-calculus. History preserving $\mu$-calculus is a first order
variant of $\mu$-calculus~\cite{Stir01,Park76}, one of the most
powerful temporal logics, which subsumes LTL, PSL, and
CTL*~\cite{ClGP99}. It was originally introduced
in~\cite{BCDDM13,BCMD*13}. Precisely, the work in~\cite{BCDDM13}
proposes a history preserving $\mu$-calculus called $\mula$ and the
work of~\cite{BCMD*13} proposes a history preserving $\mu$-calculus
called $\muladom$ (Note that the work in~\cite{BCMD*13} originally
call such logic also \mula, we use the name $\muladom$ in order to
differentiate it with the one in \cite{BCDDM13}). The different
between $\mula$ and $\muladom$ formulas is in the atomic parts of the
formulas. The former consider $\difol$ queries as the atomic
components of the formulas while the latter consider $\diecq$ queries.


 \subsection{\texorpdfstring{History Preserving $\mu$-Calculus with
     ECQ-Query (\muladom)}{History Preserving ^^ce^^bc-Calculus with ECQ-Query}}\label{subsec:muladom}


The logic $\muladom$ combines the standard temporal operators of
$\mu$-calculus with $\diecq$ queries over the states. First order
quantification is interpreted with an active domain semantics, i.e.,
it ranges over those constants 
that are explicitly present in the current ABox, and fully interacts
with temporal modalities, i.e., it applies \emph{across} states.  The
\muladom syntax is\sidetext{\ \\ \ \\ Syntax of \muladom}:
\[
\Phi ~:=~ Q ~\mid~ 
\lnot \Phi ~\mid~ 
\Phi_1 \lor \Phi_2 ~\mid~ 
\exists x.\Phi ~\mid~ 
\DIAM{\Phi} ~\mid~ 
Z ~\mid~ \mu Z.\Phi
\]
where $Q$ is a possibly open $\diecq$ query,
$Z$ is a second-order variable
denoting a predicate (of arity 0), and $\mu$ is the least fixpoint
operator, parametrized with the free variables of its bounding
formula. 
Additionally, the following standard abbreviations hold:
\begin{compactitem}
\item $\forall x.  \Phi = \neg (\exists x.\neg \Phi)$,
\item $\Phi_1 \land \Phi_2 = \neg (\neg\Phi_1 \lor \neg \Phi_2)$,
\item $\BOX{\Phi} = \neg \DIAM{\neg \Phi}$, and
\item $\nu Z. \Phi = \lnot\mu Z. \neg \Phi[Z/\neg Z]$.
\end{compactitem}


Given a KB $\tup{T, \initabox}$, we call \emph{a \muladom formula
  $\Phi$ is over $\tup{T, \initabox}$} if each query $Q$ in $\Phi$ is
a $\diecq$ query over $\tup{T, \initabox}$ (i.e., each atom in $Q$ has
either a concept, role or attribute name in $\voc(T)$ as its
predicate, and $Q$ might uses constants 
in $\adom{\initabox}$).

The semantics of \muladom formulae is defined over KB transition
systems defined as follows:

\begin{definition}[KB Transition System] \label{def:kb-ts}
  A \sidetextb{KB Transition System} \emph{KB transition system} $\ts{}$ is a tuple
  $\tup{\const,T,\stateset,s_0,\abox,\trans}$, where:
\begin{compactitem}
\item $\const$ is a countably infinite set of constants; 
\item $T$ is a \dllitea TBox;
\item $\stateset$ is a (possibly infinite) set of states;
\item $s_0 \in \stateset$ is the initial state;
\item $\abox$ is a function that, given a state $s\in \stateset$,
  returns an ABox associated to $s$;
\item $\trans \subseteq \Sigma\times\Sigma$ is a transition relation
  between pairs of states.
\end{compactitem}
\ \ 
\end{definition}

 
Given a transition system $\ts{}$, in order to assign the
meaning to \muladom formulas, the following notions are introduced:
\begin{compactitem}
\item A \emph{individual variable valuation} $\vfo$, i.e., a mapping
  from individual variables $x$~to~$\const{}$.
\item A \emph{predicate variable valuation} $\vso$, i.e., a mapping from the
  predicate variables $Z$ to a subset of $\Sigma$.
\end{compactitem}

\noindent
As for notations, since the individual variable valuation and the predicate
variable valuation are substitutions, here we also use the notation
that is defined in \Cref{sec:notation} as well (e.g., we write
$\vfo[x/c]$ to denote a valuation obtained from $\vfo$ such that
$\vfo[x/c](x) = c$, and $\vfo[x/c](y) = \vfo(y)$ if $y \neq x$, etc).



The \sidetext{Semantics of \muladom} meaning of \muladom formulas are
assigned by associating to $\ts{}$, $\vfo$ and $\vso$ an
\emph{extension function} $\MODA{\cdot}$, which maps \muladom formulas
to subsets of $\Sigma$.
The extension function $\MODA{\cdot}$ is defined inductively as
follows:
\[
  \begin{array}{r@{\ }l@{\ }l@{\ }l}
    \MODA{Q} & = &\{s \in \Sigma\mid \Ans(Q\vfo, T, \abox(s)) = \mathit{true}\}\\
   \MODA{\exists x. \Phi} & =&\{s \in \Sigma \mid\exists d. d \in
   \adom{\abox(s)} \mbox{ and } s \in \MODAX{\Phi}{[x/d]}\}\\
%
%
    \MODA{Z}  & = & V(Z) \subseteq  \Sigma\\
    \MODA{\lnot \Phi} & = & \Sigma - \MODA{\Phi}\\
%
%
    \MODA{\Phi_1 \lor \Phi_2} & = & \MODA{\Phi_1}\cup\MODA{\Phi_2}\\
    \MODA{\DIAM{\Phi}}  & = &  \{ s \in \Sigma \mid\exists s'.\ s
    \Rightarrow s' \mbox{ and } s' \in \MODA{\Phi}\}\\
%
%
    \MODA{\mu Z.\Phi} & = &\bigcap\{ \E\subseteq  \Sigma \mid
    {\MODA{\Phi}}_{[Z/\E]} \subseteq\E \} \\
%
%
 \end{array}
\]
When $\Phi$ is a closed formula, $\MODA{\Phi}$ does not depend on
$\vfo$ or $\vso$, and we denote the extension of $\Phi$ simply by
$\MOD{\Phi}$.  A closed formula $\Phi$ holds in a state $s \in \Sigma$
if $s \in \MOD{\Phi}$.  In this case, we write $\ts{},s \models \Phi$.
A closed formula $\Phi$ holds in $\ts{}$, briefly \emph{$\ts{}$
  satisfies $\Phi$}, 
if $\ts{},s_0\models \Phi$ (In this situation we write
$\ts{} \models \Phi$).
%

 \subsection{\texorpdfstring{History Preserving $\mu$-Calculus with
     FOL-Query (\mula)}{History Preserving ^^ce^^bc-Calculus with FOL-Query}}\label{subsec:mula}


The logic $\mula$ is similar to $\muladom$ except that the atomic
formulas are $\difol$ queries instead of $\diecq$ queries. Moreover,
first order quantification is interpreted with an active domain
semantics, i.e., it ranges over those constants that are explicitly
present in the current database instance. Formally the syntax of
$\mula$ is as follows\sidetext{\ \\ \ \\ Syntax of \mula}:
\[
\Phi ~:=~ Q ~\mid~ 
\lnot \Phi ~\mid~ 
\Phi_1 \lor \Phi_2 ~\mid~ 
\exists x.\Phi ~\mid~ 
\DIAM{\Phi} ~\mid~ 
Z ~\mid~ \mu Z.\Phi
\]
where $Q$ is a possibly open $\difol$ query,
and the others are the same as in $\muladom$.

Given a database instance $\dbinst$ which conforms to a database
schema $\dbschema$, we call \emph{a \mula formula $\Phi$ is over
  $\dbschema$ and $\dbinst$} if each query $Q$ in $\Phi$ is a $\difol$
query over $\dbschema$ and $\dbinst$ (i.e., the predicates of each
atom in $Q$ is a relation schema in $\dbschema$, and $Q$ might uses
constants in $\adom{\dbinst}$).

The semantics of $\mula$ formulae is defined over database transition
systems  defined as follows:
\begin{definition}[Database Transition System]\label{def:database-ts}
  A \sidetextb{Database Transition System} \emph{database transition
    system} $\ts{}$ is a tuple
  $\tup{\const, \dbschema, \dstateset, s_0, \db, \dtrans}$, where:
  \begin{compactitem}
  \item $\const$ is a countably infinite set of constants;
  \item $\R$ is a database schema;
  \item $\dstateset$ is a (possibly infinite) set of states;
  \item $s_0 \in \dstateset$ is the initial state;
  \item $\db$ is a function that, given a state $s \in \dstateset$,
    returns the database associated to $s$, which is made up of constants
    in $\const$ and conforms to $\R$;
  \item $\dtrans \subseteq \dstateset\times \dstateset$ is a
    transition relation between pairs of states.
  \end{compactitem}
\ \ 
\end{definition}


The \sidetext{Semantics of \mula} definition of the extension function
$\MODA{\cdot}$, which is used to assign meaning to \mula formula, is
the same as in \muladom except for the following
\[
  \begin{array}{r@{\ }l@{\ }l@{\ }l}
    \MODA{Q} & = &\{s \in \Sigma\mid \ANS(Q\vfo, \db(s)) = \mathit{true}\}\\
    \MODA{\exists x. \Phi} & =&\{s \in \Sigma \mid\exists d. d \in
                                \adom{\db(s)} 
                                \mbox{ and } s \in \MODAX{\Phi}{[x/d]}\}\\
  \end{array}
\]

The other notions that is defined in $\muladom$ (e.g., $\ts{}$
satisfies $\Phi$) is defined similarly as in $\mula$.


\section{Data Centric Dynamic Systems (DCDS\lowercase{s})}\label{sec:dcds}

Data Centric Dynamic Systems (DCDSs)~\cite{BCDDM13,BCDDM12} provide an
abstract model and formal foundation for various artifact-centric
systems~\cite{NiCa03,Hull08,CoHu09}.  They capture the essence of
systems in which both data and processes are first class citizens, and
thus they provide a holistic view of the system. Furthermore, a DCDS
also captures the manipulation of the data that is done by the
available processes in the system.
%
%
Here, the set $\const$ of constants denotes all possible values in the
system.  
Additionally, we consider a finite set of distinguished constants
$\iconst \subset \const$, and 
%
we also make use of a finite set $\dservcall$ of \emph{function
  symbols} that represent service calls, and can be used to inject
fresh values (constants) into the system.

\subsection{DCDSs Formalism}

Technically, a DCDS consists of:
\begin{inparaenum}[(1)]
\item the \emph{data component} which represents the data of interest
  in the application, and
\item the \emph{process component} which represents the progression
  mechanism for the DCDS.
\end{inparaenum}
In order to formally define the data component, we first introduce
several preliminaries as follows:

\begin{definition}[Equality Constraint (EC)] Given \sidetext{Equality
    Constraint (EC)} a database instance $\dbinst$ which conforms to a
  database schema $\dbschema$, an \emph{equality constraint (EC)}
  $\ec$ over $\dbschema$ and $\dbinst$ is an expression of the form
  \[
  Q(\vec{x}) \ra \bigwedge_{i=1,\dots, n} x_{i}=y_{i},
  \]
  where $Q(\vec{x})$ is a $\difol$ query over $\dbschema$ and
  $\dbinst$, and $x_{i}$ and $y_{i}$ are either a variable in
  $\vec{x}$ or a constant in $\adom{\dbinst}$.
\end{definition}

\begin{definition}[EC Satisfaction]\label{def:ec-satisfaction}
  Given \sidetextb{EC Satisfaction} a database instance $\dbinst$ and
  an equality constraint
  $\ec = Q(\vec{x}) \ra \bigwedge_{i=1,\dots, n} x_{i}=y_{i}$, we say
  \emph{$\dbinst$ satisfies $\ec$} if for each substitution
  $\sigma \in \ANS(Q, \dbinst)$, it holds that
  $x_{i}\sigma=y_{i}\sigma$.
\end{definition}

\noindent
Given a database instance $\dbinst$ which conforms to a database
schema $\dbschema$, and a set $\ecset$ of equality constraints over
$\dbschema$, we say \emph{$\dbinst$ satisfies $\ecset$} if $\dbinst$
satisfies each $\ec \in \ecset$.
Intuitively, given a database instance $\dbinst$ which conforms to a
database schema $\dbschema$, the equality constraints over $\dbinst$
and $\dbschema$ can be used to express some contraints to restrict the
relation instances (facts) that can be contained in $\dbinst$.

Now we define the data component of DCDSs as a tuple of a relational
schema, a finite set of constraints, and an initial database instance
which conforms to the given relational schema.
Formally, it is stated below:

\begin{definition}[DCDS Data Component] A \sidetext{DCDS Data
    Component} \emph{DCDS data component} is a tuple
  $\dcomp = \tup{\dbschema,\idb, \ecset}$ where:
\begin{compactitem}
\item $\dbschema$ is a \emph{database
    schema}, 
\item $\idb$ is an \emph{initial database instance} which conforms to
  the schema $\dbschema$. Intuitively, it represents the initial data
  of the system.
\item $\ecset$ is a finite set 
  of \emph{equality constraints} over $\dbschema$ and $\idb$.
\end{compactitem}
Additionally, we impose that $\idb$ satisfies each equality constraint
$\ec \in \ecset$.
\end{definition}

As uttered before, the process component constitutes the mechanism to
evolve (the data in) the system. Basically, it consists of:
\begin{compactenum}
\item \emph{Actions} which change the data from one state to another
  state, and might also issue \emph{service calls} that introduce new
  values (constants) to the system during their execution (i.e.,
  representing the interaction between the system and external
  user/environment);
\item \emph{Condition-action rules} which specify when and with which
  parameters a certain action can be executed.
\end{compactenum}
Formally, actions and condition-action rules is defined as follows.

\begin{definition}[DCDS Action]\label{def:dcds-act}
  Given \sidetext{DCDS Action} a data component
  $\dcomp = \tup{\dbschema,\idb, \ecset}$, a \emph{DCDS action $\dact$
    over $\dbschema$ and $\idb$} is an expression of the form
  \[
  \dact(p_1,\ldots,p_n): \set{e_1,\ldots,e_m},
\]
where:
\begin{compactitem}
\item $\dact$ is an \emph{action name},
\item the sequence $p_1,\ldots,p_n$ of variables are \emph{action
    parameters}, 


\item $\set{e_1,\ldots,e_m}$ is a set of \emph{DCDS action effects}
  (briefly effects).  Each effect $e_i$ has the form
  \[
  \map{q^+\land Q^-}{E},
  \]
  where:
  \begin{compactitem}
  \item $q^+\land Q^-$ is a $\difol$ query over $\dbschema$ and
    $\adom{\idb}$, that might also includes action parameters
    $\vec{p}$ as its terms and uses constants in
    $\iconst$. Additionally,
    the query $q^+$ is a UCQ, and the query $Q^-$ is an FOL query
    whose free variables are included in those of $q^+$.

  \item $E$ is a set of atoms whose predicates are relation schemas in
    $\dbschema$, which includes as terms:
    \begin{compactitem}
    \item constants in $\iconst$, 
    \item action parameters,
    \item free variables of $q^+$, and
    \item skolem terms $\dscall(\vec{t})$ (representing \emph{a
        service call}) formed by applying a function
      $\dscall\in\dservcall$ to either 
      constants in $\iconst$, 
      action parameters, or free variables of $q^+$.
    \end{compactitem}
  \end{compactitem}
\end{compactitem}
For brevity, depending on the situation, we sometimes also say that
$\dact$ is a DCDS action over $\dcomp$.
\end{definition}

\noindent
As for notation, given an action
$\dact(p_1,\ldots, p_n): \set{e_1,\ldots,e_m}$, we write
$\deff{\dact}$ to denote the set $\set{e_1,\ldots,e_m}$ of effects of
$\dact$. Later on, when it is clear from the context, a DCDS action is
simply called action (e.g., in this section we only consider DCDS
action, thus it is clear that when we say an action, it means that we
refer to a DCDS action). Similarly, for DCDS action effects, for
compactness, when it is clear, we simply call them action effects.

The intuition of the actions definition above is as follows:
\begin{compactenum}
\item Before an action $\dact(p_1,\ldots, p_n): \set{e_1,\ldots,e_m}$
  can be executed, the action parameters $p_1,\ldots, p_n$ need to be
  instantiated with constants from
  $\const$. 
\item Intuitively, in an effect of the form $\map{q^+\land Q^-}{E}$,
  the query $q^+$ selects the tuples to instantiate the atoms in $E$,
  and $Q^-$ filters away some of them.
\item The skolem terms (that might appear) in $E$ represent service
  calls, and during an action execution, they will be substituted with
  constants from $\const$ (representing the results of service call).
\end{compactenum}

\begin{definition}[DCDS Condition-Action Rule]\label{def:dcds-ca-rules}
  Given \sidetext{DCDS Condition-Action Rule} a data component
  $\dcomp = \tup{\dbschema,\idb, \ecset}$, and a set $\dactset$ of
  DCDS actions over $\dcomp$, a \emph{DCDS condition-action rule over
    $\dbschema$, $\idb$, and $\dactset$} is an expression of the form
  \[
  \carule{Q(p_1,\ldots,p_n)}{\dact(p_1,\ldots,p_n)},
  \]
  where
  \begin{compactitem}
  \item $\dact$ is an action $\dact \in \dactset$, and
  \item $Q$ is a $\difol$ query over $\dbschema$ and $\idb$ whose free
    variables are exactly the parameters of $\dact$. Additionally, $Q$
    might also uses constants in $\iconst$. 
  \end{compactitem}
  For brevity, we often also say that the above condition-action rule
  is a DCDS condition-action rule over $\dcomp$ and $\dactset$.
\end{definition}

\noindent
Later on, when it is clear from the context, a DCDS condition-action
rule is simply called condition-action
rule.  
Intuitively, the query $Q$ in the left hand side of
$\carule{Q(p_1,\ldots,p_n)}{\dact(p_1,\ldots,p_n)}$ expresses a
condition which determines when the action $\dact$ can be
executed. More precisely, the action $\dact$ is executable if the
query $Q$ is successfully evaluated. Furthermore, the query $Q$ is
also used to obtain the values (constants) to instantiate the
parameters of $\dact$ (i.e., the answers to query $Q$ is also used to
instantiate the parameters of $\dact$).

Now, we define the process component of DCDSs composed by a set of
actions and a set of condition-action rules.

\begin{definition}[DCDS Process Component] 
  Given \sidetext{DCDS Process Component} a data component
  $\dcomp = \tup{\dbschema,\idb, \ecset}$, a \emph{DCDS process
    component} over $\dcomp$ is a tuple
  $\pcomp = \tup{\dactset,\dprocset}$ where:
\begin{compactitem}
\item $\dactset$ is a finite \emph{set of actions} over $\dcomp$,
\item $\dprocset$ is a finite set of condition-action rules over
  $\dcomp$ and $\dactset$ that form the specification of the
  \emph{DCDS process} (which tells at any moment which actions can be
  executed).
\end{compactitem}
\ \ 
\end{definition}

Having the necessary ingredients
in hand (i.e., data and process component), we are now ready to
formally define a DCDS as follows:

\begin{definition}[Data Centric Dynamic System
  (DCDS)]\label{def:dcds}
  A \sidetext{Data Centric Dynamic System (DCDS)} Data Centric Dynamic
  System (DCDS) $\dcdssym = \tup{\dcomp, \pcomp}$, where
  \begin{compactitem}
  \item $\dcomp = \tup{\dbschema,\idb, \ecset}$ is a DCDS data component
    and
  \item $\pcomp = \tup{\dactset,\dprocset}$ is a DCDS process
    component over $\dcomp$.
  \end{compactitem}
  Additionally, we assume that $\adom{\idb} \subseteq \const_0$.
\end{definition}

\begin{example}\label{ex:dcds}
  Recall our running example scenario in
  \Cref{ex:example-scenario}. Here we will model such order processing
  scenario in a DCDS. In addition to this scenario, in the DCDS that
  we specify below, we do not strictly enforce that the order of the
  processing flow must be followed sequentially. For instance, the
  operation of approving an order might be followed by another order
  approval operation. However, for each specific order, we enforce
  that it is processed sequentially according to the order processing
  flow described in \Cref{ex:example-scenario} (for example, before
  the company delivers an order, the company must performs a quality
  check operation). We now specify a DCDS
  $\dcdssym = \tup{\dcomp, \pcomp}$ where the data component
  $\dcomp = \tup{\dbschema,\idb, \ecset}$ and the process component
  $\pcomp = \tup{\dactset,\dprocset}$ of $\dcdssym$ are specified
  below.

  As for the database schema, we utilize the database schema
  $\dbschema$ that is specified in \Cref{ex:dbschema-and-dbinst}. The
  initial database is specified as follows
  \[
  \begin{array}{@{}r@{}l@{}}
  \idb = \set{&\exr{ORDER}(\excon{123}, \excon{chair}, \excon{received},
                   \excon{456}, \excon{NULL}, \excon{NULL}, \excon{NULL}, \excon{NULL},
                   \excon{ecodesign}), \\
                 &\exr{ORDER}(\excon{321}, \excon{table}, \excon{approved},
                   \excon{654}, \excon{NULL}, \excon{NULL}, \excon{NULL}, \excon{NULL},
                   \excon{NULL})
                   },
  \end{array}
  \]
  and we consider empty equality constraint $\ecset = \set{}$.

  For the process component, we specify the set $\dactset$ of actions
  containing the following actions:
\begin{enumerate}

\item $\exa{approveOrder}(\exvar{x})$, which intuitively changes the
  status of an order with id $\exvar{x}$ into ``approved''. Formally
  it is
  specified as follows:
\[
\begin{array}{l}
  \exa{approveOrder}(\exvar{x}):\{ \\
  \hspace*{3mm}\begin{array}{@{}r@{}l@{}l@{}} 
                 \exists
                 \exvar{x_3}. \exr{ORDER}(&\exvar{x_1}, \ldots &
                                                                , \exvar{x_9}) \wedge \exvar{x_1} =
                                                                \exvar{x} \rightsquigarrow \{ \\
                                          &&\exr{ORDER}(\exvar{x_1}, \exvar{x_2},
                                             \excon{"approved"},\exvar{x_4},\exvar{x_5},\\
                                            &&\hspace*{15mm} \excon{NULL},\excon{NULL},\exvar{x_8},\exvar{x_9}) \\
                                          &\}, \ &
 \end{array} \\
  \hspace*{3mm}\begin{array}{@{}r@{}r@{}l@{}} 
                 \exr{ORDER}(\exvar{x_1}&,\ldots,
                                          \exvar{x_9}) \wedge& \exvar{x_1} \neq \exvar{x} \rightsquigarrow
                                                               \{  \exr{ORDER}(\exvar{x_1},\ldots, \exvar{x_9})\},
   \end{array} \\
  \hspace*{3mm}\begin{array}{@{}r@{}r@{}l@{}} 
                 \exr{DELIVERED\_ORDER}(\exvar{x_1},
                 \exvar{x_2}) \rightsquigarrow
                 \{  \exr{DELIVERED\_ORDER}(\exvar{x_1}, \exvar{x_2})\}
   \end{array} \\
\}
\end{array}
\]
Technically, the action above changes the status of the order with id
$\exvar{x}$ into $\excon{"approved"}$ while keeping the other database
entries  stay the same.

\item $\exa{prepareOrders}()$, which intuitively prepares several
  things that are needed for further processing steps of each approved
  order. For each approved order, this action prepares the design of
  the corresponding order by calling an external service
  $\exs{getDesign}(\exvar{x})$. Moreover, this action also retrieves
  the information about the corresponding designer by calling a
  service $\exs{getDesigner}(\exvar{x})$, and assign the assembling
  location for the corresponding order by calling a service
  $\exs{assignAssemblingLoc}(\exvar{x})$.  Formally it is specified as
  follows:
\[
\begin{array}{l}
  \exa{prepareOrders}():\{ \\
  \hspace*{3mm}\begin{array}{@{}r@{}l@{}l@{}l@{}} 
                 \exists \exvar{x_5}\exvar{x_8}&\exvar{x_9}.& \exr{ORDER}(&\exvar{x_1},\exvar{x_2},
                                                              \excon{"approved"},\exvar{x_4}, \ldots ,\exvar{x_9}) \rightsquigarrow \{ \\
                                               &&\exr{ORDER}(&\exvar{x_1}, \exvar{x_2}, \excon{"approved"},\exvar{x_4},
                                                               \exs{getDesigner}(\exvar{x}),
                                                               \exvar{x_6}, \exvar{x_7}, \\
                                               &&&\exs{assignAssemblingLoc}(\exvar{x}), 
                                                   \exs{getDesign}(\exvar{x})) \\
                                               &\},&&
 \end{array} \\
  \hspace*{3mm}\begin{array}{@{}r@{}r@{}l@{}} 
                 \exr{ORDER}(\exvar{x_1}&,\ldots,
                                          \exvar{x_9}) \wedge& \exvar{x_3} \neq \excon{"approved"} \rightsquigarrow
                                                               \{  \exr{ORDER}(\exvar{x_1},\ldots, \exvar{x_9})\},
   \end{array} \\
  \hspace*{3mm}\begin{array}{@{}r@{}r@{}l@{}} 
                 \exr{DELIVERED\_ORDER}(\exvar{x_1},
                 \exvar{x_2}) \rightsquigarrow
                 \{  \exr{DELIVERED\_ORDER}(\exvar{x_1}, \exvar{x_2})\}
   \end{array} \\
\}
\end{array}
\]

\item $\exa{assembleOrders}()$, which represents the step of
  assembling several components into the corresponding ordered
  furniture. This action also acquires the information about the
  assembler and the assembling location by calling external service
  calls $\exs{getAssembler}(\exvar{x})$ and
  $\exs{getAssemblingLoc}(\exvar{x})$ respectively. Additionally, this
  action only assembles every approved order that already has a
  design. Formally this action is specified as follows:
\[
\begin{array}{l}
  \exa{assembleOrders}():\{ \\
  \hspace*{3mm}\begin{array}{@{}r@{}l@{}} 
                 \exists \exvar{x_6}\exvar{x_8}.\exr{ORDER}&(\exvar{x_1},\exvar{x_2},
                                                  \excon{"approved"},\exvar{x_4},\exvar{x_5},\exvar{x_6},\exvar{x_7},\exvar{x_8},
                                                  \exvar{x_9}) \wedge \exvar{x_9}
                                                  \neq \excon{NULL}
                                                  \rightsquigarrow \{ \\
                                                &\exr{ORDER}(\exvar{x_1}, \exvar{x_2}, \excon{"assembled"},\exvar{x_4},
                                                                                                                    \exvar{x_5}, \exs{getAssembler}(\exvar{x}), \\
                                                &\hspace*{15mm}\exvar{x_7}, \exs{getAssemblingLoc}(\exvar{x}), 
                                                    \exvar{x_9} ) \\
                                                \},&
 \end{array} \\
  \hspace*{3mm}\begin{array}{@{}r@{}r@{}l@{}} 
    \exr{ORDER}(\exvar{x_1}&,\ldots,
                              \exvar{x_9}) \wedge& \exvar{x_3} \neq \excon{"approved"} \rightsquigarrow
                              \{  \exr{ORDER}(\exvar{x_1},\ldots, \exvar{x_9})\}
               \end{array} \\
  \hspace*{1.3mm}\begin{array}{r@{}l} 
                 \exr{ORDER}(\exvar{x_1},\exvar{x_2},\excon{"approved"}&,\exvar{x_4},\ldots,
                 \exvar{x_9}) \wedge \exvar{x_9} = \excon{NULL}
                 \rightsquigarrow \\
                 &\{  \exr{ORDER}(\exvar{x_1},\exvar{x_2},\excon{"approved"},\exvar{x_4},\ldots, \exvar{x_9})\},
   \end{array} \\
  \hspace*{3mm}\begin{array}{@{}r@{}r@{}l@{}} 
                 \exr{DELIVERED\_ORDER}(\exvar{x_1},
                 \exvar{x_2}) \rightsquigarrow
                 \{  \exr{DELIVERED\_ORDER}(\exvar{x_1}, \exvar{x_2})\}
   \end{array} \\
\}
\end{array}
\]

\item $\exa{checkAssembledOrders}()$, which models the quality check
  process for each assembled order. This action calls an external
  service $\exs{getQualityController}(\exvar{x})$ in order to obtain
  the quality controller officer who performs the task. Formally it is
  specified as follows:
\[
\begin{array}{l}
  \exa{checkAssembledOrders}():\{ \\
  \hspace*{3mm}\begin{array}{@{}r@{}r@{}r@{}l@{}} 
                 \exr{ORDER}(&\exvar{x_1},\exvar{x_2},
                 &\excon{"assembled"}&,\exvar{x_4},\exvar{x_5},\exvar{x_6},\excon{NULL},\exvar{x_8},
                                       \exvar{x_9})  \rightsquigarrow \{ \\
                             &&\exr{ORDER}(&\exvar{x_1}, \exvar{x_2},\excon{"assembled"},\exvar{x_4},
                                             \exvar{x_5}, \exvar{x_6}, \\
                             &&& \exs{getQualityController}(\exvar{x}),\exvar{x_8}, 
                                 \exvar{x_9} ) \} 
 \end{array} \\
  \hspace*{3mm}\begin{array}{@{}r@{}r@{}r@{}l@{}} 
                 \exr{ORDER}(&\exvar{x_1},\exvar{x_2},
                 &\excon{"assembled"}&,\exvar{x_4},\exvar{x_5},\exvar{x_6},\exvar{x_7},\exvar{x_8},
                                       \exvar{x_9}) \wedge \exvar{x_7}
                                       \neq \excon{NULL}\rightsquigarrow \{ \\
                             &&\exr{ORDER}(&\exvar{x_1}, \exvar{x_2},\excon{"assembled"},\exvar{x_4},
                                             \exvar{x_5}, \exvar{x_6},
                                             \exvar{x_7}, \exvar{x_8}, 
                                             \exvar{x_9} )  \} 
 \end{array} \\
  \hspace*{3mm}\begin{array}{@{}r@{}r@{}l@{}} 
    \exr{ORDER}(\exvar{x_1}&,\ldots,
                              \exvar{x_9}) \wedge& \exvar{x_3} \neq \excon{"assembled"} \rightsquigarrow
                              \{  \exr{ORDER}(\exvar{x_1},\ldots, \exvar{x_9})\},
   \end{array} \\
  \hspace*{3mm}\begin{array}{@{}r@{}r@{}l@{}} 
                 \exr{DELIVERED\_ORDER}(\exvar{x_1},
                 \exvar{x_2}) \rightsquigarrow
                 \{  \exr{DELIVERED\_ORDER}(\exvar{x_1}, \exvar{x_2})\}
   \end{array} \\
\}
\end{array}
\]

\item $\exa{deliverOrders}()$, which delivers each assembled order that
  has passed the quality control process. 
  Formally this action is specified as follows:
\[
\begin{array}{l}
  \exa{deliverOrders}():\{ \\
  \hspace*{3mm}\begin{array}{@{}r@{}r@{}r@{}l@{}} 
                 \exr{ORDER}(&\exvar{x_1},\exvar{x_2},
                 &\excon{"assembled"}&,\exvar{x_4},\exvar{x_5},\exvar{x_6},\excon{NULL},\exvar{x_8},
                                       \exvar{x_9})  \rightsquigarrow \{ \\
                             &&\exr{ORDER}(&\exvar{x_1}, \exvar{x_2},\excon{"assembled"},\exvar{x_4},
                                             \exvar{x_5}, \exvar{x_6},
                                             \excon{NULL},\exvar{x_8}, 
                                             \exvar{x_9} ) \\
                             &\},&&
 \end{array} \\
  \hspace*{3mm}\begin{array}{@{}r@{}r@{}r@{}l@{}} 
                 \exr{ORDER}(&\exvar{x_1},\exvar{x_2},
                 &\excon{"assembled"}&,\exvar{x_4},\exvar{x_5},\exvar{x_6},\exvar{x_7},\exvar{x_8},
                                       \exvar{x_9}) \wedge \exvar{x_7}
                                       \neq \excon{NULL}\rightsquigarrow \{ \\
                             &&\exr{ORDER}(&\exvar{x_1}, \exvar{x_2},\excon{"delivered"},\exvar{x_4},
                                             \exvar{x_5}, \exvar{x_6},
                                             \exvar{x_7}, \exvar{x_8}, 
                                             \exvar{x_9} ), \\
                             &&\exr{DELIVER}&\exr{ED\_ORDER}(\exvar{x_1}, \exs{getDeliveryDate}(\exvar{x_1}) ), \\
                             &\},&&
 \end{array} \\
  \hspace*{3mm}\begin{array}{@{}r@{}r@{}l@{}} 
    \exr{ORDER}(\exvar{x_1}&,\ldots,
                              \exvar{x_9}) \wedge& \exvar{x_3} \neq \excon{"assembled"} \rightsquigarrow
                              \{  \exr{ORDER}(\exvar{x_1},\ldots, \exvar{x_9})\},
   \end{array} \\
  \hspace*{3mm}\begin{array}{@{}r@{}r@{}l@{}} 
                 \exr{DELIVERED\_ORDER}(\exvar{x_1},
                 \exvar{x_2}) \rightsquigarrow
                 \{  \exr{DELIVERED\_ORDER}(\exvar{x_1}, \exvar{x_2})\}
   \end{array} \\
\}
\end{array}
\]
\end{enumerate}

\noindent
Furthermore, the set $\dprocset$ of condition action rules is specified as follows:\\
$\begin{array}{@{}l@{}l@{}} \bullet \ \carule{\exists
    \exvar{x_2}\exvar{x_4}\exvar{x_5}\exvar{x_6}\exvar{x_7}\exvar{x_8}
    \exvar{x_9}.\exr{ORDER}(\exvar{x_1},\exvar{x_2},
    \excon{"received"},\exvar{x_4},\ldots,\exvar{x_9})}
  {\exa{approveOrder}(\exvar{x_1})},
 \end{array}
 $ \\
$\begin{array}{@{}l@{}l@{}} \bullet \ \carule{\exists
    \exvar{x_1}\exvar{x_2}\exvar{x_4}\exvar{x_5}\exvar{x_6}\exvar{x_7}\exvar{x_8}
    \exvar{x_9}.\exr{ORDER}(\exvar{x_1},\exvar{x_2},
    \excon{"approved"},\exvar{x_4},\ldots,\exvar{x_9})}
  {\exa{prepareOrders}()},
 \end{array}
 $\\
$\begin{array}{@{}l@{}l@{}} \bullet \ \carule{\exists
    \exvar{x_1}\exvar{x_2}\exvar{x_4}\exvar{x_5}\exvar{x_6}\exvar{x_7}\exvar{x_8}
    \exvar{x_9}.\exr{ORDER}(\exvar{x_1},\exvar{x_2},
    \excon{"approved"},\exvar{x_4},\ldots,&\exvar{x_9}) \wedge
   \exvar{x_9} \neq \excon{NULL} \\ & \ }
  {\exa{assembleOrders}()},
 \end{array}
 $\\
 $\begin{array}{@{}l@{}l@{}} \bullet \ \carule{\exists
     \exvar{x_1}\exvar{x_2}\exvar{x_4}\exvar{x_5}\exvar{x_6}\exvar{x_8}
     \exvar{x_9}.\exr{ORDER}(\exvar{x_1},\exvar{x_2},
     \excon{"assembled"},\exvar{x_4},&\exvar{x_5},\exvar{x_6},\excon{NULL},\exvar{x_8},\exvar{x_9})
    \\ & \ }
   {\exa{checkAssembledOrders}()},
 \end{array}
 $\\
 $\begin{array}{@{}l@{}l@{}} \bullet \ \carule{\exists
     \exvar{x_1}\exvar{x_2}\exvar{x_4}\exvar{x_5}\exvar{x_6}\exvar{x_7}\exvar{x_8}
     \exvar{x_9}.\exr{ORDER}(\exvar{x_1},\exvar{x_2},
     \excon{"assembled"}&,\exvar{x_4},\exvar{x_5},\exvar{x_6},\exvar{x_7},\exvar{x_8},\exvar{x_9})
    \\ &\wedge\ \exvar{x_7} \neq \excon{NULL}}
   {\exa{deliverOrders}()},
 \end{array}
 $\\
 The intuition for each condition action rules above is consecutively
 presented below:
\begin{compactitem}

\item the first rule states that if there exists an order with the
  status ``received'' and has an ID $\exvar{x_1}$, then we can fire
  the execution of action $\exa{approveOrder}/1$ with the argument
  $\exvar{x_1}$ (i.e., $\exa{approveOrder}(\exvar{x_1})$).

\item The second rule says that we can execute the action
  $\exa{prepareOrders}/0$ in case there exists at least one approved order.

\item Next, the third rule encodes the condition where the execution
  of action $\exa{assembleOrders}/0$ can be fired in case there exists
  an approved order that already has a design.

\item The fourth rule indicates that if there exists an assembled
  order that has not been checked, we can execute the action
  $\exa{checkAssembledOrders}/0$

\item Finally, the last rule specifies that whenever there exists an
  assembled order that has been checked, we can execute the action $\exa{deliverOrders}/0$.

\end{compactitem}
\end{example}

\subsection{DCDSs Execution Semantics}
The semantic of DCDS is defined in terms of a possibly infinite
transition system whose states are labeled by databases and where
transitions represent the execution of actions. Such transition system
represents all possible computations that the process component can do
on the data component starting from the initial database instance
(i.e., all possible manipulations of data by actions).

During the execution, an action can issue service calls.  In
\cite{BCDDM13}, there are two kinds of service calls semantics that
are considered, namely \emph{deterministic} and
\emph{non-deterministic} service calls semantics. In the deterministic
service calls semantics, along a run of the system, whenever a service
call is issued with the same input parameters, it will return the same
value (constant). On the other hand, in the non-deterministic service
calls semantics, along a run of the system, two different issues of a
service call with the same input parameters might return distinct
results.
Here, in this thesis, we assume that the semantics of service calls is
deterministic.
%

To enforce the deterministic service calls semantics, the transition
systems of a DCDS remembers the results of previous service calls in a
so-called service call map defined as
follows.  

\begin{definition}[Service Call Map]\label{def:service-call-map}
  A \sidetext{Service Call Map} \emph{service call map} is a partial
  function
  \[
  \dscmap:\dscset\ra\const,
  \]
  where $\dscset$ is the set
  $\{f(v_1,\ldots,v_n) \mid f/n \in \dservcall \textrm{ and }
  \{v_1,\ldots,v_n\} \subseteq \const \}$
  of (skolem terms representing) \emph{service calls}.
\end{definition}
\noindent

Technically, to provide the semantics of DCDSs, we consider database
transition systems (as defined in \Cref{def:database-ts}), i.e.,
transition systems of the form
$\tup{\const, \dbschema, \dstateset, s_0, \db, \dtrans}$, where:
\begin{compactitem}
\item $\R$ is a database schema;
\item $\dstateset$ is a set of states;
\item $s_0 \in \dstateset$ is the initial state;
\item $\db$ is a function that, given a state $s \in \dstateset$,
  returns the database associated to $s$, which is made up of
  constants in $\const$ and conforms to $\R$;
\item $\dtrans \subseteq \dstateset\times \dstateset$ is a transition
  relation between pairs of states.
\end{compactitem}
%
In addition, to realize DCDSs with deterministic service calls
semantics, each state $s \in \stateset$ of the transition system
is defined as a tuple $\tup{\dbinst, \dscmap}$, where $\dbinst$ is a
database instance and $\dscmap$ is a service call map.
%
As for notations related to the service call map, we use similar
notations that is defined for substitutions in 
\Cref{sec:notation} (e.g., we write $f(c)/v \in \dscmap$ to say that a
service call map $\dscmap$ maps $f(c)$ to $v$).

In \sidetext{Action Execution Semantics in DCDS} order to define the
semantics of an \emph{action execution} in DCDS, below we define the
notion of
when an action can be executed, and
how the results of an action execution is constructed.

\begin{definition}[Executability of an Action in DCDS]\label{def:DCDS-act-executability}
  Let \sidetextb{Executability of an Action in DCDS}
  $\dcdssym = \tup{\dcomp, \pcomp}$ be a DCDS
  where $\pcomp = \tup{\dactset,\dprocset}$. Given a database instance
  $\dbinst$, an action $\dact \in \dactset$ of the form
  $\dact(\vec{p}): \set{e_1,\ldots,e_m}$,
  and a \emph{parameter substitution} $\sigma$ which substitute the
  parameters $\vec{p}$ with constants taken from $\const$.
  We say that \emph{$\dact$ is executable in $\dbinst$ with a
    parameter substitution $\sigma$}, if there exists a
  condition-action rule
  $Q(\vec{p})\mapsto\alpha(\vec{p}) \in \dprocset$ such that
  $\sigma \in \ANS(Q, \dbinst)$.  \ \ \ \ \
\end{definition}


\noindent
With a little abuse of the definition, we sometimes say that an action
\emph{$\dact$ is executable in a state $s$ with a parameter
  substitution $\sigma$} if $s = \tup{\dbinst, \dscmap}$ and $\dact$
is executable in $\dbinst$ with a parameter substitution $\sigma$.
Additionally, given an action
$\dact(p_1,\ldots, p_n): \set{e_1,\ldots,e_m}$, and a database
instance $\dbinst$, 
we say that $\sigma$ is \emph{a legal parameter assignment} for
$\dact$ in $\dbinst$ if $\dact$ is executable in $\dbinst$ with a
parameter substitution $\sigma$. Moreover, 
we write $\dact\sigma$ to denote a \emph{grounded action} that is
obtained by applying a legal parameter assignment $\sigma$ to each
$e \in \deff{\dact}$ (i.e., substituting each occurrence of $p_i$ (for
$i \in \set{1,\ldots,n}$) in $e$ with a constant in $\const$ based on
the substitution $\sigma$).

\begin{example}\label{ex:dcds-act-exec}
  Continuing our running example in \Cref{ex:dcds}, we have that the
  action $\exa{approveOrder/1}$ is executable in the state
  $s_0 = \tup{\idb, \emptyset}$ with a parameter substitution
  $\sigma$, where $\sigma$ is a substitution that substitutes the
  action parameter of $\exa{approveOrder/1}$ to $\excon{123}$. This is
  the case because we have that the query in the left hand side of the
  condition-action rule
  \[
  \begin{array}{@{}l@{}l@{}} \carule{\exists
      \exvar{x_2}\exvar{x_4}\exvar{x_5}\exvar{x_6}\exvar{x_7}\exvar{x_8}
      \exvar{x_9}.\exr{ORDER}(\exvar{x_1},\exvar{x_2},
      \excon{"received"},\exvar{x_4},\ldots,\exvar{x_9})}
    {\exa{approveOrder}(\exvar{x_1})},
  \end{array}
  \]
  is successfully evaluated over $\idb$ and give an answer
  $\excon{123}$ (i.e., $\sigma(x_1) = \excon{123}$).
%
%
  In this case, we have that $\sigma$ is the legal parameter assignment
  for $\act$ in $\idb$.
\end{example}

The execution result of a grounded action $\dact\sigma$ 
%
is captured by a function $\doo{\dbinst, \act\sigma}$ which is
formally defined as follows:

\begin{definition}[Computation of DCDS Action Execution Result]\label{def:do-dcds-action}
  Let \sidetext{Computation of DCDS Action Execution Result}
  $\dcdssym = \tup{\dcomp, \pcomp}$ be a DCDS where
  $\pcomp = \tup{\dactset,\dprocset}$.
%
  Given a database instance $\dbinst$, an action $\dact \in \dactset$,
  and a legal parameter assignment $\sigma$ for $\dact$ in $\dbinst$.
  The \emph{execution result of $\act\sigma$ in $\dbinst$} is computed
  by function $\ddoo{\dbinst, \act\sigma}$ as follows:
  \[
  \ddoo{\dbinst, \dact\sigma}= \left(\bigcup_{e_i = \map{q_i^+\land
        Q_i^-}{E} \text{ in } \deff{\dact}}
    \bigcup_{\rho\in\ANS(([q_i^+]\land Q_i^-)\sigma,\dbinst)}
    E\sigma\rho \right)
  \]
\ \ 
\end{definition}

\noindent
Intuitively, the execution result of $\act$ is obtained by collecting
all facts in $E\sigma\rho$ of each effect $\map{q^+\land Q^-}{E}$ in
$\deff{\dact}$, where the set $E\sigma\rho$ of facts of
$\map{q^+\land Q^-}{E}$ is obtained by substituting each variable in
each atom in $E$ with a constant in $\const$ based on the answers of
the query $q^+\land Q^-$ (i.e., substitution $\rho$) and the legal
parameter assignment $\sigma$.

Note that there might be some facts in $\ddoo{\dbinst, \act\sigma}$
that contain (ground) skolem terms. Intuitively, it means that some
service calls have to be issued in order to replace it with some
values (constants) and get a proper database instance as the result of
an action execution.  We \sidetext{Service Call Evaluation} denote by
$\dcalls{\ddoo{\dbinst, \act\sigma}}$ the set of such ground service
calls (ground skolem terms), and by $\deval{\dbinst,\act\sigma}$ the
set of substitutions that replace such calls with concrete constants
taken from $\const$. Specifically, $\deval{\dbinst,\act\sigma}$ is
defined as
\[
  \deval{\dbinst,\act\sigma}= \{\theta \mid \theta \mbox{ is a total function }
                                           \theta: \dcalls{\doo{\dbinst, \act\sigma}} \ra \const \}.
\]
As for notations related to the substitution
$\theta \in \deval{\dbinst,\act\sigma}$, we use similar notations,
that is defined in \Cref{sec:notation} (e.g., we write
$f(c)/v \in \theta$ to say that $\theta$ maps $f(c)$ to $v$).


Having the semantics of an action execution in place, given a DCDS
$\dcdssym = \tup{\dcomp, \pcomp}$, we employ $\ddoo{}$ and $\deval{}$
to define a transition relation $\exec{\dcdssym}$ connecting two
states through an action execution as follows.

\begin{definition}[DCDS Transition Relation $\exec{\dcdssym}$]\label{def:dcds-trans-rel}
  Let \sidetext{DCDS Transition Relation}
  $\dcdssym = \tup{\dcomp, \pcomp}$ be a DCDS
  where $\pcomp = \tup{\dactset,\dprocset}$. Given a state
  $s = \tup{\dbinst, \dscmap}$, a state
  $s' = \tup{\dbinst', \dscmap'}$, an action $\dact \in \dactset$, and
  a substitution $\sigma$. We have
  $\tup{\tup{\dbinst, \dscmap},\act\sigma, \tup{\dbinst', \dscmap'}}
  \in \exec{\dcdssym}$ if the following holds:
  \begin{compactenum}
  \item $\act$ is \emph{executable} in $s$ with legal parameter
    assignment $\sigma$;
  \item there exists $\theta \in \deval{\dbinst,\act\sigma}$ such that
    for each skolem term
    $\dscall(c) \in \domain{\dscmap} \cap \domain{\theta}$, we have
    $\dscall(c)/v \in \dscmap$ if and only if
    $\dscall(c)/v \in \theta$ (i.e., $\theta$ and $\dscmap$ ``agree''
    on the common skolem terms in their domains, in order to realize
    the deterministic service call semantics);

  \item $\dbinst' = \ddoo{\dbinst, \act\sigma}\theta$;
  \item $\dscmap' = \dscmap \cup \theta$ (i.e., updating the history
    of issued service calls).
  \end{compactenum}
  \ \
\end{definition}

\noindent
When
$\tup{\tup{\dbinst, \dscmap},\act\sigma, \tup{\dbinst', \dscmap'}} \in
\exec{\dcdssym}$, we equivalently write
\[
\tup{\dbinst, \dscmap} \dexect{\act\sigma,\ \dcdssym\ } \tup{\dbinst',
  \dscmap'}.
\]
for easiness of reading. When it is clear from the context, we also
often omit $\dcdssym$ and simply write
$\tup{\dbinst, \dscmap} \dexect{\act\sigma} \tup{\dbinst', \dscmap'}$.


The transition system $\ts{\dcdssym}$ of DCDS $\dcdssym$, which
provide the execution semantics of $\dcdssym$, is then formally
defined as follows:

\begin{definition}[DCDSs Transition System] \label{def:dcds-ts} 
  Given \sidetext{DCDSs Transition System}
  a DCDS $\dcdssym = \tup{\dcomp, \pcomp}$ with
  $\dcomp = \tup{\dbschema,\idb, \ecset}$, and
  $\pcomp = \tup{\dactset,\dprocset}$, the transition system
  $\ts{\dcdssym}$ is defined as
  $\tup{\const, \dbschema, \dstateset, s_0,\db,\dtrans}$ where
\begin{compactitem}
\item $s_0 = \tup{\idb,\emptyset}$, and
\item $\dstateset$ and $\dtrans$ are defined by simultaneous induction
  as the smallest sets satisfying the following properties:
  \begin{compactenum}
  \item $s_0 \in \dstateset$;
  \item if $\tup{\dbinst,\dscmap} \in \dstateset$, then for all
    actions $\dact \in \dactset$, for all legal parameter assignments
    $\sigma$ for $\dact$ in $\dbinst$ and
    for all $\tup{\dbinst',\dscmap'}$ such that
    \begin{compactenum}
    \item
      $\tup{\dbinst,\dscmap} \exect{\dact\sigma, \ \dcdssym\ }
      \tup{\dbinst',\dscmap'}$, and
    \item $\dbinst'$ satisfies $\ecset$,
    \end{compactenum}
    we have $\tup{\dbinst',\scmap'}\in \stateset$, and
    $\tup{\dbinst,\scmap}\trans \tup{\dbinst',\scmap'}$.
  \end{compactenum}
\end{compactitem}
\ \ 
\end{definition}
 
\noindent
Roughly speaking, the transition system of a DCDS, which provides its
execution semantics, is obtained by nondeterministically applying
every executable action starting from the initial database with
corresponding legal parameter assignments, and considering each
possible value (constant) returned by applying the involved service
calls.
Additionally, we restrict that an action with certain parameters is
executable if the database instance produced by its execution
satisfies the given equality constraints. Otherwise the action is
considered as non executable with the chosen parameters.

\begin{example}\label{ex:dcds-exec}

  Continuing our running example in \Cref{ex:dcds}, the construction
  of transition system $\ts{\dcdssym}$ of DCDS $\dcdssym$ is started
  from the initial state $s_0 = \tup{\idb, \emptyset}$ where 
  \[
  \begin{array}{@{}r@{}l@{}}
  \idb = \set{&\exr{ORDER}(\excon{123}, \excon{chair}, \excon{received},
                   \excon{456}, \excon{NULL}, \excon{NULL}, \excon{NULL}, \excon{NULL},
                   \excon{ecodesign}), \\
                 &\exr{ORDER}(\excon{321}, \excon{table}, \excon{approved},
                   \excon{654}, \excon{NULL}, \excon{NULL}, \excon{NULL}, \excon{NULL},
                   \excon{NULL})
                   }.
  \end{array}
  \]
  An example of a sucessor of state $s_0$ is a state
  $s_1 = \tup{\dbinst_1, \scmap_1}$, where
  \[
  \begin{array}{@{}r@{}l@{}}
    \dbinst_1 = \set{&\exr{ORDER}(\excon{123}, \excon{chair}, \excon{approved},
                       \excon{456}, \excon{NULL}, \excon{NULL}, \excon{NULL}, \excon{NULL},
                       \excon{ecodesign}), \\
                     &\exr{ORDER}(\excon{321}, \excon{table}, \excon{approved},
                       \excon{654}, \excon{NULL}, \excon{NULL}, \excon{NULL}, \excon{NULL},
                       \excon{NULL})
                       },\\
    \scmap_1 =\ \ &\emptyset
  \end{array}
  \]
  and $s_1$ is obtained from the execution of action
  $\exa{approveOrder/1}$ with the argument $\excon{123}$. The action
  $\exa{approveOrder/1}$ is executable with argument $\excon{123}$ in
  the state $s_0$ since we have that the query in the left hand side
  of the condition-action rule
  \[
  \begin{array}{@{}l@{}l@{}} \carule{\exists
      \exvar{x_2}\exvar{x_4}\exvar{x_5}\exvar{x_6}\exvar{x_7}\exvar{x_8}
      \exvar{x_9}.\exr{ORDER}(\exvar{x_1},\exvar{x_2},
      \excon{"received"},\exvar{x_4},\ldots,\exvar{x_9})}
    {\exa{approveOrder}(\exvar{x_1})},
  \end{array}
  \]
  is successfully evaluated and give an answer $\excon{123}$. As the
  result of executing this action, we have that now the status of
  order $\excon{123}$ become $\excon{approved}$.

  \medskip
  \noindent
  Another example of a sucessor of state $s_0$ is a state
  $s_2 = \tup{\dbinst_2, \scmap_2}$, where
  \[
  \begin{array}{@{}r@{}l@{}}
    \dbinst_2 = \set{&\exr{ORDER}(\excon{123}, \excon{chair}, \excon{received},
                       \excon{456}, \excon{NULL}, \excon{NULL}, \excon{NULL}, \excon{NULL},
                       \excon{ecodesign}), \\
                     &\exr{ORDER}(\excon{321}, \excon{table}, \excon{approved},
                       \excon{654}, \excon{bob}, \excon{NULL}, \excon{NULL}, \excon{bolzano},
                       \excon{classicdesign})
                       },\\
    \scmap_2 = \set{ &[\exs{getDesigner}(\excon{321})) \ra
                       \excon{bob}], [\exs{assignAssemblingLoc}(\excon{321})) \ra
                       \excon{bolzano}], \\
                     &    [\exs{getDesign}(\excon{321})) \ra
                       \excon{classicdesign}] }, 
  \end{array}
  \]
  and $s_2$ is obtained from the execution of action
  $\exa{prepareOrders/0}$. The action $\exa{prepareOrders/0}$ is
  executable in the state $s_0$ since we have that the query in the
  left hand side of the condition-action rule
  \[
  \begin{array}{@{}l@{}l@{}} \carule{\exists
      \exvar{x_1}\exvar{x_2}\exvar{x_4}\exvar{x_5}\exvar{x_6}\exvar{x_7}\exvar{x_8}
      \exvar{x_9}.\exr{ORDER}(\exvar{x_1},\exvar{x_2},
      \excon{"approved"},\exvar{x_4},\ldots,\exvar{x_9})}
    {\exa{prepareOrders}(\exvar{})},
  \end{array}
  \]
  is successfully evaluated. 
\end{example}

\subsection{Verification of DCDSs}
The interesting reasoning task in DCDSs is to verify whether the
transition system of a given DCDS satisfies temporal properties of
interest, specified in some first-order temporal logic. To specify the
temporal properties, the work in~\cite{BCDDM13} uses the temporal
logic history preserving $\mu$-calculus (\mula) (see \Cref{sec:mula}).
The verification problem of \mula properties over DCDSs is then
formally stated as follows:

\begin{definition}[Verification of a \mula property over a DCDS]
  Given \sidetext{Verification of a \mula Property Over a DCDS} a DCDS
  $\dcdssym$ $= \tup{\dcomp, \pcomp}$ (with
  $\dcomp = \tup{\dbschema,\idb, \ecset}$) 
%
  and a closed \mula formula $\Phi$ over $\dbschema$ and $\idb$,
  \emph{the verification of a \mula property $\Phi$ over $\dcdssym$}
  is a problem to check whether $\ts{\dcdssym} \models \Phi$.
%
\end{definition}

We also say a DCDS \emph{$\dcdssym$ satisfies a closed \mula formula
  $\Phi$} if $\ts{\dcdssym} \models \Phi$.  As studied in
\cite{BCDDM13}, it has been shown that in general the verification of
\mula over DCDS is undecidable. However, the work in~\cite{BCDDM13}
has identified some restrictions to get the decidability. Here we
briefly explain a semantic restriction that was introduced in
\cite{BCDDM13}, namely \emph{run-boundedness}.


\begin{definition}[Run of a DCDS Transition System]
  Given \sidetextb{Run of a DCDS Transition System} a DCDS $\dcdssym$,
  a \emph{run of $\ts{\dcdssym}$} is a (possibly infinite) sequence
  $s_0s_1\cdots$ of states of $\ts{\dcdssym}$ such that
  $s_i\trans s_{i+1}$, for all $i\geq 0$.
\end{definition}

\begin{definition}[Run-bounded DCDS]
  Given \sidetext{Run-bounded DCDS} a DCDS $\dcdssym$, we say
  \emph{$\dcdssym$ is run-bounded} if there exists an integer bound
  $b$ such that for every run $\pi = s_0s_1\cdots$ of $\ts{\dcdssym}$,
  we have that
  $\card{\bigcup_{s \textrm{ state of } \pi}\adom{\db(s)}} < b$.
\end{definition}


\noindent
Intuitively, run-boundedness requires that every run in the transition
system cumulatively encounters at most a bounded number of constants.
Unboundedly many constants 
can still be present in the overall system, provided that they do not
accumulate in the same run.

\begin{theorem}[Verification of \mula over run-bounded DCDS~\cite{BCDDM13}]\label{thm:verification-dcds}
  Verification of $\mula$ properties over run-bounded DCDS is
  decidable and can be reduced to finite-state model checking.
\end{theorem}

%% file: 2.chapters/3-kab.tex
\chapter{Knowledge and Action Bases (KAB\lowercase{s})}\label{ch:kab}

\ifhidecontent
 
\fi

Knowledge and Action Bases (KABs)~\cite{BCDD*12, BCMD*13} have been
proposed as a unified framework to simultaneously account for the
static and dynamic aspects of an application domain. Essentially, it
provides a semantically rich representation of the information on the
domain of interest in terms of a DL KB and a set of actions to change
such information over time, possibly introducing new objects.



Here we consider the KABs that are obtained by combining the framework
in~\cite{BCMD*13} with the action specification formalism in
\cite{MoCD14}. 
Specifically, rather than following the original
KABs~\cite{BCMD*13}, in which at each action execution the
state is reconstructed from scratch, we adopt the action formalism
in~\cite{MoCD14}, in which one specifies only the facts to add and
those to delete from the current state.
Additionally, radically different from~\cite{BCMD*13} where service
calls are not evaluated, in this work we evaluate the service calls
(in the sense that we substitute each service call with a concrete
value) when constructing the transition system. This service call
evaluation semantics is similar to the work on Data Centric Dynamic
Systems in~\cite{BCDDM13}.

%

In the following, we use \dllitea for expressing knowledge bases and
we also do not distinguish between objects and values (thus we drop
attributes).
Furthermore, we make use of a countably infinite set $\const$ of
constants, which intuitively denotes all possible values in the
system.
Additionally, we consider a finite set of distinguished constants
$\iconst \subset \const$, and 
a finite set $\servcall$ of \textit{function symbols} that represents
\textit{service calls}, which abstractly account for the injection of
fresh values (constants) from $\const$ into the system.

\section{KABs Formalism}\label{sec:kab-formalism}
In a nutshell, a KAB is composed by a \dllitea KB (see
\Cref{def:dllitea-kb}), and an action base (consisting of actions and
process) which represents the progression mechanism for KAB. In order
to formally define KABs, we first introduce the notion of KAB actions
as well as KAB process (which is specified as a finite set of
condition-action rules).



Syntactically, an action in a KAB is formalized in the following
definitions:
\begin{definition}[KAB Action]\label{def:kab-action}
  Given \sidetext{KAB Action} a KB $\tup{T, \initabox}$, a \textit{KAB
    action} $\act$ over $\tup{T, \initabox}$ is an expression of the
  form
  \[
  \act(\vec{p}):\set{e_1,\ldots,e_m},
  \]
  where
  \begin{compactitem}
  \item $\act$ is the \emph{action name},
  \item $\vec{p}$ are the \emph{action parameters}, and
  \item $\set{e_1,\ldots,e_m}$ is the set of \emph{KAB action effects}
    (briefly effects). Each effect $e_i$ has the form
    \[
    \map{[q_i^+]\land Q_i^-}{\add \facta_i, \del \factd_i}
    \]
    where:
    \begin{compactitem}
    \item $[q_i^+]\land Q_i^-$ is a $\diecq$ query over
      $\tup{T, \initabox}$, that might also includes action parameters
      $\vec{p}$ as its terms and uses constants from $\iconst$.
      Additionally, the query $[q_i^+]$ is a UCQ, and the query
      $Q_i^-$ is an ECQ whose free variables are included in those of
      $[q_i^+]$.

    \item $\facta_i$ is a set of atoms whose predicates are either
      concept or role names in $\voc(T)$ and
%
%
      each having as terms: either constants in $\iconst$, 
      action parameters $\vec{p}$, free variables of $[q_i^+]$, or
      skolem terms $\scall(\vec{t})$ (representing \emph{service
        calls}) formed by applying a function $\scall \in \servcall$
      to either
      constants in $\iconst$, 
      action parameters, or free variables of $[q^+]$. 

    \item $\factd_i$ is a set of atoms whose predicates are either
      concept or role names in $\voc(T)$, and each having as terms:
      either constants in $\iconst$,
      action parameters $\vec{p}$, or free variables of $[q_i^+]$.
    \end{compactitem}
  \end{compactitem}
  \ \
\end{definition}

\noindent

Given an action $\act$, we write $\eff{\act}$ to denote the \emph{set
  of effects in $\act$}.
For brevity, when it is clear from the context, we often only write
``action'' (resp.\ ``effect'') instead of ``KAB action'' (resp.\ ``KAB
action effect'').
Furthermore, we also often omit the ``$\add \facta$'' part (resp., the
``$\del \factd$'' part) if $\facta=\emptyset$ (resp., if $\factd=\emptyset$).

Intuitively, an action $\act$ is executed by grounding its parameters,
and then applying its effects in parallel. For each effect the form
\[
\map{[q^+]\land Q^-}{\add \facta, \del \factd}, 
\]
the query $[q^+]$ intuitively selects the values to instantiate the
atoms in $\facta$ as well as $\factd$, and $Q^-$ filters \emph{away}
some of those values\footnote{to convey this intuition, we use the
  ``$^+$'' and ``$^-$'' superscript}.
Moreover, the skolem terms that might be contained in $\facta$
represent service calls, and during an action execution, they will be
substituted with values from $\const$ (representing the results of
service calls). Intuitively, the service calls represent the
interactions between the system and the external environment.
After each atom in $\facta$ (resp.\ $\factd$) has been instantiated
and all service calls has been issued (i.e., has been substituted with
the results of service calls), it becomes a set of assertions to be
added (resp.\ deleted) to (resp.\ from) the ABox.
%
%
The update induced by $\act$ is produced by adding and removing those
assertions (the assertions to be added/deleted)
to/from the current ABox, giving higher priority to additions (i.e.,
if the same assertion is asserted to be added and deleted during the
same action execution step, then the assertion is added).
%
%
All of these intuitions of an action execution will be formalized
later when we introduce the semantics of an action execution.

  The instantiations of action parameters are determined by
  condition-action rules. Additionally, a condition-action rule also
  determines when a certain action can be executed. Formally, a
  condition-action rule is defined as follows:

\ \\ \ \\

\begin{definition}[KAB Condition-Action Rule]
  Given \sidetext{KAB Condition-Action Rule} a KB
  $\tup{T, \initabox}$, and a set $\actset$ of KAB actions over
  $\tup{T, \initabox}$, a \emph{KAB condition-action rule} over
  $\tup{T, \initabox}$ and $\actset$ is an
  expression of the form
  \[
  \carule{Q(\vec{p})}{\act(\vec{p})},
  \]
  where:
  \begin{compactitem}
  \item $\act\in \actset$ is a KAB action, and
  \item $Q(\vec{p})$ is a $\diecq$ over $\tup{T, \initabox}$, whose
    free variables are exactly the parameters of $\act$. Additionally,
    $Q(\vec{p})$ might uses constants in $\iconst$.
  \end{compactitem}
\end{definition}
\noindent
For brevity, we often simply write $\carule{Q}{\act}$ instead of
$\carule{Q(\vec{p})}{\act(\vec{p})}$. Intuitively, the query $Q$
expresses a condition when the action $\act$ can be
executed. Moreover, if $Q$ is an open query, then the answers of $Q$
are used to instantiate the parameters of $\act$.



Having the required machinery in hand, we are ready to formally define
KABs as follows.

\begin{definition}[Knowledge and Action Base]\label{def:kab}

  \ \sidetext{Knowledge and Action Base (KAB) Formalism} \\ A KAB is a
  tuple $\kabsym = \tup{T, \initabox, \actset, \procset}$ where:
  \begin{compactitem}
  \item $T$ together with $\initabox$ form a satisfiable \dllitea KB
    $\tup{T, \initabox}$, where
    \begin{compactitem}
    \item $T$ is a \emph{\dllitea TBox} that captures the intensional
      aspects of the domain of interest;
    \item $\initabox$ is the \emph{initial \dllitea ABox}, describing
      the initial configuration of data;
  \end{compactitem}
  \item $\actset$ is a finite set of \emph{KAB actions} over
    $\tup{T, \initabox}$ that evolve the ABox;
  \item $\procset$ is a finite set of \emph{KAB condition-action
      rules} over $\tup{T, \initabox}$ and $\actset$ forming a
    \emph{process} (which tells at any moment which actions can be
    executed, and with which parameters).
%
%
  \end{compactitem}
  Additionally, we assume that $\adom{A_0} \subseteq \const_0$.
\end{definition}


\noindent
%
Roughly speaking, $T$ and $\initabox$ together form the
\emph{knowledge base} while $\actset$ and $\procset$ form the
\emph{action base} which evolves the knowledge base.  We assume that
$\adom{\initabox} \subseteq \iconst$.
Intuitively, the KB maintains the information of interest.  
%
%
$\initabox$ represents the initial state of the system and,
differently from $T$, it evolves and incorporates new information from
the external world by executing actions $\actset$, according to the
sequencing established by process 
$\procset$. 

Without loss of generality, in a KAB, we assume that for each action
$\act \in \actset$ there exists at most one condition-action rule
$\carule{Q(\vec{p})}{\act(\vec{p})} \in \procset$. Notice that several
condition-action rules
$\set{\carule{Q_1(\vec{p})}{\act(\vec{p})}, \ldots,
  \carule{Q_n(\vec{p})}{\act(\vec{p})}} \subseteq \procset$
for an action $\act \in \actset$, can be compactly represented as a
single condition-action rule by taking the disjunction of each query
in the left hand side of those condition-action rule (i.e.,
$\carule{Q_1(\vec{p}) \vee \ldots \vee Q_n(\vec{p})}{\act(\vec{p})}
\in \procset$).

\begin{example}\label{ex:kab}
  To give an example for KAB, we consider the order processing
  scenario in \Cref{ex:dcds}. To model such scenario, we specify a KAB
  $\kabsym = \tup{T, \initabox, \actset, \procset}$, where $T$ is the
  same TBox as in \Cref{ex:tbox-and-abox}, and the initial ABox
  $\initabox$ is as follows:
  \[
  \initabox = \set{\exo{ReceivedOrder}(\excon{chair}),
    \exo{ApprovedOrder}(\excon{table})}
  \]
  The progression mechanism is then modeled by specifying the set
  $\actset$ of actions containing the following actions:
  \begin{enumerate}

  \item $\exa{approveOrder}(\exvar{x})$, which intuitively approves
    the order $\exvar{x}$. Technically it adds a new assertion made by
    the $\exo{ApprovedOrder}$ concept. Formally it is specified as
    follows:
    \[
    \exa{approveOrder}(\exvar{x}):\set{ \true \rightsquigarrow \add
      \set{ \exo{ApprovedOrder}(\exvar{x}) } }
    \]

  \item $\exa{prepareOrders}()$, which intuitively prepares several
    things that are needed for further processing steps of each
    approved order. For each approved order, this action prepares the
    design of the corresponding order by calling an external service
    $\exs{getDesign}(\exvar{x})$. Moreover, this action also retrieves
    the information about the corresponding designer by calling a
    service $\exs{getDesigner}(\exvar{x})$, and assign the assembling
    location for the corresponding order by calling a service
    $\exs{assignAssemblingLoc}(\exvar{x})$.  Formally it is specified
    as follows:

    $\exa{prepareOrders}():\{$ \\
    $\begin{array}{l@{}l@{}l}
       [\exo{ApprovedOrder}&(\exvar{x})] \rightsquigarrow&  \\
                           &\add \{& \\
                           &&\exo{designedBy}(\exvar{x}, \exs{getDesigner}(\exvar{x}) ), \\
                           &&\exo{Designer}(\exs{getDesigner}(\exvar{x})), \\
                           &&\exo{hasDesign}(\exvar{x}, \exs{getDesign}(\exvar{x}) ), \\
                           &&\exo{hasAssemblingLoc}(\exvar{x},\exs{assignAssemblingLoc}(\exvar{x})) \\
                           &\}& 
     \end{array}
     $\\
     $\}$

   \item $\exa{assembleOrders}()$, which represents the step of
     assembling several components into the corresponding ordered
     furniture. This action also acquires the information about the
     assembler and the assembling location by calling external service
     calls $\exs{getAssembler}(\exvar{x})$ and
     $\exs{getAssemblingLoc}(\exvar{x})$ respectively. Formally this
     action is specified as follows:

     $\exa{assembleOrders}():\{$ \\
     $\begin{array}{l@{}l@{}l} [\exo{ApprovedOrder}&(\exvar{x}) \wedge 
                                                     \exists y&.\exo{hasDesign}(\exvar{x},\exvar{y})] \rightsquigarrow \\
                                                   &\add \{& \\
                                                   &&\exo{AssembledOrder}(\exvar{x}), \\
                                                   &&\exo{assembledBy}(\exvar{x},\exs{getAssembler}(\exvar{x})), \\
                                                   &&\exo{Assembler}(\exs{getAssembler}(\exvar{x})), \\
                                                   &&\exo{hasAssemblingLoc}(\exvar{x},\exs{getAssemblingLoc}(\exvar{x})) \\
                                                   &\},& \\
                                                   &\del
                                                     \{&\exo{ApprovedOrder}(\exvar{x})\}
      \end{array}
      $\\
      $\}$

    \item $\exa{checkAssembledOrders}()$, which model the quality
      check process for each assembled order. This action calls an
      external service $\exs{getQualityController}(\exvar{x})$ in
      order to obtain the quality controller officer who performs the
      task. Formally it is specified as follows:

      $\exa{checkAssembledOrders}():\{$ \\
      $\begin{array}{l@{}l@{}l}
         [\exo{AssembledOrder}&(\exvar{x})] \rightsquigarrow& \\
                              &\add \{& \\
                              &&\exo{checkedBy}(\exvar{x},\exs{getQualityController}(\exvar{x})), \\
                              &&\exo{QualityController}(\exs{getQualityController}(\exvar{x}) ) \\
                              &\}&
       \end{array}
       $\\
       $\}$

     \item $\exa{deliverOrders}()$, which delivers each assembled
       order that has passed the quality control process. Technically,
       it changes the status of an assembled order that has been
       delivered into delivered order. Formally this action is
       specified as follows:

       $\exa{deliverOrders}():\{$ \\
       $\begin{array}{l@{}l}
          [\exo{AssembledOrder}(\exvar{x}) \wedge& \exists y. \exo{checkedBy}(\exvar{x},\exvar{y})] \rightsquigarrow \\
                                                            &\add
                                                              \set{\exo{DeliveredOrder}(\exvar{x})
                                                              }, \del
                                                              \set{\exo{AssembledOrder}(\exvar{x})
                                                              }
        \end{array}
        $\\
        $\}$

      \end{enumerate}

\noindent
Furthermore, the set $\procset$ of condition action rules is specified
as follows:
\begin{compactitem}

\item $\begin{array}{@{}l@{}l@{}}  \carule{
   [\exo{ReceivedOrder}(\exvar{x})] }
  {\exa{approveOrder}(\exvar{x})},
 \end{array}
 $ 

\item  $\begin{array}{@{}l@{}l@{}}  \carule{ [\exists
     \exvar{x}. \exo{ApprovedOrder}(\exvar{x})] }
   {\exa{prepareOrders}()},
 \end{array}
 $

\item  $\begin{array}{@{}l@{}l@{}}  \carule{ [\exists
     \exvar{x}\exvar{y}. \exo{ApprovedOrder}(\exvar{x}) \wedge
     \exo{hasDesign}(\exvar{x},\exvar{y})]} {\exa{assembleOrders}()},
 \end{array}
 $

\item  $\begin{array}{@{}l@{}l@{}}  \carule{[\exists
     \exvar{x}. \exo{AssembledOrder}(\exvar{x})] }
   {\exa{checkAssembledOrders}()},
 \end{array}
 $

\item  $\begin{array}{@{}l@{}l@{}}  \carule{ [\exists
     \exvar{x}\exvar{y}. \exo{AssembledOrder}(\exvar{x}) \wedge
     \exo{checkedBy}(\exvar{x},\exvar{y})] }
   {\exa{deliverOrders}()}.
 \end{array}
 $
\end{compactitem}
 The intuition for each condition action rules above is consecutively
 presented below:
\begin{compactitem}

\item the first rule states that if there exists a received order with
  ID $\exvar{x}$, then we can fire the execution of action
  $\exa{approveOrder}/1$ with the argument $\exvar{x}$ (i.e.,
  $\exa{approveOrder}(\exvar{x})$).

\item The second rule says that we can execute the action
  $\exa{prepareOrders}/0$ in case there exists at least one approved order.

\item Next, the third rule encodes the condition where the execution
  of action $\exa{assembleOrders}/0$ can be fired in case there exists
  an approved order that already has a design.

\item The fourth rule indicates that if there exists an assembled
  order, we can execute the action
  $\exa{checkAssembledOrders}/0$

\item Finally, the last rule specifies that whenever there exists an
  assembled order that has been checked, we can execute the action $\exa{deliverOrders}/0$.

\end{compactitem}
\end{example}

\section{KABs Standard Execution Semantics}\label{sec:kab-standard-exec-semantic}

The \emph{standard execution semantics of a KAB}
$\kabsym$ 
is given in terms of a possibly infinite-state \emph{transition
  system} whose states are labeled by knowledge bases and where
transitions represent the execution of actions. Such a transition
system represents all possible computations that the actions can do on
the knowledge base starting from the initial knowledge base (with the
corresponding initial ABox).
During the execution, an action can issue service calls.  In this thesis, we
assume that the semantics of service calls is \emph{deterministic}, i.e., along
a run of the system, whenever a service is called with the same input
parameters, it will return the same value.
To enforce this semantics, similar to DCDS (see \Cref{sec:dcds}), the
transition system of a KAB remembers the results of previous service
calls in a
service call map (see \Cref{def:service-call-map}) that is
embedded as a part of the transition system state.

Technically, to provide the semantics of KABs, we consider KB
transition systems (as in \Cref{def:kb-ts}), i.e., transition systems
of the form $\left<\const,T,\stateset,s_0,\abox,\trans\right>$, where:
\begin{compactitem}
\item $T$ is a \dllitea TBox;
\item $\stateset$ is a (possibly infinite) set of states;
\item $s_0 \in \stateset$ is the initial state;
\item $\abox$ is a function that, given a state $s\in \stateset$,
  returns an ABox associated to $s$;
\item $\trans \subseteq \stateset\times\stateset$ is a transition relation
  between pairs of states.
\end{compactitem}

%
%
%
%
%
%
\noindent
In addition, to realize the deterministic service calls semantics,
each state $s \in \stateset$ of the transition system 
is defined as a tuple $\tup{A, \scmap}$, where $\A$ is an ABox and
$\scmap$ is a service call map. Later on, a state of the form
$\tup{A, \scmap}$ is often also called a \emph{KAB state}. 

In the following, we provide the semantics of an \emph{action
  execution} \sidetext{KAB Action Execution Semantics} by defining
when an action can be executed and how to construct the result of an
action execution.
%

\begin{definition}[Executability of an Action]\label{def:kab-action-executability}
  Let \sidetextb{Executability of \\an Action}
  $\kabsym = \tup{T, \initabox, \actset, \procset}$ be a KAB.
%
  Given an ABox $A$, an action $\act \in \actset$
  of the form $\act(\vec{p}):\set{e_1,\ldots,e_m}$, 
  and a \emph{parameter substitution} $\sigma$ which substitutes the
  parameters $\vec{p}$ with values taken from $\const$.
  We say that \emph{$\act$ is executable in $A$ with a parameter
    substitution $\sigma$}, if there exists a condition-action rule
  $Q(\vec{p})\mapsto\alpha(\vec{p}) \in \procset$ such that
  $\Ans(Q\sigma, T, A)$ is $\true$.
\end{definition}



\noindent
Similar to DCDS (as in \Cref{sec:dcds}), with a little abuse of
the definition, we sometimes say that an action \emph{$\act$ is
  executable in a state $s$ with a parameter substitution $\sigma$} if
$s = \tup{A, \scmap}$ and $\act$ is executable in $A$ with a parameter
substitution $\sigma$.
Additionally, given an action
$\act(p_1,\ldots, p_n): \set{e_1,\ldots,e_m}$, and an ABox $A$, 
We say that $\sigma$ is a \sidetext{Legal Parameter Assignment and
  Grounded Action} \emph{legal parameter assignment} for $\act$
in $A$, if $\act$ is executable in $A$ with a parameter substitution
$\sigma$. Furthermore, 
we write $\dact\sigma$ to denote a \emph{grounded action} that is
obtained by applying a legal parameter assignment $\sigma$ to each
$e \in \deff{\dact}$ (i.e., substituting each occurrence of action
parameter $p_i$ (for $i \in \set{1,\ldots,n}$) in $e$ with a constant
in $\const$ based on the substitution $\sigma$).

\begin{example}\label{ex:kab-act-exec}
  Continuing our running example in \Cref{ex:kab}, the action
  $\exa{approveOrder/1}$ is executable in the state
  $s_0 = \tup{\initabox, \emptyset}$ with a parameter substitution
  $\sigma$, where $\sigma$ substitutes the action parameter of
  $\exa{approveOrder/1}$ into $\excon{chair}$. This is the case
  because we have that the query in the left hand side of the
  condition-action rule
  \[
  \begin{array}{@{}l@{}l@{}} \carule{ [\exo{ReceivedOrder}(\exvar{x})]
    } {\exa{approveOrder}(\exvar{x})},
  \end{array}
  \]
  is successfully evaluated and give an answer $\excon{chair}$ (i.e.,
  $\sigma(x) = \excon{chair}$).
\end{example}

Now we proceed to define a set of atoms to be added/deleted by a
grounded action $\act\sigma$. 

\begin{definition}[Set of Atoms to be Added]\label{def:add-kab-action}
  Let \sidetextb{Set of Atoms to be Added by an Action} $\kabsym = \tup{T, \initabox, \actset, \procset}$ be a KAB.
%
  Given an ABox $A$, an action $\act \in \actset$
  of the form $\act(\vec{p}):\set{e_1,\ldots,e_m}$ with
  $e_i = \map{[q_i^+]\land Q_i^-}{\add \facta_i, \del \factd_i}$, and
  a legal parameter assignment $\sigma$ for $\act$ in $A$.
%
%
  We define a \emph{set of atoms to be added by $\act\sigma$ w.r.t.\
    $A$} 
  as follows:
\[
\addfacts{T, A, \act\sigma} = \bigcup_{(\map{[q^+]\land Q^-}{\add \facta, \del \factd})
  \text{ in } \eff{\act}} \quad \bigcup_{\rho\ \in\ \Ans([q^+]\land Q^-\sigma,T,A)}
F^+\sigma\rho
\]
\ \ 
\end{definition}

\begin{definition}[Set of Atoms to be Deleted]\label{def:del-kab-action}
  Let \sidetextb{Set of Atoms to be Deleted by an Action}
  $\kabsym = \tup{T, \initabox, \actset, \procset}$ be a KAB.
%
Given an ABox $A$, an $\act \in \actset$ 
of the form $\act(\vec{p}):\set{e_1,\ldots,e_m}$ with
$e_i = \map{[q_i^+]\land Q_i^-}{\add \facta_i, \del \factd_i}$, and a
legal parameter assignment $\sigma$ for $\act$ in $A$.
%
%
We define a \emph{set of atoms to be deleted by $\act\sigma$ w.r.t.\
  $A$} as follows:
\[
\delfacts{T, A, \act\sigma} = \bigcup_{(\map{[q^+]\land Q^-}{\add
    \facta, \del \factd}) \text{ in } \eff{\act}} \quad \bigcup_{\rho\
  \in\ \Ans([q^+]\land Q^-\sigma,T,A)} F^-\sigma\rho
\]
\ \ 
\end{definition}


The execution result of a grounded action $\act\sigma$ 
is then captured by a function $\doo{T, A, \act\sigma}$ which is
formally defined as follows:

\begin{definition}[Computation of KAB Action Execution Result]\label{def:do-kab-action}
  Let \sidetextb{Computation of KAB Action Execution Result}
  $\kabsym = \tup{T, \initabox, \actset, \procset}$ be a KAB.
  Given an ABox $A$,
  an action $\act \in \actset$ 
  of the form $\act(\vec{p}):\set{e_1,\ldots,e_m}$ with
  $e_i = \map{[q^+]\land Q^-}{\add \facta, \del \factd}$, and a legal
  parameter assignment $\sigma$ for $\act$ in $A$.
  The \emph{execution result of $\act\sigma$ in $A$} is computed by
  function $\doo{T, A, \act\sigma}$ as follows:
  \[
  \doo{T, A, \act\sigma} = \left(A\ \setminus\ \delfacts{T, A,
      \act\sigma} \right) \cup \left(\addfacts{T, A,
      \act\sigma}\right)
  \]
  where $e = \map{[q^+]\land Q^-}{\add \facta, \del \factd}$.
\end{definition}

\noindent
Radically different from the KABs in~\cite{BCMD*13} (also different from
DCDSs), during an action execution, instead of dropping the whole ABox
and constructing a new one as the result of the action execution, in
this KABs the actions only update the corresponding current ABox
(i.e., add/delete assertions from the current ABox).
Notice that in this sense such actions are similar to STRIPS style
actions \cite{FN71} where the additions are assumed to have higher
priority than deletions, and those ``facts'' that are not affected by
the action execution stay the same. Hence, unlike in DCDSs, we do not
need to specify both the things that are changes and the things that
are not changes.
More precisely, the definition above intuitively says that the
execution result of $\act$ is obtained by first deleting from $A$ the
assertions that is obtained from the grounding of the atoms in $F^-$
and then adds the new assertions that is obtained from the grounding
of the atoms in $F^+$. The grounding of the atoms in $F^+$ and $F^-$
are obtained from all the certain answers of the query
$[q^+]\land Q^-$ over $\tup{T,A}$.

The result of $\doo{T, A, \act\sigma}$ is in general not a proper
ABox, because it could contain (ground) skolem terms, attesting that
in order to produce the ABox, some service calls have to be issued.
We \sidetext{Service Call Evaluation} denote by
$\calls{\doo{T, A, \act\sigma}}$ the set of such ground service calls,
and by $\eval{T, A,\act\sigma}$ the set of substitutions that replace
such calls with concrete values taken from $\const$. Specifically,
$\eval{T, A,\act\sigma}$ is defined as
\[
\eval{T,A,\act\sigma}= \{\theta \mid \theta \mbox{ is a total function }
                        \theta: \calls{\doo{T, A, \act\sigma}} \ra \const \}.
\]

Given a KAB $\kabsym = \tup{T, A_0, \actset, \procset}$, we employ
$\doo{}$ and $\eval{}$ to define a transition relation
$\exec{\kabsym}$ connecting two states through an action execution as
follows.


\begin{definition}[KAB Transition Relation $\exec{\kabsym}$]\label{def:KAB-exec-relation} \ \\
  Let \sidetext{KAB Transition Relation}
  $\kabsym = \tup{T, \initabox, \actset, \procset}$ be a KAB.
  Given a state $s = \tup{A, \scmap}$, a state
  $s' = \tup{A', \scmap'}$, an action $\act \in \actset$, and a
  substitution $\sigma$.
%
  We have
  $\tup{\tup{A, \scmap},\act\sigma, \tup{A', \scmap'}} \in
  \exec{\kabsym}$ if the following holds:
\begin{compactenum}
\item $\act$ is executable in $s$ with a legal parameter assignment
  $\sigma$;

\item there exists $\theta \in \eval{T, A,\act\sigma}$ such that
  $\theta$ and $\dscmap$ ``agree'' on the common skolem terms in their
  domains, in order to realize the deterministic service call
  semantics;



\item $A' = \doo{T, A, \act\sigma}\theta$;
\item $\scmap' = \scmap \cup \theta$ (i.e., updating the history of
  issued service calls).
\end{compactenum}
\ \ 
\end{definition}

\noindent
For easiness of reading, when
$\tup{\tup{A, \scmap}, \act\sigma, \tup{A', \scmap'}} \in
\exec{\kabsym}$, we equivalently write
\[
\tup{A, \scmap} \exect{\act\sigma,\ \kabsym\ } \tup{A', \scmap'}. 
\]
 When it is clear from the context, we also
often omit $\kabsym$ and just write
$\tup{A, \scmap} \exect{\act\sigma} \tup{A', \scmap'}$.


The transition system $\ts{\kabsym}$ of KAB $\kabsym$, which provide
the standard execution semantics of $\kabsym$, is then formally defined as
follows:

\begin{definition}[KABs Standard Transition
  System]\label{def:KAB-standard-ts} Given \sidetext{KABs Standard
    Transition System}
  a KAB $\kabsym = \tup{T, \initabox, \actset, \procset}$, the
  standard transition system $\ts{\kabsym}$ is defined as
  $\tup{\const, T, \stateset, s_0, \abox, \trans}$ where
\begin{itemize}
\item $s_0 = \tup{\initabox,\emptyset}$, and
\item $\stateset$ and $\trans$ are defined by simultaneous induction as the
  smallest sets satisfying the following properties:
  \begin{enumerate}
  \item $s_0 \in \stateset$;
  \item if $\tup{A,\scmap} \in \stateset$, then for all actions $\act \in
    \actset$,
    for all substitutions $\sigma$ for the parameters of $\act$ and
    for all $\tup{A',\scmap'}$ such that 
    %
    \begin{enumerate}
    \item
      $\tup{A,\scmap} \exect{\act\sigma, \ \kabsym\ }
      \tup{A',\scmap'}$ and
    \item $A'$ is $T$-consistent,
    \end{enumerate}
    we have $\tup{A',\scmap'}\in \stateset$, and
    $\tup{A,\scmap}\trans \tup{A',\scmap'}$.
  \end{enumerate}
\end{itemize}
\ \ 
\end{definition}
 
\noindent
Intuitively, the standard execution semantics for a KAB
$\kabsym = \tup{T, \initabox, \actset, \procset}$ is obtained starting
from $\initabox$ by nondeterministically applying every executable
actions with corresponding legal parameter assignments, and
considering each possible value returned by applying the involved
service calls.
%
%
The executability of an action with fixed parameters does not only
depend on the set of condition-action rules $\procset$, but also on
the $T$-consistency of the ABox produced by the execution of the
action: if the resulting ABox is $T$-inconsistent, the action is
considered as non executable with the chosen parameters.

\begin{example}\label{ex:kab-exec}
  Continuing our running example in \Cref{ex:kab}, the construction of
  transition system $\ts{\kabsym}$ of KAB $\kabsym$ in \Cref{ex:kab}
  is started from the initial state $s_0 = \tup{\initabox, \emptyset}$
  where
  \[
  \initabox = \set{\exo{ReceivedOrder}(\excon{chair}),
    \exo{ApprovedOrder}(\excon{table})}.
  \]
  An example of a sucessor of state $s_0$ is a state
  $s_1 = \tup{A_1, \scmap_1}$, where
  \[
  \begin{array}{l@{}l}
    A_1 &= \set{\exo{ApprovedOrder}(\excon{chair}),
           \exo{ApprovedOrder}(\excon{table})}, \\
    \scmap_1 &= \emptyset
  \end{array}
  \]
  and $s_1$ is obtained from the execution of action
  $\exa{approveOrder/1}$ with the argument $\excon{chair}$. The action
  $\exa{approveOrder/1}$ is executable with argument $\excon{chair}$ in
  the state $s_0$ since we have that the query in the left hand side
  of the condition-action rule
  \[
  \begin{array}{@{}l@{}l@{}} \carule{ [\exo{ReceivedOrder}(\exvar{x})]
    } {\exa{approveOrder}(\exvar{x})},
  \end{array}
  \]
  is successfully evaluated and give an answer $\excon{chair}$. As the
  result of executing this action, we now have an approved order of
  chair in state $s_1$.
\end{example}

\section{Verification of KABs}

The interesting reasoning task in KABs is to verify whether the
transition system of a given KAB satisfies temporal properties of
interest, specified in some first-order temporal logic. To specify the
temporal properties to be verified over KABs, the work
in~\cite{BCMD*13} uses the temporal logic \muladom (see Section
\Cref{subsec:muladom}).
%
%
The verification problem of \muladom properties over KABs is then
formally stated as follows:

\begin{definition}[Verification of a \muladom Property over a
  KAB]\label{def:verification-kab}
  Given \sidetext{Verification of a \muladom Formula over a KAB} a KAB
  $\kabsym = \tup{T, \initabox, \actset, \procset}$ and a closed
  $\muladom$ formula $\Phi$ over $\tup{T, \initabox}$. Let
  $\ts{\kabsym}$ be the transition system of $\kabsym$,
  \emph{the verification of a \muladom formula $\Phi$ over $\kabsym$}
  is a problem to check whether $\ts{\kabsym} \models \Phi$.
%
\end{definition}

\noindent
We also say a KAB \emph{$\kabsym$ satisfies a closed \muladom formula
  $\Phi$},
if $\ts{\kabsym} \models \Phi$. The definition above intuitively said
that given a KAB $\kabsym$ and a closed $\muladom$ formula $\Phi$, the
problem of verifying $\Phi$ over $\kabsym$ is a problem to check
whether $\Phi$ holds in the initial state of $\ts{\kabsym}$. From this
moment, we assume that the \muladom properties to be verified over
KABs $\kabsym = \tup{T, \initabox, \actset, \procset}$ are closed
\muladom formulas over $\tup{T, \initabox}$.

Here we solve the problem of $\muladom$ properties verification over
KABs by reducing it into the problem of \mula properties verification
over DCDSs. The idea of the reduction is similar to the work in
\cite{MoCD14}, which shows a reduction from Data-Aware
Commitment-Based Multiagent Systems into DCDSs.
To reduce such problem, here we do the following:
\begin{compactenum}

\item We define a generic translation $\tdcds$, that given a KAB
  $\kabsym$, produces a DCDS $\tdcds(\kabsym)$, and

\item We extend the perfect reformulation algorithm into \muladom
  formula by simply applying the algorithm only to each query in the
  corresponding \muladom formula and keeping the other parts of the
  formula unaltered.

\item We show that 
  a KAB $\kabsym$ satisfies a certain \muladom property $\Phi$ if and
  only if 
  its corresponding DCDS $\tdcds(\kabsym)$ (obtained from $\kabsym$
  via $\tdcds$) satisfies a \mula property $\Phi'$ that is obtained
  using the extended perfect reformulation algorithm for \muladom
  (i.e., by
  rewriting each query in $\Phi$ w.r.t.\ the TBox in the given KAB
  $\kabsym$ using the perfect reformulation algorithm).

\end{compactenum}




\subsection{Translating KABs into DCDSs}\label{sec:trans-kab-to-dcds}


In this section we define a generic translation which transform any
KABs into DCDSs with the aim of reducing the \muladom verification
problem over KABs into the \mula verification problem over
DCDSs. Towards this goal, we first introduce several preliminaries as
follows.

\begin{definition}[Equality Between a Database Instance and an ABox]\label{def:db-equal-kb}
  Given \sidetextb{Equality Between a Database Instance and an ABox} a
  database instance $\dbinst$ which conforms to schema $\dbschema$,
  and an ABox $A$ over $\voc(T)$, we say that \emph{$A$ is equal to
    $\dbinst$}, denoted by $A = \dbinst$, if the following hold
  \begin{compactitem}
  \item a concept name $N \in \voc(T)$ if and only if a relation
    schema $N \in \dbschema$,
  \item a role name $P \in \voc(T)$ if and only if a relation schema
    $P \in \dbschema$,
  \item a concept assertion $N(c) \in A$ if and only if a database 
    fact $N(c) \in \dbinst$,
  \item a role assertion $P(c_1,c_2) \in A$ if and only if a database
    fact $P(c_1,c_2) \in \dbinst$.
  \end{compactitem}
\ \ 
\end{definition}

\begin{definition}[Set of Deletion Effects For Concept
  Assertion]\label{def:deletion-effect}
  Given \sidetextb{Set of Deletion Effects} a
  KAB $\kabsym = \tup{T, \initabox, \actset, \procset}$, and an action
  of the form $\act(\vec{p}):\set{e_1,\ldots,e_m}$, 
%
  we define a \emph{set $\effd{N, \act}$ of deletion effects for concept assertions}
  made by concept name $N \in \voc(T)$ w.r.t.\ $\act$ as a set of effects
  constructed as follows:
  we have
  \[
  \map{[q^+
    ] \land Q^- 
  }{\del \set{N(t)}} \in \effd{N, \act}
  \]
  if there exists an effect 
  \[
  \map{[q^+]\land Q^-}{\add \facta, \del \factd} \in \eff{\act}
  \]
  such that $N(t) \in \factd$, where $t$ can be a free variable of
  $[q^+]$, 
  an action parameter (i.e., among $\vec{p}$), or a constant in
  $\adom{\initabox}$.
\end{definition}

\noindent
%
%
The case for role assertions (i.e., $\effd{P, \act}$) is defined
similarly as above. 




Having all necessary preliminaries in hand, we are ready to define a
translation $\tdcds$ that, given a KAB $\kabsym$,
produces DCDS $\tdcds(\kabsym)$
as follows.

\begin{definition}[Translation From KABs to DCDSs]\label{def:translation-kab-to-dcds}
  A \sidetext{Translation From KABs to DCDSs} \emph{translation
    $\tdcds$} is a translation that takes a KAB
  $\kabsym = \tup{T, \initabox, \actset, \procset}$ as input and
  produces DCDS $\tdcds(\kabsym) = \tup{\dcomp, \pcomp}$, where:
  \begin{compactitem}
  \item $\dcomp = \tup{\dbschema,\idb, \ecset}$ is a data component,
    where:
    \begin{compactitem}[$\bullet$]
    \item $\dbschema$ is obtained as follows: 
      \begin{compactitem}
      \item for each concept name $N \in \voc(T)$, we have $N \in \dbschema$ with
        arity 1,
      \item for each role name $P \in \voc(T)$, we have $P \in \dbschema$ with
        arity 2,
      \end{compactitem}
    \item $\idb = \initabox$, 
    \item $\ecset = \set{\qunsatfol{T} \ra \false}$,
      where $\qunsatfol{T}$
      is an FOL query defined in \Cref{def:qunsat-fol}. 
      Intuitively, here we encode the constraints in the TBox $T$
      into the equality constraints $\ecset$ in DCDS.
   \end{compactitem}
  \item $\pcomp = \tup{\dactset,\dprocset}$ is a process component,
    where 
    \begin{compactitem}[$\bullet$]
    \item $\dactset$ is obtained as follows: 
      for each $\act(\vec{p}) \in \actset$,
      we have $\dact'(\vec{p}) \in \dactset$
%
      that is constructed 
      as follows:
      \begin{compactenum}[(1)]
      \item For each effect $\map{[q^+]\land
          Q^-}{\add \facta, \del \factd} \in \eff{\act}$, \\
        we have $\map{\rew([q^+]\land
          Q^-, T)}{\facta} \in \deff{\act'} $,
      
      \item For each concept name $N \in \voc(T)$, 
        we have an effect of the form
        \[
        \map{N(w) \wedge \bigwedge_{e\ \in\ \effd{N, \act}}
          \left(\rew([q^+]\land Q^-, T) \wedge \neg(w = t) \right)
        }{\set{N(w)}}
        \]
        in $\deff{\act'}$, where
        \begin{compactitem}
        \item[-]
          $e = \map{[q^+] \land Q^- }{\del \set{N(t)}} \in
          \effd{N, \act}$,
        \item[-] $w$
          is a variable, and additionally it is neither an action
          parameter nor a free variable of any queries $[q^+]$
          in any $e \in \effd{N, \act}$.
        \end{compactitem}
        Intuitively, the effects constructed above, preserve all
        concept assertions
        that are not deleted by any deletion effect.

      \item We repeat similar construction, as in the (2), for each
        role name $P \in \voc(T)$.
      \end{compactenum}

    \item $\dprocset$ is obtained as follows: for each
      condition-action rule $Q\mapsto\act \in \procset$, we have
      $\carule{Q'}{\act'} \in \dprocset$, where $Q' = \rew(Q, T)$, and
      $\act' \in \dactset$ is obtained from $\act \in \actset$ as
      above.
  \end{compactitem}
  \end{compactitem}
\ \ 
\end{definition}


In the following, we show several interesting properties of the
translation $\tdcds$ that will be used later to reduce the problem of
\muladom verification over KABs into the problem of \mula verification
over DCDSs.
First, we show that the computation result of KAB action execution
$\act$ over a certain state $s_k$ and the computation result of DCDS
action execution $\dact'$ (which is obtained from $\act$ via $\tdcds$)
over a certain state $s_d$ produce a same result provided that the
corresponding ABox in $s_k$ and database instance in $s_d$ are equal.


\begin{lemma}\label{lem:dokab-equal-dodcds}
  Let $\kabsym = \tup{T, \initabox, \actset, \procset}$ be a KAB with
  transition system $\ts{\kabsym}$,
%
  $\tdcds(\kabsym) = \tup{\dcomp, \pcomp}$ be a DCDS obtained from
  $\kabsym$ through translation $\tdcds$ where
  $\pcomp = \tup{\dactset,\dprocset}$. Additionally, let
  $\ts{\tdcds(\kabsym)}$ 
  be transition system of $\tdcds(\kabsym)$.
  Consider a state $s = \tup{A, \scmap}$ of $\ts{\kabsym}$, a state
  $s' = \tup{\dbinst, \dscmap_d}$ of $\ts{\tdcds(\kabsym)}$, an action
  $\act \in \actset$, an action $\dact' \in \dactset$ that is obtained
  from $\act$ as in the definition of translation $\tdcds$, and a
  substitution $\sigma$ that is a legal parameter assignment for
  $\act$ in $s$ and also a legal parameter assignment for $\dact'$ in
  $s'$.
  If $A = \dbinst$, 
  then $\doo{T, A, \act\sigma} = \ddoo{\dbinst, \act'\sigma}$.
\end{lemma}
\begin{proof}
We have to show that 
\begin{compactenum}[\bf (1)]
\item 
  $N(t_1) \in \doo{T, A, \act\sigma}$ if and only if
  $N(t_1) \in \ddoo{\dbinst, \dact'\sigma}$
\item 
  $P(t_1,t_2) \in \doo{T, A, \act\sigma}$ if and only if
  $P(t_1,t_2) \in \ddoo{\dbinst, \dact'\sigma}$
\end{compactenum}
where $N$ is a concept name, $P$ is a role name, and $t_1$ (resp.\
$t_2$) is either a constant in $\const$ or a skolem term.

\smallskip
\noindent
\textbf{Proof for Case (1):} by \Cref{def:do-kab-action}, if
$N(t) \in \doo{T, A, \act\sigma}$ then we have either
\begin{compactenum}[\bf (a)]
\item $N(t_1) \in \addfacts{T, A, \act\sigma}$ (No matter whether
  $N(t_1) \in A$ or $N(t_1) \not\in A$,
  and also no matter whether $N(t_1) \in \delfacts{T, A, \act\sigma}$
  or $N(t_1) \not\in \delfacts{T, A, \act\sigma}$).
\item $N(t_1) \in A$ and $N(t_1) \not\in \delfacts{T, A, \act\sigma}$, or
%
%
\end{compactenum} 

\begin{compactitem}
\item[\textbf{Case (a):}]
By 
\Cref{def:add-kab-action}, if $N(t_1) \in \addfacts{T, A, \act\sigma}$
then there exists an effect
\[ 
\map{[q^+]\land Q^-}{\add \facta, \del \factd} \in \eff{\act} 
\]
s.t.\ $N(x) \in \facta$ where $x$ is either an action parameter, a
constant in $\adom{\initabox}$, a free variable in $[q^+]$ or a skolem
terms formed by applying a function $\scall \in \servcall$ to either
constants in $\adom{\initabox}$, action parameters, or free variables
of $[q^+]$. Moreover, the following hold:
\begin{compactitem}
%
%
\item if $x$ is a constant in $\adom{\initabox}$, then $x = t_1$
\item if $x$ is an action parameter, then $x/t_1 \in \sigma$
\item if $x$ is a free
  variable in $[q^+]$, then $x/t_1 \in \rho$, where
  $\rho \in \Ans([q^+]\land Q^-\sigma,T,A)$
\item if $x$ is a skolem term of the form $\dscall(\vec{v})$, then we
  have either $(\dscall(\vec{v})\sigma)\rho = t_1$, where
  $\rho \in \Ans([q^+]\land Q^-\sigma,T,A)$
\end{compactitem}
Now, recall that by the definition of $\tdcds$ (\Cref{def:translation-kab-to-dcds}), since
\[
\map{[q^+]\land Q^-}{\add \facta, \del \factd} \in \eff{\act}
\]
then we have 
\[ \map{\rew([q^+]\land Q^-, T)}{\facta} \in \deff{\act'}. \]
Additionally, since $A = \dbinst$, by \Cref{thm:FO-rewritability}, we have
\[
\ans([q^+]\land Q^-\sigma, T, A) = 
\ANS(\rew([q^+]\land Q^-,T)\sigma, \dbinst).
\]
Thus, by \Cref{def:do-dcds-action}
 it is easy to see that  
 $N(t_1) \in \ddoo{\dbinst, \dact'\sigma}$. 

\item[\textbf{Case (b):}] If $N(t_1) \in A$ and
  $N(t_1) \not\in \delfacts{T, A, \act\sigma}$, then we have
\begin{compactenum}
\item there exists a substitution $\rho$ such that 
  $\rho \in \Ans([N(w)], T, A)$, and $[w/t_1] \in \rho$.
\item there does not exists  $\map{[q^+] \land Q^- }{\del \set{N(t)}} \in
  \effd{N, \act}$ such that there exists $\rho \in \Ans([q^+] \land
  Q^-, T, A)$ and $t/t_1 \in \rho$.
\end{compactenum}
Now, recall that by the definition of $\tdcds$
(\Cref{def:translation-kab-to-dcds}), For each concept name
$N \in \voc(T)$, we have an effect of the form
        \[
        \map{N(w) \wedge \bigwedge_{e\ \in\ \effd{N, \act}}
          \left(\rew([q^+]\land Q^-, T) \wedge \neg(w = t) \right)
        }{\set{N(w)}}
        \]
        in $\deff{\act'}$, where
        \begin{compactitem}
        \item[-]
          $e = \map{[q^+] \land Q^- }{\del \set{N(t)}} \in
          \effd{N, \act}$,
        \item[-] $w$ is a variable, and additionally it is neither an
          action parameter nor a free variable of any queries $[q^+]$
          in any $e \in \effd{N, \act}$.

        \end{compactitem}
Additionally, since $A = \dbinst$, by \Cref{thm:FO-rewritability}, we have
%
\begin{compactenum}
\item there exists a substitution $\rho$ such that
  $\rho \in \ANS([N(w)], \dbinst)$, and $w/t_1 \in \rho$.
\item there does not exists a query $\rew([q^+]\land Q^-, T)$ (where
  $\map{[q^+] \land Q^- }{\del \set{N(t)}} \in \effd{N, \act}$) such
  that $\rho \in \ANS(\rew([q^+]\land Q^-, T), \dbinst)$ and
  $t/t_1 \in \rho$.

%
\end{compactenum}
Thus, by \Cref{def:do-dcds-action}
 we have 
 $N(t_1) \in \ddoo{\dbinst, \dact'\sigma}$. 


\end{compactitem}
The proof for the other direction for case (1) can be shown similarly.
Moreover, the proof for case (2) can be done similarly as the case
(1).
%
%
%
%
%
\end{proof}




As a consequence of \Cref{lem:dokab-equal-dodcds}, in the
following we show that the set of substitutions that replace service
calls $\eval{T,A,\act\sigma}$ is equal to
$\deval{\dbinst,\dact'\sigma}$ provided that 
\begin{inparaenum}[\it (i)]
  \item $A = \dbinst$, 
  \item $\dact'$ is obtained from $\act$ through $\tdcds$ and both of
    them are grounded with the same legal parameter assignment
    $\sigma$.
\end{inparaenum}

\begin{lemma}\label{lem:kabeval-equal-dcdseval}
  Let $\kabsym = \tup{T, \initabox, \actset, \procset}$ be a KAB with
  transition system $\ts{\kabsym}$,
%
  $\tdcds(\kabsym) = \tup{\dcomp, \pcomp}$ be a DCDS obtained from
  $\kabsym$ through translation $\tdcds$ where
  $\pcomp = \tup{\dactset,\dprocset}$. Additionally, let
  $\ts{\tdcds(\kabsym)}$ 
  be transition system of $\tdcds(\kabsym)$.
  Consider a state $s = \tup{A, \scmap}$ of $\ts{\kabsym}$, a state
  $s' = \tup{\dbinst, \dscmap_d}$ of $\ts{\tdcds(\kabsym)}$, an action
  $\act \in \actset$, an action $\dact \in \dactset$ obtained from
  $\act$ as in the definition of translation $\tdcds$, and a
  substitution $\sigma$ that is a legal parameter assignment for
  $\act$ in $s$ as well as a legal parameter assignment for $\dact'$
  in $s'$.
  If $A = \dbinst$, 
  then $\eval{T,A,\act\sigma} = \deval{\dbinst,\dact'\sigma}$.
\end{lemma}
\begin{proof}
  By \Cref{lem:dokab-equal-dodcds}, we have
  $\doo{T, A, \act\sigma} = \ddoo{\dbinst, \act'\sigma}$. Therefore by
  the definition of $\eval{T,A,\act\sigma}$ and
  $\deval{\dbinst,\dact'\sigma}$, it is easy to see that
  $\eval{T,A,\act\sigma} = \deval{\dbinst,\dact'\sigma}$.
\end{proof}

Last, we show that an equality constraints $\ecset$, that is obtained
through $\tdcds$, encode the same constraints as in the TBox $T$ (of
the corresponding KAB) that is needed for satisfiability check. As a
consequence, given an ABox $A$ and a database instance $\dbinst$ such
that $A = \dbinst$, if $A$ is $T$-consistent then $\dbinst$ satisfies
$\ecset$, and vice versa.


\begin{lemma}\label{lem:qunsat-equal-ec}
  Let $\kabsym = \tup{T, \initabox, \actset, \procset}$ be a KAB with
  transition system $\ts{\kabsym}$,
%
  $\tdcds(\kabsym) = \tup{\dcomp, \pcomp}$ be a DCDS obtained from
  $\kabsym$ through translation $\tdcds$, where
  $\dcomp = \tup{\dbschema,\idb, \ecset}$ 
  Additionally, let
  $\ts{\tdcds(\kabsym)}$
  be transition system of $\tdcds(\kabsym)$.
  Consider a state $s = \tup{A, \scmap}$ of $\ts{\kabsym}$, a state
  $s' = \tup{\dbinst, \dscmap_d}$ of $\ts{\tdcds(\kabsym)}$, 
%
  such that $A = \dbinst$, 
  then we have $\dbinst$ satisfies $\ecset$ if and only if $A$ is $T$-consistent
\end{lemma}
\begin{proof}
  Since $A$ is $T$-consistent, by \Cref{thm:qunsat}, we have
  $\ANS(\qunsatfol{T}, A) = \false$. Then, since, $\dbinst = A$, we
  have $\ANS(\qunsatfol{T}, \dbinst) = \false$. Therefore, by
  \Cref{def:ec-satisfaction}, and since
  $\ecset = \set{\qunsatfol{T} \ra \false}$,
  we have $\dbinst$ satisfies $\ecset$.
\end{proof}


\subsection{Rewriting \texorpdfstring{\muladom}{History Preserving ^^ce^^bc-Calculus with ECQ-Query} Formulas}

We extend the perfect reformulation algorithm to \muladom formulas as
follows.

\clearpage
\begin{definition}[Perfect Reformulation of \muladom Formula]\label{def:perf-ref-muladom}
  Given \sidetextb{Perfect Reformulation of \muladom Formula} a KB
  $\tup{T, A}$, and a \muladom formula $\Phi$ over $\tup{T, A}$, a
  \emph{perfect reformulation} of $\Phi$ w.r.t.\ $T$
  is $\Phi' = \rew(\Phi, T)$, where $\rew(\Phi, T)$ is inductively
  defined as follows:
\[
  \rew(\Phi, T) = 
  \left\{ 		
    \begin{array}{rcl}
      & \rew(Q, T)   &\mbox{if}\ {\Phi = Q} \\
      &  \rew(\Psi_1, T) \vee \rew(\Psi_2, T)   & \mbox{if}\ {\Phi = \Psi_1 \vee \Psi_2} \\
      &  \exists x.\rew(\Psi, T)   & \mbox{if}\ {\Phi = \exists x. \Psi} \\
      &  \DIAM{\rew(\Psi, T)}   & \mbox{if}\ {\Phi = \DIAM{\Psi}} \\
      &  \mu Z.\rew(\Psi, T)   & \mbox{if}\ {\Phi = \mu Z.\Psi} \\
    \end{array}
  \right.
\]
and $\rew(Q, T)$ is the application of the perfect reformulation
algorithm over $Q$ w.r.t.\ $T$.
\end{definition}
\noindent
The definition above intuitively said that the rewriting of \muladom
formula $\Phi$ is obtained by rewriting each query in $\Phi$ w.r.t.\
$T$ using the perfect reformulation algorithm while maintaining the
temporal operators unaltered.


\subsection{Recasting the Verification of KABs into DCDSs}



The idea to recast the problem of $\muladom$ verification over KABs
into the problem of $\mula$ verification over DCDSs is as follows:
\begin{compactenum}
\item We define a bisimulation relation namely \emph{KAB-DCDS
    Bisimulation (KD-Bisimulation)} and show some interesting
  properties of KD-Bisimulation relation which are related to
  satisfiability of \muladom formulas.

\item We show that the transition system of a KAB $\kabsym$ and the
  transition system of its corresponding DCDS $\tdcds(\kabsym)$
  (obtained through translation $\tdcds$) are bisimilar w.r.t.\ the
  KD-Bisimulation relation,
 


\item Making use the ingredients obtained from the point 1 and 2, we
  show that a KAB $\kabsym$ satisfies a \muladom property $\Phi$ if and
  only if its corresponding DCDS $\tdcds(\kabsym)$ (obtained through
  translation $\tdcds$) satisfies the \mula property $\rew(\Phi, T)$
  (where $T$ is the TBox in the given KAB $\kabsym$). 

\end{compactenum}





We define a KD-Bisimulation relation between a KB transition system
and a database transition system as follows.
%

\begin{definition}[KAB-DCDS
  Bisimulation (KD-Bisimulation)]\label{def:kab-dcds-bisimulation}
  Let \sidetext{KAB-DCDS Bisimulation (KD-Bisimulation)}
  $\ts{1} = \tup{\const, T, \stateset_1, s_{01}, \abox, \trans_1}$ be
  a KB transition system (see \Cref{def:kb-ts})
  and
  $\ts{2} = \tup{\const,\dbschema,\dstateset_2, s_{02},
    \db,\dtrans_2}$
  be a database transition system (see \Cref{def:database-ts}), with
  $\adom{\abox(s_{01})} \subseteq \const$,
and $\adom{\db(s_{02})} \subseteq \const$.
%
%
A \emph{KD-Bisimulation} between $\ts{1}$ and $\ts{2}$ is a relation
$\B \subseteq \stateset_1 \times\Sigma_2$ such that
$\tup{s_1, s_2} \in \B$ implies that:
  \begin{compactenum}
  \item $\abox(s_1) = \db(s_2)$
  \item for each $s_1'$, if $s_1 \Rightarrow_1 s_1'$ then there exists
    $s_2'$ with $ s_2 \Rightarrow_2 s_2' $ such that
    $\tup{s_1',s_2'}\in\B$.
  \item for each $s_2'$, if $ s_2 \Rightarrow_2 s_2' $ then there
    exists $s_1'$ with $s_1 \Rightarrow_1 s_1'$ such that
    $\tup{s_1',s_2'}\in\B$.
 \end{compactenum}
\ \ 
\end{definition}

Given a KB transition system
$\ts{1} = \tup{\const, T, \stateset_1, s_{01}, \abox, \trans_1}$ and a
database transition system
$\ts{2} = \tup{\const,\dbschema,\dstateset_2, s_{02}, \db,\dtrans_2}$,
we say a state $s_1 \in \stateset_1$ is \emph{KD-bisimilar} to
$s_2 \in \dstateset_2$, written $s_1 \kdbsim s_2$, if there exists a
KD-Bisimulation $\B$ between $\ts{1}$ and $\ts{2}$ such that
$\tup{s_1,s_2}\in\B$.
Moreover, a transition system $\ts{1}$ is \emph{KD-bisimilar} to
$\ts{2}$, written $\ts{1} \kdbsim \ts{2}$, if there exists a
KD-Bisimulation $\B$ between $\ts{1}$ and $\ts{2}$ such that
$\tup{s_{01},s_{02}}\in\B$.

Now, we show some important properties of KD-Bisimulation relation
w.r.t.\ satisfiability of \muladom formulas as follows.


\begin{lemma}\label{lem:kab-dcds-bisimilar-state} \ \ 
  Consider a KB transition system
  $\ts{1}$ $=$ $\tup{\const,T,\stateset_1,s_{01},\abox,\trans_1}$ and a
  database transition system
  $\ts{2}=\tup{\const,\dbschema,\dstateset_2,s_{02},\db,\dtrans_2}$
  with $\adom{\abox(s_{01})} \subseteq \const$, and
  $\adom{\db(s_{02})} \subseteq \const$. Consider also two states
  $s_1 \in \stateset_1$ and $s_2 \in \dstateset_2$ such that
  $s_1 \kdbsim s_2$. Then for every (open) \muladom formula $\Phi$, and
  for every valuations $\vfo_1$ and $\vfo_2$ that assign to each of
  its free variables a constant $c_1 \in \adom{\abox(s_1)}$ and
  $c_2 \in \adom{\db(s_2)}$, such that $c_1 = c_2$,
  we have that
  \[
  \ts{1},s_1 \models \Phi \vfo_1 \textrm{ if and only if } \ts{2},s_2
  \models \rew(\Phi, T) \vfo_2.
  \]
\end{lemma}
\begin{proof}
  The proof is then organized in three parts:
\begin{compactenum}[(1)]
\item We prove the claim for formulae of $\ladom$, obtained from
  $\muladom$ by dropping the predicate variables and the fixpoint
  constructs. $\ladom$ corresponds to a first-order variant of the
  Hennessy Milner logic, and its semantics does not depend on the
  second-order valuation.
\item We extend the results to the infinitary logic obtained by
  extending $\ladom$ with arbitrary countable disjunction.
\item We recall that fixpoints can be translated into this infinitary
  logic (cf. \cite{Stir01}), thus proving that the theorem holds for $\muladom$.
\end{compactenum}

\smallskip
\noindent
\textbf{Proof for $\ladom$.}  We proceed by induction on the structure
of $\Phi$, without considering the case of predicate variable and of
fixpoint constructs, which are not part of $\ladom$.

\smallskip
\noindent
\textit{Base case:} 
\begin{compactitem}
\item[\textbf{($\Phi = Q$).}] 
%
%
  Since $s_1 \kdbsim s_2$, we have $\abox(s_1) = \db(s_2)$. Hence, by
  \Cref{thm:FO-rew-ECQ}, for every valuations $\vfo_1$ and $\vfo_2$
  that assign to each of its free variables a constant
  $c_1 \in \adom{\abox(s_1)}$ and $c_2 \in \adom{\db(s_2)}$, such that
  $c_1 = c_2$, we have
  $\Ans(Q\vfo_1,T,\abox(s_1)) = \ANS(\rew(Q,T)\vfo_1, \abox(s_1)) =
  \ANS(\rew(Q,T)\vfo_2, \db(s_2))$. Thus we have
  \[
  \ts{1},s_1 \models Q\vfo_1 \textrm{ \ if and only if \ } \ts{2},s_2 \models
  \rew(Q, T)\vfo_2
  \]
\end{compactitem}


\smallskip
\noindent
\textit{Inductive step:}
\begin{compactitem}
\item[\textbf{($\Phi = \Psi_1 \vee \Psi_2$)}.]  
  we have $\ts{1},s_1 \models (\Psi_1\vee \Psi_2) \vfo_1$ if and only
  if either $\ts{1},s_1 \models \Psi_1 \vfo_1$ or
  $\ts{1},s_1 \models \Psi_2 \vfo_1$.  By induction hypothesis, for
  every (open) \muladom formula $\Psi$, and
  for every valuations $\vfo_1$ and $\vfo_2$ that assign to each of
  its free variables a constant $c_1 \in \adom{\abox(s_1)}$ and
  $c_2 \in \adom{\db(s_2)}$, such that $c_1 = c_2$, we have
  \begin{compactitem}
  \item
    $ \ts{1},s_1 \models \Psi_1 \vfo_1 \textrm{ if and only if }
    \ts{2},s_2 \models \rew(\Psi_1, T)\vfo_2 $, and also
  \item
    $ \ts{1},s_1 \models \Psi_2 \vfo_1 \textrm{ if and only if }
    \ts{2},s_2 \models \rew(\Psi_2, T)\vfo_2$
  \end{compactitem}
Hence, $\ts{1},s_1 \models \Psi_1 \vfo_1$ or
$\ts{1},s_1 \models \Psi_2 \vfo_1$ if and only if
$\ts{2},s_2 \models \rew(\Psi_1, T) \vfo_2$ or
$\ts{2},s_2 \models \rew(\Psi_2, T) \vfo_2$. Therefore we have
  $ \ts{1},s_1 \models (\Psi_1 \vee \Psi_2)\vfo_1 \textrm{ if and
    only if } \ts{2},s_2 \models (\rew(\Psi_1, T) \vee \rew(\Psi_2,
  T))\vfo_2 $. 
  Since
  $\rew(\Psi_1 \vee \Psi_2, T) = \rew(\Psi_1, T) \vee \rew(\Psi_2,
  T)$, we have
\[
\ts{1},s_1 \models (\Psi_1 \vee \Psi_2)\vfo_1 \textrm{ if and only if
} \ts{2},s_2 \models \rew(\Psi_1\vee \Psi_2)\vfo_2
  \]

\item[\textbf{($\Phi = \DIAM{\Psi}$)}.] Assume
  $\ts{1},s_1 \models (\DIAM{\Psi}) \vfo_1$, where $\vfo_1$ is a
  valuation that assigns to each free variable of $\Psi$ a constant
  $c_1 \in \adom{\abox(s_1)}$. Then there exists $s_1'$
  s.t.\ $s_1 \trans_1 s_1'$ and $\ts{1},s_1' \models \Psi \vfo_1$.
  Since $s_1 \kdbsim s_2$, there exists $s_2'$ such that
  $ s_2 \Rightarrow_2 s_2' $ and $s_1' \kdbsim s_2'$.
  Hence, by induction hypothesis, for every valuations $\vfo_2$ that
  assign to each free variable $x$ of $\rew(\Psi, T)$ a constant
  $c_2 \in \adom{\db(s_2)}$, such that $c_1 = c_2$,
  we have $ \ts{2},s_2' \models \rew(\Psi_1, T) \vfo_2$.  
  Consider that $ s_2 \Rightarrow_2 s_2', $ we therefore get
  $ \ts{2},s_2 \models (\DIAM{ \rew(\Psi, T)} )\vfo_2 $.  Since
  $\rew(\DIAM{\Psi}, T) = \DIAM{\rew(\Psi, T)}$, we have $\ts{2},s_2
  \models (\DIAM{ \rew(\Psi, T)})\vfo_2$. 
%
  The other direction can be shown in a similar way.

\item[\textbf{($\Phi = \exists x. \Psi$)}.]  Assume that
  $\ts{1},s_1 \models (\exists x. \Psi)\vfo'_1$, where $\vfo'_1$ is a
  valuation that assigns to each free variable of $\Psi$ a constant
  $c_1 \in \adom{\abox(s_1)}$. Then, by definition, there exists
  $c \in \adom{\abox(s_1)}$ such that $\ts{1},s_1 \models \Psi\vfo_1$,
  where $\vfo_1 = \vfo'_1[x/c]$.
  By induction hypothesis, for every valuation $\vfo_2$ that assigns
  to each free variable $y$ of $\rew(\Psi, T)$ a constant
  $c_2 \in \adom{\db(s_2)}$, such that $c_2 = c_1$ with
  $y/c_1 \in \vfo_1$, we have that
  $\ts{2},s_2 \models \rew(\Psi, T) \vfo_2$. Additionally,
  $\vfo_2 = \vfo'_2[x/c']$, where $c' \in \adom{\db(s_2)}$, and
  $c' = c$ because $\db(s_2) = \abox(s_1)$.  Hence, we get
  $\ts{2},s_2 \models (\exists x. \rew(\Psi, T))\vfo'_2$. Furthermore,
  since $\rew(\exists x. \Psi, T) = (\exists x. \rew(\Psi, T))$, we
  have $\ts{2},s_2 \models (\rew(\exists x. \Psi, T))\vfo_2'$.
  The other direction can be shown similarly.

\end{compactitem}

\smallskip
\noindent
\textbf{Extension to arbitrary countable disjunction.}  Let $\Psi$ be
a countable set of $\ladom$ formulae. Given either a KB transition
system $\ts{} = \tup{\const,\T,\stateset,s_{0},\abox,\trans}$ (or a
database transition system
$\ts{}=\tup{\const,\dbschema,\dstateset_2,s_{02},\db,\dtrans_2}$), the
semantics of $\bigvee \Psi$ is
$(\bigvee \Psi) _\vfo^{\ts{}} = \bigcup_{\psi \in \Psi}
(\psi)_\vfo^{\ts{}}$.
Therefore, given a state $s \in \Sigma$ we have
$\ts{}, s \models (\bigvee \Psi)\vfo$ if and only if there exists
$\psi \in \Psi$ such that $\ts{}, s \models \psi\vfo$. Arbitrary
countable conjunction is obtained for free because of negation.


  Now, let $\ts{1}$ $=$
  $\tup{\const,T,\stateset_1,s_{01},\abox,\trans_1}$ and
  $\ts{2}=\tup{\const,\dbschema,\dstateset_2,s_{02},\db,\dtrans_2}$.
  Consider two states $s_1 \in \stateset_1$ and $s_2 \in \stateset_2$
  such that $s_1 \kdbsim s_2$.
  By induction hypothesis, we have for every valuations $\vfo_1$ and
  $\vfo_2$ that assign to each free variable of $\bigvee \Psi$ a constant $c_1 \in \adom{\abox(s_1)}$ and
  $c_2 \in \adom{\db(s_2)}$, such that $c_2 = c_1$, we have that for
  every formula $\psi \in \Psi$, it holds
  $\ts{1}, s_1 \models \psi \vfo_1$ if and only if
  $\ts{2}, s_2 \models \rew(\psi, T)\vfo_2$.
Given the semantics of $\bigvee \Psi$ above, this implies that
$\ts{1}, s \models (\bigvee \Psi) \vfo_1$ if and only if $\ts{2}, s
\models (\bigvee \rew(\Psi, T)) \vfo_2$, where $\rew(\Psi, T) =
\{\rew(\psi, T) \mid \psi \in \Psi\}$. The proof is then obtained by
observing that $\bigvee \rew(\Psi, T) = \rew(\bigvee \Psi, T)$.

\smallskip
\noindent
\textbf{Extension to full $\muladom$.}  In order to extend the result
to the whole \muladom, we resort to the well-known result stating that
fixpoints of the $\mu$-calculus can be translated into the infinitary
Hennessy Milner logic by iterating over \emph{approximants}, where the
approximant of index $\alpha$ is denoted by $\mu^\alpha Z.\Phi$
(resp.~$\nu^\alpha Z.\Phi$) (cf. \cite{Stir01}). This is a standard
result that also holds for \muladom. In particular, approximants are
built as follows:
\[
\begin{array}{rl rl}
  \mu^0 Z.\Phi & = \false
  &  \nu^0 Z.\Phi & = \true\\
  \mu^{\beta+1} Z.\Phi & = \Phi[Z/\mu^\beta Z.\Phi]
  & \nu^{\beta+1} Z.\Phi & = \Phi[Z/\nu^\beta Z.\Phi]\\
  \mu^\lambda Z.\Phi & = \bigvee_{\beta < \lambda} \mu^\beta Z. \Phi &
  \nu^\lambda Z.\Phi & = \bigwedge_{\beta < \lambda} \nu^\beta Z. \Phi
\end{array}
\]
where $\lambda$ is a limit ordinal, and where fixpoints and their
approximants are connected by the following properties: given a
transition system $\ts{}$ and a state $s$ of $\ts{}$
\begin{compactitem}
\item $s \in \MODA{\mu Z.\Phi}$ if and only if there exists an ordinal
  $\alpha$ such that $s \in \MODA{\mu^\alpha Z.\Phi}$ and, for every
  $\beta < \alpha$, it holds that $s \notin \MODA{\mu^\beta Z.\Phi}$;
\item $s \notin \MODA{\nu Z.\Phi}$ if and only if there exists an
  ordinal $\alpha$ such that $s \notin \MODA{\nu^\alpha Z.\Phi}$ and,
  for every $\beta < \alpha$, it holds that $s \in \MODA{\nu^\beta
    Z.\Phi}$.
\end{compactitem}

\end{proof}

Having \Cref{lem:kab-dcds-bisimilar-state} in hand, we can easily show
the following important theorem which essentially says that given a KB
transition system $\ts{1}$ that is KD-bisimilar to a database
transition system $\ts{2}$, we have that $\ts{1}$ satisfies a \muladom
property $\Phi$ if and only if $\ts{2}$ satisfies a \mula property
$\Phi'$ that is obtained from $\Phi$ through perfect reformulation
algorithm.



\begin{theorem}\label{thm:kab-dcds-bisimulation-property} 
  Consider a KB transition system
  $\ts{1}=\tup{\const,T,\stateset_1,s_{01},\abox,\trans_1}$ and a
  database transition system $\ts{2}$
  such that $\ts{1} \kdbsim \ts{2}$. For every \muladom closed formula
  $\Phi$, we have:
\[
\ts{1} \models \Phi \mbox{ \ if and only if \ } \ts{2} \models \rew(\Phi, T)
\]
\end{theorem}
\begin{proof}
  Let
  $\ts{2}=\tup{\const,\dbschema,\dstateset_2,s_{02},\db,\dtrans_2}$,
  Since $\ts{1} \kdbsim \ts{2}$, then we have $s_{01} \kdbsim
  s_{02}$. Hence, by \Cref{lem:kab-dcds-bisimilar-state} we have 
  \[
  \ts{1},s_{01} \models \Phi \mbox{ \ if and only if \ } \ts{2},s_{02}
  \models \rew(\Phi, T),
  \]
  which prove the claim.
\end{proof}


Now we are going to show that the transition system of a KAB $\kabsym$
and the transition system of its corresponding DCDS $\tdcds(\kabsym)$
(obtained from $\kabsym$ via $\tdcds$) are
KD-bisimilar.  
%
As the first step, in \Cref{lem:kab-dcds-bisimilar-state}, we show
that given a state of a KAB transition system and a state of its
corresponding DCDS transition system such that those two states
contain the same data (i.e., their corresponding ABox and database
instance are equal) as well as the same service call map, we have that
they are KD-bisimilar.



\begin{lemma}\label{lem:kab-dcds-bisimilar}
  Let $\kabsym$ be a KAB with transition system $\ts{\kabsym}$, and
  let $\tdcds(\kabsym)$ be a DCDS obtained from $\kabsym$ through
  $\tdcds$ with transition system $\ts{\tdcds(\kabsym)}$. Consider a
  state $\tup{A, \scmap}$ of $\ts{\kabsym}$ and a state
  $\tup{\dbinst, \dscmap_d}$ of $\ts{\tdcds(\kabsym)}$. If
  $A = \dbinst$ and $\scmap = \dscmap_d$, then
  $\tup{A, \scmap} \kdbsim \tup{\dbinst, \dscmap_d}$.
\end{lemma}
\begin{proof}
Let
\begin{compactitem}

\item $\kabsym = \tup{T, \initabox, \actset, \procset}$, and 
  $\ts{\kabsym} = \tup{\const, T, \stateset, s_0, \abox, \trans}$,

\item $\tdcds(\kabsym) = \tup{\dcomp, \pcomp}$, 
  where $\dcomp = \tup{\dbschema,\idb, \ecset}$, and
  $\pcomp = \tup{\dactset,\dprocset}$. Additionally, let
  $\ts{\tdcds(\kabsym)} = \tup{\const,\dbschema,\dstateset,s_0,
    \db,\dtrans}$. 

\end{compactitem}

To prove the lemma, we show that for every state $\tup{A', \scmap'}$
such that $\tup{A, \scmap} \trans \tup{A', \scmap'}$, there exists a
state $\tup{\dbinst', \dscmap_d'}$ such that
\begin{compactenum}
\item $\tup{\dbinst, \dscmap_d} \trans \tup{\dbinst', \dscmap_d'}$
\item $A' = \dbinst'$
\item $\scmap' = \dscmap_d'$
\end{compactenum}

By definition of $\ts{\kabsym}$ (\Cref{def:KAB-standard-ts}), since
$\tup{A, \scmap} \trans \tup{A', \scmap'}$, then there exists
$\act \in \actset$, and a substitution $\sigma$ for parameters of
$\act$ such that
$\tup{A,\scmap} \exect{\act\sigma, \ \kabsym\ } \tup{A',\scmap'}$ and
$A'$ is $T$-consistent. Moreover, by definition of
$\exect{\act\sigma, \ \kabsym\ }$ (see \Cref{def:KAB-exec-relation}),
we have the following:

\begin{compactenum}
\item $\sigma$ is a legal parameter assignment for $\act$ in state
  $\tup{A,\scmap}$, and additionally by
  \Cref{def:kab-action-executability}, there exists a condition-action
  rule $\carule{Q}{\act} \in \procset$ such that $\Ans(Q\sigma, T, A)$
  is true;
\item there exists $\theta \in \eval{T, A,\act\sigma}$ such that
  $\theta$ and $\scmap$ ``agree'' on the common values in their
  domains;
\item $A' = \doo{T, A, \act\sigma}\theta$; and
\item $\scmap' = \scmap \cup \theta$ (i.e., updating the history of
  issued service calls).
\end{compactenum}

By definition of $\tdcds$ (\Cref{def:translation-kab-to-dcds}) we have the following:
\begin{compactitem}
\item there exists a corresponding action $\dact' \in \dactset$ which
  is obtained from $\act \in \actset$, 
\item there exists a corresponding condition-action rule
  $\carule{Q'}{\dact'} \in \dprocset$ (where $Q' = \rew(Q, T)$) which
  is obtained from $\carule{Q}{\act} \in \procset$
\end{compactitem}

Since $A = \dbinst$, by \Cref{thm:FO-rew-ECQ}, we have
\[
\Ans(Q,T,A) = \ANS(\rew(Q, T), A) = \ANS(Q', \dbinst).
\]
Hence, by \Cref{def:DCDS-act-executability}, 
$\sigma$ is also a legal parameter assignment for $\act'$ in state
$\tup{\dbinst, \dscmap_d}$.
Therefore, by \Cref{lem:dokab-equal-dodcds}, we have
$ \doo{T, A, \act\sigma} = \ddoo{ \dbinst, \act'\sigma}$. Thus, by
\Cref{lem:kabeval-equal-dcdseval}, we have
$\eval{T, A,\act\sigma} = \eval{\dbinst,\act'\sigma}$, and hence we
have $\theta \in \eval{\dbinst,\act'\sigma}$. As a consequence, we
have
\begin{compactenum}
\item
  $A' = \doo{T, A, \act\sigma}\theta = \ddoo{ \dbinst, \act'\sigma}
  \theta = \dbinst'$.
\item $\scmap' = \dscmap_d'$, because $\dscmap_d = \scmap$,
  $\scmap' = \scmap \cup \theta$, and
  $\dscmap_d' = \dscmap_d \cup \theta$.
\end{compactenum}
Furthermore, it is easy to see that we have
$\tup{\dbinst,\dscmap} \exect{\act'\sigma, \ \dcdssym\ }
\tup{\dbinst',\dscmap'}$,
and since $A' = \dbinst'$ as well as $A'$ is $T$-consistent, by
\Cref{lem:qunsat-equal-ec} we have $\dbinst'$ satisfies
$\ecset$. Hence by \Cref{def:dcds-ts}, we have
$\tup{\dbinst,\dscmap_d} \dtrans \tup{\dbinst',\dscmap_d'}$.

The other direction of bisimulation can be shown similarly.
\end{proof}

Using \Cref{lem:kab-dcds-bisimilar-state} above, now we show that
the transition system of a KAB $\kabsym$ is KD-bisimilar with the
transition system of the corresponding DCDS $\tdcds(\kabsym)$ (that is
obtained from $\kabsym$ via $\tdcds$) as follows.

\begin{theorem}\label{thm:kab-to-dcds-bisimilar-ts}
  Given a KAB $\kabsym$ with transition system $\ts{\kabsym}$, and let
  $\tdcds(\kabsym)$ be a DCDS obtained from $\kabsym$ through $\tdcds$
  with transition system $\ts{\tdcds(\kabsym)}$. 
  We have $ \ts{\kabsym} \kdbsim \ts{\tdcds(\kabsym)} $
\end{theorem}
\begin{proof}
  Let
\begin{compactitem}
\item $\kabsym = \tup{T, \initabox, \actset, \procset}$, and 
  $\ts{\kabsym} = \tup{\const, T, \stateset, s_{0k}, \abox, \trans}$,

\item $\tdcds(\kabsym) = \tup{\dcomp, \pcomp}$, where
  $\dcomp = \tup{\dbschema,\idb, \ecset}$, and
  $\pcomp = \tup{\dactset,\dprocset}$. Additionally, let
  $\ts{\tdcds(\kabsym)} = \tup{\const,\dbschema,\dstateset,s_{0d},
    \db,\dtrans}$.
\end{compactitem}
By \Cref{def:KAB-standard-ts}, we have
$s_{0k} = \tup{\initabox, \emptyset}$, and by \Cref{def:dcds-ts}, we
have $s_{0d} = \tup{\idb, \emptyset}$.  Furthermore, by the definition
of $\tdcds$ (\Cref{def:translation-kab-to-dcds}), we have
$\initabox = \idb$.  Hence by \Cref{lem:kab-dcds-bisimilar-state}, we
have $s_{0k} \kdbsim s_{0d}$.  Thus, we have
$ \ts{\kabsym} \kdbsim \ts{\tdcds(\kabsym)} $ \ \
\end{proof}


Having the fact that the transition system of a KAB $\kabsym$ is
KD-bisimilar with the transition system of the corresponding DCDS
obtained from $\kabsym$ via translation $\tdcds$, by exploiting
\Cref{thm:kab-dcds-bisimulation-property} we can finally show that the
verification of \muladom properties over KABs can be reduced to the
verification of \mula properties over DCDSs as follows.

\begin{theorem}\label{thm:verification-reduction-kab-dcds}
  Given a KAB $\kabsym$ 
  and a closed \muladom formula $\Phi$, we have
\[
\ts{\kabsym} \models \Phi \quad\mbox{if and only if}\quad
\ts{\tdcds(\kabsym)} \models \rew(\Phi, T)
\]
\end{theorem}
\begin{proof}
  By \Cref{thm:kab-to-dcds-bisimilar-ts}, we have
  $ \ts{\kabsym} \kdbsim \ts{\tdcds(\kabsym)} $. Hence, the claim is
  directly follows from \Cref{thm:kab-dcds-bisimulation-property}.
\end{proof}

\noindent
Finally, \Cref{thm:verification-reduction-kab-dcds} shows
that we can tackle the problem of \muladom verification over KABs by
reducing it into the problem of \mula verification over
DCDSs. 

\subsection{Verification of Run-Bounded KABs}

As in DCDSs, in general the verification of KABs is
undecidable. However, we can use the semantic
restriction that was originally proposed in~\cite{BCDDM13}, namely
\emph{run-boundedness} in order to gain decidability. To introduce the
notion of run-boundedness, we first define the notion of run of a KAB
transition system as follows.


\begin{definition}[Run of a KAB Transition System]\label{def:run-of-kab}
  Given \sidetextb{Run of a KAB Transition System} a KAB $\kabsym$, a \emph{run of
    $\ts{\kabsym} = \tup{\const, T, \stateset, s_{0}, \abox, \trans}$}
  is a (possibly infinite) sequence $s_0s_1\cdots$ of states of
  $\ts{\kabsym}$ such that $s_i\trans s_{i+1}$, for all $i\geq 0$.
\end{definition}

\begin{definition}[Run-bounded KAB]\label{def:run-bounded-kab}
  Given \sidetext{Run-bounded KAB} a KAB $\kabsym$, we say
  \emph{$\kabsym$ is run-bounded} if there exists an integer bound $b$
  such that for every run $\pi = s_0s_1\cdots$ of $\ts{\kabsym}$, we
  have that
  $\card{\bigcup_{s \textrm{ state of } \pi}\adom{\abox(s)}} < b$.
\end{definition}

\noindent
Intuitively, run-boundedness requires that every run in the transition
system cumulatively encounters at most a bounded number of
constants. Unboundedly many constants can still be present in the
overall system, provided that they do not accumulate in the same run.

\begin{lemma}\label{lem:tdcds-preserve-runboundedness}
  Given a run-bounded KAB $\kabsym$, the DCDS $\tdcds(\kabsym)$ 
  is run-bounded.
\end{lemma}
\begin{proof}
  Let
\begin{compactitem}
\item $\kabsym = \tup{T, \initabox, \actset, \procset}$, and
  $\ts{\kabsym} = \tup{\const, T, \stateset, s^k_{0}, \abox, \trans}$
  be its corresponding transition system,

\item $\tdcds(\kabsym) = \tup{\dcomp, \pcomp}$, 
  and
  $\ts{\tdcds(\kabsym)} = \tup{\const,\dbschema,\dstateset,s^d_{0},
    \db,\dtrans}$ be its corresponding transition system,

\item $\pi_k = s^k_0s^k_1\cdots$ be an arbitrary run of
  $\ts{\kabsym}$.
\end{compactitem}
Since, $\kabsym$ is run-bounded, we have that there exists an integer
bound $b$ such that
$\card{\bigcup_{s^k \textrm{ state of } \pi_k}\adom{\abox(s^k)}} < b$.
By \Cref{thm:kab-to-dcds-bisimilar-ts} we have that
$\ts{\kabsym} \kdbsim \ts{\tdcds(\kabsym)}$. As a consequence, there
exists a corresponding run $\pi_d = s^d_0s^d_1\cdots$ in
$\ts{\tdcds(\kabsym)}$ such that $\abox(s^k_i) = \db(s^d_i)$ (for
$i = 0, 1, \ldots$). Thus we get that there exists an integer bound
$b$ such that
$\card{\bigcup_{s^d \textrm{ state of } \pi_d}\adom{\db(s^d)}} < b$.
The proof is then completed by noticing that 
\begin{compactenum}
\item $\pi_k$ is an arbitrary run of $\ts{\kabsym}$, and
\item because $\ts{\kabsym} \kdbsim \ts{\tdcds(\kabsym)}$, for each run
  $\pi_d = s^d_0s^d_1\cdots$ in $\ts{\tdcds(\kabsym)}$ there exists a
  corresponding run $\pi_k = s^k_0s^k_1\cdots$ in $\ts{\kabsym}$ such
  that $\abox(s^k_i) = \db(s^d_i)$ (for $i = 0, 1, \ldots$).
\end{compactenum}
\ \ 
\end{proof}

\noindent
Finally, we can state the final result on verification of \muladom
over run-bounded KAB.

\begin{theorem}[Verification of \muladom over run-bounded KAB]\label{thm:verification-run-bounded-kab}
  Verification of closed $\muladom$ formulas over run-bounded KAB is
  decidable and can be reduced to finite-state model checking.
\end{theorem}
\begin{proof}
  From \Cref{thm:verification-reduction-kab-dcds} and
  \Cref{lem:tdcds-preserve-runboundedness}, we have that verification
  of closed \muladom formulas over run-bounded KAB can be reduced to
  the verification of \mula formulas over run-bounded DCDS. Then, by
  \Cref{thm:verification-dcds}, we have that verification of $\mula$
  over run-bounded DCDS is decidable and can be reduced to
  finite-state model checking.
\end{proof}


\section{Discussion: Weakly Acyclic KABs}

Studying various kind of restrictions to obtain decidability is out of
the scope of this thesis. However, in the following we provide some
discussions on the condition for obtaining decidability of
verification, in particular on the notion of \emph{weak-acyclicity}
that is a syntactic condition which implies (guarantees) run-boundedness.

As we have seen above, to get decidability of verification, in this
thesis we rely on the assumption that the considered KAB is
run-bounded.  
%
In \cite{BCDDM13}, a sufficient syntactic condition borrowed from
\emph{weak acyclicity} in data exchange \cite{FKMP05} has been
studied.
Such condition is shown to guarantee run boundedness under the
assumption that the service calls are deterministic. 

We can recast such notion of weak acyclicity that was studied in DCDSs
into KABs such that if we have a KAB $\kabsym$ is weak acyclic, then
$\kabsym$ is run bounded.
%
%
Intuitively, given a KAB $\kabsym$, this weak acyclicity test
constructs a \emph{dependency graph} tracking how the actions of
$\kabsym$ transport values from one state to the next one. To track
all the actual dependencies, every involved query is first rewritten
considering the positive inclusion assertions of the TBox. Two types
of dependencies are tracked:
%
\begin{compactenum}
\item copy of values and
\item use of values as parameters of a service call.
\end{compactenum}
%
The KAB $\K$ is said to be \emph{weakly acyclic} if there is no cyclic
chain of dependencies of the second kind. 
Intuitively, the presence of such a cycle 
could produce an infinite chain of fresh values generation
through service calls. 
Thus, such a cycle could destroy run-boundedness. In other word, a
non-weakly acyclic KAB contains at least a service that might be
called repeatedly, and each call is using fresh values that are either
directly or indirectly obtained by manipulating the previous result
that is produced by the same service.
Note that this notion of weak acyclicity is the same as the one in
\cite{BCMD*13}, that is used by \cite{BCMD*13} to get decidability of
verification. 

%% file: 2.chapters/4-gkab.tex
\chapter{Golog-KAB\lowercase{s} (GKAB\lowercase{s})}\label{ch:gkab}

\ifhidecontent
 
\fi

Knowledge and Action Bases (KABs) have been put forward as a framework
which provides a semantically rich representation of a domain that
also simultaneously takes into account the dynamic aspects of the
modeled system. However, KABs lack of a convenient way to specify
processes at a high-level of abstraction. To cope with this situation,
we enrich KABs with a high-level, compact action language inspired by
Golog~\cite{LRLLS97}. We call Golog-KAB (GKAB) the KAB enhanced with a
Golog-like programming language.
Additionally, here we also introduce 
a parametric execution semantics for GKABs, so as to elegantly
accomodate various way of updating an ABox. We will see later in the
next chapter that the parametric execution semantics allow us to
incorporate various inconsistency-aware execution semantics for GKABs.

In the following, we use \dllitea for expressing knowledge bases and
we also do not distinguish between objects and values (thus we drop
attributes).
Furthermore, we also make use of a countably infinite set $\const$ of
constants, which intuitively denotes all possible values in the
system.
Additionally, we consider a finite set of distinguished constants
$\iconst \subset \const$, and 
a finite set $\servcall$ of \textit{function symbols} that represents
\textit{service calls}, which abstractly account for the injection of
fresh values (constants) from $\const$ into the system.
The results in this chapter are published in
\cite{AS-CORR-15,AS-IJCAI-15,AS-DL-15}.

\section{GKABs Formalism}

In this section, 
%
%
we enrich KABs (cf. \Cref{ch:kab}) with a high-level action language
inspired by Golog~\cite{LRLLS97}.  This allows modelers to represent
processes in a much more intuitive and compact way.
%
In this thesis, we consider a variant 
of Golog 
that has been tailored to work on KBs based on \cite{CDLR11}.
%
%
Formally, a \emph{Golog program} is inductively defined as follows:

\begin{definition}[Golog Program]\label{def:golog-program}
  Given \sidetext{Golog Program} a set of KAB actions $\actset$ (see
  \Cref{def:kab-action}), a \emph{Golog program} $\delta$ over
  $\actset$ is an expression formed by the following grammar: \vspace*{-1.3mm}
\begin{center}
$
\begin{array}{@{}r@{\ }l@{\ }}
  \delta ::= &
  \gemptyprog ~\mid~
  \gact{Q(\vec{p})}{\act(\vec{p})} ~\mid~
  \delta_1|\delta_2  ~\mid~
  \delta_1;\delta_2 ~\mid~ \\
  &\gif{\varphi}{\delta_1}{\delta_2} ~\mid~
  \gwhile{\varphi}{\delta}
\end{array}
$\vspace*{-1.3mm}
\end{center}
where:
\begin{compactitem}
\item $\gemptyprog$ is the \emph{empty program};
\item $\gact{Q(\vec{p})}{\act(\vec{p})}$ is an \emph{atomic action
    invocation} guarded by a \diecq $Q$, such that $\act\in\actset$ is
  applied by non-deterministically substituting its parameters
  $\vec{p}$ with an answer of $Q$. Additionally, $Q(\vec{p})$ might
  uses constants in $\iconst$;
\item $\delta_1|\delta_2$ is a \emph{non-deterministic choice} between
  programs;
\item $\delta_1;\delta_2$ is \emph{sequencing};
\item $\gif{\varphi}{\delta_1}{\delta_2}$ and
  $\gwhile{\varphi}{\delta}$ are \emph{conditional} and \emph{loop}
  constructs, using a boolean ECQ $\varphi$ as condition.
\end{compactitem}
\ \ 
\end{definition}


\noindent
Notice that we are able to simulate some of other Golog program
constructs by using the constructs above. We discuss such
possibilities in \Cref{sec:discussion}.

We then define the notion of Golog-KABs as follows.

\begin{definition}[Golog-KAB]\label{def:golog-kab}
  A \sidetext{Golog-KAB} \emph{Golog-KAB (GKAB)} is a tuple
  $\gkabsym~=~\tup{T,\initabox,\actset,\ginitprog}$, where
\begin{compactitem}
\item $T$ and $\initabox$ together form a satisfiable \dllitea KB
  $\tup{T, \initabox}$, where
    \begin{compactitem}
    \item $T$ is a \emph{\dllitea TBox} that captures the intensional
      aspects of the domain of interest;
    \item $\initabox$ is the \emph{initial \dllitea ABox}, describing
      the initial configuration of data;
  \end{compactitem}
\item $\actset$ is a finite set of \emph{KAB actions} (as in
  \Cref{def:kab-action}) over $\tup{T, \initabox}$ that evolve the
  ABox;
\item $\ginitprog$ is a Golog program over $\actset$, which
  characterizes the evolution of the GKAB over time, using the atomic
  actions in $\actset$.


  \end{compactitem}
  Additionally, we assume that $\adom{A_0} \subseteq \const_0$.
\end{definition}

\noindent
The crucial difference between a GKAB and a KAB is that a GKAB
specifies its processes using a Golog program instead of a set of
condition-action
rules. 

\begin{figure}[tbp]
\centering
\includegraphics[width=1.00\textwidth]{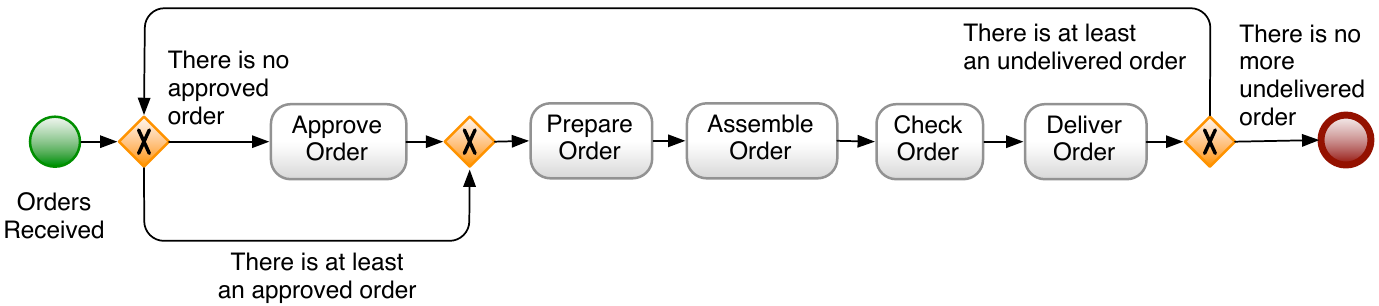}
\caption{Simple Order Processing Scenario in a Furniture
  Company \label{fig:gkab-example-bpmn}}
\end{figure}

\begin{example}\label{ex:gkab-run-ex}\textbf{An example of a GKAB.}\\
  Consider the furniture company order processing scenario described
  in \Cref{ex:example-scenario}. For the running example in this
  chapter, we slightly adjust the scenario as follows:
\begin{compactenum}
\item The order processing flows are followed strictly in a sequential
  manner (e.g., order preparation must be followed by order assembling).
\item In the beginning of the processing flow, in case there is
  already at least an approved order, the company immediately starts to
  process the order. Otherwise, the company must first approve an
  order.
\item While there is still an order that is not yet delivered, the
  processes will be repeated.
\end{compactenum}
The scenario that we consider here is visually
described in \Cref{fig:gkab-example-bpmn}.
%
  We then model this scenario by a GKAB
  $\gkabsym = \tup{T,\initabox,\actset,\ginitprog}$ where $T$,
  $\initabox$, $\actset$, and $\ginitprog$ are specified as follows.

  To capture the domain knowledge within this scenario, we consider
  the TBox $T$ that is specified in \Cref{ex:tbox-and-abox}. 
%
  As for the initial ABox, our GKAB $\gkabsym$ has the following
  initial ABox:
\[
\initabox = \set{\exo{ReceivedOrder}(\excon{chair}),
  \exo{ApprovedOrder}(\excon{table})}
\]
that basically contains a fact that there is an order of chair and
a fact that there is an approved order of table.

To model the progression mechanism in the scenario above, we consider
the set $\actset$ of actions that is specified in \Cref{ex:kab}.
The order processing flow in the scenario above is then captured by
the program $\delta$ of our GKAB $\gkabsym$, and it is specified as
follows:
\[
\delta = \gwhile{ \exists \exvar{x}.[\exo{Order}(\exvar{x})] \wedge \neg[\exo{DeliveredOrder}(\exvar{x})] }{\delta_0} 
\]
where:
\begin{compactitem}
\item $\delta_0 = \delta_1 ; \delta_2 ; \delta_3 ; \delta_4 ; \delta_5 $,
\item
  $\delta_1 = \gif{ \neg [\exists
    \exvar{x}.\exo{ApprovedOrder}(\exvar{x})] \\ \hspace*{15mm}}{
    \gact{[\exo{ReceivedOrder}(\exvar{x})]}{\exa{approveOrder}(\exvar{x})}
    \\ \hspace*{15mm}}{\gemptyprog}$,
\item $\delta_2 = \gact{\true}{\exa{prepareOrders}()}$,
\item $\delta_3 = \gact{\true}{\exa{assembleOrders}()}$,
\item $\delta_4 = \gact{\true}{\exa{checkAssembledOrders}()}$,
\item $\delta_5 = \gact{\true}{\exa{deliverOrders}()}$.
\end{compactitem}
The intuition of the program $\delta$ above is as follows:
\begin{compactitem}
\item
  $\gwhile{ \exists \exvar{x}.[\exo{Order}(\exvar{x})] \wedge
    \neg[\exo{DeliveredOrder}(\exvar{x})] }{\delta_0}$
  states the fact that as long as there exists an order that is not
  yet delivered, the program $\delta_0$, which basically processes the
  orders, will be executed.
\item $\delta_0$ specifies a consecutive sequence of programs
  ($\delta_1 ; \delta_2 ; \delta_3 ; \delta_4 ; \delta_5 $) where each
  of them captures a certain activity related to the order processing.
\item In $\delta_1$, if there does not exists any approved order, the
  program will pick up an order and then approve it by executing the
  action $\exa{approveOrder}/1$ otherwise it will perform nothing.
\item The programs $\delta_2, \delta_3, \delta_4, \delta_5$
  consecutively perform the order preparation, order assembling, order
  quality control, and order delivery. Each of them is performed by
  consecutively executing the actions $\exa{prepareOrders}/0$,
  $\exa{assembleOrders}/0$, $\exa{checkAssembledOrders}/0$, and
  $\exa{deliverOrders}/0$.
\end{compactitem}

\noindent
With a slightly abuse of notation, in order to ease the understanding,
below we provide another way to write the program above (note: we use
the curly braces (``$\{$'', ``$\}$'') to mark the scope of program
constructor):
\[
\begin{array}{l@{}l}
  &\gwhile{  \exists \exvar{x}.[\exo{Order}(\exvar{x})] \wedge
    \neg[\exo{DeliveredOrder}(\exvar{x})]  }{ \{ \\
    & \hspace*{7mm}\gif{ \neg [\exists
      \exvar{x}.\exo{ApprovedOrder}(\exvar{x})] \\
      &  \hspace*{11mm} } { \{ \gact{[\exo{ReceivedOrder}(\exvar{x})]}{\exa{approveOrder}(\exvar{x})}  \} \\
      & \hspace*{11mm} } { \{ \gemptyprog \}}; \\
    &\hspace*{7mm}\gact{\true}{\exa{prepareOrders}()};\\
    &\hspace*{7mm}\gact{\true}{\exa{assembleOrders}()};\\
    &\hspace*{7mm}\gact{\true}{\exa{checkAssembledOrders}()};\\
    &\hspace*{7mm}\gact{\true}{\exa{deliverOrders}()}\\
&\}}
\end{array}
\]
\ \ 
\end{example}

\section{GKABs Standard Execution Semantics} 

Similar to standard KABs, the execution semantics of a GKAB
$\gkabsym = \tup{T,\initabox,\actset,\ginitprog}$ is given in terms of
a possibly infinite-state KB transition system
$\ts{\gkabsym}~=~\left<\const,T,\stateset,s_0,\abox,\trans\right>$, whose
states contain ABoxes (see \Cref{def:kb-ts} for the detail of
$\ts{\gkabsym}$ components).
%
However, differently from KABs, the states we consider are tuples of
the form $\tup{A,\scmap,\delta}$, where $A$ is an ABox, $\scmap$ a
service call map (see \Cref{def:service-call-map}), and $\delta$ a
program.
%
%
Together, $A$ and $\scmap$ constitute the \emph{data-state}, which
captures the result of the actions executed so far, together with the
answers returned by service calls issued in the past. Instead,
$\delta$ is the \emph{program-state}, which represents the program
that still needs to be executed from the current data-state. Later on,
a state of the form $\tup{A, \scmap, \delta}$ is often also called a \emph{GKAB
  state}.

We adopt the functional approach by~\cite{Leve84} in defining the
semantics of action execution over GKAB $\gkabsym$, i.e., we assume
$\gkabsym$ provides two operations:
\begin{compactenum}
\item $\textsc{ask}$, to
answer queries over the current KB;
\item  $\tell$, to update the KB through an atomic action.
\end{compactenum}
%
%


By adopting the functional approach, we are able to interact with the
KB as a black box through some operations ($\ask$/$\tell$). As a result, we can
separate the reasoning over the structural/static aspect from the
reasoning over the dynamic aspect while still capturing the combined
behavior of both aspects within a system.
%
%
As a consequence, it enables us to take advantage from their existing
technique for the two kinds of reasoning (i.e., reasoning over DL KBs
and reasoning over the evolution of the system that is characterized
by action execution).
%
%
Furthermore, note that KABs \cite{BCMD*13}, which is the underlying
framework of GKABs, is also already adopting functional approach in
defining their semantics. As pointed out by \cite{BCMD*13},
combining incomplete information setting (DL KB) with the capability
to capture the dynamics of the system in one setting often brought us
into a rich setting that is fragile w.r.t. undecidability aspects
(cf. \cite{WoZa98}).
As it has been studied in the area of temporal description logics
\cite{LuWZ08,ArFr00}, that combine both temporal logics and
description logics, one crucial factor that influence the
computational complexity of such setting is the degree of interaction
between the temporal component and the DL component \cite{LuWZ08}. For
instance, the work in \cite{BaGL08,BaGL12} obtain a nice decidability result
for the combination of the Temporal Logic LTL \cite{BaKa08} and the
Description Logic $\dlalc$ \cite{BCMNP07} by limiting the interaction
of those two components compare to the works in \cite{AKLWZ07,ArLT06}
which either restrict the expressivity of the DL or the expressivity
of the temporal component. In particular, \cite{BaGL08,BaGL12} only allows
the temporal operators to be occurred in front of TBox/ABox assertions
instead of in front of any concepts/roles.
In KABs, to overcome such difficulties and obtain a robust result, the
work of \cite{BCMD*13} follows the functional approach which gives a
good control over the interaction between the DL KB and the dynamic
aspects. Similar arguments also pointed out by \cite{CDS15}.
%
%
%
%
  By adopting the functional approach, \cite{BCMD*13} obtained a
  setting that is not capturing the dynamics/evolution of each single
  model of a DL KB, but it captures the evolution of KBs in which at
  each single step of evolution we consider only the portion of
  knowledge that hold in all possible models of the KB (i.e., the
  certain answers). As a consequence, this setting is typically
  computationally easier to handle.  


Still about functional approach to knowledge representation, from the
business processes point of view (in particular data-aware business
processes), the motivation is to capture the manipulation/evolution of
data by the processes. 
%
%
Furthermore, each single operation (that manipulates the data) views a
data storage as a black box in which they can manipulate the data
inside it (i.e., by retrieving/querying the data as well as
adding/deleting the data). After the execution of such operation, as a
result, we have the data storage in which its data has been
manipulated (cf. \cite{DHPV09,Hull08,CoHu09,BCDDF11}).
Now, when we want to add the intensional domain knowledge into our
data-aware business processes systems, the information storage is
modeled by a KB that contains not only data but also the intensional
domain knowledge.
Hence, by adopting the functional approach, we can obtain a setting
where each manipulation operation views the KB as a black box and they
can simply perform some ``direct'' manipulation, i.e., retrieving
(``asking'') some objects/data from the KB as well as ``telling'' how
the KB should change. Moreover, as a result of the execution of a
manipulation operation, we have the manipulated KB. Overall, we obtain
a setting that captures the manipulation of KB by processes.




In this thesis, we consider that the $\textsc{ask}$ operator
corresponds to certain answers computation of queries. 

\begin{definition}[\textsc{ask} Operation]\label{def:ask}
  Given \sidetext{\textsc{ask} Operation} a KB $\tup{T, A}$, and a UCQ
  $q$ (resp. an ECQ $Q$), we define $\ask(q, T, A) = \ans(q, T, A)$
  (resp.\ $\ask(Q, T, A) = \Ans(Q, T, A)$).
\end{definition}
 
We proceed now to formally define $\tell$. As the first step, we
introduce several preliminaries
as follows.


\begin{definition}[Executability of an Action
  Invocation]\label{def:gkab-action-invoc-executability}
  Let \sidetextb{Executability of an Action Invocation}
  $\gkabsym = \tup{T, \initabox, \actset, \ginitprog}$ be a GKAB.
  Given an action invocation $\gact{Q(\vec{p})}{\act(\vec{p})}$ and an
  ABox $A$, we say that a substitutions $\sigma$, which substitutes the
  parameters $\vec{p}$ with constants in $\const$, is \emph{a legal
    parameter assignment for $\act$ in $A$ w.r.t.\
    $\gact{Q(\vec{p})}{\act(\vec{p})}$} if $\ask(Q\sigma, T, A)$ is
  $\true$. In this case we also say that \emph{$\act$ is executable in
    $A$ with a legal parameter assignment $\sigma$}
\end{definition}

\begin{example}
  Recall our running example in \Cref{ex:gkab-run-ex}, let
  $A = \set{\exo{ReceivedOrder}(\excon{chair})}$ be an ABox and consider the
  action invocation
  $\gact{[\exo{ReceivedOrder}(\exvar{x})]}{\exa{approveOrder}(\exvar{x})}$. In
  this case the substitution $\sigma$ that substitute $x$ with
  $\excon{chair}$ (i.e., $\sigma(x) = \excon{chair}$) is a legal
  parameter assignment for $\exa{approveOrder}$ in $A$ w.r.t.\
  $\gact{[\exo{ReceivedOrder}(\exvar{x})]}{\exa{approveOrder}(\exvar{x})}$
  since $\ask([\exo{ReceivedOrder}(\exvar{x})]\sigma, T, A)$ is $\true$. Thus,
  $\exa{approveOrder}$ is executable in $A$ with a legal parameter
  assignment $\sigma$.
\end{example}




With a slight abuse of the definition, we sometimes say that an action
\emph{$\act$ is executable in a state $s$ with a legal parameter
  assignment $\sigma$} if $s = \tup{A, \scmap}$ and $\act$ is
executable in $A$ with a legal parameter assignment $\sigma$. The
notion of grounded action $\act\sigma$ is defined similarly as in KABs
(i.e., a grounded action $\act\sigma$ is obtained by applying $\sigma$
to each $e \in \deff{\dact}$).
%



We define the sets $\addfacts{T, A, \act\sigma}$ (resp.
$\delfacts{T, A, \act\sigma}$) of atoms to be added and deleted by
$\gact{Q(\vec{p})}{\act(\vec{p})}$ w.r.t.\ $\sigma$ in $A$ similar to
the definitions in KABs (see
\Cref{def:add-kab-action,def:del-kab-action}) as follows:

\clearpage
\begin{definition}[Set of Atoms to be Added]\label{def:add-gkab-action}
  Let \sidetext{Set of Atoms to be Added By an Action} $\gkabsym = \tup{T, \initabox, \actset, \ginitprog}$ be a GKAB.
  Given an ABox $A$, an action invocation
  $\gact{Q(\vec{p})}{\act(\vec{p})}$ where $\act \in \actset$ is an
  action of the form $\act(\vec{p}):\set{e_1,\ldots,e_m}$ with
  $e_i = \map{[q_i^+]\land Q_i^-}{\add \facta_i, \del \factd_i}$, and
  a legal parameter assignment $\sigma$ for $\act$ in $A$ w.r.t.\
  $\gact{Q(\vec{p})}{\act(\vec{p})}$, 
  we define a \emph{set of atoms to be added by $\act\sigma$ w.r.t\
    $A$} 
  as follows:
\[
\addfacts{T, A, \act\sigma} = \bigcup_{(\map{[q^+]\land Q^-}{\add \facta, \del \factd})
  \text{ in } \eff{\act}} \  \bigcup_{\rho\ \in\ \ask(([q^+]\land Q^-)\sigma,T,A)}F^+\sigma\rho
\]
\ \ 
\end{definition}

\begin{definition}[Set of Atoms to be Deleted]\label{def:del-gkab-action}
  Let \sidetext{Set of Atoms to be Deleted By an Action}
  $\gkabsym = \tup{T, \initabox, \actset, \ginitprog}$ be a GKAB.
  Given an ABox $A$, an action invocation
  $\gact{Q(\vec{p})}{\act(\vec{p})}$ where $\act \in \actset$ is an
  action of the form $\act(\vec{p}):\set{e_1,\ldots,e_m}$ with
  $e_i = \map{[q_i^+]\land Q_i^-}{\add \facta_i, \del \factd_i}$, and
  a legal parameter assignment $\sigma$ for $\act$ in $A$ w.r.t.\
  $\gact{Q(\vec{p})}{\act(\vec{p})}$, 
  we define a \emph{set of atoms to be deleted by $\act\sigma$ w.r.t\
    $A$} 
  as follows:
\[
\delfacts{T, A, \act\sigma} = \bigcup_{(\map{[q^+]\land Q^-}{\add
    \facta, \del \factd}) \text{ in } \eff{\act}} \ \bigcup_{\rho\
  \in\ \ask(([q^+]\land Q^-)\sigma,T,A)} \factd\sigma\rho
\]
\ \ 
\end{definition}

\begin{example}\label{ex:gkab-add-del-assertions}
  Consider our running example in \Cref{ex:gkab-run-ex}. Let
%
  \[
  A = \set{\exo{ReceivedOrder}(\excon{chair}),\exo{ApprovedOrder}(\excon{table})}
  \]
  be an ABox. Consider the action invocation
  $\gact{\true}{\exa{assembleOrders}()}$ and a legal parameter
  assignment $\sigma$ for the action $\exa{assembleOrders}$ in $A$ 
  (in this case $\sigma$ is an empty substitution because the
  atomic invocation $\gact{\true}{\exa{assembleOrders}()}$ is guarded
  by $\true$). Then we have that
  \begin{compactitem}

  \item the set $\addfacts{T, A, \exa{assembleOrders}\sigma}$ of atoms
    to be added by $\exa{assembleOrders}\sigma$ is
    \[
    \begin{array}{rl}
      \{&\exo{AssembledOrder}(\excon{table}),
      \exo{assembledBy}(\excon{table},
      \exs{getAssembler}(\excon{table})), \\
      &\exo{Assembler}(\exs{getAssembler}(\excon{table})), \\
      &\exo{hasAssemblingLoc}(\excon{table},
        \exs{getAssemblingLoc}(\excon{table})) \ \ \},
    \end{array}
    \]

  \item the set $\delfacts{T, A, \exa{assembleOrders}\sigma}$ of atoms
    to be deleted by $\exa{assembleOrders}\sigma$ is
    $\set{\exo{ApprovedOrder}(\excon{table})}$.

  \end{compactitem}
\end{example}

In general, $\addfacts{T, A, \act\sigma}$ is not a proper set of ABox
assertions, because it could contain (ground) skolem terms
(representing service calls), to be substituted with corresponding
results.
We \sidetext{Service Call Evaluation} denote by
$\calls{\addfacts{T, A, \act\sigma}}$ the set of such ground skolem
terms in $\addfacts{T, A, \act\sigma}$, and by
$\eval{\addfacts{T, A, \act\sigma}}$ the set of substitutions that
replace such skolem terms with concrete values (constants) in
$\const$. 
Specifically, $\eval{\addfacts{T, A, \act\sigma}}$ is defined as
follows:
\[
\begin{array}{ll}
  \eval{\addfacts{T, A, \act\sigma}} = \{ \theta \mid &\theta \mbox { is
                                                        a total function, } \\
                                                      &\theta: \calls{\addfacts{T, A, \act\sigma}} \ra
                                                        \const \},
\end{array}
  \]
%
  and $\eval{\addfacts{T, A, \act\sigma}}$ is a singleton set of an
  empty substitution $\theta$ when
  $\calls{\addfacts{T, A, \act\sigma}} = \emptyset$.

  \begin{example}\label{ex:gkab-servcall}
    Continuing \Cref{ex:gkab-add-del-assertions}, we have that
    \[
    \begin{array}{l}
    \calls{\addfacts{T, A, \exa{assembleOrders}\sigma} } = \\
     \hspace*{20mm}\set{  \exs{getAssembler}(\excon{table}),
      \exs{getAssemblingLoc}(\excon{table}) }.
    \end{array}
    \]
    and $\eval{\addfacts{T, A, \exa{assembleOrders}\sigma}}$ contains
    a set of substitutions that substitute the ground skolem terms
    $\exs{getAssembler}(\excon{table})$ and
    $\exs{getAssemblingLoc}(\excon{table})$ into constants in
    $\const$. For example, a substitution $\theta$ where
    \begin{compactenum}
    \item $\theta(\exs{getAssembler}(\excon{table})) = \excon{john}$,
    \item
      $\theta(\exs{getAssemblingLoc}(\excon{table})) =
      \excon{bolzano}$, and
    \item $\set{\excon{john}, \excon{bolzano}} \subseteq \const$,
    \end{compactenum}
    is in $\eval{\addfacts{T, A, \exa{assembleOrders}\sigma}}$.
   \end{example}

  As the last step towards defining the $\tell$ relations, given two
  ABoxes $A$ and $A'$ where $A$ is assumed to be $T$-consistent, and
  two sets $\facta$ and $\factd$ of ABox assertions, we introduce
  so-called \emph{filter relation} to indicate that $A'$ is obtained
  from $A$ by adding the ABox assertions in $\facta$ and removing the
  one in $\factd$ as follows.

\begin{definition}[Filter Relation]\label{def:filter-relation}
  \ \sidetext{Filter Relation} Given a $T$-consistent ABox $A$, a
  \emph{filter relation $\filter$} is a relation that consists of
  tuples of the form $\tup{A, \facta, \factd, A'}$ such that we have
  $\emptyset \subseteq A' \subseteq ((A \setminus \factd) \cup
  \facta)$,
  where $A$ and $A'$ are ABoxes, and $\facta$ as well as $\factd$ are
  two sets of ABox assertions.
\end{definition}
\noindent
In this light, filter relations provide an abstract mechanism to
accommodate various approaches in updating an ABox.
For example, to account for inconsistencies, the filter could drop
some additional facts when producing $A'$.

Having all ingredients in hand, we are now ready to define the $\tell$
operation as follows.
\begin{definition}[$\tell$ Operation]\label{def:tell-operation}
  Given \sidetext{$\tell$ Operation} a GKAB $\gkabsym = \tup{T, \initabox, \actset, \ginitprog}$
  and a filter $\filter$, we define $\tell_\filter$ as 
  a relation over pairs of data-states
  such that we have a tuple
  $\tup{\tup{A,\scmap}, \act\sigma, \tup{A', \scmap'}} \in
  \tell_\filter$ if
\begin{compactitem}
\item $\sigma$ is a legal parameter assignment for $\act$ in $A$
  w.r.t.\ a certain action invocation
  $\gact{Q(\vec{p})}{\act(\vec{p})}$, and
\item there exists $\theta \in \eval{\addfacts{T, A, \act\sigma}}$
  such that:
  \begin{compactenum}

  \item for each skolem term
    $\dscall(c) \in \domain{\dscmap} \cap \domain{\theta}$, we have
    $\dscall(c)/v \in \dscmap$ if and only if
    $\dscall(c)/v \in \theta$ (i.e., $\theta$ and $\dscmap$ ``agree''
    on the common skolem terms in their domains, in order to realize
    the deterministic service call semantics);


  \item $\scmap' = \scmap \cup \theta$;

  \item
    $\tup{A, \addfacts{T, A, \act\sigma}\theta, \delfacts{T, A,
        \act\sigma}, A'} \in \filter$,
    where $\addfacts{T, A, \act\sigma}\theta$ denotes the 
    set of ABox assertions obtained by applying $\theta$ over each
    element of 
    $\addfacts{T, A, \act\sigma}$;
  \item $A$ and $A'$ is $T$-consistent.
  \end{compactenum}
\end{compactitem}
\ \ 
\end{definition}



\begin{example}
  Continuing \Cref{ex:gkab-servcall}, 
  let $\scmap = \emptyset$. Suppose that we have 
  \[
  \tup{A, \addfacts{T, A, \exa{assembleOrders}\sigma}\theta,
    \delfacts{T, A, \exa{assembleOrders}\sigma}, A'} \in \filter
  \]
  where     
  \begin{compactenum}
  \item $\theta(\exs{getAssembler}(\excon{table})) = \excon{john}$,
  \item
    $\theta(\exs{getAssemblingLoc}(\excon{table})) = \excon{bolzano}$,
    and
  \item $\set{\excon{john}, \excon{bolzano}} \subseteq \const$,

  \item $
    \addfacts{T, A, \exa{assembleOrders}\sigma}\theta =
    \left\{\begin{array}{l}
             \exo{AssembledOrder}(\excon{table}), \\
             \exo{assembledBy}(\excon{table}, \excon{john} ), \\
             \exo{Assembler}( \excon{john} ), \\
             \exo{hasAssemblingLoc}(\excon{table}, \excon{bolzano} )
    \end{array}\right\}
    $

  \item $A'$ is as follows:
    \[
    \begin{array}{l@{ \ }l@{ \ }l@{}l}
      A' &=& (&A \setminus \delfacts{T, A, \exa{assembleOrders}\sigma})
                \cup \addfacts{T, A, \exa{assembleOrders}\sigma}\theta\\
         &=&\{& \exo{ReceivedOrder}(\excon{chair}),\exo{AssembledOrder}(\excon{table}),
      \\
         &&&\exo{assembledBy}(\excon{table},
             \excon{john} ), \exo{hasAssemblingLoc}(\excon{table},
             \excon{bolzano} ), \\
         &&&\exo{Assembler}( \excon{john} ) \ \}.
    \end{array}
    \]

  \end{compactenum}
  Then we have
  $\tup{\tup{A,\scmap}, \exa{assembleOrders}\sigma, \tup{A', \scmap'}}
  \in \tell_\filter$
  where $\scmap' = \scmap \cup \theta$. 
  Intuitively, $\tup{A', \scmap'}$ represents the result of execution
  of $\exa{assembleOrders}\sigma$ over $\tup{A,\scmap}$ where $A'$ is
  obtained from $A$ by first deleting the facts to be deleted by
  $\exa{assembleOrders}\sigma$ and then add the facts to be added by
  $\exa{assembleOrders}\sigma$. 
\end{example}

As the last preliminary notion towards the parametric execution
semantics of GKABs, we specify when a state $\tup{A,\scmap,\delta}$
can be considered to be \emph{final} (i.e., the execution of $\delta$
can be considered completed),
written $\final{\tup{A,\scmap,\delta}}$.  This is done by defining the
set $\gfin$ of final states as follows:


\begin{definition}[Final State]\label{def:final-state-program}
  Let \sidetext{Final State}
  $\gkabsym = \tup{T, \initabox, \actset, \ginitprog}$ be a GKAB.
%
%
%
%
  We define the set $\gfin$ of \emph{final states of $\gkabsym$} as
  the least set of states
  of the form $\tup{A, \scmap, \delta}$ such that $A$ is an ABox over
  $T$, $\scmap$ is a service call map, $\delta'$ is a program over
  $\actset$, and
%
%
  the following hold:
\begin{compactenum}
\item $\final{\tup{A, \scmap, \gemptyprog}}$;
\item $\final{\tup{A, \scmap, \delta_1|\delta_2}}$ if
  $\final{\tup{A, \scmap, \delta_1}}$ or
  $\final{\tup{A, \scmap, \delta_2}}$;
\item $\final{\tup{A, \scmap, \delta_1;\delta_2}}$ if
  $\final{\tup{A, \scmap, \delta_1}}$ and
  $\final{\tup{A, \scmap, \delta_2}}$;
\item $\final{\tup{A, \scmap, \gif{\varphi}{\delta_1}{\delta_2}}}$ \\ if
  $\ask(\varphi, T, A) = \true$, and
  $\final{\tup{A, \scmap, \delta_1}}$;
\item $\final{\tup{A, \scmap, \gif{\varphi}{\delta_1}{\delta_2}}}$ \\if
  $\ask(\varphi, T, A) = \false$, and
  $\final{\tup{A, \scmap, \delta_2}}$;
\item $\final{\tup{A, \scmap, \gwhile{\varphi}{\delta}}}$ if
  $\ask(\varphi, T, A) = \false$;
\item $\final{\tup{A, \scmap, \gwhile{\varphi}{\delta}}}$ if
  $\ask(\varphi, T, A) = \true$, and
  $\final{\tup{A, \scmap, \delta}}$.
\end{compactenum}
\ \
\end{definition}

Now, given a filter relation $\filter$, we define the \emph{program
  execution relation} $\gprogtrans{\act\sigma, \filter}$, describing
how a grounded action
simultaneously evolves the data- and program-state. 

\begin{definition}[Program Execution Relation]\label{def:prog-exec-relation}
  Given \sidetext{Program Execution Relation} a GKAB
  $\gkabsym = \tup{T, \initabox, \actset, \ginitprog}$, and a filter
  relation $\filter$, we define a \emph{program execution relation}
  $\gprogtrans{\act\sigma, \filter}$ as follows:
\begin{compactenum}
\item $\tup{A, \scmap, \gact{Q(\vec{p})}{\act(\vec{p})}}
  \gprogtrans{\act\sigma, \filter}
  \tup{A', \scmap', \gemptyprog}$, \\if
  $\tup{\tup{A, \scmap}, \act\sigma, \tup{A', \scmap'}} \in
  \tell_\filter$,
  and $\sigma$ is a legal parameter assignment for $\act$ in $A$
  w.r.t.\ $\gact{Q(\vec{p})}{\act(\vec{p})}$;
\item $\tup{A, \scmap, \delta_1|\delta_2} \gprogtrans{\act\sigma,
    \filter} \tup{A', \scmap', \delta'}$, \\if $\tup{A, \scmap,
    \delta_1} \!\gprogtrans{\act\sigma, \filter}\! \tup{A', \scmap',
    \delta'}$ or $\tup{A, \scmap, \delta_2} \gprogtrans{\act\sigma,
    \filter} \tup{A', \scmap', \delta'}$;
\item $\tup{A, \scmap, \delta_1;\delta_2} \gprogtrans{\act\sigma,
    \filter} \tup{A', \scmap', \delta_1';\delta_2}$, \\if $\tup{A,
    \scmap, \delta_1} \gprogtrans{\act\sigma, \filter} \tup{A',
    \scmap', \delta_1'}$;
\item
  $\tup{A, \scmap, \delta_1;\delta_2} \gprogtrans{\act\sigma, \filter}
  \tup{A', \scmap', \delta_2'}$, \\
  if $\final{\tup{A, \scmap, \delta_1}}$, and
  $\tup{A, \scmap, \delta_2} \gprogtrans{\act\sigma, \filter} \tup{A',
    \scmap', \delta_2'}$;
\item
  $\tup{A, \scmap, \gif{\varphi}{\delta_1}{\delta_2}}
  \gprogtrans{\act\sigma, \filter}
  \tup{A', \scmap', \delta_1'}$, \\
  if $\ask(\varphi, T, A) = \true$, and
  $\tup{A, \scmap, \delta_1} \gprogtrans{\act\sigma, \filter}
  \tup{A', \scmap', \delta_1'}$;
\item $\tup{A, \scmap, \gif{\varphi}{\delta_1}{\delta_2}} \gprogtrans{\act\sigma, \filter} \tup{A', \scmap', \delta_2'}$,\\
  if $\ask(\varphi, T, A) = \false$, and
  $\tup{A, \scmap, \delta_2} \gprogtrans{\act\sigma, \filter} \tup{A',
    \scmap', \delta_2'}$;
\item $\tup{A, \scmap, \gwhile{\varphi}{\delta}}
  \gprogtrans{\act\sigma, \filter} \tup{A', \scmap', \delta';\gwhile{\varphi}{\delta}}$,\\
  if $\ask(\varphi, T, A) = \true$, and
  $\tup{A, \scmap, \delta} \gprogtrans{\act\sigma, \filter} \tup{A',
    \scmap', \delta'}$.
\end{compactenum}
\ \ 
\end{definition}


We are now defining the construction of GKABs transition systems that
is parameterized with a filter as follows.


\begin{definition}[GKAB Transition System]\label{def:gkab-ts}
  Given \sidetext{GKAB Transition System} a GKAB
  $\gkabsym = \tup{T, \initabox, \actset, \ginitprog}$ and a filter
  relation $\filter$, we define the \emph{transition system of
    $\gkabsym$ w.r.t.~$\filter$}, written $\ts{\gkabsym}^{\filter}$,
  as $\tup{\const,T,\stateset,s_0,\abox,\trans}$, where
  \begin{compactenum}
  \item $s_0 = \tup{\initabox, \emptyset, \ginitprog}$, and
  \item $\stateset$ and $\trans$ are defined by simultaneous induction
    as the smallest sets such that 
    \begin{compactenum}
    \item $s_0 \in \stateset$, and 
    \item if $\tup{A, \scmap, \delta} \in \stateset$ and
    $\tup{A, \scmap, \delta} \gprogtrans{\act\sigma, \filter} \tup{A', \scmap', \delta'}$,
    then $\tup{A', \scmap', \delta'}\in\stateset$ and $\tup{A,
      \scmap, \delta}\trans \tup{A', \scmap', \delta'}$.
  \end{compactenum}
  \end{compactenum}
\ \ 
\end{definition}
By suitably concretizing the filter relation, we can obtain
various execution semantics for GKABs.
We are now exploiting filter relations to define the standard
execution semantics of GKAB. 
%
%
In particular, we define a filter relation $\filter_S$ as follows:


\begin{definition}[Standard Filter $\filter_S$]
  Let \sidetext{Standard Filter $\filter_S$} $\gkabsym = \tup{T, \initabox, \actset, \ginitprog}$ be a GKAB.
  Given an ABox $A$, an atomic action $\act\in \actset$, a legal
  parameter assignment $\sigma$ for $\act$ in $A$ w.r.t.\ a certain
  action invocation $\gact{Q(\vec{p})}{\act(\vec{p})}$ in
  $\ginitprog$, and a service call substitution
  $\theta \in \eval{\addfacts{T, A, \act\sigma}}$, let
  $\facta = \addfacts{T, A, \act\sigma}\theta$ and
  $\factd = \delfacts{T, A, \act\sigma}$.
  We then have $\tup{A, \facta, \factd, A'} \in \filter_S$ if
  $A' = (A \setminus \factd) \cup \facta$,
%
\end{definition}

%

%
%
Filter $\filter_S$ gives rise to the \emph{standard execution
  semantics} for $\gkabsym$,
Essentially it just applies the update induced by the ground atomic
action $\act\sigma$ (giving priority to additions over deletions).
%
%
We call the GKABs adopting these semantics \emph{Standard GKABs} (\emph{S-GKABs}).

\begin{definition}[GKAB Standard Transition System]\label{def:gkab-standard-ts}
  Given \sidetext{GKAB Standard Transition System} a GKAB $\gkabsym = \tup{T, \initabox, \actset, \ginitprog}$
  and a standard filter $\filter_S$, the \emph{standard
    transition system of $\gkabsym$}, written
  $\ts{\gkabsym}^{\filter_S}$, is the transition system of $\gkabsym$
  w.r.t.\ $\filter_S$.
\end{definition}

An important observation in S-GKABs is that they reject those actions
that lead into an inconsistent state.  This is the case because a
tuple $\tup{A, \facta, \factd, A'} \in \filter_S$ might have the ABox
$A'$ $T$-inconsistent. However, each tuple
$\tup{\tup{A,\scmap}, \act\sigma, \tup{A', \scmap'}} \in
\tell_{\filter_S}$ requires that $A$ and $A'$ are $T$-consistent.

The notion of run and run-boundedness of S-GKABs transition systems is
defined similarly as in \Cref{def:run-bounded-kab,def:run-of-kab}.

\begin{example}\label{ex:gkab-execution}
  Consider our specification of GKAB
  $\gkabsym = \tup{T, \initabox, \actset, \delta}$ in
  \Cref{ex:gkab-run-ex}. Let $\gkabsym$ be an S-GKABs (i.e., it adopts
  standard execution semantics and its execution semantics is provided
  by the standard transition system as in
  \Cref{def:gkab-standard-ts}). In the following, we give the
  intuition of how a program is executed, how the system is
  progressing, and how the standard transition system is constructed. 

  The construction of the $\ts{\gkabsym}^{\filter_S}$ is started from
  the initial state $ s_0 = \tup{\initabox, \scmap_0, \delta} $,
  where
  \[
  \delta = \gwhile{ \exists \exvar{x}.[\exo{Order}(\exvar{x})] \wedge
    \neg[\exo{DeliveredOrder}(\exvar{x})] }{\delta_0}
  \]
  and $\scmap_0 = \emptyset$. Since there exists an \emph{order} in
  $\initabox$ (note that
  $\initabox = \set{\exo{ReceivedOrder}(\excon{chair}),
    \exo{ApprovedOrder}(\excon{table})}$),
  we enter the loop and execute $\delta_0$ that is a sequence of
  program $\delta_1 ; \delta_2 ; \delta_3 ; \delta_4 ; \delta_5 $.  We
  then need to execute $\delta_1$ that is basically an if-else
  conditional statement as follows:
  \[
  \begin{array}{r@{}l}
    \delta_1 = \gif{ \neg [\exists
    \exvar{x}.&\exo{ApprovedOrder}(\exvar{x})]
    \\ &}{\gact{[\exo{ReceivedOrder}(\exvar{x})]}{\exa{approveOrder}(\exvar{x})}
    \\ &}{\gemptyprog}
  \end{array}
  \]
  Since there exists an \emph{approved order} in $\initabox$ (note
  that $\exo{ApprovedOrder}(\excon{table}) \in \initabox$),  and the
  else part of $\delta_1$ is $\gemptyprog$ (i.e.,
  $\final{\tup{A_0, \scmap_0, \gemptyprog}}$), we then have
%
%
  $\final{\tup{\initabox, \scmap_0, \delta_1}}$. Then we need to execute
  $\delta_2$ that is an action invocation
  $\gact{\true}{\exa{prepareOrders}()}$. Notice that
  $\exa{prepareOrders}$ is executable in $A_0$ with a legal parameter
  assignment $\sigma$ where $\sigma$ is an empty substitution. Since
  we consider standard filter relation, we basically have the
  following:
  \begin{compactitem}

  \item
    $ \addfacts{T, A_0, \exa{prepareOrders}\sigma} = \{$ \\
    \hspace*{3mm}$\begin{array}{l}
         \exo{designedBy}( \excon{table},\exs{getDesigner}(\excon{table}) ), \\
         \exo{Designer}(\exs{getDesigner}(\excon{table})), \\
         \exo{hasDesign}( \excon{table}, \exs{getDesign}(\excon{table}) ), \\
         \exo{hasAssemblingLoc}( \excon{table}, \exs{assignAssemblingLoc}(\excon{table}) ) 
    \end{array}
    $ \\ $\}$

  \item $ \delfacts{T, A_0, \exa{prepareOrders}\sigma} = \emptyset$.

  \item $\eval{\addfacts{T, A_0, \exa{prepareOrders}\sigma}}$ contains
    infinite set of substitutions where each substitution substitutes
    the service calls $\exs{getDesigner}(\excon{table})$,
    $\exs{getDesign}(\excon{table})$, and
    $\exs{assignAssemblingLoc}(\excon{table})$ into constants in
    $\const$.

  \end{compactitem}
  and we have infinite tuple of the form
  \[
  \tup{A_0, \addfacts{T, A_0, \exa{prepareOrders}\sigma}\theta,
    \delfacts{T, A_0, \exa{prepareOrders}\sigma}, A_1} 
  \]
  in $\filter_S$, where
  \[
  A_1 = (A_0 \setminus \delfacts{T, A_0, \exa{prepareOrders}\sigma})
  \cup \addfacts{T, A_0, \exa{prepareOrders}\sigma}\theta
  \]
  and
  $\theta \in \eval{\addfacts{T, A_0, \exa{prepareOrders}\sigma}}$.
  I.e., we basically have that $A_1$ is of the form
  \[
  A_0 \cup \left\{\begin{array}{l}
         \exo{designedBy}( \excon{table},\exs{getDesigner}(\excon{table}) ), \\
         \exo{Designer}(\exs{getDesigner}(\excon{table})), \\
         \exo{hasDesign}( \excon{table}, \exs{getDesign}(\excon{table}) ), \\
         \exo{hasAssemblingLoc}( \excon{table}, \exs{assignAssemblingLoc}(\excon{table}) ) 
    \end{array}\right\}
\]
where $\exs{getDesigner}(\excon{table})$,
$\exs{getDesign}(\excon{table})$, and \\
$\exs{assignAssemblingLoc}(\excon{table})$ are arbitrarily substituted with constants
from $\const$. Furthermore, we have infinite tuple of the form 
\[
  \tup{\tup{A_0,\scmap_0}, \act\sigma, \tup{A_1, \scmap_1}}   
\]
in $\tell_{\filter_S}$ where $A_1$ is of the form as above, and
$\scmap_1$ substitutes the service calls
$\exs{getDesigner}(\excon{table})$, $\exs{getDesign}(\excon{table})$,
and $\exs{assignAssemblingLoc}(\excon{table})$ with the corresponding
constants.
Therefore, based on the definition on how the program are executed
(see \Cref{def:prog-exec-relation}), basically we have infinitely many
successors for $s_0$, each of the form
$ \tup{A_1, \scmap_1, \delta'} $ where
\begin{compactenum}
\item $\delta' = \delta_0' ; \gwhile{ \exists \exvar{x}.[\exo{Order}(\exvar{x})] \wedge
    \neg[\exo{DeliveredOrder}(\exvar{x})]  }{\delta_0}$, 
\item $\delta_0' = \delta_3 ; \delta_4 ; \delta_5$,
\item $A_1$ and $\scmap_1$ are as above.
\end{compactenum}

Next, the execution of S-GKAB is continued by applying the same
procedure to all successors of $s_0$ and so on.

\medskip
\noindent
\textbf{Rejecting Inconsistent States. \xspace}\\
We now provide an example where S-GKABs reject inconsistent states.
Continuing our example, consider a particular sucessor state of $s_0$,
namely state $s_1 = \tup{A_1, \scmap_1, \delta'}$, where
\begin{flushleft}
$\begin{array}{l@{}ll}
  \bullet \ A_1 = &\set{ & \exo{ReceivedOrder}( \excon{chair} ), 
              \exo{ApprovedOrder}( \excon{table} ), 
              \exo{designedBy}( \excon{table}, \excon{alice} ), \\
            &&\exo{Designer}( \excon{alice} ), \exo{hasDesign}( \excon{table} , \excon{ecodesign} ),\\
             &&\exo{hasAssemblingLoc}( \excon{table}, \excon{bolzano} ) \ \ }, 
\end{array}
$
\end{flushleft}
\begin{flushleft}
  $\begin{array}{l@{}ll} \bullet \ \scmap_1 =& \set{
      &[\exs{getDesigner}(\excon{table}) \ra
        \excon{alice}], [\exs{getDesign}(\excon{table}) \ra \excon{ecodesign}], \\
                                             &&[\exs{assignAssemblingLoc}(\excon{table})
                                                \ra \excon{bolzano}] \
                                                \ },
\end{array}
$
\end{flushleft}
%
%
\begin{flushleft}
$
\begin{array}{l}
\bullet \  \delta' = \delta_3 ; \delta_4 ; \delta_5 ; \gwhile{ \exists \exvar{x}.[\exo{Order}(\exvar{x})] \wedge
    \neg[\exo{DeliveredOrder}(\exvar{x})]  }{\delta_0}.
\end{array}
$
\end{flushleft}

The next step is to execute $\delta_3$ that is an action invocation of
the form $\gact{\true}{\exa{assembleOrders}()}$. 
The execution of action $\exa{assembleOrders}$ involves the service
calls $\exs{getAssembler}/1$ and $\exs{getAssemblingLoc}/1$. Thus it
is easy to see that there are infinite successor states of $s_1$ each
of the form $\tup{A_2, \scmap_2, \delta''}$, where $A_2$ is of the
form as follows:
\[
\begin{array}{l@{}l}
A_2 = (A_1 \setminus \set{&\exo{ApprovedOrder}(\excon{table})}) \cup \set{
\exo{AssembledOrder}(\excon{table}), \\
&\exo{assembledBy}( \excon{table},\exs{getAssembler}(\excon{table}) ), \\
&\exo{Assembler}(\exs{getAssembler}(\excon{table})), \\
&\exo{hasAssemblingLoc}( \excon{table},
  \exs{getAssemblingLoc}(\excon{table}) ) \ 
}
\end{array}
\]
in which $\exs{getAssemblingLoc}(\excon{table})$ as well as
$\exs{getAssembler}(\excon{table})$ are arbitrarily substituted with
constants from $\const$ by a substitution
$\theta \in \eval{\addfacts{T, A_1, \exa{assembleOrders}\sigma}}$ and
$\scmap_2 = \scmap_1 \cup \theta$. Moreover, we have
\[
\delta'' = \delta_4 ; \delta_5 ; \gwhile{   \exists \exvar{x}.[\exo{Order}(\exvar{x})] \wedge
    \neg[\exo{DeliveredOrder}(\exvar{x})]   }{\delta_0}.
\]


Now, as an example of a sucessor that causes inconsistency, consider a
possible substitution of $\exs{getAssemblingLoc}(\excon{table})$ into
``$\excon{trento}$'' and $\exs{getAssembler}(\excon{table})$ into
``$\excon{alice}$'' by a particular substitution
$\theta \in \eval{\addfacts{T, A_1, \exa{assembleOrders}\sigma}}$. We
then have a state $s_2~=~\tup{A_2, \scmap_2, \delta''}$ where
\begin{flushleft}
$\begin{array}{l@{}ll}
  \bullet \ A_2 = &\set{ & \exo{ReceivedOrder}( \excon{chair} ), 
                           \exo{designedBy}( \excon{table}, \excon{alice} ), \exo{Designer}( \excon{alice} ),\\
                  && \exo{hasDesign}( \excon{table} ,
                     \excon{ecodesign} ), \exo{hasAssemblingLoc}(  \excon{table},
                     \excon{bolzano} ), \\
                  &&\exo{AssembledOrder}(\excon{table}),  \exo{assembledBy}(
                     \excon{table},\excon{alice}  ), \\
                  &&\exo{Assembler}( \excon{alice} ), 
                  \exo{hasAssemblingLoc}( \excon{table},
                     \excon{trento} )  \ \ }, 
\end{array}
$
\end{flushleft}
\begin{flushleft}
$\begin{array}{l@{}ll}
  \bullet \  \scmap_2 =& \set{ &[\exs{getDesigner}(\excon{table}) \ra
                      \excon{alice}], [\exs{getDesign}(\excon{table}) \ra \excon{ecodesign}], \\
           &&[\exs{assignAssemblingLoc}(\excon{table}) \ra
              \excon{bolzano}], \\
           &&[\exs{getAssembler}(\excon{table}) \ra
              \excon{alice}], \\  
           &&[\exs{getAssemblingLoc}(\excon{table}) \ra
              \excon{trento}]  \ \ },
\end{array}
$
\end{flushleft}
\begin{flushleft}
$
\begin{array}{l}
  \bullet \  \delta'' = \delta_4 ; \delta_5 ; \gwhile{  \exists \exvar{x}.[\exo{Order}(\exvar{x})] \wedge
    \neg[\exo{DeliveredOrder}(\exvar{x})]   }{\delta_0}.
\end{array}
$
\end{flushleft}
Notice that the presence of assertions
$\exo{Assembler}( \excon{alice} )$ and
$\exo{Designer}( \excon{alice} )$ in $A_2$ trigger the violation of
TBox assertion $\exo{Designer} \sqsubseteq \neg \exo{Assembler}$,
because $\exo{Designer}$ and $\exo{Assembler}$ are two disjoint
concepts (i.e., a constant can not belong to both of those concepts at
the same time).
Moreover, the existence of assertions
$\exo{hasAssemblingLoc}( \excon{table}, \excon{bolzano} )$ and
$\exo{hasAssemblingLoc}( \excon{table}, \excon{trento})$ yields the
violation of TBox assertion $\funct{\exo{hasAssemblingLoc}}$ because
they make the role $\exo{hasAssemblingLoc}$ not functional (i.e.,
$\excon{table}$ has
two different ranges namely $\excon{bolzano}$ and $\excon{trento}$).
Therefore, since in this case $A_2$ is $T$-inconsistent, then we have
that $s_2$ is rejected.
\end{example}

\section{Capturing KABs within Standard GKABs}


Here we show that our S-GKABs are able to capture 
KABs, and 
show that the verification of \muladom properties over KABs can be
reduced to the verification of \muladom properties over S-GKABs.
%
%
The core idea is to invent a generic translation that transforms any
KABs into S-GKABs
%
%
such that their transition systems are ``equal'' (in the sense that
they have the same structure and each corresponding state contains the
same ABox and service call map), and hence
they should satisfy the same \muladom formulas. 


We now define the translation from KABs to S-GKABs as follows:

\begin{definition}[Translation from a KAB to an S-GKAB]
  We \sidetext{Translation from a KAB to an S-GKAB} define a
  translation $\tkabs$ that, given an KAB
  $\kabsym = \tup{T, \initabox, \actset, \procset}$,
  generates an S-GKAB
  $\tkabs(\kabsym) = \tup{T, \initabox, \actset, \ginitprog}$ in which
  program $\ginitprog$ is obtained from $\procset$ as
  \[
  \ginitprog = \gwhile{\true}{(a_1|a_2|\ldots|a_{\card{\procset}})},
  \]
  where, for each condition-action rule
  $\carule{Q_i(\vec{x})}{\act_i(\vec{x})} \in \procset$, we have
  $a_i = \gact{Q_i(\vec{x})}{\act_i(\vec{x})}$.
\end{definition}
\noindent
Intuitively, the translation produces a program that continues forever
to non-deterministically pick an executable action with parameters (as
specified by $\procset$), or stops if no action is executable.

Next, we continue our journey to recast the verification of KABs into
S-GKABs by:
\begin{compactenum}
\item formalizing the notion of ``equality'' between transition
  systems by introducing the notion of E-Bisimulation, as well as
  showing that two E-bisimilar transition systems can not be
  distinguished by \muladom properties (in
  \Cref{subsec:e-bisimulation}), and
\item showing that $\tkabs$ transforms KABs into S-GKABs such that
  their transition systems are E-bisimilar (in
  \Cref{subsec:kab-to-sgkab}).
\end{compactenum}

\subsection{E-Bisimulation}\label{subsec:e-bisimulation}

We now define the notion of \emph{E-Bisimulation} and
show that two E-bisimilar transition systems can not be distinguished
by a \muladom formula.

\begin{definition}[E-Bisimulation] \ \\
  Let \sidetext{E-Bisimulation} $\ts{1} = \tup{\const, T, \Sigma_1, s_{01}, \abox_1, \trans_1}$
  and $\ts{2} = \tup{\const, T, \Sigma_2, s_{02}, \abox_2, \trans_2}$
  be transition systems, with
  $\adom{\abox_1(s_{01})} \subseteq \const$ and
  $\adom{\abox_2(s_{02})} \subseteq \const$.  An \emph{E-Bisimulation}
  between $\ts{1}$ and $\ts{2}$ is a relation
  $\B \subseteq \Sigma_1 \times\Sigma_2$ such that
  $\tup{s_1, s_2} \in \B$ implies that:
  \begin{compactenum}
  \item $\abox_1(s_1) = \abox_2(s_2)$
  \item for each $s_1'$, if $s_1 \Rightarrow_1 s_1'$ then there exist
    $s_2'$ with $ s_2 \Rightarrow_2 s_2' $ such that
    $\tup{s_1', s_2'}\in\B$.
  \item for each $s_2'$, if $ s_2 \Rightarrow_2 s_2' $ then there
    exists $s_1'$ with $s_1 \Rightarrow_1 s_1'$, such that
    $\tup{s_1', s_2'}\in\B$.
 \end{compactenum}
\ \ 
\end{definition}

\noindent
Let $\ts{1} = \tup{\const, T, \Sigma_1, s_{01}, \abox_1, \trans_1}$
and $\ts{2} = \tup{\const, T, \Sigma_2, s_{02}, \abox_2, \trans_2}$ be
transition systems,
a state $s_1 \in \Sigma_1$ is \emph{E-bisimilar} to
$s_2 \in \Sigma_2$, written $s_1 \ebsim s_2$, if there exists an
E-Bisimulation $\B$ between $\ts{1}$ and $\ts{2}$ such that
$\tup{s_1, s_2}\in\B$.
The transition system $\ts{1}$ is \emph{E-bisimilar} to $\ts{2}$,
written $\ts{1} \ebsim \ts{2}$, if there exists an E-Bisimulation $\B$
between $\ts{1}$ and $\ts{2}$ such that $\tup{s_{01}, s_{02}}\in\B$.
%


\begin{lemma}\label{lem:e-bisimilar-ts-satisfies-same-formula}
  Consider two transition systems
  $\ts{1} = \tup{\const,T,\stateset_1,s_{01},\abox_1,\trans_1}$ and
  $\ts{2} = \tup{\const,T,\stateset_2,s_{02},\abox_2,\trans_2}$ such
  that $\ts{1} \ebsim \ts{2}$.  For every $\muladom$ closed formula
  $\Phi$, we have:
  \[
  \ts{1} \models \Phi \textrm{ if and only if } \ts{2} \models \Phi.
  \]
\end{lemma}
\begin{proof}
  The claim easily follows since 
%
  two E-bisimilar transition systems are essentially equal in terms of
  the structure and the ABoxes that are contained in each two
  bisimilar states.
\end{proof}

\subsection{Reducing the Verification of KABs to Standard GKABs}\label{subsec:kab-to-sgkab}


Here we show that the transition system of a KAB and the transition
system of its corresponding S-GKAB are E-bisimilar. Then, by using the
result from the previous subsection we can easily recast the
verification problem and hence achieve our purpose.

\begin{lemma}\label{lem:skab-to-sgkab-bisimilar-state}
  Let $\kabsym$ be a KAB with transition system
  $\ts{\kabsym}^S$, and let $\tkabs(\kabsym)$ be an
  S-GKAB with transition system $\ts{\tkabs(\kabsym)}^{\filter_S}$
  obtained through $\tkabs$.
  Consider 
  \begin{inparaenum}[]
  \item a state $\tup{A_k,\scmap_k}$ of $\ts{\kabsym}^S$ and
  \item a state $\tup{A_g,\scmap_g, \delta_g}$ of
    $\ts{\tkabs(\kabsym)}^{\filter_S}$.
  \end{inparaenum}
  If $A_k = A_g$, and $\scmap_k = \scmap_g$, then
  $\tup{A_k,\scmap_k} \ebsim \tup{A_g,\scmap_g, \delta_g}$.
\end{lemma}
\begin{proof}
Let
\begin{compactenum}
\item $\kabsym = \tup{T, \initabox, \actset, \procset}$, and \\
  $\ts{\kabsym}^S = \tup{\const, T, \stateset_k, s_{0k}, \abox_k,
    \trans_k}$,
\item $\tkabs(\kabsym) = \tup{T, \initabox, \actset, \ginitprog}$, and \\
  $\ts{\tkabs(\kabsym)}^{\filter_S} = \tup{\const, T, \stateset_g, s_{0g},
    \abox_g, \trans_g}$.
\end{compactenum}
To prove the lemma, we show that, for every state
$\tup{A_k', \scmap_k'}$ s.t.\
$\tup{A_k,\scmap_k} \trans_k \tup{A_k',\scmap_k'}$, there exists a state
$\tup{A'_g,\scmap'_g, \delta_g'}$ s.t.:
\begin{compactenum}
\item
  $\tup{A_g,\scmap_g, \delta_g} \trans_g \tup{A'_g,\scmap'_g,
    \delta_g'}$;
\item $A'_k = A_g'$;
\item $\scmap'_k = \scmap_g'$.
\end{compactenum}
By definition of $\ts{\kabsym}^S$, if
$\tup{A_k,\scmap_k} \trans \tup{A_k',\scmap_k'}$, then there exist
\begin{compactenum}
\item a condition action rule $\carule{Q(\vec{p})}{\act(\vec{p})}$, 
\item an action $\act \in \actset$ with parameters $\vec{p}$, 
\item a parameter substitution $\sigma$, and 
\item a substitution $\theta$.
\end{compactenum}
such that 
\begin{inparaenum}[\it (i)]
\item $\theta \in \eval{T,A_k,\act\sigma}$ and agrees with $\scmap_k$, 
\item $\act$ is executable in state $A_k$ with a parameter
  substitution $\sigma$,
\item $A_k' = \doo{T, A_k, \act\sigma}\theta$, and 
\item $\scmap_k' = \scmap_k \cup \theta$.
\end{inparaenum}

Now, since
$ \ginitprog = \gwhile{\true}{(a_1|a_2|\ldots|a_{\card{\procset}})}$,
and each $a_i$ is an action invocation obtained from a
condition-action rule in $\procset$, then there exists an action
invocation $a_i$ such that $a_i = \gact{Q(\vec{x})}{\act(\vec{x})}$.
Since $A_k = A_g$, and $\scmap_k = \scmap_g$, by considering how a
transition is created in the transition system of S-GKABs, it is easy
to see that there exists a state $\tup{A_g', \scmap_g', \delta_g'}$
such that
$\tup{A_g,\scmap_g, \delta_g} \trans_g \tup{A'_g,\scmap'_g,
  \delta_g'}$,
$A_g' = A_k'$, and $\scmap_g' = \scmap_k'$. Thus, the claim is proven.
\end{proof}

\begin{lemma}\label{lem:skab-to-sgkab-bisimilar-ts}
  Given a KAB $\kabsym$, we have
  $\ts{\kabsym}^S \ebsim \ts{\tkabs(\kabsym)}^{\filter_S}$
\end{lemma}
\begin{proof}
Let
\begin{compactenum}
\item $\kabsym = \tup{T, \initabox, \actset, \procset}$, and \\
  $\ts{\kabsym}^S = \tup{\const, T, \stateset_k, s_{0k}, \abox_k,
    \trans_k}$,
\item $\tkabs(\kabsym) = \tup{T, \initabox, \actset, \ginitprog}$, and \\
  $\ts{\tkabs(\kabsym)}^{\filter_S} = \tup{\const, T, \stateset_g, s_{0g},
    \abox_g, \trans_g}$.
\end{compactenum}
We have that $s_{0k} = \tup{A_0, \scmap_k}$ and
$s_{0g} = \tup{A_0, \scmap_g, \delta}$ where
$\scmap_k = \scmap_g = \emptyset$. Hence, by Lemma
\ref{lem:skab-to-sgkab-bisimilar-state}, we have
$s_{0k} \ebsim s_{0g}$. Therefore, by the definition of E-bisimulation
between two transition systems, we have
$\ts{\kabsym}^S \ebsim \ts{\tkabs(\kabsym)}^{\filter_S}$.
\end{proof}

Having Lemma~\ref{lem:skab-to-sgkab-bisimilar-ts} in hand, we can
easily show that the verification of \muladom over KABs can be
reduced to the verification of \muladom over S-GKABs by also making
use the result from the previous subsection.


\begin{theorem}
\label{thm:stog}
Verification of \muladom properties over KABs can be recast as
verification over S-GKABs.
\end{theorem}
\begin{proof}
  The proof can be easily obtained since we can translate KABs into
  S-GKABs using $\tkabs$ and then we can easily show that given a KAB
  $\kabsym$ and a closed $\muladom$ formula $\Phi$, we have
  $\ts{\kabsym}^S \models \Phi$ iff
  $\ts{\tkabs(\kabsym)}^{\filter_S} \models \Phi$ due to the fact that
  by Lemma~\ref{lem:skab-to-sgkab-bisimilar-ts}, we have that
  $\ts{\kabsym}^S \ebsim \ts{\tkabs(\kabsym)}^{\filter_S}$ and hence
  the claim is directly follows from
  Lemma~\ref{lem:e-bisimilar-ts-satisfies-same-formula}. 
\end{proof}


\section{Verification of Standard GKABs (S-GKABs)}\label{sec:verification-sgkab}

The problem definition of the \muladom formula verification over
S-GKABs is defined similarly as in KABs (see
\Cref{def:verification-kab}). 

\begin{example}
Continuing our running example, an example of \muladom properties to be
verified is as follows:
\[
\nu Z.(\forall x. \exo{Order}(x) \ra \mu Y.(\exo{DeliveredOrder}(x)
\vee \DIAM{Y})) \land \BOX{Z}
\]
Intuitively, this formula says that, along every path, it is always
true that each order $x$ will be eventually delivered.
\end{example}

Here, we solve the problem of S-GKABs verification by compiling
S-GKABs 
into  
KABs 
%
and show that the verification of \muladom formulas over S-GKAB can be
recast as verification over KAB.
%
%
(This claim is formally stated
in \Cref{thm:gtos}).
To this aim, technically we do the following:
\begin{compactenum}

\item 
  We define a special bisimulation relation between two transition
  system namely \emph{jumping bisimulation} (see
  \Cref{sec:jumping-bisimulation}). 

\item Furthermore, also in \Cref{sec:jumping-bisimulation}, we define
  a generic translation $\tforj$ that takes a \muladom formula $\Phi$
  in Negative Normal Form (NNF) as an input and produces a \muladom
  formula $\tforj(\Phi)$, and then we also show that two jumping
  bisimilar transition system can not be distinguished by any \muladom
  formula (in NNF) modulo the translation $\tforj$.

\item In \Cref{sec:transform-sgkab-to-kab}, we define a generic
  translation $\tgkab$, that given an S-GKAB $\gkabsym$, produces a
  KAB $\tgkab(\gkabsym)$. The core idea of this translation is to
  transform the given program $\delta$ and the set of actions in
  S-GKAB $\gkabsym$ into a process (a set of condition-action rules)
  and a set of KAB actions, such that all possible sequence of action
  executions that is enforced by $\delta$ can be mimicked by the
  process in KAB (which determines all possible sequence of action
  executions in KAB).

\item In the \Cref{sec:reduction-verification-sgkab-to-kab}, we show
  that the transition system of a GKAB $\gkabsym$ and the transition
  system of its corresponding KAB $\tgkab(\gkabsym)$ (obtained through
  translation $\tgkab$) are bisimilar w.r.t.\ the jumping bisimulation
  relation.

\item Making use all of the ingredients above, in the end we show that
  a GKAB $\gkabsym$ satisfies a certain \muladom formula $\Phi$ if and
  only if its corresponding KAB $\tgkab(\gkabsym)$ satisfies a
  \muladom formula $\tforj(\Phi)$ (see
  \Cref{sec:reduction-verification-sgkab-to-kab}).

\end{compactenum}

For a \sidetext{Special Markers} technical reason, 
%
we reserve some fresh concept names $\flagconceptname$,
$\noopconceptname$ and $\tmpconceptname$ (i.e., they are outside of
any TBox vocabulary), and they are not allowed to be used in any
temporal properties (i.e., in \muladom or \mula formulas). We call
them \emph{special marker concept names}.
Additionally, we make use the constants in $\const_0$ to populate
them.
%
%
We call \emph{special marker} an ABox assertion that is obtained by
applying either $\flagconceptname$, $\noopconceptname$ or
$\tmpconceptname$ to a constant in $\const_0$.
Additionally, we call \emph{flag} a special marker formed by applying
either concept name $\flagconceptname$ or $\noopconceptname$ to a
constant in $\const_0$.
Later on, we use flags as markers to impose a certain sequence of
action executions, and 
%
we use a special marker $\tmp$ (where $\tmpconst \in \const_0$) to
mark an \emph{intermediate state}.
%
%

\subsection{Jumping Bisimulation (J-Bisimulation)}\label{sec:jumping-bisimulation}

As a start towards defining the notion of J-Bisimulation, we introduce
the notion of equality modulo flag between two ABoxes as follows:

\begin{definition}[Equal Modulo Special Markers]\label{def:equal-mod-markers}
  Given \sidetextb{Equal Modulo Special Markers} a TBox $T$, two ABoxes $A_1$ and $A_2$ over $\voc(T)$ that
  might contain special markers, 
  we say \emph{$A_1$ equal to $A_2$ modulo special markers}, written
  $A_1 \eqm A_2$ (or equivalently $A_2 \eqm A_1$),
  if the following hold:
  \begin{compactitem}
  \item For each concept name $N \in \voc(T)$ (i.e., $N$ is not a
    special marker concept name), we have a concept assertion
    $N(c) \in A_1$ if and only if a concept assertion $N(c) \in A_2$,
  \item For each role name $P \in \voc(T)$, we have a role assertion
    $P(c_1,c_2) \in A_1$ if and only if a role assertion
    $P(c_1,c_2) \in A_2$.
  \end{compactitem}
\ \ 
\end{definition}

\noindent
Essentially, the definition \Cref{def:equal-mod-markers} above says
that two ABoxes are equal modulo special markers if they contain the
same ABox assertions except the special markers. Next, we show some
interesting properties related to the notion of equality modulo flag
between two ABoxes.

\begin{lemma}\label{lem:equal-ABox-imply-equal-modulo-markers}
  $A_1 = A_2$ implies $A_1 \eqm A_2$.
\end{lemma}
\begin{proof}
  Trivially true from the definition of $A_1 \eqm A_2$ above (see
  \Cref{def:equal-mod-markers}) since both $A_1$ and $A_2$ contain the
  same set of ABox assertions.
\end{proof}

\begin{lemma}\label{lem:ECQ-equal-ABox-modulo-markers}
  Given a GKAB $\gkabsym = \tup{T, \initabox, \actset, \ginitprog}$,
  two ABoxes $A_1$ and $A_2$ over $\voc(T)$ which might contain
  special markers, and an ECQ $Q$ over $\tup{T, \initabox}$ which does
  not contain any atoms whose predicates are special marker concept
  names. 
  We have that if $A_1 \eqm A_2$, then
  $\Ans(Q, T, A_1) = \Ans(Q, T, A_2)$.
\end{lemma}
\begin{proof}
  Trivially hold since 
  without considering special markers, we have that a concept
  assertion $N(c) \in A_1$ if and only if a concept assertion
  $N(c) \in A_2$, and also a role assertion $P(c_1,c_2) \in A_1$ if
  and only if a role assertion $P(c_1,c_2) \in A_2$ (i.e., we can
  consider that $A_1 = A_2$ because we do not query the special
  marker). Hence $\Ans(Q, T, A_1) = \Ans(Q, T, A_2)$.
\ \ 
\end{proof}

We now proceed to define the notion of \emph{jumping bisimulation} as
follows.
\begin{definition}[Jumping Bisimulation (J-Bisimulation)] \
  \sidetext{Jumping Bisimulation (J-Bisimulation)} \\
  Let  $\ts{1} = \tup{\const, T, \Sigma_1, s_{01}, \abox_1, \trans_1}$
  and $\ts{2} = \tup{\const, T, \Sigma_2, s_{02}, \abox_2, \trans_2}$
  be KB transition systems, with
  $\adom{\abox_1(s_{01})} \subseteq \const$
and $\adom{\abox_2(s_{02})} \subseteq \const$.
A \emph{jumping bisimulation} (J-Bisimulation) between $\ts{1}$ and
$\ts{2}$ is a relation $\B \subseteq \Sigma_1 \times\Sigma_2$ such
that $\tup{s_1, s_2} \in \B$ implies that:
  \begin{compactenum}
  \item $\abox_1(s_1) \eqm \abox_2(s_2)$
  \item for each $s_1'$, if $s_1 \Rightarrow_1 s_1'$ then there exist
    $s_2'$, $t_1, \ldots ,t_n$ (for $n \geq 0$) with
    \[
    s_2 \Rightarrow_2 t_1 \Rightarrow_2 \ldots \Rightarrow_2 t_n \Rightarrow_2 s_2'
    \] 
   such that $\tup{s_1', s_2'}\in\B$,
    $\tmp \not\in \abox_2(s_2')$ and $\tmp \in \abox_2(t_i)$ for
    $i \in \set{1, \ldots, n}$.
  \item for each $s_2'$, if 
    \[
    s_2 \Rightarrow_2 t_1 \Rightarrow_2 \ldots \Rightarrow_2 t_n \Rightarrow_2 s_2'
    \] 
    (for $n \geq 0$) with $\tmp \in \abox_2(t_i)$ for
    $i \in \set{1, \ldots, n}$ and $\tmp \not\in \abox_2(s_2')$, then
    there exists $s_1'$ with $s_1 \Rightarrow_1 s_1'$, such that
    $\tup{s_1', s_2'}\in\B$.
 \end{compactenum}
\ \ 
\end{definition}

\noindent
Given two KB transition systems
$\ts{1} = \tup{\const, T, \Sigma_1, s_{01}, \abox_1, \trans_1}$ and
$\ts{2} = \tup{\const, T, \Sigma_2, s_{02}, \abox_2, \trans_2}$,
a state $s_1 \in \Sigma_1$ is \emph{J-bisimilar} to
$s_2 \in \Sigma_2$, written $s_1 \jbsim s_2$, if there exists a
jumping bisimulation $\B$ between $\ts{1}$ and $\ts{2}$ such that
$\tup{s_1, s_2}\in\B$.
A transition system $\ts{1}$ is \emph{J-bisimilar} to $\ts{2}$,
written $\ts{1} \jbsim \ts{2}$, if there exists a jumping bisimulation
$\B$ between $\ts{1}$ and $\ts{2}$ such that
$\tup{s_{01}, s_{02}}\in\B$.

Now, in the \Cref{lem:jumping-bisimilar-states-satisfies-same-formula}
below, we 
show that two transition systems which
are J-bisimilar can not be distinguished by any \muladom formula (in
NNF) modulo a translation $\tforj$ which is defined as follows:

\begin{definition}[Translation $\tforj$]\label{def:tforj}
  We define a \emph{translation $\tforj$} that transforms an arbitrary
  \muladom formula $\Phi$ (in NNF) into another \muladom formula
  $\Phi'$ inductively by recurring over the structure of $\Phi$ as
  follows:
\[
\begin{array}{@{}l@{}ll@{}}
  \bullet\ \tforj(Q) &=& Q \\

  \bullet\ \tforj(\neg Q) &=& \neg Q \\

  \bullet\ \tforj(\Q x.\Phi) &=& \Q x. \tforj(\Phi) \\

  \bullet\ \tforj(\Phi_1 \circ \Phi_2) &=& \tforj(\Phi_1) \circ \tforj(\Phi_2) \\

  \bullet\ \tforj(\circledcirc Z.\Phi) &=& \circledcirc Z. \tforj(\Phi) \\

  \bullet\ \tforj(\DIAM{\Phi}) &=& \DIAM{\mu Z.((\tmp \wedge \DIAM{Z}) \vee
                                   (\neg \tmp \wedge \tforj(\Phi)))} \\

  \bullet\ \tforj(\BOX{\Phi}) &=& \BOX{\mu Z.((\tmp \wedge \BOX{Z} \wedge
                                  \DIAM{\top}) \vee (\neg \tmp \wedge \tforj(\Phi)))}
\end{array}
\]


\noindent
where:
\begin{compactitem}
\item $\circ$ is a binary operator ($\vee, \wedge, \ra,$ or $\lra$),
\item $\circledcirc$ is least ($\mu$) or greatest ($\nu$) fix-point operator,
\item $\Q$ is forall ($\forall$) or existential ($\exists$)
  quantifier.
\end{compactitem}
\ \ 
\end{definition}
%

\noindent
In brief, $\tforj$ translates a given formula into a formula that
skips the states in which $\tmp$ hold (i.e., bypass the intermediate
states). To better understand the translation $\tforj$, we provide
some more intuitions of it as follows:
\begin{itemize}

\item the formula
  \[
  \mu Z.((\tmp \wedge \DIAM{Z}) \vee (\neg \tmp \wedge \tforj(\Phi)))
  \]
  in the translation $\tforj(\DIAM{\Phi})$ essentially can be also
  expressed in CTL as
  \[
  \exists [\tmp\ \mathbf{U}\ (\neg \tmp \wedge \tforj(\Phi))]
  \]
  where ``$\mathbf{U}$'' is the typical CTL ``until'' operator, and
  this formula says that there exists a path in which $\tmp$ holds
  until there is a state in which $(\neg \tmp \wedge \tforj(\Phi))$
  holds (See also the translation from CTL into $\mu$-Calculus in
  \cite{BrSt07}). Combining with $\DIAM{}$, we have that intuitively
  $\tforj$ translates $\DIAM{\Phi}$ into a formula stating that there
  exists a successor state, such that from that successor state,
  $\tmp$ holds until there is a state in which
  $(\neg \tmp \wedge \tforj(\Phi))$ holds. 
%
%
  Therefore, intuitively, the translation $\tforj$ translates
  $\DIAM{\Phi}$ into a formula saying that there exists a path
  leading us into a state where $\tforj(\Phi)$ holds, and until we
  reach that state, might need to pass/skip several intermediate states
  that are marked by $\tmp$.

\item The intuition for the translation of $\BOX{\Phi}$ is similar to
  the translation of $\DIAM{\Phi}$ by also noticing that the formula
  \[
  \mu Z.((\tmp \wedge \BOX{Z} \wedge \DIAM{\top}) \vee (\neg \tmp
  \wedge \tforj(\Phi)))
  \]
  can be also expressed in CTL as
  \[
  \forall [\tmp\ \mathbf{U}\ (\neg \tmp \wedge \tforj(\Phi))].
  \]
  (See also the translation from CTL into $\mu$-Calculus in
  \cite{BrSt07}).

\item Later on, we will see that we use $\tmp$ to mark the
  intermediate states in our transition systems, and those
  intermediate states capture the intermediate results of some
  computation for generating the ``real'' successor states. Thus, we
  are not supposed to query this state and just skip these states.


\end{itemize}

\begin{lemma}\label{lem:jumping-bisimilar-states-satisfies-same-formula}
  Consider two KB transition systems
  $\ts{1} = \left<\const, T,\stateset_1,s_{01},\abox_1,\trans_1\right>$ and
  $\ts{2} = \left<\const, T,\stateset_2,s_{02},\abox_2,\trans_2\right>$,  
  two states $s_1 \in \stateset_1$ and $s_2 \in \stateset_2$ such that
  $s_1 \jbsim s_2$. Then for every formula $\Phi$ of $\muladom$ (in
  NNF), 
  and every valuations $\vfo_1$ and $\vfo_2$ that assign to each of
  its free variables a constant $c_1 \in \adom{\abox_1(s_1)}$ and
  $c_2 \in \adom{\abox_2(s_2)}$, such that $c_1 = c_2$, we have that
  \[
  \ts{1},s_1 \models \Phi \vfo_1 \textrm{ if and only if } \ts{2},s_2
  \models \tforj(\Phi) \vfo_2.
  \]
\end{lemma}
\begin{proof}
  The proof is then organized in three parts:
\begin{compactenum}[(1)]

\item We prove the claim for formulae of $\ladom$, obtained from
  $\muladom$ by dropping the predicate variables and the fixpoint
  constructs. $\ladom$ corresponds to a first-order variant of the
  Hennessy Milner logic. 

\item We extend the results to the infinitary logic obtained by extending
  $\ladom$ with arbitrary countable disjunction.

\item We recall that fixpoints can be translated into this infinitary
  logic (cf. \cite{Stir01}), thus proving that the theorem holds for
  $\muladom$.

\end{compactenum}

\smallskip
\noindent
\textbf{Proof for $\ladom$.}  We proceed by induction on the structure
of $\Phi$, without considering the case of predicate variable and of
fixpoint constructs, which are not part of $\ladom$.

\smallskip
\noindent
\textit{Base case:}
\begin{compactitem}
\item[\textbf{($\Phi = Q$)}.] Since $s_1 \jbsim s_2$, we have
  $\abox_1(s_1) \eqm \abox_2(s_2)$. Hence, since we also restrict that
  any \muladom formulas does not use special marker concept names, 
  by \Cref{lem:ECQ-equal-ABox-modulo-markers},
%
  we have $\Ans(Q, T, \abox_1(s_1)) = \Ans(Q, T, \abox_2(s_2))$.
  Hence, since $\tforj(Q) = Q$, for every valuations $\vfo_1$ and
  $\vfo_2$ that assign to each of its free variables a constant
  $c_1 \in \adom{\abox_1(s_1)}$ and $c_2 \in \adom{\abox_2(s_2)}$,
  such that $c_1 = c_2$, we have
  \[
  \ts{1},s_1 \models Q \vfo_1 \textrm{ if and only if } \ts{2},s_2
  \models \tforj(Q) \vfo_2.
  \]

\item[\textbf{($\Phi = \neg Q$)}.] Similar to the previous case.

\end{compactitem}

\smallskip
\noindent
\textit{Inductive step:}
\begin{compactitem}
\item[\textbf{($\Phi = \Psi_1 \wedge \Psi_2$)}.]  
  $\ts{1},s_1 \models (\Psi_1\wedge \Psi_2) \vfo_1$ if and only if
  either $\ts{1},s_1 \models \Psi_1 \vfo_1$ or
  $\ts{1},s_1 \models \Psi_2 \vfo_1$.  By induction hypothesis, we
  have for every valuations $\vfo_1$ and $\vfo_2$ that assign to each
  of its free variables a constant $c_1 \in \adom{\abox_1(s_1)}$ and
  $c_2 \in \adom{\abox_2(s_2)}$, such that $c_1 = c_2$, we have
\begin{compactitem}
\item
  $ \ts{1},s_1 \models \Psi_1 \vfo_1 \textrm{ if and only if }
  \ts{2},s_2 \models \tforj(\Psi_1) \vfo_2.  $, and also
\item
  $ \ts{1},s_1 \models \Psi_2 \vfo_1 \textrm{ if and only if }
  \ts{2},s_2 \models \tforj(\Psi_2) \vfo_2.  $
\end{compactitem}

Hence, $\ts{1},s_1 \models \Psi_1 \vfo_1$ and
$\ts{1},s_1 \models \Psi_2 \vfo_1$ if and only if
$\ts{2},s_2 \models \tforj(\Psi_1) \vfo_2$ and
$\ts{2},s_2 \models \tforj(\Psi_2) \vfo_2$. Therefore we have
$ \ts{1},s_1 \models (\Psi_1 \wedge \Psi_2) \vfo_1 \textrm{ if and
  only if } \ts{2},s_2 \models (\tforj(\Psi_1) \wedge \tforj(\Psi_2))
\vfo_2 $
Since
$\tforj(\Psi_1 \wedge \Psi_2) = \tforj(\Psi_1) \wedge \tforj(\Psi_2)$,
we have
\[
\ts{1},s_1 \models (\Psi_1 \wedge \Psi_2) \vfo_1 \textrm{ if and only
  if } \ts{2},s_2 \models \tforj(\Psi_1\wedge \Psi_2) \vfo_2
  \]
  The proof for the case of $\Phi = \Psi_1 \vee \Psi_2$,
  $\Phi = \Psi_1 \ra \Psi_2$, and $\Phi = \Psi_1 \lra \Psi_2$ can be
  done similarly.


\item[\textbf{($\Phi = \DIAM{\Psi}$)}.]  Assume
  $\ts{1},s_1 \models (\DIAM{\Psi}) \vfo_1$, where $\vfo_1$ is a
  valuation that assigns to each free variable of $\Psi$ a constant
  $c_1 \in \adom{\abox_1(s_1)}$. Then there exists $s_1'$ s.t.\
  $s_1 \trans_1 s_1'$ and $\ts{1},s_1' \models \Psi \vfo_1$.  Since
  $s_1 \jbsim s_2$, there exists $s_2'$, $t_1, \ldots ,t_n$ (for
  $n \geq 0$) with
    \[
    s_2 \Rightarrow_2 t_1 \Rightarrow_2 \ldots \Rightarrow_2 t_n
    \Rightarrow_2 s_2'
    \] 
    such that $s_1' \jbsim s_2'$, $\tmp \in \abox_2(t_i)$ for
    $i \in \set{1, \ldots, n}$, and $\tmp \not\in \abox_2(s_2')$.
    Hence, by induction hypothesis, for every valuations $\vfo_2$ that
    assign to each free variables $x$ of $\tforj(\Psi)$ a constant
    $c_2 \in \adom{\abox_2(s_2)}$, such that $c_1 = c_2$ and
    $x/c_1 \in \vfo_1$, we have
    $ \ts{2},s_2' \models \tforj(\Psi) \vfo_2.  $
%
%
%
%
%
    Consider that
    \[
    s_2\Rightarrow_2 t_1 \Rightarrow_2 \ldots
    \Rightarrow_2 t_n \Rightarrow_2 s_2'
    \] 
    (for $n \geq 0$), $\tmp \in \abox_2(t_i)$ for
    $i \in \set{1, \ldots, n}$, and $\tmp \not\in \abox_2(s_2')$. We
    then obtain
    \[
      \ts{2},s_2 \models (\DIAM{\mu Z.((\tmp \wedge \DIAM{Z})
                                      \vee (\neg\tmp \wedge \tforj(\Psi)))})\vfo_2.
    \]
    Since
    $\tforj(\DIAM{\Phi}) = \DIAM{\mu Z.((\tmp \wedge \DIAM{Z}) \vee
      (\neg \tmp \wedge \tforj(\Phi)))}$, we have
    \[
    \ts{2},s_2 \models \tforj(\DIAM{\Psi})\vfo_2.
    \]



    \medskip For the other direction, assume
    $\ts{2},s_2 \models \tforj(\DIAM{\Psi}) \vfo_2$, where $\vfo_2$ is
    a valuation that assigns to each free variable of
    $\tforj(\DIAM{\Psi})$ a constant $c_2 \in \adom{\abox_2(s_2)}$. By
    the definition of $\tforj$, we have
    \[
    \ts{2},s_2 \models \DIAM{\mu Z.((\tmp \wedge \DIAM{Z}) \vee (\neg
      \tmp \wedge \tforj(\Psi)))} \vfo_2
    \]

    Then there exists $s_2'$ s.t.\
    \[
    s_2 \Rightarrow_2 t_1 \Rightarrow_2 \ldots \Rightarrow_2 t_n
    \Rightarrow_2 s_2',
    \] 
    $\tmp \in \abox_2(t_i)$ for $i \in \set{1, \ldots, n}$,
    $\tmp \not\in \abox_2(s_2')$, and
    $\ts{2},s_2' \models \tforj(\Psi) \vfo_2$. Since $s_1 \jbsim s_2$,
    there exists $s_1'$, such that $s_1 \trans_1 s_1'$ and
    $s_1' \jbsim s_2'$.
    Hence, by induction hypothesis, for every valuations $\vfo_1$ that
    assign to each free variables $x$ of $\Psi$ a constant
    $c_1 \in \adom{\abox_1(s_1)}$, such that $c_1 = c_2$ and
    $x/c_2 \in \vfo_2$, we have $ \ts{1},s_1' \models \Psi \vfo_1$.
    Now, consider that $s_1 \trans_1 s_1'$, we then obtain that
    $ \ts{1},s_1 \models (\DIAM{\Psi}) \vfo_1$.

\item[\textbf{($\Phi = \BOX{\Psi}$)}.]  The proof is similar to the
  case of $\Phi = \DIAM{\Psi}$

\item[\textbf{($\Phi = \exists x. \Psi$)}.]  Assume that
  $\ts{1},s_1 \models (\exists x. \Psi)\vfo'_1$, where $\vfo'_1$ is a
  valuation that assigns to each free variable of $\Psi$ a constant
  $c_1 \in \adom{\abox_1(s_1)}$. Then, by definition, there exists
  $c \in \adom{\abox_1(s_1)}$ such that
  $\ts{1},s_1 \models \Psi\vfo_1$, where $\vfo_1 = \vfo'_1[x/c]$. By
  induction hypothesis, for every valuation $\vfo_2$ that assigns to
  each free variable $y$ of $\tforj(\Psi)$ a constant
  $c_2 \in \adom{\abox_2(s_2)}$, such that $c_1 = c_2$ and
  $y/c_1 \in \vfo_1$, we have that
  $\ts{2},s_2 \models \tforj(\Psi) \vfo_2$. Additionally, we have
  $\vfo_2 = \vfo'_2[x/c']$, where $c' \in \adom{\abox_2(s_2)}$, and
  $c' = c$ because $\abox_2(s_2) \eqm \abox_1(s_1)$.  Hence, we get
  $\ts{2},s_2 \models (\exists x. \tforj(\Psi))\vfo'_2$. Since
  $\tforj(\exists x.\Phi) = \exists x. \tforj(\Phi)$, thus we have
  $\ts{2},s_2 \models \tforj(\exists x. \Psi)\vfo'_2$.  The other
  direction can be shown similarly.

\item[\textbf{($\Phi = \forall x. \Psi$)}.]  The proof is similar to
  the case of $\Phi = \exists x. \Psi$.


\end{compactitem}

\smallskip
\noindent
\textbf{Extension to arbitrary countable disjunction.}  Let $\Psi$ be
a countable set of $\ladom$ formulae. Given a transition system
$\ts{} = \tup{\const, T,\stateset,s_{0},\abox,\trans}$, the semantics
of $\bigvee \Psi$ is
$(\bigvee \Psi) _\vfo^{\ts{}} = \bigcup_{\psi \in \Psi}
(\psi)_\vfo^{\ts{}}$.
Therefore, given a state $s \in \Sigma$ we have
$\ts{}, s \models (\bigvee \Psi)\vfo$ if and only if there exists
$\psi \in \Psi$ such that $\ts{}, s \models \psi\vfo$. Arbitrary
countable conjunction can be obtained similarly.

Now, let $\ts{1} = \tup{\const, T,\stateset_1,s_{01},\abox_1,\trans_1}$
and $\ts{2} = \tup{\const, T,\stateset_2,s_{02},\abox_2,\trans_2}$.
Consider two states $s_1 \in \stateset_1$ and $s_2 \in
\stateset_2$ such that $s_1 \jbsim s_2$.
By induction hypothesis, we have for every valuations $\vfo_1$ and
$\vfo_2$ that assign to each of its free variables a constant
$c_1 \in \adom{\abox_1(s_1)}$ and $c_2 \in \adom{\abox_2(s_2)}$, such
that $c_2 = c_1$, we have that for every formula $\psi \in \Psi$, it
holds $\ts{1}, s_1 \models \psi \vfo_1$ if and only if
$\ts{2}, s_2 \models \tforj(\psi)\vfo_2$.
Given the semantics of $\bigvee \Psi$ above, this implies that
$\ts{1}, s \models (\bigvee \Psi) \vfo_1$ if and only if
$\ts{2}, s \models (\bigvee \tforj(\Psi)) \vfo_2$, where
$\tforj(\Psi) = \{\tforj(\psi) \mid \psi \in \Psi\}$. The proof is
then obtained by observing that
$\bigvee \tforj(\Psi) = \tforj(\bigvee \Psi)$.

\smallskip
\noindent
\textbf{Extension to full $\muladom$.}  In order to extend the result
to the whole \muladom, we resort to the well-known result stating that
fixpoints of the $\mu$-calculus can be translated into the infinitary
Hennessy Milner logic by iterating over \emph{approximants}, where the
approximant of index $\alpha$ is denoted by $\mu^\alpha Z.\Phi$
(resp.~$\nu^\alpha Z.\Phi$)  (cf. \cite{Stir01}). This is a standard result that also holds
for \muladom. In particular, approximants are built as follows:
\[
\begin{array}{rl rl}
  \mu^0 Z.\Phi & = \false
  &  \nu^0 Z.\Phi & = \true\\
  \mu^{\beta+1} Z.\Phi & = \Phi[Z/\mu^\beta Z.\Phi]
  & \nu^{\beta+1} Z.\Phi & = \Phi[Z/\nu^\beta Z.\Phi]\\
  \mu^\lambda Z.\Phi & = \bigvee_{\beta < \lambda} \mu^\beta Z. \Phi &
  \nu^\lambda Z.\Phi & = \bigwedge_{\beta < \lambda} \nu^\beta Z. \Phi
\end{array}
\]
where $\lambda$ is a limit ordinal, and where fixpoints and their
approximants are connected by the following properties: given a
transition system $\ts{}$ and a state $s$ of $\ts{}$
\begin{compactitem}
\item $s \in \MODA{\mu Z.\Phi}$ if and only if there exists an ordinal
  $\alpha$ such that $s \in \MODA{\mu^\alpha Z.\Phi}$ and, for every
  $\beta < \alpha$, it holds that $s \notin \MODA{\mu^\beta Z.\Phi}$;
\item $s \notin \MODA{\nu Z.\Phi}$ if and only if there exists an
  ordinal $\alpha$ such that $s \notin \MODA{\nu^\alpha Z.\Phi}$ and,
  for every $\beta < \alpha$, it holds that $s \in \MODA{\nu^\beta
    Z.\Phi}$.
\end{compactitem}

\end{proof}

As a consequence, from
\Cref{lem:jumping-bisimilar-states-satisfies-same-formula} above, we can
easily obtain the following lemma saying that two transition systems
which are J-bisimilar can not be distinguished by any \muladom formula
(in NNF) modulo a translation $\tforj$.

\begin{lemma}\label{lem:jumping-bisimilar-ts-satisfies-same-formula}
  Consider two transition systems $\ts{1}$,
  and 
  $\ts{2}$ 
  such that $\ts{1} \jbsim \ts{2}$.  For every $\muladom$ closed
  formula $\Phi$ (in NNF)
  we have:
  \[
  \ts{1} \models \Phi \textrm{ if and only if } \ts{2} \models
  \tforj(\Phi).
  \]
\end{lemma}
\begin{proof} Let
  $\ts{1} = \tup{\const,T,\stateset_1,s_{01},\abox_1,\trans_1}$ and
  $\ts{2} = \tup{\const,T,\stateset_2,s_{02},\abox_2,\trans_2}$.
  Since by the definition we have $s_{01} \jbsim s_{02}$, we obtain
  the proof as a consequence of
  \Cref{lem:jumping-bisimilar-states-satisfies-same-formula} due to
  the fact that
  \[
  \ts{1}, s_{01} \models \Phi \textrm{ if and only if } \ts{2}, s_{02}
  \models \tforj(\Phi)
  \]
\ \ 
\end{proof}

\subsection{Transforming Standard GKABs into KABs}\label{sec:transform-sgkab-to-kab}


Here we explain how we transform S-GKABs into KABs with the aim of
reducing verification of S-GKABs into KABs. 
%
As an important observation, notice that a program is essentially a
specification of a sequence of action execution and the atomic step
within a program execution is simply an action execution.
%
Thus, the key idea of the translation is to inductively interpret a
Golog program as a structure consisting of nested processes, suitably
composed through the Golog operators.  We mark the starting and ending
point of each Golog subprogram, and use accessory assertions in the
ABox to track states corresponding to subprograms.  Each subprogram is
then inductively translated into a set of actions and a set of
condition-action rules encoding its entrance and termination
conditions.

As the first step towards defining a generic translation to compile
S-GKABs into KABs, we need to introduce the notion of sub-program and
program IDs. The purpose of the program IDs is to uniquely identify
each sub-program of a Golog program. 
%
%
%
The motivation/intuition why we need to annotate a (sub-)program with
IDs is as follows: Suppose we have a program
\[
\delta = \gact{ \true}{\act_1()} ; \gact{\true}{\act()} ; \gact{\true}{\act_2()};
\gact{\true}{\act()} ; \gact{\true}{\act_3()}
\]
which basically specifies a sequence of action execution
$\act_1, \act, \act_2, \act, \act_3$.
To translate $\delta$ into a set of KAB actions and a set of
condition-action rules that ``mimic the behavior'' of $\delta$, the
idea is to enforce such sequence by using some reserved ABox
assertions (\textit{flags}) that act as pre and post condition of each
action. Essentially, we aim to have that an action is executable when
its pre-condition holds (when the corresponding ABox assertion is
present in the KB) and after its execution, the action should make its
post-condition holds (by adding the corresponding ABox assertion into
the KB). However, since the action $\act$ occurs twice in two
different position, we cannot use a single reserved ABox assertion for
its pre or post condition. We need to differentiate the pre and post
condition of the first and the second occurrence of $\act$ since they
are different. In addition, later on we need to keep track and get the
information about pre and post condition of each
subprograms. Therefore, to differentiate those two different
occurrences of $\act$, we annotate them with IDs.



\begin{definition}[Sub-program]
  Given \sidetext{Sub-program} a program $\delta$, we define the
  notion of a \emph{sub-program} of $\delta$ inductively as follows:
  \begin{compactitem}
  \item $\delta$ is a sub-program of $\delta$,
  \item If $\delta$ is of the form $\delta_1|\delta_2$,
    $\delta_1;\delta_2$, or $\gif{\varphi}{\delta_1}{\delta_2}$, then
    \begin{compactitem}
    \item $\delta_1$ and $\delta_2$ both are sub-programs of $\delta$,
    \item each sub-program of $\delta_1$ is a sub-program of $\delta$,
    \item each sub-program of $\delta_2$ is a sub-program of $\delta$,
    \end{compactitem}
  \item  If $\delta$ is of the form $\gwhile{\varphi}{\delta_1}$
    \begin{compactitem}
    \item $\delta_1$ is a sub-program of $\delta$,
    \item each sub-program of $\delta_1$ is a sub-program of $\delta$,
    \end{compactitem}
  \end{compactitem}
\ \ 
\end{definition}
\noindent
We say a program \emph{$\delta'$ occurs in $\delta$} if $\delta'$ is a
sub-program of $\delta$. Next, we define the notion of Golog programs
with IDs, that are programs in which each of their sub-programs is
annotated with a unique ID.

\begin{definition}[Golog Program with IDs]
  Given \sidetextb{Golog Program with IDs} a set of actions $\actset$,
  a \emph{Golog program with ID} 
  over $\actset$ is an expression formed by the following grammar:
  \[
  \begin{array}{@{}r@{\ }l@{\ }}
    \tup{id, \delta} ::= &
                           \tup{id, \gemptyprog} ~\mid~
                           \tup{id, \gact{Q(\vec{p})}{\act(\vec{p})}} ~\mid~\\
                           &\tup{id, \tup{id_1, \delta_1}|\tup{id_2, \delta_2}}  ~\mid~
                           \tup{id, \tup{id_1, \delta_1};\tup{id_2, \delta_2}} ~\mid~ \\
                         &\tup{id, \gif{\varphi}{\tup{id_1, \delta_1}}{\tup{id_2, \delta_2}}} ~\mid~
                           \tup{id, \gwhile{\varphi}{\tup{id_1, \delta_1}}}
  \end{array}
  \]
  where \emph{$id$ is a program ID} 
  (that can be simply a string over some alphabets), and the rest of
  the things are the same as in \Cref{def:golog-program}.  \ \
\end{definition}

All notions related to Golog program can be defined similarly for the
Golog program with IDs, since essentially we only annotate each
sub-program with a unique ID.
In the following we define a formal translation that transforms a
Golog program into a Golog program with IDs.
As a notation, given program IDs $id$ and $id'$, we write $id.id'$ to
denote a string obtained by concatenating the strings $id$ and $id'$
consecutively.

\begin{definition}[Program ID Assignment]
  We \sidetextb{Program ID Assignment} define a translation $\tpid(\delta, id)$ that 
\begin{compactenum}
\item takes a program $\delta$ as well as a program ID $id$, and
\item produces a Golog program with ID $\tup{id, \delta_{id}}$ such that
  each sub-program of $\delta$ is associated with a unique program ID
  and occurrence matters (i.e., for each sub-program $\delta'$ of
  $\delta$ such that $\delta'$ occurs more than once in $\delta$, each
  of them has a different program ID)
\end{compactenum}
formally as follows: 

  \begin{itemize}
  \item $\tpid(\gemptyprog, id) = \tup{id, \gemptyprog}$, \\where $id'$
    is a fresh program ID.
  \item
    $\tpid(\gact{Q(\vec{p})}{\act(\vec{p})}, id) = \tup{id,
      \gact{Q(\vec{p})}{\act(\vec{p})}}$,
   \\ where $id'$ is a fresh program ID.
  \item $\tpid(\delta_1|\delta_2, id) = \tup{id, \tpid(\delta_1, id.id') |
      \tpid(\delta_2, id.id'')}$, \\ where
    $id'$ and $id''$ are fresh program IDs.
  \item $\tpid(\delta_1;\delta_2, id) = \tup{id, \tpid(\delta_1, id.id') ;
      \tpid(\delta_2, id.id'')}$, \\ where
    $id'$ and $id''$ are fresh program IDs.
  \item
    $\tpid(\gif{\varphi}{\delta_1}{\delta_2}, id) = \tup{id,
      \gif{\varphi}{\tpid(\delta_1, id.id')}{\tpid(\delta_2, id.id'')}}$,
    \\ where $id'$ and $id''$ are fresh program IDs.
  \item
    $\tpid(\gwhile{\varphi}{\delta_1}, id) = \tup{id,
      \gwhile{\varphi}{\tpid(\delta_1, id.id')}}$,
    \\ where $id'$ is a fresh program ID.
  \end{itemize}
  Given a program $\delta$, we say \emph{$\tup{id, \delta_{id}}$ is a
    program with IDs w.r.t.\ $\delta$} if
  $\tpid(\delta, id) = \tup{id, \delta_{id}}$ where $id$ is a fresh
  program ID and $\delta_{id}$ is a program with ID.
\end{definition}

\begin{definition}[Program ID Retrieval function]
  Let \sidetextb{Program ID Retrieval function} $\delta$ be a Golog
  program and $\tup{id, \delta_{id}}$ be its corresponding program
  with IDs w.r.t.\ $\delta$, we define a function $\pid$ that
  \begin{compactenum}
  \item maps each sub-program of $\tup{id, \delta}$ into its unique
    ID. I.e., for each sub-program $\tup{id', \delta'}$ of
    $\tup{id, \delta}$, we have $\pid(\tup{id', \delta'}) = id'$, and
  \item additionally, for a technical reason related to the
    correctness proof of our translation from S-GKABs to KABs, for
    each action invocation
    $\tup{id_\act, \gact{Q(\vec{p})}{\act(\vec{p})}}$ that is a
    sub-program of $\tup{id, \delta_{id}}$, we also have
    $\pid(\tup{id_\act.\gemptyprog, \gemptyprog}) =
    id_\act.\gemptyprog$
    (where $id_\act.\gemptyprog$ is a new ID simply obtained by
    concatenating $id_\act$ with a string $\gemptyprog$).
\end{compactenum}
\ \ 
\end{definition}

\noindent
For simplicity of the presentation, from now on we assume that every
program is associated with ID. 
Note that every program without ID can be transform into
a program with ID as above.
Moreover we will not write the ID that
is attached to a (sub-)program, and when it is clear from the context,
we simply write \emph{$\pid(\delta')$}, instead of
$\pid(\tup{id, \delta'})$, to denote the \emph{unique program ID of a
  sub-program $\delta'$ of $\delta$} that is based on its occurrence
in $\delta$.

We now proceed to define a translation $\tgprog$ that, given a Golog
program $\delta$, produces a set of condition-action rules and a set
of actions that mimic the behavior of $\delta$.
Note that the translation that we present below might not be the only
way to translate a program into the set of condition-action rules and
actions.
Before we formally define $\tgprog$, in \Cref{ex:intuition-tgprog} we
briefly illustrate the idea of how we translate a program and also how
we use the flags to make the resulting condition-action rules mimic
the program behavior (Note that the purpose of
\Cref{ex:intuition-tgprog} is only to give the ideas, and there are
some simplification w.r.t. the translation $\tgprog$ that we define
later).

\begin{example}\label{ex:intuition-tgprog}
Consider a program
\[
\delta = \gact{Q_1(\vec{x})}{\act_1(\vec{x})} ; \gact{Q_2(\vec{y})}{\act_2(\vec{y})}
\]
and a state $s_0 = \tup{\initabox, \scmap_0, \delta}$. Suppose we have
the following run:
\[
\tup{\initabox, \scmap_0, \delta} \trans \tup{A_1, \scmap_1, \delta_1}
\trans \tup{A_2, \scmap_2, \gemptyprog}
\]
where $\delta_1 = \gemptyprog ;
\gact{Q(\vec{y})}{\act_2(\vec{y})}$. Notice that the execution of
$\act_1$ above change $A_0$ into $A_1$ and $\scmap_0$ into $\scmap_1$
(similarly for the execution of $\act_2$).
Now, to mimic the run above within KAB, we do the following:
\begin{enumerate}

\item we use three flags namely $\flagconceptname(c_1)$,
  $\flagconceptname(c_2)$, and $\flagconceptname(c_3)$.

\item we translate $\act_1$ into $\act_1'$ such that
  \[
  \eff{\act_1'} = \eff{\act_1} \cup \set{\map{\true}{ \add \set{ \flagconceptname(c_2) },
                                    \del \set{  \flagconceptname(c_1)  } }}
  \]
  intuitively, the action $\act_1'$ do the same thing as the action
  $\act_1$ except that it deletes the ABox assertion
  $\flagconceptname(c_1)$ and adds the ABox assertion
  $\flagconceptname(c_2)$.

\item we translate $\act_2$ into $\act_2'$ such that
  \[
  \eff{\act_2'} = \eff{\act_2} \cup \set{\map{\true}{ \add \set{ \flagconceptname(c_3) },
                                    \del \set{  \flagconceptname(c_2)  } }}
  \]
  the intuition for $\act_2'$ is similar to $\act_1'$.

\item we introduce the following condition-action rules:
  \begin{compactenum}
  \item
    $\carule{Q_1(\vec{x}) \wedge \flagconceptname(c_1)}{\act_1'(\vec{x})}$.
  \item
    $\carule{Q_2(\vec{y}) \wedge \flagconceptname(c_2)}{\act_2'(\vec{y})}$.
  \end{compactenum}
  intuitively, the first rule enforce that $\act_1$ is executable when
  $Q_1$ is successfully evaluated over the current KB and
  $\flagconceptname(c_1)$ is in the current ABox.
  Similarly, the second condition-action rule require that in order to
  have $\act_2'$ executable, $\flagconceptname(c_2)$ must be in the
  current ABox and $Q_2$ must be successfully evaluated. Since at the
  end of the execution of $\act_1'$ it adds the assertion
  $\flagconceptname(c_2)$, it is easy to see that the flags here
  enforce the sequence of action execution as in the specified program.

\item we add $\flagconceptname(c_1)$ into the ABox in the state where
  the execution of $\delta$ begin. I.e., we have now the state
  $\tup{\initabox \cup \set{\flagconceptname(c_1)} }$. As an
  intuition, later on when we translate S-GKABs into KABs, we add the
  pre condition flag of the corresponding program of S-GKABs into the
  initial state of KABs.

\end{enumerate}

\noindent
Hence, we can now have the following run in KAB:
\[
\tup{\initabox \cup \set{\flagconceptname(c_1)}, \scmap_0} \trans
\tup{A_1 \cup \set{\flagconceptname(c_2)}, \scmap_1} \trans \tup{A_2
  \cup \set{\flagconceptname(c_3)}, \scmap_2}
\]
where the run above is obtained by sequentially executing $\act_1'$
and $\act_2'$.

Roughly speaking, $\flagconceptname(c_1)$ (resp.\
$\flagconceptname(c_2)$) is the pre (resp.\ post) condition of the
program $\gact{Q_1(\vec{x})}{\act_1(\vec{x})}$. Generalizing the idea,
we can say that $\flagconceptname(c_1)$ (resp.\
$\flagconceptname(c_3)$) is the pre (resp.\ post) condition of the
program $\delta$. Thus, later on we will see that the translation
$\tgprog$ will translate a program $\delta$ recursively over the
sub-programs $\delta'$ of $\delta$ and each sub-program has their own
pre/post condition.
\end{example}

As learned from \Cref{ex:intuition-tgprog}, the important intuition to
create the translation $\tgprog$ is that a program is essentially
specifying a particular sequence of action execution. Now,
generalizing the idea in \Cref{ex:intuition-tgprog}, the translation
$\tgprog$ basically takes as inputs:
\begin{compactitem}
\item a Golog program $\delta$,
\item two flags (i.e., special ABox assertions) $\pre$ and $\post$.
\end{compactitem}
and produces the corresponding set of condition-action rules
$\procset$ and actions $\actset$ that mimic the execution of the
program $\delta$. Moreover, the flags $\pre$ and $\post$ are used by
$\procset$ and $\actset$ to mark the start and end of a run $\tau$ in
KAB that is ``induced by'' $\procset$ (together with $\actset$) and
mimics the run $\tau'$ that is ``induced by'' $\delta$ in S-GKAB.
%
%
%
The translation $\tgprog$ works recursively over the structure of the
given program $\delta$, while at the same time for each sub-program
$\delta'$ of $\delta$, it produces and accumulates the corresponding
set of actions and condition-action rules for $\delta'$ in order to
produce the whole set of actions $\actset$ and condition-action rules
$\procset$ that mimic the given program $\delta$.
%
%
Formally, the translation $\tgprog$ 
is
defined 
as follows:

\begin{definition}[Program Translation]\label{def:prog-translation}
%
%
A \sidetext{Program Translation} program translation $\tgprog$ that takes as inputs:
\begin{compactenum}[(i)]
\item A program $\delta$ over a set of actions $\actset$,
\item Two flags (i.e., two special ABox assertions which will be used
  as markers indicating the start and the end of the execution of a
  program $\delta$).
\end{compactenum}
and produces as outputs:
\begin{compactenum}[(i)]

\item $\ppre$ is a function that maps the ID of $\delta$ (as well as
  the IDs of all sub-programs of $\delta$) to the flag (called
  \emph{start flag}) that acts as a marker indicating the start of the
  run induced by the corresponding set of condition-action rules and
  actions obtained from $\delta$ (as well as all of its sub-programs),

\item $\ppost$ is a function that maps the ID of $\delta$ (as well as
  the IDs of all sub-programs of $\delta$) to the flag
  (called \emph{end flag}) that acts as a marker indicating the end of
  the run induced by the corresponding set of condition-action rules
  and actions obtained from $\delta$ (as well as all of its
  sub-programs),

\item $\procset$ is a process (a set of condition-action rules),

\item $\actset'$ is a set of actions.

\end{compactenum}
I.e.,
$\tgprog(\pre, \delta, \post) = \tup{\ppre, \ppost, \procset,
  \actset'}$, where $\pre$ and $\post$ are flags.
Formally, $\tgprog(\pre, \delta, \post)$ is inductively defined over
the structure of a program $\delta$ as follows:

 \begin{enumerate}

 \item For the case of $\delta = \gemptyprog$ (i.e., $\delta$ is an
   empty program):
   \[
   \tgprog(\pre, \gemptyprog, \post) = 
   \tup{ \ppre, \ppost, \set{\carule{\pre} {\act_\gemptyprog()}},
     \set{\act_\gemptyprog} },
   \] 
   where
   \begin{compactitem}
   \item $\ppre = \set{\tap{\pid(\gemptyprog) \ra \pre}}$,
   \item $\ppost = \set{\tap{\pid(\gemptyprog) \ra \post}}$,
   \item $\act_\gemptyprog$ is of the form
     $ \act_\gemptyprog():\set{\map{\true}{ \add \set{\post, \tmp},
         \del \set{\pre} }}; $
   \end{compactitem}


 \item For the case of $\delta = \gact{Q(\vec{p})}{\act(\vec{p})}$
   (i.e., $\delta$ is an action invocation) with
   $\pid(\gact{Q(\vec{p})}{\act(\vec{p})}) = id_\act$:
  \[
  \tgprog(\pre, \gact{Q(\vec{p})}{\act(\vec{p})}, \post) = \tup{\ppre,
    \ppost, \procset, \actset'},
  \]
  where
  \begin{compactitem}
  \item
    $\ppre = \set{\tap{\pid(\gact{Q(\vec{p})}{\act(\vec{p})}) \ra \pre}}
    \cup \ppre'$,
  \item
    $\ppost = \set{\tap{\pid(\gact{Q(\vec{p})}{\act(\vec{p})}) \ra \post}}
    \cup \ppost'$,
  \item
    $\procset = \set{\carule{Q(\vec{p}) \wedge \pre}{\act'(\vec{p})}}
    \cup \procset'$,
  \item $\actset' = \set{\act'} \cup \actset''$,
    where
  \begin{center}
    $\begin{array}{ll} \eff{\act'} = &\eff{\act} \cup \\
       &\set{\map{\true}{ \add \set{\post}, \del \set{\pre, \tmp} }}
         \cup \\
       &\set{\map{\noopconcept{x}}{\del \noopconcept{x} }},
  \end{array}
  $\end{center}
\item
  $\tgprog(\post, \gemptyprog, \post) = \tup{\ppre', \ppost',
    \procset', \actset''}$, where $\pid(\gemptyprog) = id_\act.\gemptyprog$
%
\end{compactitem}


\item For the case of $\delta = \delta_1 | \delta_2$ (i.e., $\delta$
  is a non-deterministic choice between programs):
  \[
  \tgprog(\pre, \delta_1|\delta_2, \post) = \tup{\ppre, \ppost, \procset, \actset},
  \]
  where
  \begin{compactitem}
  \item $ \procset = \set{
      \carule{\pre}{\gamma_{\delta_1}()},
      \carule{\pre}{\gamma_{\delta_2}()}}
    \cup \procset_1 \cup \procset_2$,
  \item $\actset = \actset_1~\cup~\actset_2~\cup~\set{\gamma_{\delta_1}, \gamma_{\delta_2}}$, where
    \begin{compactitem}
    \item $\gamma_{\delta_1}():\set{\map{\true} {\add \set{\flagconcept{c_1}, \tmp}, \del \set{\pre} }}$,
    \item $\gamma_{\delta_2}() : \set{\map{\true}{\add \set{\flagconcept{c_2}, \tmp}, \del \set{\pre} }}$,
    \end{compactitem}
\item
  $\ppre = \set{\tap{\pid(\delta_1|\delta_2) \ra \pre}} \cup \ppre_1 \cup
  \ppre_2$,

\item
  $\ppost = \set{\tap{\pid(\delta_1|\delta_2) \ra \post}} \cup \ppost_1 \cup
  \ppost_2$,

  \item $\tgprog(\flagconcept{c_1}, \delta_1, \post) = \tup{\ppre_1, \ppost_1, \procset_1, \actset_1}$,
  \item $\tgprog(\flagconcept{c_2}, \delta_2, \post) = \tup{\ppre_2, \ppost_2, \procset_2, \actset_2}$,
  \item $c_1, c_2 \in \const_0$ are fresh constants;
  \end{compactitem}

%
\item
  $\tgprog(\pre, \delta_1;\delta_2, \post) = \tup{\ppre, \ppost,
    \procset_1 \cup \procset_2, \actset_1 \cup \actset_2}$, where
\begin{compactitem}
\item
  $\ppre = \set{\tap{\pid(\delta_1;\delta_2) \ra \pre}} \cup \ppre_1 \cup
  \ppre_2$,
\item
  $\ppost = \set{\tap{\pid(\delta_1;\delta_2) \ra \post}} \cup \ppost_1 \cup
  \ppost_2$,
\item $\tgprog(\pre, \delta_1, \flagconcept{c}) =
  \tup{\ppre_1, \ppost_1, \procset_1, \actset_1}$,
\item $\tgprog(\flagconcept{c}, \delta_2, \post) =
  \tup{\ppre_2, \ppost_2, \procset_2, \actset_2}$, 
\item $c \in \const_0$ is a fresh constant;
\end{compactitem}


\item
  $\tgprog(\pre, \gif{\varphi}{\delta_1}{\delta_2}, \post) =
  \tup{\ppre, \ppost, \procset,\actset}$, where
\begin{compactitem}
\item
  $\ppre = \set{\tap{\pid(\gif{\varphi}{\delta_1}{\delta_2}) \ra \pre}} \cup
  \ppre_1 \cup \ppre_2$,
\item
  $\ppost = \set{\tap{\pid(\gif{\varphi}{\delta_1}{\delta_2}) \ra \post}}
  \cup \ppost_1 \cup \ppost_2$,

\item
  $ \procset = \set{
    \carule{\pre \wedge \varphi}{\gamma_{if}()},
    \carule{\pre \wedge \neg \varphi}{\gamma_{else}()}}
  \cup \procset_1 \cup \procset_2$,
\item $\actset = \actset_1~\cup~\actset_2~\cup~\set{\gamma_{if},
    \gamma_{else}}$, where
\begin{compactitem}
\item $\gamma_{if}():\set{\map{\true}
    {\add \set{\flagconcept{c_1}, \tmp}, \del
      \set{\pre} }}$,
\item
  $\gamma_{else}() : \set{\map{\true}{\add \set{\flagconcept{c_2},
        \tmp}, \del \set{\pre} }}$, 
\end{compactitem}
\item
  $\tgprog(\flagconcept{c_1}, \delta_1, \post) = \tup{\ppre_1,
    \ppost_1, \procset_1, \actset_1}$,

\item
  $\tgprog(\flagconcept{c_2}, \delta_2, \post) = \tup{\ppre_2,
    \ppost_2, \procset_2, \actset_2}$,

\item $c_1, c_2 \in \const_0$ are fresh constants;
\end{compactitem}

\item $\tgprog(\pre, \gwhile{\varphi}{\delta}, \post) =
  \tup{\ppre, \ppost, \procset, \actset}$, where 
\begin{compactitem}
\item
  $\ppre = \set{ \tap{\pid(\gwhile{\varphi}{\delta}) \ra \pre}} \cup
  \ppre'$

\item
  $\ppost = \set{ \tap{\pid(\gwhile{\varphi}{\delta}) \ra \post}} \cup
  \ppost'$


\item $ \procset~=~\procset'\cup~ \procset_{loop}$, where
  $\procset_{loop}$ contains:
\begin{compactitem}
\item
  $ \carule{\pre \wedge \varphi \wedge \neg \noopconcept{noop}
  }{\gamma_{doLoop}()}$,
\item
  $ \carule{\pre \wedge (\neg \varphi \vee \noopconcept{noop}) }
  {\gamma_{endLoop}()}$,


\end{compactitem}

\item $\actset~=~\actset'~\cup \actset_{loop}$, where $\actset_{loop}$
  contains the following:
\begin{compactitem}


\item $\gamma_{doLoop}(): \set{\true  \rightsquigarrow\\
    \hspace*{10mm} \add \set{\flagconcept{lStart}, \noopconcept{noop},
      \tmp }, \del \set{\pre}}$,
\item $\gamma_{endLoop}(): \set{\true  \rightsquigarrow \\
    \hspace*{10mm} \add \set{\post, \tmp }, \del \set{\pre,
      \noopconcept{noop}}}$,

\end{compactitem}

\item
  $\tgprog(\flagconcept{lStart}, \delta, \pre) =
  \tup{\ppre', \ppost', \procset', \actset'}$,


\item $noop, lStart \in \const_0$ are fresh constants.
\end{compactitem}
%
%
%
%
%
%
%
\end{enumerate}

\ \ 
\end{definition}


For compactness reason, in the following we often simply write
$\ppre(\delta)$ to abbreviate the notation $\ppre(\pid(\delta))$ that
essentially returns the start flag of a program with program ID
$\pid(\delta)$. Similarly for $\ppost(\delta)$. 
To give the intuition of the translation $\tgkab$ 
as well as to show some of its properties, we
first introduce 
the notion when a state of an S-GKAB is mimicked by a state of
a KAB. Roughly speaking, a state $\tup{A_g, \scmap_g, \delta_g}$ of an
S-GKAB is mimicked by a state $\tup{A_k, \scmap_k}$ of a KAB if
\begin{compactenum}
\item The corresponding ABox of those states contain the same
  assertions except the special markers. I.e., they are equal modulo
  special markers (cf. \Cref{def:equal-mod-markers}),
\item They both have the same service call map, and
\item The ABox $A_k$ contains the start flag of the program
  $\delta_g$.
\end{compactenum}
This notion is formally defined as follows.

\begin{definition}\label{def:mimic-state}
  Let \sidetextb{Mimicked States}
  $\gkabsym = \tup{T, \initabox, \actset, \delta}$ be an S-GKAB with
  transition system $\ts{\gkabsym}^{\filter_S}$, and
  $\tgkab(\gkabsym) = \tup{T, \initabox', \actset', \procset'}$ be a
  KAB with transition system $\ts{\tgkab(\gkabsym)}$ obtained from
  $\gkabsym$ through $\tgkab$.
  Consider two states $\tup{A_g,\scmap_g, \delta_g}$ of
  $\ts{\gkabsym}^{\filter_S}$ and $\tup{A_k,\scmap_k}$ of
  $\ts{\tgkab(\gkabsym)}$. 
We say \emph{$\tup{A_g,\scmap_g, \delta_g}$ is mimicked by
  $\tup{A_k,\scmap_k}$} (or equivalently \emph{$\tup{A_k,\scmap_k}$
  mimics $\tup{A_g,\scmap_g, \delta_g}$}), written
$\tup{A_g,\scmap_g, \delta_g} \mimic \tup{A_k,\scmap_k}$, if
\begin{compactenum}
\item $A_k \eqm A_g$,
\item $\scmap_k = \scmap_g$, and 
\item $\ppre(\delta_g) \in A_k$. 
\end{compactenum}
\ \ 
\end{definition}


\noindent
The intuition of each recursive step in the translation $\tgprog$
is then explained as follows:
\begin{itemize}
\item For the case of $\delta = \gemptyprog$, given a program
  $\gemptyprog$, a start flag $\pre$, and an end flag $\post$, the
  translation $\tgprog$ do the following:
  \begin{compactenum}

  \item it produces a function $\ppre$ such that
    $\ppre(\pid(\gemptyprog)) = \pre$,

  \item it produces a function $\ppost$ such that
    $\ppost(\pid(\gemptyprog)) = \post$,

  \item it produces a singleton set of actions containing an action
    $\act_\gemptyprog$ that adds (resp.\ deletes) the end flag $\post$
    (resp.\ the start flag $\pre$).

  \item Additionally, $\act_\gemptyprog$ also adds the assertion
    $\tmp$. The idea is to make the state generated by this action
    ignored (during the verification). Because in S-GKABs, when we
    have a program $\gemptyprog$, they essentially do not make any
    transition and the program execution can be considered completed
    (cf. \Cref{def:final-state-program}). Basically, it is also the
    reason why we introduce the translation $\tforj$
    (cf. \Cref{def:tforj}) that bypass some intermediate states
    (states containing $\tmp$).
  \end{compactenum}
  
  As a further intuition for the translation of the case
  $\delta = \gemptyprog$, consider a program 
    \[
    \delta = \gact{\true}{\act_1()}; \delta_2 ; \gact{\true}{\act_3()},
    \]
    where $\delta_2 = \gact{\true}{\act_2()}\ |\ \gemptyprog$.
    Thus, due to the use of non-deterministic choice construct, there
    are two possible sequences of action execution namely:
    \begin{compactenum}[\bf (1)]
    \item $\act_1$ and then followed by $\act_2$ and $\act_3$.
    \item $\act_1$ and then followed by $\act_3$.
    \end{compactenum}
    To emulate such situation in KAB, 
    \begin{compactitem}
    \item for \textbf{(1)}, the post-condition of $\act_1$
      should be the pre-condition for $\act_2$, 
    \item for \textbf{(2)}, the post-condition of $\act_1$ should be
      the pre-condition for $\act_3$.
    \end{compactitem}
    Hence, when we translate $\act_1$, it is not clear which end flag
    that $\act_1$ should add at the end of its execution. Therefore,
    it is one of the reason why we translate $\gemptyprog$ into an
    action $\act_\gemptyprog$ that only change the flag.
    Note that it is also aligned with the general intuition of the
    translation $\tgprog$ in which for each sub-program there will be
    a corresponding start flag and end flag.
    Thus, for the case \textbf{(2)} above, we then have an action
    $\act_\gemptyprog$ that bridges the execution of $\act_1$ and
    $\act_3$. Essentially, the translation $\tgprog$ will translate
    $\act_1$ such that at the end of its execution it adds an end flag
    that is also the start flag of $\delta_2$ and then, no matter
    which choice that is taken in $\delta_2$, at the end of execution
    of $\delta_2$, it will add an end flag that is also the start flag
    of $\act_3$ (i.e., in the case of choosing $\gemptyprog$, the
    action $\act_\gemptyprog$ will put the start flag of $\act_3$ so
    that $\act_3$ can be fired, however the generated state by
    $\act_\gemptyprog$ will be marked as an intermediate state by the
    assertion $\tmp$ and hence it will be ignored during the
    verification).
%
%
    Similarly for the case of a program sequence, as an intuition,
    consider the
    program 
    \[
    \delta = \gemptyprog ; \gact{\true}{\act_1()} ;
    \gact{\true}{\act_2()}.
    \]
    Let $A$ be an ABox and $\scmap$ be a service call map, then we
    have
    \[
    \tup{A, \scmap, \delta} \gprogtrans{\act_1\sigma, \filter_S}
    \tup{A', \scmap', \delta'}
    \]
    where $\delta' = \gact{\true}{\act_2()}$. However, following the
    general idea of translation $\tgprog$, especially for the case of
    program sequence $\delta = \delta_1 ; \delta_2$, each sub-program
    will be translated into a set of condition-action rules and
    actions, and each sub-program will have the corresponding start
    (resp.\ end) flag that will drive the execution of actions in
    KAB. In the case of $\delta = \delta_1 ; \delta_2$, the end flag
    of $\delta_1$ should be the start flag of $\delta_2$. Therefore,
    when $\delta_1$ is $\gemptyprog$, we need to change the flag from
    the start flag of $\delta_1$ into the end flag of $\delta_1$. This
    is also one of the reason why we translate $\gemptyprog$ this way.

  \item For the case of $\delta = \gact{Q(\vec{p})}{\act(\vec{p})}$,
    given a program $\gact{Q(\vec{p})}{\act(\vec{p})}$, a start flag
    $\pre$, and an end flag $\post$, the idea of translation $\tgprog$
    for this case is as follows:
  \begin{compactitem}

  \item the translation $\tgprog$ produces a function $\ppre$ (resp.\
    $\ppost$) such that
    $\ppre(\pid(\gact{Q(\vec{p})}{\act(\vec{p})})) = \pre$ (resp.\
    $\ppost(\pid(\gact{Q(\vec{p})}{\act(\vec{p})})) = \post$),


  \item we translate the action $\act$ into an action $\act'$ such
    that $\act'$ does the same thing as $\act$ except that it also
    does the following:
    \begin{compactitem}
    \item  $\act'$ adds the end flag $\post$ and deletes the start
      flag $\pre$.
    \item $\act'$ deletes all flags made by the concept name
      $\noopconceptname$. This deletion is related to the translation
      of the while loop construct. In the semantics of while loop, as
      it is explained in the beginning of this chapter, for any ABox
      $A$ and service call map $\scmap$, we have that
      $\final{\tup{A, \scmap, \gwhile{\varphi}{\delta'}}}$ if
      $\ask(\varphi, T, A) = \true$, and
      $\final{\tup{A, \scmap, \delta'}}$. I.e., when $\delta'$ is
      considered to be completed, the whole loop can be considered to
      be completed as well.
%
%
%
      To check this, we make use the flags made by the concept name
      $\noopconceptname$. The idea is that if there is still a flag
      made by $\noopconceptname$ at the end of the loop, then it means
      that there is no action that is executed. Therefore, here we
      translate an action such that it clears all flags made by
      $\noopconceptname$ to give a sign that there is an action that
      is executed.
    \end{compactitem}

  \item we create a condition-action rule
    $\carule{Q(\vec{p}) \wedge \pre}{\act'(\vec{p})}$. So, the idea is
    that $\act'$ can be executed when $\pre$ is in the current ABox
    and at the end of the execution of $\act'$, we have $\post$
    in the ABox.

  \item we also recursively call the translation
    $\tgprog(\post, \gemptyprog, \post)$ where
    $\pid(\gemptyprog) = id_\act.\gemptyprog$. 
    This step is needed for some part of the correctness proof of the
    translation $\tgprog$.
    In particular, later on we will show that given a GKAB state
    $s_1 = \tup{A_g, \scmap_g, \delta_g}$ and a KAB state
    $s_2 = \tup{A_k, \scmap_k}$ such that $A_k$ mimics $A_g$, if $s_1$
    can reach a state $s_1' = \tup{A_g', \scmap_g', \delta_g'}$, then
    there exists a state $s_2' = \tup{A_k', \scmap_k'}$ such that
    $s_2$ can reach $s_2'$ and $s_2'$ mimics $s_1'$. Thus, for the
    base case of the proof (the case of action invocation), since we
    have
    \[
    \tup{A_g,\scmap_g, \gact{Q(\vec{x})}{\act(\vec{x})}} \gprogtrans{\act\sigma, \filter_S}
    \tup{A'_g,\scmap'_g, \gemptyprog},
    \]
    then we need to have the start flag of $\gemptyprog$ in
    $A_k'$. Therefore we need to translate $\gemptyprog$ (where
    $\pid(\gemptyprog) = id_\act.\gemptyprog$) such that it has the
    start flag $\post$ (that is also the end flag of $\act$). In
    addition, since we don't need to change further the flag, we call
    the translation $\tgprog$ for $\gemptyprog$ with the end flag
    $\post$.

  \end{compactitem}


\item For the case of $\delta = \delta_1 ; \delta_2$, given a program
  $\delta_1 ; \delta_2$, a start flag $\pre$, and an end flag $\post$,
  the translation $\tgprog$ do the following:
  \begin{compactitem}

  \item it introduces a fresh flag $\flagconceptname(c)$ that become
    the end flag of $\delta_1$ and also the start flag of
    $\delta_2$. The intuition is that $\flagconceptname(c)$
    bridges/connects the run in the KAB that emulates $\delta_1$ and
    the run in the KAB that emulates $\delta_2$.

  \item it recursively translates $\delta_1$ and $\delta_2$ using the
    translation $\tgprog$.  For translating $\delta_1$, the
    translation $\tgprog$ is fired with the start flag $\pre$ and the
    end flag $\flagconceptname(c)$, while for translating $\delta_2$,
    the translation $\tgprog$ is fired with the start flag
    $\flagconceptname(c)$ and the end flag $\post$. The idea is to
    enforce that $\delta_2$ will be executed after $\delta_1$ has been
    successfully executed.

  \item it produces a set of condition-action rules (resp.\ a set of
    actions) that is obtained by merging the sets of condition-action
    rules (resp.\ the sets of actions) that are obtained from
    recursively translating $\delta_1$ and $\delta_2$

  \item it produces a function $\ppre$ such that
%
  $\ppre = \set{\tap{\pid(\delta_1;\delta_2) \ra \pre}} \cup \ppre_1 \cup
  \ppre_2$,
  where $\ppre_1$ (resp.\ $\ppre_2$) is obtained from the translation
  of $\delta_1$ (resp.\ $\delta_2$). Intuitively, $\ppre$ is obtained
  by mapping the ID of the current program into the given start flag
  $\pre$ while also accumulating the mapping between IDs and start
  flags obtained from the translation of all of its sub-program.

  \item the intuition for $\ppost$ is similar to $\ppre$.

  \end{compactitem}

\item For the case of $\delta = \delta_1 | \delta_2$, given a program
  $\delta_1 | \delta_2$, a start flag $\pre$ and an end flag $\post$,
  the translation $\tgprog$ do the following:
  \begin{compactitem}
  \item it introduces two flags $\flagconceptname(c_1)$ and
    $\flagconceptname(c_2)$. One for the start flag of $\delta_1$ and
    one for the start flag of $\delta_2$. Two corresponding actions
    also generated in order to put either $\flagconceptname(c_1)$ or
    $\flagconceptname(c_2)$ into the Abox (i.e., to mark the start of
    the run in the KAB that emulates the run in the S-GKAB that is
    induced by either $\delta_1$ or $\delta_2$).
  \item it recursively translates $\delta_1$ with the start flag
    $\flagconceptname(c_1)$ and the end flag $\post$. Similarly for
    $\delta_2$ except that the start flag is $\flagconceptname(c_2)$
  \item it constructs $\ppre$, $\ppost$, $\procset$, and $\actset$ by
    accumulating the results from translating $\delta_1$ and
    $\delta_2$ (similar to the previous construct).
  \end{compactitem}  
  Furthermore, the reason why we need to introduce two fresh flags for
  the start flag of $\delta_1$ and $\delta_2$ is to enforce that each
  sub-program has a unique start flag. Later on, we will see that it
  is important because as a step to show the correctness of the
  translation $\tgprog$, we show that the behavior of the transition
  system will be the ``same'' starting from a GKAB state
  $s_g = \tup{A_g,\scmap_g,\delta_1|\delta_2}$ and a KAB state $s_k = \tup{A_k,\scmap_k}$ in
  which $s_k$ mimics $s_g$.
%
%
  Now, suppose that we do not invent two fresh flags for the
  translation of $\delta_1$ as well as $\delta_2$, and we simply use
  $\pre$ as their start flag, then we will have that
  $\delta_1 | \delta_2$, $\delta_1$, and $\delta_2$ are having the
  same start flag. Hence it is easy to see that the property that we
  want to show above cannot be proven. As an intuition, suppose that
  the execution of $\delta_1$ will be stuck while the execution of
  $\delta_2$ is not. Then we cannot say that the behavior of the
  transition system will be the ``same'' starting from
  $s_1 = \tup{A_g,\scmap_g,\delta_1}$ and $s_2 = \tup{A_k,\scmap_k}$
  where $s_2$ mimics $s_1$.
%
  Because $\tup{A_g,\scmap_g,\delta_1}$ will be stuck but
  $\tup{A_k,\scmap_k}$ will be able to continue the execution by
  emulating the execution of $\delta_2$ and this is possible since in
  any case the start flag of $\delta_2$ is the same as the start flag
  of $\delta_1$ and $A_k$ contains such start flag.

\item For the case of $\delta =
  \gif{\varphi}{\delta_1}{\delta_2}$, the idea is as follows:
  \begin{compactitem}
  \item we first introduce two fresh flags that will be used as the
    start flags of $\delta_1$ and $\delta_2$.

  \item to check whether $\varphi$ is successfully evaluated over
    the current KB or not, we introduce two condition-action rules as
    follows:
    \begin{compactenum}
    \item $ \carule{\pre \wedge \varphi}{\gamma_{if}()}$,
    \item $ \carule{\pre \wedge \neg \varphi}{\gamma_{else}()} $
  \end{compactenum}
  in case $\varphi$ is successfully evaluated, the condition of the
  first rule will be satisfied, and then the action $\gamma_{if}$
  will be fired. As a consequence, $\gamma_{if}$ will add the start
  flag for $\delta_1$. 
  In case $\varphi$ is not successfully evaluated, the condition of
  the second rule will be satisfied, and then the action
  $\gamma_{else}$ will be fired. As a consequence, $\gamma_{else}$
  will add the start flag for $\delta_2$.

\item similar to the other previous construct, in this case the
  translation $\tgprog$ also recursively translates the program
  $\delta_1$ and $\delta_2$ each with their own start flag but both of
  them have the same end flag.

\item note that the states generated by either $\gamma_{if}$ or
  $\gamma_{else}$ will be just intermediate states that will be
  ignored during the verification.

\item the idea for the construction of $\ppre$, $\ppost$, $\procset$,
  and $\actset$, that involves accumulating the results from
  translating $\delta_1$ and $\delta_2$, is similar to the other
  previous construct.

\end{compactitem}


\item For the case of $\delta = \gwhile{\varphi}{\delta'}$, the idea is
  as follows: 
  \begin{compactitem}

  \item when we translate a loop, we introduce a fresh flag
    $\flagconceptname(lStart)$ that is used to mark the situation when
    we enter the body of the corresponding loop.

  \item we introduce an action $\gamma_{doLoop}$ that adds the flag
    $\flagconceptname(lStart)$. 

  \item based on the semantics of the loop construct,
    as it is explained in the beginning of this chapter, for any ABox
    $A$ and service call map $\scmap$, we have that
    $\final{\tup{A, \scmap, \gwhile{\varphi}{\delta'}}}$ if
    $\ask(\varphi, T, A) = \true$, and
    $\final{\tup{A, \scmap, \delta'}}$. I.e., when $\delta'$ is
    considered to be completed, the whole loop can be considered to be
    completed.
%
%
    To mimic such situation in KABs, when we translate a loop, we
    introduce a fresh flag $\noopconceptname(noop)$ that will be also
    added by $\gamma_{doLoop}$. The idea is that the flag
    $\noopconceptname(noop)$ should be added when we start mimicking
    the computation of the loop in KABs and it should be deleted in
    case there is an action that is executed within the run that
    emulates the body of the loop.  Otherwise, if the flag
    $\noopconceptname(noop)$ remains in the ABox until the end of the
    run that mimics the loop computation, then it means that there is
    no action execution (Recall that we translate an action such that
    it will deletes all flags made by $\noopconceptname$).

  \item the program $\delta'$ is recursively translated by $\tgprog$
    with the start flag $\flagconceptname(lStart)$ and the end flag
    $\pre$. The reason of using $\pre$ as the end flag of $\delta'$ is
    to enforce repetition. So, when the corresponding run in KAB have
    finished mimicking the execution of $\delta'$, it should check
    whether the guard of the loop still holds or not. In case yes, it
    should emulate $\delta'$ again. Hence, by having $\pre$ as the end
    flag, the execution will be back to the beginning again and it
    will check the guard of the loop again.

  \item we also introduce a condition-action rule
    \[
    \carule{\pre \wedge \varphi \wedge \neg \noopconcept{noop}}{\gamma_{doLoop}()}
    \]
    that basically guards the execution of $\gamma_{doLoop}$ such that
    $\gamma_{doLoop}$ will be executed when the following hold:
    \begin{compactenum}
    \item when we start to (re-)enter the loop (marked by $\pre$),
    \item when the guard of the loop $\varphi$ is satisfied, and
    \item when there is no assertion $\noopconcept{noop}$.
    \end{compactenum}
    The reason why we need to check the presence of
    $\noopconcept{noop}$ is because we want to check whether there was
    any action execution within the body of the loop or not (Notice
    that this condition-action rule will be re-evaluated at the end of
    a loop).
%

  \item we also introduce an action $\gamma_{endLoop}$ that adds the
    flag $\post$ and makes us leave the loop. The execution of
    $\gamma_{endLoop}$ is then guarded by the condition-action rule
    \[
    \carule{\pre \wedge (\neg \varphi \vee \noopconcept{noop}) }
    {\gamma_{endLoop}()}
    \]
    in which its left hand side will be checked when we (re-)enter the
    loop. This rule says that $\gamma_{endLoop}()$ will be fired when
    the guard of the loop $\varphi$ is not satisfied or when there is
    $\noopconcept{noop}$ (i.e., there is no need to re-execute the
    loop).

  \item the idea for the construction of $\ppre$, $\ppost$,
    $\procset$, and $\actset$ that involves accumulating the results
    from translating $\delta'$ is similar to the other previous
    construct.

\end{compactitem}

\end{itemize}

Below we show an important Lemma about the function $\ppre$ and
$\ppost$ that will be used quite often later when we reduce the
verification of S-GKABs into KABs.


%
\begin{lemma}\label{lem:program-pre-post}
  Given a program $\delta$ over a set $\actset$ of actions.  We have
\[
\tgprog(\pre, \delta, \post) = \tup{\ppre, \ppost, \procset, \actset}
\textrm{ \ if and only if \ } \ppre(\delta) = \pre \mbox{ and }
\ppost(\delta) = \post 
\]
\end{lemma}
\begin{proof}
  Directly follows from the definition of $\tgprog$.
\end{proof}
%
%

Having $\tgprog$ in hand, we define a translation $\tgkab$ that
compile S-GKABs into KABs as follows.

\begin{definition}[Translation from S-GKABs to KABs]
  We \sidetext{Translation from S-GKABs to KABs} define a translation
  $\tgkab$ that takes an S-GKAB
  $\gkabsym = \tup{T, \initabox, \actset, \ginitprog}$ as the input
  and produces a KAB
  $\tgkab(\gkabsym) = \tup{T, \initabox', \actset', \procset'}$ s.t.\
  \begin{compactitem}
  \item $\initabox' = \initabox \cup \set{\flagconcept{start}}$, and
  \item
    $\tgprog(\flagconcept{start}, \ginitprog, \flagconcept{end}) =
    \tup{\ppre, \ppost, \procset', \actset'}$.
  \end{compactitem}
\ \ 
\end{definition}

Next, we define the notion of temp adder/deleter action as follows.

\begin{definition}[Temp Marker Adder Action]\label{def:tmp-adder-action}
  Let \sidetextb{Temp Marker Adder Action}
  $\gkabsym = \tup{T, \initabox, \actset, \delta}$ be an S-GKAB and
  $\tgkab(\gkabsym) = \tup{T, \initabox', \actset', \procset'}$ be the
  corresponding KAB obtained from $\gkabsym$ via $\tgkab$.
  An action $\act \in \actset$ is \emph{a temp adder action of
    $\tgkab(\gkabsym)$} if there exists an effect $e \in \eff{\act}$
  of the form $ \map{[q^+]\land Q^-}{\add \facta, \del \factd} $ such
  that $\tmp \in \facta$.
  We write $\actsettmpa$ to denote the set of temp adder actions
  of $\tgkab(\gkabsym)$.
\end{definition}

\begin{definition}[Temp Marker Deleter Action]\label{def:tmp-deleter-action}
  Let \sidetextb{Temp Marker Deleter Action}
  $\gkabsym = \tup{T, \initabox, \actset, \delta}$ be an S-GKAB and
  $\tgkab(\gkabsym) = \tup{T, \initabox', \actset', \procset'}$ be the
  corresponding KAB obtained from $\gkabsym$ via $\tgkab$.
  An action $\act \in \actset$ is \emph{a temp deleter action of
    $\tgkab(\gkabsym)$} if there exists an effect $e \in \eff{\act}$
  of the form $ \map{[q^+]\land Q^-}{\add \facta, \del \factd} $ such
  that $\tmp \in \factd$.
  We write $\actsettmpd$ to denote the set of temp deleter
  actions of $\tgkab(\gkabsym)$.
\end{definition}

\noindent
Roughly speaking, a temp adder action is an action that adds the ABox
assertion $\tmp$. Similarly, a temp deleter action is an action that
removes the ABox assertion $\tmp$. In the following, we show several
important properties of temp adder/deleter action that will be used
later for reducing the verification of S-GKABs into KABs.



\begin{lemma}\label{lem:action-set-separation}
  Let $\gkabsym$ be an S-GKAB,
  $\tgkab(\gkabsym) = \tup{T, \initabox', \actset', \procset'}$ be the
  corresponding KAB obtained from $\gkabsym$ via $\tgkab$, and
  $\actsettmpa$ (resp.\ $\actsettmpd$) be a set of temp adder
  (resp.\ deleter) actions of $\tgkab(\gkabsym)$. We have that
  $\actset' = \actsettmpa \uplus \actsettmpd$.
\end{lemma}
\begin{proof}
  Trivially true by observing
  \Cref{def:prog-translation,def:tmp-adder-action,def:tmp-deleter-action}.
\end{proof}

\begin{lemma}\label{lem:temp-state-produced-by-temp-act}
  Let $\gkabsym$ be an S-GKAB,
  $\tgkab(\gkabsym) = \tup{T, \initabox', \actset', \procset'}$ be the
  corresponding KAB (with transition system $\ts{\tgkab(\gkabsym)}$)
  obtained from $\gkabsym$ via $\tgkab$, and $\actsettmpa$ be a set of
  temp adder actions of $\tgkab(\gkabsym)$.
  Consider a state $\tup{A_k, \scmap_k}$ of $\ts{\tgkab(\gkabsym)}$,
  if there exists a state $\tup{A_k', \scmap_k'}$ such that
  $\tup{A_k, \scmap_k} \exect{\act\sigma} \tup{A_k', \scmap_k'}$, and
  $\tmp \in A_k'$ then 
  $\sigma$ is an empty substitution,
  $\act \in \actsettmpa$,
  $\act$ does not involve any service calls,
  $A_k' \eqm A_k$ and $\scmap_k' = \scmap_k$.
\end{lemma}
\begin{proof}
%
  Since $\tmp \in A_k'$, then by \Cref{def:tmp-adder-action} and
  \Cref{lem:action-set-separation} we must have
  $\act \in \actsettmpa$. By the definition of translation $\tgprog$
  (see \Cref{def:prog-translation}), any actions in $\actsettmpa$ does
  not involve service calls and only do a manipulation on special
  markers. Thus, it is easy to see that $A_k' \eqm A_k$ and
  $\scmap_k' = \scmap_k$.
\end{proof}

The following lemma basically says that if there is a transition from
a state $s$ to $s'$ that is obtained by execution an action $\act'$
and $\tmp \not\in \abox(s_k')$, then the action $\act'$ must be an
action that deletes $\tmp$ and it must be obtained from a translation
of an atomic action invocation.

\begin{lemma}\label{lem:non-temp-state-produced-by-normal-action}
  Let $\gkabsym = \tup{T, \initabox, \actset, \ginitprog}$ be an S-GKAB,
  $\tgkab(\gkabsym) = \tup{T, \initabox', \actset', \procset'}$ be the
  corresponding KAB (with transition system $\ts{\tgkab(\gkabsym)}$)
  obtained from $\gkabsym$ via $\tgkab$, and $\actsettmpa$ be a set of
  temp adder actions of $\tgkab(\gkabsym)$.
  Consider a state $\tup{A_k, \scmap_k}$ of $\ts{\tgkab(\gkabsym)}$,
  if there exists a state $\tup{A_k', \scmap_k'}$ such that
  $\tup{A_k, \scmap_k} \exect{\act'\sigma} \tup{A_k', \scmap_k'}$, and
  $\tmp \not\in A_k'$ then $\act' \in \actsettmpd$, and there
  exists action invocation $\gact{Q(\vec{p})}{\act(\vec{p})}$ in the
  sub-proram of $\ginitprog$ such that $\act'$ is obtained from the
  translation of $\gact{Q(\vec{p})}{\act(\vec{p})}$ via $\tgprog$.
\end{lemma}
\begin{proof}
%
  Since $\tmp \not\in A_k'$, then by \Cref{def:tmp-deleter-action}
  and \Cref{lem:action-set-separation} we must have
  $\act' \in \actsettmpd$. By the definition of translation $\tgprog$
  (see \Cref{def:prog-translation}), $\act'$ must be obtained from the
  translation of an action invocation
  $\gact{Q(\vec{p})}{\act(\vec{p})}$ in the sub-program of
  $\ginitprog$.
\end{proof}


The following lemma shows that given two action invocations that has
different program ID, we have that their start flags are
different. I.e., any actions invocations that occur in a different
place inside a certain program will have different start
flag. This claim is formalized below.

\begin{lemma}\label{lem:action-invocation-unique-start-flag}
  Let $\gkabsym = \tup{T, \initabox, \actset, \ginitprog}$ be an
  S-GKAB,
  $\tgkab(\gkabsym) = \tup{T, \initabox', \actset', \procset'}$ be the
  corresponding KAB (with transition system $\ts{\tgkab(\gkabsym)}$)
  obtained from $\gkabsym$ via $\tgkab$, and $\actsettmpa$ be a set of
  temp adder actions of $\tgkab(\gkabsym)$.
  Consider two action invocations
  $\gact{Q_1(\vec{x})}{\act_1(\vec{x})}$ and
  $\gact{Q_2(\vec{y})}{\act_2(\vec{y})}$ that are sub-programs of
  $\ginitprog$. We have that
  $\pid(\gact{Q_1(\vec{x})}{\act_1(\vec{x})}) \neq
  \pid(\gact{Q_2(\vec{y})}{\act_2(\vec{y})})$
  if and only if
  $\ppre(\pid(\gact{Q_1(\vec{x})}{\act_1(\vec{x})})) \neq
  \ppre(\pid(\gact{Q_2(\vec{y})}{\act_2(\vec{y})}))$.
\end{lemma}
\begin{proof}
  Trivially true from the definition of translation $\tgprog$ (see
  \Cref{def:prog-translation}).
\end{proof}

We now proceed to show a property of translation $\tgkab$ that is
related to the final states of S-GKABs transition system. Roughly
speaking, we show that given a final state
$s_g = \tup{A_g,\scmap_g, \delta_g}$ of an S-GKAB transition system
and a state $s_k = \tup{A_k,\scmap_k}$ of its corresponding KAB
transition system such that $s_k$ mimics $s_g$ (i.e.,
$s_g \mimic s_k$),
%
%
we have that there exists a state $s_k'$ that is reachable from $s_k$
(possibly through some intermediate states) and we have that
$\ppost(\delta_g)$ is in the ABox that is contained in $s_k'$ (i.e.,
$\ppost(\delta_g) \in
\abox(s_k')$). 

\begin{lemma}\label{lem:final-state-add-transition}
  Given
  \begin{inparaenum}[]
  \item an S-GKAB $\gkabsym$, and 
  \item a KAB $\tgkab(\gkabsym)$ 
    that is obtained from $\gkabsym$ through $\tgkab$.
\end{inparaenum}
Consider the states
\begin{inparaenum}[]
\item $s_g = \tup{A_g,\scmap_g, \delta_g}$ of $\ts{\gkabsym}^{\filter_S}$
  and
\item $s_k = \tup{A_k,\scmap_k}$ of $\ts{\tgkab(\gkabsym)}$.
\end{inparaenum}
If 
\begin{inparaitem}[]
\item 
  $s_g$ is a final state, and
\item $s_g \mimic s_k$,
\end{inparaitem}
then there exists states $\tup{A_i, m_k}$ and actions $\act_i$ (for
$i \in \set{1, \ldots, n}$, and $n \geq 0$)
such that
\begin{compactitem}
\item
  $\tup{A_k, m_k} \exect{\act_1\sigma} \tup{A_1,
    m_k}\exect{\act_2\sigma} \cdots \exect{\act_n\sigma} \tup{A_n,
    m_k}$ \\(with an empty substitution $\sigma$),
\item $\tmp \in A_i$ (for $i \in \set{1, \ldots, n}$),
\item $\ppost(\delta_g) \in A_n$, 
\item $\ppre(\delta_g) \not\in A_n$ (if
  $\ppost(\delta_g) \neq \ppre(\delta_g)$),
\item $A_n \eqm A_g$, and
\item if $\noopconcept{c} \in A_k$ (where $c \in \const_0$), then
  $\noopconcept{c} \in A_i$ (for $i \in \set{1, \ldots, n}$).
\end{compactitem}
\end{lemma}
\begin{proof}
  We proof the claim by induction over the definition of a final
  state. The rough idea about the correctness of this claim is as
  follows:
  \begin{compactitem}

  \item Consider $\delta_g=\gemptyprog$ (note that it is the base case
    of the definition of final state).  In this case, we have that the
    translation of $\gemptyprog$ by $\tgprog$ produces an action that
    simply delete the $\ppre(\gemptyprog)$ from $A_k$ and then add
    $\ppost(\gemptyprog)$.  Let $A_n$ be the resulting ABox, then it
    is easy to see that $A_n \eqm A_g$.

  \item Since $s_g \mimic s_k$, based on the idea of translation
    $\tgprog$, we have that the action execution starting from $s_k$
    should mimic the execution of program $\delta_g$ starting from
    $s_g$. Since $s_g$ is a final state, it makes no transition to any
    other states. Hence, to mimic the state $s_g$, generalizing from
    the translation of $\gemptyprog$, $s_k$ should make transitions to
    the state $s_k'$ in which $\ppost(\delta_g) \in
    \abox(s_k')$.
    Moreover, since $s_g$ is a final state, by the definition of final
    states as well as the translation $\tgprog$, we have that those
    actions and condition-action rules that mimic $\delta_g$ only
    manipulate the flags. Hence it is easy to see that
    $\abox(s_k') \eqm A_g$.

  \end{compactitem}
%
The detail proof is as follows:
  Let
  \begin{compactitem}
  \item $\gkabsym = \tup{T, \initabox, \actset, \delta}$, and 
    $\ts{\gkabsym}^{\filter_S} = \tup{\const, T, \stateset_g, s_{0g},
      \abox_g, \trans_g}$,

  \item $\tgkab(\gkabsym) = \tup{T, \initabox', \actset', \procset'}$,
    and 
    $\ts{\tgkab(\gkabsym)} = \tup{\const, T, \stateset_k, s_{0k},
      \abox_k, \trans_k}$
    where
    \begin{compactitem}[$\bullet$]
    \item $\initabox' = \initabox \cup \set{\flagconcept{start}}$, and
    \item
      $\tgprog(\flagconcept{start}, \ginitprog, \flagconcept{end}) = \tup{\ppre, \ppost, \procset', \actset'}$.
    \end{compactitem}
\end{compactitem}
We show the claim by induction over the definition of a final
state 
as follows:

\smallskip
\noindent
\textbf{Base case:}
\begin{compactitem}
\item[\textbf{[$\delta_g = \gemptyprog$]}.] Since
  $\tup{A_g,\scmap_g, \gemptyprog} \mimic \tup{A_k,\scmap_k}$, then by
  \Cref{def:mimic-state} we have $\ppre(\gemptyprog) \in A_k$.  By
  the definition of $\tgprog$, we have a 0-ary action
  $\act_\gemptyprog()$ where
  \begin{compactitem}
  \item $\carule{\ppre(\gemptyprog)} {\act_\gemptyprog()}$, and
  \item
    $\eff{\act_\gemptyprog} = \set{\map{\true}{\add
        \set{\ppost(\gemptyprog), \tmp}, \del \set{\ppre(\gemptyprog)}
      }}$
  \end{compactitem}
  Hence, by observing how an action is executed and the result of an
  action execution is constructed, we easily obtain that there exists
  $\tup{A_1, m_k}$ such that
  \begin{compactitem}
  \item
    $\tup{A_k, m_k} \exect{\act_{\gemptyprog}\sigma} \tup{A_1, m_k}$
    (with an empty substitution $\sigma$),
  \item $\tmp \in A_1$,
  \item $\ppost(\gemptyprog) \in A_1$,
  \item $\ppre(\gemptyprog) \not\in A_1$ 
    (if $\ppre(\gemptyprog) \neq \ppost(\gemptyprog)$), and
  \item $A_1 \eqm A_g$.
  \end{compactitem}
  Additionally, it is also true that if $\noopconcept{c} \in A_k$ (for
  a constant $c \in \const_0$), then $\noopconcept{c} \in A_1$,
  because, by the definition of $\tgprog$, the action
  $\act_{\gemptyprog}$ does not delete any concept made by concept
  names $\noopconceptname$ and only actions that are obtained from the
  translation of an action invocation delete such kind of concept
  assertions. Therefore the claim is proven for this case.
\end{compactitem}

\smallskip
\noindent
\textbf{Inductive cases:}
\begin{itemize}

\item[\textbf{[$\delta_g = \delta_1|\delta_2$]}.] Since
  $\tup{A_g, \scmap_g, \delta_1|\delta_2}$ is a final state, then by
  the definition of final state (see \Cref{def:final-state-program})
  we have either
  \begin{compactenum}[\bf (1)]
  \item $\tup{A_g, \scmap_g, \delta_1}$ is a final state, or
  \item $\tup{A_g, \scmap_g, \delta_2}$ is a final state.
  \end{compactenum}
  For compactness of the proof, here we only show the case
  \textbf{(1)}. The case \textbf{(2)} can be done similarly.
  Since
  $\tup{A_g,\scmap_g, \delta_1|\delta_2} \mimic \tup{A_k,\scmap}$,
  then $\ppre(\delta_1|\delta_2) \in A_k$.
  By the definition of $\tgkab$, 
  we have
  \begin{compactitem}
  \item
    $\carule{\ppre(\delta_1|\delta_2)}{\gamma_{\delta_1}()} \in
    \procset$,
  \item
    $\gamma_{\delta_1}():\set{\map{\true} {\add \set{\ppre(\delta_1), \tmp},
        \del \set{ \ppre(\delta_1|\delta_2) } }}$,
  \item $\ppost(\delta_1|\delta_2) = \ppost(\delta_1)$.
  \end{compactitem}
  By observing how an action is executed as well as how the result of
  an action execution is constructed, we have that there exists
  $A_{k+1}$ such that 
  \begin{compactitem}
  \item
    $\tup{A_k,\scmap} \exect{\gamma_{\delta_1}\sigma}
    \tup{A_{k+1},\scmap}$,
  \item $\set{\ppre(\delta_1), \tmp} \subseteq A_{k+1}$,
  \end{compactitem}
  Thus it is easy to see that
  $\tup{A_g,\scmap_g, \delta_1} \mimic \tup{A_{k+1},\scmap_k}$.
  Then, since $\tup{A_g, \scmap_g, \delta_1}$ is a final state, by
  induction hypothesis, 
  it is easy to see that the claim is proven.

\item[\textbf{[$\delta_g = \delta_1;\delta_2$]}.] Since
  $\tup{A_g, \scmap_g, \delta_1;\delta_2}$ is a final state, then by
  the definition of final state (see \Cref{def:final-state-program})
  we have that $\tup{A_g, \scmap_g, \delta_1}$ is a final state and
  $\tup{A_g, \scmap_g, \delta_2}$ is a final state.
  Since
  $\tup{A_g,\scmap_g, \delta_1;\delta_2} \mimic \tup{A_k,\scmap_k}$,
  then $\ppre(\delta_1;\delta_2) \in A_k$.  By the definition of
  $\tgkab$ 
  we have that $\ppre(\delta_1; \delta_2) = \ppre(\delta_1)$,
  $\ppost(\delta_1) = \ppre(\delta_2)$, and
  $\ppost(\delta_2) = \ppost(\delta_1; \delta_2)$.
  By induction hypothesis, there exists states $\tup{A_i, m_k}$, and
  actions $\act_i$, (for $i \in \set{1, \ldots, l}$, and $n \geq 0$)
such that
\begin{compactitem}
\item
  $\tup{A_k, m_k} \exect{\act_1\sigma} \tup{A_1,
    m_k}\exect{\act_2\sigma} \cdots
  \exect{\act_l\sigma} \tup{A_l, m_k}$
\\
  (with an empty substitution $\sigma$),
\item $\tmp \in A_i$ (for $i \in \set{1, \ldots, l}$),
\item $\ppost(\delta_1) \in A_l$,
\item $\ppre(\delta_1) \not\in A_l$ (if
  $\ppre(\delta_1) \neq \ppost(\delta_1)$),
\item $A_l \eqm A_g$,
\item if $\noopconcept{c} \in A_k$ (where $c \in \const_0$), then
  $\noopconcept{c} \in A_i$ (for $i \in \set{1, \ldots, l}$).
\end{compactitem}

Now, since $A_l \eqm A_g$, $\scmap_k = \scmap_g$,
$\ppre(\delta_2) \in A_l$, then we have
$\tup{A_g, \scmap_g, \delta_2} \mimic \tup{A_l, \scmap_k}$.  Hence, by
induction hypothesis again, there exists states $\tup{A_i, m_k}$, and
actions $\act_i$ (for $i \in \set{l+1, \ldots, n}$, and $n \geq 0$)
such that
\begin{compactitem}
\item $\tup{A_l, m_k}\exect{\act_{l+1}\sigma} \tup{A_{l+1},
    m_k}\exect{\act_{l+2}\sigma} \cdots \exect{\act_n\sigma} \tup{A_n, m_k}$\\
  (with an empty substitution
  $\sigma$),
\item $\tmp \in A_i$ (for $i \in \set{l+1, \ldots, n}$),
\item $\ppost(\delta_2) \in A_n$,
\item $\ppre(\delta_2) \not\in A_n$ (if $\ppre(\delta_2) \neq \ppost(\delta_2)$),
\item $A_n \eqm A_g$,
\item if $\noopconcept{c} \in A_l$ (where $c \in \const_0$), \\ then
  $\noopconcept{c} \in A_i$ (for $i \in \set{l+1, \ldots, n}$).
\end{compactitem}
Therefore, it is easy to see that the claim is proven.

\item[\textbf{[$\delta_g = \gif{\varphi}{\delta_1}{\delta_2}$]}.]
  Since
  $\tup{A_g, \scmap_g, \gif{\varphi}{\delta_1}{\delta_2}}$ is a final state,
  then by the definition of final state (see \Cref{def:final-state-program}) we have either
\begin{compactenum}[\bf (1)]
\item $\tup{A_g, \scmap_g, \delta_1}$ is a final state and
  $\ask(\varphi, T, A) = \true$, or 
\item $\tup{A_g, \scmap_g, \delta_2}$ is a final state and
  $\ask(\varphi, T, A) = \false$.
\end{compactenum}
For compactness of the proof, here we only show the case
\textbf{(1)}. The case \textbf{(2)} can be done similarly. 
Now, since
$\tup{A_g,\scmap_g, \gif{\varphi}{\delta_1}{\delta_2}} \mimic
\tup{A_k,\scmap_k}$, then $\ppre(\gif{\varphi}{\delta_1}{\delta_2})
\in A_k$. 
By the definition of $\tgkab$, 
we have
\begin{compactitem}
\item $\carule{\varphi \wedge \ppre(\gif{\varphi}{\delta_1}{\delta_2})}{\gamma_{if}()} \in \procset$,
\item $\gamma_{if}():\set{\map{\true} {\add \set{\ppre(\delta_1),
        \tmp},$ \\
      \hspace*{25mm}
      $\del \set{\ppre(\gif{\varphi}{\delta_1}{\delta_2})} }}$,
\item $\ppost(\gif{\varphi}{\delta_1}{\delta_2}) = \ppost(\delta_1)$.
\end{compactitem}
  By observing how an action is executed as well as how the result of
  an action execution is constructed, we have that there exists
  $A_{k+1}$ such that 
  \begin{compactitem}
  \item
    $\tup{A_k,\scmap} \exect{\gamma_{if}\sigma}
    \tup{A_{k+1},\scmap}$,
  \item $\set{\ppre(\delta_1), \tmp} \subseteq A_{k+1}$,
  \end{compactitem}
  Thus it is easy to see that
  $\tup{A_g,\scmap_g, \delta_1} \mimic \tup{A_{k+1},\scmap_k}$.
  Then, since $\tup{A_g, \scmap_g, \delta_1}$ is a final state, by
  induction hypothesis, 
  it is easy to see that the claim is proven.


\item[\textbf{[$\delta_g = \gwhile{\varphi}{\delta}$]}.]  Since
  $\tup{A_g, \scmap_g, \gwhile{\varphi}{\delta}}$ is a final state,
  then by the definition of final state (see
  \Cref{def:final-state-program}) we have either
\begin{compactenum}[\bf (1)]
\item $\ask(\varphi, T, A) = \false$, or
\item $\tup{A_g, \scmap_g, \delta}$ is a final state and
  $\ask(\varphi, T, A) = \true$.
\end{compactenum}
%
%
\begin{compactitem}

\item[\textbf{Proof for the case (1)}:] Now, since
  $\tup{A_g,\scmap_g, \gwhile{\varphi}{\delta}} \mimic
  \tup{A_k,\scmap_k}$,
  then $\ppre(\gwhile{\varphi}{\delta}) \in A_k$.  By the definition
  of $\tgkab$, 
  we have
%
%
%
\begin{compactitem}[$\bullet$]
\item
  $ \carule{\ppre(\gwhile{\varphi}{\delta}) \wedge (\neg \varphi \vee
    \noopconcept{noop}) }{\gamma_{endLoop}()}$, 
\item
  $\gamma_{endLoop}(): \set{\true  \rightsquigarrow \\
    \hspace*{10mm} \add \set{\ppost(\gwhile{\varphi}{\delta}), \tmp }, \\
    \hspace*{10mm} \del \set{\ppre(\gwhile{\varphi}{\delta}),
      \noopconcept{noop}}}$,
\end{compactitem}

Then, by induction hypothesis, and also by observing how an action is
executed as well as the result of an action execution is constructed,
it is easy to see that the claim is proved.

\item[\textbf{Proof for the case (2)}:] Now, since
  $\tup{A_g,\scmap_g, \gwhile{\varphi}{\delta}} \mimic
  \tup{A_k,\scmap_k}$,
  then $\ppre(\gwhile{\varphi}{\delta}) \in A_k$.  By the definition
  of $\tgkab$, 
  we have
\smallskip
\begin{itemize}[$\bullet$]
\item
  $ \carule{\ppre(\gwhile{\varphi}{\delta})
    \wedge \varphi \wedge \neg \noopconcept{noop}
  }{\gamma_{doLoop}()}$,

\item
  $ \carule{\ppre(\gwhile{\varphi}{\delta}) \wedge (\neg \varphi \vee
    \noopconcept{noop}) }{\gamma_{endLoop}()}$, 

\item 
$\gamma_{endLoop}(): \set{\true  \rightsquigarrow \\
    \hspace*{10mm} \add \set{\ppost(\gwhile{\varphi}{\delta}), \tmp }, \\
    \hspace*{10mm} \del \set{\ppre(\gwhile{\varphi}{\delta}),
      \noopconcept{noop}}}$,

\item
  $\gamma_{doLoop}(): \set{\true  \rightsquigarrow\\
    \hspace*{10mm} \add \set{\ppre(\delta), \noopconcept{noop},
      \tmp }, \\
    \hspace*{10mm} \del \set{\ppre(\gwhile{\varphi}{\delta}) }}$.

\end{itemize}
\smallskip


Hence, it is easy to see that we have
\[
\tup{A_k,\scmap_k} \exect{\gamma_{doLoop} \sigma} \tup{A'_k,\scmap_k}
\]
where $\sigma$ is an empty substitution, and
$\set{\tmp, \ppre(\delta), \noopconcept{noop}} \subseteq A_k'$. Hence
$\tup{A_g, \scmap_g, \delta} \mimic \tup{A_k',\scmap_k}$. Since
$\tup{A_g, \scmap_g, \delta}$ is a final state and
$\tup{A_g, \scmap_g, \delta} \mimic \tup{A_k',\scmap_k}$, by induction
hypothesis, then there exists states $\tup{A_i, m_k}$, and actions
$\act_i$ (for $i \in \set{1, \ldots, n}$, and $n \geq 0$)
such that
\smallskip
\begin{itemize}[$\bullet$]
\item
  $\tup{A_k', m_k} \exect{\act_1\sigma} \tup{A_1,
    m_k}\exect{\act_2\sigma} \cdots \exect{\act_n\sigma} \tup{A_n,
    m_k}$ \\(with an empty substitution $\sigma$),
\item $\tmp \in A_i$ (for $i \in \set{1, \ldots, n}$),
\item $\ppost(\delta) \in A_n$, 
\item $\ppre(\delta) \not\in A_n$ (if
  $\ppost(\delta) \neq \ppre(\delta)$),
\item $A_n \eqm A_g$, and
\item if $\noopconcept{c} \in A'_k$ (where $c \in \const_0$), then
  $\noopconcept{c} \in A_i$ (for $i \in \set{1, \ldots, n}$).
\end{itemize}
\smallskip Hence we have
$\set{\ppost(\delta), \noopconcept{noop}, \tmp} \subseteq A_n$. Now,
since by the definition of $\tgprog$ we have that
$\ppost(\delta) = \ppre(\gwhile{\varphi}{\delta})$, then the action
$\gamma_{endLoop}$ is executable in $A_n$ (notice that we do not care
whether $\ask(\varphi, T, A) = \false$, or
$\ask(\varphi, T, A) = \true$ because $\noopconcept{noop} \in
A_n$). Hence we have
\[
\tup{A_n,\scmap_k} \exect{\gamma_{endLoop} \sigma} \tup{A'_n,\scmap_k}
\]
with $\set{\tmp, \ppost(\gwhile{\varphi}{\delta})} \subseteq A'_n$,
and $\noopconcept{noop} \not\in A'_n$ (which is fine since
$\noopconcept{noop} \not\in A_k$). Thus we have that the claim
is proven.
Intuitively, the idea for the proof of this case is that since
$\tup{A_g, \scmap_g, \delta}$ is a final state, there is no action
executed and no one removes the flag made by concept name
$\noopconceptname$. In that situation, for the second iteration, no
matter whether $\varphi$ (the guard of the loop) holds or not, we can
exit the loop and additionally keeping all assertions in the ABox
(except the special markers) stay the same. Essentially it reflects
the situation that in the corresponding S-GKAB, there is no transition
(since $\tup{A_g, \scmap_g, \delta}$ is a final state).
\end{compactitem}
\end{itemize}
\end{proof}

\subsection{Reducing the Verification of S-GKABs into KABs}\label{sec:reduction-verification-sgkab-to-kab}


In this section we show that the transition system of an S-GKAB is
J-bisimilar to the transition system of its corresponding KAB that is
obtained via translation $\tgkab$. Then, taking the advantage of
J-Bisimulation property, we essentially show that we can reduce the
verification of \muladom formulas over S-GKABs into a verification
over KABs.


As a start, below we show that given a state $s_1$ of an S-GKAB
transition system, and a state $s_2$ of its corresponding KAB
transition system such that $s_2$ mimics $s_1$, we have that if $s_1$
reaches $s_1'$ in one step, then it implies that there exists $s_2'$
reachable from $s_2$ (possibly through some intermediate states
$s^t_1, \ldots, s^t_n$ that contain $\tmp$) and $s_2'$ mimics $s_1'$.

\begin{lemma}\label{lem:prog-exec-bsim}
  Let
  \begin{inparaitem}[]
  \item $\gkabsym$ be an S-GKAB with transition system
    $\ts{\gkabsym}^{\filter_S}$,
  \item $\tgkab(\gkabsym)$ be a KAB (obtained from $\gkabsym$ through
    $\tgkab$) with transition system $\ts{\tgkab(\gkabsym)}$.
  \end{inparaitem}
  Consider two states
\begin{inparaenum}[]
\item $\tup{A_g,\scmap_g, \delta_g}$ of $\ts{\gkabsym}^{\filter_S}$,
  and
\item $\tup{A_k,\scmap_k}$ of $\ts{\tgkab(\gkabsym)}$ 
\end{inparaenum}
such that $\tup{A_g,\scmap_g, \delta_g} \mimic \tup{A_k,\scmap_k} $. 
For every state $\tup{A'_g,\scmap'_g, \delta_g'}$ such that
\[
\tup{A_g,\scmap_g, \delta_g} \gprogtrans{\alpha\sigma, \filter_S}
\tup{A'_g,\scmap'_g, \delta_g'}
\]
(for a certain action $\act$, and a legal parameter assignment
$\sigma$), 
there exist states $\tup{A'_k, m'_k}$, $\tup{A_i^t, m_k}$ (for
$i \in \set{1, \ldots, n}$, where $n \geq 0$), and actions $\act'$, $\act_i$
(for $i \in \set{1, \ldots, n}$, where $n \geq 0$)
such that
\begin{compactitem}
\item
  $\tup{A_k, m_k} \exect{\act_1\sigma_e} \tup{A_1^t, m_k}
  \exect{\act_2\sigma_e} \cdots \exect{\act_n\sigma_e} \tup{A_n^t,
    m_k} \exect{\act'\sigma} \tup{A'_k, m'_k}$ where
  \begin{compactitem}
  \item $\sigma_e$ is an empty substitution, 
  \item $\act'$ is obtained from $\act$ through $\tgprog$,
  \item $\tmp \in A_i^t$ (for $i \in \set{1, \ldots, n}$), $\tmp \not\in A_k'$, 
  \end{compactitem}
\item 
  $\tup{A'_g,\scmap'_g, \delta'_g} \mimic \tup{A'_k,\scmap'_k} $.
\end{compactitem}
\end{lemma}
\begin{proof}
Let 
  \begin{compactitem}
  \item $\gkabsym = \tup{T, \initabox, \actset, \delta}$, and
    $\ts{\gkabsym}^{\filter_S} = \tup{\const, T, \stateset_g, s_{0g},
      \abox_g, \trans_g}$,
  \item $\tgkab(\gkabsym) = \tup{T, \initabox', \actset', \procset'}$
    and
    $\ts{\tgkab(\gkabsym)} = \tup{\const, T, \stateset_k, s_{0k},
      \abox_k, \trans_k}$ where
    \begin{compactenum}
    \item $\initabox' = \initabox \cup \set{\flagconcept{start}}$, and
    \item
      $\tgprog(\flagconcept{start}, \ginitprog, \flagconcept{end}) =
      \tup{\ppre, \ppost, \procset', \actset'}$.
    \end{compactenum}
\end{compactitem}
We prove by induction over the structure of $\delta$.

\smallskip
\noindent
\textbf{Base case:}
\begin{enumerate}
\item[\textbf{[$\delta_g = \gemptyprog$]}.] Since
  $\tup{A_g,\scmap_g, \gemptyprog}$ is a final state, then there does
  not exists $\tup{A'_g,\scmap'_g, \delta'_g}$ such that
  \[
  \tup{A_g,\scmap_g, \delta_g} \gprogtrans{\act\sigma, \filter_S}
  \tup{A'_g,\scmap'_g, \delta'_g},
  \]
  Hence, we do not need to show anything.

\item[\textbf{[$\delta_g = \gact{Q(\vec{p})}{\act(\vec{p})}$]}.] For
  compactness of the presentation, let
  $a = \gact{Q(\vec{p})}{\act(\vec{p})}$.  Since
  $\tup{A_g,\scmap_g, a} \mimic \tup{A_k,\scmap_k} $, then
  $\ppre(a) \in A_k$, by 
  the definition of $\tgkab$, we have:
  \begin{compactitem}
  \item
    $\carule{Q(\vec{p}) \wedge \ppre(a)}{\act'(\vec{p})} \in
    \procset'$,
  \item $\act' \in \actset'$. 
  \end{compactitem}
Since 
\[
\tup{A_g,\scmap_g, a}
\gprogtrans{\act\sigma, \filter_S} 
\tup{A'_g,\scmap'_g, \gemptyprog},
\]
then $\sigma \in \ask(Q, T, A_g)$. Since $A_k \eqm A_g$, and $Q$ does
not use any special marker concept names, by
\Cref{lem:ECQ-equal-ABox-modulo-markers} and \Cref{def:ask} we have
$\ask(Q, T, A_g) = \Ans(Q, T, A_k)$ and hence
$\sigma \in \Ans(Q, T, A_k)$.
Now, since $\ppre(a) \in A_k$, then $\act'$ is executable in $A_k$
with legal parameter assignment $\sigma$. Additionally, considering
\[
\begin{array}{l@{}l} \eff{\act'} = \eff{\act} &\cup
                                                   \set{\map{\true}{ \add \set{\ppost(a)}, \del \set{\ppre(a), \tmp} }}\\
                                                 &\cup
                                                   \set{\map{\noopconcept{x}}{\del
                                                   \noopconcept{x} }},
  \end{array}
\]
by \Cref{def:add-gkab-action,def:add-kab-action}, we have
$\addfacts{T, A_g, \act\sigma} = \addfacts{T, A_k, \act'\sigma}
\setminus \set{\ppost(a)}$,
and hence
$\calls{\addfacts{T, A_g, \act\sigma}} = \calls{\addfacts{T, A_k,
    \act'\sigma}}$.  Thus 
we have $\theta \in \calls{\addfacts{T, A_k, \act'\sigma}}$.
Now, since $\scmap_g' = \theta \cup \scmap_g$, $\scmap_k = \scmap_g$
and $\theta \in \calls{\addfacts{T, A_k, \act'\sigma}}$, we can
construct $\scmap_k' = \theta \cup \scmap_k$. Therefore it is easy to
see that there exists $\tup{A'_k, \scmap'_k}$, such that
\[
\tup{A_k, \scmap_k} \exect{\act'\sigma} \tup{A'_k, \scmap'_k}
\]
(with service call substition $\theta$) and $A'_g \eqm A_k'$ (by
considering how $A_k'$ is constructed), $\scmap'_g = \scmap'_k$. By
the definition of $\tgprog$ (in the translation of an action
invocation) we also have $\ppre(\gemptyprog) \in A_k'$. Thus the claim
is proven.
\end{enumerate}

\smallskip
\noindent
\textbf{Inductive case:}
\begin{enumerate}
\item[\textbf{[$\delta_g = \delta_1|\delta_2$]}.] Since 
\[
\tup{A_g,\scmap_g, \delta_1|\delta_2} \gprogtrans{\act\sigma,
  \filter_S} \tup{A'_g,\scmap'_g, \delta'},
\]
then, there are two cases, that is either
\begin{compactenum}[\bf (1)]
\item
  $\tup{A_g,\scmap_g, \delta_1} \gprogtrans{\act\sigma, \filter_S}
  \tup{A'_g,\scmap'_g, \delta'}$, or
\item
  $\tup{A_g,\scmap_g, \delta_2} \gprogtrans{\act\sigma, \filter_S}
  \tup{A'_g,\scmap'_g, \delta'}$.
\end{compactenum}
Here we only give the derivation for the first case, the second case
is similar. Since
$\tup{A_g,\scmap_g, \delta_1|\delta_2} \mimic \tup{A_k,\scmap_k} $,
then $A_g \eqm A_k$, $\scmap_g = \scmap_k$, and
$\ppre(\delta_1|\delta_2) \in A_k$.  By the definition of $\tgkab$ and
\Cref{lem:program-pre-post}, we have
\begin{compactitem}
\item
  $\carule{\ppre(\delta_1|\delta_2)}{\gamma_{\delta_1}()} \in
  \procset'$
\item $\gamma_{\delta_1} \in \actset'$, where
\begin{center}  
$ \gamma_{\delta_1}():\set{\map{\true} {\add \set{\ppre(\delta_1),
        \tmp}, \del \set{\ppre(\delta_1|\delta_2)} } \ }, $
\end{center}
\end{compactitem}
Since $\ppre(\delta_1|\delta_2) \in A_k$, it is easy to see that 
\[
\tup{A_k, m_k} \exect{\gamma_{\delta_1}\sigma_t}\tup{A_t, m_k}
\]
where $\sigma_t$ is an empty substitution,
$\set{\ppre(\delta_1), \tmp} \in A_t$, and $A_t \eqm A_k$.
Since
$A_t \eqm A_k$ and $A_g \eqm A_k$, it is easy to see that
$A_g \eqm A_t$. Since $A_g \eqm A_t$, $\scmap_g = \scmap_k$, and
$\ppre(\delta_1) \in A_t$, then we have
$\tup{A_g,\scmap_g, \delta_1} \mimic \tup{A_t,\scmap_k} $. 
Therefore, since
$\tup{A_g, \scmap_g, \delta_1} \gprogtrans{\act\sigma, \filter_S}
\tup{A_g', \scmap_g', \delta_1'}$
and $\tup{A_g,\scmap_g, \delta_1} \mimic \tup{A_t,\scmap_k} $, by
induction hypothesis, it is easy to see that the claim is proven for
this case.

\item[\textbf{[$\delta_g = \delta_1;\delta_2$]}.] There are two cases:
\begin{compactenum}[\bf (1)]
\item
  $\tup{A_g,\scmap_g, \delta_1;\delta_2} \gprogtrans{\act\sigma,
    \filter_S} \tup{A'_g,\scmap'_g, \delta_1';\delta_2}, $
\item
  $\tup{A_g,\scmap_g, \delta_1;\delta_2} \gprogtrans{\act\sigma,
    \filter_S} \tup{A'_g,\scmap'_g, \delta_2'}, $
\end{compactenum}

\smallskip
\noindent
\textbf{Case (1).} Since
$\tup{A_g,\scmap_g, \delta_1;\delta_2} \gprogtrans{\act\sigma,
  \filter_S} \tup{A'_g,\scmap'_g, \delta_1';\delta_2}$, then we have
\[
\tup{A_g,\scmap_g, \delta_1} \gprogtrans{\act\sigma, \filter_S}
\tup{A'_g,\scmap'_g, \delta_1'},
\]
Since
$\tup{A_g,\scmap_g, \delta_1;\delta_2} \mimic \tup{A_k,\scmap_k} $,
then $A_g \eqm A_k$, $\scmap_g = \scmap_k$, and
$\ppre(\delta_1;\delta_2) \in A_k$.  By the definition of $\tgkab$ and
\Cref{lem:program-pre-post}, it is easy to see that
$\ppre(\delta_1;\delta_2) = \ppre(\delta_1)$, and hence because
$\ppre(\delta_1;\delta_2) \in A_k$, we have $\ppre(\delta_1) \in A_k$.
Now, since $A_g \eqm A_k$, $\scmap_g = \scmap_k$,
$\ppre(\delta_1) \in A_k$, then we have
$\tup{A_g,\scmap_g, \delta_1} \mimic \tup{A_k,\scmap_k} $. Thus, since
we also have
$\tup{A_g,\scmap_g, \delta_1} \gprogtrans{\act\sigma, \filter_S}
\tup{A'_g,\scmap'_g, \delta_1'}$,
by using induction hypothesis we have that the claim is proven.

\smallskip
\noindent
\textbf{Case (2).}  Since
$\tup{A_g,\scmap_g, \delta_1;\delta_2} \gprogtrans{\act\sigma,
  \filter_S} \tup{A'_g,\scmap'_g, \delta_2'} $,
then we have $\tup{A_g,\scmap_g, \delta_1}$ is a final state, and
\[
\tup{A_g,\scmap_g, \delta_2} \gprogtrans{\act\sigma, \filter_S}
\tup{A'_g,\scmap'_g, \delta_2'},
\]

Since $\tup{A_g,\scmap_g, \delta_1}$ is a final state and
$\tup{A_g,\scmap_g, \delta_g} \mimic \tup{A_k,\scmap_k}$, by
\Cref{lem:final-state-add-transition}, there exist states
$\tup{A_i, m_k}$ and actions $\act_i$ (for $i \in \set{1, \ldots, n}$, and
$n \geq 0$)
such that
\begin{compactitem}
\item
  $\tup{A_k, m_k} \exect{\act_1\sigma} \tup{A_1,
    m_k}\exect{\act_2\sigma} \cdots \exect{\act_n\sigma} \tup{A_n,
    m_k}$ \\ (with empty an empty substitution $\sigma$),
\item $\tmp \in A_i$ (for $i \in \set{1, \ldots, n}$),
\item $\ppost(\delta_1) \in A_n$, $\ppre(\delta_1) \not\in A_n$, 
   and 
\item $A_n \eqm A_g$.
\end{compactitem}
Since
$\tup{A_g,\scmap_g, \delta_1;\delta_2} \mimic \tup{A_k,\scmap_k} $,
then $A_g \eqm A_k$, $\scmap_g = \scmap_k$, and
$\ppre(\delta_1;\delta_2) \in A_k$.  
By the definition of $\tgkab$ and
\Cref{lem:program-pre-post}, it is easy to see that
\begin{compactitem}
\item $\ppre(\delta_1;\delta_2) = \ppre(\delta_1)$, 
\item $\ppost(\delta_1) = \ppre(\delta_2)$, 
\item $\ppost(\delta_1;\delta_2) = \ppost(\delta_2)$, 
\end{compactitem}
Hence, because $\ppost(\delta_1) \in A_n$, and
$\ppost(\delta_1) = \ppre(\delta_2)$, we have
$\ppre(\delta_2) \in A_n$. 
Now, since $A_g \eqm A_n$, $\scmap_g = \scmap_k$,
$\ppre(\delta_2) \in A_n$, then we have
$\tup{A_g,\scmap_g, \delta_2} \mimic \tup{A_n,\scmap_k} $. Thus, since
we also have
$\tup{A_g,\scmap_g, \delta_2} \gprogtrans{\act\sigma, \filter_S}
\tup{A'_g,\scmap'_g, \delta_2'}$,
by using induction hypothesis we have that the claim is proven.

\item[\textbf{[$\delta_g = \gif{\varphi}{\delta_1}{\delta_2}$]}.]
  There are two cases:
\begin{compactenum}[\bf (1)]
\item
  $\tup{A_g, \scmap_g, \gif{\varphi}{\delta_1}{\delta_2}}
  \gprogtrans{\alpha\sigma, \filter_S} \tup{A_g', \scmap_g',
    \delta_1'}$,
\item
  $\tup{A_g, \scmap_g, \gif{\varphi}{\delta_1}{\delta_2}}
  \gprogtrans{\alpha\sigma, \filter_S} \tup{A_g', \scmap_g',
    \delta_2'}$.
\end{compactenum}
Here we only consider the first case. The second case is similar. 

\noindent
\smallskip \textbf{Case (1).} Since
$\tup{A_g, \scmap_g, \gif{\varphi}{\delta_1}{\delta_2}}
\gprogtrans{\alpha\sigma, \filter_S} \tup{A_g', \scmap_g',
  \delta_1'}$, then we have
\[
\tup{A_g, \scmap_g, \delta_1} \gprogtrans{\act\sigma, \filter_S}
\tup{A_g', \scmap_g', \delta_1'}.
\] 
with $\ask(\varphi, T, A_g) = \true$.  
Since
$\tup{A_g,\scmap_g, \gif{\varphi}{\delta_1}{\delta_2}} \mimic
\tup{A_k,\scmap_k} $,
then $A_g \eqm A_k$, $\scmap_g = \scmap_k$, and
$\ppre(\gif{\varphi}{\delta_1}{\delta_2}) \in A_k$.
By the definition of $\tgkab$ and \Cref{lem:program-pre-post}, we have
\begin{compactitem}
\item
  $\carule{\ppre(\gif{\varphi}{\delta_1}{\delta_2}) \wedge
    \varphi}{\gamma_{if}()} \in \procset'$
\item $\gamma_{if} \in \actset'$, where \\
  $
\begin{array}{ll}
\gamma_{if}():\set{\map{\true} {&\add \set{\ppre(\delta_1), \tmp}, \\
      &\del \set{\ppre(\gif{\varphi}{\delta_1}{\delta_2})} }},
\end{array}
$
\end{compactitem}
Since $A_k \eqm A_g$, $\ask(\varphi, T, A_g) = \true$, and $\varphi$
does not use any special marker concept names, by
\Cref{lem:ECQ-equal-ABox-modulo-markers} and \Cref{def:ask} we have
$\Ans(\varphi, T, A_k) = \true$.
Now, since $\ppre(\gif{\varphi}{\delta_1}{\delta_2}) \in A_k$, and
$\Ans(\varphi, T, A_k) = \true$, it is easy to see that
\[
\tup{A_k, m_k} \exect{\gamma_{if}\sigma_t}\tup{A_t, m_k}
\]
where $\sigma_t$ is an empty substitution,
$\set{\ppre(\delta_1), \tmp} \in A_t$, and $A_t \eqm A_k$. Since
$A_t \eqm A_k$ and $A_g \eqm A_k$, it is easy to see that
$A_g \eqm A_t$. Since $A_g \eqm A_t$, $\scmap_g = \scmap_k$, and
$\ppre(\delta_1) \in A_t$, then we have
$\tup{A_g,\scmap_g, \delta_1} \mimic \tup{A_t,\scmap_k} $. 
Thus, since
$\tup{A_g, \scmap_g, \delta_1} \gprogtrans{\act\sigma, \filter_S}
\tup{A_g', \scmap_g', \delta_1'}$
and $\tup{A_g,\scmap_g, \delta_1} \mimic \tup{A_t,\scmap_k} $, by
induction hypothesis, it is easy to see that the claim is proven for
this case.

\item[\textbf{[$\delta_g = \gwhile{\varphi}{\delta}$]}.]  Since
\[  
\tup{A_g, \scmap_g,
  \gwhile{\varphi}{\delta}} \gprogtrans{\act\sigma,
  \filter_S} \tup{A_g', \scmap_g', \delta';\gwhile{\varphi}{\delta}},
\]
then we have $\ask(\varphi, T, A) = \true$ and
\[
\tup{A_g, \scmap_g, \delta} \gprogtrans{\act\sigma, \filter_S}
\tup{A_g', \scmap_g', \delta'}.
\]
Since
$\tup{A_g,\scmap_g, \gwhile{\varphi}{\delta}} \mimic
\tup{A_k,\scmap_k}$,
then $A_g \eqm A_k$, $\scmap_g = \scmap_k$, and
$\ppre(\gwhile{\varphi}{\delta}) \in A_k$.
By the definition of $\tgkab$ and \Cref{lem:program-pre-post}, we have
\begin{compactitem}
\item
  $ \carule{(\ppre(\gwhile{\varphi}{\delta}) \vee \ppost(\delta)) \wedge \varphi \wedge \neg
    \noopconcept{noop} }{\gamma_{doLoop}()} \in \procset'$,
\item $\gamma_{doLoop}(): \set{\true  \rightsquigarrow
\add \set{\ppre(\delta), \noopconcept{noop},
      \tmp }, \\
    \hspace*{33mm} \del \set{ \ppre(\gwhile{\varphi}{\delta}),
      \ppost(\delta) }}$,
\end{compactitem}
Since $A_k \eqm A_g$, $\ask(\varphi, T, A_g) = \true$, and $\varphi$
does not use any special marker concept names, by
\Cref{lem:ECQ-equal-ABox-modulo-markers} and \Cref{def:ask} we have
$\Ans(\varphi, T, A_k) = \true$. Additionally, it is easy to see from
the definition of $\tgprog$ that $\noopconcept{noop} \not\in A_k$.
Now, since $\ppre(\gwhile{\varphi}{\delta}) \in A_k$,
$\Ans(\varphi, T, A_k) = \true$, and $\noopconcept{noop} \not\in A_k$,
it is easy to see that
\[
\tup{A_k, m_k} \exect{\gamma_{doLoop}\sigma_t}\tup{A_t, m_k}
\]
%
%
where $\sigma_t$ is an empty substitution,
$\set{\ppre(\delta), \tmp} \in A_t$, and $A_t \eqm A_k$. Since
$A_t \eqm A_k$ and $A_g \eqm A_k$, it is easy to see that
$A_g \eqm A_t$. Since $A_g \eqm A_t$, $\scmap_g = \scmap_k$, and
$\ppre(\delta) \in A_t$, then we have
$\tup{A_g,\scmap_g, \delta} \mimic \tup{A_t,\scmap_k} $.
Thus, since
$\tup{A_g, \scmap_g, \delta} \gprogtrans{\act\sigma, \filter_S}
\tup{A_g', \scmap_g', \delta'}$
and $\tup{A_g,\scmap_g, \delta} \mimic \tup{A_t,\scmap_k} $, by
induction hypothesis, 
there exist states $\tup{A'_k, m'_k}$, $\tup{A_i^t, m_k}$ (for
$i \in \set{1, \ldots, n}$, where $n \geq 0$), and actions $\act'$, $\act_i$
(for $i \in \set{1, \ldots, n}$, where $n \geq 0$)
such that
\begin{compactitem}
\item
  $\tup{A_t, m_k} \exect{\act_1\sigma_e} \tup{A_1^t, m_k}
  \exect{\act_2\sigma_e} \cdots \exect{\act_n\sigma_e} \tup{A_n^t,
    m_k} \exect{\act'\sigma} \tup{A'_k, m'_k}$ where
  \begin{compactitem}
  \item $\sigma_e$ is an empty substitution, 
  \item $\act'$ is obtained from $\act$ through $\tgprog$,
  \item $\tmp \in A_i^t$ (for $i \in \set{1, \ldots, n}$), $\tmp \not\in A_k'$, 
  \end{compactitem}
\item 
  $\tup{A'_g,\scmap'_g, \delta'} \mimic \tup{A'_k,\scmap'_k} $.
\end{compactitem}
The proof for this case is then completed by also observing that by
the definition of program execution relation (see
\Cref{def:prog-exec-relation}), we have that we repeat the loop at the
end of the execution of program $\delta$, and this situation is
captured in the definition of $\tgprog$ by having that
$\ppost(\delta) = \ppre(\gwhile{\varphi}{\delta})$.
\end{enumerate}
\ \ 
\end{proof}

We now proceed to show another crucial lemma for showing the
bisimulation between S-GKAB transition system and the transition
system of its corresponding KAB that is obtained via
$\tgprog$. Basically, we show that given a state $s_1$ of an S-GKAB
transition system, and a state $s_2$ of its corresponding KAB
transition system such that $s_2$ mimics $s_1$, 
we have that if $s_2$ reaches $s_2'$ (possibly through some
intermediate states $s^t_1, \ldots, s^t_n$ that contains $\tmp$), then
$s_1$ reaches $s_1'$ in one step and $s_1'$ is mimicked by $s_2'$.


\begin{lemma}\label{lem:kab-transition-bsim}
  Let
  \begin{inparaitem}[]
  \item $\gkabsym  = \tup{T, \initabox, \actset, \delta}$ be an S-GKAB with transition system
    $\ts{\gkabsym}^{\filter_S}$,
  \item $\tgkab(\gkabsym) = \tup{T, \initabox', \actset', \procset'}$
    be a KAB (obtained from $\gkabsym$ through $\tgkab$) with
    transition system $\ts{\tgkab(\gkabsym)}$,
  \end{inparaitem}
  where
  \begin{compactenum}
  \item $\initabox' = \initabox \cup \set{\flagconcept{start}}$, and
  \item
    $\tgprog(\flagconcept{start}, \ginitprog, \flagconcept{end}) =
    \tup{\ppre, \ppost, \procset', \actset'}$.
  \end{compactenum}
    Consider two states
  \begin{inparaenum}[]
  \item $\tup{A_g,\scmap_g, \delta_g}$ of $\ts{\gkabsym}^{\filter_S}$,
    and
  \item $\tup{A_k,\scmap_k}$ of $\ts{\tgkab(\gkabsym)}$ 
  \end{inparaenum}
  such that $\tup{A_g,\scmap_g, \delta_g} \mimic \tup{A_k,\scmap_k}$.
  For every state $\tup{A'_k, m'_k}$
%
  such that
  \begin{compactitem}
  \item there exist $\tup{A_i^t, m_k}$ $($for $i \in \set{1,\ldots, n}$,
    $n \geq 0$$)$, and 
  \item either we have
\[\tup{A_k,
      m_k} \exect{\act_1\sigma_e} \tup{A_1^t, m_k}
    \exect{\act_2\sigma_e} \cdots \exect{\act_n\sigma_e} \tup{A_n^t,
      m_k} \exect{\act'\sigma} \tup{A'_k, m'_k}\]
    if $n > 0$ or we have
    \[\tup{A_k, m_k} \exect{\act'\sigma} \tup{A'_k, m'_k}\]
    if $n = 0$ where
    \begin{compactitem}
    \item $\sigma_e$ is an empty substitution,
    \item $\act_i \in \actsettmpa$ $($for $i \in \set{1, \ldots, n}$$)$,
    \item $\act' \in \actsettmpd$,
    \item $\tmp \in A_i^t$ $($for $i \in \set{1,\ldots, n}$$)$,
      $\tmp \not\in A_k'$,
    \end{compactitem}
  \end{compactitem}
  then there exists a state $\tup{A'_g,\scmap'_g, \delta_g'}$ such
  that
  \begin{compactitem}
  \item
    $ \tup{A_g,\scmap_g, \delta_g} \gprogtrans{\act\sigma,
      \filter_S} \tup{A'_g,\scmap'_g, \delta_g'}, $
  \item $\act'$ is obtained from the translation of a certain action
    invocation $\gact{Q(\vec{p})}{\act(\vec{p})}$ via $\tgprog$, 
  \item 
    $\tup{A'_g,\scmap'_g, \delta'_g} \mimic \tup{A'_k,\scmap'_k} $.
  \end{compactitem}
\ \ 
\end{lemma}
\begin{proof}
  Let
\begin{compactitem}
\item
  $\ts{\gkabsym}^{\filter_S}\!=\!\tup{\const, T, \stateset_g, s_{0g},
    \abox_g, \trans_g}$, and
\item
  $\ts{\tgkab(\gkabsym)}\!=\!\tup{\const, T, \stateset_k, s_{0k},
    \abox_k, \trans_k}$.
\end{compactitem}
%
  We prove by induction over the structure of $\delta$.

\smallskip
\noindent
\textbf{Base case:}
\begin{enumerate}
\item[\textbf{[$\delta_g = \gemptyprog$]}.] Since
  $\delta_g
  = \gemptyprog$, by the definition of
  $\tgprog$,
  there must not exist $\tup{A'_k,
    m'_k}$, $\tup{A_i^t, m_k}$ (for $i \in \set{1,\ldots, n}$, and $n
  \geq 0$) such that either
\[\tup{A_k,
      m_k} \exect{\act_1\sigma_e} \tup{A_1^t, m_k}
    \exect{\act_2\sigma_e} \cdots \exect{\act_n\sigma_e} \tup{A_n^t,
      m_k} \exect{\act'\sigma} \tup{A'_k, m'_k}\] if $n > 0$ or
    \[\tup{A_k, m_k} \exect{\act'\sigma} \tup{A'_k, m'_k}\]
    if $n = 0$ 
  where
    \begin{compactitem}
    \item $\sigma_e$ is an empty substitution,
    \item $\act_i \in \actsettmpa$ (for $i \in \set{1,\ldots, n}$),
    \item $\act' \in \actsettmpd$,
    \item $\tmp
      \in A_i^t$ (for $i \in \set{1,\ldots, n}$), $\tmp \not\in A_k'$,
    \end{compactitem}
    Hence, the base case is trivially true.
    The reason is that the translation $\tgprog$
    translates empty programs into an action that only adds $\tmp$
    and changes the flag.
%
%
    In fact, $\tup{A_g,\scmap_g,
      \delta_g}$ is a final state, then there does not exists
    $\tup{A'_g,\scmap'_g, \delta'_g}$ such that
  \[
  \tup{A_g,\scmap_g, \delta_g} \gprogtrans{\act\sigma, \filter_S}
  \tup{A'_g,\scmap'_g, \delta'_g},
  \]

  \item[\textbf{[$\delta_g = \gact{Q(\vec{p})}{\act(\vec{p})}$]}.]
    Assume that there exist states $\tup{A'_k, m'_k}$,
    $\tup{A_i^t, m_k}$ (for $i \in \set{1,\ldots, n}$, and $n \geq 0$) such
    that either
    \[\tup{A_k, m_k} \exect{\act_1\sigma_e} \tup{A_1^t, m_k}
    \exect{\act_2\sigma_e} \cdots \exect{\act_n\sigma_e} \tup{A_n^t,
      m_k} \exect{\act'\sigma} \tup{A'_k, m'_k}\] if $n > 0$ or
    \[\tup{A_k, m_k} \exect{\act'\sigma} \tup{A'_k, m'_k}\]
    if $n = 0$
    where
    \begin{compactitem}
    \item $\sigma_e$ is an empty substitution,
    \item $\act_i \in \actsettmpa$ (for $i \in \set{1,\ldots, n}$),
    \item $\act' \in \actsettmpd$,
    \item $\tmp \in A_i^t$ (for $i \in \set{1,\ldots, n}$),
      $\tmp \not\in A_k'$,
    \end{compactitem}
    Additionally, w.l.o.g., let $\theta$ be the corresponding
    substitution that evaluates service calls in the transition
    \[
    \tup{A_n^t, m_k} \exect{\act'\sigma} \tup{A'_k, m'_k}
    \]
    (Note that we consider $A_n^t = A_k$ if $n = 0$). Now, since
    $\tup{A_g,\scmap_g, \gact{Q(\vec{p})}{\act(\vec{p})} } \mimic
    \tup{A_k,\scmap_k} $,
    then $\ppre(\gact{Q(\vec{p})}{\act(\vec{p})}) \in A_k$.  Moreover,
    since we also have $\delta_g = \gact{Q(\vec{p})}{\act(\vec{p})}$,
    by 
    the definition of $\tgkab$,
    \Cref{lem:non-temp-state-produced-by-normal-action,lem:action-invocation-unique-start-flag},
    we have that $\act'$ must be obtained from
    $\gact{Q(\vec{p})}{\act(\vec{p})}$ and hence we have that
    $\carule{Q(\vec{p}) \wedge
      \ppre(\gact{Q(\vec{p})}{\act(\vec{p})})}{\act'(\vec{p})} \in
    \procset'$.

  Now, by our assumption above 
  and by \Cref{lem:temp-state-produced-by-temp-act}, we have that
  $A_k \eqm A_1^t$, $A_i^t \eqm A_{i+1}^t$ (for
  $i \in \set{1,\ldots, n-1}$) 
  and hence we have $A_g \eqm A_n^t$. Then, since $A_g \eqm A_n^t$,
  and $Q$ does not use any special marker concept names, by
  \Cref{lem:ECQ-equal-ABox-modulo-markers} and \Cref{def:ask} we have
  $\ask(Q, T, A_g) = \Ans(Q, T, A_n^t)$ and hence
  $\sigma \in \ask(Q, T, A_g)$.
  Additionally, considering
  \[
  \begin{array}{l@{}l@{}l} \eff{\act'} = \eff{\act} &\cup
    \set{\map{\true}{ &\add
                        \set{\ppost(\gact{Q(\vec{p})}{\act(\vec{p})})}, \\
                                                    &&\del
                                                       \set{\ppre(\gact{Q(\vec{p})}{\act(\vec{p})}), \tmp} }}\\
                                                    &\cup
                                                      \set{\map{\noopconcept{x}&}{\del
                                                                                 \noopconcept{x}
                                                                                 }},
  \end{array}
  \]
  by \Cref{def:add-gkab-action,def:add-kab-action}, we have
   \[
   \addfacts{T, A_g, \act\sigma} = \addfacts{T, A_n^t, \act'\sigma}
   \setminus \ppost(\gact{Q(\vec{p})}{\act(\vec{p})}),
   \]
   and hence
   $\calls{\addfacts{T, A_g, \act\sigma}} = \calls{\addfacts{T, A_n^t,
       \act'\sigma}}$
   and $\theta \in \calls{\addfacts{T, A_g, \act\sigma}}$.
   Since $\scmap_k' = \theta \cup \scmap_k$, $\scmap_k = \scmap_g$ and
   $\theta \in \calls{\addfacts{T, A_g, \act\sigma}}$, we can
   construct $\scmap_g' = \theta \cup \scmap_g$. 
   Thus, it is easy to see that there exists
   $\tup{A'_g,\scmap'_g, \gemptyprog}$ such that
  \[
  \tup{A_g,\scmap_g, \gact{Q(\vec{p})}{\act(\vec{p})}  } \gprogtrans{\act\sigma, \filter_S}
  \tup{A'_g,\scmap'_g, \gemptyprog},
  \]
  (with service call substition $\theta$) and $A'_g \eqm A_k'$ (by
  considering how $A_k'$ is constructed), $\scmap'_g = \scmap'_k$.
  Moreover, by the definition of $\tgprog$ (in the translation of an
  action invocation) we also have $\ppre(\gemptyprog) \in A_k'$
  (because $\ppost(\gact{Q(\vec{p})}{\act(\vec{p})}) = \ppre(\gemptyprog)$). Thus
  the claim is proven.
\end{enumerate}

\smallskip
\noindent
\textbf{Inductive case:}
\begin{enumerate}
\item[\textbf{[$\delta_g = \delta_1|\delta_2$]}.]  Assume that there
  exist states $\tup{A'_k, m'_k}$, $\tup{A_i^t, m_k}$ (for
  $i \in \set{1,\ldots, n}$, and $n \geq 0$) such that 
    \[\tup{A_k, m_k} \exect{\act_1\sigma_e} \tup{A_1^t, m_k}
    \exect{\act_2\sigma_e} \cdots \exect{\act_n\sigma_e} \tup{A_n^t,
      m_k} \exect{\act'\sigma} \tup{A'_k, m'_k}\] 
  where
    \begin{compactitem}
    \item $\sigma_e$ is an empty substitution,
    \item $\act_i \in \actsettmpa$ (for $i \in \set{1,\ldots, n}$), 
    \item $\act' \in \actsettmpd$, 
    \item $\tmp \in A_i^t$ (for $i \in \set{1,\ldots, n}$),
      $\tmp \not\in A_k'$,
    \end{compactitem}
%
    Now, by the definition of $\tgprog$ on the translation of a
    program of the form $\delta_1 | \delta_2$ and
    $\gact{Q(\vec{p})}{\act(\vec{p})}$, there exists
    $j \in \set{1,\ldots,n}$ such that
    $\ppre(\delta_1|\delta_2) \in A_{j-1}^t$, and either
    \begin{compactenum}
    \item $\act_j = \gamma_{\delta_1}$, and
      $\ppre(\delta_1) \in A_j^t$, or
    \item $\act_j = \gamma_{\delta_2}$, and
      $\ppre(\delta_2) \in A_j^t$.
    \end{compactenum}
    where $\gamma_{\delta_1}$ and $\gamma_{\delta_2}$ are the actions
    obtained from the translation of $\delta_1 | \delta_2$ by
    $\tgprog$, and note that we consider $A_k = A_{j-1}^t$ if $j = 1$.
    Now, by our assumption above and by
    \Cref{lem:temp-state-produced-by-temp-act}, we have that
    $A_k \eqm A_j^t$, and hence it is easy to see that we also have
    $A_g \eqm A_j^t$. Thus, essentially we have
    \begin{center}
      $\tup{A_{j-1}^t, m_k} \exect{\act_j\sigma_e} \tup{A_{j}^t, m_k}
      \exect{\act_{j+1}\sigma_e} \cdots \exect{\act_n\sigma_e}
      \tup{A_n^t, m_k} \exect{\act'\sigma} \tup{A'_k, m'_k}$,
  \end{center}
  and 
  \begin{compactenum}
  \item if $\act_j = \gamma_{\delta_1}$, then
    $\tup{A_g, \scmap_g, \delta_1} \mimic \tup{A_j^t,\scmap_k}$
    (because $A_g \eqm A_j^t$, $\scmap_g = \scmap_k$,
    $\ppre(\delta_1) \in A_j^t$), otherwise
  \item if $\act_j = \gamma_{\delta_2}$, then
    $\tup{A_g, \scmap_g, \delta_2} \mimic \tup{A_j^t,\scmap_k}$
    (because $A_g \eqm A_j^t$, $\scmap_g = \scmap_k$,
    $\ppre(\delta_2) \in A_j^t$).
  \end{compactenum}
  Therefore by induction hypothesis, 
  it is easy to see that the claim is proven.

\item[\textbf{[$\delta_g = \delta_1;\delta_2$]}.]  Assume that there
  exist states $\tup{A'_k, m'_k}$, $\tup{A_i^t, m_k}$ (for
  $i \in \set{1,\ldots, n}$, and $n \geq 0$) such that
  \begin{center}
    $\tup{A_k, m_k} \exect{\act_1\sigma_e} \tup{A_1^t, m_k}
    \exect{\act_2\sigma_e} \cdots \exect{\act_n\sigma_e} \tup{A_n^t,
      m_k} \exect{\act'\sigma} \tup{A'_k, m'_k}$
  \end{center}
  where
    \begin{compactitem}
    \item $\sigma_e$ is an empty substitution,
    \item $\act_i \in \actsettmpa$ (for $i \in \set{1,\ldots, n}$), 
    \item $\act' \in \actsettmpd$, 
    \item $\tmp \in A_i^t$ (for $i \in \set{1,\ldots, n}$),
      $\tmp \not\in A_k'$,
    \end{compactitem}
    By the definition of $\tgprog$ on the translation of
    $\delta_1 ; \delta_2$ and $\gact{Q(\vec{p})}{\act(\vec{p})}$, as
    well as \Cref{lem:final-state-add-transition}, then there are two cases:

%
    \begin{compactenum}[\bf (a)]
%
    \item The case when $\tup{A_g, \scmap_g, \delta_1}$ is not a final
      state. Either
      \begin{compactitem}
        
      \item $\ppre(\delta_1) \in A_k$ and
        $\ppre(\delta_1) \not\in A_j^t$ for $j \in \set{1,\ldots,n}$,
        or


      \item there exists $j \in \set{1,\ldots,n}$ such that
        $\ppre(\delta_1) \in A_j^t$, and
        $\ppre(\delta_1) \not\in A_l^t$ for
        $l \in \set{j+1,\ldots,n}$. 

      \end{compactitem}
    \item The case when $\tup{A_g, \scmap_g, \delta_1}$ is a final
      state. There exists $j \in \set{1,\ldots,n-1}$ and
      $l \in \set{j+1,\ldots,n}$ such that
      $\ppre(\delta_1) \in A_j^t$, $\ppost(\delta_1) \in A_{l}^t$,
      $\ppre(\delta_2) \in A_{l}^t$,
      $\ppost(\delta_1) =
      \ppre(\delta_2)$. 
    \end{compactenum}
%
%
    Now, by our assumption above and by
    \Cref{lem:temp-state-produced-by-temp-act}, we have that
    \begin{compactenum}
    \item[- \textbf{For the case} \textbf{(a)}:] either 
      \begin{itemize}
      \item $A_g \eqm A_k$ and thus we have that
        $\tup{A_g, \scmap_g, \delta_1} \mimic \tup{A_k,\scmap_k}$, or 
      \item $A_k \eqm A_j^t$, and hence it is easy to see that we also
        have $A_g \eqm A_j^t$. Thus we have that
        $\tup{A_g, \scmap_g, \delta_1} \mimic \tup{A_j^t,\scmap_k}$.
    \end{itemize}
    
    \item[- \textbf{For the case} \textbf{(b)}:] $A_k \eqm A_l^t$, and hence it is
      easy to see that we also have $A_g \eqm A_l^t$. Thus we have
      that $\tup{A_g, \scmap_g, \delta_2} \mimic
      \tup{A_l^t,\scmap_k}$. 
    \end{compactenum}
    Therefore by induction hypothesis, it is easy to see that the
    claim is proven. 

\item[\textbf{[$\delta_g = \gif{\varphi}{\delta_1}{\delta_2}$]}.]
  Assume that there exist states $\tup{A'_k, m'_k}$,
  $\tup{A_i^t, m_k}$ (for $i \in \set{1,\ldots, n}$, and $n \geq 0$)
  such that
  \begin{center}
    $\tup{A_k, m_k} \exect{\act_1\sigma_e} \tup{A_1^t, m_k}
    \exect{\act_2\sigma_e} \cdots \exect{\act_n\sigma_e} \tup{A_n^t,
      m_k} \exect{\act'\sigma} \tup{A'_k, m'_k}$
  \end{center}
  where
    \begin{compactitem}
    \item $\sigma_e$ is an empty substitution,
    \item $\act_i \in \actsettmpa$ (for $i \in \set{1,\ldots, n}$), 
    \item $\act' \in \actsettmpd$, 
    \item $\tmp \in A_i^t$ (for $i \in \set{1,\ldots, n}$),
      $\tmp \not\in A_k'$,
    \end{compactitem}
    By the definition of $\tgprog$ on the translation of a program of
    the form $\gif{\varphi}{\delta_1}{\delta_2}$ and
    $\gact{Q(\vec{p})}{\act(\vec{p})}$, there exists
    $j \in \set{1,\ldots,n-1}$ such that
    $\ppre(\gif{\varphi}{\delta_1}{\delta_2}) \in A_{j-1}^t$ (note
    that we consider $A_k = A_{j-1}^t$ if $j = 1$) and either
    \begin{compactenum}
    \item $\act_j = \gamma_{if}$, $\ppre(\delta_1) \in A_j^t$, and
      $\Ans(\varphi, T, A_{j-1}^t) = \true$, or
    \item $\act_j = \gamma_{else}$, $\ppre(\delta_2) \in A_j^t$, and
      $\Ans(\varphi, T, A_{j-1}^t)  = \false$.
    \end{compactenum}
    where $\gamma_{if}$ and $\gamma_{else}$ are the actions obtained
    from the translation of $\gif{\varphi}{\delta_1}{\delta_2}$ by
    $\tgprog$. 
    Now, by our assumption above and by
    \Cref{lem:temp-state-produced-by-temp-act}, we have that
    $A_k \eqm A_j^t$, and hence it is easy to see that we also have
    $A_g \eqm A_j^t$. Thus, essentially we have
    \begin{center}
      $\tup{A_{j-1}^t, m_k} \exect{\act_j\sigma_e} \tup{A_{j}^t, m_k}
      \exect{\act_{j+1}\sigma_e} \cdots \exect{\act_n\sigma_e}
      \tup{A_n^t, m_k} \exect{\act'\sigma} \tup{A'_k, m'_k}$,
  \end{center}
  and
  \begin{compactenum}
  \item if $\act_j = \gamma_{if}$, then
    $\tup{A_g, \scmap_g, \delta_1} \mimic \tup{A_j^t,\scmap_k}$
    (because $A_g \eqm A_j^t$, $\scmap_g = \scmap_k$,
    $\ppre(\delta_1) \in A_j^t$), otherwise
  \item if $\act_j = \gamma_{else}$, then
    $\tup{A_g, \scmap_g, \delta_2} \mimic \tup{A_j^t,\scmap_k}$
    (because $A_g \eqm A_j^t$, $\scmap_g = \scmap_k$,
    $\ppre(\delta_2) \in A_j^t$).
  \end{compactenum}
  Therefore by induction hypothesis, 
  it is easy to see that the claim is proven. 

\item[\textbf{[$\delta_g = \gwhile{\varphi}{\delta_1}$]}.]  Assume
  that there exist states $\tup{A'_k, m'_k}$, $\tup{A_i^t, m_k}$ (for
  $i \in \set{1,\ldots, n}$, and $n \geq 0$) such that
  \begin{center}
    $\tup{A_k, m_k} \exect{\act_1\sigma_e} \tup{A_1^t, m_k}
    \exect{\act_2\sigma_e} \cdots \exect{\act_n\sigma_e} \tup{A_n^t,
      m_k} \exect{\act'\sigma} \tup{A'_k, m'_k}$
  \end{center}
  where
    \begin{compactitem}
    \item $\sigma_e$ is an empty substitution,
    \item $\act_i \in \actsettmpa$ (for $i \in \set{1,\ldots, n}$), 
    \item $\act' \in \actsettmpd$, 
    \item $\tmp \in A_i^t$ (for $i \in \set{1,\ldots, n}$),
      $\tmp \not\in A_k'$,
    \end{compactitem}
    By the definition of $\tgprog$ on the translation of a program of
    the form $\gwhile{\varphi}{\delta_1}$ and
    $\gact{Q(\vec{p})}{\act(\vec{p})}$, there exists
    $j \in \set{1,\ldots,n-1}$ such that
    \begin{compactitem}
    \item $\act_j
      = \gamma_{doLoop}$
      ($\gamma_{doLoop}$
      is the action obtained during the translation of
      $\gwhile{\varphi}{\delta_1}$ by $\tgprog$),
    \item $\ppre(\gwhile{\varphi}{\delta_1})
      \in A_{j-1}^t$ (where $A_{j-1}^t = A_k$ when $j = 1$), and
    \item $\ppre(\delta_1) \in A_{j}^t$.
    \end{compactitem}
    Now, by our assumption above and by
    \Cref{lem:temp-state-produced-by-temp-act}, we have that $A_k
    \eqm A_j^t$, and hence it is easy to see that $A_g \eqm
    A_j^t$. Thus, essentially we have
    \begin{center}
      $\tup{A_{j-1}^t,
        m_k} \exect{\act_j\sigma_e} \tup{A_{j}^t, m_k}
      \exect{\act_{j+1}\sigma_e} \cdots \exect{\act_n\sigma_e}
      \tup{A_n^t, m_k} \exect{\act'\sigma} \tup{A'_k, m'_k}$,
  \end{center}
  and $\tup{A_g, \scmap_g, \delta_1} \mimic \tup{A_j^t,\scmap_k}$
  (because $A_g \eqm A_j^t$, $\scmap_g = \scmap_k$,
  $\ppre(\delta_1) \in A_j^t$).
  Therefore by induction hypothesis, it is easy to see that the claim
  is proven. 
\end{enumerate}
\end{proof}

Now we will show that given a state $s_g$ of an S-GKAB transition
system and a state $s_k$ of its corresponding KAB transition system
such that $s_g$ is mimicked by $s_k$, then we have $s_g$ and $s_k$ are
J-bisimilar. Formally this claim is stated and shown below.

\begin{lemma}\label{lem:sgkab-to-kab-bisimilar-state}
  Let $\gkabsym$ be an S-GKAB with transition system
  $\ts{\gkabsym}^{\filter_S}$, and let $\tgkab(\gkabsym)$ be a KAB
  with transition system $\ts{\tgkab(\gkabsym)}$ obtained from
  $\gkabsym$ through $\tgkab$.  Consider
\begin{inparaenum}[]
\item a state $\tup{A_g,\scmap_g, \delta_g}$ of
  $\ts{\gkabsym}^{\filter_S}$ and
\item a state $\tup{A_k,\scmap_k}$ of $\ts{\tgkab(\gkabsym)}$.
\end{inparaenum}
If $\tup{A_g,\scmap_g, \delta_g} \mimic \tup{A_k,\scmap_k} $ then
$\tup{A_g,\scmap_g, \delta_g} \jbsim \tup{A_k,\scmap_k}$.
\end{lemma}
\begin{proof}
Let 
\begin{compactitem}
\item $\gkabsym = \tup{T, \initabox, \actset, \ginitprog}$ and
  $\ts{\gkabsym}^{\filter_S} = \tup{\const, T, \stateset_g, s_{0g},
    \abox_g, \trans_g}$, 
\item $\tgkab(\gkabsym) = \tup{T, \initabox', \actset', \procset'}$,  and
  $\ts{\tgkab(\gkabsym)} = \tup{\const, T, \stateset_k, s_{0k},
    \abox_k, \trans_k}$
\end{compactitem}
We have to show:
\begin{enumerate}[\bf (1)]
\item for every state $\tup{A'_g,\scmap'_g, \delta'_g}$ such that
  $\tup{A_g,\scmap_g, \delta_g} \trans_g \tup{A'_g,\scmap'_g,
    \delta'_g},$
  there exist states $\tup{A'_k,\scmap'_k}$,
  $\tup{A_1^t,\scmap_k} \ldots \tup{A_n^t,\scmap_k}$ (for $n \geq 0$)
  with
  \[
  \tup{A_k,\scmap_k} \trans_k \tup{A_1^t,\scmap_k} \trans_k \ldots
  \trans_k \tup{A_n^t,\scmap_k} \trans_k \tup{A'_k,\scmap'_k}
  \]
  such that:
  \begin{compactenum}
  \item $\tmp \not\in A'_k$, $\tmp \in A^t_i$ for
    $i \in \set{1, \ldots, n}$, and
  \item 
    $\tup{A'_g,\scmap'_g, \delta'_g} \mimic \tup{A'_k,\scmap'_k} $.
  \end{compactenum}

\item for every state $\tup{A'_k,\scmap'_k}$ such that there exist
  states $\tup{A_1^t,\scmap_1} \ldots \tup{A_n^t,\scmap_n}$ (for
  $n \geq 0$) and
  \[
  \tup{A_k,\scmap_k} \trans_k \tup{A_1^t,\scmap_1} \trans_k \ldots
  \trans_k \tup{A_n^t,\scmap_n} \trans_k \tup{A'_k,\scmap'_k}
  \]
  where $\tmp \not\in A'_k$, and $\tmp \in A^t_i$ for
  $i \in \set{1, \ldots, n}$,
  then there exists a state $\tup{A'_g,\scmap'_g, \delta'_g}$ with
  $\tup{A_g,\scmap_g, \delta_g} \trans_g \tup{A'_g,\scmap'_g,
    \delta'_g}$
  such that 
  $\tup{A'_g,\scmap'_g, \delta'_g} \mimic \tup{A'_k,\scmap'_k} $.

\end{enumerate}

\begin{enumerate}

\item[\textbf{Proof for (1):}] 
Assume
$ \tup{A_g,\scmap_g, \delta_g} \trans \tup{A'_g,\scmap'_g, \delta'_g},
$ then by \Cref{def:gkab-ts} we have
\[
\tup{A_g,\scmap_g, \delta_g} \gprogtrans{\act\sigma, \filter_S}
\tup{A'_g,\scmap'_g, \delta'_g}.
\]
Additionally, by \Cref{def:prog-exec-relation,def:tell-operation} we
have that $A'_g$ is $T$-consistent.
By \Cref{lem:prog-exec-bsim}, there exist states $\tup{A_i^t, m_k}$,
and actions $\act_i$ (for $i \in \set{1, \ldots, n}$, where $n \geq 0$)
such that
\begin{compactitem}
\item
  $\tup{A_k, m_k} \exect{\act_1\sigma_e} \tup{A_1^t, m_k}
  \exect{\act_2\sigma_e} \cdots \exect{\act_n\sigma_e} \tup{A_n^t, m_k}
  \exect{\act'\sigma} \tup{A'_k, m'_k}$ where
  \begin{compactitem}
  \item $\sigma_e$ is an empty substitution, 
  \item $\act'$ is obtained from $\act$ through $\tgprog$,
  \item $\tmp \in A_i^t$ (for $i \in \set{1, \ldots, n}$), $\tmp \not\in A_k'$
  \end{compactitem}
\item 
  $\tup{A'_g,\scmap'_g, \delta'_g} \mimic \tup{A'_k,\scmap'_k} $,
\end{compactitem}
Additionally, since $A'_g$ is $T$-consistent and $A'_g \eqm A_k'$ then
$A'_k$ is $T$-consistent. 
As a consequence, we have that the claim is easily proven, since by
\Cref{def:KAB-standard-ts}, we have
\[
\begin{array}{l}
  \tup{A_k, m_k} \trans_k \tup{A_1^t, m_k} \trans_k
  \cdots \trans_k \tup{A_n^t, m_k} \trans_k \tup{A'_k,
  m'_k}
\end{array}
\]
where 
\begin{compactenum}
\item $\tmp \not\in A'_k$, and $\tmp \in A^t_i$ for
  $i \in \set{1, \ldots, n}$, and
\item 
  $\tup{A'_g,\scmap'_g, \delta'_g} \mimic \tup{A'_k,\scmap'_k} $,
\end{compactenum}

\item[\textbf{Proof for (2):}] 
%
  Assume
  \[
  \tup{A_k,\scmap_k} \trans_k \tup{A_1^t,\scmap_k} \trans_k \ldots
  \trans_k \tup{A_n^t,\scmap_k} \trans_k \tup{A'_k,\scmap'_k}
  \]
  where $n \geq 0$, $\tmp \not\in A'_k$, and $\tmp \in A^t_i$ for
  $i \in \set{1, \ldots, n}$. By \Cref{def:KAB-standard-ts}, we
  have 
  \[
  \tup{A_k, m_k} \exect{\act_1\sigma_1} \tup{A_1^t, m_1}
  \exect{\act_2\sigma_2} \cdots \exect{\act_n\sigma_n} \tup{A_n^t,
    m_n} \exect{\act'\sigma'} \tup{A'_k, m'_k}.
  \]
  For some actions $\act'$, $\act_i$ (for $i \in \set{1,\ldots, n}$), and
  substitutions $\sigma'$, $\sigma_i$ (for $i \in \set{1,\ldots, n}$). 
  Let $\actsettmpa$ (resp.\ $\actsettmpd$) be the set of temp
  adder (resp.\ deleter) actions of $\tgkab(\gkabsym)$, since
  $\tmp \not\in A'_k$, and $\tmp \in A^t_i$ for
  $i \in \set{1, \ldots, n}$, by
  \Cref{lem:temp-state-produced-by-temp-act,lem:non-temp-state-produced-by-normal-action},
  we have that
\begin{compactitem}
\item $\sigma_i$ is an empty substitution (for
  $i \in \set{1,\ldots, n}$).
\item $\act_i \in \actsettmpa$ (for $i \in \set{1,\ldots, n}$), and
\item $\act' \in \actsettmpd$, 
\item there exists an action invocation
  $\gact{Q(\vec{p})}{\act(\vec{p})}$ that is a sub-program of
  $\ginitprog$ such that $\act'$ is obtained from the translation of
  $\gact{Q(\vec{p})}{\act(\vec{p})}$ by $\tgprog$,
\item $\act_i$ (for $i \in \set{1,\ldots, n}$) does not involve any
  service calls, and hence $m_i = m_{i+1}$ (for $i \in \set{1,\ldots, n-1}$).
\end{compactitem}

Therefore by \Cref{lem:kab-transition-bsim}, then there exists a
state $\tup{A'_g,\scmap'_g, \delta_g'}$ such that
\begin{compactitem}
\item
  $ \tup{A_g,\scmap_g, \delta_g} \gprogtrans{\alpha\sigma, \filter_S}
  \tup{A'_g,\scmap'_g, \delta_g'}, $
  \item $\act'$ is obtained from the translation of a certain action
    invocation $\gact{Q(\vec{p})}{\act(\vec{p})}$ via $\tgprog$.
  \item 
  $\tup{A'_g,\scmap'_g, \delta'_g} \mimic \tup{A'_k,\scmap'_k} $.
\end{compactitem}
Since
$\tup{A_g,\scmap_g, \delta_g} \gprogtrans{\act\sigma, \filter_S}
\tup{A'_g,\scmap'_g, \delta'_g}$,
by \Cref{def:gkab-ts}, we have that
$ \tup{A_g,\scmap_g, \delta_g} \trans \tup{A'_g,\scmap'_g, \delta'_g}
$.
Thus it is easy that the claim is proven since we also have that
$\tup{A'_g,\scmap'_g, \delta'_g} \mimic \tup{A'_k,\scmap'_k} $.
\end{enumerate}
\ \ 
\end{proof}
%

Having \Cref{lem:sgkab-to-kab-bisimilar-state} in hand, we can easily
show that given an S-GKAB, its transition system is J-bisimilar to the
transition system of its corresponding KAB that is obtained via the
translation $\tgkab$.

\begin{lemma}\label{lem:sgkab-to-kab-bisimilar-ts}
  Given an S-GKAB $\gkabsym$, let $\tgkab(\gkabsym)$ be the KAB
  obtained from $\gkabsym$ by applying the translation $\tgkab$, we
  have $\ts{\gkabsym}^{\filter_S} \jbsim \ts{\tgkab(\gkabsym)} $
\end{lemma}
\begin{proof}
Let
\begin{compactenum}
\item $\gkabsym = \tup{T, \initabox, \actset, \ginitprog}$, and 
  $\ts{\gkabsym}^{\filter_S} = \tup{\const, T, \stateset_g, s_{0g},
    \abox_g, \trans_g}$,
\item $\tgkab(\gkabsym) = \tup{T, \initabox', \actset', \procset'}$, and
  $\ts{\tgkab(\gkabsym)} = \tup{\const, T, \stateset_k,
    s_{0k}, \abox_k, \trans_k}$ 
\end{compactenum}
We have that $s_{0g} = \tup{A_0, \scmap_g, \delta_g}$ and
$s_{0k} = \tup{A'_0, \scmap_k}$ where
$\scmap_g = \scmap_k = \emptyset$.
Since $A_0' = A_0 \cup \set{\flagconcept{start}}$, and
$\flagconceptname$ is a special vocabulary outside the vocabulary of
$T$, hence $A'_0 \eqm A_0$. 
Now, by \Cref{lem:program-pre-post}, we have
$\ppre(\ginitprog) = \flagconcept{start}$ and
$\ppost(\ginitprog) = \flagconcept{end}$. 
Furthermore, since $\flagconcept{start} \in A_0'$, then we have
$s_{0g} \mimic s_{0k}$. Hence by
\Cref{lem:sgkab-to-kab-bisimilar-state}, we have
$s_{0g} \jbsim s_{0k}$. Therefore,
we have
$\ts{\gkabsym}^{\filter_S} \jbsim \ts{\tgkab(\gkabsym)} $.

\end{proof}

Having all of these machinery in hand, we are now ready to show that
the verification of \muladom properties over S-GKABs can be recast as
verification over KAB as follows.

\begin{theorem}\label{thm:gtos}
  Given an S-GKAB $\gkabsym$ and a closed $\muladom$ property $\Phi$
  in NNF, let $\tgkab(\gkabsym)$ be the KAB obtained from $\gkabsym$
  by applying the translation $\tgkab$, we have
\begin{center}
  $\ts{\gkabsym}^{\filter_S} \models \Phi$ if and only if
  $\ts{\tgkab(\gkabsym)} \models \tforj(\Phi)$
\end{center}
\end{theorem}
\begin{proof}
  By \Cref{lem:sgkab-to-kab-bisimilar-ts}, we have that
  $\ts{\gkabsym}^{\filter_S} \jbsim \ts{\tgkab(\gkabsym)}$.
  Hence, by
  \Cref{lem:jumping-bisimilar-ts-satisfies-same-formula}, we have
  that for every $\muladom$ property $\Phi$
\[
\ts{\gkabsym}^{\filter_S} \models \Phi \textrm{ if and only if }
\ts{\tgkab(\gkabsym)} \models \tforj(\Phi)
\]
\end{proof}



\subsection{Verification of Run-Bounded S-GKABs}

An interesting property of the translation $\tgkab$ is that it
preserves run-boundedness. 

\begin{lemma}\label{lem:run-bounded-preservation-sgkab}
  Let $\gkabsym$ be an arbitrary S-GKAB and $\tgkab(\gkabsym)$ be its
  corresponding KAB obtained through the translation $\tgkab$. We have
  that $\gkabsym$ is run-bounded if and only if $\tgkab(\gkabsym)$ is
  run-bounded.
\end{lemma}
\begin{proof}
  Let $\ts{\gkabsym}^{\filter_S}$ be the transition system of
  $\gkabsym$, and $\ts{\tgkab(\gkabsym)}$ be the transition system
  of $\tgkab(\gkabsym)$.
  The proof is then easily obtained since
  \begin{compactitem}

  \item only a bounded number of new constants are introduced when
    emulating the Golog program with KAB condition-action rules and
    actions. This fact can be easily seen by observing the following:
    \begin{compactitem}

    \item Given a Golog program $\delta$ there are only finitely many
      sub-programs (i.e., it only yields finite number of program IDs).

    \item As it can be seen from the definition of program translation
      $\tgprog$ (cf. \Cref{def:prog-translation}), which recursively
      translates each sub-program, for each case of the sub-program
      translation, we only introduce finitely many fresh constants.

    \end{compactitem}

  \item by Lemma~\ref{lem:sgkab-to-kab-bisimilar-ts}, we have that
    $\ts{\gkabsym}^{\filter_S} \jbsim \ts{\tgkab(\gkabsym)} $.  Thus,
    basically they are ``equivalent'' modulo intermediate states
    (states containing $\tmp$), and each two bisimilar states are
    equivalent modulo special markers.

  \end{compactitem}
\ \ 
\end{proof}

\noindent
Now, we can easily acquire the following result on verification of
\muladom properties over run-bounded S-GKABs.

\begin{theorem}[Verification of Run-Bounded S-GKABs]\label{thm:ver-run-bounded-sgkab}
  Verification of closed $\muladom$ formulas over a run-bounded S-GKAB is
  decidable and can be reduced to finite-state model checking.
\end{theorem}
\begin{proof}
  From \Cref{thm:gtos} and \Cref{lem:run-bounded-preservation-sgkab},
  we have that verification of closed \muladom formulas over
  run-bounded S-GKABs can be reduced to the verification of \muladom
  formulas over run-bounded KABs. Then, by
  \Cref{thm:verification-run-bounded-kab}, we have that verification
  of \muladom over run-bounded KABs are decidable and can be reduced
  to finite-state model checking.
\end{proof}

\section{Discussion}\label{sec:discussion}

As we have seen above, here we only consider some constructs of
typical Golog program (cf. \cite{LRLLS97,DeLL00}). However, we are
able to simulate some other Golog program constructs within GKABs. In
this section, we discuss some of those possibilities.

Nondeterministic iteration (i.e., $\delta^*$) is seamlessly supported
by G-KABs, using the core set of constructs we considered.  In fact,
$\delta^*$ can be simulated as
%
%
\[
\gwhile{\true}{\delta_1}
\]
where $\delta_1 = \delta\ \! |\ \! \gemptyprog$. Notice that
essentially the meaning of $\delta^*$ is that we execute $\delta$ zero
or more times. Thus, in the while loop above, by having
$\delta\ \! |\ \! \gemptyprog$ within the body of the loop we can
either choose:
\begin{compactenum}
\item to execute $\delta$ (and it can be done as much as we
  want).
\item to consider the program as completed and move to next
  instruction (Note that by the definition of final state, we have
  that
  \[
  \final{\tup{A, \scmap, \gwhile{\true}{\delta_1}}}
  \]
  for any ABox $A$ and service call map $\scmap$).
\end{compactenum}

Regarding the test construct (i.e., $\varphi ?$), based on the
semantics of test construct in \cite{LRLLS97,DeLL00,DeLL97}, roughly
speaking, the meaning of the test construct $\varphi ?$ is that when
the test $\varphi$ is satisfied, then the system makes a transition
into a state where the data stay the same but the remaining program to
be executed become $\gemptyprog$. I.e., within our setting we can
extend our program execution relation definition by adding the
following:
\[
\tup{A, \scmap, \varphi?}  \gprogtrans{\act\sigma, \filter} \tup{A,
  \scmap, \gemptyprog}, \mbox{ if } \ask(\varphi, T, A) = \true;
\]
it is easy to see that such construct can be simulated by
\[
\gif{\varphi}{ \delta_1 }{ \delta_2  }
\]
where:
\begin{compactitem}
\item $\delta_1 = \gact{\true}{\act_{DoNothing}()}$, and
  $\act_{DoNothing}$ is an action that does not change the ABox,
\item $\delta_2 = \gact{\false}{\act_{Block}()}$, and $\act_{Block}$
  is any action (the idea is just to block the execution and basically
  it will be blocked in any case since the guard is $\false$).
\end{compactitem}
The idea of the program above is as follows:
\begin{compactitem}

\item in case $\varphi$ is satisfied, the system will execute the
  action $\act_{DoNothing}$ which will not change the data but the
  remaining program to be executed becomes $\gemptyprog$ (i.e., the
  system can progress further to execute the next program
  instruction).

\item in case $\varphi$ is not satisfied, the system execution will be
  blocked since the action $\act_{Block}$ in any case can not be
  executed.

\end{compactitem}

Moreover, we can also simulate another semantics of test construct
$\varphi ?$ as in \cite{ZC14, ClLa08}. According to \cite{ZC14,
  ClLa08}, the program made by a test construct is considered to be
completed when the test is satisfied. Technically the state is
considered to be final state (i.e., completed) when the test is
satisfied. Therefore, within our setting, we can extend our definition
of final state by adding the following:
\[
\final{\tup{A, \scmap, \varphi ?}}, \mbox{ if } \ask(\varphi, T, A) = \true;
\]
In this case, we can simulate such construct as follows:
\[
\gif{\varphi}{ \gemptyprog }{ \gact{\false}{\act_{Block}()} }
\]
where $\act_{Block}$ is any action.
The idea of the program above is as follows:
\begin{compactitem}

\item in case $\varphi$ is satisfied, since the ``if case'' goes to
  $\gemptyprog$, then according to the definition of final states, the
  corresponding state that has the program
  $\gif{\varphi}{ \gemptyprog }{ \delta }$ can be considered as a
  final state (i.e., completed).

\item in case the $\varphi$ is not satisfied, the system execution
  will be blocked since the action $\act_{Block}$ in any case can not
  be executed because it is guarded with $\false$.

\end{compactitem}

About the pick construct, so far our GKABs only consider the pick
construct that ranges over an action. However, one might easily extend
it into the pick construct that ranges over a program, and GKABs with
such extension can actually be simulated by our GKABs that only
consider the pick construct that ranges over an action. We provide the
ideas below.

Basically, we can extend our definition of Golog program
(cf. \Cref{def:golog-program}) by allowing the following pick
construct:
\[
\gact{Q(\vec{x})}{\delta[\vec{x}]}
\]
that 
%
\begin{compactenum}
\item picks a tuple $\vec{c}$ in the answer of $Q(\vec{x})$,
\item instantiates the rest of the program $\delta$ by substituting
  $\vec{x}$ with $\vec{c}$, and
\item then executes $\delta$.
\end{compactenum}
Notice that each $x \in \vec{x}$ that occurs in a query within
$\delta$ must be a free variable (recall that the idea of pick is to
pick a constant to instantiate the action).
We can extend our definition of the program execution relation
(cf. \Cref{def:prog-exec-relation}) by adding the following:
\begin{compactitem}

\item
  $\tup{A, \scmap, \gact{Q(\vec{x})}{\delta[\vec{x}] }}
  \gprogtrans{\act\sigma, \filter} \tup{A', \scmap',
    \delta'[\vec{x} / \vec{c}] }$,
  \\if $\vec{c} \in \ask(Q(\vec{x}), T, A)$, and
  $\tup{A, \scmap, \delta[\vec{x}/\vec{c}]} \gprogtrans{\act\sigma,
    \filter} \tup{A', \scmap', \delta'[\vec{x}/\vec{c}]}$,
%
%
\end{compactitem}
where $\delta[\vec{x}/\vec{c}]$ means that we substitute the variables
$\vec{x}$ in $\delta$ with $\vec{c}$.

We first illustrate the idea of translating GKABs with such extension
into our GKABs as follows: Consider the program:
\[
\delta = \gact{Q(x_1, x_2)}{\delta_1[x_1, x_2]}
\]
where
\[
\delta_1 = \gact{ Q_1( x_1, x_2, x_3 ) }{ \act_1( x_1, x_2, x_3 ) } ;
\gact{ Q_2(x_1 , x_4) }{ \act_2(x_1, x_4) }.
\]
Let $\tup{a, b}$ be an answer of $Q(x_1, x_2)$ (i.e., $x_1$ (resp.\
$x_2$) is substituted by $a$ (resp.\ $b$)). Hence, because we need to
substitute each occurrence of $x_1$ (resp.\ $x_2$) within $\delta$
with $a$ (resp.\ $b$), we basically have the following:
\begin{compactitem}

\item to instantiate the parameters of $\act_1$, we must only consider
  those answers of $Q_1$ in which $x_1$ (resp.\ $x_2$) is substituted
  by $a$ (resp.\ $b$);

\item similarly, to instantiate the parameters of $\act_2$, we must
  only consider those answers of $Q_2$ in which $x_1$ is substituted
  by $a$.

\end{compactitem}
To simulate such situation, the idea is as follows:
\begin{compactenum}
\item We use some temporary ABox assertions to keep the answer of $Q$
  that is used to instantiate $x_1$ and $x_2$ in $\delta'$. In our
  example, we can introduce two fresh concepts $V^Q_{x_1}$ and
  $V^Q_{x_2}$ to store the picked values for $x_1$ and $x_2$.

\item We change each action invocation within $\delta'$ such that when
  they want to get a value for $x_1$ (resp.\ $x_2$), they should only
  consider the value in $V^Q_{x_1}$ (resp.\ $V^Q_{x_2}$). To do this,
  we can just conjunct each query in each action invocation with the
  query that retrieves values from $V^Q_{x_1}$ and $V^Q_{x_2}$ (i.e.,
  $Q^{x}(x_1, x_2) = V^Q_{x_1}(x_1) \wedge
  V^Q_{x_2}(x_2)$).
  Essentially, we can translate $\delta_1$ into the following:
  \[
  \begin{array}{ll}
    \delta_1' = &\gact{ Q_1( x_1, x_2, x_3 ) \wedge Q^{x}(x_1, x_2) }{ \act_1( x_1, x_2, x_3 ) } ;\\
                &\gact{ Q_2(x_1 , x_4) \wedge Q^{x}(x_1, x_2)  }{ \act_2(x_1, x_4) }.
  \end{array}
  \]
  Note that in the second atomic action invocation above (i.e., that
  execute $\act_2$), we need to abuse our definition of atomic action
  invocation because we pick more values than what is needed in the
  action parameters of $\act_2$ (i.e., we have an atomic action
  invocation of the form $\gact{Q(\vec{y})}{\act({\vec{x}})}$ where
  $\vec{x} \subseteq \vec{y}$). However, considering the form of the
  query $Q_x$, it is easy to see that we can do post processing in our
  translation to remove unnecessary query. I.e., we can simply
  translate $\delta_1'$ into
  \[
  \begin{array}{ll}
    \delta_1'' = &\gact{ Q_1( x_1, x_2, x_3 ) \wedge V^Q_{x_1}(x_1) \wedge V^Q_{x_2}(x_2) }{ \act_1( x_1, x_2, x_3 ) } ;\\
                 &\gact{ Q_2(x_1 , x_4) \wedge V^Q_{x_1}(x_1)}{ \act_2(x_1, x_4) }.
  \end{array}
  \]

\item At the end of the execution of $\delta_1'$ we need to delete
  those temporary ABox assertions that was used to store the picked
  values (i.e., in our example it means that those ABox assertions
  that are made by $V^Q_{x_1}$ and $V^Q_{x_2}$).

\item Hence, to sum up those ideas above, basically we translate
  $\delta$ into $\delta'$ as follows:
\[
\delta' =  \gact{ Q(x_1,x_2) }{ \act_{pick}(x_1,x_2)} \ ;\  \delta_1'\ ;\ \gact{\true }{ \act_{del}() }
\]
where 
\begin{compactitem}[]

\item $\bullet$ $\eff{\act_{pick}} = \set{\map{\true}{\add \set{V^Q_{x_1}(x_1),
        V^Q_{x_2}(x_2), \tmp} } }$,


\item $
  \begin{array}{ll}
    \hspace*{-1.7mm}\bullet\ \eff{\act_{del}} = \set{  
    &\map{ \true }{\add \set{ \tmp } },  \\
    &\map{V^Q_{x_1}(x_1)}{\del \set{V^Q_{x_1}(x_1)} },  \\
    &\map{V^Q_{x_2}(x_2)}{\del \set{V^Q_{x_2}(x_2)} } \ \ }
\end{array}
$

\end{compactitem}
The idea is that the action $\act_{pick}$ stores the information about
the picked values for $x_1$ and $x_2$, while $\act_{del}$ deletes such
information when it is not useful anymore.

\item Note that we need to use the intermediate states (i.e., states
  marked by $\tmp$). Therefore, we also need to translate each atomic
  action invocation such that it deletes $\tmp$ and we also need to
  translate each formula to be verified such that it ignores the
  states marked by $\tmp$.

\end{compactenum}

Note that such idea also work properly in the case where we have
nested pick that ranges over program and might pick a constant for a
particular variable more than once. To illustrate the idea, consider
the program:
\[
\delta = \gact{Q(x_1, x_2)}{\delta_1[x_1, x_2]}
\]
where
\[
\delta_1 = \gact{ Q_1( x_1, x_2, x_3 ) }{ \act_1( x_1, x_2, x_3 ) } \
; \ \gact{ Q_2(x_1 , x_4) }{ \delta_2[x_1, x_4] }.
\]
and 
\[
\delta_2 = \gact{ Q_3( x_1, x_4, x_5 ) }{ \act_3( x_1, x_4, x_5 ) }.
\]
We can then translate $\delta$ into $\delta'$ as follows:
\[
\delta' =  \gact{ Q(x_1,x_2) }{ \act_{pick}(x_1,x_2)}\ ;\ \delta_1'\ ;\ \gact{\true }{ \act_{del}() }
\]
where 
\begin{itemize}[]

\item $\bullet$ $\eff{\act_{pick}} = \set{\map{\true}{\add \set{V^Q_{x_1}(x_1),
        V^Q_{x_2}(x_2), \tmp} } }$,


\item $
  \begin{array}{ll}
    \hspace*{-1.7mm}\bullet\ \eff{\act_{del}} = \set{  
    &\map{ \true }{\add \set{ \tmp } },  \\
    &\map{V^Q_{x_1}(x_1)}{\del \set{V^Q_{x_1}(x_1)} },  \\
    &\map{V^Q_{x_2}(x_2)}{\del \set{V^Q_{x_2}(x_2)} } \ \ },
\end{array}
$


\item $
\begin{array}{ll}
  \hspace*{-1.7mm}\bullet\ \delta_1' = &\gact{ Q_1( x_1, x_2, x_3 ) \wedge Q^{x}(x_1, x_2) }{
                                         \act_1( x_1, x_2, x_3 ) } ; \\
                                       &\gact{ Q_2(x_1,x_4) \wedge Q^{x}(x_1, x_2)}{
                                         \act'_{pick}(x_1,x_4)} \ ;\ 
                                         \delta_4\ ;\ \gact{\true }{
                                         \act'_{del}() }, \mbox{ where}
\end{array}
$
\begin{compactitem}[]
\item $\bullet$ $Q^{x}(x_1, x_2) = V^Q_{x_1}(x_1) \wedge V^Q_{x_2}(x_2)$




\item $\bullet$ $\eff{\act'_{pick}} = \set{\map{\true}{\add \set{V^{Q_2}_{x_1}(x_1),
        V^{Q_2}_{x_4}(x_4), \tmp} } }$,


\item $
  \begin{array}{ll}
    \hspace*{-1.7mm}\bullet\ \eff{\act'_{del}} = \set{  
    &\map{ \true }{\add \set{ \tmp } },  \\
    &\map{V^{Q_2}_{x_1}(x_1)}{\del \set{V^{Q_2}_{x_1}(x_1)} },  \\
    &\map{V^{Q_2}_{x_4}(x_4)}{\del \set{V^{Q_2}_{x_4}(x_4)} } \ \ },
\end{array}
$

\item $\bullet$
  $\delta_4 = \gact{ Q_3( x_1, x_4, x_5 ) \wedge Q^{x}(x_1, x_2) \wedge
    Q_2^{x}(x_1, x_4) }{ \act_3( x_1, x_4, x_5 ) }$, with

\begin{compactitem}
\item  $Q_2^{x}(x_1, x_4) = V^{Q_2}_{x_1}(x_1) \wedge V^{Q_2}_{x_4}(x_4)$
\end{compactitem}


\end{compactitem}

\end{itemize}

Generalizing the idea above, we now proceed to provide a systematic
way to translate GKABs with such extension into our GKABs by defining
a translation $\tpick$ that takes as inputs:
\begin{compactenum}
\item a program $\delta$ (that might contain pick that ranges over a program), and
\item a query $Q$
\end{compactenum}
and produces a program $\delta'$, in which each pick construct ranges
over an action. I.e., we have $\tpick(\delta, Q) = \delta'$. Formally,
the translation $\tpick$ is inductively defined over the structure of
a program $\delta$ as follows:

\begin{enumerate}


\item For the case of $\delta = \gemptyprog$, we have:
  \[
  \tpick (\gemptyprog, Q_\pi(\vec{y}) ) = \gemptyprog,
  \]

\item For the case of $\delta = \gact{Q(\vec{x})}{\act(\vec{x})}$, we
  have:
  \[
  \tpick (\gact{Q(\vec{x})}{\act(\vec{x})}, Q_\pi(\vec{y}) ) =
  \gact{Q(\vec{x}) \wedge Q_\pi(\vec{y})}{\act'(\vec{x})},
  \]
  where $\eff{\act'} = \eff{\act} \cup \set{ \map{ \true }{\del \set{ \tmp } } }$

\item For the case of $\delta = \gact{ Q(\vec{x}) }{ \delta[\vec{x}] }$, we
  have:
  \[
  \tpick ( \gact{ Q(\vec{x}) }{ \delta[\vec{x}] }, Q_\pi(\vec{y}) ) =
  \delta_{pick}\ ;\ \tpick (\delta, Q_\pi(\vec{y}) \wedge
  Q_\pi'(\vec{x}))\ ;\ \delta_{del},
  \]
  where
  \begin{compactitem}
  \item
    $\delta_{pick} = \gact{ Q(\vec{x}) \wedge Q_\pi(\vec{y}) }{
      \act_{pick}(\vec{x})}$, where
    \begin{compactitem}
    \item for each $x \in \vec{x}$, we have
      $\set{\map{\true}{\add \set{V^Q_{x}(x)} } } \in
      \eff{\act_{pick}}$, 
    \item $V^Q_{x}$ is a fresh concept name.
    \item
      $\set{\map{\true}{\add \set{\tmp} } } \in
      \eff{\act_{pick}}$
   \end{compactitem}

  \item $\delta_{del} = \gact{\true}{ \act_{del}() }$, where
    \begin{compactitem}
    \item for each $x \in \vec{x}$, we have
      $\set{\map{V^Q_{x}(x)}{\del \set{V^Q_{x}(x)} } } \in
      \eff{\act_{del}}$, and
    \item
      $\set{\map{\true}{\add \set{\tmp} } } \in
      \eff{\act_{del}}$
    \end{compactitem}
    
  \item $Q_\pi'(\vec{x}) = \bigwedge_{x \in \vec{x}} V^Q_{x}(x)$

  \end{compactitem}

\item For the case of $\delta = \delta_1 | \delta_2$, we
  have:
  \[
\tpick (\delta_1 | \delta_2, Q_\pi(\vec{y}) ) = 
\tpick (\delta_1, Q_\pi(\vec{y}) )\ |\ 
\tpick (\delta_2, Q_\pi(\vec{y}) ),
\]

\item For the case of $\delta = \delta_1 ; \delta_2$, we
  have:
\[  
\tpick (\delta_1 ; \delta_2, Q_\pi(\vec{y}) ) = 
\tpick (\delta_1, Q_\pi(\vec{y}) )\ ;\ 
\tpick (\delta_2, Q_\pi(\vec{y}) ),
\]

\item For the case of $\delta = \gif{\varphi}{\delta_1}{\delta_2}$, we
  have:
\[
\tpick (\gif{\varphi}{\delta_1}{\delta_2}, Q_\pi(\vec{y}) ) =
\gif{\varphi}{\tpick (\delta_1 , Q_\pi(\vec{y}) )}{\tpick (\delta_2 ,
  Q_\pi(\vec{y}) )}
\]

\item For the case of $\delta = \gwhile{\varphi}{\delta}$, we
  have:
  \[
  \tpick (\gwhile{\varphi}{\delta}, Q_\pi(\vec{y}) ) =
  \gwhile{\varphi}{\tpick (\delta , Q_\pi(\vec{y}) )}
  \]

\end{enumerate}

Hence, given a GKAB $\gkabsym = \tup{T, \initabox, \actset, \delta}$
where $\delta$ might uses the pick construct that ranges over a
program, we can translate it into a GKAB
$\gkabsym' = \tup{T, \initabox, \actset, \delta'}$ where
$\delta' = \tpick(\delta, \true)$ and each pick construct in $\delta'$
ranges over a single action.


Concerning Golog procedures, first of all, it is easy to see that
GKABs can simulate non-recursive procedures. The idea is to simply
unfold each procedure call with its definition, until all procedure
calls have been removed from the program.
However, it is also possible to show that GKABs can also simulate
arbitrary recursive Golog procedures, by explicitly implementing the
stack used by the recursion, and making use of loops. This is fully in
line with the standard results in computation and programming language
theory showing that recursion can be encoded by means of while loops
(see, e.g., \cite{LS99,AK82,HK92}). The details of this translation
still need to be worked out, but one can draw inspiration for this
from work that establishes a generic translation from Golog programs
(possibly containing recursive procedures) into basic action theories
\cite{FBM08}.  Notice that, due to the need to explicitly represent
and maintain the stack of recursive calls, even when the GKAB is
run-bounded, the resulting KAB might not be so.

%% file: 2.chapters/5-ia-gkab.tex
\chapter{Inconsistency-Aware Golog-KAB\lowercase{s}
  (I-GKAB\lowercase{s})}\label{ch:ia-gkab}

\ifhidecontent
 
\fi

We have seen KABs in \Cref{ch:kab} as well as its extension to GKABs
in \Cref{ch:gkab} which
%
%
provide a sophisticated framework that captures the evolution of KBs
by actions. However so far they treat inconsistency in a simplistic
way. Basically, they handle inconsistency by rejecting inconsistent
states produced through action execution. In general, this is not
satisfactory, since the inconsistency may affect just a small portion
of the entire KB, and should be treated in a more careful way.
%



As a motivating example, consider our \Cref{ex:gkab-execution}. Recall
that we have a state $s_1$ containing an ABox
\begin{flushleft}
$
\begin{array}{l@{}ll}
\ A_1 = &\set{ & \exo{ReceivedOrder}( \excon{chair} ), \exo{ApprovedOrder}(
  \excon{table} ),
  \exo{designedBy}( \excon{table}, \excon{alice} ), \\
  &&\exo{Designer}( \excon{alice} ), \exo{hasDesign}( \excon{table} , \excon{ecodesign} ),\\
  &&\exo{hasAssemblingLoc}( \excon{table}, \excon{bolzano} ) \ \ }.
\end{array}
$
\end{flushleft}
As explained in \Cref{ex:gkab-execution}, one possible sucessor of
state $s_1$ is the state $s_2$ that contains a $T$-inconsistent ABox
$A_2$ (note that $T$ is specified in \Cref{ex:tbox-and-abox}), and
\begin{flushleft}
$\begin{array}{l@{}ll}
  A_2 = &\set{ & \exo{ReceivedOrder}( \excon{chair} ), 
                           \exo{designedBy}( \excon{table}, \excon{alice} ), 
                 {\underline{\exo{ Designer } ( \excon{alice} ) }},\\
                  && \exo{hasDesign}( \excon{table} , \excon{ecodesign} ), 
                     {\underline{\exo{ hasAssemblingLoc }(  \excon{table}, \excon{bolzano} )} }, \\
                  &&\exo{AssembledOrder}(\excon{table}),  \exo{assembledBy}(
                     \excon{table},\excon{alice}  ), \\
                  &&{\underline{\exo{ Assembler }( \excon{alice} )}}, 
                  {\underline{\exo{hasAssemblingLoc}( \excon{table}, \excon{trento} )}}  \ \ }. \\
\end{array}$
\end{flushleft}
Basically, the ABox $A_2$ is obtained from the execution of action
$\exa{assembleOrders}/0$. 
This action execution introduces the assertions
$\exo{Assembler}( \excon{alice} )$ and
$\exo{hasAssemblingLoc}( \excon{table}, \excon{trento})$ that,
together with another assertions (see the underlined assertions),
cause an inconsistency due to the violation of TBox assertions
$\exo{Designer} \sqsubseteq \neg \exo{Assembler}$ and
$\funct{\exo{hasAssemblingLoc}}$.
%
For the violation of
$\exo{Designer} \sqsubseteq \neg \exo{Assembler}$, notice that the
assembler and the designer are assigned from outside of our system
(e.g., other department), and the information is brought into the
system by calls to external services (i.e.,
$\exs{getAssembler}(\excon{table})$ and
$\exs{getAssemblingLoc}(\excon{table})$). Hence, we do not have
control over the employee task assignment that is done outside of our
system. It could also be the case that by the time of assigning the
assembler, $\excon{alice}$ has been moved from the design department
into the assembling department, or it could also be the case that
$\excon{alice}$ is replacing her friend who works as an assembler.
Thus, in this situation, it might be desirable to repair the
inconsistency (e.g., by removing either
$\exo{Designer}(\excon{alice})$ or $\exo{Assembler}(\excon{alice})$)
instead of rejecting the inconsistent state and block the system
evolution.
For the violation of $\funct{\exo{hasAssemblingLoc}}$, observe that
the assertion $\exo{hasAssemblingLoc}(\excon{table}, \excon{bolzano})$
is introduced by the action $\exa{prepareOrders}/0$ and the value for
the assembling location is obtained from the service call
$\exs{assignAssemblingLoc}(\excon{table})$. Therefore, it could be the
case that during the preparation phase, the assembling location is
assigned to $\excon{bolzano}$ but in the reality, due to some
unpredicted situation, the assembling is done in $\excon{trento}$ or
there is just an error, and we do not have any control over it. Thus,
it is also desirable in this situation to repair the inconsistency
(e.g., by either removing
$\exo{hasAssemblingLoc}(\excon{table}, \excon{bolzano})$ or
$\exo{hasAssemblingLoc}(\excon{table}, \excon{trento})$) instead of
rejecting the inconsistent state and block the system.


Starting from all of these observation, here we leverage on the
research about instance-level evolution of knowledge bases
\cite{Wins90,EiGo92,FMKPA08,CKNZ10b}, and, in particular, on the
notion of knowledge base repair~\cite{LLRRS10,LLRRS11}, in order to
make GKABs inconsistency-aware.
In particular we exploit filter relations in GKABs to define three
inconsistency-aware semantics that incorporate the different
repair-based approaches in order to handle the inconsistencies.

As in KABs and GKABs, in the following we use \dllitea for expressing
KBs and we also do not distinguish between objects and values (thus we
drop attributes).
Moreover we make use of a countably infinite set $\const$ of
constants, which intuitively denotes all possible values in the
system.
Additionally, we consider a finite set of distinguished constants
$\iconst \subset \const$, and
a finite set $\servcall$ of \textit{function symbols} that represents
\textit{service calls}, which abstractly account for the injection of
fresh values (constants) from $\const$ into the system.

The corresponding publications of the results presented in this
chapter are
\cite{AS-CORR-15,AS-IJCAI-15,AS-DL-15,AS-IJCAI-13,AS-CORR-13a,AS-DL-13}.

\section{Inconsistency Management in DL KBs}
\label{sec:inconsistency-management-dl}
We open this chapter by elaborating some inconsistency management
approaches in DL KBs. Basically, retrieving certain answers from a KB
makes sense only if the KB is consistent: if it is not, then each
query returns all possible tuples of constants of the ABox. In a
dynamic setting where the ABox evolves over time, consistency is a too
strong requirement, and in fact a number of approaches have been
proposed to handle the instance-level evolution of KBs, managing
inconsistency when it arises. Such approaches typically follow one of
the two following two strategies:
\begin{compactenum}
\item inconsistencies are kept in the KBs, but the semantics of query
  answering is refined to take this into account (\emph{consistent
    query answering}~\cite{Bert06});
\item the extensional part of an inconsistent KB is (minimally)
  \emph{repaired} so as to remove inconsistencies, and certain answers
  are then applied over the curated KB.
\end{compactenum}

Here, we follow the approach
that 
focuses on repair-based approaches.
We then recall the basic notions related to inconsistency management
via repair, distinguishing approaches that repair an ABox and those
that repair an update.

\subsection{ABox repairs}
Starting from the seminal work in
\cite{EiGo92}, in~\cite{LLRRS10} two approaches for repairing KBs are
proposed: \emph{ABox repair} (AR) and \emph{intersection ABox repair}
(IAR). 
Here we use these approaches 
to handle inconsistency in KABs, and are respectively called
\emph{bold-repair} (\emph{b-repair}) and \emph{certain-repair}
(\emph{c-repair}).

\begin{definition}[Bold-repair]
  Given \sidetext{B-repair} an ABox $A$ and a TBox $T$, a \emph{b-repair of an ABox $A$
    w.r.t.\ $T$} is an ABox $A'$ such that
  \begin{compactenum}
  \item $A' \subseteq A$,
  \item $A'$ is $T$-consistent, and
  \item there does not exists $A''$ such that
    $A' \subset A'' \subseteq A$ and $A''$ is $T$-consistent.
  \end{compactenum}
  We also call $A'$ a \emph{maximal T-consistent subset of $A$}.
\end{definition}

\noindent
We denote by $\arset{T,A}$ the \emph{set of all b-repairs} of $A$ w.r.t.\
$T$.

\begin{definition}[Certain-repair]\label{def:c-rep}
  Given \sidetext{C-repair} an ABox $A$ and a TBox $T$, a
  \emph{c-repair of $A$ w.r.t.\ $T$} is the (unique) set
  $\iarset{T,A} = \cap_{A_i \in\arset{T,A}} A_i$ of ABox assertions,
  obtained by intersecting all b-repairs of $A$ w.r.t.\ $T$.
\end{definition}


\begin{example}\label{ex:b-rep-and-c-rep}
  Continuing \Cref{ex:gkab-execution}, recall that we have an
  inconsistent ABox $A_2$ w.r.t.\ TBox $T$, where $T$ is specified in
  \Cref{ex:tbox-and-abox}, and
  \begin{flushleft}
    $\begin{array}{l@{}ll} A_2 = &\set{ & \exo{ReceivedOrder}( \excon{chair}
        ), \exo{designedBy}( \excon{table}, \excon{alice} ),
                                          {\underline{\exo{ Designer } ( \excon{alice} ) }},\\
                                 && \exo{hasDesign}( \excon{table} ,
                                    \excon{ecodesign} ),
                                    {\underline{\exo{ hasAssemblingLoc }(  \excon{table}, \excon{bolzano} )} }, \\
                                 &&\exo{AssembledOrder}(\excon{table}),
                                    \exo{assembledBy}(
                                    \excon{table},\excon{alice}  ), \\
                                 &&{\underline{\exo{ Assembler }(
                                    \excon{alice} )}},
                                    {\underline{\exo{hasAssemblingLoc}( \excon{table}, \excon{trento} )}}  \ \ }. \\
     \end{array}$
   \end{flushleft}
   \medskip
   \textbf{Example of Bold-repair.}\xspace\\
   In this case, we have
   $\arset{T,A_2} = \set{A_2^1, A_2^2, A_2^3, A_2^4}$, where
   \begin{flushleft}
    $\begin{array}{l@{}ll} A_2^1 = &\set{ & \exo{ReceivedOrder}( \excon{chair}
        ), \exo{designedBy}( \excon{table}, \excon{alice} ), \\
                                 && \exo{hasDesign}( \excon{table} ,
                                    \excon{ecodesign} ), \exo{AssembledOrder}(\excon{table}),
                                    \\
                                 &&\exo{assembledBy}(
                                    \excon{table},\excon{alice}  ), {\underline{\exo{ Assembler }(
                                    \excon{alice} )}}, \\
                                    &&{\underline{\exo{hasAssemblingLoc}( \excon{table}, \excon{trento} )}}  \ \ }. \\
     \end{array}$
   \end{flushleft}
   \begin{flushleft}
    $\begin{array}{l@{}ll} A_2^2 = &\set{ & \exo{ReceivedOrder}( \excon{chair}
        ), \exo{designedBy}( \excon{table}, \excon{alice} ),
                                            {\underline{\exo{ Designer } ( \excon{alice} ) }},\\
                                   && \exo{hasDesign}( \excon{table} ,
                                      \excon{ecodesign} ), \exo{AssembledOrder}(\excon{table}),\\
                                   && \exo{assembledBy}( \excon{table},\excon{alice} ),
                                      {\underline{\exo{hasAssemblingLoc}( \excon{table}, \excon{trento} )}}  \ \ }. \\
     \end{array}$
   \end{flushleft}
   \begin{flushleft}
     $\begin{array}{l@{}ll} A_2^3 = &\set{ & \exo{ReceivedOrder}(
         \excon{chair}
                                             ), \exo{designedBy}( \excon{table}, \excon{alice} ), \\
                                    &&\exo{hasDesign}( \excon{table} ,
                                       \excon{ecodesign} ),
                                       {\underline{\exo{
                                       hasAssemblingLoc }(
                                       \excon{table}, \excon{bolzano}
                                       )} }, \\
                                    &&\exo{AssembledOrder}(\excon{table}),
                                       \exo{assembledBy}(
                                       \excon{table},\excon{alice}  ), \\
                                    &&{\underline{\exo{ Assembler }(
                                       \excon{alice} )}}            \ \ }. \\
     \end{array}$
   \end{flushleft}
   \begin{flushleft}
    $\begin{array}{l@{}ll} A_2^4 = &\set{ & \exo{ReceivedOrder}( \excon{chair}
        ), \exo{designedBy}( \excon{table}, \excon{alice} ),
                                          {\underline{\exo{ Designer } ( \excon{alice} ) }},\\
                                 && \exo{hasDesign}( \excon{table} ,
                                    \excon{ecodesign} ),
                                    {\underline{\exo{ hasAssemblingLoc }(  \excon{table}, \excon{bolzano} )} }, \\
                                 &&\exo{AssembledOrder}(\excon{table}),
                                    \exo{assembledBy}(
                                    \excon{table},\excon{alice}  )  \ \ }. \\
     \end{array}$
   \end{flushleft}

   In this scenario, a plausible justification when we drop
   $\exo{Designer}(\excon{alice})$ and keep
   $\exo{Assembler}(\excon{alice})$ (i.e., $A_2^1$ and $A_2^3$) is
   that it could be the case that $\excon{alice}$ is just moved from
   the Design Department into the Assembling Department. For the other
   way around, a possible explanation of dropping
   $\exo{Assembler}(\excon{alice})$ and keeping
   $\exo{Designer}(\excon{alice})$ (i.e., $A_2^2$ and $A_2^4$) is
   because it could be the case that due to some exceptional condition
   $\excon{alice}$ is just substituting her friend, but she is still a
   designer.
   Moreover, a possible justification of dropping
   $\exo{hasAssemblingLoc}(\excon{table}, \excon{bolzano})$ and
   keeping $\exo{hasAssemblingLoc}(\excon{table}, \excon{trento})$
   (i.e., $A_2^1$ and $A_2^2$) is that it could be the case there are
   some exceptional conditions (e.g., some disasters) such that the
   assembling place must be changed from the one that has been
   planned. On the other hand, a plausible explanation when we drop
   $\exo{hasAssemblingLoc}(\excon{table}, \excon{trento})$ and keep
   $\exo{hasAssemblingLoc}(\excon{table}, \excon{bolzano})$ (i.e.,
   $A_2^3$ and $A_2^4$) is that there is just a mistake in recording
   the assembling place but the reality is still aligned as it is
   planned.

   \medskip
   \noindent
   \textbf{Example of Certain-repair.}\xspace\\
   Regarding c-repair of $A_2$, we have the following
  \begin{flushleft}
    $\begin{array}{ll@{ \ }l@{}l} 
       \iarset{T,A_2} &= &\cap&_{A_i
                               \in\arset{T,A_2}} A_i \\
       &=&& A_2^1 \cap
           A_2^2 \cap A_2^3 \cap A_2^4\\
       &= &\set{ & \exo{ReceivedOrder}( \excon{chair}
                                 ), \exo{designedBy}( \excon{table}, \excon{alice} ),
       \\
                        &&& \exo{hasDesign}( \excon{table} ,
                           \excon{ecodesign} ),
                           \exo{AssembledOrder}(\excon{table}), \\
                        &&&\exo{assembledBy}(
                           \excon{table},\excon{alice}  )         \ \ }. \\
     \end{array}$
   \end{flushleft}
%
%
%
   The intuition of this approach is that we only keep those facts
   that are certainly correct (i.e., do not involve in any inconsistency).
\end{example}

\subsection{Inconsistency in KB evolution}
In a setting where the KB is subject to instance-level evolution, b-
and c-repairs are computed agnostically from the updates: each update
is committed, and only secondly the obtained ABox is repaired if
inconsistent. In~\cite{CKNZ10b}, a so-called \emph{bold semantics} is
proposed to apply the notion of repair to the update itself.
Specifically, the bold semantics is defined over a consistent KB
$\tup{T, A}$ and an instance-level update that comprises two ABoxes
$F^-$ and $F^+$, respectively containing those assertions that have to
be deleted from and then added to $A$. It is assumed that $F^+$ is
$T$-consistent, and that new assertions have ``priority'': if an
inconsistency arises, newly introduced assertions are preferred to
those already present in $A$.
%

\begin{definition}[Bold-Evolution of an ABox]\label{def:bold-evol-abox-icma}
  Let \sidetext{Bold-Evolution} $T$ be a TBox, $A$ an ABox, $\facta$ a
  set of ABox assertions to be added, and $\factd$ a set of ABox
  assertions to be deleted. An \emph{evolution of an ABox $A$ w.r.t.\
    $T$ by $\facta$ and $\factd$}, written $\evol(T, A, \facta, \factd)$, is
  an ABox $A_e = \facta \cup A'$, where
\begin{compactenum}
\item $A' \subseteq (A \setminus \factd)$,
\item $\facta \cup A'$ is $T$-consistent, and
\item there does not exists $A''$ such that
  $A' \subset A'' \subseteq (A \setminus \factd)$ and
  $\facta \cup A''$ is $T$-consistent.
\end{compactenum}
\ \ 
\end{definition}
%
%

\begin{example}\label{ex:b-evol}
  Consider our \Cref{ex:gkab-execution}. Recall that we have an ABox
\begin{flushleft}
$
\begin{array}{l@{}ll}
\ A_1 = &\set{ & \exo{ReceivedOrder}( \excon{chair} ), \exo{ApprovedOrder}(
  \excon{table} ),
  \exo{designedBy}( \excon{table}, \excon{alice} ), \\
  &&\exo{Designer}( \excon{alice} ), \exo{hasDesign}( \excon{table} , \excon{ecodesign} ),\\
  &&\exo{hasAssemblingLoc}( \excon{table}, \excon{bolzano} ) \ \ }.
\end{array}
$
\end{flushleft}
Consider that we have 
\begin{flushleft}
  $\begin{array}{l@{}l} 
     \facta = \set{
      &\exo{AssembledOrder}(\excon{table}), \exo{assembledBy}(
      \excon{table}, \excon{alice}
        ), \exo{Assembler}(\excon{alice}), \\
      &\exo{hasAssemblingLoc}( \excon{table},
        \excon{trento} ) \ }
\end{array}
$
\end{flushleft}
\begin{flushleft}
  $\begin{array}{l@{}l} 
     \factd = \set{ &\exo{ApprovedOrder}(\excon{table})}
\end{array}
$
\end{flushleft}
we then have the following
\begin{flushleft}
  $\begin{array}{l@{}l@{}l} \evol(T, A_1, F^+, F^-) = &\set{ &
      \exo{ReceivedOrder}( \excon{chair}
                                                               ), \exo{designedBy}( \excon{table}, \excon{alice} ), \\
                                                      &&
                                                         \exo{hasDesign}(
                                                         \excon{table}
                                                         ,
                                                         \excon{ecodesign} ), \\
                                                      &&\exo{AssembledOrder}(\excon{table}),
                                                         \exo{assembledBy}(
                                                         \excon{table},\excon{alice}  ), \\
                                                      &&{\underline{\exo{
                                                         Assembler }(
                                                         \excon{alice}
                                                         )}},
                                                         {\underline{\exo{hasAssemblingLoc}( \excon{table}, \excon{trento} )}}  \ \ }. \\
     \end{array}$
   \end{flushleft}
   Since $\exo{ Assembler }( \excon{alice} )$ and
   $\exo{hasAssemblingLoc}( \excon{table}, \excon{trento} )$ are new
   facts, when an inconsistency arises after adding these new facts,
   we keep these two facts and throw away the other facts that
   together with these two facts cause an inconsistency.

   As an intuition, this repair mechanism assumes that the new facts
   are more correct or reliable. Thus we keep them and throw the old
   ones. In this scenario, it could be the case that $\excon{alice}$
   has been moved from the Design Department to the Assembling
   Department. Therefore, the new fact that $\excon{alice}$ is an
   assembler is correct and the fact that $\excon{alice}$ is a
   designer is obsolete but the system just have not throw it
   away. Similarly, regarding the fact that the assembling is
   performed in $\excon{trento}$, it could be the case that due to
   some reasons, they change the assembling place from the one that
   has been planned.
\end{example}

\section{Inconsistency-Aware Semantics for GKAB.}\label{sec:ia-gkab}
Given a GKAB $\gkabsym = \tup{T, \initabox, \actset, \ginitprog}$, we
exploit filter relations to define
three inconsistency-aware semantics that incorporate the repair-based
approaches reviewed in \Cref{sec:inconsistency-management-dl}. 
In particular, we introduce 3 filter relations 
$\filter_B$, $\filter_C$, and $\filter_E$, as follows. 

\begin{definition}[B-repair Filter $\filter_B$]\label{def:b-rep-filter}
  A \sidetext{B-repair Filter $\filter_B$} \emph{B-repair Filter
    $\filter_B$} is a relation that consists of tuples of the form
  $\tup{A, \facta, \factd, A'}$ such that
  $A' \in \arset{T, (A \setminus \factd) \cup \facta}$, where $A$ and
  $A'$ are ABoxes, and $\facta$ as well as $\factd$ are two sets of
  ABox assertions.
\end{definition}


\noindent
Filter $\filter_B$ gives rise to the \emph{b-repair execution semantics} for
GKABs, 
where inconsistent ABoxes are repaired by non-deterministically
picking a b-repair. Precisely, transition systems which provide the
b-repair execution semantics for GKABs is defined as follows.

\begin{definition}[GKAB B-Transition System]
  Given \sidetext{GKAB B-Transition System} a GKAB
  $\gkabsym$ 
  and a b-repair filter $\filter_B$, the \emph{b-transition system of
    $\gkabsym$}, written $\ts{\gkabsym}^{\filter_B}$, is the
  transition system of $\gkabsym$ w.r.t.\ $\filter_B$.
\end{definition}

\noindent
We call \emph{B-GKABs} the GKABs adopting this semantics.

\begin{example}\label{ex:bgkab-execution}
  As an example for B-GKABs, let
  $\gkabsym = \tup{T, \initabox, \actset, \delta}$ be a B-GKAB where
  $T$, $\initabox$, $\actset$, and $\delta$ are as in
  \Cref{ex:gkab-run-ex}. Similar to \Cref{ex:gkab-execution},
  executing $\gkabsym$ starting from $s_0$, we have
  $s_1 = \tup{A_1, \scmap_1, \delta'}$ as a reachable state from
  $s_0$, where
  \begin{flushleft}
    $\begin{array}{l@{}ll} 
       \bullet \ A_1 = &\set{ & \exo{ReceivedOrder}(
        \excon{chair} ), \exo{ApprovedOrder}( \excon{table} ),
                                                    \exo{designedBy}( \excon{table}, \excon{alice} ), \\
                                           &&\exo{Designer}( \excon{alice} ), \exo{hasDesign}( \excon{table} , \excon{ecodesign} ),\\
                                           &&\exo{hasAssemblingLoc}( \excon{table}, \excon{bolzano} ) \ \ }, 
     \end{array}$
   \end{flushleft}
  \begin{flushleft}
    $\begin{array}{l@{}ll} 
       \bullet \  \scmap_1 =& \set{ &[\exs{getDesigner}(\excon{table}) \ra
                                      \excon{alice}], [\exs{getDesign}(\excon{table}) \ra \excon{ecodesign}], \\
                                           &&[\exs{assignAssemblingLoc}(\excon{table}) \ra
                                              \excon{bolzano}]  \ \ },
     \end{array}$ 
   \end{flushleft}
   \begin{flushleft}
     $     
     \begin{array}{l}
       \bullet \  \delta' = \delta_3 ; \delta_4 ; \delta_5 ; \gwhile{ [\exists
       \exvar{x}.\exo{Order}(\exvar{x})] }{\delta_0}.
     \end{array}
     $
   \end{flushleft}
   As explained in \Cref{ex:gkab-execution}, the next step is to
   execute $\delta_3$ that is an action invocation of the form
   $\gact{\true}{\exa{assembleOrders}()}$ and it involves the service
   calls $\exs{getAssembler}/1$ and $\exs{getAssemblingLoc}/1$.
   Thus, it is easy to see that there are infinite successor states of
   $s_1$ each of the form $\tup{A_2', \scmap_2, \delta''}$ where
   \[
   \begin{array}{r@{}l}
     A_2' \in \arset{T, (A_1 \setminus
     \set{ \ &\exo{ApprovedOrder}(\excon{table})}) \cup \\
     \set{ \ &\exo{AssembledOrder}(\excon{table}), \\
          &\exo{assembledBy}( \excon{table},\exs{getAssembler}(\excon{table}) ), \\
          &\exo{Assembler}(\exs{getAssembler}(\excon{table})), \\
          &\exo{hasAssemblingLoc}( \excon{table},
            \exs{getAssemblingLoc}(\excon{table}) ) \ 
            }\\
     }
   \end{array}
   \]
   in which $\exs{getAssemblingLoc}(\excon{table})$ as well as
   $\exs{getAssembler}(\excon{table})$ are arbitrarily substituted
   with constants from $\const$ by a substitution
   $\theta \in \eval{\addfacts{T, A_1, \exa{assembleOrders}\sigma}}$,
   and $\scmap_2 = \scmap_1 \cup \theta$. Moreover, we have
   \[
   \delta'' = \delta_4 ; \delta_5 ; \gwhile{  \exists \exvar{x}.[\exo{Order}(\exvar{x})] \wedge
    \neg[\exo{DeliveredOrder}(\exvar{x})]   }{\delta_0}.
   \]

   Now, as an example of sucessor states of $s_1$, consider a possible substitution
   of $\exs{getAssemblingLoc}(\excon{table})$ into
   ``$\excon{trento}$'' and $\exs{getAssembler}(\excon{table})$ into
   ``$\excon{alice}$'' by a particular substitution
   $\theta \in \eval{\addfacts{T, A_1,
       \exa{assembleOrders}\sigma}}$. We then have states
\begin{compactitem}
\item $s_2^1~=~\tup{A_2^1, \scmap_2, \delta''}$ 
\item $s_2^2~=~\tup{A_2^2, \scmap_2, \delta''}$ 
\item $s_2^3~=~\tup{A_2^3, \scmap_2, \delta''}$ 
\item $s_2^4~=~\tup{A_2^4, \scmap_2, \delta''}$
\end{compactitem}
as some sucessor states of $s_1$ where
$\arset{T, A_2} = \set{A_2^1, A_2^2, A_2^3, A_2^4}$,
\begin{flushleft}
  $\begin{array}{l@{}ll} A_2 = &\set{ & \exo{ReceivedOrder}( \excon{chair} ),
      \exo{designedBy}( \excon{table}, \excon{alice} ),
                                        {\underline{\exo{ Designer } ( \excon{alice} ) }},\\
                               && \exo{hasDesign}( \excon{table} ,
                                  \excon{ecodesign} ),
                                  {\underline{\exo{ hasAssemblingLoc }(  \excon{table}, \excon{bolzano} )} }, \\
                               &&\exo{AssembledOrder}(\excon{table}),
                                  \exo{assembledBy}(
                                  \excon{table},\excon{alice}  ), \\
                               &&{\underline{\exo{ Assembler }(
                                  \excon{alice} )}},
                                  {\underline{\exo{hasAssemblingLoc}( \excon{table}, \excon{trento} )}}  \ \ }, \\
   \end{array}$
 \end{flushleft}
 and $A_2^1, A_2^2, A_2^3, A_2^4$ are the same as in the example of
 b-repair in
 \Cref{ex:b-rep-and-c-rep}. 
\end{example}

Now we proceed further to define the c-repair filter as follows.

\begin{definition}[C-repair Filter $\filter_C$]\label{def:c-rep-filter}
  A \sidetext{C-repair Filter $\filter_C$} \emph{C-repair Filter
    $\filter_C$} is a relation that consists of tuples of the form
  $\tup{A, \facta, \factd, A'}$ such that
  $A' = \iarset{T, (A \setminus \factd) \cup \facta}$, where $A$ and
  $A'$ are ABoxes, and $\facta$ as well as $\factd$ are two sets of
  ABox assertions.
\end{definition}


\noindent
Filter $f_C$ gives rise to the \emph{c-repair execution semantics} for
GKABs, 
where inconsistent ABoxes are repaired by computing their
unique c-repair.
The transition systems which provide the c-repair execution semantics
for GKABs is defined as follows.

\begin{definition}[GKAB C-Transition System]
  Given \sidetext{GKAB C-Transition System} a GKAB
  $\gkabsym$ 
  and a c-repair filter $\filter_C$, the \emph{c-transition system of
    $\gkabsym$}, written $\ts{\gkabsym}^{\filter_C}$, is the
  transition system of $\gkabsym$ w.r.t.\ $\filter_C$.
\end{definition}

\noindent
We call \emph{C-GKABs} the GKABs adopting this semantics. 

\begin{example}\label{ex:cgkab-execution}
  Similar to \Cref{ex:bgkab-execution}, let
  $\gkabsym = \tup{T, \initabox, \actset, \delta}$ be a C-GKAB where
  $T$, $\initabox$, $\actset$, and $\delta$ are as in
  \Cref{ex:gkab-run-ex}. Same as \Cref{ex:bgkab-execution},
  we have that $s_1 = \tup{A_1, \scmap_1, \delta'}$ is a reachable
  state from the initial state $s_0$ of $\gkabsym$, where $A_1$,
  $\scmap_1$, $\delta'$ are the same as in \Cref{ex:bgkab-execution}.
  The next step is to execute $\delta_3$ that is an action invocation
  of the form $\gact{\true}{\exa{assembleOrders}()}$ and it involves
  the service calls $\exs{getAssembler}/1$ and
  $\exs{getAssemblingLoc}/1$.
  Thus, it is easy to see that there are infinite successor states of
  $s_1$, each of the form $\tup{A_2', \scmap_2, \delta''}$ where
   \[
   \begin{array}{r@{}l}
     A_2' = \iarset{T, (A_1 \setminus
     \set{ \ &\exo{ApprovedOrder}(\excon{table})}) \cup \\
     \set{ \ &\exo{AssembledOrder}(\excon{table}), \\
          &\exo{assembledBy}( \excon{table},\exs{getAssembler}(\excon{table}) ), \\
          &\exo{Assembler}(\exs{getAssembler}(\excon{table})), \\
          &\exo{hasAssemblingLoc}( \excon{table},
            \exs{getAssemblingLoc}(\excon{table}) ) \ 
            }\\
     }
   \end{array}
   \]
   in which $\exs{getAssemblingLoc}(\excon{table})$ as well as
   $\exs{getAssembler}(\excon{table})$ are arbitrarily substituted
   with constants from $\const$ by a substitution
   $\theta \in \eval{\addfacts{T, A_1, \exa{assembleOrders}\sigma}}$,
   and $\scmap_2 = \scmap_1 \cup \theta$.  Moreover, we have
   \[
   \delta'' = \delta_4 ; \delta_5 ; \gwhile{    \exists \exvar{x}.[\exo{Order}(\exvar{x})] \wedge
    \neg[\exo{DeliveredOrder}(\exvar{x})]   }{\delta_0}.
   \]

   Now, as an example of a successor, consider a possible substitution
   of $\exs{getAssemblingLoc}(\excon{table})$ into
   ``$\excon{trento}$'' and $\exs{getAssembler}(\excon{table})$ into
   ``$\excon{alice}$'' by a particular substitution
   $\theta \in \eval{\addfacts{T, A_1, \exa{assembleOrders}\sigma}}$.
   We then have a state $s_2~=~\tup{A_2', \scmap_2, \delta''}$ where
   $A_2'$ is the same as $\iarset{T, A_2}$ in the example of c-repair
   in \Cref{ex:b-rep-and-c-rep}.
\end{example}

Next, we define the evolution filter as follows.

\begin{definition}[B-evol Filter $\filter_E$]\label{def:bevol-filter}
  A \sidetext{B-evol Filter $\filter_E$} \emph{B-evol Filter
    $\filter_E$} is a relation that consists of tuples of the form
  $\tup{A, \facta, \factd, A'}$ such that
  $A' = \evol(T, A, \facta, \factd)$ and $\facta$ is $T$-consistent,
  where $A$ and $A'$ are ABoxes, and $\facta$ as well as $\factd$ are
  two sets of ABox assertions.
\end{definition}


\noindent
Filter $f_E$ gives rise to the \emph{b-evol execution semantics} for
GKABs
where for updates leading to inconsistent
ABoxes, their unique bold-evolution is computed.  
The transition systems which provide the b-evol execution semantics
for GKABs is defined as follows.

\begin{definition}[GKAB E-Transition System]
  Given \sidetext{GKAB E-Transition System} a GKAB
  $\gkabsym$ 
  and a b-evol filter $\filter_E$, the \emph{e-transition system of
    $\gkabsym$}, written $\ts{\gkabsym}^{\filter_E}$, is the
  transition system of $\gkabsym$ w.r.t.\ $\filter_E$.
\end{definition}

\noindent
We call \emph{E-GKABs} the GKABs adopting this semantics.  
%
%
We group these three forms of GKABs (i.e., B-GKABs, C-GKABs, E-GKABs)
under the umbrella of \emph{inconsistency-aware GKABs (I-GKABs)}. The
definition of \muladom verification over I-GKABs is usual, i.e.,
similar to the case of KABs (see also \Cref{def:verification-kab}).

\begin{example}\label{ex:egkab-execution}
  Similar to \Cref{ex:bgkab-execution} and \Cref{ex:cgkab-execution},
  let $\gkabsym = \tup{T, \initabox, \actset, \delta}$ be an E-GKAB
  where $T$, $\initabox$, $\actset$, and $\delta$ are as in
  \Cref{ex:gkab-run-ex}. Same as
  \Cref{ex:bgkab-execution,ex:cgkab-execution},
  we have that $s_1 = \tup{A_1, \scmap_1, \delta'}$ is a reachable
  state from the initial state $s_0$ of $\gkabsym$, where $A_1$,
  $\scmap_1$, $\delta'$ are the same as in
  \Cref{ex:bgkab-execution,ex:cgkab-execution}.  The next step is to
  execute $\delta_3$ that is an action invocation of the form
  $\gact{\true}{\exa{assembleOrders}()}$ and it involves the service
  calls $\exs{getAssembler}/1$ and $\exs{getAssemblingLoc}/1$.
  It is easy to see that there are infinite successor states of $s_1$,
  each of the form $\tup{A_2', \scmap_2, \delta''}$ where
  $A_2' = \evol(T, A_1, F^+, F^-)$ with 
\begin{flushleft}
$
\begin{array}{l@{}ll}
\ A_1 = &\set{ & \exo{ReceivedOrder}( \excon{chair} ), \exo{ApprovedOrder}(
  \excon{table} ),
  \exo{designedBy}( \excon{table}, \excon{alice} ), \\
  &&\exo{Designer}( \excon{alice} ), \exo{hasDesign}( \excon{table} , \excon{ecodesign} ),\\
  &&\exo{hasAssemblingLoc}( \excon{table}, \excon{bolzano} ) \ \ }.
\end{array}
$
\end{flushleft}
\begin{flushleft}
$\begin{array}{l@{}l} 
   \facta = \set{
   &\exo{AssembledOrder}(\excon{table}), 
   \exo{assembledBy}( \excon{table},\exs{getAssembler}(\excon{table}) ), \\
   &\exo{Assembler}(\exs{getAssembler}(\excon{table})), \\
   &\exo{hasAssemblingLoc}( \excon{table},
     \exs{getAssemblingLoc}(\excon{table}) )
     }\\
\end{array}
$
\end{flushleft}
\begin{flushleft}
$\begin{array}{l@{}l} \factd = \set{
    &\exo{ApprovedOrder}(\excon{table})}
\end{array}
$
\end{flushleft}
in which $\exs{getAssemblingLoc}(\excon{table})$ as well as
$\exs{getAssembler}(\excon{table})$ are arbitrarily substituted with
constants from $\const$ by a substitution
$\theta \in \eval{\addfacts{T, A_1, \exa{assembleOrders}\sigma}}$, and
$\scmap_2 = \scmap_1 \cup \theta$ (note that $\facta$
and $\factd$ are the set of assertions to be added and deleted by
$\exa{assembleOrders}/0$).  Moreover, we have
\[
\delta'' = \delta_4 ; \delta_5 ; \gwhile{  \exists \exvar{x}.[\exo{Order}(\exvar{x})] \wedge
    \neg[\exo{DeliveredOrder}(\exvar{x})]  }{\delta_0}.
\]

Now, consider a possible substitution of
$\exs{getAssemblingLoc}(\excon{table})$ into ``$\excon{trento}$'' and
$\exs{getAssembler}(\excon{table})$ into ``$\excon{alice}$'' by a
particular substitution
$\theta \in \eval{\addfacts{T, A_1, \exa{assembleOrders}\sigma}}$.  We
then have a state $s_2~=~\tup{A_2', \scmap_2, \delta''}$ where $A_2'$
is the same as $\evol(T, A_1, F^+, F^-)$ in the example of
bold-evolution in \Cref{ex:b-evol}.
\end{example}

\section{Compilation of Inconsistency Management}
\label{sec:compilation-inconsistency-management}

In this section we show that all inconsistency-aware variants of GKABs
introduced in \Cref{sec:ia-gkab} can be reduced to S-GKABs. In
particular, we show that verification of \muladom formulas over
I-GKABs can be reduced to the verification of \muladom over S-GKABs.

Our general strategy is to show that S-GKABs are sufficiently
expressive to incorporate the repair-based approaches of
\Cref{sec:inconsistency-management-dl}, so that an action
executed under a certain inconsistency semantics can be compiled into
a Golog program that applies the action with the standard semantics,
and then explicitly handles the inconsistency, if needed. 

For compactness of presentation, we will use some abbreviations for
query atoms similar to \Cref{def:abbreviation-query} as follows.

\begin{definition}[Q-UNSAT-ECQ Query Abbreviations]\label{def:qunsat-ecq-abbreviation}
  We \sidetextb{Q-UNSAT-ECQ Query Abbreviations} define several
  abbreviations for ECQ queries as follows:
\[
\begin{array}{lll}
  \qunsatf(\funct{Z},x,y,z) &=& [Z(x,y)] \land [Z(x,z)] \land \neg [y = z];\\
  \qunsatn(B_1 \ISA \neg B_2, x) &=& [B_1(x)] \land [B_2(x)]; \\
  \qunsatn(R_1 \ISA \neg R_2,x,y) &=& [R_1(x,y)] \land [R_2(x,y)]; \\
\end{array}
\]
 \ \ 
\end{definition}
%
%
\noindent
Similar to the FOL query $\qunsatfol{T}$ (as in
\Cref{def:qunsat-fol}), we can define an ECQ $\qunsatecq{T}$
\sidetext{$\qunsatecq{T}$} for checking the satisfiability of a
\dllitea KB by making use the query abbreviations in
\Cref{def:qunsat-ecq-abbreviation} above.


\begin{theorem}[\cite{CDLLR13}]\label{thm:qunsat-ecq}
  Given a KB $\tup{T,A}$, we have $\Ans(\qunsatecq{T}, T, A) = \true$
  if and only if $A$ is $T$-inconsistent.
\qedw
\end{theorem}

\noindent
From now on,\sidetext{Abbreviations for ABox assertion} we also use
the following abbreviations to compactly express various ABox
assertions:
  \begin{compactitem}
  \item Let $B$ be a basic concept, an ABox assertion $B(c)$ denotes
    \begin{compactitem}
    \item $N(c)$ if $B = N$,
    \item $P(c,c')$ if $B = \exists P$,
    \item $P(c',c)$ if $B = \exists P^-$,
    \end{compactitem}
    where `$c'$' is a constant;
  \item Let $R$ be a basic role, an ABox assertion $R(c_1, c_2)$
    denotes
    \begin{compactitem}
    \item $P(c_1, c_2)$ if $R = P$,
    \item $P(c_2, c_1)$ if $R = P^-$.
    \end{compactitem}
  \end{compactitem}

As the last preliminaries before we proceed to compile each I-GKABs
into S-GKABs, we introduce some notions related to violations of
negative inclusion assertion (resp. functionality assertions).

\begin{definition}[Violation of a Negative Inclusion Assertion]
  Let \sidetextb{Violation of a Negative Inclusion Assertion}
  $\tup{T,A}$ be a KB, and $T \models B_1 \sqsubseteq \neg B_2$.  We
  say \emph{$B_1 \sqsubseteq \neg B_2$ is violated} if there exists a
  constant $c$ such that $\set{B_1(c), B_2(c)} \subseteq A$. In this
  situation, we also say that \emph{$B_1(c)$ (resp. $B_2(c)$) violates
    $B_1 \sqsubseteq \neg B_2$}. Similarly for roles.
\end{definition}

\begin{definition}[Violation of a Functionality Assertion]
  Let \sidetextb{Violation of a Functionality Assertion} $\tup{T,A}$
  be a KB, and $\funct{R} \in T$.  We say \emph{$\funct{R}$ is
    violated} if there exists constants $c, c_1, c_2$ such that
  $\set{R(c,c_1), R(c,c_2)} \subseteq A$ and $c_1 \neq c_2$. In this
  situation, we also say that \emph{$R(c,c_1)$ (resp. $R(c,c_2)$)
    violates $\funct{R}$}.
\end{definition}


For a technical reason, to encode I-GKABs into S-GKABs, we reserve a
special ABox assertion $\tmp$, where $\tmpconst \in \const_0$ and
$\tmpconceptname$ is a reserved 
concept name (i.e., outside of any TBox vocabulary). In particular, we
use $\tmp$ to distinguish \emph{stable} states, where an atomic action
can be applied, from intermediate states used by the S-GKABs to
(incrementally) remove inconsistent assertions from the ABox.
Stable/repair states are marked by the absence/presence of $\tmp$. As
in \Cref{sec:verification-sgkab}, here the ABox assertion $\tmp$ is
often also called special marker.

\subsection{From B-GKABs to Standard GKABs}\label{sec:BGKABToSGKAB}



As a preliminary towards defining the translation from B-GKAB to
S-KAB, we first define the notion of b-repair actions and b-repair
atomic action invocations which in the end will be used to form a
b-repair program. The main purpose of the b-repair program is to mimic
the computation of b-repair in B-GKAB and thus we can mimic the whole
computation in B-GKAB inside S-GKAB.

\begin{definition}[B-Repair Actions and Atomic Action Invocations]
  G\sidetext{B-Repair Actions and Atomic Action Invocations}iven a
  TBox $T$
  we define the \emph{set $\actset_b^T$ of b-repair actions} over $T$
  and the set $\setinvocation_b^T$ of \emph{b-repair atomic action
    invocations} over $T$ 
  as follows:

\begin{compactenum}
\item For each functionality assertion $\funct{R} \in T$, we include
  in $\actset_b^T$ and $\setinvocation_b^T$ respectively:
  \begin{compactitem}
  \item
    $\act_{F}(x,y):\set{\map{R(x,z) \wedge \neg [z = y]}{\del \set{R(
          x, z)}}} \in \actset_b^T $, and
  \item
    $\gact{\exists z.\qunsatf(\funct{R}, x, y, z)}{\act_{F}(x,y)} \in
    \setinvocation_b^T$.
  \end{compactitem}
  Essentially, the atomic action invocation and action above together
  repair an inconsistency related to $\funct{R}$ by removing all
  tuples causing the inconsistency, except one.

\item For each negative concept inclusion $B_1 \ISA \neg B_2$ such
  that $T \models B_1 \ISA \neg B_2$, we include in $\actset_b^T$ and
  $\setinvocation_b^T$ respectively:
  \begin{compactitem}
  \item
    $\act_{B_1}(x):\set{\map{B_1(x)}{\del \set{B_1(x)}}} \in
    \actset_b^T$,\footnote{Note:
      if $B_1 = \SOMET{P}$, then we have that $\act_{B_1}$ is of the
      form
      $\act_{B_1}(x):\set{\map{P(x,y)}{\del \set{P(x,y)}}} \in
      \actset_b^T$
      (Similarly for the case of $B_1 = \SOMET{P^-}$).} and
  \item
    $\gact{\qunsatn(B_1 \ISA \neg B_2, x)}{\act_{B_1}(x)} \in
    \setinvocation_b^T$.
  \end{compactitem}
  Basically, the atomic action invocation and action above together
  repair an inconsistency related to $B_1 \ISA \neg B_2$ by removing a
  constant that is both in $B_1$ and $B_2$ from $B_1$.

\item For each negative role inclusion $R_1 \ISA \neg R_2$ such that
  $T \models R_1 \ISA \neg R_2$, we include in $\actset_b^T$ and
  $\setinvocation_b^T$ respectively:
  \begin{compactitem}
  \item
    $\act_{R_1}(x,y):\set{\map{R_1(x, y)}{\del \set{R_1(x, y)}}} \in
    \actset_b^T$, and
  \item
    $\gact{\qunsatn(R_1 \ISA \neg R_2, x, y)}{\act_{R_1}(x,y)} \in
    \setinvocation_b^T$.
  \end{compactitem}
  The atomic action invocation and action above repair an
  inconsistency related to $R_1 \ISA \neg R_2$ by removing constants
  that are in both $R_1$ and $R_2$ from $R_1$.
\end{compactenum}
\ \ 
\end{definition}

The B-repair program is then defined below by employing the b-repair atomic
action invocations as well  as the b-repair actions.

\begin{definition}[B-Repair Program]\label{def:brepair-prog}
  Given \sidetext{B-Repair Program} a B-GKAB
  $\gkabsym = \tup{T, \initabox, \actset, \ginitprog}$. Let
  $\actset_b^T$ be a set of b-repair actions over $T$,
  and $\setinvocation_b^T = \set{a_1,\ldots,a_n}$ be a set of b-repair
  atomic action invocations 
  over $T$. 
  We then define the \emph{b-repair program over } $T$ 
  as follows:
\[
\delta_b^T =\gwhile{\qunsatecq{T}}{\delta_{r}}
\]
where $\delta_r = a_{1}|a_{2}|\ldots|a_n$. 
\end{definition}

Intuitively, a repair program $\delta_b^T$ iterates while the ABox is
inconsistent, and at each iteration, non-deterministically picks one
of the sources of inconsistency, and removes one or more assertions
causing it. Consequently, the loop is guaranteed to terminate, in a
state that corresponds to one of the b-repairs of the initial ABox.

We now define a translation function $\tgprogb$ which basically
concatenates each action invocation with a b-repair program in order
to simulate the action executions in B-GKABs.  Additionally, the
translation function $\tgprogb$ also serves as a one-to-one
correspondence (bijection) between the original and the translated
program (as well as between the sub-program).

\begin{definition}[Program Translation $\tgprogb$]
  Given \sidetextb{Program Translation $\tgprogb$} a B-GKAB
  $\gkabsym = \tup{T, \initabox, \actset,
    \ginitprog}$, 
  we define a \emph{translation $\tgprogb$} that translates a program
  $\delta$ into a program $\delta'$ inductively as follows:
\[
\begin{array}{@{}l@{}l@{}}
  \tgprogb(\gact{Q(\vec{p})}{\act(\vec{p})}) &=  
                                               \gact{Q(\vec{p})}{\act'(\vec{p})};\delta^T_b ; \gact{\true}{\act^-_{tmp}()}\\
  \tgprogb(\gemptyprog) &= \gemptyprog \\
  \tgprogb(\delta_1|\delta_2) &= \tgprogb(\delta_1)|\tgprogb(\delta_2) \\
  \tgprogb(\delta_1;\delta_2) &= \tgprogb(\delta_1);\tgprogb(\delta_2) \\
  \tgprogb(\gif{\varphi}{\delta_1}{\delta_2}) &= \gif{\varphi}{\tgprogb(\delta_1)}{\tgprogb(\delta_2)} \\
  \tgprogb(\gwhile{\varphi}{\delta}) &= \gwhile{\varphi}{\tgprogb(\delta)}
\end{array}
\]
  where
  \begin{compactitem}
  \item $\act^-_{temp}() :\set{\map{\true}{\del \set{\tmp}}}$,
  \item $\delta^T_b$ is b-repair program over $T$, 
  \item $\act'$ is an action obtained from
    $\act(\vec{p}):\set{e_1, \ldots, e_m}$ such that we have
    $\act'(\vec{p}):\set{e_1, \ldots, e_m, e_{temp}}$, where
    \[
    e_{temp} = \map{\true}{\add \set{\tmp}}.
    \]
  \end{compactitem}
\ \ 
\end{definition}
\noindent

Having all of the machinery above, we are ready to define a
translation $\tgkabb$ that, given a B-GKAB, produces an
S-GKAB as follows:

\begin{definition}[Translation from B-GKAB to S-GKAB]
  We \sidetext{Translation from B-GKAB to S-GKAB} define a translation
  $\tgkabb$ that, given a B-GKAB
  $\gkabsym = \tup{T, \initabox, \actset, \ginitprog}$, produces an
  S-GKAB
  $\tgkabb(\gkabsym) = \tup{T_p, \initabox, \actset' \cup \actset_b^T
    \cup \set{\act^-_{temp}},
    \ginitprog'}$,
  where 
\begin{compactitem}
\item $T_p$ is the positive inclusion assertions of $T$ (see \Cref{def:dllitea-tbox}),
\item $\actset'$ is obtained from $\actset$ such that for each
  $\act \in \actset$ of the form
  $\act(\vec{p}):\set{e_1, \ldots, e_m}$, we have $\act' \in \actset'$
  of the form $\act'(\vec{p}):\set{e_1, \ldots, e_m, e_{temp}}$, where
    \[
    e_{temp} = \map{\true}{\add \set{\tmp}}.
    \]
\item $\actset_b^T$ is a set of b-repair actions over $T$,
\item $\act^-_{temp}$ is an action of the form
  $\act^-_{temp}() :\set{\map{\true}{\del \set{\tmp}}}$,
\item $\ginitprog' = \tgprogb(\ginitprog)$.
\end{compactitem}
\ \ 
\end{definition}

In the translation above, only the positive inclusion assertions $T_p$
of the original TBox $T$ are maintained (guaranteeing that the S-GKAB
$\tgkabb(\gkabsym)$ never encounters inconsistency).
Moreover, the translation of the program $\tgprogb$ concatenates each
original action invocation with a corresponding ``repair''
phase. Obviously, this means that when an inconsistent ABox is
produced, a single transition in B-GKAB $\gkabsym$ corresponds to a
sequence of transitions in S-GKAB $\tgkabb(\gkabsym)$.
Hence, we need to translate the given \muladom formula $\Phi$ to be
verified over B-GKAB $\gkabsym$ 
%
into a corresponding formula over S-GKAB $\tgkabb(\gkabsym)$.
This is done by first obtaining formula $\Phi' = \nnf(\Phi)$, where
$\nnf(\Phi)$ denotes the \emph{negation normal form} of $\Phi$. Then
translating $\nnf(\Phi)$ using the translation $\tforj$ as in \Cref{def:tforj}.
%
%
Intuitively, $\tforj$ translates every sub-formula of $\Phi$ of the
form $\DIAM{\Psi}$ becomes
$\DIAM{\mu Z.((\tmp \wedge \DIAM{Z}) \vee (\neg\tmp \wedge
  \tforb(\Psi)))}$,
so as to translate a next-state condition over $\gkabsym$ into
reachability of the next stable state over
$\tgkabb(\gkabsym)$. Similarly for $\BOX{\Psi}$.

We will show later that $\ts{\gkabsym}^{\filter_B} \models \Phi$ if
and only if $\ts{\tgkabb(\gkabsym)}^{\filter_S} \models \tforj(\Phi)$.
Thus, we can show that the verification of \muladom over B-GKABs can
be reduced to the corresponding verification over S-GKABs.

\subsubsection{Termination and Correctness of B-repair Program}\label{sec:term-corr-brep-prog}

Towards recasting the \muladom verification over B-GKABs into S-GKABs,
in this section we show that the b-repair program is always terminate
and produces the same result as the result of b-repair over a
knowledge base. To this aim, we first need introduce some
preliminaries. As a start, we define the notion of a set of
inconsistent ABox assertions as follows.

\begin{definition}[Set of Inconsistent ABox Assertions]
  Given \sidetext{Set of Inconsistent ABox Assertions} a KB
  $\tup{T, A}$, we define the set $\inc(A)$ containing all ABox
  assertions that participate in the inconsistencies w.r.t.\ $T$ as
  the smallest set satisfying the following:
  \begin{compactenum}
  \item For each TBox assertion $B_1 \sqsubseteq \neg B_2$ such that
    $T \models B_1 \sqsubseteq \neg B_2$, we have
    $B_1(c) \in \inc(A)$, if $B_1(c) \in A$ and there exists
    $B_2(c) \in A$. 
%
  \item For each TBox assertion $R_1 \sqsubseteq \neg R_2$ such that
    $T \models R_1 \sqsubseteq \neg R_2$, we have
    $R_1(c_1, c_2) \in \inc(A)$ if $R_1(c_1, c_2) \in A$ and there
    exists $R_2(c_1, c_2) \in A$.
  \item For each functional assertion $\funct{R} \in T$, we have
    $R(c_1, c_2) \in \inc(A)$, if $R(c_1, c_2) \in A$ and there exists
    $R(c_1, c_3) \in A$ such that $c_2 \neq c_3$.
%
  \end{compactenum}
\ \ 
\end{definition}

\begin{lemma}\label{lem:card-inc-abox}
  Given a TBox $T$ and an ABox $A$, we have $\card{\inc(A)} = 0$ if
  and only if $A$ is $T$-consistent.
\end{lemma}
\begin{proof}
  Trivially follows from the definition. Since there is no ABox
  assertion violating any functionality or negative inclusion
  assertions.
\end{proof}


In the following, we show that an execution of b-repair action always
reduces the number of ABox assertions that participate in the
inconsistency (i.e., $\card{\inc(A)}$).


\begin{lemma}\label{lem:brepair-action-decrease}
  Let $\gkabsym = \tup{T, \initabox, \actset, \ginitprog}$ be a
  B-GKAB,
  $\tgkabb(\gkabsym)$ 
  be an S-GKAB $($with transition system
  $\ts{\tgkabb(\gkabsym)}^{\filter_S}$$)$ obtained from $\gkabsym$
  through $\tgkabb$, and $\actset_b^T$ be the set of b-repair action
  over $T$.
  Consider a $T$-inconsistent ABox $A$, a service call map $\scmap$,
  an arbitrary b-repair action $\act \in \actset_b^T$, and a legal
  parameter assignment $\sigma$ for $\act$. If
  $(\tup{A,\scmap}, \act\sigma, \tup{A', \scmap'}) \in
  \tell_{\filter_S}$, then $\card{\inc(A)} > \card{\inc(A')}$.
%
\end{lemma}
\begin{proof}
  Intuitively, the correctness of the claim can be seen by observing
  that each action in the set $\actset_b^T$
  of b-repair action over $T$
  only removes ABox assertions that participate in an
  inconsistency. 
%
%
We proof the claim by reasoning over all cases of b-repair actions as
follows:
\begin{compactitem}
\item[\textbf{Case 1:}] \textit{The actions obtained from functionality
    assertion $\funct{R} \in T_f$.} \\
  Let $\act_{F}$ be such action and has the following form:
  \[
  \act_{F}(x,y):\set{\map{R(x,z) \wedge \neg [z = y]}{\del \set{R(x,z)}}}.
  \]
  Suppose, $\act_{F}$ is executable in $A$ with a legal parameter
  assignment $\sigma$.  Since we have
  \[
  \gact{\exists z.\qunsatf(\funct{R}, x, y, z)}{\act_{F}(x,y)} \in
  \setinvocation_b^T,
  \]
  then there exists $c \in \adom{A}$ and
  $\set{c_1, c_2, c_3, \ldots, c_n} \subseteq \adom{A}$ such that
  $\set{R(c,c_1), R(c,c_2), \ldots, R(c,c_n)} \subseteq A$ where
  $n \geq 2$.  W.l.o.g.\ let $\sigma$ substitutes $x$ to $c$, and $y$
  to $c_1$, then we have
  $(\tup{A,\scmap}, \act\sigma, \tup{A', \scmap}) \in
  \tell_{\filter_S}$,
  where $A' = A \setminus \set{R(c,c_2), \ldots, R(c,c_n)}$. Therefore
  we have $\card{\inc(A)} > \card{\inc(A')}$.

\item[\textbf{Case 2:}] \textit{The actions obtained from negative
    concept $B_1 \ISA \neg B_2$ such that
    $T \models B_1 \ISA \neg B_2$.}  Let $\act_{B_1}$ be such action
  and has the following form:
  \[
  \act_{B_1}(x):\set{\map{B_1(x)}{\del \set{B_1(x)}}}.
  \]
  Suppose, $\act_{B_1}$ is executable in $A$ with a legal parameter
  $\sigma$.  Since we have
  \[
  \gact{\qunsatn(B_1 \ISA \neg B_2, x)}{\act_{B_1}(x)} \in
  \setinvocation_b^T,
  \]
  then there exists $c \in \adom{A}$ such that
  $\set{B_1(c), B_2(c)} \subseteq A$.  W.l.o.g.\ let $\sigma$
  substitutes $x$ to $c$, then we have
  $(\tup{A,\scmap}, \act\sigma, \tup{A', \scmap}) \in
  \tell_{\filter_S}$,
  where $A' = A \setminus \set{B_1(c)}$. Therefore we have
  $\card{\inc(A)} > \card{\inc(A')}$.

\item[\textbf{Case 3:}] \textit{The actions obtained from negative
    role
    inclusion $R_1 \ISA \neg R_2$ s.t.\ $T \models R_1 \ISA \neg
    R_2$.} 
  The proof is similar to the case 2.
\end{compactitem}

\end{proof}

As the next preliminaries, in the following we define the notion of a
program execution trace as well as the notion when such a trace is
called terminating. Moreover, we also define the notion of program
execution result in the case of terminating program execution trace.

\begin{definition}[Program Execution Trace]\label{def:prog-exec-trace}
  Let \sidetextb{Program Execution Trace}
  $\ts{\gkabsym}^{\filter} = \tup{\const, T, \stateset, s_0, \abox,
    \trans}$
  be the transition system of a GKAB
  $\gkabsym = \tup{T, \initabox, \actset, \ginitprog}$. Given a state
  $\tup{A_1, \scmap_1, \delta_1}$, a \emph{program execution trace
    $\pi$ induced by $\delta$ on $\tup{A_1, \scmap_1, \delta_1}$
    w.r.t.\ filter $\filter$} is a (possibly infinite) sequence of
  states of the form
\[
\pi = \tup{A_1, \scmap_1, \delta_1} \gexectrans \tup{A_2, \scmap_2,
  \delta_2} \gexectrans \tup{A_3, \scmap_3, \delta_3} \gexectrans
\cdots
\]
s.t.  $\tup{A_i, \scmap_i, \delta_i} \gprogtrans{\alpha_i\sigma_i,
  \filter} \tup{A_{i+1}, \scmap_{i+1}, \delta_{i+1}}$ for $i\geq1$.
\ \ 
\end{definition}

\begin{definition}[Terminating Program Execution Trace]\label{def:term-prog-exec-trace}
  Let \sidetextb{Terminating Program Execution Trace}
  $\ts{\gkabsym}^{\filter} = \tup{\const, T, \stateset, s_0, \abox,
    \trans}$
  be the transition system of a GKAB
  $\gkabsym = \tup{T, \initabox, \actset, \ginitprog}$. Given a state
  $\tup{A_1, \scmap_1, \delta_1}$, and a program execution trace $\pi$
induced by $\delta_1$ on $\tup{A_1, \scmap_1, \delta_1}$, we call
\emph{$\pi$ terminating} if
\begin{compactenum}[(1)]
\item $\tup{A_1, \scmap_1, \delta_1}$ is a final state, or
\item if $\tup{A_1, \scmap_1, \delta_1}$ is not a final state, then there
  exists a state $\tup{A_n, \scmap_n, \delta_n}$ s.t.\ we have the
  following finite program execution trace
\[
  \pi = 
  \tup{A_1,\scmap_1, \delta_1} \gexectrans \tup{A_2,\scmap_2,
    \delta_2} \gexectrans \cdots \gexectrans \tup{A_n, \scmap_n,
    \delta_n}.
\]
where $\tup{A_i, \scmap_i, \delta_i}$ (for
$i \in \set{1,\ldots, n-1}$) are not final states, and
$\tup{A_n, \scmap_n, \delta_n}$ is a final state.
\end{compactenum}
In the situation (1) (resp.\ (2)), we call the ABox $A_1$ (resp.\
$A_n$) \emph{the result of executing $\delta_1$ on
  $\tup{A_1, \scmap_1, \delta_1}$} w.r.t.\ filter
$\filter$. Additionally, we also say that $\pi$ is the \emph{program
  execution trace that produces $A_1$ (resp.\ $A_n$)}.
\end{definition}
We write $\progres(A_1, \scmap_1, \delta_1)$ to denote the set of all
ABoxes that is the result of executing $\delta_1$ on
$\tup{A_1, \scmap_1, \delta_1}$ w.r.t.\ filter $\filter$. Note that
given a state $\tup{A_1, \scmap_1, \delta_1}$, it is possible to have
several terminating program execution traces.
%
%
Intuitively, a program execution trace is a sequence of states which
captures the computation of the program as well as the evolution of
the system states by the program. Additionally, it is terminating if
at some point it reaches a final state.

We now proceed to show the termination of b-repair program as follows:

\begin{lemma}\label{lem:bprog-termination}
  Let $\gkabsym = \tup{T, \initabox, \actset, \ginitprog}$ be a
  B-GKAB,
  $\tgkabb(\gkabsym)$ 
  be an S-GKAB $($with transition system
  $\ts{\tgkabb(\gkabsym)}^{\filter_S}$$)$
  obtained from $\gkabsym$
  through $\tgkabb$,
  and $\delta^T_{b}$
  be a b-repair program over $T$.
  We have that $\delta^T_{b}$
  is always terminate. I.e., given a state $\tup{A,
    \scmap, \delta^T_{b}}$ of
  $\ts{\tgkabb(\gkabsym)}^{\filter_S}$,
  every program execution trace induced by $\delta^T_{b}$
  on $\tup{A,
    \scmap, \delta^T_{b}}$ w.r.t.\ filter $\filter_S$ is terminating.
%
%
\end{lemma}
\begin{proof}
  Roughly speaking, the claim is obtained due to the fact that at each
  step of the execution of b-repair program $\delta^T_{b}$,
  we have that the number of ABox assertions that cause inconsistency
  are always decreasing (cf.\ \Cref{lem:brepair-action-decrease}). Hence,
  since the execution of $\delta^T_{b}$
  never adds a new ABox assertion and there are only finitely many
  ABox assertions in the current ABox $A$,
  at some point the program $\delta^T_{b}$
  will be terminated (when there is no more ABox assertions that cause
  inconsistency).
%
%
  Technically, we show the claim by dividing the proof into two cases:

\smallskip
\noindent
\textbf{Case 1: $A$ is $T$-consistent. }\\
Trivially true, since $\ask(\qunsatecq{T}, T, A) = \false$, we have
$\tup{A, \scmap, \delta^T_b}$ is a final state, by the definition.

\smallskip
\noindent
\textbf{Case 2: $A$ is $T$-inconsistent. }\\
Given a state $\tup{A, \scmap, \delta^T_{b}}$ such that $A$ is
$T$-inconsistent, w.l.o.g.\ let
\[
\pi = \tup{A, \scmap, \delta^T_{b}} \gexectrans \tup{A_1, \scmap,
  \delta_1} \gexectrans \tup{A_2, \scmap, \delta_2} \gexectrans \cdots
\]
be an arbitrary program execution trace induced by $\delta^T_{b}$ on
$\tup{A, \scmap, \delta^T_{b}}$ w.r.t.\ filter $\filter_S$. Notice
that the service call map $\scmap$ always stay the same since every
b-repair action $\act \in \actset^T_b$ (which is the only action that
might appears in $\delta^T_b$) does not involve any service
calls. Now, we have to show that eventually there exists a state
$\tup{A_n, \scmap, \delta_n}$, such that
\[
\pi = \tup{A, \scmap, \delta^T_{b}} \gexectrans \tup{A_1, \scmap,
  \delta_1} \gexectrans \cdots \gexectrans \tup{A_n, \scmap, \delta_n}
\]
and $\tup{A_n, \scmap, \delta_n}$ is a final state. 
By \Cref{lem:brepair-action-decrease}, we have that
\[
\card{\inc(A)} > \card{\inc(A_1)} > \card{\inc(A_2)} > \cdots
\]
Additionally, due to the following facts:
\begin{compactenum}[(1)]
\item Since we assume that every concepts (resp.\ roles) are
  satisfiable, inconsistency can only be caused by
  \begin{compactenum}
  \item pair of assertions $B_1(c)$ and $B_2(c)$ (resp.\
    $R_1(c_1,c_2)$ and $R_2(c_1,c_2)$) that violate a negative
    inclusion assertion $B_1 \sqsubseteq \neg B_2$ (resp.\
    $R_1 \sqsubseteq \neg R_2$) such that \
    $T \models B_1 \sqsubseteq \neg B_2$ (resp.\
    $T \models R_1 \sqsubseteq \neg R_2$), or
  \item $n$-number role assertions
    \[
    R(c, c_1), R(c, c_2), \ldots, R(c, c_n)
    \]
    that violate a functionality assertion $\funct{R} \in T$.
  \end{compactenum}
\item To deal with both source of inconsistency in the point (1):
  \begin{compactenum}
  \item we consider all negative concept inclusions $B_1
    \sqsubseteq \neg B_2$
    such that $T \models B_1 \sqsubseteq \neg B_2$
    when constructing the b-repair actions $\actset^T_b$ (i.e., we
    saturate the negative inclusion assertions w.r.t.\ $T$ obtaining
    all derivable negative inclusion assertions from $T$). Moreover,
    for each negative concept inclusion $B_1 \sqsubseteq \neg B_2$
    such that $T \models B_1 \sqsubseteq \neg B_2$, we have an action
    which removes the ABox assertion $B_1(c)$ (for a certain constant
    $c$) in case $B_1 \sqsubseteq \neg B_2$ is violated. Similarly for
    negative role inclusions. 
    
%
  \item we consider all functionality assertions $\funct{R} \in T$
    when constructing the b-repair actions $\actset^T_b$, and each
    $\act_F \in \actset^T_b$ removes all role assertions that violates
    $\funct{R}$, except one.
  \end{compactenum}
\item Observe that $\ask(\qunsatecq{T}, T, A_n) = \true$ as long as
  $\card{\inc(A)} > 0$ (for any ABox $A$). Moreover, in such
  situation, by construction of $\setinvocation_b^T$, there always
  exists an executable action $\act \in \actset^T_b$ (Observe that
  $\qunsatecq{T}$ is a disjunction of every ECQ $Q$ that guard every
  corresponding atomic action invocation
  $\gact{Q(\vec{p})}{\act(\vec{p})} \in \setinvocation_b^T$ of each
  $\act \in \actset^T_b$ where each of its free variables are
  existentially quantified).
\end{compactenum}
As a consequence, eventually there exists $A_n$ such that
$\card{\inc(A_n)} = 0$. Hence by \Cref{lem:card-inc-abox} $A_n$ is
$T$-consistent. Therefore $\ask(\qunsatecq{T}, T, A_n) = \false$, and
$\tup{A_n, \scmap, \delta_n}$ is a final state.

\end{proof}

We now proceed to show the correctness of the b-repair program. I.e.,
showing that a b-repair program produces exactly the result of a
b-repair operation over the given (inconsistent) KB. As the first
step, we will show that every ABoxes produced by the b-repair program
is a maximal $T$-consistent subset of the given input ABox as
follows. 

\begin{lemma}\label{lem:characteristic-bprog-result}
  Let $\gkabsym = \tup{T, \initabox, \actset, \ginitprog}$ be a
  B-GKAB,
  $\tgkabb(\gkabsym)$ 
  be an S-GKAB $($with transition system
  $\ts{\tgkabb(\gkabsym)}^{\filter_S}$$)$
  obtained from $\gkabsym$
  through $\tgkabb$,
  and $\delta^T_{b}$
  be a b-repair program over $T$.
  Consider an ABox $A$,
  and a service call map $\scmap$.
  if $A'
  \in \progres(A, \scmap, \delta^T_b)$ then
  $A'$ is a maximal $T$-consistent subset of $A$.
%
\end{lemma}
\begin{proof} 
  Let $A' \in \progres(A, \scmap, \delta^T_b)$. We have to show that
\begin{compactenum}[(1)]
\item $A' \subseteq A$
\item $A'$ is $T$-consistent
\item There does not exists $A''$ such that $A' \subset A'' \subseteq
  A$ and $A''$ is $T$-consistent.
\end{compactenum}
We divide the proof into two cases:
\begin{compactenum}
\item[\textbf{Case 1: $A$ is $T$-consistent.}]~Trivially true, because
  $\ask(\qunsatecq{T}, T, A) = \false$, hence $\tup{A, \scmap,
    \delta^T_{b}}$ is a final state and $A \in \progres(A, \scmap,
  \delta^T_b)$. Thus, $A$ trivially satisfies the condition (1) - (3).

\item[\textbf{Case 2: $A$ is $T$-inconsistent.}]~Let
\[
\pi = \tup{A, \scmap, \delta^T_{b}} \gexectrans \tup{A_1, \scmap,
  \delta_1} \gexectrans \cdots \gexectrans \tup{A', \scmap, \delta'}
\]
be the corresponding program execution trace that produces $A'$ (This
trace should exists because $A' \in \progres(A, \scmap, \delta^T_b)$).

\smallskip
\noindent
\textbf{For condition (1).}~
Trivially true from the construction of b-repair program $\delta^T_b$.
Since, each step of the program always and only removes some ABox
assertions and also by recalling \Cref{lem:brepair-action-decrease}
that we have
\[
\card{\inc(A)} > \card{\inc(A_1)} > \card{\inc(A_2)} > \cdots
\]

\smallskip
\noindent
\textbf{For condition (2).}~Since the b-repair program $\delta^T_b$ is
terminated at a final state $\tup{A', \scmap, \delta'}$ where
$\ask(\qunsatecq{T}, T, A') = \false$, hence $A'$ is $T$-consistent.

\smallskip
\noindent
\textbf{For condition (3).}~Suppose by contradiction that there exists $A''$
s.t. $A' \subset A'' \subseteq A$ and $A''$ is $T$-consistent. Recall
that in \dllitea, since we assume that every concepts (resp.\ roles)
are satisfiable, inconsistency is only caused by
\begin{compactenum}[\it (i)]
\item pair of assertions $B_1(c)$ and $B_2(c)$ (resp.\ $R_1(c_1,c_2)$
  and $R_2(c_1,c_2)$) that violate a negative inclusion assertion $B_1
  \sqsubseteq \neg B_2$ (resp.\ $R_1 \sqsubseteq \neg R_2$) s.t.\ $T
  \models B_1 \sqsubseteq \neg B_2$ (resp.\ $T \models R_1 \sqsubseteq
  \neg R_2$), or
\item $n$-number role assertions 
  \[
  R(c, c_1), R(c, c_2), \ldots, R(c, c_n)
  \]
  that violate a functionality assertion $\funct{R} \in T$.
\end{compactenum}
However, by the construction of b-repair program $\delta^T_b$,
we have that each action $\act \in \actset^T_b$ is executable when
there is a corresponding inconsistency (detected by each guard $Q$ of
each corresponding atomic action invocation
$\gact{Q(\vec{p})}{\act(\vec{p})} \in \setinvocation_b^T$) and each
action only either
\begin{compactenum}[\it (i)]
\item removes one of the pair of assertions that violate a negative
  inclusion assertion, or
\item removes $n-1$ role assertions among $n$ role assertions that
  violate a functionality assertion.
\end{compactenum}
Hence, if $A''$ exists, then there exists an ABox assertion that
should not be removed, but then $A''$ is $T$-inconsistent. Thus, we
have a contradiction. Hence, there does not exists $A''$ such that
$A' \subset A'' \subseteq A$ and $A''$ is $T$-consistent.
\end{compactenum}
\ \ 
\end{proof}

From \Cref{lem:characteristic-bprog-result}, we can show that
every ABox that is produced by b-repair program is in the set of
b-repair of the given (inconsistent) KB. Formally it is stated below:

\begin{lemma}\label{lem:brep-is-bprog}
  Let $\gkabsym = \tup{T, \initabox, \actset, \ginitprog}$ be a
  B-GKAB,
  $\tgkabb(\gkabsym)$ 
  be an S-GKAB $($with transition system
  $\ts{\tgkabb(\gkabsym)}^{\filter_S}$$)$
  obtained from $\gkabsym$
  through $\tgkabb$,
  and $\delta^T_{b}$
  be a b-repair program over $T$.
  Consider an ABox $A$
  and a service call map $\scmap$.
  If $A'
  \in \progres(A, \scmap, \delta^T_b)$ then $A' \in \arset{T,A}$.
%
\end{lemma}
\begin{proof}
  By \Cref{lem:characteristic-bprog-result} and the definition of
  $\arset{T,A}$.
\end{proof}

In order to complete the proof that a b-repair program produces
exactly all b-repair results of the given (inconsistent) KB, we will
show that every b-repair result of the given (inconsistent) KB is
produced by the b-repair program
below. 

\begin{lemma}\label{lem:bprog-is-brep}
  Let $\gkabsym = \tup{T, \initabox, \actset, \ginitprog}$ be a
  B-GKAB,
  $\tgkabb(\gkabsym)$ 
  be an S-GKAB $($with transition system
  $\ts{\tgkabb(\gkabsym)}^{\filter_S}$$)$
  obtained from $\gkabsym$
  through $\tgkabb$,
  and $\delta^T_{b}$
  be a b-repair program over $T$.
  Consider an ABox $A$
  and a service call map $\scmap$.
  If $A'
  \in \arset{T,A}$, then $A' \in \progres(A, \scmap, \delta^T_b)$
%
\end{lemma}
\begin{proof}
  We divide the proof into two cases:
  \begin{compactenum}
  \item[\textbf{Case 1: $A$ is $T$-consistent.}]~Trivially true,
    because $\arset{T,A}$ is a singleton set containing $A$ and since
    $\ask(\qunsatecq{T}, T, A) = \false$, we have
    $\tup{A, \scmap, \delta^T_{b}}$ is a final state and hence
    $\progres(A, \scmap, \delta^T_b)$ is also a singleton set
    containing $A$.

\item[\textbf{Case 2: $A$ is $T$-inconsistent.}]~
%
%
  Let $A_1$ be an arbitrary ABox in $\arset{T,A}$, we have to show
that there exists $A_2 \in \progres(A, \scmap, \delta^T_b)$ such that $A_2 = A_1$.
%
%


Now, consider an arbitrary concept assertion $N(c) \in A_1$ (resp.\
role assertion $P(c_1,c_2) \in A_1$), we have to show that
$N(c) \in A_2$ (resp.\ $P(c_1,c_2) \in A_2$).
%
%
%
For compactness reason, here we only consider the case for $N(c)$ (the
case for $P(c_1,c_2)$ is similar). Now we have to consider two cases:
    \begin{compactenum}[\bf (a)]
    \item $N(c)$ does not violate any negative concept inclusion
      assertion,
    \item $N(c)$, together with another assertion, violate a negative
      concept inclusion assertion. 
    \end{compactenum}
    The proof is as follows:
    \begin{compactitem}

    \item[Case \textbf{(a)}:] It is easy to see that there exists
      $A_2 \in \progres(A, \scmap, \delta^T_b)$ such that
      $N(c) \in A_2$ because by construction of $\delta^T_b$, every
      action $\act \in \actset^T_b$ never deletes any assertion that
      does not violate any negative inclusion.

    \item[Case \textbf{(b)}:] Due to the fact about the source of
      inconsistency in \dllitea, there exists 
      \begin{compactenum}[i.]
      \item $N(c) \in A$,
      \item a negative inclusion $N \sqsubseteq \neg B$ (such that
        $T \models N \sqsubseteq \neg B$), and
      \item $B(c) \in A$.
      \end{compactenum}
      Since $N(c) \in A_1$, then there exists $A_1' \in \arset{T,A}$
      such that $B(c) \in A_1'$.  Now, it is easy to see from the
      construction of b-repair program $\delta^T_b$ that we have two
      actions in $\actset^T_b$ 
      that one removes only $N(c)$ from $A$ and the other removes only
      $B(c)$ from $A$. Hence, w.l.o.g.\ we must have
      $\set{A_2, A_2'} \subseteq \progres(A, \scmap, \delta^T_b)$ such that
      $N(c) \in A_2$ but $N(c) \not\in A_2'$ and $B(c) \not\in A_2$
      but $B(c) \in A_2'$.

    \end{compactitem}

    Now, since $N(c)$ is an arbitrary assertion in $A$, by the two
    cases above, and also considering that the other case can be
    treated similarly, we have that
    $A_2 \in \progres(A, \scmap, \delta^T_b)$, where $A_2 = A_1$.

\end{compactenum}
\ \ 
\end{proof}

As a consequence of \Cref{lem:bprog-is-brep,lem:brep-is-bprog}, we
finally show the correctness of b-repair program (i.e., it produces
the same result as the result of b-repair over KB) as follows.

\begin{theorem}\label{thm:bprog-equal-brep}
  Let $\gkabsym = \tup{T, \initabox, \actset, \ginitprog}$ be a
  B-GKAB,
  $\tgkabb(\gkabsym)$ 
  be an S-GKAB $($with transition system
  $\ts{\tgkabb(\gkabsym)}^{\filter_S}$$)$
  obtained from $\gkabsym$
  through $\tgkabb$,
  and $\delta^T_{b}$
  be a b-repair program over $T$.
  Consider an ABox $A$
  and a service call map $\scmap$, 
  we have that $\progres(A, \scmap, \delta^T_b) =\arset{T,A}$.
%
\end{theorem}
\begin{proof}
  Direct consequence of \Cref{lem:bprog-is-brep,lem:brep-is-bprog}.
\end{proof}

\subsubsection{Recasting the Verification of B-GKABs Into S-GKABs}

To show that the verification of \muladom over B-GKABs can be recast
as verification over S-GKABs, we make use the J-Bisimulation relation
defined in \Cref{sec:jumping-bisimulation} with a slight modification
that two J-bisimilar states $s_1$ and $s_2$ should have
$\abox(s_1) = \abox(s_2)$ instead of $\abox(s_1) \eqm \abox(s_2)$.  It
is easy to see that
\Cref{lem:jumping-bisimilar-states-satisfies-same-formula,lem:jumping-bisimilar-ts-satisfies-same-formula}
still hold for this small modification.
We now aim to show that given a B-GKAB $\gkabsym$, its transition
system $\ts{\gkabsym}^{\filter_B}$ is J-bisimilar to the transition
system $\ts{\tgkabb(\gkabsym)}^{\filter_S}$ of S-GKAB
$\tgkabb(\gkabsym)$ that is obtained via the translation $\tgkabb$.
As a consequence, we have that both transition systems
$\ts{\gkabsym}^{\filter_B}$ and $\ts{\tgkabb(\gkabsym)}^{\filter_S}$
can not be distinguished by any \muladom (in NNF) modulo the
translation $\tforj$ (as in \Cref{def:tforj}).


\begin{lemma}\label{lem:bgkab-to-sgkab-bisimilar-state}
  Let $\gkabsym$ be a B-GKAB with transition system
  $\ts{\gkabsym}^{\filter_B}$, and let $\tgkabb(\gkabsym)$ be an
  S-GKAB with transition system $\ts{\tgkabb(\gkabsym)}^{\filter_S}$
  obtained through $\tgkabb$.
  Consider 
\begin{inparaenum}[]
\item a state $\tup{A_b,\scmap_b, \delta_b}$ of
  $\ts{\gkabsym}^{\filter_B}$ and
\item a state $\tup{A_s,\scmap_s, \delta_s}$ of
  $\ts{\tgkabb(\gkabsym)}^{\filter_S}$.
\end{inparaenum}
If  
\begin{inparaenum}[]
\item $A_s = A_b$, $\scmap_s = \scmap_b$ and
\item $\delta_s = \tgprogb(\delta_b)$,
\end{inparaenum}
then
$\tup{A_b,\scmap_b, \delta_b} \jbsim \tup{A_s,\scmap_s, \delta_s}$.
\end{lemma}
\begin{proof}
Let 
\begin{compactitem}
\item $\gkabsym = \tup{T, \initabox, \actset, \ginitprog}$ and
  $\ts{\gkabsym}^{\filter_B} = \tup{\const, T, \stateset_b, s_{0b},
    \abox_b, \trans_b}$,
\item
  $\tgkabb(\gkabsym) = \tup{T_s, \initabox, \actset_s,
    \ginitprog_{s}}$ and
  $\ts{\tgkabb(\gkabsym)}^{\filter_S} = \tup{\const, T_s, \stateset_s,
    s_{0s}, \abox_s, \trans_s}$.
%
\end{compactitem}
We have to show the following: for every state
$\tup{A''_b,\scmap''_b, \delta''_b}$ such that
$\tup{A_b,\scmap_b, \delta_b} \trans \tup{A''_b,\scmap''_b,
  \delta''_b}$,
there exists states $t_1, \ldots, t_n$, and
$\tup{A''_s,\scmap''_s, \delta''_s}$ such that:
\begin{compactenum}[\bf (a)]
\item $s \trans_s t_1 \trans_s \ldots \trans_s t_n \trans_s s''$,
  where $s = \tup{A_s,\scmap_s, \delta_s}$,
  $s'' = \tup{A''_s,\scmap''_s, \delta''_s}$, $n \geq 0$,
  $\tmp \not\in A''_s$, and
  $\tmp \in \abox_s(t_i)$ for $i \in \set{1, \ldots, n}$;
\item $A''_s = A''_b$;
\item $\scmap''_s = \scmap''_b$;
\item $\delta''_s = \tgprogb(\delta''_b)$.
\end{compactenum}

By definition of $\ts{\gkabsym}^{\filter_B}$, 
Since $\tup{A_b,\scmap_b, \delta_b} \trans \tup{A''_b,\scmap''_b,
  \delta''_b}$, we have $\tup{A_b,\scmap_b, \delta_b}
\gprogtrans{\act\sigma_b, \filter_B} \tup{A''_b,\scmap''_b,
  \delta''_b}$.
Hence, by the definition of $\gprogtrans{\act\sigma_b, \filter_B}$, we
have:
\begin{compactitem}


\item
  $\tup{\tup{A_b, \scmap_b}, \act\sigma_b, \tup{A''_b, \scmap''_b}}
  \in \tell_{\filter_B}$, and

\item $\sigma_b$ is a legal parameter assignment for $\act$ in $A_b$
  w.r.t.\ $\gact{Q(\vec{p})}{\act(\vec{p})}$ (i.e.,
  $\ask(Q\sigma_b, T, A_b) = \true$).

\end{compactitem}
Since
$\tup{\tup{A_b, \scmap_b}, \act\sigma_b, \tup{A''_b, \scmap''_b}} \in
\tell_{\filter_B}$,
by the definition of $\tell_{\filter_B}$, there exists
$\theta_b \in \eval{\addfacts{T, A_b, \act\sigma_b}}$ such that
\begin{compactitem}
\item $\theta_b$ and $\scmap_b$ agree on the common values in their
  domains.
\item $\scmap''_b = \scmap_b \cup \theta_b$.
\item
  $\tup{A_b, \addfacts{T, A_b, \act\sigma_b}\theta_b, \delfacts{T,
      A_b, \act\sigma_b}, A_b''} \in \filter_B$.
\item $A''_b$ is $T$-consistent.
\end{compactitem}
Since
$\tup{A_b, \addfacts{T, A_b, \act\sigma_b}\theta_b, \delfacts{T, A_b,
  \act\sigma_b}, A_b''} \in \filter_B$,
by the definition of $\filter_B$, there exists $A'_b$ such that
$A''_b \in \arset{T, A'_b}$, and
$A'_b = (A_b \setminus \delfacts{T, A_b, \act\sigma_b}) \cup
\addfacts{T, A_b, \act\sigma_b}\theta_b$.

Since $\delta_s = \tgprogb(\delta_b)$, by the definition of
$\tgprogb$, we have that
\[
\begin{array}{l}
  \tgprogb(\gact{Q(\vec{p})}{\act(\vec{p})}) =
  \gact{Q(\vec{p})}{\act'(\vec{p})} ; \delta^T_b ; \gact{\true}{\act^-_{tmp}()}
\end{array}
\]
Hence, the next executable sub-program on state
$\tup{A_s,\scmap_s, \delta_s}$ is
\[
\delta_s' = 
\gact{Q(\vec{p})}{\act'(\vec{p})}; \delta^T_b ;
\gact{\true}{\act^-_{tmp}()}.
\]
Now, since 
\begin{compactitem}
\item $\act' \in \actset_s$ is obtained from $\act \in \actset$
  through $\tgkabb$,
\item the translation $\tgkabb$ transform $\act$ into $\act'$ without
  changing its parameters, 
\item $\sigma_b$ maps parameters of $\act \in \actset$ to constants
  in $\adom{A_b}$, and 
\item $A_b = A_s$
\end{compactitem}
we can construct
$\sigma_s$ such that $\sigma_s = \sigma_b$.
%
Moreover, we also know that the certain answers computed over $A_b$
are the same to those computed over $A_s$.
Hence, $\act' \in \actset_s$ is executable in $A_s$ with legal
parameter assignment $\sigma_s$.
Now, since we have $\scmap_s = \scmap_b$, we can construct
$\theta_s$ such that $\theta_s = \theta_b$. 
Hence, we have the following:
\begin{compactitem}
\item $\theta_s$ and $\scmap_s$ agree on the common values in their
  domains.
\item $\scmap''_s = \theta_s \cup \scmap_s = \theta_b \cup \scmap_b = \scmap_b''$.
\end{compactitem}
Let
$A_s' = (A_s \setminus \delfacts{T_s, A_s, \act'\sigma_s}) \cup
\addfacts{T_s, A_s, \act'\sigma_s}\theta_s$,
as a consequence, we have that
$\tup{A_s, \addfacts{T_s, A_s, \act'\sigma_s}\theta_s, \delfacts{T_s,
    A_s, \act'\sigma_s}, A_s'} \in \filter_S$.
Since $A_s = A_b$, $\sigma_s = \sigma_b$ and $\theta_s = \theta_b$,
it follows that
\begin{compactitem}
\item
  $\delfacts{T_s, A_s, \act'\sigma_s} = \delfacts{T, A_b,
    \act\sigma_b}$, and
\item
  $\addfacts{T_s, A_s, \act'\sigma_s}\theta_s \setminus \tmp =
  \addfacts{T, A_b, \act\sigma_b}\theta_b$.
\end{compactitem}
Hence, by the construction of $A_s'$ and $A_b'$ above, we have
$A_b' = A_s' \setminus \tmp$.  
By the definition of $\tgkabb$, we have $T_s = T_p$ (i.e., only
positive inclusion assertion of $T$), hence $A'_s$ is
$T_s$-consistent. Thus, by the definition of $\tell_{\filter_s}$, we
have
$\tup{\tup{A_s,\scmap_s}, \act'\sigma_s, \tup{A'_s, \scmap''_s}} \in
\tell_{\filter_s}$.
Moreover, we have
\[
\tup{A_s, \scmap_s, \gact{Q(\vec{p})}{\act'(\vec{p})};\delta_0}
\gprogtrans{\act'\sigma_s, \filter_s} \tup{A_s', \scmap_s'', \delta_0}
\]
where $\delta_0 = \delta^T_b ; \gact{\true}{\act^-_{tmp}()}$.

Now, we need to show that the rest of program in $\delta_s'$ that
still need to be executed (i.e., $\delta_0$)
will bring us into a state $\tup{A_s'', \scmap_s'', \delta_s''}$
s.t. the claim \textbf{(a)} - \textbf{(e)} are proved.
Since $\delta_0$ does not involve any service calls, w.l.o.g.\ let
\[
\pi = \tup{A_s', \scmap_s'', \delta_0} \gexectrans \tup{A_1, \scmap_s'',
  \delta_1} \gexectrans \cdots 
\]
be a program execution trace induced by $\delta_0$ on
$\tup{A_s', \scmap_s'', \delta_0}$.
By \Cref{lem:bprog-termination} and \Cref{thm:bprog-equal-brep}, we
have that
\begin{compactitem}
\item $\delta^T_b$ is always terminate,
\item $\delta^T_b$ produces an ABox $A_n$ such that
  $A_n \in \arset{T, A_s'}$,
\end{compactitem}
additionally, by the construction of $\delta^T_b$ and
$\act^-_{tmp}$, we have that 
\begin{compactitem}
\item $\delta^T_b$ never deletes $\tmp$, and 
\item $\act^-_{tmp}$ only deletes $\tmp$ from the corresponding
  ABox,
\end{compactitem}
therefore, there exists a state
$\tup{A_s'', \scmap_s'', \delta_{n+1}}$ such that we have the
following program execution trace
\[
\pi = \tup{A_s', \scmap_s'', \delta_0} \gexectrans \tup{A_1,
  \scmap_s'', \delta_1} \gexectrans \cdots \cdots \gexectrans
\tup{A_n, \scmap_s'', \delta_n} \gexectrans \tup{A_s'', \scmap_s'',
  \delta_{n+1}}
\]
where 
\begin{compactitem}
\item $\tmp \not\in A''_s$, 
\item $\tmp \in A'_s$, $\tmp \in A_i$ (for $1 \leq i \leq n$), 
\item $\tup{A_s'', \scmap_s'', \delta_{n+1}}$ is a final state,
\item $A_n \in \arset{T, A'_s}$ (by \Cref{thm:bprog-equal-brep}),
\item $A_s'' \in \arset{T, A'_b}$ (Because
  $A'_b = A'_s \setminus \tmp$, $A_s'' = A_n \setminus \tmp$,
  $A_n \in \arset{T, A'_s}$, and $\tmp$ is a special marker).
\end{compactitem}
Since $\tup{A_s'', \scmap_s'', \delta_{n+1}}$ is a final state, we
have finished executing $\delta_s'$, and by the definition of
$\tgprogb$ the rest of the program to be executed is
$\delta_s'' = \tgprogb(\delta_b'')$.

Therefore, we have shown that there exists $s'', t_1, \ldots, t_n$
(for $n \geq 0$) such that
  \[
  s \trans_s t_1 \trans_s \ldots \trans_s t_n \trans_s s''
  \]
where
\begin{compactitem} 
\item $s = \tup{A_s,\scmap_s, \delta_s}$,
  $s'' = \tup{A''_s,\scmap''_s, \delta''_s}$,
\item $\tmp \not\in A''_s$, and 
\item 
  $\tmp \in \abox_2(t_i)$ for $i \in \set{1, \ldots, n}$;
\item $A''_s = A''_b$
\end{compactitem}
The other direction of bisimulation relation can be proven similarly.

\end{proof}

Having \Cref{lem:bgkab-to-sgkab-bisimilar-state} in hand, we can
easily show that given a B-GKAB $\gkabsym$, its transition system
$\ts{\gkabsym}^{\filter_B}$ is J-bisimilar to the transition
$\ts{\tgkabb(\gkabsym)}^{\filter_S}$ of S-GKAB $\tgkabb(\gkabsym)$
(which is obtained via the translation $\tgkabb$).

\begin{lemma}\label{lem:bgkab-to-sgkab-bisimilar-ts}
  Given a B-GKAB $\gkabsym$, we have
  $\ts{\gkabsym}^{\filter_B} \jbsim
  \ts{\tgkabb(\gkabsym)}^{\filter_S}$
\end{lemma}
\begin{proof}
Let
\begin{compactenum}
\item $\gkabsym = \tup{T, \initabox, \actset, \ginitprog_b}$ and
  $\ts{\gkabsym}^{\filter_B} = \tup{\const, T, \stateset_b, s_{0b},
    \abox_b, \trans_b}$, 
\item
  $\tgkabb(\gkabsym) = \tup{T_s, \initabox, \actset_s, \ginitprog_s}$,
  and 
%
  $\ts{\tgkabb(\gkabsym)}^{\filter_S} = \tup{\const, T_s, \stateset_s,
    s_{0s}, \abox_s, \trans_s}$.
\end{compactenum}
We have that $s_{0b} = \tup{A_0, \scmap_b, \delta_b}$ and
$s_{0s} = \tup{A_0, \scmap_s, \delta_s}$ where
$\scmap_b = \scmap_s = \emptyset$. By the definition of $\tgkabb$, we
also have $\delta_s = \tgprogb(\delta_b)$. Hence, by
\Cref{lem:bgkab-to-sgkab-bisimilar-state}, we have
$s_{0b} \jbsim s_{0s}$. Therefore, by the definition of
J-bisimulation, we have
$\ts{\gkabsym}^{\filter_B} \jbsim \ts{\tgkabb(\gkabsym)}^{\filter_S}
$.

\end{proof}

With all of these machinery in hand, we are now ready to show that the
verification of \muladom over B-GKABs can be recast as verification
over S-GKAB as follows.


\begin{theorem}\label{thm:bgkab-to-sgkab}
  Given a B-GKAB $\gkabsym$ and a closed $\muladom$ formula $\Phi$ in
  NNF,
\begin{center}
  $\ts{\gkabsym}^{\filter_B} \models \Phi$ iff
  $\ts{\tgkabb(\gkabsym)}^{\filter_S} \models \tforj(\Phi)$
\end{center}
\end{theorem} 
\begin{proof}
  By \Cref{lem:bgkab-to-sgkab-bisimilar-ts}, we have that
  $\ts{\gkabsym}^{\filter_B} \jbsim
  \ts{\tgkabb(\gkabsym)}^{\filter_S}$.
  Hence, the claim is directly follows from
  \Cref{lem:jumping-bisimilar-ts-satisfies-same-formula}.
\end{proof}


\subsection{From C-GKABs to Standard GKABs}\label{CGKABToSGKAB}

Making inconsistency management for C-GKABs explicit requires just a
single action, which removes all ABox assertions that are involved in
some form of inconsistency. To this aim, 
we define a 0-ary c-repair action $\act^T_c$, as follows:

\begin{definition}[C-Repair Action]\label{def:c-rep-act}
  Given \sidetext{C-Repair Action} a TBox $T$, we define a 0-ary (i.e., has no action parameters)
  \emph{c-repair action $\act^T_c$} over $T$, where $\eff{\act^T_c}$
  is the smallest set containing the following effects:
  \begin{compactitem}
  \item for each functionality assertion $\funct{R} \in T$, we have 
\begin{center}
$   
 \map{\qunsatf(\funct{R}, x, y, z)} {\del \set{R(x, y),R(x,
          z)}} \in \eff{\act^T_c}
$\\ \ \\
\end{center}
  \item for each negative concept inclusion assertion $B_1 \ISA \neg B_2$ such that  $T
    \models B_1 \ISA \neg B_2$, we have 
\begin{center}
  $ \map{\qunsatn(B_1 \ISA \neg B_2, x)} {\del \set{B_1(x), B_2(x)} }
  \in \eff{\act^T_c}; \footnote{Note: if $B_1 =
    \SOMET{P_1}$ (resp.\ $B_2 =
    \SOMET{P_2}$), then with a little abuse of notation, specifically
    for this case, we have that the atom $B_1(x)$ (resp.\
    $B_2(x)$) denotes $P_1(x,y)$ (resp.\
    $P_2(x,y)$).  For instance, for the assertion
    $\SOMET{P_1} \ISA \neg
    N$, 
    we have
    $\map{P_1(x,y) \land N(x)} {\del \set{P_1(x,y), N(x)} } \in
    \eff{\act^T_c}$. Similarly when $B_1 = \SOMET{P_1^-}$ (resp.\
    $B_2 = \SOMET{P_2^-}$).}
  $\\ \ \\
\end{center}
  \item 
    for each negative role inclusion assertion $R_1 \ISA \neg R_2$
    such that $T \models R_1 \ISA \neg R_2$, we have
\begin{center}
$   
    \map{\qunsatn(R_1 \ISA \neg R_2, x, y)} {\del \set{R_1(x, y),
          R_2(x, y)}} \in \eff{\act^T_c}.
$\\ 
\end{center}
  \item $\map{\true} {\del \set{\tmp} \in \eff{\act^T_c}}.$

  \end{compactitem}
\ \ 
\end{definition}

\noindent
Roughly speaking, the c-repair action deletes every ABox assertion
that involves in inconsistency. Technically, the first three effects
of the c-repair action above will delete the ABox assertions that are
obtained by populating the atoms in the right hand side of the effects
with every constant that satisfies the unsatisfiability query in the
left hand side of the effects.
Notice that all effects are guarded by queries that extract only
constants involved in an inconsistency. Hence, other assertions are
kept unaltered, which also means that $\act^T_c$ is a no-op when
applied over a $T$-consistent ABox. The effect in the last line above
is used to flush the marker for an intermediate state.
%

%
%

We now define a translation function $\tgprogc$ that essentially
concatenates each action invocation with a c-repair action in order
to simulate the action executions in C-GKABs.  Additionally, the
translation function $\tgprogc$ also serves as a one-to-one
correspondence (bijection) between the original and the translated
program (as well as between the sub-program).

\begin{definition}[Program Translation $\tgprogc$]
  Given \sidetextb{Program Translation $\tgprogc$} a C-GKAB
  $\gkabsym = \tup{T, \initabox, \actset,
    \ginitprog}$, 
  we define a \emph{translation $\tgprogc$} that translates a program
  $\delta$ into a program $\delta'$ inductively as follows:
  \[
\begin{array}{@{}l@{}l@{}}
  \tgprogc(\gact{Q(\vec{p})}{\act(\vec{p})}) &=  
                                               \gact{Q(\vec{p})}{\act'(\vec{p})};\gact{\true}{\act^T_c()}\\
  \tgprogc(\gemptyprog) &= \gemptyprog \\
  \tgprogc(\delta_1|\delta_2) &= \tgprogc(\delta_1)|\tgprogc(\delta_2) \\
  \tgprogc(\delta_1;\delta_2) &= \tgprogc(\delta_1);\tgprogc(\delta_2) \\
  \tgprogc(\gif{\varphi}{\delta_1}{\delta_2}) &= \gif{\varphi}{\tgprogc(\delta_1)}{\tgprogc(\delta_2)} \\
  \tgprogc(\gwhile{\varphi}{\delta}) &= \gwhile{\varphi}{\tgprogc(\delta)}
\end{array}
\]
where 
\begin{compactitem}
\item action $\act'$ is obtained from
  $\act \in \actset$, such that
  \[
  \eff{\act'} = \eff{\act} \cup \set{\map{\true}{\add \set{\tmp} } }
  \]
 \item $\act^T_c$ is a c-repair action over $T$.  
 \end{compactitem}
\ \ 
\end{definition}
\noindent

To transform a C-GKAB into the corresponding S-GKAB, We define a
translation $\tgkabc$ that, given a C-GKAB, generates an S-GKAB as
follows.

\begin{definition}[Translation from C-GKAB to S-GKAB]
  We \sidetext{Translation from C-GKAB to S-GKAB} define a translation
  $\tgkabc$ that, given a C-GKAB
  $\gkabsym = \tup{T, \initabox, \actset, \ginitprog}$, generates an
  S-GKAB
  $\tgkabc(\gkabsym) = \tup{T_p, \initabox, \actset' \cup
    \set{\act^T_c}, \ginitprog'}$, where
  \begin{compactitem}
  \item $T_p$ is the positive inclusion assertions of $T$ (see \Cref{def:dllitea-tbox}),
  \item $\actset'$ is obtained from $\actset$ such that for each $\act
    \in \actset$, we have $\act' \in \actset'$ and
    \[
    \eff{\act'} = \eff{\act} \cup \set{\map{\true}{\add \set{\tmp} } }
    \]
  \item $\act^T_c$ is a c-repair action over $T$,
  \item $\ginitprog' = \tgprogc(\ginitprog)$. 
\end{compactitem}
 \ \ 
\end{definition}

As for B-GKABs, the translation above only maintains positive
inclusion assertions of TBox $T$. Moreover, intuitively the new
program obtained from the translation above
%
%
attests that each transition in C-GKAB $\gkabsym$ corresponds to a
sequence of two transitions in S-GKAB $\tgkabc(\gkabsym)$: the first
mimics the action execution, while the second computes the c-repair of
the obtained ABox. 


A \muladom property $\Phi$ over C-GKAB $\gkabsym$ can then be recast
as a corresponding property over $\tgkabc(\gkabsym)$ that simply
substitutes each subformula $\DIAM{\Psi}$ of $\Phi$ with
$\DIAM{\DIAM{\Psi}}$ (similarly for $\BOX{\Phi}$). Formally we define
such formula translation as follows:

\begin{definition}[Translation $\tford$]\label{def:tdup}
  We \sidetext{Translation $\tford$} define a \emph{translation
    $\tford$} that takes a \muladom formula $\Phi$ as an input and
  produces a new \muladom formula $\tford(\Phi)$ by recurring over the
  structure of $\Phi$ as follows:
  \begin{compactitem}
  \item $\tford(Q) = Q$
  \item $\tford(\neg \Phi) = \neg \tford(\Phi)$ 
  \item $\tford(\exists x.\Phi) = \exists x. \tford(\Phi)$ 
  \item $\tford(\Phi_1 \vee \Phi_2) = \tford(\Phi_1) \vee \tford(\Phi_2)$ 
  \item $\tford(\mu Z.\Phi) = \mu Z. \tford(\Phi)$ 
  \item $\tford(\DIAM{\Phi}) = \DIAM{\DIAM{\tford(\Phi)}}$
  \end{compactitem}
\ \ 
\end{definition}

With these two translations at hand, we will show later that
$\ts{\gkabsym}^{\filter_C} \models \Phi$ if and only if
$\ts{\tgkabc(\gkabsym)}^{\filter_S} \models \tford(\Phi)$ which
consequently means that the verification of \muladom over C-GKABs can
be reduced to the corresponding verification over S-GKABs.
The core idea of the proof is to use a certain bisimulation relation
in which two bisimilar transition systems 
can not be distinguished by \muladom properties modulo the
formula translation $\tford$. Then, we show that the transition system
of a C-GKAB is bisimilar to the transition system of its corresponding
S-GKAB w.r.t.\ this bisimulation relation. 


\subsubsection{Skip-one Bisimulation (S-Bisimulation)}\label{sec:s-bsim}
As a start towards reducing the verification of C-GKABs into S-GKABs,
we define the notion of \emph{skip-one bisimulation} as follows.
\begin{definition}[Skip-one Bisimulation (S-Bisimulation)] \
  \sidetext{Skip-one Bisimulation (S-Bisimulation)} \\
  Let $\ts{1} = \tup{\const, T, \stateset_1, s_{01}, \abox_1, \trans_1}$
  and $\ts{2} = \tup{\const, T, \stateset_2, s_{02}, \abox_2, \trans_2}$
  be KB transition systems, with
  $\adom{\abox_1(s_{01})} \subseteq \const$
  and $\adom{\abox_2(s_{02})} \subseteq \const$.
  A \emph{skip-one bisimulation} (S-Bisimulation) between $\ts{1}$ and
  $\ts{2}$ is a relation $\B \subseteq \Sigma_1 \times\Sigma_2$ such
  that $\tup{s_1, s_2} \in \B$ implies that:
  \begin{compactenum}
  \item $\abox_1(s_1) = \abox_2(s_2)$
  \item for each $s_1'$, if $s_1 \Rightarrow_1 s_1'$ then there exists
    $t$, and $s_2'$ with
    \[
    s_2 \Rightarrow_2 t \Rightarrow_2 s_2'
    \] 
    such that $\tup{s_1', s_2'}\in\B$, $\tmp \not\in \abox_2(s_2')$
    and $\tmp \in \abox_2(t)$.
  \item for each $s_2'$, if 
    \[
    s_2 \Rightarrow_2 t \Rightarrow_2 s_2'
    \] 
    with $\tmp \in \abox_2(t)$ 
    and $\tmp \not\in \abox_2(s_2')$, then there exists $s_1'$ with
    $s_1 \Rightarrow_1 s_1'$, such that $\tup{s_1', s_2'}\in\B$.
 \end{compactenum}
\ \ 
\end{definition}

\noindent
Let $\ts{1} = \tup{\const, T, \Sigma_1, s_{01}, \abox_1, \trans_1}$
and $\ts{2} = \tup{\const, T, \Sigma_2, s_{02}, \abox_2, \trans_2}$ be
KB transition systems,
a state $s_1 \in \Sigma_1$ is \emph{S-bisimilar} to
$s_2 \in \Sigma_2$, written $s_1 \sbsim s_2$, if there exists an
S-bisimulation relation $\B$ between $\ts{1}$ and $\ts{2}$ such that
$\tup{s_1, s_2}\in\B$.
A transition system $\ts{1}$ is \emph{S-bisimilar} to $\ts{2}$,
written $\ts{1} \sbsim \ts{2}$, if there exists an S-bisimulation
relation $\B$ between $\ts{1}$ and $\ts{2}$ such that
$\tup{s_{01}, s_{02}}\in\B$.

Now, we advance further to show that two S-bisimilar transition
systems can not be distinguished by any \muladom formula modulo the
translation $\tford$.


\begin{lemma}\label{lem:sbisimilar-state-satisfies-same-formula}
  Consider two KB transition systems
  $\ts{1} = \tup{\const, T,\stateset_1,s_{01},\abox_1,\trans_1}$ and
  $\ts{2} = \tup{\const, T,\stateset_2,s_{02},\abox_2,\trans_2}$, 
  two states $s_1 \in \stateset_1$ and $s_2 \in \stateset_2$ such that
  $s_1 \sbsim s_2$. Then for every formula $\Phi$ of
  $\muladom$, 
  and every valuations $\vfo_1$ and $\vfo_2$ that assign to each of
  its free variables a constant $c_1 \in \adom{\abox_1(s_1)}$ and
  $c_2 \in \adom{\abox_2(s_2)}$, such that $c_1 = c_2$, we have that
  \[
  \ts{1},s_1 \models \Phi \vfo_1 \textrm{ if and only if } \ts{2},s_2
  \models \tford(\Phi) \vfo_2.
  \]
\end{lemma}
\begin{proof}
  Similar to
  \Cref{lem:jumping-bisimilar-states-satisfies-same-formula}, we
  divide the proof into three parts:
  \begin{compactenum}[(1)]
  \item First, we obtain the proof of the claim for formulae of
    $\ladom$
  \item Second, we extend the results to the infinitary logic obtained
    by extending $\ladom$ with arbitrary countable disjunction.
  \item Last, we recall that fixpoints can be translated into this
    infinitary logic, thus proving that the theorem holds for
    $\muladom$.
\end{compactenum}
Since the step (2) and (3) are similar to the proof of
\Cref{lem:jumping-bisimilar-states-satisfies-same-formula}, here we
only highlight some interesting cases of the proof for the step (1) as
follow (the other cases of step (1) can be shown similarly):

\smallskip
\noindent
\textbf{Proof for $\ladom$.}  

\smallskip
\noindent
\textit{Base case:}
\begin{compactitem}
\item[\textbf{($\Phi = Q$)}.] Since $s_1 \sbsim s_2$, we have
  $\abox_1(s_1) = \abox_2(s_2)$, and hence
  $\Ans(Q, T, \abox_1(s_1)) = \Ans(Q, T, \abox_2(s_2))$.
  Since $\tforj(Q) = Q$, for every valuations $\vfo_1$ and $\vfo_2$
  that assign to each of its free variables a constant
  $c_1 \in \adom{\abox_1(s_1)}$ and $c_2 \in \adom{\abox_2(s_2)}$,
  such that $c_1 = c_2$, we have
  \[
  \ts{1},s_1 \models Q \vfo_1 \textrm{ if and only if } \ts{2},s_2
  \models \tford(Q) \vfo_2.
  \]
\end{compactitem}

\smallskip
\noindent
\textit{Inductive step:}
\begin{compactitem}
\item[\textbf{($\Phi = \neg\Psi$)}.]  By Induction hypothesis, for
  every valuations $\vfo_1$ and $\vfo_2$ that assign to each of its
  free variables a constant $c_1 \in \adom{\abox_1(s_1)}$ and
  $c_2 \in \adom{\abox_2(s_2)}$, such that $c_2 = c_1$, we have that
  $\ts{1},s_1 \models \Psi \vfo_1$ if and only if
  $\ts{2},s_2 \models \tford(\Psi) \vfo_1$. Hence,
  $\ts{1},s_1 \not\models \Psi \vfo_1$ if and only if
  $\ts{2},s_2 \not\models \tford(\Psi) \vfo_2$. By definition,
  $\ts{1},s_1 \models \neg \Psi \vfo_1$ if and only if
  $\ts{2},s_2 \models \neg \tford(\Psi) \vfo_2$.  Hence, by the
  definition of $\tford$, we have
  $\ts{1},s_1 \models \neg \Psi \vfo_1$ if and only if
  $\ts{2},s_2 \models \tford(\neg \Psi) \vfo_2$.

\item[\textbf{($\Phi = \DIAM{\Psi}$)}.]  Assume $\ts{1},s_1 \models
  (\DIAM{\Psi}) \vfo_1$, then there exists $s_1'$ s.t.\ $s_1 \trans_1
  s_1'$ and $\ts{1},s_1' \models \Psi \vfo_1$. Since $s_1 \sbsim
  s_2$, there exist $t$ and $s_2'$ s.t.\
    \[
    s_2 \trans_2 t \trans_2 s_2'
    \] 
    and $s_1' \sbsim s_2'$.
    Hence, by induction hypothesis, for every valuations $\vfo_2$ that
    assign to each free variables $x$ of $\tford(\Psi)$ a constant $c_2 \in
    \adom{\abox_2(s_2)}$, such that $c_2 = c_1$ with $x/c_1 \in
    \vfo_1$, we have
    \[
    \ts{2},s_2' \models \tford(\Psi_1) \vfo_2.
    \]
    Since 
    $ s_2 \trans_2 t \trans_2 s_2', $ therefore we get
    \[
    \ts{2},s_2 \models ( \DIAM{\DIAM{\tford(\Psi)}} )\vfo_2.
    \]
    Since $\tford(\DIAM{\Phi}) = \DIAM{\DIAM{\tford(\Phi)}} $, we
    therefore have
    \[
    \ts{2},s_2 \models \tford(\DIAM{\Psi} )\vfo_2. 
    \]
    The other direction can be shown in a similar way.

\end{compactitem}

\end{proof}

Having Lemma~\ref{lem:sbisimilar-state-satisfies-same-formula} in
hand, we can easily show that two S-bisimilar transition systems can
not be distinguished by any \muladom formulas modulo translation
$\tford$.

\begin{lemma}\label{lem:sbisimilar-ts-satisfies-same-formula}
  Consider two transition systems
  $\ts{1}$ 
  and
  $\ts{2}$ 
  such that $\ts{1} \sbsim \ts{2}$.  For every closed \muladom formula
  $\Phi$, we have:
  \[
  \ts{1} \models \Phi \textrm{ if and only if } \ts{2} \models
  \tford(\Phi)
  \]
\end{lemma}
\begin{proof} Let
  $\ts{1} = \tup{\const_1, T, \stateset_1, s_{01}, \abox_1, \trans_1}$
  and
  $\ts{2} = \tup{\const_2, T, \stateset_2, s_{02}, \abox_2,
    \trans_2}$. 
  By the definition of S-bisimilar transition system we have that
  $s_{01} \sbsim s_{02}$. Thus we obtain the proof as a consequence of
  \Cref{lem:sbisimilar-state-satisfies-same-formula}, due to the fact
  that
  \[ \ts{1}, s_{01} \models \Phi \textrm{ if and only if } \ts{2},
  s_{02} \models \tford(\Phi)
  \]
\end{proof}

\subsubsection{Important Properties of C-Repair and C-Repair Actions.}

To the aim of reducing verification of C-GKABs into S-GKABs, we
now aiming to show an important property of c-repair and c-repair
action, namely we show that c-repair action produces the same results
as the computation of c-repair.
To this aim, we first show some important properties of b-repair,
c-repair and also c-repair action.
%
As a start, below we show that for every pair of ABox assertions that
violates a certain negative inclusion assertion, each of them will be
contained in two different ABoxes in the result of b-repair.

\begin{lemma}\label{lem:disjoint-brep}
  Let $T$ be a TBox, and $A$ be an ABox. For every negative concept
  inclusion assertion $B_1 \sqsubseteq \neg B_2$ such that
  $T \models B_1 \sqsubseteq \neg B_2$ and $B_1 \neq B_2$, if 
%
  $\set{B_1(c), B_2(c)} \subseteq A$ (for any constant
  $c \in \const$), then there exist $A' \in \arset{T, A}$ such that
  \begin{inparaenum}[\it (i)]
  \item $B_1(c) \in A'$,
  \item $B_2(c) \not\in A'$.
  \end{inparaenum}
  (Similarly for the case of negative role inclusion assertion
  $R_1 \sqsubseteq \neg R_2$ s.t.\
  $T \models R_1 \sqsubseteq \neg R_2$).
\end{lemma}
\begin{proof}
  Suppose by contradiction $\set{B_1(c), B_2(c)} \subseteq A$, and
  there does not exist $A' \in \arset{T, A}$ such that $B_1(c) \in A'$
  and $B_2(c) \not\in A'$.
%
  Since in \dllitea the violation of negative concept inclusion
  $B_1 \sqsubseteq \neg B_2$ 
  is only caused by a pair of assertions $B_1(c)$ and $B_2(c)$ (for
  any constant $c \in \const$) and by the definition of
  $\arset{T, A}$, it contains all maximal $T$-consistent subsets of
  $A$,
  then there should be a $T$-consistent ABox $A' \in \arset{T, A}$
  such that $B_1(c) \in A'$ and $B_2(c) \not\in A'$ that is obtained
  by just removing $B_2(c)$ from $A$ and keep $B_1(c)$ (otherwise we
  will not have all maximal $T$-consistent subsets of $A$ in
  $\arset{T, A}$, which contradicts the definition of $\arset{T, A}$
  itself).
%
%
  Hence, we have a contradiction 
  Thus, there exists $A' \in \arset{T, A}$ such that
  \begin{inparaenum}[\it (i)]
  \item $B_1(c) \in A'$,
  \item $B_2(c) \not\in A'$.
  \end{inparaenum}
  The proof for the case of negative role inclusion is similar.
\end{proof}

Similarly for the case of functionality assertion, below we show that
for each role assertion that violates a functional assertion, there
exists an ABox in the set of b-repair results that contains only this
role assertion but not the other role assertions that together they
violate the corresponding functional assertion.

\begin{lemma}\label{lem:disjoint-brep-for-funct}
  Given a TBox $T$, and an ABox $A$, for every functional assertion
  $\funct{R}$, if $\set{R(c,c_1), R(c,c_2), \ldots, R(c,c_n)} \subseteq A$ (for
  any constants $\set{c, c_1, c_2, \ldots, c_n} \subseteq \const$),
  then there exist $A' \in \arset{T, A}$ such that
\begin{inparaenum}[\it (i)]
\item $R(c,c_1) \in A'$, 
\item $R(c,c_2) \not\in A', \ldots, R(c,c_n) \not\in A'$,
\end{inparaenum}
\end{lemma}
\begin{proof}
  Similar to the proof of \Cref{lem:disjoint-brep}.
\end{proof}

Below, we show that the result of c-repair does not contain any ABox
assertions that, together with another ABox assertions, violate a
negative inclusion assertion. Intuitively, this fact is obtained by
using Lemma \ref{lem:disjoint-brep} which said that for every pair of
ABox assertions that violates some negative inclusion assertions, each
of them will be contained in two different ABoxes in the results of
b-repair. As a consequence, we have that both of them are not in the
result of c-repair when we compute the intersection of all of b-repair
results.

\begin{lemma}\label{lem:inconsistent-assertion-not-in-crep}
  Given a TBox $T$, and an ABox $A$, for every negative concept
  inclusion assertion $B_1 \sqsubseteq \neg B_2$ such that
  $T \models B_1 \sqsubseteq \neg B_2$ and $B_1 \neq B_2$, if there
  exists $B_1(c) \in A$ (for any constant $c \in \const{}$) such that
  $B_1(c)$ violates $B_1 \sqsubseteq \neg B_2$,
%
%
  then $B_1(c)
  \not\in \iarset{T, A}$.  (Similarly for the
  case of negative role inclusion assertion).
\end{lemma}
\begin{proof}
  Let $\set{B_1(c), B_2(c)} \subseteq A$ (for any constant
  $c \in \const$), then by \Cref{lem:disjoint-brep}, there exist
  $A' \in \arset{T, A}$ and $A'' \in \arset{T, A}$ such that
\begin{inparaenum}[\it (i)]
\item $B_1(c) \in A'$, 
\item $B_2(c) \not\in A'$,
\item $B_2(c) \in A''$, and 
\item $B_1(c) \not\in A''$.
\end{inparaenum}
By \Cref{def:c-rep}, $\iarset{T, A} = \cap_{A_i \in\arset{T,A}} A_i$.
Since $\set{B_1(c), B_2(c)} \not\subseteq A' \cap A''$, then we have
that 
$\set{B_1(c), B_2(c)} \not\subseteq \iarset{T, A}$. Thus
$B_1(c) \not\in \iarset{T, A}$.
The proof for the case of negative role inclusion is similar.
\end{proof}

Similarly, below we show that the result of c-repair does not contain
any role assertions that, together with another role assertions,
violate a functional assertion. The intuition of the proof is similar
to the proof of
Lemma~\ref{lem:inconsistent-assertion-not-in-crep}. I.e., they are
thrown away when we compute the intersection of all of b-repair
results.

\begin{lemma}\label{lem:inconsistent-assertion-not-in-crep-for-functional}
  Given a TBox $T$, and an ABox $A$, for every functionality assertion
  $\funct{R} \in T$, 
  if there exists $R(c, c_1) \in A$ $($for some constants
  $\set{c, c_1} \subseteq \const$$)$
  such that $R(c,
  c_1)$ violates $\funct{R}$, then $R(c, c_1) \not\in A'$.
%
\end{lemma}
\begin{proof}
Similar to the proof of \Cref{lem:inconsistent-assertion-not-in-crep}
\end{proof}

Now, in the two following Lemmas we show a property of a c-repair
action, namely that a c-repair action deletes all ABox assertions
that, together with another ABox assertions, violate a negative
inclusion or a functionality assertion.

\begin{lemma}\label{lem:inconsistent-assertion-not-in-act-crep}
  Given a TBox $T$, and an ABox $A$, a service call map $\scmap$ and a
  c-repair action $\act^T_c$. 
%
  If
  $(\tup{A,\scmap}, \act^T_c\sigma, \tup{A', \scmap}) \in
  \tell_{\filter_S}$ (where $\sigma$ is an empty substitution), 
  then for every negative concept inclusion assertion
  $B_1 \sqsubseteq \neg B_2$ such that
  $T \models B_1 \sqsubseteq \neg B_2$ and $B_1 \neq B_2$, if there
  exists $B_1(c) \in A$ (for any constant $c \in \const$) such that
  $B_1(c)$ violates $B_1 \sqsubseteq \neg B_2$,
%
%
  then $B_1(c) \not\in A'$. 
  (Similarly for the case of the negative role inclusion assertion).
\end{lemma}
\begin{proof}
  Since $T \models B_1 \sqsubseteq \neg B_2$, by the definition of
  $\act^T_c$, we have
\[
\map{\qunsatn(B_1 \ISA \neg B_2, x)} {\set{\del \set{ B_1(x), B_2(x)} } }
\in \eff{\act^T_c}.
\]
Since, $(\tup{A,\scmap}, \act^T_c\sigma, \tup{A', \scmap}) \in
\tell_{\filter_S}$, by the definition of $\tell_{\filter_S}$, we have
$\tup{A, \addfacts{T, A, \act^T_c\sigma}\theta, \delfacts{T, A, \act^T_c\sigma}, A'}
\in \filter_S$.
Now, it is easy to see that by the definition of filter $\filter_S$
and $\delfacts{T, A, \act^T_c\sigma}$, we have $B_1(c) \not\in
A'$. 

\end{proof}

\begin{lemma}\label{lem:inconsistent-assertion-not-in-act-crep-for-functional}
  Given a TBox $T$, an ABox $A$, a service call map $\scmap$ and a
  c-repair action $\act^T_c$.
%
  If
  $(\tup{A,\scmap}, \act^T_c\sigma, \tup{A', \scmap}) \in
  \tell_{\filter_S}$ (where $\sigma$ is an empty substitution)
  then for every functional assertion $\funct{R} \in T$, if there
  exists $R(c, c_1) \in A$ $($for some constants
  $\set{c, c_1} \subseteq \const$$)$ such that $R(c, c_1)$ violates
  $\funct{R}$, then $R(c, c_1) \not\in A'$.
%
\end{lemma}
\begin{proof}
  Similar to the proof of
  \Cref{lem:inconsistent-assertion-not-in-act-crep}.
\end{proof}

Next, we show that every Abox assertion that does not violate any
TBox assertions will appear in all results of a b-repair.

\begin{lemma}\label{lem:all-consistent-assertion-in-brep}
  Given a TBox $T$, and an ABox $A$, for every concept assertion
  $C(c) \in A$ (for any constant $c \in \const$) such that
  $C(c) \not\in \inc(A)$,
  it holds that for every $A' \in \arset{T, A}$, we have
  $C(c) \in A'$.
  (Similarly for role assertion).
\end{lemma}
\begin{proof}
  Suppose by contradiction there exists $C(c) \in A$ such that
  $C(c) \not\in \inc(A)$ and there exists $A' \in \arset{T, A}$, such
  that $C(c) \not\in A'$. Then, since $C(c) \not\in \inc(A)$, there
  exists $A''$ such that $A' \subset A'' \subseteq A$ and $A''$ is
  $T$-consistent (where $A'' = A' \cup \set{C(c)}$. Hence,
  $A' \not\in \arset{T, A}$.  Thus we have a contradiction. Therefore
  the claim is proven. The proof for the case of role assertion can be
  done similarly.
\end{proof}

From the previous Lemma, we can show that every ABox assertion that
does not violate any TBox assertion will appear in the c-repair
results.

\begin{lemma}\label{lem:all-consistent-assertion-in-crep}
  Given a TBox $T$, and an ABox $A$, for every concept assertion
  $C(c) \in A$ s.t.\ $C(c) \not\in \inc(A)$
  we have that $C(c) \in \iarset{T, A}$.
  (Similarly for role assertion).
\end{lemma}
\begin{proof}
  Let $C(c) \in A$ be any arbitrary concept assertion s.t.\
  $C(c) \not\in \inc(A)$. By
  \Cref{lem:all-consistent-assertion-in-brep}, for every
  $A' \in \arset{T, A}$, we have $C(c) \in A'$. Hence, since
  $\iarset{T, A} = \cap_{A_i \in\arset{T,A}} A_i$, we have
  $C(c) \in \iarset{T, A}$. The proof for the case of role assertion
  can be done similarly.
\end{proof}

Finally, below we can show that the c-repair action is correctly mimicked the
c-repair computation. I.e., they produce the same results.

\begin{theorem}\label{thm:cact-equal-crep}
  Given a TBox $T$, an ABox $A$, a service call map $\scmap$ and a
  c-repair action $\act^T_c$.
  Let $A_1 = \iarset{T,A}$, and
  $\tup{\tup{A,\scmap}, \act^T_c\sigma, \tup{A_2, \scmap}} \in
  \tell_{\filter_S}$
  where $\sigma$ is an empty substitution, then we have $A_1 = A_2$.
\end{theorem}
\begin{proof}
  The core idea of the proof is as follows:
  \begin{inparaenum}[]
  \item since c-repair take the intersection of all b-repairs, and the
    behavior of b-repair is to choose one assertion among those
    conflicting assertions, we have that basically c-repair throw away
    all assertions that involve in making inconsistency.
 \item As it can be seen from the construction of c-repair action
    $\act^T_c$, such an action basically also throw away all
    assertions that involve in making inconsistency.
  \item Moreover, both c-repair and c-repair action do nothing
    regarding those ABox assertions that do not involve in any inconsistency.
  \end{inparaenum}
%
%
  Technically, the claim is proven by the fact that $\act^T_c$ never
  deletes any ABox assertion that does not involve in any source of
  inconsistency, and also by using
  \Cref{lem:all-consistent-assertion-in-crep,lem:inconsistent-assertion-not-in-act-crep,lem:inconsistent-assertion-not-in-act-crep-for-functional,lem:inconsistent-assertion-not-in-crep-for-functional,lem:inconsistent-assertion-not-in-crep}.
\end{proof}

\subsubsection{Reducing the Verification of C-GKABs Into S-GKABs}

In the following two Lemmas, we aim to show that the transition
systems of a C-GKAB and its corresponding S-GKAB (obtained through
$\tgkabc$) are S-bisimilar.


\begin{lemma}\label{lem:cgkab-to-sgkab-bisimilar-state}
  Let $\gkabsym$ be a C-GKAB with transition system
  $\ts{\gkabsym}^{\filter_C}$, and let $\tgkabc(\gkabsym)$ be an
  S-GKAB with transition system $\ts{\tgkabc(\gkabsym)}^{\filter_S}$
  obtained through
  $\tgkabc$. 
  Consider
  \begin{inparaenum}[]
  \item a state $\tup{A_c,\scmap_c, \delta_c}$ of
    $\ts{\gkabsym}^{\filter_C}$ and
  \item a state $\tup{A_s,\scmap_s, \delta_s}$ of
    $\ts{\tgkabc(\gkabsym)}^{\filter_S}$.
  \end{inparaenum}
If  
\begin{inparaenum}[]
\item $A_s = A_c$, $\scmap_s = \scmap_c$ and
\item $\delta_s = \tgprogc(\delta_c)$,
\end{inparaenum}
then
$\tup{A_c,\scmap_c, \delta_c} \sbsim \tup{A_s,\scmap_s, \delta_s}$.
\end{lemma}
\begin{proof}
Let 
\begin{compactenum}
\item $\gkabsym = \tup{T, \initabox, \actset, \ginitprog}$, and
  $\ts{\gkabsym}^{\filter_C} = \tup{\const, T, \stateset_c, s_{0c},
    \abox_c, \trans_c}$,
\item
  $\tgkabc(\gkabsym) = \tup{T_s, \initabox, \actset_s,
    \ginitprog_{s}}$, and
  $\ts{\tgkabc(\gkabsym)}^{\filter_S} = \tup{\const, T_s,
    \stateset_s, s_{0s}, \abox_s, \trans_s}$.
\end{compactenum}
Now, we have to show the following: For every state
$\tup{A''_c,\scmap''_c, \delta''_c}$ such that
  \[
  \tup{A_c,\scmap_c, \delta_c} \trans \tup{A''_c,\scmap''_c,
    \delta''_c},
  \]
  there exists states $\tup{A'_s,\scmap'_s, \delta'_s}$ and
  $\tup{A''_s,\scmap''_s, \delta''_s}$ such that:
\begin{compactenum}[\bf (a)]
\item we have
  $ \tup{A_s,\scmap_s, \delta_s} \trans_s \tup{A'_s,\scmap'_s,
    \delta'_s} \trans_s \tup{A''_s,\scmap''_s, \delta''_s} $
\item $A''_s = A''_c$;
\item $\scmap''_s = \scmap''_c$;
\item $\delta''_s = \tgprogc(\delta''_c)$.
\end{compactenum}

By definition of $\ts{\gkabsym}^{\filter_C}$, 
since $\tup{A_c,\scmap_c, \delta_c} \trans \tup{A''_c,\scmap''_c,
  \delta''_c}$, we have $\tup{A_c,\scmap_c, \delta_c}
\gprogtrans{\alpha\sigma_c, \filter_C} \tup{A''_c,\scmap''_c,
  \delta''_c}$.
Furthermore, by the definition of
$\gprogtrans{\act\sigma_c, \filter_C}$, we have that:
\begin{compactitem}


\item
  $\tup{\tup{A_c, \scmap_c}, \act\sigma_c, \tup{A''_c, \scmap''_c}}
  \in \tell_{\filter_C}$,
  and 

\item $\sigma_c$ is a legal parameter assignment for $\act$ in $A_c$
  w.r.t.\ $\gact{Q(\vec{p})}{\act(\vec{p})}$ (i.e.,
  $\ask(Q\sigma_c, T, A_c) = \true$).

\end{compactitem}
Since
$\tup{\tup{A_c, \scmap_c}, \act\sigma_c, \tup{A''_c, \scmap''_c}} \in
\tell_{\filter_C}$,
by the definition of $\tell_{\filter_C}$, there exists
$\theta_c \in \eval{\addfacts{T, A_c, \act\sigma_c}}$ such that
\begin{compactitem}
\item $\theta_c$ and $\scmap_c$ agree on the common values in their domains.
\item $\scmap''_c = \scmap_c \cup \theta_c$.
\item
  $(A_c, \addfacts{T, A_c, \act\sigma_c}\theta_c, \delfacts{T, A_c,
    \act\sigma_c}, A_c'') \in \filter_C$.
\item $A''_c$ is $T$-consistent.
\end{compactitem}
Since
$\tup{A_c, \addfacts{T, A_c, \act\sigma_c}\theta_c, \delfacts{T, A_c,
    \act\sigma_c}, A_c''} \in \filter_C$,
by the definition of $\filter_C$, there exists $A'_c$ such that
$A''_c \in \iarset{T, A'_c}$ where
$A'_c = (A_c \setminus \delfacts{T, A_c, \act\sigma_c}) \cup
\addfacts{T, A_c, \act\sigma_c}\theta_c$.
Furthermore, since $\delta_s = \tgprogc(\delta_c)$, by the definition of
$\tgprogc$, we have that
\[
\begin{array}{l}
  \tgprogc(\gact{Q(\vec{p})}{\act(\vec{p})}) =
  \gact{Q(\vec{p})}{\act'(\vec{p})}; \gact{\true}{\act^T_c()}.
\end{array}
\]
Hence, we have that the next executable part of program on state
$\tup{A_s,\scmap_s, \delta_s}$ is
\[
\gact{Q(\vec{p})}{\act'(\vec{p})}; \gact{\true}{\act^T_c()}.
\]
%

Now, since $\sigma_c$ maps parameters of $\act \in \actset$ to
constants in $\adom{A_c}$, and $A_c = A_s$, we can construct
$\sigma_s$ mapping parameters of $\act' \in \actset_s$ to constants in
$\adom{A_s}$ such that $\sigma_c = \sigma_s$.
Moreover, since $A_s = A_c$, the certain answers computed over $A_c$
are the same to those computed over $A_s$.
%
%
Hence, $\act' \in \actset_s$ is executable in $A_s$ with (legal
parameter assignment) $\sigma_s$.
Now, since 
we have $\scmap_s= \scmap_c$, then we can construct $\theta_s$ such
that $\theta_s = \theta_c$.
%
%
Hence, we have the following:
\begin{compactitem}
\item $\theta_s$ and $\scmap_s$ agree on the common values in their
  domains.
 \item $\scmap'_s = \theta_s \cup \scmap_s = \theta_c \cup \scmap_c =\scmap_c''$.
\end{compactitem}
Let
$A_s' = (A_s \setminus \delfacts{T_s, A_s, \act'\sigma_s}) \cup
\addfacts{T_s, A_s, \act'\sigma_s}\theta_s$,
as a consequence, we have
$\tup{A_s, \addfacts{T_s, A_s, \act'\sigma_s}\theta_s, \delfacts{T_s,
    A_s, \act'\sigma_s}, A_s'} \in \filter_S$.
Since $A_s = A_c$, $\theta_s = \theta_c$, and $\sigma_s = \sigma_c$, it follows that
\begin{compactitem}
\item
  $\delfacts{T_s, A_s, \act'\sigma_s} = \delfacts{T, A_c, \act\sigma_c}$,
  and
\item
  $\addfacts{T_s, A_s, \act'\sigma_s}\theta_s \setminus \set{\tmp} =
  \addfacts{T, A_c, \act\sigma_c}\theta_c$.
\end{compactitem}
Hence, by the construction of $A_s'$ and $A_c'$ above, we also have
$A_s' \setminus \set{\tmp} = A_c'$.
By the definition of $\tgkabc$, we have $T_s = T_p$ (i.e., only
positive inclusion assertion of $T$), hence $A'_s$ is
$T_s$-consistent. Thus, by the definition of $\tell_{\filter_s}$, we
have
$\tup{\tup{A_s,\scmap_s}, \act'\sigma_s, \tup{A'_s, \scmap'_s}} \in
\tell_{\filter_s}$.
Moreover, we have 
\[
\tup{A_s, \scmap_s, \gact{Q(\vec{p})}{\act'(\vec{p})};\delta_0}
\gprogtrans{\act'\sigma_s, \filter_s} \tup{A_s', \scmap_s', \delta_0}
\]
where $\delta_0 = \gact{\true}{\act^T_c()}$. 
Now, it is easy to see that
 \[
 \tup{A_s', \scmap_s', \gact{\true}{\act^T_c()}}
 \gprogtrans{\act^T_c\sigma, \filter_s} \tup{A_s'', \scmap_s'',
   \gemptyprog}
 \]
 where 
\begin{compactitem}
\item $\scmap_s'' = \scmap_s'$ (since $\act^T_c$ does not involve
 any service call), 
\item $\sigma$ is empty substitution (because $\act^T_c$
 is a 0-ary action),
\item $\tup{A_s'', \scmap_s'', \gemptyprog}$ is a final state,
\item $\tmp \not\in A_s''$,
\item $A_s'' = \iarset{T, A_s' \setminus \set{\tmp}}$ (by \Cref{thm:cact-equal-crep}, and
  by considering that $\tmp$ is only a special marker).
\end{compactitem}
Since $A_s' \setminus \set{\tmp} = A_c'$,
$A_s'' = \iarset{T, A_s' \setminus \set{\tmp}}$, and
$A_c'' = \iarset{T, A_c'}$, then it is easy to see that
$A_s'' = A_c''$.  Moreover, since
$\tup{A_s'', \scmap_s'', \gemptyprog}$ is a final state, we have
successfully finished executing 
\[
\gact{Q(\vec{p})}{\act'(\vec{p})}; \gact{\true}{\act^T_c()},
\]
and by the definition of $\tgprogc$
the rest of the program to be executed is
$\delta_s'' = \tgprogc(\delta_c'')$. Thus, we have
\[
\tup{A_s,\scmap_s, \delta_s} \trans_s \tup{A'_s,\scmap'_s, \delta'_s}
\trans_s \tup{A''_s,\scmap''_s, \delta''_s}
\]
where
\begin{compactenum}[\bf (a)]
\item $A''_s = A''_c$;
\item $\scmap''_s = \scmap''_c$;
\item $\delta''_s = \tgprogc(\delta''_c)$.
\end{compactenum}

The other direction of bisimulation relation can be proven in a
similar way.

\end{proof}

Having \Cref{lem:cgkab-to-sgkab-bisimilar-state} in hand, we can
easily show that given a C-GKAB, its transition system is S-bisimilar
to the transition of its corresponding S-GKAB that is obtained via
the translation $\tgkabc$ as follows.

\begin{lemma}\label{lem:cgkab-to-sgkab-bisimilar-ts}
  Given a C-GKAB $\gkabsym$, we have
  $\ts{\gkabsym}^{\filter_C} \sbsim
  \ts{\tgkabc(\gkabsym)}^{\filter_S}$
\end{lemma}
\begin{proof}
Let
\begin{compactenum}
\item $\gkabsym = \tup{T, \initabox, \actset, \ginitprog_c}$, and
  $\ts{\gkabsym}^{\filter_C} = \tup{\const, T, \stateset_c, s_{0c},
    \abox_c, \trans_c}$,
\item
  $\tgkabc(\gkabsym) = \tup{T_s, \initabox, \actset_s, \ginitprog_s}$,
  and 
  $\ts{\tgkabc(\gkabsym)}^{\filter_S} = \tup{\const, T_s,
    \stateset_s, s_{0s}, \abox_s, \trans_s}$.
\end{compactenum}
We have that $s_{0c} = \tup{A_0, \scmap_c, \delta_c}$ and
$s_{0s} = \tup{A_0, \scmap_s, \delta_s}$ where
$\scmap_c = \scmap_s = \emptyset$. By the definition of $\tgprogc$ and
$\tgkabc$, we also have $\delta_s = \tgprogc(\delta_c)$. Hence, by
\Cref{lem:cgkab-to-sgkab-bisimilar-state}, we have
$s_{0c} \sbsim s_{0s}$. Therefore, by the definition of
S-bisimulation, we have
$\ts{\gkabsym}^{\filter_C} \sbsim \ts{\tgkabc(\gkabsym)}^{\filter_S}$.
\ \ 
\end{proof}

Finally, we are now ready to show that the verification of \muladom
formulas over C-GKABs can be recast as verification of \muladom
formulas over S-GKAB as follows.

\begin{theorem}\label{thm:cgkab-to-sgkab}
  Given a C-GKAB $\gkabsym$ and a closed \muladom property $\Phi$,
\begin{center}
  $\ts{\gkabsym}^{\filter_C} \models \Phi$ iff
  $\ts{\tgkabc(\gkabsym)}^{\filter_S} \models \tford(\Phi)$
\end{center}
\end{theorem} 
\begin{proof}
  By \Cref{lem:cgkab-to-sgkab-bisimilar-ts}, we have that
  $\ts{\gkabsym}^{\filter_C} \sbsim
  \ts{\tgkabc(\gkabsym)}^{\filter_S}$.
  Hence, by \Cref{lem:sbisimilar-ts-satisfies-same-formula}, it is
  easy to see that the claim is proved.
\end{proof}

\subsection{From E-GKABs to Standard GKABs}\label{EGKABToSGKAB}

Differently from the case of B-GKABs and C-GKABs, the case of E-GKABs
pose two challenges while translating it into S-GKABs:
\begin{compactenum}
\item when applying an atomic action (and managing the possibly
  arising inconsistency) it is necessary to distinguish those
  assertions that are newly introduced by the action from those
  already present in the system;
\item the evolution semantics can be applied only if the assertions to
  be added are consistent with the TBox, and hence an additional check
  is required to abort the action execution if this is not the case.
\end{compactenum}
To this aim, given a TBox $T$, we duplicate concepts and roles in $T$,
introducing a fresh concept name $N^{n}$ for every concept name $N$ in
$T$ (similarly for roles). The key idea is to insert those constants
that are added to $N$ also in $N^{n}$, so as to trace that they are
part of the update.

The first issue described above is then tackled by compiling the bold
evolution semantics into a 0-ary \emph{evolution action} $\act^T_e$ as
follows:

\begin{definition}[Evolution Action]\label{def:evolution-act}
  Given \sidetext{Evolution Action} a TBox $T$, we define an
  \emph{evolution action} $\act^T_e$ over $T$ as a 0-ary (i.e., has no
  action parameters), where $\eff{\act^T_e}$ is the smallest set of
  effects containing:

\begin{compactitem}

\item for each assertion $\funct{R} \in T$, we have
\begin{center}
$
\map{\exists z.\qunsatf(\funct{R}, x, y, z) \land R^{n}(x, y)} {\del
  \set{R(x, z)}} \in \eff{\act^T_e},
$
\end{center}

\item for each negative concept inclusion assertion $B_1 \ISA \neg
  B_2$ such that $T \models B_1 \ISA \neg B_2$, we have
\begin{center}
$
  \map{\qunsatn(B_1 \ISA \neg B_2, x) \land B_1^{n}(x)} \del
  \set{B_2(x)} \in \eff{\act^T_e},
$  \footnote{Note: if $B_1 = \SOMET{P_1}$ (resp.\ $B_2 = \SOMET{P_2}$
    and $B_1^{n} = \SOMET{P_1^n}$), then with a little abuse of
    notation, specifically for this case, we have that the atom
    $B_1(x)$ (resp.\ $B_2(x)$ and $B_1^{n}(x)$) denotes $P_1(x,y)$
    (resp.\ $P_2(x,y)$ and $P_1^n(x,y)$).  For instance, for the
    assertion $\SOMET{P_1} \ISA \neg
    N$, 
    we have
    $\map{P_1(x,y) \land N(x) \land P_1^n(x,y)} {\del \set{N(x)} } \in
    \eff{\act^T_e}$.
    Similarly when $B_1 = \SOMET{P_1^-}$ (resp.\ $B_2 = \SOMET{P_2^-}$
    and $B_1^{n} = \SOMET{P_1^{n-}}$).}

\end{center}

\item for each negative role inclusion assertion $R_1 \ISA \neg R_2$
  such that $T \models R_1 \ISA \neg R_2$, we have:
\begin{center}
$
\map{\qunsatn(R_1 \ISA \neg R_2, x, y) \wedge R_1^{n}(x, y)} {\del
  \set{R_2(x, y)}} \in \eff{\act^T_e},
$
\end{center}

\item for each concept name $N \in \voc(T)$, we have: 
\begin{center}
$
\map{N^{n}(x)}{\del \set{N^{n}(x)}} \in \eff{\act^T_e},
$
\end{center}

\item for each role name $P \in \voc(T)$, we have:
\begin{center}
$
\map{P^{n}(x, y)}{\del \set{P^{n}(x, y)}} \in \eff{\act^T_e}.
$
\end{center}

\item $\map{\true} {\del \set{\tmp} \in \eff{\act^T_e}}.$

\end{compactitem}
\ \ 
\end{definition}

\noindent
Those effects in $\eff{\act^T_e}$ mirror those of
\Cref{CGKABToSGKAB}, with the difference that they
asymmetrically remove old assertions when inconsistency arises. The
last two bullets guarantee that the content of concept and role names
tracking the newly added assertions are flushed.

We now define a translation function $\tgproge$ that essentially
concatenates each action invocation with an evolution action in order
to simulate the action executions in E-GKABs.  Additionally, the
translation function $\tgproge$ also serves as a one-to-one
correspondence (bijection) between the original and the translated
program (as well as between the sub-program).

\begin{definition}[Program Translation $\tgproge$]
  Given \sidetextb{Program Translation $\tgproge$} an E-GKAB
  $\gkabsym = \tup{T, \initabox, \actset,
    \ginitprog}$, 
  we define a \emph{translation $\tgproge$} that translates a program
  $\delta$ into a program $\delta'$ inductively as follows:
  \[
\begin{array}{@{}l@{}l@{}}
  \tgproge(\gact{Q(\vec{p})}{\act(\vec{p})}) &=  
                                               \gact{Q(\vec{p})}{\act'(\vec{p})};\gact{\true}{\act^T_e()}\\
  \tgproge(\gemptyprog) &= \gemptyprog \\
  \tgproge(\delta_1|\delta_2) &= \tgproge(\delta_1)|\tgproge(\delta_2) \\
  \tgproge(\delta_1;\delta_2) &= \tgproge(\delta_1);\tgproge(\delta_2) \\
  \tgproge(\gif{\varphi}{\delta_1}{\delta_2}) &= \gif{\varphi}{\tgproge(\delta_1)}{\tgproge(\delta_2)} \\
  \tgproge(\gwhile{\varphi}{\delta}) &= \gwhile{\varphi}{\tgproge(\delta)}
\end{array}
\]
where 
\begin{compactitem}
\item action $\act'(\vec{p})$ is obtained from
  $\act(\vec{p}) \in \actset$, such that for each effect
  \[
  \map{[q^+]\land Q^-}{\add \facta, \del \factd} \in \eff{\act}
  \]
  we have
  \[
  \map{[q^+]\land Q^-}{\add \facta \cup {\facta}^n, \del \factd} \in
  \eff{\act'}
  \]
  where ${\facta}^n$
  duplicates $\facta$ by using the vocabulary for newly introduced
  atoms and additionally, we also have 
  \[
  \map{\true}{\add \set{\tmp} } \in \eff{\act'} 
  \]

\item $\act^T_e$ is an evolution action over $T$.
\end{compactitem}
\ \ 
\end{definition}

We then define a translation $\tgkabe$ that transforms an E-GKAB to an
S-GKAB as follows.

\begin{definition}[Translation from E-GKAB to S-GKAB]\label{def:translation-egkab-to-sgkab}
  We \sidetext{Translation from E-GKAB to S-GKAB} define a translation
  $\tgkabe$ that, given an E-GKAB
  $\gkabsym = \tup{T, \initabox, \actset, \ginitprog}$, generates an
  S-GKAB
  $\tgkabe(\gkabsym) = \tup{T_p \cup T^n, \initabox, \actset' \cup
    \set{\act^T_e}, \ginitprog'}$, where:
\begin{compactitem}
\item $T_p$ is the positive inclusion assertions of $T$ (see \Cref{def:dllitea-tbox}),
\item $T^n$ is obtained from $T$ by renaming each concept name $N$ in
  $T$ into $N^n$ (similarly for roles). In this way, the original
  concepts/roles are only subject in $\tgkabe(\gkabsym)$ to the
  positive inclusion assertions of $T$ (i.e., $T_p$), while
  concepts/roles tracking newly inserted assertions are subject also
  to negative constraints. This blocks the generation of the successor
  state when the assertions to be added to the current ABox are
  $T$-inconsistent.
\item $\actset'$ is obtained by translating each action in
  $\act(\vec{p}) \in \actset$ into action $\act'(\vec{p})$, such that
  for each effect 
  \[
  \map{[q^+]\land Q^-}{\add \facta, \del \factd} \in \eff{\act}
  \]
  we have
  \[
  \map{[q^+]\land Q^-}{\add \facta \cup {\facta}^n, \del \factd} \in
  \eff{\act'}
  \]
  where ${\facta}^n$
  duplicates $\facta$ by using the vocabulary for newly introduced
  atoms and additionally, we also have 
  \[
  \map{\true}{\add \set{\tmp} } \in \eff{\act'} 
  \]
%
  \item $\ginitprog' = \tgproge(\ginitprog)$. 
%
\end{compactitem}
\ \ 
\end{definition}

By exploiting the same \muladom translation used in
\Cref{CGKABToSGKAB} (i.e., the translation $\tford$ in
\Cref{def:tdup}), we will show later that
$\ts{\gkabsym}^{\filter_E} \models \Phi$ if and only if
$\ts{\tgkabe(\gkabsym)}^{\filter_S} \models \tford(\Phi)$. Hence
reducing the \muladom verification over E-GKABs to S-GKABs.
The strategy of the proof is similar to the reduction from the
verification of C-GKABs into the verification of S-GKABs in
\Cref{CGKABToSGKAB}. I.e., to show that the transition system of an
E-GKAB is S-bisimilar to the transition system of its corresponding
S-GKAB, and hence they can not be distinguish by any \muladom formulas
modulo translation $\tford$.


\subsubsection{Recasting the Verification of E-GKABs Into S-GKABs}

As the first step, we show an important property of the filter
$\filter_E$ (which is also a property of $\evol$
operator). Specifically, we show that every ABox assertion in the
evolution result is either a new assertion or it was already in the
original ABox and it was not deleted as well as did not violate any
TBox constraints (together with another ABox assertions). Formally the
claim is stated below.

\begin{lemma}\label{lem:evol-prop}
  Given
  \begin{inparaenum}[]
  \item a TBox $T$,
  \item a $T$-consistent ABox $A$,
  \item a $T$-consistent set $\facta$ of ABox assertions to be added,
    and
  \item a set $\factd$ of ABox assertions to be deleted
  \end{inparaenum}
  such that $A_e = \evol(T, A, \facta, \factd)$, 
  we have $N(c) \in A_e$
  if and only if either
  \begin{compactenum}
  \item $N(c) \in \facta$, or
  \item $N(c) \in (A \setminus \factd)$ and there does not exists
    $B(c) \in \facta$ such that
    $T \models N \sqsubseteq \neg B$.
  \end{compactenum}
  (Similarly for the case of role assertion with the corresponding
  violation of negative role inclusion or functional assertion).
\end{lemma}
\begin{proof}
  The intuition of the correctness of this claim is simply obtained
  from the definition of bold-evolution operator itself
  (cf. \Cref{def:bold-evol-abox-icma}).  I.e.,
  \[
  \evol(T, A, \facta, \factd) = \facta \cup A'
  \]
  where 
  \begin{compactitem}
  \item $\facta$ is a set of newly added assertions,
  \item $A'$ is a set of ABox assertions that is obtained from $A$ by
    throwing away those assertions that are either also belongs to
    $\factd$ or have a conflict with some assertions in $\facta$. 
  \end{compactitem}
  Hence, it is easy to see that an assertion is belong to the result
  of the application of bold-evolution if it is either
  \begin{compactenum}
  \item a newly added assertion, or
  \item an assertion that is not deleted and doesn't have a conflict
    with the newly added assertion.
  \end{compactenum}
%
%
  Technically, the proof is as follows:

\begin{compactitem} 
\item[``$\Lora$'':]
  Assume $N(c) \in A_e$, since
  $A_e = \evol(T, A, \facta, \factd)$, 
  by the definition of $\evol(T, A, \facta, \factd)$,
  we have $A_e = \facta \cup A'$, where
  \begin{compactenum}
  \item $A' \subseteq (A \setminus \factd)$,
  \item $\facta \cup A'$ is $T$-consistent, and
  \item there does not exists $A''$ such that
    $A' \subset A'' \subseteq (A \setminus \factd)$ and
    $\facta \cup A''$ is $T$-consistent.
  \end{compactenum}
  Hence, we have either
  \begin{compactenum}[(1)]
  \item $N(c) \in \facta$, or
  \item $N(c) \in A'$. 
  \end{compactenum}
  For the case (2), as a consequence:
  \begin{compactitem}
  \item Since $N(c) \in A'$ and $A' \subseteq (A \setminus \factd)$ it
    follows that $N(c) \in (A \setminus \factd)$.
  \item Since $F^+ \cup A'$ is $T$-consistent, then we have that there
    does not exists $B(c) \in \facta$ s.t.\
    $T \models N \sqsubseteq \neg B$.
  \end{compactitem}
  Thus, the claim is proven.

\smallskip
\item[``$\Lola$'':]
We divide the proof into two parts:
\begin{compactenum}[(1)]

\item Assume $N(c) \in \facta$. Then simply by the definition of
  $\evol(T, A, \facta, \factd)$, we have $N(c) \in A_e$.

\item Supposed by contradiction we have that
  $N(c) \in (A \setminus \factd)$ and there does not exists
  $B(c) \in \facta$ s.t.\ $T \models N \sqsubseteq \neg B$, and
  $N(c) \not\in A_e$. Since $N(c) \not\in A_e$, by the definition of
  $\evol(T, A, \facta, \factd)$, we have that $N(c) \not\in \facta$
  and $N(c) \not\in A'$ in which $A'$ should satisfies the following:
  \begin{compactitem}
  \item $A' \subseteq (A \setminus \factd)$,
  \item $\facta \cup A'$ is $T$-consistent, and
  \item there does not exists $A''$ such that
    $A' \subset A'' \subseteq (A \setminus \factd)$ and $\facta\cup A''$ is
    $T$-consistent.
  \end{compactitem}
  But then we have a contradiction since there exists
  $A'' = A' \cup \set{N(c)}$ such that
  $A' \subset A'' \subseteq (A \setminus \factd)$ and $\facta\cup A''$
  is $T$-consistent. 
  Hence, we must have $N(c) \in A_e$.
\end{compactenum}
\end{compactitem}

\end{proof}

Now we show a property of evolution action $\act^T_e$ which says that
every ABox assertion in the result of the execution of $\act^T_e$ is
either a newly added assertion, or an old assertion that does not
violate any TBox constraints. Precisely we state this property below.

\begin{lemma}\label{lem:eact-prop}
  Given 
\begin{compactitem}
\item an E-GKAB
  $\gkabsym = \tup{T, \initabox, \actset_e, \ginitprog_e}$ with
  transition system $\ts{\gkabsym}^{\filter_E}$,
  and 
\item an S-GKAB
  $\tgkabe(\gkabsym) = \tup{T_s, \initabox, \actset_s, \ginitprog_s}$
  $($with transition system $\ts{\tgkabe(\gkabsym)}^{\filter_S}$$)$
  that is obtained from $\gkabsym$ through $\tgkabe$,
  where $T_s = T_p \cup T^n$.
\end{compactitem}
Let $\tup{A, \scmap, \delta}$ be any state in
$\ts{\tgkabe(\gkabsym)}^{\filter_S}$, $\act' \in \actset_s$ be any
action, $A$ is $T_s$-consistent and does not contain any ABox
assertions constructed from $\voc(T^n)$ and we have:
\[
\tup{A, \scmap, \delta} 
\gprogtrans{\act'\sigma, \filter_S} 
\tup{A', \scmap', \delta'} 
\gprogtrans{\act^T_e\sigma', \filter_S} 
\tup{A'', \scmap'', \delta''}
\]
for
\begin{compactitem}
\item a particular legal parameter assignment $\sigma$
\item an empty substitution $\sigma'$,
\item a particular service call evaluation
  $\theta \in \eval{\addfacts{T, A, \act'\sigma}}$ that agree with
  $\scmap$ on the common values in their domains.
\end{compactitem}
We have $N(c) \in A''$
if and only if $N$ is not in the vocabulary of TBox $T^n$ and either
\begin{compactenum}
\item $N(c) \in \addfacts{T_s, A, \act'\sigma}\theta$, or
\item $N(c) \in (A \setminus \delfacts{T_s, A, \act'\sigma})$ and there
  does not exists $B(c) \in \addfacts{T_s, A, \act'\sigma}\theta$ such
  that $T \models N \sqsubseteq \neg B$.
\end{compactenum}
(Similarly for the case of role assertion with the corresponding
violation of negative role inclusion or functional assertion).
\end{lemma}
\begin{proof} 
First of all, it will
not be the case that $N(c)$ is $\tmp$ since in any case $\act^T_e$
deletes $\tmp$.  The proof of this Lemma is then divided into two
parts as follows:
\begin{compactitem}
\item[``$\Lora$'':] Assume $N(c) \in A''$, since the evolution action
  $\act^T_e$ only
\begin{compactenum}
\item removes old assertions when inconsistency arises,
\item flushes every ABox assertions constructed by the vocabulary of
  $T^n$,
\end{compactenum}
then we have the following:
\begin{compactenum}
\item $N$ is not in the vocabulary of TBox $T^n$ (otherwise it will
  be flushed by $\act^T_e$)
\item $N(c) \in A'$ (because $\act^T_e$ never introduce a new ABox
  assertion),
\item if there exists $B(c) \in A'$ such that
  $T \models N \sqsubseteq \neg B$, then $B(c) \not\in A''$,
  $B^n(c) \not\in A'$, and $N^n(c) \in A'$ (i.e., if
  $N(c) \in A'$ violates a negative inclusion assertion, $N(c)$
  must be a newly added ABox assertion, otherwise it will be deleted
  by $\act^T_e$).
\end{compactenum}
Now, since $A$ and $A'$ are $T_s$-consistent (because
$\tup{A, \scmap, \delta} \gprogtrans{\act'\sigma, \filter_S} \tup{A',
  \scmap', \delta'} $),
then $\addfacts{T, A, \act'\sigma}\theta$ is $T_s$-consistent.  Hence
we have either
\begin{compactenum}
\item $N(c) \in \addfacts{T, A, \act'\sigma}\theta$ (and there does
  not exists $B(c)$ such that
  $B(c) \in \addfacts{T, A, \act'\sigma}\theta$, and
  $T \models N \sqsubseteq \neg B$), or
\item $N(c) \in (A \setminus \delfacts{T, A, \act'\sigma})$ and
  there does not exists 
  $B(c) \in \addfacts{T, A, \act'\sigma}\theta$ such that
  $T \models N \sqsubseteq \neg B$ (otherwise we have
  $\set{N(c), B(c), B^n(c)} \subseteq A'$ and then $N(c)$ will
  be deleted by $\act_e^T$).
\end{compactenum}
Therefore, the claim is proved.

\smallskip
\item[``$\Lola$'':]
  Assume $N$
  is not in the vocabulary of TBox $T^n$,
  then we divide the proof into two parts:
\begin{compactenum}

\item Assume $N(c) \in \addfacts{T, A, \act'\sigma}\theta$. Then, by
  the construction of $\act'$ and the definition of
  $\gprogtrans{\act'\sigma, \filter_S}$, it is easy to see that
  $\set{N(c), N^n(c)} \subseteq A'$. Therefore $N(c) \in A''$ (by construction of
  $\act^T_e$).

\item Assume $N(c) \in (A \setminus \delfacts{T, A, \act'\sigma})$ and
  there does not exists $B(c) \in \addfacts{T, A, \act'\sigma}\theta$
  such that $T \models N \sqsubseteq \neg B$. Hence, by the definition
  of $\gprogtrans{\act'\sigma, \filter_S}$, we have $N(c) \in A'$.
  Moreover, because $N(c) \in A'$ does not violate any negative
  inclusion assertions, by construction of $\act^T_e$, we also simply
  have $N(c) \in A''$.
\end{compactenum}

\end{compactitem}

\end{proof}


Next, in the following two Lemmas we aim to show that the transition
system of an E-GKAB is S-bisimilar to the transition system of its
corresponding S-GKAB that is obtained from translation $\tgkabe$.

\begin{lemma}\label{lem:egkab-to-sgkab-bisimilar-state}
  Let $\gkabsym = \tup{T, \initabox, \actset, \ginitprog}$ be an E-GKAB with transition system
  $\ts{\gkabsym}^{\filter_E}$, and let $\tgkabe(\gkabsym) = \tup{T_s, \initabox, \actset_s, \ginitprog_s}$ be an
  S-GKAB with transition system $\ts{\tgkabe(\gkabsym)}^{\filter_S}$
  obtained through $\tgkabe$. 
%
  Consider
\begin{inparaenum}[]
\item a state $\tup{A_e,\scmap_e, \delta_e}$ of
  $\ts{\gkabsym}^{\filter_E}$ and
\item a state $\tup{A_s,\scmap_s, \delta_s}$ of
  $\ts{\tgkabc(\gkabsym)}^{\filter_S}$.
\end{inparaenum}
If  
\begin{inparaenum}[]
\item $A_s = A_e$, $\scmap_s = \scmap_e$, $A_s$ is $T_s$-consistent
  and 
\item $\delta_s = \tgproge(\delta_e)$, 
\end{inparaenum}
then
$\tup{A_e,\scmap_e, \delta_e} \sbsim \tup{A_s,\scmap_s, \delta_s}$.
\end{lemma}
\begin{proof}
  For the simplicity of the proof, here we ignore the presence of
  $\tmp$ that acts as a special marker (that marks an intermediate
  state). The important thing to observe is that $\tmp$ is always
  added to the intermediate state (where we need to execute the
  evolution action) but then it will be deleted after
  that. 
  Now, let
\begin{inparaitem}[]
\item 
  $\ts{\gkabsym}^{\filter_E} = \tup{\const, T, \stateset_e, s_{0e},
    \abox_e, \trans_e}$, and  
\item
%
  $\ts{\tgkabe(\gkabsym)}^{\filter_S} = \tup{\const, T_s,
    \stateset_s, s_{0s}, \abox_s, \trans_s}$. 
\end{inparaitem}
We have to show the following: for every state
$\tup{A''_e,\scmap''_e, \delta''_e}$ such that
  \[
  \tup{A_e,\scmap_e, \delta_e} \trans \tup{A''_e,\scmap''_e,
    \delta''_e},
  \]
  there exist states $\tup{A'_s,\scmap'_s, \delta'_s}$ and
  $\tup{A''_s,\scmap''_s, \delta''_s}$ such that:
\begin{compactenum}[\bf (a)]
\item
  $ \tup{A_s,\scmap_s, \delta_s} \trans_s \tup{A'_s,\scmap'_s,
    \delta'_s} \trans_s \tup{A''_s,\scmap''_s, \delta''_s} $
\item $A''_s = A''_e$;
\item $\scmap''_s = \scmap''_e$;
\item $\delta''_s = \tgproge(\delta''_e)$.
\end{compactenum}
By definition of $\ts{\gkabsym}^{\filter_E}$, 
since $\tup{A_e,\scmap_e, \delta_e} \trans \tup{A''_e,\scmap''_e,
  \delta''_e}$, we have $\tup{A_e,\scmap_e, \delta_e}
\gprogtrans{\act\sigma_e, \filter_E} \tup{A''_e,\scmap''_e,
  \delta''_e}$.
Hence, by the definition of $\gprogtrans{\act\sigma_e, \filter_E}$, we
have:
\begin{compactitem}


\item
  $\tup{\tup{A_e, \scmap_e}, \act\sigma_e, \tup{A''_e, \scmap''_e}}
  \in \tell_{\filter_E}$, and 

\item $\sigma_e$ is a legal parameter assignment for $\act$ in $A_e$
  w.r.t.\ $\gact{Q(\vec{p})}{\act(\vec{p})}$ (i.e.,
  $\ask(Q\sigma_e, T, A_e) = \true$).

\end{compactitem}
Since
$\tup{\tup{A_e, \scmap_e}, \act\sigma_e, \tup{A''_e, \scmap''_e}} \in
\tell_{\filter_E}$,
by the definition of $\tell_{\filter_E}$, there exists
$\theta_e \in \eval{\addfacts{T ,A_e, \act\sigma_e}}$ such that
\begin{compactitem}
\item $\theta_e$ and $\scmap_e$ agree on the common values in their domains.
\item $\scmap''_e = \scmap_e \cup \theta_e$.
\item
  $\tup{A_e, \addfacts{T, A_e, \act\sigma_e}\theta_e, \delfacts{T, A_e,
    \act\sigma_e}, A_e''} \in \filter_E$.
\item $A''_e$ is $T$-consistent.
\end{compactitem}
Since
$\tup{A_e, \addfacts{T, A_e, \act\sigma_e}\theta_e, \delfacts{T, A_e,
    \act\sigma_e}, A_e''} \in \filter_E$,
by the definition of $\filter_E$, we have
\begin{compactitem}
\item $\addfacts{T, A_e, \act\sigma_e}\theta_e$ is $T$-consistent.
\item $ A_e'' = \evol(T, A_e, \addfacts{T, A_e, \act\sigma_e}\theta_e,
\delfacts{T, A_e, \act\sigma_e})$.
\end{compactitem}
Furthermore, since $\delta_s = \tgproge(\delta_e)$, by the definition of
$\tgproge$, we have that
\[
\begin{array}{l}
  \tgproge(\gact{Q(\vec{p})}{\act(\vec{p})}) =  \gact{Q(\vec{p})}{\act'(\vec{p})}; \gact{\true}{\act^T_e()}
\end{array}
\]
Hence, the part of program that we need to execute on state
$\tup{A_s,\scmap_s, \delta_s}$ is
\[
\gact{Q(\vec{p})}{\act'(\vec{p})}; \gact{\true}{\act^T_e()}.
\]
%
%

Now, since: 
\begin{compactitem}
\item $\act' \in \actset_s$ is obtained from $\act \in \actset$
through $\tgkabe$,
%
%
\item the translation $\tgkabe$ transform $\act$ into $\act'$ without
  changing its parameters, 
\item $\sigma_e$ maps parameters of $\act \in \actset$ to constants
  in $\adom{A_e}$
\end{compactitem}
then we can construct $\sigma_s$ mapping parameters of
$\act' \in \actset_s$ to constants in $\adom{A_s}$ such that
$\sigma_s = \sigma_e$
%
Moreover, since $A_s = A_e$, we know that the certain answers computed
over $A_e$ are the same to those computed over $A_s$. Hence
%
%
$\act' \in \actset_s$ is executable in $A_s$ with (legal parameter
assignment) $\sigma_s$.
Furthermore, since $\scmap_s= \scmap_e$, then we can construct
$\theta_s$, such that $\theta_s = \theta_e$. 
%
%
Hence, we have the following:
\begin{compactitem}
\item $\theta_s$ and $\scmap_s$ agree on the common
values in their domains. 
\item
  $\scmap'_s = \theta_s \cup \scmap_s = \theta_e \cup \scmap_e =
  \scmap_e''$.
\end{compactitem}
Let
$A_s' = (A_s \setminus \delfacts{T_s, A_s, \act'\sigma_s}) \cup
\addfacts{T_s, A_s, \act'\sigma_s}\theta_s$,
as a consequence, we have
$\tup{A_s, \addfacts{T_s, A_s, \act'\sigma_s}\theta_s, \delfacts{T_s,
    A_s, \act'\sigma_s}, A_s'} \in \filter_S$.

Since $A_s = A_e$, $\theta_s = \theta_e$, and $\sigma_s = \sigma_e$,
it follows that
\begin{compactitem}
\item $\delfacts{T, A_e, \act\sigma_e}  =  \delfacts{T_s, A_s, \act'\sigma_s}$.
\item $N(c) \in \addfacts{T, A_e, \act\sigma_e}\theta_e$ if and only
  if $N(c), N^n(c) \in \addfacts{T_s, A_s, \act'\sigma_s}\theta_s$.
\item $P(c_1,c_2) \in \addfacts{T, A_e, \act\sigma_e}\theta_e$ if and
  only if
  $P(c_1,c_2), P^n(c_1,c_2) \in \addfacts{T_s, A_s,
    \act'\sigma_s}\theta_s$.
\end{compactitem}



\noindent
As a consequence, since $\addfacts{T, A_e, \act\sigma_e}\theta_e$ is
$T$-consistent, then we have
$\addfacts{T_s, A_s, \act'\sigma_s}\theta_s$ is $T_s$-consistent.
Moreover, 
because $A_s$ is $T_s$-consistent,
$\addfacts{T_s, A_s, \act'\sigma_s}\theta_s$ is $T_s$-consistent, and
also considering how $A_s'$ is constructed, we then have $A_s'$ is
$T_s$-consistent.
Thus we have
$\tup{\tup{A_s,\scmap_s}, \act'\sigma_s, \tup{A_s', \scmap_s'}} \in
\tell_{\filter_S}$, and we also have
\[
\tup{A_s,\scmap_s,
  \gact{Q(\vec{p})}{\act'(\vec{p})};\gact{\true}{\act^T_e()}}
\gprogtrans{\act'\sigma_s, \filter_S} \tup{A_s', \scmap_s',
  \gact{\true}{\act^T_e()}}.
\]
It is easy to see that we have
\[
\begin{array}{@{}l@{}l@{}}
  \tup{A_s', \scmap_s', \gact{\true}{\act^T_e()}}
  \gprogtrans{\act^T_e\sigma'_s, \filter_S} 
  \tup{A_s'', \scmap_s'', \gemptyprog}
\end{array}
\]
 where 
\begin{compactitem}
\item $\tup{A_s'', \scmap_s'', \gemptyprog}$ is a final state
\item $\sigma'_s$ is empty legal parameter assignment (because $\act^T_e$ is
  0-ary action).
\item $\scmap_s'' = \scmap_e''$, (due to the fact that $\act^T_e$ does
  not involve any service call (i.e., $\scmap_s'' = \scmap_s'$) and
  $\scmap_s' = \scmap_e''$).
\end{compactitem}
Additionally, by the definition of $\tgproge$, we have
$\delta''_s = \tgproge(\delta''_e)$ as the rest of the program to be
executed (because $\tup{A_s'', \scmap_s'', \gemptyprog}$ is a final
state). Hence, we have
\[
\tup{A_s,\scmap_s, \delta_s} \trans_s 
\tup{A'_s,\scmap'_s, \delta'_s} \trans_s 
\tup{A''_s,\scmap''_s, \delta''_s}
\]
%
To complete the proof, we obtain $A_s'' = A_e''$ simply as a
consequence of the following facts:
\begin{compactenum}
\item $A_s = A_e$;

\item $\delfacts{T, A_e, \act\sigma_e}  =  \delfacts{T_s, A_s, \act'\sigma_s}$.
\item $N(c) \in \addfacts{T, A_e, \act\sigma_e}\theta_e$ if and only
  if $N(c), N^n(c) \in \addfacts{T_s, A_s, \act'\sigma_s}\theta_s$.
\item $P(c_1,c_2) \in \addfacts{T, A_e, \act\sigma_e}\theta_e$ if and
  only if
  $P(c_1,c_2), P^n(c_1,c_2) \in \addfacts{T_s, A_s,
    \act'\sigma_s}\theta_s$.

\item   By \Cref{lem:evol-prop}, we have $N(c) \in A''_e$ if and only
  if either
\begin{compactitem}[$\bullet$]
\item $N(c) \in \addfacts{T, A_e, \act\sigma_e}\theta_e$, or
\item $N(c) \in (A_e \setminus \delfacts{T, A_e, \act\sigma_e})$ and
  there does not exists
  $B(c) \in \addfacts{T, A_e, \act\sigma_e}\theta_e$ such that
  $T \models N \sqsubseteq \neg B$;
\end{compactitem}

\item By \Cref{lem:eact-prop}, we have $N(c') \in A''_s$ if and only
  if $N$ is not in the vocabulary of TBox $T^n$ and either
\begin{compactitem}[$\bullet$]
\item $N(c') \in \addfacts{T_s, A_s, \act'\sigma_s}\theta_s$, or
\item $N(c') \in (A_s \setminus \delfacts{T_s, A_s, \act\sigma_s})$ and
  there does not exists
  $B(c') \in \addfacts{T_s, A_s, \act'\sigma_s}\theta_s$ such that
  $T \models N \sqsubseteq \neg B$.
\end{compactitem}

\item $\act^T_e$ flushes all ABox assertions made by using
  $\voc(T^n)$.


\end{compactenum}
The other direction of bisimulation relation can be proven
in a similar way.

\end{proof}

Having \Cref{lem:egkab-to-sgkab-bisimilar-state} in hand, we can
easily show that given an E-GKAB, its transition system is S-bisimilar
to the transition of its corresponding S-GKAB that is obtained via the
translation $\tgkabe$.

\begin{lemma}\label{lem:egkab-to-sgkab-bisimilar-ts}
  Given an E-GKAB $\gkabsym$, we have
  $\ts{\gkabsym}^{\filter_E} \sbsim \ts{\tgkabe(\gkabsym)}^{\filter_S}
  $
\end{lemma}
\begin{proof}
Let
\begin{compactenum}
\item $\gkabsym = \tup{T, \initabox, \actset, \ginitprog_e}$ and
  $\ts{\gkabsym}^{\filter_E} =
  \tup{\const, T, \stateset_e, s_{0e}, \abox_e, \trans_e}$,
\item
  $\tgkabe(\gkabsym) = \tup{T_s, \initabox, \actset_s, \ginitprog_s}$
  and %
  $\ts{\tgkabe(\gkabsym)}^{\filter_S} = \tup{\const, T_s, \stateset_s,
    s_{0s}, \abox_s, \trans_s}$)
\end{compactenum}
We have that $s_{0e} = \tup{A_0, \scmap_e, \delta_e}$ and
$s_{0s} = \tup{A_0, \scmap_s, \delta_s}$ where
$\scmap_e = \scmap_s = \emptyset$. 
By the definition of $\tgproge$ and $\tgkabe$, we also have
$\delta_s = \tgproge(\delta_e)$.  Hence, by
\Cref{lem:egkab-to-sgkab-bisimilar-state}, we have
$s_{0e} \sbsim s_{0s}$. Therefore, by the definition of
S-Bisimulation, we have
$\ts{\gkabsym}^{\filter_E} \sbsim \ts{\tgkabe(\gkabsym)}^{\filter_S}$.
\end{proof}

Having all of the ingredients in hand, we are now ready to show that
the verification of \muladom properties over E-GKABs can be recast as
verification over S-GKAB as follows.

\begin{theorem}\label{thm:egkab-to-sgkab}
  Given an E-GKAB $\gkabsym$ and a closed $\muladom$ property $\Phi$,
\begin{center}
  $\ts{\gkabsym}^{\filter_E} \models \Phi$ iff
  $\ts{\tgkabe(\gkabsym)}^{\filter_S} \models \tford(\Phi)$
\end{center}
\end{theorem} 
\begin{proof}
  By \Cref{lem:egkab-to-sgkab-bisimilar-ts}, we have that
  $\ts{\gkabsym}^{\filter_E} \sbsim
  \ts{\tgkabe(\gkabsym)}^{\filter_S}$.
  Hence, by \Cref{lem:sbisimilar-ts-satisfies-same-formula}, we have
  that the claim is proven.
\end{proof}

\subsection{Putting It All Together: Verification of I-GKABs}


To sum up, we state the result of I-GKABs verification as follows:

\begin{theorem}
\label{thm:itos}
Verification of \muladom properties over I-GKABs can be recast as
verification over S-GKABs.
\end{theorem}
\begin{proof}
  As a consequence of
  \Cref{thm:bgkab-to-sgkab,thm:cgkab-to-sgkab,thm:egkab-to-sgkab}, we
  essentially show that the verification of \muladom properties over
  I-GKABs can be recast as verification over S-GKABs since we can
  recast the verification of \muladom properties over B-GKABs,
  C-GKABs, and E-GKABs as verification over S-GKABs.
\end{proof}

From \Cref{thm:itos,thm:gtos}, we get our next major result that
verification of all inconsistency-aware variants of GKABs introduced
in \Cref{sec:ia-gkab} can be compiled into verification of KABs by
concatenating the two translations from I-GKABs to S-GKABs, and then
to KABs.
\begin{theorem}
\label{thm:itoverys}
Verification of \muladom properties over I-GKABs can be recast as
verification over KABs.
\end{theorem}
\begin{proof}
  The proof is easily obtained from the \Cref{thm:itos,thm:gtos},
  since by \Cref{thm:itos} we can recast the verification of \muladom
  over I-GKABs as verification over S-GKABs and then by
  \Cref{thm:gtos} we can recast the verification of \muladom over
  S-GKABs as verification over KABs. Thus combining those two
  ingredients, we can reduce the verification of \muladom over I-GKABs
  into the corresponding verification of \muladom over KABs.
\end{proof}

\subsection{Verification of Run-bounded I-GKABs}

We now aim to show that the reductions from I-GKABs to S-GKABs
preserve run-boundedness. 

\begin{lemma}\label{lem:run-bounded-preservation-bgkab}
  Let $\gkabsym$ be a B-GKAB and $\tgkabb(\gkabsym)$ be its
  corresponding S-GKAB. We have $\gkabsym$ is run-bounded if and only
  if $\tgkabb(\gkabsym)$ is run-bounded.
\end{lemma}
\begin{proof}
  Let
  \begin{compactenum}
  \item $\gkabsym = \tup{T, \initabox, \actset, \ginitprog}$ and
    $\ts{\gkabsym}^{\filter_B} $ be its transition system.
  \item $\ts{\tgkabb(\gkabsym)}^{\filter_S}$ be the transition system
    of $\tgkabb(\gkabsym)$.
  \end{compactenum}
  The proof is easily obtained due to the following facts:
  \begin{compactitem}
  \item the translation $\tgkabb$ essentially only appends each action
    invocation in $\delta$ with some additional programs to manage
    inconsistency.
  \item the actions introduced to manage inconsistency never inject
    new constants, but only remove facts causing inconsistency,
  \item by \Cref{lem:bgkab-to-sgkab-bisimilar-ts}, we have that
    $\ts{\gkabsym}^{\filter_B} \jbsim \ts{\tgkabb(\gkabsym)}^{\filter_S}$. Thus,
    basically they are equivalent modulo intermediate states (states
    containing $\tmp$).
  \end{compactitem}
\end{proof}

\begin{lemma}\label{lem:run-bounded-preservation-cgkab}
  Let $\gkabsym$ be a C-GKAB and $\tgkabc(\gkabsym)$ be its
  corresponding S-GKAB. We have $\gkabsym$ is run-bounded if and only
  if $\tgkabc(\gkabsym)$ is run-bounded.
\end{lemma}
\begin{proof}
  Similar to the proof of
  Lemma~\ref{lem:run-bounded-preservation-bgkab} but using the
  S-Bisimulation.
\end{proof}

\begin{lemma}\label{lem:run-bounded-preservation-egkab}
  Let $\gkabsym$ be a E-GKAB and $\tgkabe(\gkabsym)$ be its
  corresponding S-GKAB. We have $\gkabsym$ is run-bounded if and only
  if $\tgkabe(\gkabsym)$ is run-bounded.
\end{lemma}
\begin{proof}
  Similar to the proof of
  Lemma~\ref{lem:run-bounded-preservation-bgkab} but using the
  S-Bisimulation.
\end{proof}

Finally, we show the result on the verification of \muladom properties
over run-bounded I-GKABs as follows.

\begin{theorem}
  Verification of \muladom properties over run-bounded I-GKABs is
  decidable, and reducible to standard $\mu$-calculus finite-state
  model checking.
\end{theorem}
\begin{proof}
  By
  \Cref{lem:run-bounded-preservation-bgkab,lem:run-bounded-preservation-cgkab,lem:run-bounded-preservation-egkab},
  the translation from I-GKABs to S-GKABs preserves run-boundedness.
%
%
  Thus, the claim follows by combining \Cref{thm:itos} and
  \Cref{thm:ver-run-bounded-sgkab}.
\end{proof}

\section{Back From Standard to Inconsistency-aware GKABs}\label{sec:cap-sgkabs-to-igkabs}

So far we have seen how we can reduce the verification of I-GKABs into
S-GKABs. In this section we show the other direction of reduction,
i.e., we show that we can also recast the verification of S-GKABs into
I-GKABs. In particular, we show how we reduce the verification of
S-GKABs into B-GKABs. The reductions from S-GKABs into C-GKABs and
E-GKABs can be done similarly.
The following sections are then organized as follows: first we explain
how we translate S-GKABs into B-GKABs. 
Then, 
we show that the transition system of S-GKABs and B-GKABs that is
obtained from our translation are S-bisimilar. As a consequence,
utilizing the result in
\Cref{lem:sbisimilar-ts-satisfies-same-formula}, we show that we can
reduce the verification of S-GKABs into B-GKABs.

The main challenge of this reduction is how to make B-GKABs mimic the
standard execution semantics such that they stop evolving (instead of
doing the repair) when there is an inconsistency. 
To deal with this, the core strategy of the translation from S-GKABs
into B-GKABs is as follows:

\begin{compactenum}
\item We 
  throw away all of the negative inclusion assertions as well as the
  functionality assertions from the current TBox and keep only the
  positive inclusion assertion. The purpose of this step is to avoid
  the changes by the repair mechanism when there is an
  inconsistency. Note that any ABox will be consistent with the TBox
  that has only positive inclusion assertions.
%

\item We delegate the inconsistency check to a certain action that
  checks the violation of negative inclusion assertions as well as the
  functionality assertions. 
  To do this, we exploit the fact that the inconsistency can be
  checked through query answering.


\end{compactenum}

Before defining the translation from S-GKABs into B-GKABs, we first
introduce the translation for program in S-GKABs. Particularly, we
define a translation function $\tgprogsb$ that essentially
concatenates each action invocation with an action invocation that
checks the inconsistency.  Additionally, the translation function
$\tgprogsb$ also serves as a one-to-one correspondence (bijection)
between the original and the translated program (as well as between
the sub-program).

\clearpage
\begin{definition}[Program Translation $\tgprogsb$]
  Given \sidetextb{Program Translation $\tgprogsb$} a set of actions $\actset$, a program $\delta$ over $\actset$,
  and a TBox $T$, we define a \emph{translation $\tgprogsb$} which
  translates a program into a program inductively as follows:
\[
\begin{array}{@{}l@{}l@{}}
  \tgprogsb(\gact{Q(\vec{p})}{\act(\vec{p})}) &=  
                                                \gact{Q(\vec{p})
                                                }{\act'(\vec{p})};\gact{\neg\qunsatecq{T}}{\act_\bot()}\\
  \tgprogsb(\gemptyprog) &= \gemptyprog \\
  \tgprogsb(\delta_1|\delta_2) &= \tgprogsb(\delta_1)|\tgprogsb(\delta_2) \\
  \tgprogsb(\delta_1;\delta_2) &= \tgprogsb(\delta_1);\tgprogsb(\delta_2) \\
  \tgprogsb(\gif{\varphi}{\delta_1}{\delta_2}) &= \gif{\varphi}{\tgprogsb(\delta_1)}{\tgprogsb(\delta_2)} \\
  \tgprogsb(\gwhile{\varphi}{\delta}) &= \gwhile{\varphi}{\tgprogsb(\delta)}
\end{array}
\]
where 
\begin{compactitem}
\item action $\act'(\vec{p})$ is obtained from
  $\act(\vec{p}) \in \actset$, such that
  \[
  \eff{\act'} = \eff{\act} \cup \set{\map{\true}{\add \set{\tmp} } }
  \]

\item $\qunsatecq{T}$ is a boolean Q-UNSAT-ECQ over $T$ (similar to
  Q-UNSAT-FOL in \Cref{def:qunsat-fol}) that is used to check the
  inconsistency. It will be evaluated to true if the ABox is
  $T$-inconsistent.

\item $\act_\bot$ is a 0-ary action of the form
  \[
  \act_{\bot}():\set{\map{\true}{\del \set{\tmp} }}
  \]
\end{compactitem}
\ \ 
\end{definition}

Finally, to compile an S-GKAB into the corresponding B-GKAB, we define
a translation $\tgkabsb$ that, given an S-GKAB, generates a B-GKAB as
follows.

\begin{definition}[Translation from S-GKAB to B-GKAB]
  We \sidetext{Translation from S-GKAB to B-GKAB} define a translation
  $\tgkabsb$ that, given an S-GKAB
  $\gkabsym = \tup{T, \initabox, \actset, \ginitprog}$, generates a
  B-GKAB
  $\tgkabsb(\gkabsym) = \tup{T_p, \initabox, \actset' \cup
    \set{\act_\bot}, \ginitprog'}$, where
  \begin{compactitem}
  \item $T_p$ is the positive inclusion assertions of $T$ (see \Cref{def:dllitea-tbox}),
  \item $\actset'$ is obtained from $\actset$ as follows: for each
    $\act \in \actset$, we have $\act' \in \actset$, where
    $ \eff{\act'} = \eff{\act} \cup \set{\map{\true}{\add \set{\tmp} }
    }$
  \item $\act_\bot$ is an action of the form $\act_{\bot}():\set{\map{\true}{\del \set{\tmp} }}$,
  \item $\ginitprog' = \tgprogsb(\ginitprog)$.
\end{compactitem}
 \ \ 
\end{definition}

\noindent
Essentially, the translation above only keeps the positive inclusion
assertions in order to prevent the repair. Moreover, it delegates the
inconsistency checks into query answering.


To the aim of reducing the verification of S-GKABs into B-GKABs, in
the following owe show that the transition system of an S-GKAB is
S-bisimilar to the transition system of its corresponding B-GKAB that
is obtained via the translation $\tgkabsb$.

\begin{lemma}\label{lem:sgkab-to-bgkab-bisimilar-state}
  Let $\gkabsym$ be an S-GKAB with transition system
  $\ts{\gkabsym}^{\filter_S}$, and let $\tgkabsb(\gkabsym)$ be a
  B-GKAB with transition system $\ts{\tgkabsb(\gkabsym)}^{\filter_B}$
  obtained through
  $\tgkabsb$. 
  Consider
  \begin{inparaenum}[]
  \item a state $\tup{A_s,\scmap_s, \delta_s}$ of
    $\ts{\gkabsym}^{\filter_S}$ and
  \item a state $\tup{A_b,\scmap_b, \delta_b}$ of
    $\ts{\tgkabsb(\gkabsym)}^{\filter_B}$.
  \end{inparaenum}
If  
\begin{inparaenum}[]
\item $A_s = A_b$, $\scmap_s = \scmap_b$ and
\item $\delta_s = \tgprogsb(\delta_b)$,
\end{inparaenum}
then
$\tup{A_s,\scmap_s, \delta_s} \sbsim \tup{A_b,\scmap_b, \delta_b}$.
\end{lemma}
\begin{proof}
  Similar line of reasoning as in the proof of
  \Cref{lem:bgkab-to-sgkab-bisimilar-state,lem:cgkab-to-sgkab-bisimilar-state,lem:egkab-to-sgkab-bisimilar-state}
  can be applied here. The important different are as follows:
\begin{compactitem}
\item Each state in the transition system of B-GKAB is always
  consistent, because we throw away all negative inclusion assertions
  as well as functionality assertions from the TBox and only keep the
  positive inclusion assertions. As a consequence, the repair
  mechanism in B-GKAB will not change anything.

\item Each action invocation in the program in the given S-GKAB is
  translated in such a way that it will always be followed by the
  action invocation
  $\gact{\neg\qunsatecq{T}}{\act_\bot()}$. 
  Note that $\qunsatecq{T}$ will be evaluated to true when the
  corresponding ABox is $T$-inconsistent.
  Therefore, the inconsistency check is basically delegated to the
  evaluation of the query that acts as the guard of the action
  $\act_\bot$ and it is triggered after each action execution.
  Moreover, the action $\act_\bot$ will not be executed if the
  previous action execution leads into an inconsistent state w.r.t.\
  the TBox $T$. Thus, it is easy to see that when an action execution
  in S-GKAB is blocked because it leads into a $T$-inconsistent state,
  then the corresponding action execution in B-GKAB will not lead into
  a new state without $\tmp$ as well. However, when an execution in
  S-GKAB leads into a new $T$-consistent state, the corresponding
  action execution in B-GKAB will be followed by the execution of
  $\act_\bot$ and it leads into a new state without $\tmp$.


\end{compactitem}
\ \ 
\end{proof}

Having \Cref{lem:sgkab-to-bgkab-bisimilar-state} in hand, we can
easily show that given an S-GKAB, its transition system is S-bisimilar
to the transition of its corresponding B-GKAB that is obtained via the
translation $\tgkabsb$ as follows.

\begin{lemma}\label{lem:sgkab-to-bgkab-bisimilar-ts}
  Given an S-GKAB $\gkabsym$, we have
  $\ts{\gkabsym}^{\filter_S} \sbsim
  \ts{\tgkabsb(\gkabsym)}^{\filter_B}$
\end{lemma}
\begin{proof}
Let
\begin{compactenum}
\item $\gkabsym = \tup{T, \initabox, \actset, \ginitprog_s}$, and
  $\ts{\gkabsym}^{\filter_S} = \tup{\const, T, \stateset_s, s_{0s},
    \abox_s, \trans_s}$,
\item
  $\tgkabsb(\gkabsym) = \tup{T_b, \initabox, \actset_b, \ginitprog_b}$,
  and 
  $\ts{\tgkabsb(\gkabsym)}^{\filter_B} = \tup{\const, T_b,
    \stateset_b, s_{0b}, \abox_b, \trans_b}$.
\end{compactenum}
We have that $s_{0s} = \tup{A_0, \scmap_s, \delta_s}$ and
$s_{0b} = \tup{A_0, \scmap_b, \delta_b}$ where
$\scmap_s = \scmap_b = \emptyset$. By the definition of $\tgprogsb$
and $\tgkabsb$, we also have $\delta_b = \tgprogsb(\delta_s)$. Hence,
by \Cref{lem:sgkab-to-bgkab-bisimilar-state}, we have
$s_{0s} \sbsim s_{0b}$. Therefore, by the definition of
S-bisimulation, we have
$\ts{\gkabsym}^{\filter_S} \sbsim
\ts{\tgkabsb(\gkabsym)}^{\filter_B}$.  \ \
\end{proof}

Finally, we now able to show that the verification of \muladom
formulas over S-GKABs can be recast as verification of \muladom
formulas over B-GKABs as follows.

\begin{theorem}\label{thm:sgkab-to-bgkab}
  Given an S-GKAB $\gkabsym$ and a closed \muladom property $\Phi$,
\begin{center}
  $\ts{\gkabsym}^{\filter_S} \models \Phi$ \ iff \
  $\ts{\tgkabsb(\gkabsym)}^{\filter_B} \models \tforsb(\Phi)$
\end{center}
\end{theorem} 
\begin{proof}
  By \Cref{lem:sgkab-to-bgkab-bisimilar-ts}, we have that
  $\ts{\gkabsym}^{\filter_S} \sbsim
  \ts{\tgkabsb(\gkabsym)}^{\filter_B}$.
  Hence, by \Cref{lem:sbisimilar-ts-satisfies-same-formula}, it is
  easy to see that the claim is proven.
\end{proof}

\section{Discussion: Extended Inconsistency-Aware Golog-KABs}

In the following we elaborate some possible extensions of
I-GKABs. 
First, we discuss an I-GKABs extension that enables us to keep track
some information regarding inconsistencies and hence it gives us
finer-grained insights concerning inconsistencies. 
Second, we elaborate a possibility to allow dynamic selection of
repair mechanisms and hence enable us incorporate several repair
mechanisms within one system.

\subsection{Tracking Inconsistencies}
As we have seen, I-GKABs (B-GKABs, C-GKABs, E-GKABs) enhance GKABs
with inconsistency-handling mechanisms by adopting repair-based
semantics. However, despite dealing with possible repairs when some
action step produces a $T$-inconsistent ABox, they do not explicitly
track whether a repair has been actually enforced, nor do they provide
finer-grained insights about which TBox assertions were involved in
the inconsistency.
As a brief discussion, here we elaborate a possible extension of
I-GKABs so as to equip the transition system of I-GKABs with these
additional information. We will also see that such information enable
us to have a more fine-grained analysis over the system
evolution. Furthermore, we also give an intuition that the
verification of GKABs with such extended inconsistency-aware semantics
can be reduced to the corresponding verification of S-GKABs.

The idea of extending the inconsistency-aware semantics for GKABs is
elaborated step by step as follows:
\begin{enumerate}


\item We assume w.l.o.g.\ that $\iconst$ contains one distinguished
  constant per TBox assertion in $T$, 

\item We introduce a function $\lab$, that, given a TBox assertion,
  returns the corresponding constant.

\item We then define the set $\viol(A,T)$ of constants labeling TBox
  assertions in $T$ violated by $A$, as follows:
  \[
  \begin{array}{ll@{\ }l@{}}
    \viol(A,T) = &\{ d \in \const \mid \funct{Z} &\in T_f,\  
                                                   d = \lab(\funct{Z}) \text{ and } \\
                 &&A \models \exists xyz.\qunsatf(\funct{Z},x,y,z) \ \}\  \cup\\
                 &\{ d \in \const \mid B_1 \ISA \neg B_2 &\in T_n,\ 
                                                            d = \lab(B_1 \ISA \neg B_2) \text{ and } \\
                 &&A \models \rew(\exists x.\qunsatn(B_1 \ISA \neg
                    B_2,x), T)\}\  \cup\\
                 &\{ d \in \const \mid R_1 \ISA \neg R_2 &\in T_n,\ 
                                                            d = \lab(R_1 \ISA \neg R_2) \text{ and } \\
                 &&A \models \rew(\exists x.\qunsatn(R_1 \ISA \neg
                    R_2,x), T)\}
  \end{array}
  \]

\item We then employ the information provided by the set $\viol(A,T)$
  to decorate the states that is produced in each transition. This is
  done by utilizing a fresh concept $\violcon$ that keeps track of the
  labels of violated TBox assertions. Formally, we adjust the
  definitions of b-repair filter (\Cref{def:b-rep-filter}), c-repair
  filter (\Cref{def:c-rep-filter}), and e-repair filter
  (\Cref{def:bevol-filter}) as follows:
  \begin{itemize}

  \item A \emph{B-repair Filter $\filter_B$} is a relation that
    consists of tuples of the form $\tup{A, \facta, \factd, A''}$ such
    that $A' \in \arset{T, (A \setminus \factd) \cup \facta}$ and
    $A'' = A' \cup \set{ \violcon(d) \mid d \in \viol((A \setminus
      \factd) \cup \facta,T) }$.

  \item A \emph{C-repair Filter $\filter_C$} is a relation that
    consists of tuples of the form $\tup{A, \facta, \factd, A''}$ such
    that $A' = \iarset{T, (A \setminus \factd) \cup \facta}$ and
    $A'' = A' \cup \set{ \violcon(d) \mid d \in \viol((A \setminus
      \factd) \cup \facta,T) }$.

  \item A \emph{B-evol Filter $\filter_E$} is a relation that consists
    of tuples of the form $\tup{A, \facta, \factd, A''}$ such that
    $A' = \evol(T, A, \facta, \factd)$, $\facta$ is $T$-consistent, and
    $A'' = A' \cup \set{ \violcon(d) \mid d \in \viol((A \setminus
      \factd) \cup \facta,T) }$.

  \end{itemize}
  Additionally, notice that we also need to flush away each existing
  ABox assertion made by the concept $\violcon$ whenever we want to
  generate a new state. This is necessary in order to make sure that
  the information about violated TBox assertions from the previous
  states are not augmented to the new states. It is easy to see that
  we can achieve such purpose by modifying the definition of the
  filters above a little bit.
\end{enumerate}

We have seen how we can decorate each state in the transition systems
of I-GKABs such that they contain information about violated TBox
assertion. Now, with this machinery in hand, observe that we can do
finer-grained analysis over the system evolution by exploiting the
concept $\violcon$. For instance we can now verify the following
\muladom properties:
\begin{compactitem}
\item $\nu Z.(\neg \exists l. \textsf{Viol}(l)) \wedge \BOX{Z}$.  It
  says that no state of the system is violating the TBox constraints;
\item
  $\nu Z.(\forall l.  \textsf{Viol}(l) \ra (\mu Y. (\nu W. \neg
  \textsf{Viol}(l) \wedge \BOX{W}) \vee \DIAM{Y}) ) \wedge \BOX{Z}$.
  It says that, in all states, whenever a certain TBox assertion $t$
  is violated, there exists a run that reaches a state starting from
  which $t$ will never be violated anymore.
\end{compactitem}

We now proceed to give the intuition that the verification of I-GKABs
(B-GKABs, C-GKABs, and E-GKABs) with such kind of extension can be
reduced to the corresponding verification of S-GKABs as follows:
\begin{compactenum}
\item Since we can detect the violated TBox assertions through query
  answering, we can simply construct an action in which each of its
  effects detects the violation of a particular TBox assertion, and
  then when a certain TBox assertion is violated, this action adds the
  corresponding assertion made by the concept $\violcon$ and the
  corresponding label of the violated TBox assertion.
\item We concat each action invocation with the action that marks the
  violated TBox assertions and then we concat them with the repair
  program. With this approach, we can simulate the computation of
  extended inconsistency-aware GKABs inside S-GKABs.
\item As for translating the temporal properties, similar approach can
  be followed. For the case of extended C-GKABs and E-GKABs we might
  need to triplicate the modal operator instead of just duplicating
  it.
\end{compactenum}
Last, note that it is easy to see that our extended
inconsistency-aware semantics can easily capture S-GKABs. Similar
approach as in \Cref{sec:cap-sgkabs-to-igkabs} can be followed. 

\subsection{Dynamic Selection of Repair Mechanisms}

In the I-GKABs framework that has been presented so far, each of them
only employ one kind of repair mechanism when there is an
inconsistency (e.g., B-GKABs always apply b-repair when there is an
inconsistency). However, it might be desirable to employ different
kind of repair mechanisms in a different situation within a
system. For example, when specifying the program, the designer of the
system might know that in case there is an inconsistency after the
execution of a particular action, it is better to apply a certain
repair mechanism instead of another repair mechanisms. In general, it
might also desirable to use different way of updating the ABox for
different action.

To deal with this, we can extend our Golog program
(cf. \Cref{def:golog-program}) such that we can have \emph{annotated
  atomic action invocation} that is atomic action invocation annotated
with the desired way of updating the ABox as well as the preferred
repair mechanism. Technically, it is the usual atomic action invocation
annotated with a filter relation written as follows:
\[
\gact{Q(\vec{p})}{\act(\vec{p})}:\filterb
\]
where $\filterb$ is the desired filter relation. Furthermore, we can
refine the notion of program execution relation
(cf. \Cref{def:prog-exec-relation}) by adding the following:

 \begin{compactitem}
 \item
   $\tup{A, \scmap, \gact{Q(\vec{p})}{\act(\vec{p})}:\filterb}
   \gprogtrans{\act\sigma, \filter} \tup{A', \scmap', \gemptyprog}$
   \\
   if
   $\tup{A, \scmap, \gact{Q(\vec{p})}{\act(\vec{p})}}
   \gprogtrans{\act\sigma, \filterb} \tup{A', \scmap', \gemptyprog}$


\end{compactitem}

\noindent
Note that the additional condition for the program execution relation
above updates $A$ into $A'$ by employing filter $\filterb$ instead of
the given filter $\filter$. Essentially we define the relation
$\gprogtrans{\act\sigma, \filter}$ by also utilizing the relation
$\gprogtrans{\act\sigma, \filterb}$. As an example, one might specify
the following atomic action invocation:
\[
\gact{\true}{\act()}:\filter_C
\]
which requires that the update by the action $\act$ should be done
using the c-repair filter $\filter_C$.

Regarding verification, it can be shown that the verification of
I-GKABs with such kind of extension can be reduced to verification of
S-GKABs as follows. The intuition is that we just need to translate
each annotated action invocation into an action invocation that is
concatenated with the corresponding program that is suitable with the
preferred way of updating the ABox. For instance, to simulate
$\gact{\true}{\act()}:\filter_C$ inside S-GKAB, we can concat
$\gact{\true}{\act()}$ with c-repair program.

Now, instead of just annotating some particular atomic action
invocations, one might want to annotate a program with a preferred way
of updating an ABox. Thus, we can extend our Golog program further
with the following construct:
\[
\delta:\filterb
\]
where $\filterb$ is a filter relation and $\delta$ is a golog program
that does not contain any annotated atomic action invocation.  We can
then extend the program execution relation as follows:
\[
\tup{A, \scmap,\delta :\filterb} \gprogtrans{\act\sigma, \filter}
\tup{A', \scmap', \delta' : \filterb} \mbox{ \ if \ } \tup{A, \scmap,
  \delta} \gprogtrans{\act\sigma, \filterb} \tup{A', \scmap', \delta'}
\]
However, notice that the construct $\delta:\filterb$ essentially can
be translated into the previous setting (i.e., the setting with
annotated atomic action invocation) by translating each atomic action
invocation $\gact{Q(\vec{p})}{\act(\vec{p})}$ within
$\delta : \filterb$ into an annotated atomic action invocation
$\gact{Q(\vec{p})}{\act(\vec{p})} : \filterb$. Therefore, in the end
the verification in this setting can also be reduced to the
verification of S-GKABs.

%% file: 2.chapters/6-cs-gkab.tex
\chapter{Embracing Contexts into Golog-KAB\lowercase{s}}\label{ch:cs-gkab}

\ifhidecontent
 
\fi

In GKABs (as well as in KABs), the intensional knowledge about the
domain, expressed in terms of a DL TBox, is assumed to be fixed along
the evolution of the system, i.e., independent of the actual state.
However, this assumption is in general too restrictive, since specific
knowledge might hold or be applicable only in specific,
\emph{context-dependent} circumstances.  Ideally, one should be able
to form statements that are known to be true in certain cases, but not
necessarily in all. For instance, in our simple order processing
scenario (cf.\ \Cref{ex:example-scenario,ex:gkab-run-ex}), in the
normal situation an employee should only act as either a designer or
an assembler. However, during the peak season, a designer might also
work as an assembler.
In addition, the needs of having flexible business processes that are
able to adapt themselves according to the situation (context) 
also has been identified in the area of business process modeling
\cite{RRF08,RRIG09}. For example, in our simple order processing
scenario, under the peak season, the company might prefer to outsource
the quality control operation instead of performing such action by
themselves.

In this chapter, we enrich GKABs with contextual information, making
use of different context dimensions.  On the one hand, context is
determined by the environment using context-changing actions that make
use of the current state of the KB and the current context.  On the
other hand, it affects the set of TBox assertions that are relevant at
each time point, and that have to be considered when processing
queries posed over GKABs.

We follow here the approach of~\cite{BaKP12,CePe14}, and introduce
\emph{contextualized TBoxes}, in which each inclusion assertion is
adorned with context information that determines under which
circumstances the inclusion assertion is considered to hold.  The
relation among contexts is described by means of a lattice
in~\cite{BaKP12} and by means of a directed acyclic graph
in~\cite{CePe14}. In our case, we represent context using a finite set
of context dimensions, each characterized by a finite set of domain
values that are organized in a tree structure. If for a context
dimension $d$, a value $v_2$ is placed below $v_1$ in the tree (i.e.,
$v_2$ is a descendant of $v_1$), then the context associated to $v_1$
is considered to be more general than the one for $v_2$, and hence
whenever context dimension $d$ is in value $v_2$, it is also in value
$v_1$.

Starting from this representation of contexts, we enrich GKABs towards
\emph{Context-Sensitive GKABs} (\csgkabs), by representing the
intensional information about the domain using a contextualized TBox,
in place of an ordinary one.  Moreover, the action component of GKABs,
which specifies how the states of the system evolve, is extended in
\csgkabs with \emph{context changing actions}.  Such actions determine
values for context dimensions in the new state, based on the data and
the context in the current state.  In addition, also regular
state-changing actions can query, besides the state, also the context,
and hence be enabled or disabled according to the context.
Notably, we show that verification of a very rich temporal logic,
which can be used to query the system evolution, contexts, and data,
is decidable for run-bounded \csgkabs. 



For the setting, as in KABs and GKABs, in the following we use
\dllitea for expressing KBs and we also do not distinguish between
objects and values (thus we drop attributes).
Moreover we make use of a countably infinite set $\const$ of
constants, which intuitively denotes all possible values in the
system.
Additionally, we consider a finite set of distinguished constants
$\iconst \subset \const$, and
a finite set $\servcall$ of \textit{function symbols} that represents
\textit{service calls}, which abstractly account for the injection of
fresh values (constants) from $\const$ into the system.

The core results in this chapter are published in
\cite{AS-JELIA-14,AS-ARCOE-14}

\section{Context Formalization} 

Following~\cite{McCa93}, we formalize context as a mathematical
object.  Basically, we follow the approach in~\cite{SeHo12} of
contextualizing knowledge bases by adopting the metaphor of
considering context as a box~\cite{BoGS13,GiBo97}.  Specifically, this
means that the knowledge represented by the TBox (together with the
ABox) in a certain context is affected by the values of parameters
used to characterize the context itself.

\begin{definition}[Tree-shaped Value Domain]
  Given \sidetextb{Tree-shaped Value Domain} a variable $d$, a
  \emph{tree-shaped finite value domain} of $d$ is a pair
  $\tup{\cdom[d], \cover[d]}$ where:
  \begin{compactitem}
  \item $\cdom[d]$ is a finite set of domain values, and
  \item $\cover[d]$ is a binary relation between values in $\cdom[d]$
  \end{compactitem}
  and it holds that
  \begin{compactenum}
  \item There exists exactly one value $v \in \cdom[d]$ such that
    there does not exists $v' \in \cdom[d]$ and $v \cover[d] v'$ (in
    this case we say that $v$ is a \emph{root}),
  \item For each $v \in \cdom[d]$, if $v$ is not a root, then there
    exists exactly one value $v' \in \cdom[d]$ such that
    $v \cover[d] v'$, and 
  \item There is no cycle (i.e., there does not exist
    $v_1, v_2, \ldots, v_n$ such that
    $v_1 \cover[d] v_2, v_2 \cover[d] v_3, \ldots, v_n \cover[d]
    v_1$).
  \end{compactenum}
 \ \ 
\end{definition}


\noindent
In the definition above, intuitively the condition 1 to 3 impose that
the binary relation $\cover[d]$ relates the values in $\cdom[d]$ such
that they form a tree structure.
As notation, we denote the domain value in the root of the tree
with $\topv[d]$.  Intuitively, $\topv[d]$ is the most general value in
the tree-shaped value hierarchy of $\cdom[d]$.

We now proceed to define the notion of context dimension and context dimension assignment
which are the crucial ingredients for formalizing the notion of
context.

\begin{definition}[Context Dimension]
  A \sidetext{Context Dimension} \emph{context dimension} is a variable $d$ that has its own
  \emph{tree-shaped finite value domain} $\tup{\cdom[d], \cover[d]}$.
\end{definition}

\begin{definition}[Context Dimension Assignment]
  Let \sidetext{Context Dimension Assignment} $d$ be a context
  dimension with a tree-shaped finite value domain
  $\tup{\cdom[d], \cover[d]}$. A \emph{context dimension assignment},
  denoted by $\cval{d}{v}$, is the assignment of value
  $v \in \cdom[d]$ into the context dimension $d$.  
\end{definition}

\noindent
Intuitively, a \emph{context dimension assignment} $\cval{d}{v}$
denotes the fact that a context dimension $d$ is in value $v$.
%

From this moment, except for \Cref{sec:cap-sgkab-to-scsgkabs}, for the
technical development of this chapter we fix a set 
\[
\cdimset = \{d_1,\ldots,d_n\}
\]
of context dimensions. Each context dimension $d_i \in \cdimset$ comes
with its own tree-shaped finite value domain
$\tup{\cdom[d_i],\cover[d_i]}$, where $\cdom[d_i]$ represents the
finite set of domain values, and $\cover[d_i]$ represents the
predecessor relation forming the tree.

The notion of context is then formally defined as follows: 

\begin{definition}[Context]
  A \sidetext{Context} \emph{context} $\ctx$ over a set $\cdimset$ of
  context dimensions is defined as a set
  $\set{\cval{d_1}{v_1},\ldots,\cval{d_n}{v_m}}$ of context dimension
  assignments such that for each context dimension $d\in \cdimset$,
  there exists exactly one assignment $\cval{d}{v}\in\ctx$.
\end{definition}

\noindent
Informally, a context represents a particular situation and is
characterized by the assignment of a particular value into each
context dimension.

\begin{figure}[bp]
\centering
\includegraphics[width=1.00\textwidth]{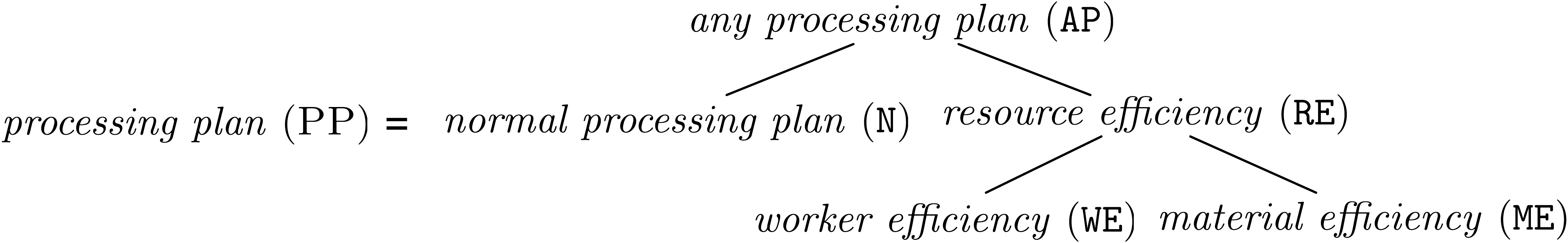}
\caption{Value domain of the context dimension
  \emph{processing plan} (\exc{PP})} \label{fig:csgkab-pp-cdim}
\end{figure}

\begin{figure}[bp]
\centering
\includegraphics[width=0.77\textwidth]{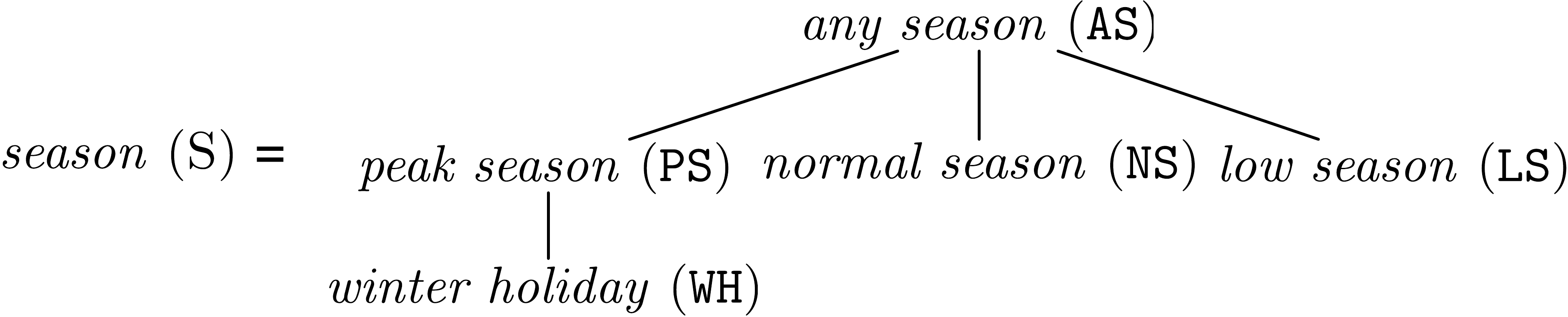}
\caption{Value domain of the context dimension
  \emph{season} (\exc{S})} \label{fig:csgkab-s-cdim}
\end{figure}

\begin{example}\label{ex:context-formalization}
  Recall our simple order processing scenario in
  \Cref{ex:gkab-run-ex}. Here we
  extend this running example into the case where some contextual
  information come into play.
  Consider the following situations:
  \begin{compactitem}
  \item Under a certain processing plan or during a particular season,
    the company might prefer to perform a certain operation compared to
    the other operation (e.g., During peak season, instead of doing
    the quality check by themselves, the company might outsource the
    operation).
  \item Within a specific season or when a particular processing plan
    is applied, our domain knowledge might changes.

  \end{compactitem}
  To model this situation, in this scenario we consider the set of
  context dimensions $\cdimset = \set{ \exc{PP} , \exc{S} }$, where
  \exc{PP} stands for \emph{processing plan}, and \exc{S} stands for
  \emph{season}. Both context dimensions \exc{PP} and \exc{S} are used
  later to characterized some contexts (some particular
  situations). We then define $\cdom[\exc{PP}]$ as well as
  $\cdom[\exc{S}]$ as follows:
%
\begin{itemize}
\item   $\cdom[\exc{PP}] = \{\exv{WE}, \exv{ME}, \exv{RE}, \exv{N},
  \exv{AP}\}$, where
  \begin{compactenum}
  \item \exv{WE} stands for \emph{worker efficiency},
  \item \exv{ME} stands for \emph{material efficiency},
  \item\exv{RE} stands for \emph{resource efficiency},
  \item \exv{N} stands for \emph{normal processing plan}, and
  \item\exv{AP} stands for \emph{any processing plan}.
  \end{compactenum}
  and
  \begin{compactenum}
  \item $\exv{WE} \cover[\exc{PP}] \exv{RE} $,
  \item $\exv{ME} \cover[\exc{PP}] \exv{RE} $,
  \item $\exv{RE} \cover[\exc{PP}] \exv{AP} $,
  \item $\exv{N} \cover[\exc{PP}] \exv{AP} $.
  \end{compactenum}
  To give more intuition, the value domain of
  $\exc{PP}$ 
  is visually described in \Cref{fig:csgkab-pp-cdim}. As an example,
  $\exv{WE} \cover[\exc{PP}] \exv{RE} $ means that \emph{worker
    efficiency} is a form of \emph{resource efficiency}.
%
\item   $\cdom[\exc{S}] = \{\exv{WH}, \exv{PS}, \exv{LS}, \exv{NS},
  \exv{AS}\}$, where
  \begin{compactenum}
  \item \exv{WH} stands for \emph{winter holiday},
  \item \exv{PS} stands for \emph{peak season},
  \item \exv{LS} stands for \emph{low season},
  \item\exv{NS} stands for \emph{normal season},
  \item\exv{AS} stands for \emph{any season}.
  \end{compactenum}
  and
  \begin{compactenum}
  \item $\exv{WH} \cover[\exc{S}] \exv{PS} $,
  \item $\exv{PS} \cover[\exc{S}] \exv{AS} $,
  \item $\exv{NS} \cover[\exc{S}] \exv{AS} $,
  \item $\exv{LS} \cover[\exc{S}] \exv{AS} $.
  \end{compactenum}
\end{itemize}
To give more intuition, the value domain of $\exc{S}$ is visually
described in \Cref{fig:csgkab-s-cdim}.

An example of a context over the set
$\cdimset = \set{ \exc{PP} , \exc{S} }$ of context dimensions is the
context
  \[
  \ctx = \set{\cval{\exc{PP}}{\exv{N}}, \cval{\exc{S}}{\exv{NS}}},
  \]
  which essentially encode the context (situation) where the normal
  processing plan is applied and the season is normal. Technically,
  the context $\ctx$ is characterized by the assignment of the value
  $\exv{N}$ into $\exc{PP}$ and the value $\exv{N}$ into $\exc{S}$.

\end{example}

To predicate over contexts, we introduce a \emph{context expression
  language} $\ctxlang$ over $\cdimset$, which corresponds to
propositional logic where the propositional letters are context
dimension assignments over $\cdimset$.  Formally, it is defined as
follows:


\begin{definition}[Context Expression Language]
  The \sidetext{Context Expression Language} syntax of \emph{context
    expression language} $\ctxlang$ is as follows:
\[
  \ctxe ~::=~ \cval{d}{v} ~\mid~ \ctxe\land \ctxe' ~\mid~ \lnot \ctxe
\]
where $d\in \cdimset$, and $v \in \cdom[d]$.  
\end{definition}

\noindent
We call a formula expressed in $\ctxlang$ a \emph{context
  expression}. For the semantics of $\ctxlang$, we adopt the standard
propositional logic semantics and the usual abbreviations.  The notion
of \emph{satisfiability} and \emph{model} are as usual.

Observe that a context
$\ctx=\{\cval{d_1}{v_1}, \ldots, \cval{d_n}{v_m}\}$, being a set of
(atomic) formulas in $\ctxlang$, can be considered as a propositional
theory.
The semantics of value domain of each context dimension in $\cdimset$
can also be characterized by an $\ctxlang$ theory.  Specifically, we
define the \emph{value domain theory $\ctxth$ of $\cdimset$} as an
$\ctxlang$ theory below:

\begin{definition}[Context Dimension Value
  Domain Theory]
  We \sidetextb{Context Dimension Value Domain Theory} define a \emph{value domain
    theory $\ctxth$} of $\cdimset$ as the smallest set of context
  expressions satisfying the following conditions: For every context
  dimension $d \in \cdimset$, we have:
\begin{compactitem}
\item For all values $v_1,v_2\in\cdom[d]$ such that $v_1\cover v_2$,
  we have that $\Phi_{\cdimset}$ contains the expression
  $\cval{d}{v_1} \ra \cval{d}{v_2}$.  Intuitively, this states that the
  value $v_2$ is more general than $v_1$, and hence, whenever we have
  $\cval{d}{v_1}$ we can infer that $\cval{d}{v_2}$.
\item For all values $v_1,v_2,v\in\cdom[d]$ s.t.\ $v_1\cover v$ and
  $v_2\cover v$, we have that $\Phi_{\cdimset}$ contains the expression
  $\cval{d}{v_1} \ra\neg \cval{d}{v_2}$.  Intuitively, this expresses that
  sibling values $v_1$ and $v_2$ are disjoint.
\end{compactitem}
\ \ 
\end{definition}

\noindent
In the following, we write $\ctxth$ to denote the value domain theory
of $\cdimset$.




\section{Contextualizing Knowledge Bases}


Essentially, we define a \emph{context-sensitive knowledge base} (CKB)
over the set $\cdimset$ of context dimensions as a standard DL
knowledge base in which the TBox assertions are contextualized. 

\begin{definition}[Contextualized TBox]
  A \sidetext{Contextualized TBox} \emph{contextualized TBox} $\ctbox$
  over $\cdimset$ is a finite set of assertions of the form
  $\tup{t:\varphi}$, where $t$ is a usual TBox assertion and $\varphi$
  is a context expression over $\cdimset$.
\end{definition}

\noindent
Intuitively, $\tup{t:\varphi}$ expresses that the TBox assertion $t$
holds in all those contexts satisfying $\varphi$, taking into account
the theory $\Phi_{\cdimset}$.
%
%
%
Similar to the usual TBox, given a contextualized TBox $\ctbox$, we
write $\voc(\ctbox)$ to denote the vocabulary of TBox $\ctbox$,
independently from the context.
As a remark, the idea of our contextualized TBox is inspired by
\cite{CePe14} (see the notion of V-TBox in \cite{CePe14}).

\begin{definition}[Contextualized KB]
  A \sidetext{Contextualized KB} \emph{contextualized KB} is a tuple
  $\tup{\ctbox,A}$ where $\ctbox$ is a contextualized TBox and $A$ is
  the usual ABox over $\voc(\ctbox)$.
\end{definition}

We now define the notion of a KB in context $\ctx$ as follows:

\begin{definition}[KB Under the Context $\ctx$]
  Given \sidetext{KB Under the Context $\ctx$} a CKB $\tup{\ctbox, A}$
  and a context $\ctx$, both over $\cdimset$, we define the \emph{KB
    under the context $\ctx$} as the KB $\tup{\ctbox^{\ctx},A}$, where
  $\ctbox^{\ctx} =\{t\mid \tup{t:\varphi} \in \ctbox \text{ and } \ctx
  \cup \ctxth \models \varphi\}$.
  Additionally, in this case we say that $\ctbox^{\ctx}$ is
  \emph{contextualized TBox $\ctbox$ under the context $\ctx$}.
\end{definition}

\begin{example}\label{ex:contextualized-tbox}
  Continuing our example, in a normal situation, to enforce the
  segregation of duties, \emph{designer} and \emph{assembler} must be
  different. However, in the situation (context) where we have either
  \emph{peak season} ($\cval{\exc{S}}{\exv{PS}}$) or the company wants
  to promote \emph{worker efficiency} ($\cval{\exc{PP}}{ \exv{WE}}$),
  each \emph{designer} is also an \emph{assembler}. In
  addition, the other assertions hold in any situation (context).
  This situation can be encoded in a contextualized TBox $\ctbox$
  containing the following assertions: \\
\begin{align*}
  \tup{ \exo{ApprovedOrder} \sqsubseteq \exo{Order} &: \cval{\exc{PP}}{\exv{AP}} \wedge \cval{\exc{S}}{\exv{AS}} }\\
  \tup{\exo{AssembledOrder} \sqsubseteq \exo{Order} &: \cval{\exc{PP}}{\exv{AP}} \wedge \cval{\exc{S}}{\exv{AS}} }\\
  \tup{\exo{DeliveredOrder} \sqsubseteq \exo{Order} &: \cval{\exc{PP}}{\exv{AP}} \wedge \cval{\exc{S}}{\exv{AS}} }\\
  \tup{\exo{ReceivedOrder} \sqsubseteq \exo{Order} &: \cval{\exc{PP}}{\exv{AP}} \wedge \cval{\exc{S}}{\exv{AS}} }\vspace*{2mm}\\
  \tup{\exo{Designer} \sqsubseteq \exo{Employee} &: \cval{\exc{PP}}{\exv{AP}} \wedge \cval{\exc{S}}{\exv{AS}} }\\
  \tup{\exo{Assembler} \sqsubseteq \exo{Employee} &: \cval{\exc{PP}}{\exv{AP}} \wedge \cval{\exc{S}}{\exv{AS}} }\\
  \tup{\exo{QualityController} \sqsubseteq \exo{Employee} &: \cval{\exc{PP}}{\exv{AP}} \wedge \cval{\exc{S}}{\exv{AS}} }\vspace*{2mm}\\
  \tup{\exo{Designer} \sqsubseteq \neg \exo{Assembler} &:                                                                 
                                                         \cval{\exc{PP}}{\exv{N}}
                                                         \wedge 
                                                         (\cval{\exc{S}}{\exv{NS}}
                                                         \vee 
                                                         \cval{\exc{S}}{\exv{LS}})
                                                         }\\
  \tup{\exo{Designer} \sqsubseteq \exo{Assembler} &: \cval{\exc{PP}}{\exv{WE}} \vee \cval{\exc{S}}{\exv{PS}} }\\
   \tup{\exo{Designer} \sqsubseteq \neg \exo{QualityController} &: \cval{\exc{PP}}{\exv{AP}} \vee \cval{\exc{S}}{\exv{AS}} }\\
  \tup{\exo{Assembler} \sqsubseteq \neg \exo{QualityController} &: \cval{\exc{PP}}{\exv{AP}} \wedge \cval{\exc{S}}{\exv{AS}} }\vspace*{2mm}\\
%
%
  \tup{\SOMET{\exo{assembledBy}^-}\sqsubseteq \exo{Employee} &: \cval{\exc{PP}}{\exv{AP}} \wedge \cval{\exc{S}}{\exv{AS}} }\\
  \tup{\SOMET{\exo{assembledBy}}\sqsubseteq \exo{Order} &: \cval{\exc{PP}}{\exv{AP}} \wedge \cval{\exc{S}}{\exv{AS}} }\\
  \tup{\SOMET{\exo{designedBy}^-}\sqsubseteq \exo{Employee} &: \cval{\exc{PP}}{\exv{AP}} \wedge \cval{\exc{S}}{\exv{AS}} }\\
  \tup{\SOMET{\exo{designedBy}}\sqsubseteq \exo{Order} &: \cval{\exc{PP}}{\exv{AP}} \wedge \cval{\exc{S}}{\exv{AS}} }\\
  \tup{\SOMET{\exo{checkedBy}^-}\sqsubseteq \exo{Employee} &: \cval{\exc{PP}}{\exv{AP}} \wedge \cval{\exc{S}}{\exv{AS}} }\\
  \tup{\SOMET{\exo{checkedBy}}\sqsubseteq \exo{Order} &: \cval{\exc{PP}}{\exv{AP}} \wedge \cval{\exc{S}}{\exv{AS}} }\vspace*{2mm}\\
\tup{\SOMET{\exo{hasAssemblingLoc}^-}\sqsubseteq \exo{Location} &: \cval{\exc{PP}}{\exv{AP}} \wedge \cval{\exc{S}}{\exv{AS}} }\\
    \tup{\SOMET{\exo{hasAssemblingLoc}}\sqsubseteq \exo{Order} &: \cval{\exc{PP}}{\exv{AP}} \wedge \cval{\exc{S}}{\exv{AS}} }\vspace*{2mm}\\
  \tup{\SOMET{\exo{hasDesign}^-}\sqsubseteq \exo{Design} &: \cval{\exc{PP}}{\exv{AP}} \wedge \cval{\exc{S}}{\exv{AS}} }\\
\tup{\SOMET{\exo{hasDesign}}\sqsubseteq \exo{Order} &: \cval{\exc{PP}}{\exv{AP}} \wedge \cval{\exc{S}}{\exv{AS}} }\vspace*{2mm}\\
   \tup{\funct{\exo{hasAssemblingLoc}} &: \cval{\exc{PP}}{\exv{AP}} \wedge \cval{\exc{S}}{\exv{AS}} }\\
   \tup{\funct{\exo{hasDesign}} &: \cval{\exc{PP}}{\exv{AP}} \wedge \cval{\exc{S}}{\exv{AS}} }
\end{align*}
%
Given a CKB $\tup{\ctbox, A}$, and a context
$\ctx = \set{\cval{\exc{PP}}{\exv{N}}, \cval{\exc{S}}{\exv{NS}}}$, we
have that $\tup{\ctbox^{\ctx}, A}$ is a KB under the context $\ctx$
where $\ctbox^{\ctx}$ containing the same TBox assertions as in
\Cref{ex:tbox-and-abox}. As another example, consider the context
$\ctx' = \set{\cval{\exc{PP}}{\exv{WE}}, \cval{\exc{S}}{\exv{PS}}}$,
we have that $\tup{\ctbox^{\ctx'}, A}$ is a KB under the context
$\ctx'$ where $\ctbox^{\ctx'}$ containing the same TBox assertions as
in \Cref{ex:tbox-and-abox} except that it contains
$\exo{Designer} \sqsubseteq \exo{Assembler}$ instead of
$\exo{Designer} \sqsubseteq \neg \exo{Assembler}$.
\end{example}
  
\section{Contextualizing Golog-Program}\label{sec:contextualized-golog-program}

As the presence of contexts
influence the condition when an action can be executed, we now lift
our Golog program variant (in \Cref{def:golog-program}) into
Contextualized Golog program as follows:

\begin{definition}[Contextualized Golog Program]\label{def:context-golog-program}
  Given \sidetext{Contextualized Golog Program} a set of KAB actions
  $\actset$, a \emph{contextualized Golog program} $\delta$ over
  $\actset$ is an expression formed by the following grammar:
\[
\begin{array}{@{}r@{\ }l@{\ }}
  \delta ::= &
  \gemptyprog ~\mid~
  \gactc{Q(\vec{p})}{\ctxe}{\act(\vec{p})} ~\mid~
  \delta_1|\delta_2  ~\mid~
  \delta_1;\delta_2 ~\mid~ \\
  &\gif{\varphi}{\delta_1}{\delta_2} ~\mid~
  \gwhile{\varphi}{\delta}
\end{array}
\]
where:
\begin{compactitem}
\item $\gemptyprog$, $\delta_1|\delta_2$, $\delta_1;\delta_2$,
  $\gif{\varphi}{\delta_1}{\delta_2}$ and $\gwhile{\varphi}{\delta}$
  are the same as in \Cref{def:golog-program}.
\item $\gactc{Q(\vec{p})}{\ctxe}{\act(\vec{p})}$ is a
  \emph{context-sensitive atomic action invocation} guarded by a
  \diecq $Q$ and context expression $\ctxe$, such that
  $\act\in\actset$ is executable when $\ctxe$ is satisfied by the
  current context (taking into account the theory $\Phi_{\cdimset}$),
  and it is executed by non-deterministically substituting its
  parameters $\vec{p}$ with an answer of $Q$;
\end{compactitem}
\ \ 
\end{definition}

\section{Context-Sensitive Golog-KABs (\csgkabs)}


We now enhance GKABs with context-related information, introducing
in particular \emph{Context-Sensitive GKABs} (\csgkabs), which
consist of:
\begin{compactenum}
\item a context-sensitive knowledge base (CKB), which maintains the information of interest,
\item an action base, which characterizes the system evolution, and
\item context information that evolves over time, capturing changing
  circumstances.
\end{compactenum}
Differently from GKABs, where the TBox is fixed a-priori and remains
rigid during the evolution of the system, in \csgkabs the TBox changes
depending on the current context.  Alongside the evolution mechanism
for data borrowed from GKABs, \csgkabs include also a progression
mechanism for the context itself, giving raise to a system in which
data and context evolve simultaneously.



\begin{definition}[Context-Sensitive GKABs (\csgkabs)]\label{def:csgkabs-formalism}
  A \sidetext{Context-Sensitive GKABs (\csgkabs)} \csgkab is a tuple
  $\csgkabsym = \tup{\ctbox, \initabox, \actset, \ginitprog, \initctx,
    \ctxprocset}$ where:
  \begin{compactitem}
  \item $\ctbox$ is a \dllitea \emph{contextualized TBox} capturing
    the domain of interest.

  \item $\initabox$ and $\actset$ are as in a GKAB.

  \item $\ginitprog$ is a contextualized Golog program over $\actset$,
    which characterizes the evolution of the GKAB over time, using the
    atomic actions in $\actset$.


  \item $\initctx$ is the initial context over $\cdimset$.


  \item $\ctxprocset$ is a finite set of context-evolution rules, each
    of which determines the configuration of the new context depending
    on the current context and data. Each \emph{context-evolution
      rule} has the form $\tup{Q, \ctxe} \mapsto C_{new}$, where:
      \begin{compactenum}
      \item $Q$ is a boolean ECQ over $\ctbox$,
      \item $\ctxe$ is a context expression, and
      \item $C_{new}$ is a finite set of context dimension assignments
        such that for each context dimension $d\in \cdimset$, there
        exists \emph{at most one} context dimension assignment
        $\cval{d}{v} \in C_{new}$. In the execution, if a context
        variable is not assigned by $C_{new}$, it maintains the
        assignment of the previous state.
      \end{compactenum}
\end{compactitem}
\ \ 
\end{definition}

\begin{example}\label{ex:csgkab}\textbf{An example of a \csgkab.}\\
  Continuing our running example, consider the scenario in which
  either during the \emph{peak season} ($\cval{\exc{S}}{\exv{PS}}$) or
  when the company wants to promote \emph{worker efficiency}
  ($\cval{\exc{PP}}{ \exv{WE}}$), the company outsource the quality
  control task.
%
%
  To model this scenario, we specify a \csgkab
  $\csgkabsym = \tup{\ctbox, \initabox, \actset, \ginitprog, \initctx,
    \ctxprocset}$
  where $\ctbox$ is the same as contextualized TBox in
  \Cref{ex:contextualized-tbox}, $\initabox$ is the same as the one in
  \Cref{ex:gkab-run-ex}, $\actset$ is the same as in
  \Cref{ex:kab} except that we augment an additional action
  that essentially performs the quality control by outsourcing
  it. This action obtains the quality controller by calling a service
  $\exs{getQCService}/1$ and it is specified as follows:
  \begin{flalign*}
    &\exa{outsourceQualityCheck}():\{ \\
    &\quad\begin{array}{l@{}l@{}l}
            [\exo{AssembledOrder}&(\exvar{x})] \rightsquigarrow& \\
                                 &\add \set{&
                                              \exo{checkedBy}(\exvar{x},\exs{getQCService}(\exvar{x}))
                                              }
          \end{array}\\
          &\}.
 \end{flalign*}

\noindent 
To realize the flow of order processing in the scenario above, the
initial program $\ginitprog$ is specified as follows:
\[
\delta = \gwhile{  \exists \exvar{x}.[\exo{Order}(\exvar{x})] \wedge
    \neg[\exo{DeliveredOrder}(\exvar{x})]  }{\delta_0} 
\]
where:
\begin{compactitem}
\item $\delta_0 = \delta_1 ; \delta_2 ; \delta_3 ; \delta_4 ; \delta_5 $
\item $\delta_1 = \gif{ \neg [\exists
    \exvar{x}.\exo{ApprovedOrder}(\exvar{x})] \\
  \hspace*{8mm}}{ \gactc{\exo{ReceivedOrder}(\exvar{x})}{ \cval{\exc{PP}}{\exv{AP}} \wedge
      \cval{\exc{S}}{\exv{AS}} }{\exa{approveOrder}(\exvar{x})} \\
  \hspace*{8mm}}{\gemptyprog}$,
\item
  $\delta_2 = \gactc{\true}{\cval{\exc{PP}}{\exv{AP}} \wedge
    \cval{\exc{S}}{\exv{AS}}}{\exa{prepareOrders}()}$,
\item
  $\delta_3 = \gactc{\true}{\cval{\exc{PP}}{\exv{AP}} \wedge
    \cval{\exc{S}}{\exv{AS}}}{\exa{assembleOrders}()}$,
\item
  $\delta_4 = 
  \gactc{\true}{ \neg ( \cval{\exc{PP}}{\exv{RE}} \vee
    \cval{\exc{S}}{\exv{PS}} ) }{\exa{checkAssembledOrders}()} \ | \\
  \hspace*{9mm}\gactc{\true}{ \cval{\exc{PP}}{\exv{RE}} \vee
    \cval{\exc{S}}{\exv{PS}} }{\exa{outsourceQualityCheck}()}$,
\item
  $\delta_5 = \gactc{\true}{\cval{\exc{PP}}{\exv{AP}} \wedge
    \cval{\exc{S}}{\exv{AS}}}{\exa{deliverOrder}()}$.
\end{compactitem}
Note that the program above is similar to the program in
\Cref{ex:gkab-run-ex}. The intuition of the program is also similar,
except that here the program is decorated with context expressions
that act as additional guards for each action invocation. The most
different one is $\delta_4$. In $\delta_4$, 
depending on the context, we have a choice whether the quality check
will be performed by the company or by outsourcing it. In particular,
within the \emph{peak season} ($\cval{\exc{S}}{\exv{PS}}$) or when the
company wants to promote \emph{resource efficiency}
($\cval{\exc{PP}}{ \exv{RE}}$), the company outsource the quality
control task. Additionally, note that we also need to consider the
value domain theory. For instance, when the season is \emph{winter
  holiday} (i.e., $\cval{\exc{S}}{\exv{WH}}$) together with the value
domain theory it will implies that the season is peak season
($\cval{\exc{S}}{\exv{PS}}$), and hence the company will also
outsource the quality control task. Similarly,
$\cval{\exc{PP}}{ \exv{WE}}$ and $\cval{\exc{PP}}{ \exv{ME}}$ will
imply $\cval{\exc{PP}}{ \exv{RE}}$. On the other hand, when neither
$\cval{\exc{S}}{\exv{PS}}$ nor $\cval{\exc{PP}}{ \exv{RE}}$ are
implied by the current context together with the value domain theory,
the company will perform the quality control by themselves.
%

\noindent
With a slightly abuse of notation, below we provide another way to
write the program above by also making use curly braces (``$\{$'',
``$\}$'') for marking the scope of program operators: 
\begin{align*}
&\gwhile{  \exists \exvar{x}.[\exo{Order}(\exvar{x})] \wedge
    \neg[\exo{DeliveredOrder}(\exvar{x})]  }{ \{ \\
&  \hspace*{5mm}\gif{ \neg [\exists
    \exvar{x}.\exo{ApprovedOrder}(\exvar{x})] \\
&    \hspace*{9mm}}{\{ 
    \gactc{\exo{ReceivedOrder}(\exvar{x})}{\cval{\exc{PP}}{\exv{AP}} \wedge \cval{\exc{S}}{\exv{AS}}}{\exa{approveOrder}(\exvar{x})}  
    \} \\
 &   \hspace*{9mm}}{\{ 
    \gemptyprog 
    \}}; \\
&  \hspace*{5mm}
  \gactc{\true}{\cval{\exc{PP}}{\exv{AP}} \wedge \cval{\exc{S}}{\exv{AS}}}{\exa{prepareOrders}()};\\
& \hspace*{5mm}
 \gactc{\true}{\cval{\exc{PP}}{\exv{AP}} \wedge \cval{\exc{S}}{\exv{AS}}}{\exa{assembleOrders}()};\\
& \hspace*{2mm}
\begin{array}{l@{\ }l}
  \ \{ &\gactc{\true}{ 
         \neg  (\cval{\exc{PP}}{\exv{RE}} \vee \cval{\exc{S}}{\exv{PS}}  ) }{\exa{checkAssembledOrders}()} \ | \\  
         &\gactc{\true}{\cval{\exc{PP}}{\exv{RE}} \vee
           \cval{\exc{S}}{\exv{PS}}}{\exa{outsourceQualityCheck}()} \};
\end{array} \\
&  \hspace*{5mm}
  \gactc{\true}{\cval{\exc{PP}}{\exv{AP}} \wedge \cval{\exc{S}}{\exv{AS}}}{\exa{deliverOrder}()}\\
\}}
\end{align*}

\noindent
As for the initial Context, we consider the following Context:
\[
\initctx = \set{\cval{\exc{PP}}{\exv{N}}, \cval{\exc{S}}{\exv{NS}}}.
\]
which capture the context (situation) where the normal processing plan
is applied and the season is normal.

\noindent
Furthermore, we have the set $\ctxprocset$ of context-evolution rules
containing the following rules:
\begin{compactenum}

\item $\carulex{\true}{\cval{\exc{S}}{\exv{PS}}}{\cval{\exc{S}}{\exv{NS}}}$. 

\item $\carulex{\true}{\cval{\exc{PP}}{\exv{N}} \wedge
\cval{\exc{S}}{\exv{NS}} }{ \cval{\exc{PP}}{\exv{WE}},
\cval{\exc{S}}{\exv{PS}} }$. 

\item $\carulex{\true}{\cval{\exc{PP}}{\exv{RE}} \wedge
\cval{\exc{S}}{\exv{PS}} }{ \cval{\exc{PP}}{\exv{N}},
\cval{\exc{S}}{\exv{NS}} }$. 

\item $\carulex{\true}{\cval{\exc{S}}{\exv{AS}} }{ }$.

\item
  $\carulex{\exists \exvar{x}.[\exo{Order}(\exvar{x})] \wedge
    \neg[\exo{DeliveredOrder}(\exvar{x})]}{\cval{\exc{PP}}{\exv{WE}}
    \wedge \cval{\exc{S}}{\exv{PS}} }{ }$.

\end{compactenum}
The intuition of each rule above is as follows:
\begin{compactenum}
\item The first rule models the transition from \emph{peak season}
  ($\cval{\exc{S}}{\exv{PS}}$) to \emph{normal season}
  ($\cval{\exc{S}}{\exv{NS}}$), independently from the data.
\item The second rule models the transition from the situation where
  the processing plan is normal ($\cval{\exc{PP}}{\exv{N}}$) and the
  season is also normal ($\cval{\exc{S}}{\exv{NS}}$) into the
  situation where the season is peak ($\cval{\exc{S}}{\exv{PS}}$) and
  the worker efficiency ($\cval{\exc{PP}}{\exv{WE}}$) processing plan
  is applied.
\item The third rule models the transition from the situation where
  the company promotes resource efficiency
  ($\cval{\exc{PP}}{\exv{RE}}$) and the season is peak
  ($\cval{\exc{S}}{\exv{PS}}$) into the situation where the season is
  normal ($\cval{\exc{S}}{\exv{NS}}$) and the normal processing plan
  is applied ($\cval{\exc{PP}}{\exv{N}}$).
\item The fourth rule represents the transition where the context stay
  the same, independently from the current data and context.  This is
  the case because the right hand side of the rule is an empty set,
  and hence it does not change the assignment of any context
  dimensions. Additionally, the context expression
  $\cval{\exc{S}}{\exv{AS}}$ will be entailed no matter which value is
  assigned to the context dimension $\exc{S}$.
\item The fifth rule represents the context-evolution where the context stay
  the same, given that
  \begin{inparaenum}[\it (i)]
  \item the current processing plan is worker efficiency,
  \item the current season is peak season, and
  \item there exists an order that is not yet delivered.
  \end{inparaenum}
\end{compactenum}
\end{example}

\subsection{\csgkabs Standard Execution Semantics}

As before, we are interested in verifying temporal properties over the
evolution of \csgkabs, in particular ``robust'' properties that the
system is required to guarantee independently from context changes.
Towards this goal, we define the execution semantics of \csgkabs in
terms of a possibly infinite-state transition system that
simultaneously captures all possible evolutions of the system as well
as all possible context changes.

Each state in the transition system 
of a \csgkab is a tuple 
$\tup{A, \scmap, \ctx, \delta}$, where 
%
%
$A$ is an ABox maintaining the current data, $\scmap$ is a service
call map accounting for the service call results obtained so far,
$\ctx$ is the current context, and $\delta$ is a program. The context
selects which are the axioms of the contextual TBox that currently
hold, in turn determining the current KB.
Specifically, we introduce the notion of context-sensitive transition
system in order to provide the semantics of \csgkab as follows:

\begin{definition}[Context-Sensitive Transition System]\label{def:cs-trans-sys}
  A \sidetext{Context-Sensitive Transition System}
  \emph{context-sensitive transition system} is a tuple
  $\ts{\csgkabsym} = \tup{\const, \ctbox, \stateset, s_0, \abox,
    \cntx, \trans}$, where:
\begin{compactenum}
\item $\ctbox$ is a contextualized TBox;
\item $\stateset$ is a set of states;
\item $s_0 \in \stateset$ is the initial state;
\item $\abox$ is a function that, given a state $s\in\stateset$,
  returns the ABox associated to $s$;
\item $\cntx$ is a function that, given a state $s\in\stateset$,
  returns the context associated to $s$;
\item $\trans \subseteq \stateset \times \stateset$ is a transition
  relation between pairs of states.
\end{compactenum}
 \ \ 
\end{definition}

Starting from the initial state $s_0$, $\ts{\csgkabsym}$ accounts for
all the possible (simultaneous) data and context transitions. 
Technically, we revise the notion of executability of action
invocation for GKABs by taking into account context expressions as follows: 
%
%
Let
$\csgkabsym$ 
  be a \csgkab, 
  given a context-sensitive action invocation
  $\gactc{Q(\vec{p})}{\ctxe}{\act(\vec{p})}$, we say that a
  substitution $\sigma$, which substitutes the parameters $\vec{p}$
  with constants in $\const$, is \emph{a legal parameter assignment
    for $\act$ in $A$ w.r.t.\ context $\ctx$ and action invocation}
  $\gactc{Q(\vec{p})}{\ctxe}{\act(\vec{p})}$ if
  $\ask(Q\sigma, \ctbox^{\ctx}, A)$ is $\true$.


To capture all possible context changes in a certain state, we define
the notion of $\ctxchg$ relation that essentially captures all
possible changes of a context based on the current ABox, the current
context and the available context-evolution rules as follows.

\ \\ 
\begin{definition}[Context Change Relation $\ctxchg$]\label{def:ctx-chg-relation}
  Given \sidetext{Context Change Relation} a \csgkab
  $\csgkabsym = \tup{\ctbox, \initabox, \actset, \ginitprog, \initctx,
    \ctxprocset}$,
  we define $\ctxchg$ relation of $\csgkabsym$ such that given an ABox
  $A$, two contexts $\ctx$ and $\ctx'$, we have a tuple
  $\tup{A, \ctx, \ctx'} \in \ctxchg$ if there exists a
  context-evolution rule $\tup{Q, \ctxe} \mapsto C_{new}$ in
  $\ctxprocset$ s.t.:
  \begin{compactenum}
  \item $\ask(Q, \ctbox^C, A)$ is $\true$;
  \item $C \cup \ctxth \models \ctxe$;
  \item for every context dimension $d\in\cdimset$ s.t.\
    $\cval{d}{v} \in C_{new}$, \\we have $\cval{d}{v} \in C'$;
  \item for every context dimension $d\in\cdimset$ s.t.\
    $\cval{d}{v} \in C$, and there does not exist any $v_2$ s.t.\
    $\cval{d}{v_2} \in C_{new}$, we have $\cval{d}{v} \in C'$.
  \end{compactenum}
\ \ 
\end{definition}

\begin{example}
  Consider our running example. We have
  \[
  \tup{A, \set{\cval{\exc{PP}}{\exv{N}}, \cval{\exc{S}}{\exv{NS}}},
    \set{\cval{\exc{PP}}{\exv{WE}}, \cval{\exc{S}}{\exv{PS}}} } \in
  \ctxchg
  \]
  where $A$ is any ABox. In this case, the context-evolution rule that
  changes the values of the context dimensions is
  \[
  \carulex{\true}{\cval{\exc{PP}}{\exv{N}} \wedge
    \cval{\exc{S}}{\exv{NS}} }{ \cval{\exc{PP}}{\exv{WE}},
    \cval{\exc{S}}{\exv{PS}} }
  \]
%
%
As another example, we have
\[
\tup{A, \set{\cval{\exc{PP}}{\exv{N}}, \cval{\exc{S}}{\exv{WH}}},
  \set{\cval{\exc{PP}}{\exv{N}}, \cval{\exc{S}}{\exv{NS}}} } \in
\ctxchg
\]
where $A$ is any ABox. In this case, the context-evolution rule that
changes the values of the context dimensions is
\[
\carulex{\true}{ \cval{\exc{S}}{\exv{PS}} }{ \cval{\exc{S}}{\exv{NS}} }
\]
Notice that in this example, the reason why the rule above is
applicable is because \emph{winter holiday} entail \emph{peak season}.
\end{example}

Next, in order to take into account the presence of the context, we
need to redefine the notion of 
\begin{inparaenum}[\it (i)]
\item filter relation, 
\item \tell operation, 
\item final states, 
and 
\item program execution relation.  
\end{inparaenum}
%
%
For the refinement of the filter relation, because the TBox changes
along with the context evolution, and also because sometimes we need
the TBox w.r.t.\ the current context in order to construct the new
ABox, thus we need to incorporate the corresponding context
information inside the filter relation. Therefore, we then refine the
filter relation as follows:

\begin{definition}[Context-sensitive Filter Relation]\label{def:cs-filter-rel}
  A \sidetext{Context-sensitive Filter Relation $\csfilter$}
  \emph{Context-sensitive Filter Relation $\csfilter$} is a relation
  that consists of tuples of the form
  $\tup{A, \facta, \factd, \ctx, A'}$ such that
  $\emptyset \subseteq A' \subseteq ((A \setminus \factd) \cup
  \facta)$,
  where $A$ and $A'$ are ABoxes, $\ctx$ is a context, and $\facta$ as
  well as $\factd$ are two sets of ABox assertions.
\end{definition}


\noindent
Roughly speaking, the filter relation indicates that the new ABox $A'$
is constructed based on the current ABox $A$, the set of assertions to
be added/deleted $\facta$/$\factd$ and also the context $\ctx$. The
context $\ctx$ in $\csfilter$ will be mainly used later when we
incorporate the inconsistency handling mechanism that based on
repair. Essentially, we need the context $\ctx$ during the repair in
order to determine in which context we should do the repair and hence
also determine which TBox assertions that we need to use.
From now on, unless explicitly stated, for brevity, we simply say
filter to refer to context-sensitive filter relation.

The \tell operation is then refined into \cstell operation as follows.

\begin{definition}[Context-sensitive $\cstell$ Operation]\label{def:cs-tell-operation}
  Given \sidetext{Context-sensitive $\cstell$} a \csgkab
  $\csgkabsym = \tup{\ctbox, \initabox, \actset, \ginitprog, \initctx,
    \ctxprocset}$
  and a filter $\csfilter$, we define $\cstell_{\csfilter}$ as
  a relation 
  such that we have a tuple
  $\tup{\tup{A,\scmap, \ctx}, \act\sigma, \tup{A', \scmap', \ctx'}}
  \in \cstell_{\csfilter}$ if
  \begin{compactenum}

  \item $\sigma$ is a legal parameter assignment for $\act$ in $A$
    w.r.t.\ context $\ctx$ and a certain action invocation
    $\gactc{Q(\vec{p})}{\ctxe}{\act(\vec{p})}$,



  \item $\tup{A, \ctx, \ctx'} \in \ctxchg$,

  \item there exists
    $\theta \in \eval{\addfacts{\ctbox^{\ctx}, A, \act\sigma}}$ such
    that:
    \begin{compactenum}

    \item for each skolem term
      $\dscall(c) \in \domain{\dscmap} \cap \domain{\theta}$, we have
      $\dscall(c)/v \in \dscmap$ if and only if
      $\dscall(c)/v \in \theta$ (i.e., $\theta$ and $\dscmap$
      ``agree'' on the common skolem terms in their domains, in order
      to realize the deterministic service call semantics);


    \item $\scmap' = \scmap \cup \theta$;

    \item
      $\tup{A, \addfacts{\ctbox^{\ctx}, A, \act\sigma}\theta,
        \delfacts{\ctbox^{\ctx}, A, \act\sigma}, \ctx', A'} \in
      \csfilter$;

    \item $A$ is $\ctbox^\ctx$-consistent, and $A'$ is
      $\ctbox^{\ctx'}$-consistent.

    \end{compactenum}
  \end{compactenum}
  \ \
\end{definition}

\noindent
Essentially, instead of only capturing the changes of ABox and service
call map by an action, the $\cstell$ operation also capture the
changes of the contexts. In addition, the inconsistency check is
performed with respect to the new context.
%
%
%
%
Next, we refine the notion of final states as follows.

\begin{definition}[Final State]\label{def:cs-final-state-program}
  Let \sidetextb{Final State}
  $\csgkabsym = \tup{\ctbox, \initabox, \actset, \ginitprog, \initctx,
    \ctxprocset}$
  be a \csgkab with transition system $\ts{\csgkabsym}$.
  We define when a state $\tup{A, \scmap, \ctx, \delta}$ of
  $\ts{\csgkabsym}$ is \emph{a final state}, written
  $\final{\tup{A, \scmap, \ctx, \delta}}$, as follows:
\begin{compactenum}
\item $\final{\tup{A, \scmap, \ctx, \gemptyprog}}$;
\item $\final{\tup{A, \scmap, \ctx, \delta_1|\delta_2}}$ if
  $\final{\tup{A, \scmap, \ctx, \delta_1}}$ or
  $\final{\tup{A, \scmap, \ctx, \delta_2}}$;
\item $\final{\tup{A, \scmap, \ctx, \delta_1;\delta_2}}$ if
  $\final{\tup{A, \scmap, \ctx, \delta_1}}$ and
  $\final{\tup{A, \scmap, \ctx, \delta_2}}$;
\item $\final{\tup{A, \scmap, \ctx, \gif{\varphi}{\delta_1}{\delta_2}}}$ \\ if
  $\ask(\varphi, T, A) = \true$, and
  $\final{\tup{A, \scmap, \ctx, \delta_1}}$;
\item $\final{\tup{A, \scmap, \ctx, \gif{\varphi}{\delta_1}{\delta_2}}}$ \\if
  $\ask(\varphi, T, A) = \false$, and
  $\final{\tup{A, \scmap, \ctx, \delta_2}}$;
\item $\final{\tup{A, \scmap, \ctx, \gwhile{\varphi}{\delta}}}$ if
  $\ask(\varphi, T, A) = \false$;
\item $\final{\tup{A, \scmap, \ctx, \gwhile{\varphi}{\delta}}}$ if
  $\ask(\varphi, T, A) = \true$, and
  $\final{\tup{A, \scmap, \ctx, \delta}}$.
\end{compactenum}
\ \
\end{definition}

Having the refined filter relation, 
$\cstell$ operation, and also the refined final states, we now proceed
to refine the \emph{program execution relation}
$\gprogtrans{\act\sigma, \csfilter}$, which describes how a grounded
action
simultaneously evolves the contexts as well as data- and
program-state. 

\ \\

\begin{definition}[Context-sensitive Program Execution Relation]\label{def:cs-prog-exec-relation}
  Given \sidetext{Context-sensitive Program Execution Relation} a
  \csgkab
  $\csgkabsym = \tup{\ctbox, \initabox, \actset, \ginitprog, \initctx,
    \ctxprocset}$,
  and a filter relation $\csfilter$, we define a
  \emph{context-sensitive program execution relation}
  $\gprogtrans{\act\sigma, \csfilter}$ as follows:
\begin{compactenum}
\item $\tup{A, \scmap, \ctx, \gactc{Q(\vec{p})}{\ctxe}{\act(\vec{p})} }
  \gprogtrans{\act\sigma, \csfilter}
  \tup{A', \scmap', \ctx', \gemptyprog}$, \\if the following hold:
  \begin{compactenum}
  \item
    $\tup{\tup{A, \scmap, \ctx}, \act\sigma, \tup{A', \scmap', \ctx'}}
    \in \cstell_{\csfilter}$,
  \item $\sigma$ is a legal parameter assignment for $\act$ in $A$
    w.r.t.\ context $\ctx$ and action invocation
    $\gactc{Q(\vec{p})}{\ctxe}{\act(\vec{p})}$, 
  \item $\ctx \cup \ctxth \models \ctxe$.
  \end{compactenum}

%
\item $\tup{A, \scmap, \ctx, \delta_1|\delta_2} \gprogtrans{\act\sigma,
    \csfilter} \tup{A', \scmap', \ctx', \delta'}$, \\if $\tup{A, \scmap, \ctx,
    \delta_1} \!\gprogtrans{\act\sigma, \csfilter}\! \tup{A', \scmap', \ctx',
    \delta'}$ or $\tup{A, \scmap, \delta_2} \gprogtrans{\act\sigma,
    \csfilter} \tup{A', \scmap', \ctx', \delta'}$;
\item $\tup{A, \scmap, \ctx, \delta_1;\delta_2} \gprogtrans{\act\sigma,
    \csfilter} \tup{A', \scmap', \ctx', \delta_1';\delta_2}$, \\if $\tup{A,
    \scmap, \ctx, \delta_1} \gprogtrans{\act\sigma, \csfilter} \tup{A',
    \scmap', \ctx', \delta_1'}$;
\item
  $\tup{A, \scmap, \ctx, \delta_1;\delta_2} \gprogtrans{\act\sigma, \csfilter}
  \tup{A', \scmap', \ctx', \delta_2'}$, \\
  if $\final{\tup{A, \scmap, \ctx, \delta_1}}$, and
  $\tup{A, \scmap, \ctx, \delta_2} \gprogtrans{\act\sigma, \csfilter} \tup{A',
    \scmap', \ctx', \delta_2'}$;
\item
  $\tup{A, \scmap, \ctx, \gif{\varphi}{\delta_1}{\delta_2}}
  \gprogtrans{\act\sigma, \csfilter}
  \tup{A', \scmap', \ctx', \delta_1'}$, \\
  if $\ask(\varphi, T, A) = \true$, and
  $\tup{A, \scmap, \ctx, \delta_1} \gprogtrans{\act\sigma, \csfilter}
  \tup{A', \scmap', \ctx', \delta_1'}$;
\item $\tup{A, \scmap, \ctx, \gif{\varphi}{\delta_1}{\delta_2}} \gprogtrans{\act\sigma, \csfilter} \tup{A', \scmap', \ctx', \delta_2'}$,\\
  if $\ask(\varphi, T, A) = \false$, and
  $\tup{A, \scmap, \ctx, \delta_2} \gprogtrans{\act\sigma, \csfilter}
  \tup{A', \scmap', \ctx', \delta_2'}$;
\item $\tup{A, \scmap, \ctx, \gwhile{\varphi}{\delta}}
  \gprogtrans{\act\sigma, \csfilter} \tup{A', \scmap', \ctx', \delta';\gwhile{\varphi}{\delta}}$,\\
  if $\ask(\varphi, T, A) = \true$, and
  $\tup{A, \scmap, \ctx, \delta} \gprogtrans{\act\sigma, \csfilter} \tup{A',
    \scmap', \ctx', \delta'}$.
\end{compactenum}
\ \ 
\end{definition}

We are now defining the construction of \csgkabs transition systems that
is parameterized with filter as follows.

\begin{definition}[\csgkab Transition System]\label{def:cs-gkab-ts}
  Given \sidetext{\csgkab Transition System} a \csgkab
  $\csgkabsym = \tup{\ctbox, \initabox, \actset, \ginitprog, \initctx,
    \ctxprocset}$,
  and a filter relation $\csfilter$, we define the \emph{transition
    system of $\csgkabsym$ w.r.t.~$\csfilter$}, written
  $\ts{\csgkabsym}^{\csfilter}$, as
  $\tup{\const, \ctbox, \stateset, s_0, \abox, \cntx, \trans}$, where
  \begin{compactenum}
  \item $s_0 = \tup{\initabox, \emptyset, \ctx_0, \ginitprog}$, and
  \item $\stateset$ and $\trans$ are defined by simultaneous induction
    as the smallest sets such that 
    \begin{compactenum}
    \item $s_0 \in \stateset$, and 
    \item if $\tup{A, \scmap, \ctx, \delta} \in \stateset$ and
      $\tup{A, \scmap, \ctx, \delta} \gprogtrans{\act\sigma, \csfilter}
      \tup{A', \scmap', \ctx', \delta'}$,
      then $\tup{A', \scmap', \ctx', \delta'}\in\stateset$ and
      $\tup{A, \scmap, \ctx, \delta}\trans \tup{A', \scmap', \ctx',
        \delta'}$.
  \end{compactenum}
  \end{compactenum}
 \ \ 
\end{definition}
As in GKABs, by suitably concretizing the filter relation, we can obtain various
execution semantics for \csgkabs.
We are now exploiting filter relations to define the standard
execution semantics of \csgkab.
Particularly, we define a context-sensitive standard filter relation
$\csfilter_S$ as follows:

\begin{definition}[Context-sensitive Standard Filter $\csfilter_S$]
  Let \sidetext{Context-sensitive Standard Filter $\csfilter_S$}
  $\csgkabsym = \tup{\ctbox, \initabox, \actset, \ginitprog, \initctx,
    \ctxprocset}$
  be a \csgkab, $A$ and $A'$ be ABoxes over $\voc(T)$, $\facta$ be a
  set of ABox assertions over $\voc(T)$ to be added, $\factd$ be a set
  of ABox assertions over $\voc(T)$ to be deleted, and $\ctx$ be a
  context,
  we then have $\tup{A, \facta, \factd, \ctx, A'} \in \csfilter_S$ if
  $A' = (A \setminus \factd) \cup \facta$,
\end{definition}


Filter $\csfilter_S$ gives rise to the \emph{standard execution
  semantics} for $\csgkabsym$. 
%
%
We call the \csgkabs adopting these semantics \emph{\scsgkabs}.

\begin{definition}[\scsgkabs Standard Transition System]
  Given \sidetext{\scsgkabs Standard Transition System} a \csgkab
  $\csgkabsym = \tup{\ctbox, \initabox, \actset, \ginitprog, \initctx,
    \ctxprocset}$
  and a standard filter $\csfilter_S$, the \emph{standard transition
    system of $\csgkabsym$}, written $\ts{\csgkabsym}^{\csfilter_S}$,
  is the transition system of $\csgkabsym$ w.r.t.\ $\csfilter_S$.
\end{definition}

The notion of run and run-boundedness of \scsgkabs transition systems
is defined similarly as in \Cref{def:run-bounded-kab,def:run-of-kab}.

\begin{example}\label{ex:exec-csgkab}
  Continuing our running example, let the \csgkab $\csgkabsym$
  specified in \Cref{ex:csgkab} be an \scsgkab. Consider a state
  $s = \tup{A, \scmap, \ctx, \delta}$ where:
  \begin{flushleft}
    $\begin{array}{l@{}ll} 
       \bullet \ A = &\set{ & \exo{ReceivedOrder}(
        \excon{chair} ), \exo{ApprovedOrder}( \excon{table} ),
                                                    \exo{designedBy}( \excon{table}, \excon{alice} ), \\
                                           &&\exo{Designer}( \excon{alice} ), \exo{hasDesign}( \excon{table} , \excon{ecodesign} ),\\
                                           &&\exo{hasAssemblingLoc}( \excon{table}, \excon{bolzano} ) \ \ }, 
     \end{array}$ 
   \end{flushleft}
  \begin{flushleft}
    $\begin{array}{l@{}ll} 
       \bullet \  \scmap =& \set{ &[\exs{getDesigner}(\excon{table}) \ra
                                      \excon{alice}], [\exs{getDesign}(\excon{table}) \ra \excon{ecodesign}], \\
                                           &&[\exs{assignAssemblingLoc}(\excon{table}) \ra
                                              \excon{bolzano}]  \ \ },
     \end{array}$ 
   \end{flushleft}
   \begin{flushleft}
     $     
     \begin{array}{l}
       \bullet \  \ctx = \set{\cval{\exc{PP}}{\exv{N}}, \cval{\exc{S}}{\exv{NS}}},
     \end{array}
     $
   \end{flushleft}
   \begin{flushleft}
     $     
     \begin{array}{l}
       \bullet \  \delta = \delta_3 ; \delta_4 ; \delta_5 ; \gwhile{
       \exists \exvar{x}.[\exo{Order}(\exvar{x})] \wedge
       \neg[\exo{DeliveredOrder}(\exvar{x})]   }{\delta_0}.
     \end{array}
     $
   \end{flushleft}
   Observe that the state $s$ is a reachable state from the initial
   state $s_0$ in the transition system
   $\ts{\csgkabsym}^{\csfilter_S}$ of $\csgkabsym$. From the state
   $s$, we have a possible successor state $s' = \tup{A', \scmap',
     \ctx', \delta'}$ with 
  \begin{flushleft}
    $\begin{array}{l@{}ll} 
       \bullet \ A' = &\set{ & \exo{ReceivedOrder}( \excon{chair} ), 
                               \exo{designedBy}( \excon{table}, \excon{alice} ), \exo{Designer}( \excon{alice} ), \\
                      &&\exo{hasDesign}( \excon{table} ,
                         \excon{ecodesign} ), \exo{hasAssemblingLoc}( \excon{table},
                         \excon{bolzano} ), \\
                      &&\exo{AssembledOrder}(\excon{table}),  \exo{assembledBy}(
                         \excon{table},\excon{bob}  ),  \exo{Assembler}( \excon{bob} ) \ \ },
     \end{array}$ 
   \end{flushleft}
  \begin{flushleft}
   $     
    \begin{array}{l@{}ll} 
       \bullet \  \scmap' =& \set{ &[\exs{getDesigner}(\excon{table}) \ra
                                      \excon{alice}], [\exs{getDesign}(\excon{table}) \ra \excon{ecodesign}], \\
                      &&[\exs{assignAssemblingLoc}(\excon{table}) \ra
                         \excon{bolzano}], \\
                      &&[\exs{getAssembler}(\excon{table}) \ra
                         \excon{bob}], \\  
                      &&[\exs{getAssemblingLoc}(\excon{table}) \ra
                         \excon{bolzano}]  \ \ }, 
     \end{array}$ 
   \end{flushleft}
  \begin{flushleft}
   $     
     \begin{array}{l}
       \bullet \  \ctx' = \set{\cval{\exc{PP}}{\exv{WE}}, \cval{\exc{S}}{\exv{PS}}},
     \end{array}
     $
   \end{flushleft}
  \begin{flushleft}
   $     
     \begin{array}{l}
       \bullet \  \delta' = \delta_4 ; \delta_5 ; \gwhile{ \exists \exvar{x}.[\exo{Order}(\exvar{x})] \wedge
    \neg[\exo{DeliveredOrder}(\exvar{x})]  }{\delta_0}.
     \end{array}
     $
   \end{flushleft}
   The state $s'$ is obtained from the execution of action invocation
   \[
   \gactc{\true}{\cval{\exc{PP}}{\exv{AP}} \wedge
     \cval{\exc{S}}{\exv{AS}}}{\exa{assembleOrders}()}
   \]
   where the context is changing from $\ctx$ to $\ctx'$ due to the
   application of the following context evolution rule:
   \[
   \carulex{\true}{\cval{\exc{PP}}{\exv{N}} \wedge
     \cval{\exc{S}}{\exv{NS}} }{ \cval{\exc{PP}}{\exv{WE}},
     \cval{\exc{S}}{\exv{PS}} }.
   \]
   Next, in $\delta_4$ there are choices between doing the quality
   check internally or by outsourcing it.
   \[
   \begin{array}{rl}
     \delta_4 =& \gactc{\true}{ \neg ( \cval{\exc{PP}}{\exv{RE}} \vee
     \cval{\exc{S}}{\exv{PS}} ) }{\exa{checkAssembledOrders}()} \ | \\
     &\gactc{\true}{ \cval{\exc{PP}}{\exv{RE}} \vee
     \cval{\exc{S}}{\exv{PS}} }{\exa{outsourceQualityCheck}()}
   \end{array}
   \]
   Since the current context is $\ctx'$, and
   $\ctx' \cup \ctxth \models \cval{\exc{PP}}{\exv{RE}} \vee
   \cval{\exc{S}}{\exv{PS}}$,
   here we can execute $\exa{outsourceQualityCheck} /0$. In this case
   one plausible successor of $s'$ is
   $s'' = \tup{A'', \scmap'', \ctx'', \delta''}$ with 
  \begin{flushleft}
    $\begin{array}{l@{}ll} 
       \bullet \ A'' = &\set{ & \exo{ReceivedOrder}( \excon{chair} ), 
                                \exo{designedBy}( \excon{table}, \excon{alice} ), \exo{Designer}( \excon{alice} ), \\
                       &&\exo{hasDesign}( \excon{table} ,
                          \excon{ecodesign} ), \exo{hasAssemblingLoc}( \excon{table},
                          \excon{bolzano} ), \\
                       &&\exo{AssembledOrder}(\excon{table}),  \exo{assembledBy}(
                          \excon{table},\excon{bob}  ),
                          \exo{Assembler}( \excon{bob} ), \\
                       &&\exo{checkedBy}(
                          \excon{table},\excon{qccompany}  ) \ \ }, 
     \end{array}$
   \end{flushleft}
  \begin{flushleft}
   $\begin{array}{l@{}ll} 
       \bullet \  \scmap'' =& \set{ &[\exs{getDesigner}(\excon{table}) \ra
                                      \excon{alice}], [\exs{getDesign}(\excon{table}) \ra \excon{ecodesign}], \\
                       &&[\exs{assignAssemblingLoc}(\excon{table}) \ra
                          \excon{bolzano}], \\
                       &&[\exs{getAssembler}(\excon{table}) \ra
                          \excon{bob}], \\  
                       &&[\exs{getAssemblingLoc}(\excon{table}) \ra
                          \excon{bolzano}], \\  
                       &&[\exs{getQCService}(\excon{table}) \ra
                          \excon{qccompany}]  \ \ }, 
     \end{array}$ 
   \end{flushleft}
   \begin{flushleft}
       $
     \begin{array}{l}
       \bullet \  \ctx'' = \set{\cval{\exc{PP}}{\exv{WE}}, \cval{\exc{S}}{\exv{PS}}},
     \end{array}
     $
   \end{flushleft}
   \begin{flushleft}
       $
     \begin{array}{l}
       \bullet \  \delta'' = \delta_5 ; \gwhile{  \exists \exvar{x}.[\exo{Order}(\exvar{x})] \wedge
    \neg[\exo{DeliveredOrder}(\exvar{x})]  }{\delta_0}.
     \end{array}
     $
   \end{flushleft}
   In this case, the value of each context dimension stay the same due
   to the application of the following context evolution rule:
   $\carulex{\true}{\cval{\exc{S}}{\exv{AS}} }{ }$.

\medskip
\noindent
\textbf{Rejecting Inconsistent State in \scsgkabs. \xspace}
Recall our state $s$ above, one of its sucessor state is the state
$s'$. Now, consider a state
$s''' = \tup{A''', \scmap''', \ctx''', \delta'''}$ with $A''' = A'$,
$\scmap''' = \scmap'$,
$\ctx''' = \set{\cval{\exc{PP}}{\exv{N}}, \cval{\exc{S}}{\exv{NS}}}$,
and $\delta''' = \delta'$.  In this case, similar to $s'$, the
execution of on the state $s$
   \[
   \gactc{\true}{\cval{\exc{PP}}{\exv{AP}} \wedge
     \cval{\exc{S}}{\exv{AS}}}{\exa{assembleOrders}()}
   \]
   might leads us into $s'''$ where the value of each context
   dimension stay the same due to the application of the following
   context evolution rule:
   \[
   \carulex{\true}{\cval{\exc{S}}{\exv{AS}} }{ }.
   \]
   Though $s'$ and $s'''$ only differ on their contexts, we have that
   $s'$ is consistent while $s'''$ is inconsistent due to the
   following TBox assertions:
   \begin{align*}
     \tup{\exo{Designer} \sqsubseteq \neg \exo{Assembler} &:                                                                 
                                                            \cval{\exc{PP}}{\exv{N}}
                                                            \wedge 
                                                            (\cval{\exc{S}}{\exv{NS}}
                                                            \vee 
                                                            \cval{\exc{S}}{\exv{LS}})
                                                            }\\
     \tup{\exo{Designer} \sqsubseteq \exo{Assembler} &: \cval{\exc{PP}}{\exv{WE}} \vee \cval{\exc{S}}{\exv{PS}} }.
   \end{align*}
   From this example, we have seen that depending on the context, some
   domain constraint might be activated or not. Thus, the context
   change influence the consistency of a state.
   In the case the construction of transition system
   $\ts{\csgkabsym}^{\csfilter_S}$ will reject the generation of
   $s'''$.
\end{example}

\section{Verifying Temporal Properties over Standard \csgkab}

We are now interested in verifying whether the evolution of an
\scsgkab $\csgkabsym$, which is represented by $\ts{\csgkabsym}$,
complies with some temporal properties.
%

\subsection{Context-Sensitive FO-Variant of \texorpdfstring{$\mu$}{^^ce^^bc}-calculus}

As previous, in order to specify temporal properties to be verified
over \scsgkabs, we use a first-order variant of $\mu$-calculus
\cite{Stir01,Park76}.
In particular, we introduce the language \mulcs of
\emph{context-sensitive temporal properties}, which is based on
\muladom (see \Cref{subsec:muladom}).
Basically, we exploit ECQs to query the states, and support a
first-order quantification across states, where the quantification
ranges over the constants in the current active domain.  Additionally,
we augment \mulcs with context expressions, which allows us to check
also context information while querying states.
Formally, \mulcs is defined as follows \sidetext{Syntax of \mulcs}:
\[
\Phi ~:=~ Q ~\mid~ \ctxe ~\mid~ \lnot \Phi ~\mid~ \Phi_1 \lor \Phi_2
~\mid~ \exists x.\Phi ~\mid~ \DIAM{\Phi}
~\mid~ Z
~\mid~ \mu Z.\Phi
\]
where $\ctxe$ is a context expression over $\ctxlang$ and the rest are
the same as in \muladom (see \Cref{subsec:muladom}). 
Similar to \muladom, let $\tup{\ctbox, \initabox}$ be a contextualized
KB, we call \emph{a \mulcs formula $\Phi$ is over
  $\tup{\ctbox, \initabox}$} if each query $Q$ in $\Phi$ is a query
over $\tup{\ctbox, \initabox}$ (i.e., each atom in $Q$ only use the
vocabulary from $\voc(\ctbox)$ and might uses constants in
$\adom{\initabox}$).

The \sidetext{Semantics of \mulcs} semantics of \mulcs is also defined
over a (possibly infinite) transition system
$\ts{} = \tup{\const, \ctbox, \stateset, s_0, \abox, \cntx, \trans}$.
Similar to \muladom, 
%
%
given a transition system $\ts{}$, in order to assign the meaning to
\mulcs formulas, we introduce an individual variable valuation $\vfo$,
i.e., a mapping from individual variables $x$ to $\const$, and a
predicate variable valuation $\vso$, i.e., a mapping from predicate
variables $Z$ to subsets of $\stateset$.
The semantics of \mulcs follows the
standard $\mu$-calculus semantics, except for the semantics of queries
and of quantification. 
We assign meaning to \mulcs formulas by associating to $\ts{}$, $\vfo$
and $\vso$ an \emph{extension function} $\MODA{\cdot}$, which maps
\mulcs formulas to subsets of $\stateset$.
The extension function $\MODA{\cdot}$ is defined inductively as follows:
\[
  \begin{array}{r@{\ }l@{\ }l@{\ }l}
    \MODA{Q} &=&\{s\in\stateset \mid
                 \Ans(Q\vfo,\ctbox^{\cntx(s)},\abox(s)) = \mathit{true}\}\\
    \MODA{\ctxe} &=&\{s\in\stateset \mid \cntx(s)\cup\ctxth \models \ctxe\}\\
    \MODA{\exists x. \Phi} &=&\{s\in\stateset \mid\exists d.d\in\adom{\abox(s)}
      \text{ and } s \in \MODAX{\Phi}{[x/d]}\}\\
    \MODA{Z}  &=& V(Z) \subseteq  \stateset\\
    \MODA{\lnot \Phi} &=& \stateset - \MODA{\Phi}\\
%
%
    \MODA{\Phi_1 \lor \Phi_2} &=& \MODA{\Phi_1}\cup\MODA{\Phi_2}\\
%
%
    \MODA{\DIAM{\Phi}}  &=&  \{s\in\stateset \mid\exists s'.\
      s \Rightarrow s' \text{ and } s'\in\MODA{\Phi}\}\\
%
%
    \MODA{\mu Z.\Phi} &=&
      \bigcap\{\E\subseteq\stateset \mid {\MODA{\Phi}}_{[Z/\E]} \subseteq\E \}
%
  \end{array}
\]
where $Q\vfo$ is the query obtained from $Q$ by substituting its free variables
according to $\vfo$.
For a closed formula $\Phi$ (for which $\MODA{\Phi}$ does not depend on $\vfo$
or $\vso$), we denote with $\MOD{\Phi}$ the extension of $\Phi$ in $\ts{}$, and
we say that $\Phi$ holds in a state $s\in\stateset$ if $s\in\MOD{\Phi}$.
%
In this case, we write $\ts{},s \models \Phi$.
Furthermore, a closed formula $\Phi$ holds in $\ts{}$, briefly
\emph{$\ts{}$ satisfies $\Phi$}, 
if $\ts{},s_0\models \Phi$ (In this situation we write
$\ts{} \models \Phi$).

\begin{example}
In our running example, the property \vspace*{-1mm}
\begin{center}
$
\nu Z.(\forall x. \exo{Order}(x)
\wedge \cval{\exc{S}}{\exv{PS}} \ra \mu Y.(\exo{DeliveredOrder}(x) \vee
\DIAM{Y})) \wedge \BOX{Z}
$
\end{center}\vspace*{-1mm}
%
%
checks that along every path, it is always true that every customer
order in the peak season will be eventually delivered, independently
on how the context and the state evolve.
\end{example}


\subsection{Verification of Standard \csgkabs}

The problem definition of the \mulcs formula verification over
\scsgkabs is defined similarly as in KABs (see
\Cref{def:verification-kab}). Precisely it is defined as follows:

\begin{definition}[Verification of a \mulcs Property over an
  \scsgkab]\label{def:verification-csgkab}
  \ \sidetext{Verification of a \mulcs Formula over an \scsgkab} Given
  an \scsgkab
  $\csgkabsym = \tup{\ctbox, \initabox, \actset, \ginitprog, \initctx,
    \ctxprocset}$
  and a closed $\mulcs$ formula $\Phi$ over $\tup{\ctbox, \initabox}$.  Let
  $\ts{\csgkabsym}$ be the transition system of $\csgkabsym$,
  \emph{the verification of a \mulcs formula $\Phi$ over $\csgkabsym$}
  is a problem to check whether $\ts{\csgkabsym} \models \Phi$.
%
\end{definition}

\noindent
We solve this problem by compiling \scsgkabs into S-GKABs and show
that the verification of \mulcs formulas over \scsgkabs can be recast
as verification over S-GKAB (This claim is formally stated in
\Cref{thm:ver-scsgkab-to-sgkab}).
Technically, we do the following:
\begin{compactenum}
\item We define a generic translation $\tgkabcs$ (in
  \Cref{sec:transf-scsgkabs-to-sgkabs}), that given an \scsgkabs
  $\csgkabsym$, produces an S-GKAB $\tgkabcs(\csgkabsym)$.

\item We define a generic translation $\tcsmula$ (in
  \Cref{sec:transf-scsgkabs-to-sgkabs}) that takes a \mulcs formula
  $\Phi$ as an input and produces a \muladom formula $\tcsmula(\Phi)$.

\item We introduce a certain bisimulation relation  in which given a
  context-sensitive transition system $\ts{1}$ and a KB transition
  system $\ts{2}$ such that they are bisimilar w.r.t.\ this
  bisimulation relation, we have that $\ts{1}$ satisfy a \mulcs
  formula $\Phi$ if and only if $\ts{2}$ satisfy the \muladom formula
  $\tcsmula(\Phi)$.

\item We show that the transition system of an \scsgkab $\csgkabsym$
  and the transition system of the corresponding S-GKAB
  $\tgkabcs(\csgkabsym)$ (that is obtained from $\csgkabsym$ via the
  translation $\tgkabcs$) are bisimilar w.r.t.\ the bisimulation
  relation introduced in the previous step.

\end{compactenum}

\subsubsection{Transforming \scsgkabs into S-GKABs}\label{sec:transf-scsgkabs-to-sgkabs}

Recall that we fix a set
\[
\cdimset = \{d_1,\ldots,d_n\}
\]
of context dimensions. Each context dimension $d_i \in \cdimset$ has
its own tree-shaped finite value domain
$\tup{\cdom[d_i],\cover[d_i]}$, where $\cdom[d_i]$ represents the
finite set of domain values, and $\cover[d_i]$ represents the
predecessor relation forming the tree. 
%



The idea of our translation $\tgkabcs$, that translates \scsgkabs
$\csgkabsym$ into an S-GKAB $\tgkabcs(\csgkabsym)$, is as follows:

  \begin{compactenum}
  \item We capture the context information using ABox assertions.  
    Precisely, each context dimension assignment is internally
    captured by an ABox assertion. Thus, for each context dimension
    assignment $\cval{d_i}{v_j}$ we reserve two fresh concept names
    $\cdcc_i^{v_j}$ and $\cdcq_i^{v_j}$ in order to represent it as an
    ABox assertion.  Such kind of concept name is called
    \sidetext{Context Dimension Concept Name}\emph{context dimension
      concept name}.
    The reason for introducing two different concepts for representing
    a context dimension assignment is to simplify the correctness
    proof of the reduction. The idea is to make a separation between
    the following situations:
    \begin{compactenum}
    \item when we need to reason using only the current context
      information (in this case we use $\cdcq_i^{v_j}$).
    \item when we need to use the current context information together
      with the value domain semantics of each context dimension value
      domain (in this case we use $\cdcc_i^{v_j}$).
  \end{compactenum}
  Furthermore, we reserve a special constant $\ctxconst \in \iconst$
  to populate such kind of concept name. We call \sidetext{Context
    ABox Assertion} \emph{context ABox assertion} an ABox assertion
  made by context dimension concepts. Furthermore, the semantics of context
  dimensions value domains is captured inside the TBox.
%


\item The context expressions are captured by ECQs queries which use
  context dimension concepts as its vocabulary.

  \item We simulate the context-evolution rules by actions.

  \item To check the inconsistency, since the TBox assertions that is
    needed to check inconsistency are determined based on the context,
    we introduce several special actions to check inconsistency which
    also taking into account the context. To this aim, for a technical
    reason, we reserve a fresh concept name $\inccon$ and we will use
    the TBox assertion $\inccon \sqsubseteq \neg \inccon$ to prevent
    the generation of an inconsistent state,
    and we also use $\ctxconst$ to populate such concept.
%

  \item For translating the program, the idea is to concatenate each
    action execution with non-deterministic choice of actions that
    change context and the action that checks the
    inconsistency. Furthermore, since the changes of context requires
    the original ABox, we don't materialize the result of an action
    execution directly after its execution, instead we just mark the
    assertions that should be added/deleted,
  and materialized it during the execution of the action that evolves
  context.  To this aim, for each concept name $N \in \voc(\ctbox)$,
  we introduce two fresh concept name $N^a$ and $N^d$ to keep track
  the temporary information about ABox assertions to be added/deleted
  before we materialize the update (similarly for roles).

\item We also use a special marker $\tmp$, made by a reserved concept
  name $\tmpconceptname$ and a constant $\tmpconst$, to mark
  intermediate states (the states where we still need to change the
  context and check the inconsistency). We call a stable state the
  state except the intermediate states (i.e., the states that does not
  contain $\tmp$).


\end{compactenum}

Now, in order to reduce the verification of \mulcs over \scsgkab into
the verification of \muladom over S-GKAB, we first need to introduce
several preliminaries below.

\begin{definition}[Set of All Possible Contexts Over 
  $\cdimset$]\label{def:set-all-possible-context} \ \sidetext{Set of All Possible Contexts}
%
%
  We define \emph{the set of all possible contexts over $\cdimset$} as
  a set $\ctxall{\cdimset}$ of contexts such that
  $\ctx \in \ctxall{\cdimset}$ if $\ctx$ is a context
  over $\cdimset$.
\end{definition}

\noindent
Roughly speaking, the definition above stated that the set
$\ctxall{\cdimset}$ contains all possible context over $\cdimset$.



We now proceed to define a TBox which capture the semantics of context
dimensions value domains in terms of TBox assertions. I.e., this TBox
capture the value domain theory $\ctxth$ of $\cdimset$. 

\begin{definition}[TBox obtained from  a Set of Context Dimensions $\cdimset$]\label{def:tbox-from-ctxdim}
  \ \sidetext{TBox obtained from a Set of Context Dimensions
    $\cdimset$} We define a \emph{TBox obtained from a set of context
    dimensions $\cdimset$} as a \dllitea TBox $T_\cdimset$ such that:
    \begin{compactitem}
    
    \item For each $d_i \in \cdimset$, and for all values
      $v_1,v_2\in\cdom[d_i]$ such that $v_1\cover v_2$, we have that
      $T_\cdimset$ contains $\cdcc_i^{v_1} \sqsubseteq \cdcc_i^{v_2}$,
      where $\cdcc_i^{v_1}$ and $\cdcc_i^{v_2}$ are fresh concept
      names each representing the context dimension assignment
      $\cval{d_i}{v_1}$ and $\cval{d_i}{v_2}$ respectively.
      Intuitively, this states that the value $v_2$ is more general
      than $v_1$, and hence, whenever we have $\cval{d}{v_1}$ we can
      infer that $\cval{d}{v_2}$.

    \item For each $d_i \in \cdimset$, and for all values
      $v_1,v_2,v\in\cdom[d_i]$ such that $v_1\cover v$ and
      $v_2\cover v$, we have that $T_\cdimset$ contains
      $\cdcc_i^{v_1} \sqsubseteq \neg \cdcc_i^{v_2}$, where
      $\cdcc_i^{v_1}$ and $\cdcc_i^{v_2}$ are fresh concept names each
      representing the context dimension assignment $\cval{d_i}{v_1}$
      and $\cval{d_i}{v_2}$ respectively.
      Intuitively, this expresses that sibling values $v_1$ and $v_2$
      are disjoint.
    \end{compactitem}
\ \ 
\end{definition}

\noindent
Notice that in the definition above we use the context dimension
concepts that are used to reason in the situation where we take into
account the value domain semantics. Since the value domain semantics
is encoded only using those concept names, it will be ignored when we
make a query using the context dimension concepts that are used
to reason without considering the value domain semantics.

We now define a query that represents a context as follows:

\begin{definition}[A Query That Represents Context $\ctx$]\label{def:query-represent-context}
\ \sidetext{A Query That Represents Context $\ctx$}  
Given a context
  \[
  \ctx = \set{\cval{d_1}{v_1},\ldots,\cval{d_n}{v_m}},
  \]
  the \emph{query that represents context $\ctx$} is a boolean CQ
  \[
  q_\ctx = \cdcq_1^{v_1}(\ctxconst) \wedge \ldots \wedge \cdcq_n^{v_m}(\ctxconst)
  \]
\ \ 
\end{definition}

\noindent
Note that in the definition above we use the context dimension concepts
that are used to reason without considering the value domain
semantics.  Later on, we will see that the query above can be used to
check the current context.

To capture the context information within an ABox, we define the
notion of a set of ABox assertions representing a context as follows:

\begin{definition}[Set of ABox Assertions Representing a Context]\label{def:abox-context}
  \ \sidetext{Set of ABox Assertions Representing a Context} Let
  \[
  \ctx = \set{\cval{d_1}{v_1},\ldots,\cval{d_n}{v_m}}
  \]
  be a context, the \emph{set of ABox assertions representing the
    context $\ctx$} is a set $A_\ctx$ of context ABox assertions as
  follows:
  \[
  A_\ctx = \set{\cdcc_1^{v_1}(\ctxconst), \ldots,
    \cdcc_n^{v_m}(\ctxconst), \cdcq_1^{v_1}(\ctxconst), \ldots,
    \cdcq_n^{v_m}(\ctxconst)}.
  \]
  where each $\cdcc_i^{v_j}$ (resp.\ $\cdcq_i^{v_j}$) is a context
  dimension concept name.
\end{definition}

When we compile \scsgkabs into S-GKABs, we represent a context
expression as a query and it is done as follows:

\begin{definition}[A Query That Represents A Context Expressions]\label{def:query-rep-context-exp}
  \ \sidetext{A Query that Represents a Context Expressions} Given a
  context expressions $\ctxe$, \emph{the query that represents the
    context expressions $\ctxe$} is a boolean ECQ query $q_{\ctxe}$
  obtained by replacing each occurrence of context dimension
  assignment of the form $\cval{d_i}{v_j}$ with an atom
  $\cdcc_i^{v_j}(\ctxconst)$, where $\cdcc_i^{v_j}$ (resp.\
  $\ctxconst$) is the reserved concept name (resp.\ reserved constant)
  explained before.
\end{definition}

\noindent
Notice that in the definition above we use the context dimension
concepts that are used to reason in the situation where we take into
account the value domain semantics. 
In the following lemma we show the ``correctness'' of our mechanism in
expressing a context expression as a query.

\begin{lemma}\label{lem:ctx-exp-and-query-about-it}
  Given a context $\ctx$, and a context expression $\ctxe$. Let
  $q_{\ctxe}$ be the query that represents the context expressions
  $\ctxe$. We have that 
\[
\ctx \cup \ctxth \models \ctxe \mbox{ if and only if } \Ans(q_{\ctxe},
T_\cdimset, A_\ctx) = \true
\]
\end{lemma}
\begin{proof}
  Trivially true by observing
  \Cref{def:abox-context,def:query-rep-context-exp,def:tbox-from-ctxdim}.
\end{proof}

We now proceed to describe how we compile queries when we transform
\scsgkabs into S-GKABs such that they will be answered using the
correct TBox w.r.t.\ the corresponding context.


\begin{definition}[Contextually Compiled Query]\label{def:contextually-compiled-query}
  Given \sidetext{Contextually Compiled Query} an ECQ $Q$, a
  \emph{contextually compiled query} of $Q$ w.r.t.\ $\cdimset$ is a
  query $Q_\ctxb$ of the form
    \[
    Q_\ctxb =  \left( \bigvee_{\ctx \in \ctxall{\cdimset}}  \left(q_\ctx \wedge
        \rew(Q, \ctbox^\ctx) \right) \right) 
    \]
    where 
    $q_\ctx$ is the query obtained from the context $\ctx$.
\end{definition}

\noindent
In the following lemma we show the ``correctness'' of our contextually
compiled query.

\begin{lemma}\label{lem:correctness-contextually-compiled-query}
  Given a contextualized KB $\tup{\ctbox, A}$, an ECQ $Q$ over
  $\voc(\ctbox)$ and a context $\ctx$ over $\cdimset$. Let $Q_\ctxb$
  be the contextually compiled query of $Q$ w.r.t.\ $\cdimset$,
  $A_\ctx$ be the set of ABox assertions representing the context
  $\ctx$, and $T_\cdimset$ be the TBox obtained from $\cdimset$. We
  have
  \[
  \Ans(Q,\ctbox^{\ctx}, A) = \Ans(Q_\ctxb, T_\cdimset, A \cup
  A_\ctx)
  \]
\end{lemma}
\begin{proof}
  Let
  $Q_\ctxb = \left( \bigvee\limits_{\ctx \in \ctxall{\cdimset}}
    \left(q_\ctx \wedge \rew(Q, \ctbox^\ctx) \right) \right)$
  where $q_\ctx$ is the query obtained from the context $\ctx$.
  By \Cref{thm:FO-rew-ECQ}, we have
$\Ans(Q,\ctbox^{\ctx}, A) = \ANS(\rew(Q,\ctbox^{\ctx}), A)$,
and
$\Ans(Q_\ctxb, T_\cdimset, A \cup A_\ctx) = \ANS(\rew(Q_\ctxb,
T_\cdimset), A \cup A_\ctx)$.
Since $Q_\ctxb$ doesn't use any vocabulary from $\voc(T_\cdimset)$
then $\rew(Q_\ctxb, T_\cdimset) = Q_\ctxb$. Now we have to show that 
$\ANS(\rew(Q,\ctbox^{\ctx}), A) = \ANS(Q_\ctxb, A \cup A_\ctx)$.
By construction of $Q_\ctxb$ and $q_\ctx$ (see
\Cref{def:contextually-compiled-query,def:query-represent-context}) we
have that 
\begin{compactenum}
\item there exists $\ctx \in \ctxall{\cdimset}$ such that
$\ANS(q_\ctx, A \cup A_\ctx) = \true$ and 
\item for all $\ctx' \in \ctxall{\cdimset}$ such that
  $\ctx' \neq \ctx$ we have
  $\ANS(q_{\ctx'}, A \cup A_{\ctx}) = \false$.
\end{compactenum}
Thus, now we only need to show that
\[  
\ANS(\rew(Q,\ctbox^{\ctx}), A) = \ANS(q_\ctx \wedge \rew(Q,
\ctbox^\ctx), A \cup A_\ctx).
\] 
The proof is then easily completed by observing that $Q$ doesn't use
any context dimension concepts, hence we can ignore $A_\ctx$.
%
%
\end{proof}

Recall that in S-GKABs we simulate the action execution and the
context change in two different steps. Furthermore, since the context
change is conducted after the action execution and it requires the
original ABox, we can not immediately update the ABox after the action
execution. Therefore, now we introduce the notion of delayed action
that does not add/deleted ABox assertions immediately, but only adds
markers about which ABox assertions that should be added or
deleted. Essentially when we compile \scsgkabs into S-GKABs, we will
see it later that we translate each action into a delayed action.

\begin{definition}[Delayed Action]\label{def:delayed-action}
  Given \sidetext{Delayed Action} an action
  $\act(\vec{p}):\set{e_1,\ldots,e_n}$ with
  $e_i = \map{[q_i^+]\land Q_i^-}{\add \facta_i, \del \factd_i}$, a
  \emph{delayed action} obtained from $\act$ is an action
  $\act'(\vec{p}):\set{e'_1,\ldots,e'_n, e_{temp}}$, where 
\begin{compactitem}
\item $e_i'$ (for $i \in \set{1, \ldots, n}$) is obtained from $e_i$
  such that \\
  $e'_i = \map{  \mbox{Q}^i_{\ctxb} }{\add {\facta_i}' \cup
    {\factd_i}'
  }$ where:
  \begin{itemize}

  \item $\mbox{Q}^i_{\ctxb}$ is contextually compiled query of
    $[q_i^+]\land Q_i^-$ w.r.t.\ $\cdimset$.

  \item for each atom $N(t) \in \facta_i$ (resp.\
    $P(t_1, t_2) \in \facta_i$), \\ we have $N^a(t) \in {\facta_i}'$
    (resp.\ $P^a(t_1, t_2) \in {\facta_i}'$).


  \item for each atom $N(t) \in \factd_i$ (resp.\
    $P(t_1, t_2) \in \factd_i$), \\ we have $N^d(t) \in {\factd_i}'$
    (resp.\ $P^d(t_1, t_2) \in {\factd_i}'$).


  \end{itemize}
\item $e_{temp} = \set{\map{\true}{\add \set{\tmp} } }$

\end{compactitem}
\ \ 
\end{definition}

Utilizing the notion of delayed action above as well as the other
notion introduced earlier, in the following we present how we
translate a context-sensitive action invocation into the usual action
invocation in S-GKABs. Essentially we express a query as a
contextually compiled query, we transform a context expression into
the corresponding query, and we translate an action into delayed
action.

\begin{definition}[Action Invocation Obtained From Context-Sensitive
  Atomic Action Invocation]\label{def:action-invoc-from-cs-action-invoc}
  An \sidetextb{Action Invocation Obtained From Context-Sensitive
    Atomic Action Invocation} \emph{action invocation obtained from
    context-sensitive atomic action invocation}
  $\gactc{Q(\vec{p})}{\ctxe}{\act(\vec{p})}$ is an action invocation
$\gact{Q'(\vec{p})}{\act'(\vec{p})}$, where
\begin{compactitem}

\item $Q' = Q_\ctxb \wedge q_{\ctxe}$
  where $Q_\ctxb$ is contextually compiled query of $Q$ (see
  \Cref{def:contextually-compiled-query}), and $q_{\ctxe}$ is the
  query that represents the context expression $\ctxe$ (see
  \Cref{def:query-rep-context-exp}).


\item $\act'$ is a delayed action obtained from $\act$ (see
  \Cref{def:delayed-action}).

\end{compactitem}
\ \ 
\end{definition}

To mimic the context-evolution rules within S-GKABs, we translate them
into action invocations as follows:

\begin{definition}[Action and Action Invocation Obtained From
  Context-\\evolution Rule]
\label{def:action-and-action-invocation-obtained-from-context-evolution-rule}
Let \sidetext{Action and Action Invocation Obtained From
  Context-evolution Rule}
$\csgkabsym = \tup{\ctbox, \initabox, \actset, \ginitprog, \initctx,
  \ctxprocset}$
be a \csgkab.  An \emph{action invocation obtained from a
  context-evolution rule} $\tup{Q, \ctxe} \mapsto C_{new}$ in
$\ctxprocset$, is an action invocation $\gact{Q'}{\act_{\ctx}()}$
where
  \begin{compactenum}
  \item $Q' = Q_\ctxb \wedge q_{\ctxe}$
    where $Q_\ctxb$ is contextually compiled query of $Q$, and
    $q_{\ctxe}$ is the query obtained from the context expression
    $\ctxe$.

  \item $\act_\ctx$ is a 0-ary action obtained from
    $\tup{Q, \ctxe} \mapsto C_{new}$ as follows:


  \begin{compactenum}[(a)]

  \item For each $[d_i \mapsto v_j] \in C_{new}$, we have:
    \begin{enumerate}[(i)]
    \item $\map{\true}{\add \set{\cdcc_i^{v_j}(\ctxconst), \cdcq_i^{v_j}(\ctxconst)  } }$ in
      $\eff{\act_\ctx}$, and 
    \item 
      $\map{\true}{\del \set{ \cdcc_i^{v_k}(\ctxconst),
          \cdcq_i^{v_k}(\ctxconst) } }$
      in $\eff{\act_\ctx}$ for every $v_k\in\cdom[d_i]$ such that
      $v_k \neq v_j$.
    \end{enumerate}
    The intuition is that the effect constructed in the step (i)
    assigns a new value $v_j$ into $d_i$ while the effects constructed
    in the step (ii) delete the old value of $d_i$.

\item For each concept name $N \in \voc(\ctbox)$, we have
  \begin{enumerate}[(i)]
  \item $\map{N^a(x)}{\add \set{N(x) }, \del \set{N^a(x) } }$ in
    $\eff{\act_\ctx}$, and 
  \item $\map{N^d(x)}{\del \set{N(x), N^d(x)} }$ in
    $\eff{\act_\ctx}$. 
  \end{enumerate}
  Intuitively, the effects constructed in this step concretely
  add/delete the ABox assertions that was marked to be added/deleted,
  and additionally delete all of those markers.
  
\item Similarly for the role names, we create the same effect as in
  the step (b) above. 
 
\end{compactenum}
In this case we say that $\act_{\ctx}$ is \emph{an action obtained
  from the context-evolution rule $\tup{Q, \ctxe} \mapsto C_{new}$}.
\end{compactenum}
\ \ 
\end{definition}

%
%

\noindent
Given a set of context-evolution rules $\ctxprocset$, we write
$\ctxproc$ to denote the set of action invocations obtained from all
context-evolution rules in $\ctxprocset$. 

In order to simulate the non-deterministic choice of context-evolution
rule that change context inside the S-GKABs, we introduce the notion
of context-change program as follows.

\begin{definition}[Context-Change Program]\label{def:ctx-chg-prog}
  Let \sidetext{Context-Change Program} $\ctxprocset$ be a set of
  context-evolution rules and $\ctxproc$ be the set of action
  invocations obtained from $\ctxprocset$, we define
  \emph{context-change program} as follows:
  \[
  \delta_{\ctxprocset} = \act_1 | \ldots | \act_{\card{\ctxproc}}
  \]
  where $\act_i \in \ctxproc$.
\end{definition}

To check the consistency of the resulting state after the action
execution and the context change, we introduce the notion of
context-sensitive consistency check action below.

\begin{definition}[Context-Sensitive Consistency Check Action]\label{def:cs-cons-chk-act}
  Let \sidetext{Context-Sensitive Consistency Check Action}
  $\tup{\ctbox, A}$ be a contextualized KB, we define a
  \emph{context-sensitive consistency check action}
  $\act^{\ctbox}_\bot$ over $\ctbox$ as a 0-ary (i.e., has no action
  parameters), where $\eff{\act^{\ctbox}_\bot}$ is the smallest set of
  effects containing:

\begin{itemize}

\item for each functionality assertion $\tup{\funct{R} : \ctxe} \in
  \ctbox$, we have \\
$
\map{q_{\ctxe} \wedge \exists x, y, z.\qunsatf(\funct{R}, x, y, z) } {\add
  \set{\inccon(\ctxconst)} } \mbox{ in } \eff{\act^{\ctbox}_\bot},
$

\item for each negative concept inclusion assertion
  $\tup{B_1 \ISA \neg B_2 : \ctxe} \in \ctbox$, and for each
  $\ctx \in \ctxall{\cdimset}$ we have \\
  $ \map{q_{\ctxe} \wedge q_\ctx \wedge \rew(\exists x.\qunsatn(B_1
    \ISA \neg B_2, x), \ctbox^\ctx) } $\\
  \hspace*{75mm}$ \add \set{\inccon(\ctxconst)} \mbox{ in }
  \eff{\act^{\ctbox}_\bot}, $

\item for each negative role inclusion assertion
  $\tup{R_1 \ISA \neg R_2 : \ctxe} \in \ctbox$, and for each
  $\ctx \in \ctxall{\cdimset}$ we have \\
  $ \map{q_{\ctxe} \wedge q_\ctx \wedge \rew(\exists x, y.\qunsatn(R_1
    \ISA \neg R_2, x, y), \ctbox^\ctx) }$ \\
  \hspace*{75mm}$\add \set{\inccon(\ctxconst)} \mbox{ in } \eff{\act^{\ctbox}_\bot},
  $

\item additionally, we have $\map{\true}{\del \tmp}$ in
  $\eff{\act^{\ctbox}_\bot}$
\end{itemize}
where $q_{\ctxe}$ is a query that represents the context expression
$\ctxe$, $q_\ctx$ is a query that represents context $\ctx$, and we
also make use the abbreviations of FOL query in
\Cref{def:qunsat-abbreviation}.
\end{definition}


\noindent
For brevity, in this section we simply say consistency check action
instead of context-sensitive consistency check action.

We now define a translation function $\tgprogcs$ that essentially
concatenates each action invocation with a program that
non-deterministically choose an action that changes the context, and
also an action that checks the inconsistency.  Additionally, the
translation function $\tgprogcs$ also serves as a one-to-one
correspondence (bijection) between the original and the translated
program (as well as between the sub-program).

\begin{definition}[Program Translation $\tgprogcs$]
  Given \sidetextb{Program Translation $\tgprogcs$} a \scsgkab
  $\csgkabsym = \tup{\ctbox, \initabox, \actset, \ginitprog, \initctx,
    \ctxprocset}$,
  we define a \emph{translation $\tgprogcs$} which translates the
  program $\ginitprog$ inductively as follows:
\begin{compactitem}
\item
  $\tgprogcs(\gactc{Q(\vec{p})}{\ctxe}{\act(\vec{p})}) =
  \gact{Q'(\vec{p})}{\act'(\vec{p})} \ ;\ \delta_{\ctxprocset}\ ;\
  \gact{\true}{\act^{\ctbox}_\bot()}$
                                                      \item
                                                        $\tgprogcs(\gemptyprog)
                                                        = \gemptyprog$
                                                      \item   $\tgprogcs(\delta_1|\delta_2) = \tgprogcs(\delta_1)|\tgprogcs(\delta_2)$ 

                                                      \item $\tgprogcs(\delta_1;\delta_2) = \tgprogcs(\delta_1);\tgprogcs(\delta_2)$ 

                                                      \item
                                                        $\tgprogcs(\gif{\varphi}{\delta_1}{\delta_2})
                                                        =
                                                        \gif{\varphi}{\tgprogcs(\delta_1)}{\tgprogcs(\delta_2)}$
                                                      \item
                                                        $
                                                        \tgprogcs(\gwhile{\varphi}{\delta})
                                                        =
                                                        \gwhile{\varphi}{\tgprogcs(\delta)}$
\end{compactitem}
%
where 
\begin{compactitem}

\item $\gact{Q'(\vec{p})}{\act'(\vec{p})}$ is an action invocation
  obtained from $\gactc{Q(\vec{p})}{\ctxe}{\act(\vec{p})}$ (see
  \Cref{def:action-invoc-from-cs-action-invoc}),

\item $\delta_{\ctxprocset}$ is a context-change program obtained from
  $\ctxprocset$ (see \Cref{def:ctx-chg-prog}),

\item $\act^{\ctbox}_\bot$ is a consistency check action (see
  \Cref{def:cs-cons-chk-act}).

\end{compactitem}
 \ \ 
\end{definition}
\noindent

To transform \scsgkabs into the corresponding S-GKABs, we define a
translation $\tgkabcs$ that, given an \scsgkab, generates an S-GKAB as
follows.

\begin{definition}[Translation from \scsgkab to S-GKAB]
  \ \sidetext{Translation from \scsgkab to S-GKAB}
%
%
  We define a translation $\tgkabcs$ that, given an \scsgkab
  $\csgkabsym = \tup{\ctbox, \initabox, \actset, \ginitprog, \initctx,
    \ctxprocset}$,
  produces an S-GKAB
  $\tgkabcs(\csgkabsym) = \tup{T', \initabox', \actset', \ginitprog'}$,
  where
\begin{compactitem}

\item $T' = \set{\inccon \sqsubseteq \neg \inccon} \cup T_\cdimset$,
  where $T_\cdimset$ is a TBox obtained from a set of context
  dimensions $\cdimset$,

\item $\initabox' = \initabox \cup A_{\initctx}$ (where $A_{\initctx}$
  is an ABox obtained from $\initctx$),

\item $\actset' = \actset_\act \cup \actset_\ctx$ where:

\begin{compactitem}
\item $\actset_\act$ is obtained from $\actset$ such that for each
  action $\act \in \actset$, we have $\act' \in \actset_\act$ where
  $\act'$ is a delayed action obtained from $\act$ (see \Cref{def:delayed-action}).

\item $\actset_\ctx$ is obtained from $\ctxprocset$ such that for each
  context-evolution rule $\tup{Q, \ctxe} \mapsto C_{new}$ in
  $\ctxprocset$, we have $\act_\ctx \in \actset_\ctx$ where
  $\act_\ctx$ is \emph{an action obtained from the context-evolution
    rule $\tup{Q, \ctxe} \mapsto C_{new}$} (see
  \Cref{def:action-and-action-invocation-obtained-from-context-evolution-rule}).
\end{compactitem}

\item $\ginitprog' = \tgprogcs(\ginitprog)$. 

\end{compactitem}
\ \ 
\end{definition}

A \mulcs property $\Phi$ over \csgkabs $\csgkabsym$ can then be recast
as a corresponding \muladom property over S-GKAB
$\tgkabcs(\csgkabsym)$ by simply substituting each subformula
$\DIAM{\Psi}$ of $\Phi$ with $\DIAM{\DIAM{\DIAM{\Psi}}}$ (similarly
for $\BOX{\Phi}$). Formally we define such formula translation as
follows:
\begin{definition}[Translation $\tfort$]\label{def:ttrip}
  We \sidetext{\mulcs Formula Translation $\tfort$} define a \emph{translation
    $\tfort$} that takes a \mulcs formula $\Phi$ as an input and
  produces a new \mulcs formula $\tfort(\Phi)$ by recurring over the
  structure of $\Phi$ as follows:
  \[
  \begin{array}{lll}
    \bullet\ \tfort(Q) &=& Q_\ctxb \\
    \bullet\ \tfort(\ctxe) &=& q_{\ctxe} \\
    \bullet\ \tfort(\neg \Phi) &=& \neg \tfort(\Phi) \\
    \bullet\ \tfort(\exists x.\Phi) &=& \exists x. \tfort(\Phi) \\
    \bullet\ \tfort(\Phi_1 \vee \Phi_2) &=& \tfort(\Phi_1) \vee \tfort(\Phi_2) \\
    \bullet\ \tfort(\mu Z.\Phi) &=& \mu Z. \tfort(\Phi) \\
    \bullet\ \tfort(\DIAM{\Phi}) &=& \DIAM{\DIAM{\DIAM{\tfort(\Phi)}}} 
  \end{array}
  \]
  where $Q_\ctxb$ is a contextually compiled query of $Q$ (see
  \Cref{def:contextually-compiled-query}), and $q_{\ctxe}$ is the
  query that represents the context expression $\ctxe$ (see
  \Cref{def:query-rep-context-exp}).
\end{definition}

With 
this translation in hand, we will show later that
$\ts{\csgkabsym}^{\csfilter_S} \models \Phi$ if and only if
$\ts{\tgkabcs(\csgkabsym)}^{\csfilter_S} \models \tfort(\Phi)$, which
consequently means that the verification of \mulcs over \scsgkabs can
be reduced to the corresponding verification over S-GKABs.
The core idea of the proof is to use a certain bisimulation relation
in which two bisimilar transition systems (w.r.t.\ this bisimulation
relation) can not be distinguished by \mulcs properties modulo the
formula translation $\tfort$. Then, we show that the transition system
of an \scsgkab is bisimilar to the transition system of its corresponding
S-GKAB w.r.t.\ this bisimulation relation. 

\subsubsection{Skip-two Bisimulation (ST-Bisimulation)}\label{sec:skip-two-st-bisimulation}

Towards defining the notion of ST-Bisimulation, we
introduce the notion of contextually equal between a state of
context-sensitive transition system and a state of KB transition
system as follows:


\begin{definition}[Contextually Equal State]\label{def:contextually-equal}
  \ \sidetext{Contextually Equal State} \\ Let
  $\ts{1} = \tup{\const, \ctbox, \stateset_1, s_{01}, \abox_1, \cntx,
    \trans_1}$
  be context-sensitive transition system, and
  $\ts{2} = \tup{\const, T, \stateset_2, s_{02}, \abox_2, \trans_2}$
  be KB transition systems
  .
%
%
  Consider two states $s_1 \in \stateset_1$ and $s_2 \in \stateset_2$,
  we say $s_1$ is \emph{contextually equal} to $s_2$, written
  $s_1 \eqc s_2$ if $\abox_1(s_1) \cup A_{\cntx(s_1)} = \abox_2(s_2)$.
\end{definition}

\noindent
Intuitively, two contextually equal states contain the same data/facts
in the ABox and also have the same context information (although they
are encoded in a different way).
%
%
We then define the notion of ST-Bisimulation as follows:
\begin{definition}[Skip-two Bisimulation (ST-Bisimulation)] \
  \sidetext{Skip-two Bisimulation (ST-Bisimulation)} \\
  Let
  $\ts{1} = \tup{\const, \ctbox, \stateset_1, s_{01}, \abox_1, \cntx,
    \trans_1}$
  be a context-sensitive transition system, and
  $\ts{2} = \tup{\const, T, \stateset_2, s_{02}, \abox_2, \trans_2}$
  be a KB transition system, with
  $\adom{\abox_1(s_{01})} \subseteq \const$ and
  $\adom{\abox_2(s_{02})} \subseteq \const$.  A \emph{skip-two
    bisimulation} (ST-Bisimulation) between $\ts{1}$ and $\ts{2}$ is a
  relation $\B \subseteq \Sigma_1 \times\Sigma_2$ such that
  $\tup{s_1, s_2} \in \B$ implies that:
  \begin{compactenum}
  \item $s_1 \eqc s_2$
  \item for each $s_1'$, if $s_1 \Rightarrow_1 s_1'$ then there exists
    $t_1$, $t_2$, and $s_2'$ with
    \[
    s_2 \Rightarrow_2 t_1 \Rightarrow_2 t_2 \Rightarrow_2 s_2'
    \] 
    such that $\tup{s_1', s_2'}\in\B$, $\tmp \not\in \abox_2(s_2')$
    and $\tmp \in \abox_2(t_i)$ for $i \in \set{1, 2}$.

  \item for each $s_2'$, if 
    \[
    s_2 \Rightarrow_2 t_1 \Rightarrow_2 t_2 \Rightarrow_2 s_2'
    \] 
    with $\tmp \in \abox_2(t_i)$ for $i \in \set{1, 2}$ and
    $\tmp \not\in \abox_2(s_2')$, then there exists $s_1'$ with
    $s_1 \Rightarrow_1 s_1'$, such that $\tup{s_1', s_2'}\in\B$.
 \end{compactenum}
\ \ 
\end{definition}

\noindent
Let
$\ts{1} = \tup{\const, \ctbox, \stateset_1, s_{01}, \abox_1, \cntx,
  \trans_1}$
be a context-sensitive transition system, and
$\ts{2} = \tup{\const, T, \stateset_2, s_{02}, \abox_2, \trans_2}$ be
a KB transition system,
a state $s_1 \in \stateset_1$ is \emph{ST-bisimilar} to
$s_2 \in \stateset_2$, written $s_1 \stbsim s_2$, if there exists an
ST-bisimulation relation $\B$ between $\ts{1}$ and $\ts{2}$ such that
$\tup{s_1, s_2}\in\B$.
A transition system $\ts{1}$ is \emph{ST-bisimilar} to $\ts{2}$,
written $\ts{1} \stbsim \ts{2}$, if there exists an ST-bisimulation
relation $\B$ between $\ts{1}$ and $\ts{2}$ such that
$\tup{s_{01}, s_{02}}\in\B$.


In the following two lemmas we show some important properties of
ST-bisimilar states and transition systems that will be useful later
to show that we can recast the verification of \scsgkabs into S-GKABs.

\begin{lemma}\label{lem:stbisimilar-state-satisfies-same-formula}
  Let
  $\ts{1} = \tup{\const, \ctbox, \stateset_1, s_{01}, \abox_1, \cntx,
    \trans_1}$
  be a context-sensitive transition system, and
  $\ts{2} = \tup{\const, T, \stateset_2, s_{02}, \abox_2, \trans_2}$
  be a KB transition system. 
  Consider two states $s_1 \in \stateset_1$ and $s_2 \in \stateset_2$
  such that $s_1 \stbsim s_2$. Then for every formula $\Phi$ of
  $\mulcs$, and every valuations $\vfo_1$ and $\vfo_2$ that assign to
  each of its free variables a constant $c_1 \in \adom{\abox_1(s_1)}$
  and $c_2 \in \adom{\abox_2(s_2)}$, such that $c_1 = c_2$, we have
  that
  \[
  \ts{1},s_1 \models \Phi \vfo_1 \textrm{ if and only if } \ts{2},s_2
  \models \tfort(\Phi) \vfo_2.
  \]
\end{lemma}
\begin{proof}
  In general, the proof is similar to the proof of
  \Cref{lem:sbisimilar-state-satisfies-same-formula}. Thus, here we
  only highlight some interesting cases in the induction.
\medskip
\begin{compactitem}
\item[$\bullet$ Case of \textbf{$\Phi = Q$}:] Since $s_1 \stbsim s_2$,
  we have $s_1 \eqc s_2$. Hence, by
  \Cref{def:contextually-equal}, 
  we have 
  $\abox_1(s_1) \cup A_{\cntx(s_1)} = \abox_2(s_2)$, and furthermore,
  by \Cref{lem:correctness-contextually-compiled-query}, 
  we have
  \[
  \Ans(Q,\ctbox^{\cntx(s_1)}, \abox(s_1)) = \Ans(Q_\ctxb, T_\cdimset,
  \abox(s_2))
  \]
  Since $\tfort(Q) = Q_\ctxb$, it is easy to see that for every
  valuations $\vfo_1$ and $\vfo_2$ that assign to each of its free
  variables a constant $c_1 \in \adom{\abox_1(s_1)}$ and
  $c_2 \in \adom{\abox_2(s_2)}$, such that $c_1 = c_2$, we have
  \[
  \ts{1},s_1 \models Q \vfo_1 \textrm{ if and only if } \ts{2},s_2
  \models \tfort(Q) \vfo_2.
  \]

\item[$\bullet$ Case of \textbf{$\Phi = \DIAM{\Psi}$}:]  Assume $\ts{1},s_1 \models
  (\DIAM{\Psi}) \vfo_1$, then there exists $s_1'$ s.t.\ $s_1 \trans_1
  s_1'$ and $\ts{1},s_1' \models \Psi \vfo_1$. Since $s_1 \stbsim
  s_2$, there exist $t_1$, $t_2$ and $s_2'$ s.t.\
    \[
    s_2 \trans_2 t_1 \trans_2 t_2 \trans_2 s_2'
    \] 
    and $s_1' \stbsim s_2'$.
    Hence, by induction hypothesis, for every valuations $\vfo_2$ that
    assign to each free variables $x$ of $\tfort(\Psi)$ a constant $c_2 \in
    \adom{\abox_2(s_2)}$, such that $c_2 = c_1$ with $x/c_1 \in
    \vfo_1$, we have
    \[
    \ts{2},s_2' \models \tfort(\Psi_1) \vfo_2.
    \]
    Since 
    $ s_2 \trans_2 t_1 \trans_2 t_2 \trans_2 s_2', $ therefore we get
    \[
    \ts{2},s_2 \models ( \DIAM{\DIAM{\DIAM{\tfort(\Psi)}}} )\vfo_2.
    \]
    Since $\tfort(\DIAM{\Phi}) = \DIAM{\DIAM{\DIAM{\tfort(\Phi)}}} $, we
    therefore have
    \[
    \ts{2},s_2 \models \tfort(\DIAM{\Psi} )\vfo_2. 
    \]
    The other direction can be shown in a similar way.

\end{compactitem}

\end{proof}


\begin{lemma}\label{lem:stbisimilar-ts-satisfies-same-formula}
  Consider a context-sensitive transition system
  $\ts{1}$ 
  and a KB transition system
  $\ts{2}$ 
%
  such that $\ts{1} \stbsim \ts{2}$.  For every closed \mulcs formula
  $\Phi$, we have:
  \[
  \ts{1} \models \Phi \textrm{ if and only if } \ts{2} \models
  \tfort(\Phi)
  \]
\end{lemma}
\begin{proof} Let
  $\ts{1} = \tup{\const, \ctbox, \stateset_1, s_{01}, \abox_1, \cntx,
    \trans_1}$,
  and
  $\ts{2} = \tup{\const, T, \stateset_2, s_{02}, \abox_2, \trans_2}$.
  By the definition of ST-bisimilar transition system we have that
  $s_{01} \stbsim s_{02}$. Thus, we obtain the proof as a consequence
  of \Cref{lem:stbisimilar-state-satisfies-same-formula}, due to the
  fact that
  \[ \ts{1}, s_{01} \models \Phi \textrm{ if and only if } \ts{2},
  s_{02} \models \tfort(\Phi)
  \]
\end{proof}

\subsubsection{Reducing the Verification of \scsgkabs into S-GKABs}

We now step forward to show that we can recast the verification of
\scsgkabs into S-GKABs. We open this section by showing that the
transition systems of an \scsgkab and its corresponding S-GKAB
(obtained through $\tgkabcs$) are ST-bisimilar. The following two
lemmas are aimed to show this fact.



\begin{lemma}\label{lem:scsgkab-to-sgkab-bisimilar-state}
  Let $\csgkabsym$ be an \scsgkab with transition system
  $\ts{\csgkabsym}^{\csfilter_S}$, and let $\tgkabcs(\csgkabsym)$ be
  its corresponding S-GKAB $($with transition system
  $\ts{\tgkabcs(\csgkabsym)}^{\filter_S}$$)$ obtained through
  $\tgkabcs$. 
  Consider
  \begin{inparaenum}[]
  \item a state $s_{cx} = \tup{A_{cx},\scmap_{cx}, \ctx, \delta_{cx}}$ of
    $\ts{\csgkabsym}^{\csfilter_S}$ and
  \item a state $s_s = \tup{A_s,\scmap_s, \delta_s}$ of
    $\ts{\tgkabcs(\csgkabsym)}^{\filter_S}$.
  \end{inparaenum}
If  
\begin{inparaenum}[]
\item $s_{cx} \eqc s_s$, $\scmap_{cx} = \scmap_s$ and
\item $\delta_s = \tgprogcs(\delta_{cx})$,
\end{inparaenum}
then
$\tup{A_{cx},\scmap_{cx}, \ctx, \delta_{cx}} \stbsim \tup{A_s,\scmap_s, \delta_s}$.
\end{lemma}
\begin{proof}
  For the simplicity of the proof, here we ignore the presence of the
  ABox assertion $\tmp$ that acts as a special marker and marks the
  intermediate states. The important thing to observe is that $\tmp$
  is always added to the intermediate state (where we still need to
  change the context and do the inconsistency check) but then it will
  be deleted after that.  Furthermore, the presence of $\tmp$ also
  distinguish the intermediate and stable state. 
%
%
  Now, let
  \begin{compactenum}
\item
  $\csgkabsym = \tup{\ctbox, \initabox, \actset, \ginitprog, \initctx,
    \ctxprocset}$,
  and \\
  $\ts{\csgkabsym}^{\csfilter_S} = \tup{\const, \ctbox, \stateset_\ctxb, s_{0\ctxb},
    \abox_\ctxb, \cntx, \trans_\ctxb}$,
\item   $\tgkabcs(\csgkabsym) = \tup{T', \initabox', \actset',
    \ginitprog'}$ and 
  $\ts{\tgkabcs(\csgkabsym)}^{\filter_S} = \tup{\const, T',
    \stateset_s, s_{0s}, \abox_s, \trans_s}$.
\end{compactenum}
Now, we have to show the following: For every state
$\tup{A'''_{cx},\scmap'''_{cx}, \ctx''', \delta'''_{cx}}$ such that
  \[
  \tup{A_{cx},\scmap_{cx}, \ctx, \delta_{cx}} \trans \tup{A'''_{cx},\scmap'''_{cx},
    \ctx''', \delta'''_{cx}},
  \]
  there exists states $\tup{A'_s,\scmap'_s, \delta'_s}$, 
  $\tup{A''_s,\scmap''_s, \delta''_s}$, and $\tup{A'''_s,\scmap'''_s, \delta'''_s}$ such that:
\begin{compactenum}[\bf (a)]
\item we have
  $ \tup{A_s,\scmap_s, \delta_s} \trans_s \tup{A'_s,\scmap'_s,
    \delta'_s} \trans_s \tup{A''_s,\scmap''_s, \delta''_s}  \trans_s \tup{A'''_s,\scmap'''_s, \delta'''_s} $
\item
  $\tup{A'''_{cx},\scmap'''_{cx}, \ctx''', \delta'''_{cx}} \eqc
  \tup{A'''_s,\scmap'''_s, \delta'''_s}$
\item $\scmap'''_s = \scmap'''_{cx}$;
\item $\delta'''_s = \tgprogc(\delta'''_{cx})$.
\end{compactenum}

\noindent
By definition of $\ts{\csgkabsym}^{\csfilter_S}$, 
since
$ \tup{A_{cx},\scmap_{cx}, \ctx, \delta_{cx}} \trans \tup{A'''_{cx},\scmap'''_{cx},
  \ctx''', \delta'''_{cx}}$,
we have
$\tup{A_{cx},\scmap_{cx}, \ctx, \delta_{cx}} \gprogtrans{\alpha\sigma_{cx},
  \filter_C} \tup{A'''_{cx},\scmap'''_{cx}, \ctx''', \delta'''_{cx}}$.
Hence, by the definition of $\gprogtrans{\act\sigma_{cx}, \filter_C}$, we
have that:
\begin{compactitem}

\item
  $\tup{\tup{A_{cx}, \scmap_{cx}, \ctx}, \act\sigma_{cx}, \tup{A'''_{cx}, \scmap'''_{cx}, \ctx'''}}
  \in \cstell_{\csfilter_S}$, and 

\item $\sigma_{cx}$ is a legal parameter assignment for $\act$ in
  $A_{cx}$ w.r.t.\ context $\ctx$ and action invocation
  $\gactc{Q(\vec{p})}{\ctxe}{\act(\vec{p})}$ (i.e.,
  $\ask(Q\sigma_{cx}, \ctbox^\ctx, A_{cx}) = \true$). Notice that
  w.l.o.g.  $\gactc{Q(\vec{p})}{\ctxe}{\act(\vec{p})}$ is the next
  instruction that should be executed in $\delta_{cx}$.



\item $\ctx \cup \ctxth \models \ctxe$.

\end{compactitem}
Since
  $\tup{\tup{A_{cx}, \scmap_{cx}, \ctx}, \act\sigma_{cx}, \tup{A'''_{cx}, \scmap'''_{cx}, \ctx'''}}
  \in \cstell_{\csfilter_S}$,
by the definition of $\cstell_{\csfilter_S}$, we have:
\begin{compactenum}




  \item $\tup{A_{cx}, \ctx, \ctx'''} \in \ctxchg$,

  \item there exists
    $\theta_{cx} \in \eval{\addfacts{\ctbox^{\ctx}, A_{cx}, \act\sigma_{cx}}}$ such
    that:
    \begin{compactenum}

\item $\theta_{cx}$ and $\scmap_{cx}$ agree on the common values in their domains.



    \item $\scmap_{cx}''' = \scmap_{cx} \cup \theta_{cx}$;

    \item
      $\tup{A_{cx}, \addfacts{\ctbox^{\ctx}, A_{cx}, \act\sigma_{cx}}\theta_{cx},
        \delfacts{\ctbox^{\ctx}, A_{cx}, \act\sigma_{cx}}, \ctx''', A_{cx}'''} \in
      \csfilter_S$;

    \item $A_{cx}$ is $\ctbox^\ctx$-consistent, and $A_{cx}'''$ is
      $\ctbox^{\ctx'''}$-consistent.

    \end{compactenum}
\end{compactenum}

Since
$\tup{A_{cx}, \addfacts{\ctbox^{\ctx}, A_{cx}, \act\sigma_{cx}}\theta_{cx},
  \delfacts{\ctbox^{\ctx}, A_{cx}, \act\sigma_{cx}}, \ctx''', A_{cx}'''} \in
\csfilter_S$,
by the definition of $\csfilter_S$, we have
$A_{cx}''' = (A_{cx} \setminus \delfacts{\ctbox^{\ctx}, A_{cx}, \act\sigma_{cx}})
\cup \addfacts{\ctbox^{\ctx}, A_{cx}, \act\sigma_{cx}}\theta_{cx}$.
Furthermore, since $\delta_s = \tgprogcs(\delta_{cx})$, by the definition of
$\tgprogcs$, we have that
\[
\begin{array}{l}
  \tgprogcs(\gactc{Q(\vec{p})}{\ctxe}{\act(\vec{p})}) =  
                                               \gact{Q'(\vec{p})}{\act'(\vec{p})}
                                                        \ ;\ \delta_{\ctxprocset}\ ;\ \gact{\true}{\act^{\ctbox}_\bot()}\\
\end{array}
\]
Hence, the next executable part of the program on state
$\tup{A_s,\scmap_s, \delta_s}$ is
\[
\gact{Q'(\vec{p})}{\act'(\vec{p})} \ ;\ \delta_{\ctxprocset}\ ;\
\gact{\true}{\act^{\ctbox}_\bot()}.
\]




\noindent
Notice that 
\begin{itemize}
\item $\gact{Q'(\vec{p})}{\act'(\vec{p})}$ is obtained from
  $\gactc{Q(\vec{p})}{\ctxe}{\act(\vec{p})}$, and
\item since $s_{cx} \eqc s_s$, then we have that
  $A_s = A_{cx} \cup A_\ctx$, where $A_\ctx$ is the set of ABox
  assertion that represents the context $\ctx$ (see
  \Cref{def:abox-context}).
\end{itemize}
Thus, by \Cref{def:action-invoc-from-cs-action-invoc}, we have that
$Q' = Q_\ctxb \wedge q_{\ctxe}$. 
Since $\ctx \cup \ctxth \models \ctxe$ and $q_{\ctxe}$ only use
context dimension concept, by \Cref{lem:ctx-exp-and-query-about-it},
it is easy to see that $\Ans(q_{\ctxe}, T', A_s) = \true$.
Furthermore, by \Cref{lem:correctness-contextually-compiled-query}, we
have that $\Ans(Q,\ctbox^{\ctx}, A_{cx}) = \Ans(Q_\ctxb, T', A_s)$.
Therefore, now we can construct $\sigma_s$ that maps parameters of
$\act'$ to constants in $\adom{A_s}$ such that $\sigma_c = \sigma_s$.

Now, since we have $\scmap_s= \scmap_{cx}$, then we can construct
$\theta_s$ such that $\theta_s = \theta_{cx}$.
Hence, we have the following:
\begin{compactitem}
\item $\theta_s$ and $\scmap_s$ agree on the common values in their
  domains.
 \item $\scmap'''_s = \theta_s \cup \scmap_s = \theta_{cx} \cup \scmap_{cx} =\scmap_{cx}'''$.
\end{compactitem}
Now, let $A_s' = A_s \cup \addfacts{T', A_s, \act'\sigma_s}\theta_s$,
i.e., $A_s'$ captures the result of the execution of action $\act'$
and by the definition of delayed action $\act'$ (see
\Cref{def:delayed-action}). Considering the form of $T'$, it is easy
to see that $A_s'$ is $T'$-consistent. Thus, by the definition of
$\tell_{\filter_s}$, we have
$\tup{\tup{A_s,\scmap_s}, \act'\sigma_s, \tup{A'_s, \scmap'''_s}} \in
\tell_{\filter_s}$.
Moreover, we have 
\[
\tup{A_s, \scmap_s,\gact{Q(\vec{p})}{\act'(\vec{p})};\delta_0}
\gprogtrans{\act'\sigma_s, \filter_s} \tup{A_s', \scmap_s''',
  \delta_0}
\]
where
$\delta_0 = \delta_{\ctxprocset}\ ;\
\gact{\true}{\act^{\ctbox}_\bot()}$,
and $A_{cx} \cup A_\ctx \subseteq A_s'$ (notice that, by construction,
$\act'$ only adds new ABox assertions). W.l.o.g., let
$A_s' = A_{cx} \cup A_{\act'} \cup A_\ctx$ (i.e., $A_{\act'}$ represents
the set of ABox assertions that was just added by $\act'$).

Now, since $\tup{A_{cx}, \ctx, \ctx'''} \in \ctxchg$, by
\Cref{def:ctx-chg-relation}, there exists a context-evolution rule
$\tup{Q, \ctxe} \mapsto C_{new}$ in $\ctxprocset$ s.t.:
  \begin{compactenum}
  \item $\ask(Q, \ctbox^C, A_{cx})$ is $\true$;
  \item $C \cup \ctxth \models \ctxe$;
  \item for every context dimension $d\in\cdimset$ s.t.\
    $\cval{d}{v} \in C_{new}$, \\we have $\cval{d}{v} \in C'''$;
  \item for every context dimension $d\in\cdimset$ s.t.\
    $\cval{d}{v} \in C$, and there does not exist any $v_2$ s.t.\
    $\cval{d}{v_2} \in C_{new}$, we have $\cval{d}{v} \in C'''$.
  \end{compactenum}
  Additionally, by the definition of $\delta_{\ctxprocset}$ (see
  \Cref{def:ctx-chg-prog}), it is easy to see that there exists
  $A''_s$ such that we have
  \[
  \tup{A'_s, \scmap_s''', \delta_0} \gprogtrans{\act_\ctx\sigma_s',
    \filter_s} \tup{A''_s, \scmap_s''', \delta_1}
  \]
  where $\sigma_s'$ is an empty substitution,
  $A''_s = A_{cx}''' \cup A_{\ctx'''}$, and
  $\delta_1 = \gact{\true}{\act^{\ctbox}_\bot()}$.

  Now, notice that $\act^{\ctbox}_\bot$ only change the ABox when
  there is an inconsistency, since $A_{cx}'''$ is
  $\ctbox^{\ctx'''}$-consistent, it is easy to see that we have
  \[
  \tup{A''_s, \scmap_s''', \delta_1} \gprogtrans{\act^{\ctbox}_\bot\sigma_s'',
    \filter_s} \tup{A'''_s, \scmap_s''', \delta_s'''}
  \]
  where $A'''_s = A''_s$, and
  $\tup{A'''_{cx},\scmap'''_{cx}, \ctx''', \delta'''_{cx}} \eqc
  \tup{A'''_s,\scmap'''_s, \delta'''_s}$.

  The other direction of bisimulation relation can be proven in a
  similar way.

\end{proof}

Having \Cref{lem:scsgkab-to-sgkab-bisimilar-state} in hand, we can
easily show that given an \scsgkab, its transition system is ST-bisimilar
to the transition of its corresponding S-GKAB that is obtained via
the translation $\tgkabcs$ as follows.

\begin{lemma}\label{lem:scsgkab-to-sgkab-bisimilar-ts}
  Given an \scsgkab $\csgkabsym$ with transition system
  $\ts{\csgkabsym}^{\csfilter_S}$, let $\tgkabcs(\csgkabsym)$ be the
  corresponding S-GKAB $($with transition system
  $\ts{\tgkabcs(\csgkabsym)}^{\filter_S}$$)$ obtained from $\csgkabsym$
  via $\tgkabcs$.
  We have
  $\ts{\csgkabsym}^{\csfilter_S} \stbsim
  \ts{\tgkabcs(\csgkabsym)}^{\filter_S}$.   
\end{lemma}
\begin{proof}
Let
\begin{compactenum}
\item
  $\csgkabsym = \tup{\ctbox, \initabox, \actset, \ginitprog, \initctx,
    \ctxprocset}$,
  and \\
  $\ts{\csgkabsym}^{\csfilter_S} = \tup{\const, \ctbox, \stateset_\ctxb, s_{0\ctxb},
    \abox_\ctxb, \cntx, \trans_\ctxb}$,
\item   $\tgkabcs(\csgkabsym) = \tup{T', \initabox', \actset',
    \ginitprog'}$ and 
  $\ts{\tgkabcs(\csgkabsym)}^{\filter_S} = \tup{\const, T_s,
    \stateset_s, s_{0s}, \abox_s, \trans_s}$.
\end{compactenum}
We have that
$s_{0\ctxb} = \tup{\initabox, \scmap_{cx}, \ctx_0, \delta}$ and
$s_{0s} = \tup{\initabox', \scmap_s, \delta'}$ where
$\scmap_{cx} = \scmap_s = \emptyset$. By the definition of $\tgprogcs$
and $\tgkabcs$, we also have $s_{0\ctxb} \eqc s_{0s}$, and
$\delta' = \tgprogcs(\delta)$. Hence, by
\Cref{lem:scsgkab-to-sgkab-bisimilar-state}, we have
$s_{0\ctxb} \stbsim s_{0s}$. Therefore, by the definition of
ST-bisimulation, we have
$\ts{\csgkabsym}^{\csfilter_S} \stbsim
\ts{\tgkabcs(\csgkabsym)}^{\filter_S}$.  \ \
\end{proof}

Having all of these machinery in hand, we are now ready to show that
the verification of \mulcs properties over \scsgkabs can be recast as
verification of \muladom over S-GKABs as follows.

\begin{theorem}\label{thm:ver-scsgkab-to-sgkab}
  Given an \scsgkab $\csgkabsym$ and a closed $\mulcs$ property
  $\Phi$, we have
\begin{center}
  $\ts{\csgkabsym}^{\csfilter_S} \models \Phi$ if and only if
  $ \ts{\tgkabcs(\csgkabsym)}^{\filter_S}\models \tfort(\Phi)$
\end{center}
\end{theorem}
\begin{proof}
  By \Cref{lem:scsgkab-to-sgkab-bisimilar-ts}, we have that
  $\ts{\csgkabsym}^{\csfilter_S} \stbsim
  \ts{\tgkabcs(\csgkabsym)}^{\filter_S}$.
  Hence, by \Cref{lem:stbisimilar-ts-satisfies-same-formula}, we
  have that for every $\mulcs$ property $\Phi$
\[
\ts{\csgkabsym}^{\csfilter_S} \models \Phi \textrm{ if and only if }
\ts{\tgkabcs(\csgkabsym)}^{\filter_S}\models \tfort(\Phi)
\]
\end{proof}


\subsection{Verification of Run-Bounded Standard \csgkabs}

An interesting property of the translation $\tgkabcs$ is that it
preserves run-boundedness. 

\begin{lemma}\label{lem:run-bounded-preservation-scsgkab}
  Let $\csgkabsym$ be an \scsgkab and $\tgkabcs(\csgkabsym)$ be its
  corresponding S-GKAB. We have $\csgkabsym$ is run-bounded if and
  only if $\tgkabcs(\csgkabsym)$ is run-bounded.
\end{lemma}
\begin{proof}
  Let
  \begin{compactenum}

  \item $\ts{\csgkabsym}^{\csfilter_S} $ be the transition system of
    $\csgkabsym$.

  \item $\ts{\tgkabcs(\csgkabsym)}^{\filter_S}$ be the transition system
    of $\tgkabcs(\csgkabsym)$.
  \end{compactenum}
  The proof is easily obtained due to the following facts:

  \begin{compactitem}

  \item the program that is used to simulate the context evolution
    does not inject unbounded number of new constants. In fact, we
    only reserve a constant $\ctxconst$ to simulate the context (i.e.,
    to construct the ABox assertions that represent the context
    dimension assignments).

  \item similarly, the action that is used to check the inconsistency
    does not introduce unbounded number of new constants.

  \item by \Cref{lem:scsgkab-to-sgkab-bisimilar-ts}, we have that
    $\ts{\csgkabsym}^{\csfilter_S} \stbsim
    \ts{\tgkabcs(\csgkabsym)}^{\filter_S}$.
    Thus, basically they are ``equivalent'' modulo intermediate states
    (states containing $\tmp$), and each two bisimilar states are
    equivalent modulo context ABox assertions.

  \end{compactitem}

\end{proof}

\noindent
Now, we can easily acquire the following result on verification of
\mulcs properties over run-bounded \scsgkabs.

\begin{theorem}[Verification of Run-Bounded \scsgkabs]\label{thm:ver-run-bounded-scsgkab}
  Verification of closed $\mulcs$ formulas over a run-bounded \scsgkab
  is decidable and can be reduced to finite-state model checking.
\end{theorem}
\begin{proof}
  From \Cref{thm:ver-scsgkab-to-sgkab},
  \Cref{lem:run-bounded-preservation-scsgkab}, and \Cref{thm:gtos} we
  have that verification of closed \mulcs formulas over run-bounded
  \scsgkabs can be reduced to the verification of \muladom formulas
  over run-bounded KABs. Then, by
  \Cref{thm:verification-run-bounded-kab}, we have that verification
  of \muladom over run-bounded DCDS is decidable and can be reduced to
  finite-state model checking.
\end{proof}

\section{Capturing Standard GKABs within Standard \csgkabs}\label{sec:cap-sgkab-to-scsgkabs}

So far we have seen that we can compile \scsgkabs into S-GKABs and
recast the verification of \scsgkabs into S-GKABs. Now, we show that
we can actually capture S-GKABs within \scsgkabs. As a consequence, we
have that S-GKABs and \scsgkabs are essentially reducible to each
other in terms of verification.

The idea to capture S-GKABs within \scsgkabs is as follows:
\begin{compactitem}
\item We introduce only a single context dimension and it has only a
  single possible value. Thus we basically can only have one possible
  context.
\item We transform the TBox in the given S-GKAB into a contextualized
  TBox such that each assertion holds in our only one possible
  context.
\item We introduce a single context-evolution rule that never change
  the context (keep the context stay the same).
\item The initial context will be our only one possible context.
\item Each action invocation in the program of the given S-GKAB is
  translated into a context-sensitive action invocation where the
  corresponding context expression is our only one possible context
  dimension assignment. Thus this context expression will not affect
  the action execution since it will always holds.
\end{compactitem}
%

To formalize the ideas above, in the following we fix a set $\cdimset$
of context dimension containing only a single context dimension $d$
(i.e., $\cdimset = \set{d}$). Moreover, $d \in \cdimset$ has a tree
shaped finite value domain $\tup{\cdom[d],\cover[d]}$ where $\cdom[d]$
contains only a single value $\topv[d]$ (i.e., $\cdom[d] = \topv[d]$).

We now introduce the translation for program in S-GKABs. In
particular, we define a translation function $\tgprogscs$ that
basically replaces each action invocation with a context-sensitive
action invocation in which its context expression always holds in any
context.
Additionally, the translation function $\tgprogsb$ also serves as a
one-to-one correspondence (bijection) between the original and the
translated program (as well as between the sub-program).

\begin{definition}[Program Translation $\tgprogscs$]
  Given \sidetextb{Program Translation $\tgprogscs$} a set of actions
  $\actset$, a program $\delta$ over $\actset$, and a TBox $T$, we
  define a \emph{translation $\tgprogscs$} which translates a program
  into a program inductively as follows:
\[
\begin{array}{@{}l@{}l@{}}
  \tgprogscs(\gact{Q(\vec{p})}{\act(\vec{p})}) &=  
                                                \gactc{Q(\vec{p})}{\ctxe}{\act(\vec{p})}\\
  \tgprogscs(\gemptyprog) &= \gemptyprog \\
  \tgprogscs(\delta_1|\delta_2) &= \tgprogscs(\delta_1)|\tgprogscs(\delta_2) \\
  \tgprogscs(\delta_1;\delta_2) &= \tgprogscs(\delta_1);\tgprogscs(\delta_2) \\
  \tgprogscs(\gif{\varphi}{\delta_1}{\delta_2}) &= \gif{\varphi}{\tgprogscs(\delta_1)}{\tgprogscs(\delta_2)} \\
  \tgprogscs(\gwhile{\varphi}{\delta}) &= \gwhile{\varphi}{\tgprogscs(\delta)}
\end{array}
\]
where $\ctxe = \set{\cval{d}{\topv[d]}}$
\end{definition}

Having the necessary ingredients, we define the following translation
that transform S-GKABs into \scsgkabs as follows.

\begin{definition}[Translation from S-GKAB to \scsgkab]\label{def:trans-sgkab-scsgkab}
  \ \sidetext{Translation from S-GKAB to \scsgkab}
  We define a translation $\tgkabscs$ that, given an S-GKAB
  $\gkabsym = \tup{T, \initabox, \actset, \ginitprog}$, produces
  an \scsgkab
  $\tgkabscs(\gkabsym) = \tup{\ctbox, \initabox, \actset, \ginitprog',
    \initctx, \ctxprocset}$, where
\begin{compactitem}

\item $\ctbox$ is obtained from $T$ such that for each TBox assertion
  $t \in T$ we have $\tup{t:\varphi}$ where
  $\varphi = \cval{d}{\topv[d]}$,

\item $\ginitprog' = \tgprogscs(\ginitprog)$. 

\item $\initctx = \set{\cval{d}{\topv[d]}}$, 

\item $\ctxprocset = \set{\tup{\true, \cval{d}{\topv[d]}} \mapsto
    \set{\cval{d}{\topv[d]}} }$

\end{compactitem}
\ \ 
\end{definition}

We now proceed to show that given an S-GKAB $\gkabsym$, and \muladom
formula $\Phi$, we have that $\ts{\gkabsym}^{\filter_S} \models \Phi$
if and only if $ \ts{\tgkabscs(\gkabsym)}^{\csfilter_S}\models
\Phi$. The strategy is as follows:
\begin{compactenum}

\item Recall the notion of E-Bisimulation in
  \Cref{subsec:e-bisimulation}. Here we use a similar notion of
  bisimulation except that now the bisimulation relation is defined
  between a KB transition system and a context-sensitive transition
  system. However, the bisimulation condition are kept the
  same. Therefore, for brevity, here we do not redefine a new
  bisimulation relation. All notions related to E-Bisimulation that
  was introduced in \Cref{subsec:e-bisimulation} can be seamlessly
  adjusted into this setting.

\item Later we show that given an S-GKAB, its transition system is
  E-bisimilar to the transition system of its corresponding \scsgkab
  that is obtained through $\tgkabscs$.

\item Thus, utilizing \Cref{lem:e-bisimilar-ts-satisfies-same-formula}
  (except that now we consider a KB transition system and a
  context-sensitive transition system) and also by considering that
  \mulcs without context expression is the same as \muladom, we can
  easily recast the verification of S-GKABs into \scsgkabs.

\end{compactenum}


In the following two lemmas we aim to show that given an S-GKAB, its
transition system is E-bisimilar to the transition system of its
corresponding \scsgkab that is obtained through $\tgkabscs$.

\begin{lemma}\label{lem:sgkab-to-scsgkab-bisimilar-state}
  Let $\gkabsym$ be an S-GKAB with transition system
  $\ts{\gkabsym}^{\filter_S}$, and let $\tgkabscs(\gkabsym)$ be its
  corresponding \scsgkab $($with transition system
  $\ts{\tgkabscs(\gkabsym)}^{\csfilter_S}$$)$
  obtained through
  $\tgkabscs$. 
  Consider
  \begin{inparaenum}[]
  \item a state $s_s = \tup{A_s,\scmap_s, \delta_s}$ of
    $\ts{\gkabsym}^{\filter_S}$, and  
%
  \item a state $s_{cx} = \tup{A_{cx},\scmap_{cx}, \ctx, \delta_{cx}}$ of
    $\ts{\tgkabscs(\gkabsym)}^{\csfilter_S}$.
  \end{inparaenum}
  If
  \begin{inparaenum}[]
  \item $A_{cx} = A_s$,
  \item $\scmap_{cx} = \scmap_s$, and
  \item $\delta_{cx} = \tgprogscs(\delta_{s})$,
  \end{inparaenum}
  then
  $\tup{A_{cx},\scmap_{cx}, \ctx, \delta_{cx}} \ebsim
  \tup{A_s,\scmap_s, \delta_s}$.
\end{lemma}
\begin{proof}
  Now, let
  \begin{compactenum}
  \item $\gkabsym = \tup{T, \initabox, \actset, \ginitprog}$ and
    $\ts{\gkabsym}^{\filter_S}
    = \tup{\const, T, \stateset_s, s_{0s}, \abox_s, \trans_s}$.
  \item $\tgkabscs(\gkabsym)
    = \tup{\ctbox, \initabox, \actset, \ginitprog', \initctx,
      \ctxprocset}$,
    and \\
    $\ts{\tgkabscs(\gkabsym)}^{\csfilter_S}
    = \tup{\const, \ctbox, \stateset_\ctxb, s_{0\ctxb}, \abox_\ctxb,
      \cntx, \trans_\ctxb}$,
\end{compactenum}
Now, we have to show the following: For every state $
\tup{A'_s,\scmap'_s, \delta'_s}$ such that $\tup{A_s,\scmap_s, \delta_s}
\trans_s \tup{A'_s,\scmap'_s, \delta'_s} $
there exists
$\tup{A'_{cx},\scmap'_{cx}, \ctx', \delta'_{cx}}$ such that
\begin{compactenum}[\bf (a)]
\item we have $
  \tup{A_{cx},\scmap_{cx}, \ctx, \delta_{cx}} \trans
  \tup{A'_{cx},\scmap'_{cx}, \ctx', \delta'_{cx}}, $
\item $A'_s = A'_{cx}$
\item $\scmap'_s = \scmap'_{cx}$;
\item $\delta'_{cx} = \tgprogscs(\delta'_{s})$.
\end{compactenum}
The proof can be easily obtained by considering that within \scsgkab,
by the definition of $\tgkabscs$
(see \Cref{def:trans-sgkab-scsgkab}), it is easy to see that the
following hold:
\begin{compactitem}

\item There is only one possible context that is $\ctx
  = \set{\cval{d}{\topv[d]}}$.

\item The initial context is $\initctx = \set{\cval{d}{\topv[d]}}$ and
  it stays the same along the system evolution because we only have a
  single context evolution rule
  $\tup{\true, \cval{d}{\topv[d]}} \mapsto \set{\cval{d}{\topv[d]}}$
  that essentially never change the context.  Thus, the TBox stay the
  same for all states. Moreover, all of the TBox assertions hold in
  our only one possible context since for each
  $\tup{t:\varphi} \in \ctbox$ we have $\varphi = \cval{d}{\topv[d]}$.
  Therefore, basically the situation of the TBox is the same as in the
  original S-GKAB.

\item Each context-sensitive action invocation in $\delta'$ has a
  context expression $\varphi = \cval{d}{\topv[d]}$. Thus, basically
  we can ignore it since it will always satisfied due to all of the
  facts above. Due to this fact, each context-sensitive action
  invocation in $\delta'$ is the same as the usual action invocation
  in $\delta$.

\item By the definition of execution semantics of \scsgkabs and
  S-GKAB, they have the same way in updating an ABox and both of them
  reject each action execution that leads into an inconsistent state.

\end{compactitem}
\end{proof}

\begin{lemma}\label{lem:sgkab-to-scsgkab-bisimilar-ts}
  Given an S-GKAB $\gkabsym$, we have
  $\ts{\gkabsym}^{\filter_S} \ebsim
  \ts{\tgkabscs(\gkabsym)}^{\csfilter_S}$
\end{lemma}
\begin{proof}
Let
\begin{compactenum}
\item $\gkabsym = \tup{T, \initabox, \actset, \ginitprog}$ and
  $\ts{\gkabsym}^{\filter_S} = \tup{\const, T,
    \stateset_s, s_{0s}, \abox_s, \trans_s}$.
\item
  $\tgkabscs(\gkabsym) = \tup{\ctbox, \initabox, \actset, \ginitprog',
    \initctx, \ctxprocset}$,
  and \\
  $\ts{\tgkabscs(\gkabsym)}^{\csfilter_S} = \tup{\const, \ctbox,
    \stateset_\ctxb, s_{0\ctxb}, \abox_\ctxb, \cntx, \trans_\ctxb}$,
\end{compactenum}
We have that $s_{0s} = \tup{\initabox, \scmap_s, \delta}$ and
$s_{0\ctxb} = \tup{\initabox, \scmap_{cx}, \ctx_0, \delta'}$ where
$\scmap_s = \scmap_{cx} = \emptyset$. By the definition of
$\tgprogscs$ and $\tgkabscs$, we also have that their initial ABoxes
are the same, and $\delta' = \tgprogscs(\delta)$. Hence, by
\Cref{lem:sgkab-to-scsgkab-bisimilar-state}, we have
$s_{0s} \ebsim s_{0\ctxb}$. Therefore, by the definition of
E-bisimulation, we have
$\ts{\gkabsym}^{\filter_S} \ebsim
\ts{\tgkabscs(\gkabsym)}^{\csfilter_S}$.
\end{proof}

Having all machinery in hand, we now show that the verification of
\muladom properties over S-GKAB can be recast as verification over
\scsgkab as follows.



\begin{theorem}
  \label{thm:ver-sgkab-to-scsgkab}
Verification of closed \muladom properties over S-GKABs can be recast
as verification over \scsgkabs.
\end{theorem}
\begin{proof}
  By \Cref{lem:sgkab-to-scsgkab-bisimilar-ts}, we have that
  $\ts{\gkabsym}^{\filter_S} \ebsim
  \ts{\tgkabscs(\gkabsym)}^{\csfilter_S}$.
  Hence, by \Cref{lem:e-bisimilar-ts-satisfies-same-formula} (but
  consider that now it is between a KB transition system and a
  context-sensitive transition system), for every $\muladom$ property
  $\Phi$, we have that
\[
\ts{\gkabsym}^{\filter_S} \models \Phi \textrm{ if and only if }
\ts{\tgkabscs(\gkabsym)}^{\csfilter_S}\models \Phi
\]
Hence, by using the translation $\tgkabscs$ we can easily transform an
S-GKAB into an \scsgkab and then the claim is easily follows due to
the fact above. 
\end{proof}

\section{Discussion}

In \Cref{sec:contextualized-golog-program} we have seen how we
incorporate contextual information as an additional information that
influence the flow of program execution. In particular, we lift the
usual atomic action invocations into context-sensitive atomic action
invocations that are not only guarded by queries but also with context
expressions (see \Cref{def:context-golog-program}). Now, observe that
we can actually also extend the conditional and loop constructs (i.e.,
$\gif{\varphi}{\delta_1}{\delta_2}$ and $\gwhile{\varphi}{\delta}$)
such that they also incorporate contextual information. Formally, we
can easily augment context expressions as an additional ``if
condition'' (resp.\ ``loop guard'') as follows:
\[
\begin{array}{c}
\gif{(\varphi,\ctxe)}{\delta_1}{\delta_2}\\
\gwhile{(\varphi,\ctxe)}{\delta}
\end{array}
\]
Furthermore we can also easily adjust the execution semantics
concerning the two constructs such that they are
context-sensitive. For the conditional construct
$\gif{(\varphi,\ctxe)}{\delta_1}{\delta_2}$, we can require that
$\delta_1$ is executable in case $\varphi$ is successfully evaluated
over the current KB and $\ctxe$ is entailed by the corresponding
current context together with the corresponding value domain
theory. Similarly, for the loop construct
$\gwhile{(\varphi,\ctxe)}{\delta}$, we can require that the program
$\delta$ will be executed as long as $\varphi$ is successfully
evaluated over the current KB and $\ctxe$ is entailed by the
corresponding current context together with the corresponding value
domain theory.



Interestingly, with the extensions above, we can still reduce the
verification of \scsgkabs into the corresponding verification of
S-GKABs. This can be done easily since we can emulate the context
expression as a query. Thus we can follow the similar way of
simulating context-sensitive atomic action invocation inside
S-GKABs. Another interesting point is that we can also easily simulate
S-GKABs inside \scsgkabs. Similar approach as in
\Cref{sec:cap-sgkab-to-scsgkabs} can be followed.

One might also extend the framework further by introducing a construct
that contextualized a program in general. I.e., introduce the
following construct:
\[
\ctxe:\delta
\]
There are actually two possible ways in defining the semantics of such
construct:
\begin{enumerate} 
\item The first semantics would be constraining that the whole program
  $\delta$ can only be executed as long as the current context $\ctx$
  together with the value domain theory $\ctxth$ entail $\ctxe$.
\item The second semantics would be constraining that we can start to
  execute the program $\delta$ if the current context $\ctx$ together
  with the value domain theory $\ctxth$ entail $\ctxe$. 
\end{enumerate}
The different between those two semantics is that in the second
semantics, the checking whether the context expression $\ctxe$ holds
or not is only done when we want to start to execute the program
$\delta$, while in the first semantics, the checking whether the
context expression $\ctxe$ holds or not is done along the execution of
$\delta$ (i.e., each atomic action invocation within $\delta$ can only
be executed if the context expression $\ctxe$ holds).

More formally, to adopt the first semantics, we can extend the
definition of context-sensitive program execution relation
(cf. \Cref{def:cs-prog-exec-relation}) by adding the following:
\begin{compactitem}
\item
  $\tup{A, \scmap, \ctx, \ctxe:\delta } \gprogtrans{\act\sigma,
    \csfilter} \tup{A', \scmap', \ctx', \ctxe:\delta'}$,
  if \\ \hspace*{5mm}
  \begin{inparaenum}[]
  \item
    $\tup{A, \scmap, \ctx, \delta} \gprogtrans{\act\sigma,
      \csfilter} \tup{A', \scmap', \ctx', \delta'}$, and 
  \item $\ctx \cup \ctxth \models \ctxe$.
  \end{inparaenum}
\end{compactitem}
On the other hand to adopt the second semantics, we can extend the
definition of context-sensitive program execution relation
(cf. \Cref{def:cs-prog-exec-relation}) by adding the following:
\begin{compactitem}
\item
  $\tup{A, \scmap, \ctx, \ctxe:\delta } \gprogtrans{\act\sigma,
    \csfilter} \tup{A', \scmap', \ctx', \delta'}$, if \\ \hspace*{5mm}
  \begin{inparaenum}[]
  \item
    $\tup{A, \scmap, \ctx, \delta} \gprogtrans{\act\sigma, \csfilter}
    \tup{A', \scmap', \ctx', \delta'}$, and
  \item $\ctx \cup \ctxth \models \ctxe$.
  \end{inparaenum}
\end{compactitem}

Notice that no matter whether we adopt the first or the second
semantics, it can still be shown that verification of \scsgkabs can be
reduced to verification of S-GKABs. 
For the first semantics, notice that a program expression
$\ctxe:\delta$ can be translated into the standard contextualized
Golog program introduced before
(cf. \Cref{def:context-golog-program}) by ``distributing'' the context
expressions $\ctxe$ into each atomic action invocation in $\delta$. By
doing this, each atomic action invocation will be constrained by the
context expressions $\ctxe$.
For the second semantics, we can emulate the construct $\ctxe:\delta$
by using the contextualized ``if'' construct introduced
above. Essentially, having $\ctxe:\delta$ is the same as having
\[
\gif{(\true,\ctxe)}{\delta}{\gact{\false}{\act()}}.
\]


%% file: 2.chapters/7-cs-ia-gkab.tex
\chapter{Inconsistency-Aware Context-Sensitive
 GKAB\lowercase{s}}\label{ch:cs-ia-gkab}

\ifhidecontent
 
\fi

In \Cref{ch:ia-gkab} we have seen how GKABs can be lifted into
inconsistency-aware GKABs such that they handle inconsistency in a
more sophisticated way. Moreover, in \Cref{ch:cs-gkab} we have also
seen an extension of GKABs into context-sensitive GKABs which taking
into account the contextual information during the evolution of the
system. Now, in this chapter we blend those two extensions and
introduce the so called Inconsistency-aware Context-sensitive GKABs
(\icgkabs).

Technically, to obtain \icgkabs, we start from \csgkabs (that has been
introduced in \Cref{ch:cs-gkab}) and then lift it into
inconsistency-aware \csgkabs using a similar approach as the way how
we get into inconsistency-aware GKABs in \Cref{ch:ia-gkab}. I.e., we
exploit the parametric execution semantics of \csgkabs and obtain
inconsistency-aware \csgkabs by introducing various kind of filters
that incorporate the inconsistency handling mechanisms and simply plug
them in into \csgkabs.


In this chapter, we also tackle the problem of verifying \mulcs
properties over \icgkabs. Similar to how we deal with the verification
problem of I-GKABs, in this chapter we show how we can reduce the
verification of \icgkabs into S-GKABs.
%
%
However, due to the presence of context, the TBox might change
depending on the context. Therefore, the repair of the inconsistency
should take into account the context and it must be performed based on
the TBox assertions that ``hold'' within the corresponding
context. Hence, when it comes to transforming \icgkabs into S-GKABs,
we can not simply directly re-use all of the tools that we use to
reduce the verification of I-GKABs into the verification of
S-GKABs. The repair program (that simulates the repair computation)
need to work based on the context. 
Thus, the challenge is how to make the repair program
context-sensitive such that it performs the repair w.r.t.\ the
TBox under the new context. I.e., the repair program (might) always
need to be adjusted ``on the fly'' based on the context.

Similar to KABs and GKABs, in the following we use \dllitea for
expressing KBs and we also do not distinguish between objects and
values (thus we drop attributes).
Moreover we make use of a countably infinite set $\const$ of
constants, which intuitively denotes all possible values in the
system.
Additionally, we consider a finite set of distinguished constants
$\iconst \subset \const$, and
a finite set $\servcall$ of \textit{function symbols} that represents
\textit{service calls}, which abstractly account for the injection of
fresh values (constants) from $\const$ into the system.
Additionally, for technical development of this chapter, except for
\Cref{sec:cap-sgkabs-to-icsgkabs}, we fix a set
$ \cdimset = \{d_1,\ldots,d_n\} $ of context dimensions. Each context
dimension $d_i \in \cdimset$ has its own tree-shaped finite value
domain $\tup{\cdom[d_i],\cover[d_i]}$, where $\cdom[d_i]$ represents
the finite set of domain values, and $\cover[d_i]$ represents the
predecessor relation forming the tree.

\section{The Inconsistency-aware Context-sensitive Execution
  Semantics}\label{sec:icgkabs-execsem}

Similar to \Cref{sec:ia-gkab}, the inconsistency-aware
context-sensitive execution semantics for \csgkabs are obtained by
simply exploiting the context-sensitive filter relations (in
\Cref{def:cs-filter-rel}) to define three inconsistency-aware
semantics that incorporate the repair-based approaches reviewed in
\Cref{sec:inconsistency-management-dl}.  In particular, we introduce
three context sensitive filter relations namely $\csfilter_B$,
$\csfilter_C$, and $\csfilter_E$.
%
For brevity, from this moment we often simply say filter to refer to
context-sensitive filter relation.

As the first one, we define B-repair Context-sensitive Filter
$\csfilter_B$ as follows:
\begin{definition}[B-repair Context-sensitive Filter $\csfilter_B$]\label{def:b-repair-cs-filter}
  A \sidetext{B-repair Context-sensitive Filter $\csfilter_B$}
  \emph{B-repair Context-sensitive Filter $\csfilter_B$} is a relation
  that consists of tuples of the form
  $\tup{A, \facta, \factd, \ctx, A'}$ such that
  $A' \in \arset{\ctbox^\ctx, (A \setminus \factd) \cup \facta}$,
  where $A$ and $A'$ are ABoxes, $\ctx$ is a context, and $\facta$ as
  well as $\factd$ are two sets of ABox assertions.
\end{definition}



\noindent
Employing the b-repair context-sensitive filter $\csfilter_B$ into
\csgkabs gives us \bicgkabs that is a \csgkabs with \emph{b-repair
  execution semantics}, i.e., where inconsistent ABoxes are repaired
by non-deterministically picking a b-repair. Formally, we define the
transition system which provide the b-repair execution semantics for
\csgkabs as follows.

\begin{definition}[\csgkab B-Transition System]
  Given \sidetext{\csgkab B-Transition System} a \csgkab
  $\csgkabsym$ 
  and a b-repair context sensitive filter $\csfilter_B$, the
  \emph{b-transition system of $\csgkabsym$}, written
  $\ts{\csgkabsym}^{\csfilter_B}$, is the transition system of
  $\csgkabsym$ w.r.t.\ $\csfilter_B$ (see also \Cref{def:cs-gkab-ts}).
\end{definition}

\noindent
We call \emph{\bicgkabs}the \csgkabs adopting this semantics.

\begin{example}
  Let the \csgkab $\csgkabsym$ specified in \Cref{ex:csgkab} be a
  \bicgkab.  Consider the state $s = \tup{A, \scmap, \ctx, \delta}$
  where:
  \begin{flushleft}
    $\begin{array}{l@{}ll} 
       \bullet \ A = &\set{ & \exo{ReceivedOrder}(
        \excon{chair} ), \exo{ApprovedOrder}( \excon{table} ),
                                                    \exo{designedBy}( \excon{table}, \excon{alice} ), \\
                                           &&\exo{Designer}( \excon{alice} ), \exo{hasDesign}( \excon{table} , \excon{ecodesign} ),\\
                                           &&\exo{hasAssemblingLoc}( \excon{table}, \excon{bolzano} ) \ \ }, 
     \end{array}$ 
   \end{flushleft}
  \begin{flushleft}
    $\begin{array}{l@{}ll} 
       \bullet \  \scmap =& \set{ &[\exs{getDesigner}(\excon{table}) \ra
                                      \excon{alice}], [\exs{getDesign}(\excon{table}) \ra \excon{ecodesign}], \\
                                           &&[\exs{assignAssemblingLoc}(\excon{table}) \ra
                                              \excon{bolzano}]  \ \ },
     \end{array}$ 
   \end{flushleft}
   \begin{flushleft}
     $     
     \begin{array}{l}
       \bullet \  \ctx = \set{\cval{\exc{PP}}{\exv{WE}}, \cval{\exc{S}}{\exv{PS}}},
     \end{array}
     $
   \end{flushleft}
   \begin{flushleft}
     $     
     \begin{array}{l}
       \bullet \  \delta = \delta_3 ; \delta_4 ; \delta_5 ; \gwhile{  \exists \exvar{x}.[\exo{Order}(\exvar{x})] \wedge
    \neg[\exo{DeliveredOrder}(\exvar{x})]   }{\delta_0}.
     \end{array}
     $
   \end{flushleft}
   Note that the state $s$ is a reachable state from the initial state
   $s_0$ in the transition system $\ts{\csgkabsym}^{\csfilter_B}$ of
   $\csgkabsym$. 
%
 %
   One possible successor state of $s$ is a state
   $s_1 = \tup{A_1, \scmap_1, \ctx_1, \delta_1}$ with
  \begin{flushleft}
    $\begin{array}{l@{}ll} 
       \bullet \ A_1 = &\set{ & \exo{ReceivedOrder}( \excon{chair} ), 
                               \exo{designedBy}( \excon{table}, \excon{alice} ), \exo{Designer}( \excon{alice} ), \\
                      &&\exo{hasDesign}( \excon{table} ,
                         \excon{ecodesign} ), \exo{hasAssemblingLoc}( \excon{table},
                         \excon{bolzano} ), \\
                      &&\exo{AssembledOrder}(\excon{table}),  \exo{assembledBy}(
                         \excon{table},\excon{alice}  ) 
                         \ \ },
     \end{array}$ 
   \end{flushleft}
  \begin{flushleft}
   $     
    \begin{array}{l@{}ll} 
       \bullet \  \scmap_1 =& \set{ &[\exs{getDesigner}(\excon{table}) \ra
                                      \excon{alice}], [\exs{getDesign}(\excon{table}) \ra \excon{ecodesign}], \\
                      &&[\exs{assignAssemblingLoc}(\excon{table}) \ra
                         \excon{bolzano}], \\
                      &&[\exs{getAssembler}(\excon{table}) \ra
                         \excon{alice}], \\  
                      &&[\exs{getAssemblingLoc}(\excon{table}) \ra
                         \excon{bolzano}]  \ \ }, 
     \end{array}$ 
   \end{flushleft}
  \begin{flushleft}
   $     
     \begin{array}{l}
       \bullet \  \ctx_1 = \set{\cval{\exc{PP}}{\exv{N}}, \cval{\exc{S}}{\exv{NS}}},
     \end{array}
     $
   \end{flushleft}
  \begin{flushleft}
   $     
     \begin{array}{l}
       \bullet \  \delta_1 = \delta_4 ; \delta_5 ; \gwhile{  \exists \exvar{x}.[\exo{Order}(\exvar{x})] \wedge
    \neg[\exo{DeliveredOrder}(\exvar{x})]   }{\delta_0}.
     \end{array}
     $
   \end{flushleft}
   The state $s_1$ is obtained from the execution of action invocation
   \[
   \gactc{\true}{\cval{\exc{PP}}{\exv{AP}} \wedge
     \cval{\exc{S}}{\exv{AS}}}{\exa{assembleOrders}()}
   \]
   where the context is changing from $\ctx$ to $\ctx'$ due to the
   application of the following context evolution rule:
   \[
   \carulex{\true}{\cval{\exc{PP}}{\exv{RE}} \wedge
     \cval{\exc{S}}{\exv{PS}} }{ \cval{\exc{PP}}{\exv{N}},
     \cval{\exc{S}}{\exv{NS}} }.
   \]
   Furthermore, we have that 
   \[
   \begin{array}{r@{}l}
     A_1 \in \arset{\ctbox^{\ctx_1}, (A \setminus
     \set{ \ &\exo{ApprovedOrder}(\excon{table})}) \cup \\
     \set{ \ &\exo{AssembledOrder}(\excon{table}),  \exo{assembledBy}( \excon{table},\excon{alice} ), \\
             &\exo{Assembler}(\excon{alice}),  \exo{hasAssemblingLoc}( \excon{table},\excon{bolzano} ) \ 
               } \ \ 
     }
   \end{array}
   \]
\end{example}

To obtain c-repair execution semantics, now we proceed to define the
c-repair context-sensitive filter as follows.

\begin{definition}[C-repair Context-sensitive Filter $\csfilter_C$]\label{def:c-repair-cs-filter}
  A \sidetext{C-repair Context-sensitive Filter $\csfilter_C$}
  \emph{C-repair Context-sensitive Filter $\csfilter_C$} is a relation
  that consists of tuples of the form
  $\tup{A, \facta, \factd, \ctx, A'}$ such that
  $A' = \iarset{\ctbox^\ctx, (A \setminus \factd) \cup \facta}$, where
  $A$ and $A'$ are ABoxes, $\ctx$ is a context, and $\facta$ as well
  as $\factd$ are two sets of ABox assertions.
\end{definition}


\noindent
Filter $\csfilter_C$ gives rise to the \emph{c-repair execution
  semantics} for \csgkabs, where inconsistent ABoxes are repaired
by computing their unique c-repair.
The transition systems which provide the c-repair execution semantics
for \csgkabs is then defined as follows.

\begin{definition}[\csgkab C-Transition System]
  Given \sidetext{\csgkab C-Transition System} a \csgkab
  $\csgkabsym$ 
  and a c-repair filter $\csfilter_C$, the \emph{c-transition system
    of $\csgkabsym$}, written $\ts{\csgkabsym}^{\csfilter_C}$, is the
  transition system of $\csgkabsym$ w.r.t.\ $\csfilter_C$ (see also
  \Cref{def:cs-gkab-ts}).
\end{definition}

\noindent
We call \emph{\cicgkabs}the \csgkabs adopting this semantics.

\begin{example}
  Let the \csgkab $\csgkabsym$ specified in \Cref{ex:csgkab} be a \\
  \cicgkab.  Consider the state $s = \tup{A, \scmap, \ctx, \delta}$
  where:
  \begin{flushleft}
    $\begin{array}{l@{}ll} 
       \bullet \ A = &\set{ & \exo{ReceivedOrder}(
        \excon{chair} ), \exo{ApprovedOrder}( \excon{table} ),
                                                    \exo{designedBy}( \excon{table}, \excon{alice} ), \\
                                           &&\exo{Designer}( \excon{alice} ), \exo{hasDesign}( \excon{table} , \excon{ecodesign} ),\\
                                           &&\exo{hasAssemblingLoc}( \excon{table}, \excon{bolzano} ) \ \ }, 
     \end{array}$ 
   \end{flushleft}
  \begin{flushleft}
    $\begin{array}{l@{}ll} 
       \bullet \  \scmap =& \set{ &[\exs{getDesigner}(\excon{table}) \ra
                                      \excon{alice}], [\exs{getDesign}(\excon{table}) \ra \excon{ecodesign}], \\
                                           &&[\exs{assignAssemblingLoc}(\excon{table}) \ra
                                              \excon{bolzano}]  \ \ },
     \end{array}$ 
   \end{flushleft}
   \begin{flushleft}
     $     
     \begin{array}{l}
       \bullet \  \ctx = \set{\cval{\exc{PP}}{\exv{WE}}, \cval{\exc{S}}{\exv{PS}}},
     \end{array}
     $
   \end{flushleft}
   \begin{flushleft}
     $     
     \begin{array}{l}
       \bullet \  \delta = \delta_3 ; \delta_4 ; \delta_5 ; \gwhile{
       \exists \exvar{x}.[\exo{Order}(\exvar{x})] \wedge
       \neg[\exo{DeliveredOrder}(\exvar{x})]  }{\delta_0}.
     \end{array}
     $
   \end{flushleft}
   Note that the state $s$ is a reachable state from the initial state
   $s_0$ in the transition system $\ts{\csgkabsym}^{\csfilter_C}$ of
   $\csgkabsym$. 
   One possible successor state of $s$ is a state
   $s_1 = \tup{A_1, \scmap_1, \ctx_1, \delta_1}$ with
  \begin{flushleft}
    $\begin{array}{l@{}ll} 
       \bullet \ A_1 = &\set{ & \exo{ReceivedOrder}( \excon{chair} ), 
                                \exo{designedBy}( \excon{table},
                                \excon{alice} ),  \exo{hasDesign}( \excon{table} ,
                                \excon{ecodesign} ), \\
                       &&\exo{AssembledOrder}(\excon{table}),  \exo{assembledBy}(
                          \excon{table},\excon{alice}  ) 
                          \ \ },
     \end{array}$ 
   \end{flushleft}
  \begin{flushleft}
   $     
    \begin{array}{l@{}ll} 
       \bullet \  \scmap_1 =& \set{ &[\exs{getDesigner}(\excon{table}) \ra
                                      \excon{alice}], [\exs{getDesign}(\excon{table}) \ra \excon{ecodesign}], \\
                      &&[\exs{assignAssemblingLoc}(\excon{table}) \ra
                         \excon{bolzano}], \\
                      &&[\exs{getAssembler}(\excon{table}) \ra
                         \excon{alice}], \\  
                      &&[\exs{getAssemblingLoc}(\excon{table}) \ra
                         \excon{trento}]  \ \ }, 
     \end{array}$ 
   \end{flushleft}
  \begin{flushleft}
   $     
     \begin{array}{l}
       \bullet \  \ctx_1 = \set{\cval{\exc{PP}}{\exv{N}}, \cval{\exc{S}}{\exv{NS}}},
     \end{array}
     $
   \end{flushleft}
  \begin{flushleft}
   $     
     \begin{array}{l}
       \bullet \  \delta_1 = \delta_4 ; \delta_5 ; \gwhile{  \exists \exvar{x}.[\exo{Order}(\exvar{x})] \wedge
    \neg[\exo{DeliveredOrder}(\exvar{x})]   }{\delta_0}.
     \end{array}
     $
   \end{flushleft}
   The state $s_1$ is obtained from the execution of action invocation
   \[
   \gactc{\true}{\cval{\exc{PP}}{\exv{AP}} \wedge
     \cval{\exc{S}}{\exv{AS}}}{\exa{assembleOrders}()}
   \]
   where the context is changing from $\ctx$ to $\ctx'$ due to the
   application of the following context evolution rule:
   \[
   \carulex{\true}{\cval{\exc{PP}}{\exv{RE}} \wedge
     \cval{\exc{S}}{\exv{PS}} }{ \cval{\exc{PP}}{\exv{N}},
     \cval{\exc{S}}{\exv{NS}} }.
   \]
   Furthermore, we have that 
   \[
   \begin{array}{r@{}l}
     A_1 = \iarset{\ctbox^{\ctx_1}, (A \setminus
     \set{ \ &\exo{ApprovedOrder}(\excon{table})}) \cup \\
     \set{ \ &\exo{AssembledOrder}(\excon{table}),  \exo{assembledBy}( \excon{table},\excon{alice} ), \\
             &\exo{Assembler}(\excon{alice}),  \exo{hasAssemblingLoc}( \excon{table},\excon{trento} ) \ 
               } \ \ 
     }
   \end{array}
   \]
\end{example}

Next, we define the evolution filter which handle the inconsistency
using the bold-evolution mechanism as in
\Cref{def:bold-evol-abox-icma}.

\begin{definition}[B-evol Context-sensitive Filter $\csfilter_E$]\label{def:evol-cs-filter}
  A \sidetext{B-evol Context-sensitive Filter $\csfilter_E$}
  \emph{B-evol Context-sensitive Filter $\csfilter_E$} is a relation
  that consists of tuples of the form
  $\tup{A, \facta, \factd, \ctx, A'}$ such that
  $A' = \evol(\ctbox^\ctx, A, \facta, \factd)$, and $\facta$ is
  $\ctbox^\ctx$-consistent, where $A$ and $A'$ are ABoxes, $\ctx$ is a
  context, and $\facta$ as well as $\factd$ are two sets of ABox
  assertions.
\end{definition}


\noindent
Filter $\csfilter_E$ gives rise to the \emph{b-evol execution
  semantics} for \csgkabs, where for updates leading to inconsistent
ABoxes, their unique bold-evolution is computed.
Notice that by combining the definition of $\cstell$ (see
\Cref{def:cs-tell-operation}) and filter $\csfilter_E$, we basically
assume that the new ABox assertions are consistent with the TBox under
the new context (i.e., after the context change). This means that we
assume that the update is always accepted/applicable since it gives
the system the new information.
The transition systems which provide the b-evol execution semantics
for \csgkabs is defined as follows.

\begin{definition}[\csgkab E-Transition System]\label{def:e-cs-ts}
  Given \sidetext{\csgkab E-Transition System} a \csgkab
  $\csgkabsym$ 
  and a e-repair filter $\csfilter_E$, the \emph{e-transition system
    of $\csgkabsym$}, written $\ts{\csgkabsym}^{\csfilter_E}$, is the
  transition system of $\csgkabsym$ w.r.t.\ $\csfilter_E$.
\end{definition}
\noindent
We call \emph{\eicgkabs}the \csgkabs adopting this semantics.
%
%
We group these three forms of \csgkabs (i.e., \bicgkabs, \cicgkabs,
\eicgkabs) under the umbrella of \emph{inconsistency-aware \csgkabs
  (\icgkabs)}. The definition of \mulcs verification over \icgkabs is
as usual, i.e., similar to the case of \csgkabs (see
\Cref{def:verification-csgkab}).

\begin{example}
  Let the \csgkab $\csgkabsym$ specified in \Cref{ex:csgkab} be an \\
  \eicgkab.  Consider the state $s = \tup{A, \scmap, \ctx, \delta}$
  where:
  \begin{flushleft}
    $\begin{array}{l@{}ll} 
       \bullet \ A = &\set{ & \exo{ReceivedOrder}(
        \excon{chair} ), \exo{ApprovedOrder}( \excon{table} ),
                                                    \exo{designedBy}( \excon{table}, \excon{alice} ), \\
                                           &&\exo{Designer}( \excon{alice} ), \exo{hasDesign}( \excon{table} , \excon{ecodesign} ),\\
                                           &&\exo{hasAssemblingLoc}( \excon{table}, \excon{bolzano} ) \ \ }, 
     \end{array}$ 
   \end{flushleft}
  \begin{flushleft}
    $\begin{array}{l@{}ll} 
       \bullet \  \scmap =& \set{ &[\exs{getDesigner}(\excon{table}) \ra
                                      \excon{alice}], [\exs{getDesign}(\excon{table}) \ra \excon{ecodesign}], \\
                                           &&[\exs{assignAssemblingLoc}(\excon{table}) \ra
                                              \excon{bolzano}]  \ \ },
     \end{array}$ 
   \end{flushleft}
   \begin{flushleft}
     $     
     \begin{array}{l}
       \bullet \  \ctx = \set{\cval{\exc{PP}}{\exv{WE}}, \cval{\exc{S}}{\exv{PS}}},
     \end{array}
     $
   \end{flushleft}
   \begin{flushleft}
     $     
     \begin{array}{l}
       \bullet \  \delta = \delta_3 ; \delta_4 ; \delta_5 ; \gwhile{
   \exists \exvar{x}.[\exo{Order}(\exvar{x})] \wedge
    \neg[\exo{DeliveredOrder}(\exvar{x})]    }{\delta_0}.
     \end{array}
     $
   \end{flushleft}
   Note that the state $s$ is a reachable state from the initial state
   $s_0$ in the transition system $\ts{\csgkabsym}^{\csfilter_E}$ of
   $\csgkabsym$. 
   One possible successor state of $s$ is a state
   $s_1 = \tup{A_1, \scmap_1, \ctx_1, \delta_1}$ with
  \begin{flushleft}
    $\begin{array}{l@{}ll} 
       \bullet \ A_1 = &\set{ & \exo{ReceivedOrder}( \excon{chair} ), 
                                \exo{designedBy}( \excon{table},
                                \excon{alice} ),  \exo{hasDesign}( \excon{table} ,
                                \excon{ecodesign} ), \\
                       &&\exo{AssembledOrder}(\excon{table}),  \exo{assembledBy}(
                          \excon{table},\excon{alice}  ), \\
                       &&\exo{Assembler}(\excon{alice}), \exo{hasAssemblingLoc}( \excon{table},
                          \excon{trento} ) 
                         \ \ },
     \end{array}$ 
   \end{flushleft}
  \begin{flushleft}
   $     
    \begin{array}{l@{}ll} 
       \bullet \  \scmap_1 =& \set{ &[\exs{getDesigner}(\excon{table}) \ra
                                      \excon{alice}], [\exs{getDesign}(\excon{table}) \ra \excon{ecodesign}], \\
                      &&[\exs{assignAssemblingLoc}(\excon{table}) \ra
                         \excon{bolzano}], \\
                      &&[\exs{getAssembler}(\excon{table}) \ra
                         \excon{alice}], \\  
                      &&[\exs{getAssemblingLoc}(\excon{table}) \ra
                         \excon{trento}]  \ \ }, 
     \end{array}$ 
   \end{flushleft}
  \begin{flushleft}
   $     
     \begin{array}{l}
       \bullet \  \ctx_1 = \set{\cval{\exc{PP}}{\exv{N}}, \cval{\exc{S}}{\exv{NS}}},
     \end{array}
     $
   \end{flushleft}
  \begin{flushleft}
   $     
     \begin{array}{l}
       \bullet \  \delta_1 = \delta_4 ; \delta_5 ; \gwhile{ \exists \exvar{x}.[\exo{Order}(\exvar{x})] \wedge
    \neg[\exo{DeliveredOrder}(\exvar{x})] }{\delta_0}.
     \end{array}
     $
   \end{flushleft}
   The state $s_1$ is obtained from the execution of action invocation
   \[
   \gactc{\true}{\cval{\exc{PP}}{\exv{AP}} \wedge
     \cval{\exc{S}}{\exv{AS}}}{\exa{assembleOrders}()}
   \]
   where the context is changing from $\ctx$ to $\ctx'$ due to the
   application of the following context evolution rule:
   \[
   \carulex{\true}{\cval{\exc{PP}}{\exv{RE}} \wedge
     \cval{\exc{S}}{\exv{PS}} }{ \cval{\exc{PP}}{\exv{N}},
     \cval{\exc{S}}{\exv{NS}} }.
   \]
   Furthermore, we have that
   $A_1 = \evol(\ctbox^{\ctx_1}, A, \facta, \factd)$ where $\facta$
   and $\factd$ are the set of assertions to be added
   and deleted by the action $\exa{assembleOrders}/0$ as follows:
\begin{flushleft}
$\begin{array}{l@{}l} 
   \facta = \set{
   &\exo{AssembledOrder}(\excon{table}), 
   \exo{assembledBy}( \excon{table},\excon{alice} ), \\
   &\exo{Assembler}(\excon{alice}), \exo{hasAssemblingLoc}( \excon{table},
     \excon{trento} )
     }\\
\end{array}
$
\end{flushleft}
\begin{flushleft}
$\begin{array}{l@{}l} \factd = \set{
    &\exo{ApprovedOrder}(\excon{table})}
\end{array}
$
\end{flushleft}
\end{example}

\section[From Inconsistency-aware Context-sensitive to
  Standard GKABs]{From Inconsistency-aware Context-sensitive GKABs to
  Standard GKABs}

In this section we show that all \icgkabs introduced in
\Cref{sec:icgkabs-execsem} (i.e., \bicgkabs, \cicgkabs, and \eicgkabs)
can be compiled into S-GKABs. In particular, we show that verification
of \mulcs formulas over \icgkabs can be reduced to the verification of
\muladom over S-GKABs. To this aim, here we combine our results in
\Cref{ch:ia-gkab} and \Cref{ch:cs-gkab}.

For a technical reason, we reserve a special ABox assertion $\tmp$,
where $\tmpconst \in \const_0$ and $\tmpconceptname$ is a
reserved 
concept name (i.e., outside of any TBox vocabulary). Basically, we use
$\tmp$ to distinguish \emph{stable} states, where an atomic action can
be applied, from intermediate states used by the S-GKABs to mimics the
context evolution as well as (incrementally) remove inconsistent
assertions from the ABox. Stable/intermediate states are marked by the
absence/presence of $\tmp$. As before, here the ABox assertion $\tmp$
is often also called special marker.

Similar to \Cref{sec:transf-scsgkabs-to-sgkabs}, since the changes of
context requires the original ABox and we do it after the action
execution, we do not materialize the result of an action execution
directly after its execution, instead we just mark the assertions that
should be added/deleted,
and concretize it during the execution of the action that evolves
context.  To this aim, for each concept name $N \in \voc(\ctbox)$, we
introduce two fresh concept name $N^a$ and $N^d$ to keep track the
temporary information about ABox assertions to be added/deleted before
we materialize the update (similarly for roles). 
Consecutively, we call such kind of concept names \emph{added} and
\emph{deleted} \emph{fact marker} concept names.
Additionally, when we compile \eicgkabs into S-GKABs, we also use
added fact marker concept names 
to mark the information about newly added
assertions.

In order to mimic the context evolution within S-GKABs, we simply
adopt our approach in \Cref{sec:transf-scsgkabs-to-sgkabs}. Thus,
similar to \Cref{sec:transf-scsgkabs-to-sgkabs}, for each context
dimension assignment $\cval{d_i}{v_j}$ we reserve two fresh concept
names $\cdcc_i^{v_j}$ and $\cdcq_i^{v_j}$ in order to represent it as
an ABox assertion. Similarly, this kind of concept name is also called
\emph{context dimension concept}.

The following sections is then organized as follows: First we show
how we compile \bicgkabs into S-GKABs and show that the verification
of \mulcs over \bicgkabs can be reduced into the verification of
\muladom over S-GKABs. After that, we also show similar results for
\cicgkabs and \eicgkabs. Last, we close the tour by exhibiting our
core result that the verification of \icgkabs can be reduced into the
verification of S-GKABs.



\subsection{From \bicgkabs into
  Standard GKABs}\label{sec:from-bicgkabs-to-sgkabs}

We start this section by exhibiting how we translate \bicgkabs into
S-GKABs such that the resulting S-GKAB simulates the evolution of the
given \bicgkab and also we will show how we translate the \mulcs
formulas into \muladom formulas with the aim to reduce the
verification of \bicgkabs into S-GKABs.
While defining the translation from \bicgkabs into S-GKABs, we also
introduce the notion of context-sensitive b-repair program that is
used to simulate the b-repair computation inside S-GKABs.
After that, we continue to show the termination and the correctness of
context-sensitive b-repair program by lifting the results in
\Cref{sec:term-corr-brep-prog}. 
%
The journey is then continued by introducing a certain bisimulation
relation that will be used to prove the reduction of the \mulcs
verification over \bicgkabs into the verification of $\muladom$ over
S-GKABs. Last, we close this section by presenting the proof that we
can recast the verification of \bicgkabs into the corresponding
verification of S-GKABs.

\subsubsection{Translating \bicgkabs into S-GKABs}

Essentially, a single transition in the transition
system of \bicgkabs do the following:
\begin{compactenum}
\item Update the ABox based on the executed action.
\item Change the context.
\item Apply the b-repair to the newly updated ABox. Additonally, the
  b-repair is applied w.r.t.\ the TBox under the new context. 
\end{compactenum}
Therefore, in order to mimic the evolution of \bicgkabs inside S-GKABs, we do the
following:
\begin{compactenum}
\item First, to simulate the ABox and context evolution, we adopt the
  way how we simulate the ABox and context evolution of \scsgkabs
  inside S-GKABs as in \Cref{sec:transf-scsgkabs-to-sgkabs} but
  dropping the inconsistency check,
\item Then, to simulate the b-repair computation, we adopt the way how
  we simulate b-repair computation when we compile B-GKABs into
  S-GKABs (see \Cref{sec:BGKABToSGKAB}). However, we can not use the
  b-repair program as in \Cref{def:brepair-prog} directly, because the
  TBox is changing depending on the context. Thus, the challenge is
  how to make the b-repair program context-sensitive such that it is
  always do the b-repair w.r.t.\ the TBox under the new context. I.e.,
  the b-repair program (might) always need to be adjusted ``on the
  fly'' based on the 
  current context.
\item Last, combining the two steps above sequentially gives us an
  S-GKAB that mimics the evolution of the given \bicgkab.
\end{compactenum}

As a preliminary towards defining the translation from \bicgkabs to
S-KAB, we first define the notion of context-sensitive b-repair
actions and context-sensitive b-repair atomic action invocations which
will be used to define a context-sensitive b-repair program. The main
purpose of introducing the context-sensitive b-repair program is to
mimic the computation of b-repair in \bicgkabs while also taking into
account the context, and thus we can mimic the whole computation in
\bicgkabs inside S-GKAB.

\begin{definition}[Context-sensitive B-Repair Actions and Atomic Action Invocations]\label{def:cs-brep-act-brep-actinv}
  G\sidetext{Context-sensitive B-Repair Actions and Atomic Action
    Invocations}iven a Contextualized TBox $\ctbox$, let
  $\ctxall{\cdimset}$ be the set of all possible context (see
  \Cref{def:set-all-possible-context}).
  We define the \emph{set $\actset_b^{\ctbox}$ of b-repair actions} over $\ctbox$
  and the set $\setinvocation_b^{\ctbox}$ of \emph{b-repair atomic
    action invocations} over $\ctbox$ 
  as follows: for each context $\ctx \in \ctxall{\cdimset}$, we have:
  \begin{compactenum}
  \item For each functionality assertion $\funct{R} \in \ctbox^\ctx$,
    we include in $\actset_b^{\ctbox}$ and $\setinvocation_b^{\ctbox}$
    respectively:
    \begin{compactitem}
    \item
      $\act_{F}(x,y):\set{\map{R(x,z) \wedge \neg [z = y]}{\del
          \set{R( x, z)}}} \in \actset_b^{\ctbox} $, and
    \item
      $\gact{ (q_\ctx \wedge \exists z.\qunsatf(\funct{R}, x, y, z)) }{\act_{F}(x,y)}
      \in \setinvocation_b^{\ctbox}$.
    \end{compactitem}
    Essentially, the atomic action invocation and action above
    together repair an inconsistency related to $\funct{R}$ by
    removing all tuples causing the inconsistency, except one.

  \item For each negative concept inclusion $B_1 \ISA \neg B_2$ such
    that $\ctbox^\ctx \models B_1 \ISA \neg B_2$, we include in $\actset_b^{\ctbox}$
    and $\setinvocation_b^{\ctbox}$ respectively:
    \begin{compactitem}
    \item
      $\act_{B_1}(x):\set{\map{\true}{\del \set{B_1(x)}}} \in
      \actset_b^{\ctbox}$, and
    \item
      $\gact{ (q_\ctx \wedge \qunsatn(B_1 \ISA \neg B_2, x)) }{\act_{B_1}(x)} \in
      \setinvocation_b^{\ctbox}$.
    \end{compactitem}
    Basically, the atomic action invocation and action above together
    repair an inconsistency related to $B_1 \ISA \neg B_2$ by removing
    a constant that is both in $B_1$ and $B_2$ from $B_1$.

  \item For each negative role inclusion $R_1 \ISA \neg R_2$ such that
    $\ctbox^\ctx \models R_1 \ISA \neg R_2$, we include in
    $\actset_b^{\ctbox}$ and $\setinvocation_b^{\ctbox}$ respectively:
    \begin{compactitem}
    \item
      $\act_{R_1}(x,y):\set{\map{\true}{\del \set{R_1(x, y)}}} \in
      \actset_b^{\ctbox}$, and
    \item
      $\gact{ (q_\ctx \wedge \qunsatn(R_1 \ISA \neg R_2, x,
        y)) }{\act_{R_1}(x,y)} \in \setinvocation_b^{\ctbox}$.
    \end{compactitem}
  The atomic action invocation and action above repair an
  inconsistency related to $R_1 \ISA \neg R_2$ by removing constants
  that is both in $R_1$ and $R_2$ from $R_1$.
\end{compactenum}
\ \ 
\end{definition}

\noindent
As the b-repair program in \Cref{sec:BGKABToSGKAB}, we will see later
that the context-sensitive b-repair program essentially will
non-deterministically choose the context-sensitive b-repair atomic
action invocations while there is still an inconsistency. Therefore,
in order to specify the guard of the while loop (i.e., to check
whether there is still an inconsistency), we need to define the notion
of context-sensitive Q-UNSAT, by leveraging on the usual notion of
Q-UNSAT, as follows.

\begin{definition}[Context-sensitive
  Q-UNSAT-ECQ]\label{def:cs-q-unsat-ecq}
  Given \sidetext{Context-sensitive Q-UNSAT-ECQ} a contextualized TBox
  $\ctbox$, a \emph{context-sensitive Q-UNSAT-ECQ} over $\ctbox$
  is a boolean FOL query $\csqunsatecq{{\ctbox}}$ of the following
  form: 
    \[
    \csqunsatecq{{\ctbox}} =  \left( \bigvee_{\ctx \in
        \ctxall{\cdimset}}  \left(q_\ctx \wedge \qunsatecq{{\ctbox^{\ctx}}} \right) \right) 
    \]
    where 
    \begin{compactitem}
    \item $\qunsatecq{{\ctbox^{\ctx}}}$ is Q-UNSAT-ECQ over TBox
      $\ctbox^{\ctx}$ (Note that $\ctbox^{\ctx}$ is $\ctbox$ under the
      context $\ctx$),
    \item $\ctxall{\cdimset}$ is the set of all possible context as in
      \Cref{def:set-all-possible-context},
    \item $q_\ctx$ is the query obtained from the context $\ctx$.
    \end{compactitem}
\end{definition}

We now proceed to define the context-sensitive b-repair program by
utilizing the context-sensitive b-repair atomic action invocations as
well as the context-sensitive b-repair actions that has been
introduced before.

\begin{definition}[Context-sensitive B-Repair
  Program]\label{def:cs-brepair-prog}
  Given \sidetext{Context-sensitive B-Repair Program} a \bicgkabs
  $\csgkabsym = \tup{\ctbox, \initabox, \actset, \ginitprog, \initctx,
    \ctxprocset}$.
  Let $\actset_b^{\ctbox}$ be a set of context-sensitve b-repair
  actions over $\ctbox$, and
  $\setinvocation_b^{\ctbox} = \set{a_1,\ldots,a_n}$ be a set of
  context-sensitive b-repair atomic action invocations over $\ctbox$.
  We then define the \emph{context-sensitve b-repair program over}
  $\ctbox$ 
  as follows:
  \[
  \delta_b^{\ctbox} =\gwhile{\csqunsatecq{\ctbox}}{\delta_{r}}
  \]
where $\delta_r = a_{1}|a_{2}|\ldots|a_n$.
\end{definition}

As the last step before we formally define the translation of
\bicgkabs into S-GKABs, below we define the program translation that
will be used to translate the program in \bicgkabs.  Basically we
define a translation function $\tgprogbcs$ that concatenates each
action invocation with a program that non-deterministically choose an
action that changes the context, and then concatenates them with the
context-sensitive b-repair program.  Additionally, the translation
function $\tgprogbcs$ also serves as a one-to-one correspondence
(bijection) between the original and the translated program (as well
as between the sub-program).

\clearpage

\begin{definition}[Program Translation $\tgprogbcs$]
  Given \sidetextb{Program Translation $\tgprogbcs$} a \bicgkabs
  $\csgkabsym = \tup{\ctbox, \initabox, \actset, \ginitprog, \initctx,
    \ctxprocset}$
  we define a \emph{translation $\tgprogbcs$} which translates a
  program $\delta$ into a program $\delta'$ inductively as follows:
\[
\begin{array}{@{}l@{}l@{}}
  \tgprogbcs(\gactc{Q(\vec{p})}{\ctxe}{\act(\vec{p})}) &=  
                                                         \gact{Q'(\vec{p})}{\act'(\vec{p})}
                                                         ; \delta_{\ctxprocset}
                                                         ; \delta^{\ctbox}_b
                                                         ; \gact{\true}{\act^-_{tmp}()}\\
  \tgprogbcs(\gemptyprog) &= \gemptyprog \\
  \tgprogbcs(\delta_1|\delta_2) &= \tgprogbcs(\delta_1)|\tgprogbcs(\delta_2) \\
  \tgprogbcs(\delta_1;\delta_2) &= \tgprogbcs(\delta_1);\tgprogbcs(\delta_2) \\
  \tgprogbcs(\gif{\varphi}{\delta_1}{\delta_2}) &= \gif{\varphi}{\tgprogbcs(\delta_1)}{\tgprogbcs(\delta_2)} \\
  \tgprogbcs(\gwhile{\varphi}{\delta}) &= \gwhile{\varphi}{\tgprogbcs(\delta)}
\end{array}
\]
where 
\begin{compactitem}
\item $\gact{Q'(\vec{p})}{\act'(\vec{p})}$ is an action invocation
  obtained from $\gactc{Q(\vec{p})}{\ctxe}{\act(\vec{p})}$ as in
  \Cref{def:action-invoc-from-cs-action-invoc},
\item $\delta_{\ctxprocset}$ is a context-change program obtained from
  $\ctxprocset$ as in
  \Cref{def:ctx-chg-prog}, 
\item $\delta^{\ctbox}_b$ is a context-sensitive b-repair program over
  $\ctbox$ as in \Cref{def:cs-brepair-prog},
\item $\act^-_{temp}() :\set{\map{\true}{\del \set{\tmp}}}$.
\end{compactitem}
\ \ 
\end{definition}

Having all of the machinery in hand, we are ready to define a
translation $\tgkabbcs$ that, given a \bicgkab, produces an S-GKAB
as follows:

\begin{definition}[Translation from \bicgkab to S-GKAB]
  We \sidetext{Translation from \bicgkab to S-GKAB} define a
  translation $\tgkabbcs$ that, given a \bicgkab
  $\csgkabsym = \tup{\ctbox, \initabox, \actset, \ginitprog, \initctx,
    \ctxprocset}$,
  produces an S-GKAB
  $\tgkabbcs(\csgkabsym) = \tup{T_\cdimset, \initabox \cup
    A_{\initctx}, \actset', \ginitprog'}$, where
\begin{compactitem}

\item $T_\cdimset$ is a TBox obtained from a
  set of context dimensions $\cdimset$ (see
  \Cref{def:tbox-from-ctxdim}),

\item 
  $A_{\initctx}$ is an ABox obtained from $\initctx$ (see
  \Cref{def:abox-context}),

\item
  $\actset' = \actset_\act \cup \actset_\ctx \cup \actset_b^{\ctbox}
  \cup \set{\act^-_{temp}}$ where:

\begin{compactitem}
\item $\actset_\act$ is obtained from $\actset$ such that for each
  action $\act \in \actset$, we have $\act' \in \actset_\act$ where
  $\act'$ is a delayed action obtained from $\act$ (see
  \Cref{def:delayed-action}),

\item $\actset_\ctx$ is obtained from $\ctxprocset$ such that for each
  context-evolution rule $\tup{Q, \ctxe} \mapsto C_{new}$ in
  $\ctxprocset$, we have $\act_\ctx \in \actset_\ctx$ where
  $\act_\ctx$ is \emph{an action obtained from the context-evolution
    rule $\tup{Q, \ctxe} \mapsto C_{new}$} (see
  \Cref{def:action-and-action-invocation-obtained-from-context-evolution-rule}),

\item $\actset_b^{\ctbox}$ is the set of context-sensitive b-repair
  actions over $\ctbox$ (see \Cref{def:cs-brep-act-brep-actinv}),

\item $\act^-_{temp}$ is an action of the form
  $\act^-_{temp}() :\set{\map{\true}{\del \set{\tmp}}}$.

\end{compactitem}

\item $\ginitprog' = \tgprogbcs(\ginitprog)$. 
\end{compactitem}
\ \ 
\end{definition}

The \mulcs property $\Phi$ over a \bicgkab $\csgkabsym$ can then be
recast as a corresponding property over an S-GKAB
$\tgkabbcs(\csgkabsym)$ using the following formula translation:
\begin{definition}[Translation $\tforjcs$]\label{def:tforjcs}
  We \sidetextb{Translation $\tforjcs$} define a \emph{translation
    $\tforjcs$} that transforms an arbitrary \mulcs formula $\Phi$ (in
  NNF) into another \muladom formula $\Phi'$ inductively by recurring
  over the structure of $\Phi$ as follows:
$
  \begin{array}{@{}l@{}ll@{}}
    \bullet\ \tforjcs(Q) &=& Q_\ctxb \\

    \bullet\ \tforjcs(\ctxe) &=& q_{\ctxe} \\

    \bullet\ \tforjcs(\neg Q) &=& \neg Q_\ctxb \\

    \bullet\ \tforjcs(\Q x.\Phi) &=& \Q x. \tforjcs(\Phi) \\

    \bullet\ \tforjcs(\Phi_1 \circ \Phi_2) &=& \tforjcs(\Phi_1) \circ \tforjcs(\Phi_2) \\

    \bullet\ \tforjcs(\circledcirc Z.\Phi) &=& \circledcirc Z. \tforjcs(\Phi) \\

    \bullet\ \tforjcs(\DIAM{\Phi}) &=& \DIAM{\mu Z.((\tmp \wedge \DIAM{Z}) \vee
                                     (\neg \tmp \wedge \tforjcs(\Phi)))} \\

    \bullet\ \tforjcs(\BOX{\Phi}) &=& \BOX{\mu Z.((\tmp \wedge \BOX{Z} \wedge
                                    \DIAM{\top}) \vee (\neg \tmp \wedge \tforjcs(\Phi)))}
  \end{array}
  $
  \noindent
  where:
  \begin{compactitem}
  \item $\circ$ is a binary operator ($\vee, \wedge, \ra,$ or $\lra$),
  \item $\circledcirc$ is least ($\mu$) or greatest ($\nu$) fix-point
    operator,
  \item $\Q$ is forall ($\forall$) or existential ($\exists$)
    quantifier.
  \end{compactitem}
\ \ 
\end{definition}

\noindent
Having those two translations in hand, we show it later that
$\ts{\csgkabsym}^{\csfilter_B} \models \Phi$ if and only if
$\ts{\tgkabbcs(\csgkabsym)}^{\filter_S} \models \tforjcs(\Phi)$ which
consequently means that the verification of \mulcs over \bicgkabs can
be reduced to the corresponding verification of \muladom over S-GKABs.
%

\subsubsection{Termination and Correctness of Context-sensitive
  B-repair Program}\label{sec:term-corr-cs-brep-prog}

In this section we aim to show the termination and correctness of
context-sensitive b-repair program by essentially lifting the result
in \Cref{sec:term-corr-brep-prog}. 
%
%
%
First, we lift the result about the termination of b-repair program
into the case of context-sensitive b-repair program as follows:

\begin{lemma}\label{lem:csbprog-termination}
  Let
  $\csgkabsym = \tup{\ctbox, \initabox, \actset, \ginitprog, \initctx,
    \ctxprocset}$
  be a \bicgkab, $\tgkabbcs(\csgkabsym)$ be an S-GKAB $($with
  transition system
  $\ts{\tgkabbcs(\csgkabsym)}^{\filter_S}$$)$
  obtained from $\csgkabsym$
  through $\tgkabbcs$,
  and $\delta^{\ctbox}_{b}$
  be a context-sensitive b-repair program over $\ctbox$.
  We have that $\delta^{\ctbox}_{b}$
  is always terminate. I.e., given a state $\tup{A,
    \scmap,
    \delta^{\ctbox}_{b}}$ of
  $\ts{\tgkabbcs(\csgkabsym)}^{\filter_S}$,
  every program execution trace induced by $\delta^{\ctbox}_{b}$
  on $\tup{A,
    \scmap, \delta^{\ctbox}_{b}}$ w.r.t.\ filter
  $\filter_S$ is terminating.
\end{lemma}
\begin{proof}
  Similar to the proof of \Cref{lem:bprog-termination} except that we
  need to accommodate the presence of context that is encoded as ABox
  assertions. All of the supporting lemmas to prove
  \Cref{lem:bprog-termination} can be also easily lifted to the
  context-sensitive case in order to support the proof of this
  lemma. The important argument in this proof is that each step of the
  program $\delta^{\ctbox}_{b}$
  always reduce the number of assertions that is participated in
  making the inconsistency. Thus, at some point when there is no more
  inconsistency, the loop will be exited.
\end{proof}

\noindent
Furthermore, we can also lift the correctness result of b-repair
program into the case of context-sensitive b-repair program
below. Essentially, we show that the context-sensitive b-repair
program produces the same result as the result of b-repair over KB.

\bigskip
\begin{theorem}\label{thm:cs-bprog-equal-brep}
  Let
  $\csgkabsym = \tup{\ctbox, \initabox, \actset, \ginitprog, \initctx,
    \ctxprocset}$
  be a \bicgkab, $\tgkabbcs(\csgkabsym)$ be an S-GKAB $($with
  transition system
  $\ts{\tgkabbcs(\csgkabsym)}^{\filter_S}$$)$
  obtained from $\csgkabsym$
  through $\tgkabbcs$,
  and $\delta^{\ctbox}_{b}$
  be a context-sensitive b-repair program over $\ctbox$.
  Consider an ABox $A$,
  a service call map $\scmap$,
  and a context $\ctx$.
  We have that $\progres(A
  \cup A_\ctx, \scmap, \delta^{\ctbox}_b) =
  \arset{\ctbox^{\ctx},A}$, where
  $A_\ctx$ is a set of ABox assertions representing the context $\ctx$.
%
\end{theorem}
\begin{proof}
  Similar to the proof of \Cref{thm:bprog-equal-brep} by also
  observing the following:
  \begin{compactitem}
  \item All supporting lemmas to prove \Cref{thm:bprog-equal-brep} can
    be also easily lifted to the context-sensitive case in order to
    support the proof of this lemma.

  \item The set $\arset{\ctbox^{\ctx},A}$
    of all b-repairs is computed w.r.t.\ the TBox
    $\ctbox^{\ctx}$ (i.e., $\ctbox$ under the context $\ctx$).

  \item The presesence of context determines which b-repair action
    invocation in the context-sensitive b-repair program
    $\delta^{\ctbox}_b$ that is executable and hence influence the
    execution flow of the program $\delta^{\ctbox}_b$. Moreover,
    $\delta^{\ctbox}_b$ is executed under the context $\ctx$.
    Therefore, by construction of $\delta^{\ctbox}_b$, the
    context-sensitive b-repair action invocations that are executed
    are only those that is related to the TBox assertion in
    $\ctbox^{\ctx}$. Thus, the repair is done based on the TBox
    $\ctbox$ under the context $\ctx$.



  \end{compactitem}
\end{proof}

\subsubsection{Context-sensitive Jumping Bisimulation (\cjbsimabr-Bisimulation)}

In this section we introduce the notion of Context-sensitive Jumping
Bisimulation (\cjbsimabr-Bisimulation) by leveraging on the notion of
Jumping Bisimulation as in (see \Cref{sec:jumping-bisimulation}) and
the notion of Skip-two Bisimulation (see
\Cref{sec:skip-two-st-bisimulation}). Furthermore, here we also show
some properties related to \cjbsimabr-Bisimulation.

\begin{definition}[Context-sensitive Jumping Bisimulation
  (\cjbsimabr-Bisimulation)] Let \sidetext{Context-sensitive Jumping
    Bisimulation (\cjbsimabr-Bisimulation)}
  $\ts{1} = \tup{\const, \ctbox, \stateset_1, s_{01}, \abox_1, \cntx,
    \trans_1}$
  be a context-sensitive transition system, and
  $\ts{2} = \tup{\const, T, \stateset_2, s_{02}, \abox_2, \trans_2}$
  be a KB transition system, with
  $\adom{\abox_1(s_{01})} \subseteq \const$ and
  $\adom{\abox_2(s_{02})} \subseteq \const$.  A
  \emph{context-sensitive jumping bisimulation}
  (\cjbsimabr-Bisimulation) between $\ts{1}$ and $\ts{2}$ is a
  relation $\B \subseteq \Sigma_1 \times\Sigma_2$ such that
  $\tup{s_1, s_2} \in \B$ implies that:
  \begin{compactenum}
  \item $s_1 \eqc s_2$, i.e., $s_1$ and $s_2$ are contextually equal
    (see \Cref{def:contextually-equal}),

  \item for each $s_1'$, if $s_1 \Rightarrow_1 s_1'$ then there exist
    $s_2'$, $t_1, \ldots ,t_n$ (for $n \geq 0$) with
    \[
    s_2 \Rightarrow_2 t_1 \Rightarrow_2 \ldots \Rightarrow_2 t_n
    \Rightarrow_2 s_2'
    \] 
    such that $\tup{s_1', s_2'}\in\B$, $\tmp \not\in \abox_2(s_2')$
    and $\tmp \in \abox_2(t_i)$ for $i \in \set{1, \ldots, n}$,

  \item for each $s_2'$, if 
    \[
    s_2 \Rightarrow_2 t_1 \Rightarrow_2 \ldots \Rightarrow_2 t_n
    \Rightarrow_2 s_2'
    \] 
    (for $n \geq 0$) with $\tmp \in \abox_2(t_i)$ for
    $i \in \set{1, \ldots, n}$ and $\tmp \not\in \abox_2(s_2')$, then
    there exists $s_1'$ with $s_1 \Rightarrow_1 s_1'$, such that
    $\tup{s_1', s_2'}\in\B$.
 \end{compactenum}
\ \ 
\end{definition}

\noindent
Let
$\ts{1} = \tup{\const, \ctbox, \stateset_1, s_{01}, \abox_1, \cntx,
  \trans_1}$
be a context-sensitive transition system, and
$\ts{2} = \tup{\const, T, \stateset_2, s_{02}, \abox_2, \trans_2}$ be
a KB transition system,
a state $s_1 \in \stateset_1$ is \emph{\cjbsimabr-bisimilar} to
$s_2 \in \stateset_2$, written $s_1 \cjbsim s_2$, if there exists a
\cjbsimabr-bisimulation relation $\B$ between $\ts{1}$ and $\ts{2}$
such that $\tup{s_1, s_2}\in\B$.
A transition system $\ts{1}$ is \emph{\cjbsimabr-bisimilar} to
$\ts{2}$, written $\ts{1} \cjbsim \ts{2}$, if there exists a
\cjbsimabr-bisimulation relation $\B$ between $\ts{1}$ and $\ts{2}$
such that $\tup{s_{01}, s_{02}}\in\B$.

In the following lemmas we show some important properties of
\cjbsimabr-bisimilar states and transition systems that will be useful later to show that
we can recast the verification of \bicgkabs into S-GKABs.

\begin{lemma}\label{lem:cjbisimilar-state-satisfies-same-formula}
  Let
  $\ts{1} = \tup{\const, \ctbox, \stateset_1, s_{01}, \abox_1, \cntx,
    \trans_1}$
  be a context-sensitive transition system, and
  $\ts{2} = \tup{\const, T, \stateset_2, s_{02}, \abox_2, \trans_2}$
  be a KB transition system. 
  Consider two states $s_1 \in \stateset_1$ and $s_2 \in \stateset_2$
  such that $s_1 \cjbsim s_2$. Then for every formula $\Phi$ of
  $\mulcs$, and every valuations $\vfo_1$ and $\vfo_2$ that assign to
  each of its free variables a constant $c_1 \in \adom{\abox_1(s_1)}$
  and $c_2 \in \adom{\abox_2(s_2)}$, such that $c_1 = c_2$, we have
  that
  \[
  \ts{1},s_1 \models \Phi \vfo_1 \textrm{ if and only if } \ts{2},s_2
  \models \tforjcs(\Phi) \vfo_2.
  \]
\end{lemma}
\begin{proof}
  Similar to the combination of the proof of
  \Cref{lem:stbisimilar-state-satisfies-same-formula,lem:jumping-bisimilar-states-satisfies-same-formula}.
\end{proof}

\begin{lemma}\label{lem:cjbisimilar-ts-satisfies-same-formula}
  Consider a context-sensitive transition system
  $\ts{1}$, 
  and a KB transition system
  $\ts{2}$ 
%
  such that $\ts{1} \cjbsim \ts{2}$.  For every closed \mulcs formula
  $\Phi$, we have:
  \[
  \ts{1} \models \Phi \textrm{ if and only if } \ts{2} \models
  \tforjcs(\Phi)
  \]
\end{lemma}
\begin{proof} Let
  $\ts{1} = \tup{\const, \ctbox, \stateset_1, s_{01}, \abox_1, \cntx,
    \trans_1}$,
  and
  $\ts{2} = \tup{\const, T, \stateset_2, s_{02}, \abox_2, \trans_2}$.
  By the definition of \cjbsimabr-bisimilar transition system we have
  that $s_{01} \cjbsim s_{02}$. Thus, we obtain the proof as a
  consequence of \Cref{lem:cjbisimilar-state-satisfies-same-formula},
  due to the fact that
  \[ \ts{1}, s_{01} \models \Phi \textrm{ if and only if } \ts{2},
  s_{02} \models \tforjcs(\Phi)
  \]
\end{proof}

\subsubsection{Reducing the Verification of \bicgkabs into S-GKABs}

In order to show that we can recast the verification of \bicgkabs into
S-GKABs, in the following two lemmas we show that the transition
system of a \bicgkabs $\csgkabsym$ is \cjbsimabr-bisimilar with the
transition system of the corresponding S-GKAB $\tgkabbcs(\csgkabsym)$
that is obtained via translation $\tgkabbcs$.

\begin{lemma}\label{lem:bicgkab-to-sgkab-bisimilar-state}
  Let $\csgkabsym$ be a \bicgkab with transition system
  $\ts{\csgkabsym}^{\csfilter_B}$, and let $\tgkabbcs(\csgkabsym)$ be
  its corresponding S-GKAB $($with transition system
  $\ts{\tgkabbcs(\csgkabsym)}^{\filter_S}$$)$
  obtained through
  $\tgkabbcs$. 
  Consider
  \begin{inparaenum}[]
  \item a state $s_{cx} = \tup{A_{cx},\scmap_{cx}, \ctx, \delta_{cx}}$ of
    $\ts{\csgkabsym}^{\csfilter_B}$ and
  \item a state $s_s = \tup{A_s,\scmap_s, \delta_s}$ of
    $\ts{\tgkabbcs(\csgkabsym)}^{\filter_S}$.
  \end{inparaenum}
If  
\begin{inparaenum}[]
\item $s_{cx} \eqc s_s$, $\scmap_{cx} = \scmap_s$ and
\item $\delta_s = \tgprogbcs(\delta_{cx})$,
\end{inparaenum}
then
$\tup{A_{cx},\scmap_{cx}, \ctx, \delta_{cx}} \cjbsim \tup{A_s,\scmap_s, \delta_s}$.
\end{lemma}
\begin{proof}
  The proof is similar to the combination of the proof for
  \Cref{lem:scsgkab-to-sgkab-bisimilar-state,lem:bgkab-to-sgkab-bisimilar-state},
  by also considering the following:
  \begin{compactenum}
    
  \item The \bicgkabs do the b-repair over the updated ABox and under
    the new context.

  \item The context-sensitive b-repair program is executed after the
    ABox has been changed and the context has been updated.

  \item By \Cref{thm:cs-bprog-equal-brep}, we have that the result of
    the context-sensitive b-repair program is the same as the result
    of the b-repair computation.


  \end{compactenum}

\end{proof}

\begin{lemma}\label{lem:bicgkab-to-sgkab-bisimilar-ts}
  Given a \bicgkab $\csgkabsym$, we have
  $\ts{\csgkabsym}^{\csfilter_B} \cjbsim
  \ts{\tgkabbcs(\csgkabsym)}^{\filter_S}$
\end{lemma}
\begin{proof}
Let
\begin{compactenum}
\item
  $\csgkabsym = \tup{\ctbox, \initabox, \actset, \ginitprog, \initctx,
    \ctxprocset}$,
  and \\
  $\ts{\csgkabsym}^{\csfilter_B} = \tup{\const, \ctbox, \stateset_\ctxb, s_{0\ctxb},
    \abox_\ctxb, \cntx, \trans_\ctxb}$,
\item   $\tgkabbcs(\csgkabsym) = \tup{T', \initabox', \actset',
    \ginitprog'}$ and 
  $\ts{\tgkabbcs(\csgkabsym)}^{\filter_S} = \tup{\const, T',
    \stateset_s, s_{0s}, \abox_s, \trans_s}$.
\end{compactenum}
We have that
$s_{0\ctxb} = \tup{\initabox, \scmap_{cx}, \ctx_0, \delta}$ and
$s_{0s} = \tup{\initabox', \scmap_s, \delta'}$ where
$\scmap_{cx} = \scmap_s = \emptyset$. By the definition of
$\tgprogbcs$ and $\tgkabbcs$, we also have $s_{0\ctxb} \eqc s_{0s}$,
and $\delta' = \tgprogbcs(\delta)$. Hence, by
\Cref{lem:bicgkab-to-sgkab-bisimilar-state}, we have
$s_{0\ctxb} \cjbsim s_{0s}$. Therefore, by the definition of
\cjbsimabr-bisimulation, we have
$\ts{\csgkabsym}^{\csfilter_B} \cjbsim
\ts{\tgkabbcs(\csgkabsym)}^{\filter_S}$.  \ \
\end{proof}

Next, we show that the verification of \mulcs properties over
\bicgkabs can be recast as verification of \muladom over S-GKABs as
follows.

\begin{theorem}\label{thm:bicgkab-to-sgkab}
  Given an \bicgkab $\csgkabsym$ and a closed $\mulcs$ property
  $\Phi$, we have
\begin{center}
  $\ts{\csgkabsym}^{\csfilter_B} \models \Phi$ if and only if
  $ \ts{\tgkabbcs(\csgkabsym)}^{\filter_S}\models \tforjcs(\Phi)$
\end{center}
\end{theorem}
\begin{proof}
  By \Cref{lem:bicgkab-to-sgkab-bisimilar-ts}, we have that
  $\ts{\csgkabsym}^{\csfilter_B} \cjbsim
  \ts{\tgkabbcs(\csgkabsym)}^{\filter_S}$.
  Hence, by \Cref{lem:cjbisimilar-ts-satisfies-same-formula}, we have
  that for every $\mulcs$ property $\Phi$
\[
\ts{\csgkabsym}^{\csfilter_B} \models \Phi \textrm{ if and only if }
\ts{\tgkabbcs(\csgkabsym)}^{\filter_S}\models \tforjcs(\Phi)
\]
\end{proof}



\subsection{From \cicgkabs into Standard GKABs}\label{sec:from-cicgkabs-to-sgkabs}

Similar to the case of \bicgkabs, a single transition in \cicgkabs
essentially update the ABox and context, and then apply c-repair to
the newly updated ABox. Thus, to mimic the evolution of \cicgkabs
inside S-GKABs, we first adopt our approach in
\Cref{sec:transf-scsgkabs-to-sgkabs} to simulate the ABox and context
evolution of \cicgkabs within S-GKABs. Then, we adopt our approach in
\Cref{CGKABToSGKAB} to simulate the c-repair computation. As in the
case of transforming \bicgkabs into S-GKABs, we can not re-use the
notion of c-repair action in \Cref{def:c-rep-act} directly because the
TBox is evolving based on the context. Thus, we need to make the
c-repair action able to adapt its behavior based on the context. I.e.,
we require that the c-repair action do the repair based on the TBox
assertions that ``hold'' under the current context.  To deal with
this, we introduce a so called context-sensitive c-repair action that
only consider those TBox assertions that ``hold'' under the current
context and removes all ABox assertions that are involved in some form
of inconsistency.

\subsubsection{Translating \cicgkabs into S-GKABs}

As the first step towards translating \cicgkabs into S-GKABs, in the
following we introduce the notion of context-sensitive c-repair action.

\begin{definition}[Context-sensitive C-Repair Action]\label{def:cs-c-repair-act}
  Given \sidetext{Context-sensitive C-Repair Action} a Contextualized
  TBox $\ctbox$, let $\ctxall{\cdimset}$ be the set of all possible
  contexts (see \Cref{def:set-all-possible-context}). We define a
  0-ary (i.e., has no action parameters) \emph{context-sensitive
    c-repair action $\act^{\ctbox}_c$} over $\ctbox$, where
  $\eff{\act^{\ctbox}_c}$ is the smallest set containing the following
  effects: For each context $\ctx \in \ctxall{\cdimset}$, we have:
  \begin{compactitem}
  \item for each functionality assertion $\funct{R} \in \ctbox^\ctx$, we have 
    \[
    \map{q_\ctx \wedge \qunsatf(\funct{R}, x, y, z)} {\del \set{R(x, y),R(x,
          z)}} \in \eff{\act^{\ctbox}_c}
    \]
  \item for each negative concept inclusion assertion $B_1 \ISA \neg B_2$ such that  $\ctbox^\ctx
    \models B_1 \ISA \neg B_2$, we have 
    \[
    \map{q_\ctx \wedge \qunsatn(B_1 \ISA \neg B_2, x)} {\del \set{B_1(x),
          B_2(x)} } \in \eff{\act^{\ctbox}_c};
    \]
  \item 
    for each negative role inclusion assertion $R_1 \ISA \neg R_2$
    such that $\ctbox^\ctx \models R_1 \ISA \neg R_2$, we have
    \[
    \map{q_\ctx \wedge \qunsatn(R_1 \ISA \neg R_2, x, y)} {\del \set{R_1(x, y),
          R_2(x, y)}} \in \eff{\act^{\ctbox}_c}.
    \]
  \item $\map{\true} {\del \set{\tmp}} \in \eff{\act^{\ctbox}_c}.$

  \end{compactitem}
\ \ 
\end{definition}

As the last preliminary before we formally define the translation of
\cicgkabs into S-GKABs, in the following we define the program
translation that will be used to translate the program in \cicgkabs.
Essentially we define a translation function $\tgprogccs$ that
concatenates each action invocation with a program that
non-deterministically choose an action that changes the context, and
then concatenates them with the context-sensitive c-repair action.
Additionally, the translation function $\tgprogccs$ also serves as a
one-to-one correspondence (bijection) between the original and the
translated program (as well as between the sub-program).

\begin{definition}[Program Translation $\tgprogccs$]
  Given \sidetextb{Program Translation $\tgprogccs$} a \cicgkabs
  $\csgkabsym = \tup{\ctbox, \initabox, \actset, \ginitprog, \initctx,
    \ctxprocset}$we
  define a \emph{translation $\tgprogccs$} that translates a program
  $\delta$ into a program $\delta'$ inductively as follows:
  \[
\begin{array}{@{}l@{}l@{}}
  \tgprogccs(\gactc{Q(\vec{p})}{\ctxe}{\act(\vec{p})})  &=  
                                                 \gact{Q'(\vec{p})}{\act'(\vec{p})}
                                                 ;\delta_{\ctxprocset}
                                                 ;\gact{\true}{\act^{\ctbox}_c()}\\
  \tgprogccs(\gemptyprog) &= \gemptyprog \\
  \tgprogccs(\delta_1|\delta_2) &= \tgprogccs(\delta_1)|\tgprogccs(\delta_2) \\
  \tgprogccs(\delta_1;\delta_2) &= \tgprogccs(\delta_1);\tgprogccs(\delta_2) \\
  \tgprogccs(\gif{\varphi}{\delta_1}{\delta_2}) &= \gif{\varphi}{\tgprogccs(\delta_1)}{\tgprogccs(\delta_2)} \\
  \tgprogccs(\gwhile{\varphi}{\delta}) &= \gwhile{\varphi}{\tgprogccs(\delta)}
\end{array}
\]
where 
\begin{compactitem}
\item $\gact{Q'(\vec{p})}{\act'(\vec{p})}$ is an action invocation
  obtained from $\gactc{Q(\vec{p})}{\ctxe}{\act(\vec{p})}$ (see 
  \Cref{def:action-invoc-from-cs-action-invoc}),
\item $\delta_{\ctxprocset}$ is a context-change program obtained from
  $\ctxprocset$ as in
  \Cref{def:ctx-chg-prog}, 
\item $\act^{\ctbox}_c$ is a context-sensitive c-repair action over
  $\ctbox$ as in \Cref{def:cs-c-repair-act}.
 \end{compactitem}
\ \ 
\end{definition}

Having all of the machinery in hand, we are ready to define a
translation $\tgkabccs$ that, given a \cicgkab, produces an S-GKAB
as follows:

\begin{definition}[Translation from \cicgkab to S-GKAB]
  We \sidetext{Translation from \cicgkab to S-GKAB} define a
  translation $\tgkabccs$ that, given a \cicgkab
  $\csgkabsym = \tup{\ctbox, \initabox, \actset, \ginitprog, \initctx,
    \ctxprocset}$,
  produces an S-GKAB
  $\tgkabccs(\csgkabsym) = \tup{T_\cdimset, \initabox \cup
    A_{\initctx}, \actset', \ginitprog'}$, where
\begin{compactitem}

\item $T_\cdimset$ is a TBox obtained from a
  set of context dimensions $\cdimset$ (see
  \Cref{def:tbox-from-ctxdim}),

\item 
  $A_{\initctx}$ is an ABox obtained from $\initctx$ (see
  \Cref{def:abox-context}),

\item
  $\actset' = \actset_\act \cup \actset_\ctx \cup
  \set{\act^{\ctbox}_c}$ where:

\begin{compactitem}
\item $\actset_\act$ is obtained from $\actset$ such that for each
  action $\act \in \actset$, we have $\act' \in \actset_\act$ where
  $\act'$ is a delayed action obtained from $\act$ (see
  \Cref{def:delayed-action}),

\item $\actset_\ctx$ is obtained from $\ctxprocset$ such that for each
  context-evolution rule $\tup{Q, \ctxe} \mapsto C_{new}$ in
  $\ctxprocset$, we have $\act_\ctx \in \actset_\ctx$ where
  $\act_\ctx$ is \emph{an action obtained from the context-evolution
    rule $\tup{Q, \ctxe} \mapsto C_{new}$} (see
  \Cref{def:action-and-action-invocation-obtained-from-context-evolution-rule}),

\item $\act^{\ctbox}_c$ is a context-sensitive c-repair action over
  $\ctbox$ (see \Cref{def:cs-c-repair-act})

\end{compactitem}

\item $\ginitprog' = \tgprogccs(\ginitprog)$. 
\end{compactitem}
\ \ 
\end{definition}

The \mulcs property $\Phi$ over \cicgkabs $\csgkabsym$ can then be
recast as a corresponding property over S-GKABs
$\tgkabccs(\csgkabsym)$ using the following formula translation
$\tfort$ (see \Cref{def:ttrip}).
Utilizing those two translations, later we show that
$\ts{\csgkabsym}^{\csfilter_C} \models \Phi$ if and only if
$\ts{\tgkabccs(\csgkabsym)}^{\filter_S} \models \tfort(\Phi)$. As a
consequence, we have that the verification of \mulcs over \cicgkabs
can be reduced to the corresponding verification of \muladom over
S-GKABs.

\subsubsection{Reducing the Verification of \cicgkabs into S-GKABs}

In this subsection, we show that we can reduce the verification of
\cicgkabs into S-GKABs. To this aim, in the following we first show
that the transition system of a \cicgkabs $\csgkabsym$ is ST-bisimilar
with the transition system of the corresponding S-GKAB
$\tgkabccs(\csgkabsym)$ that is obtained via translation
$\tgkabccs$. 

\begin{lemma}\label{lem:cicgkab-to-sgkab-bisimilar-state}
  Let $\csgkabsym$ be a \cicgkab with transition system
  $\ts{\csgkabsym}^{\csfilter_C}$, and let $\tgkabccs(\csgkabsym)$ be
  its corresponding S-GKAB $($with transition system
  $\ts{\tgkabccs(\csgkabsym)}^{\filter_S}$$)$
  obtained through
  $\tgkabccs$. 
  Consider
  \begin{inparaenum}[]
  \item a state $s_{cx} = \tup{A_{cx},\scmap_{cx}, \ctx, \delta_{cx}}$ of
    $\ts{\csgkabsym}^{\csfilter_C}$ and
  \item a state $s_s = \tup{A_s,\scmap_s, \delta_s}$ of
    $\ts{\tgkabccs(\csgkabsym)}^{\filter_S}$.
  \end{inparaenum}
If  
\begin{inparaenum}[]
\item $s_{cx} \eqc s_s$, $\scmap_{cx} = \scmap_s$ and
\item $\delta_s = \tgprogccs(\delta_{cx})$,
\end{inparaenum}
then
$\tup{A_{cx},\scmap_{cx}, \ctx, \delta_{cx}} \stbsim \tup{A_s,\scmap_s, \delta_s}$.
\end{lemma}
\begin{proof}
  The proof is similar to the combination of the proof for
  \Cref{lem:scsgkab-to-sgkab-bisimilar-state,lem:cgkab-to-sgkab-bisimilar-state},
  by also considering the following:
  \begin{compactenum}
    
  \item The \cicgkabs do the c-repair over the updated ABox and under
    the new context.

  \item The context-sensitive c-repair action is executed after the
    ABox has been changed and the context has been updated.

  \item The different with the S-GKABs that capture \scsgkabs is that
    after changing the context and materializing the ABox changes,
    instead of executing an action that check for inconsistency, the
    S-GKABs that capture \cicgkabs execute the c-repair action which
    performs c-repair computation.

  \item Similar to \Cref{thm:cact-equal-crep}, we can also easily show
    the correctness of context-sensitive c-repair action. The
    important observation is that the context-sensitive c-repair
    action do the repair based on the context, i.e., it only consider
    those assertion in the TBox that ``hold'' under the corresponding
    context.

  \end{compactenum}
\end{proof}

\begin{lemma}\label{lem:cicgkab-to-sgkab-bisimilar-ts}
  Given a \cicgkab $\csgkabsym$, we have
  $\ts{\csgkabsym}^{\csfilter_C} \stbsim
  \ts{\tgkabccs(\csgkabsym)}^{\filter_S}$
\end{lemma}
\begin{proof}
Let
\begin{compactenum}
\item
  $\csgkabsym = \tup{\ctbox, \initabox, \actset, \ginitprog, \initctx,
    \ctxprocset}$,
  and \\
  $\ts{\csgkabsym}^{\csfilter_C} = \tup{\const, \ctbox, \stateset_\ctxb, s_{0\ctxb},
    \abox_\ctxb, \cntx, \trans_\ctxb}$,
\item   $\tgkabccs(\csgkabsym) = \tup{T', \initabox', \actset',
    \ginitprog'}$ and 
  $\ts{\tgkabccs(\csgkabsym)}^{\filter_S} = \tup{\const, T',
    \stateset_s, s_{0s}, \abox_s, \trans_s}$.
\end{compactenum}
We have that $s_{0\ctxb} = \tup{\initabox, \scmap_c, \ctx_0, \delta}$
and $s_{0s} = \tup{\initabox', \scmap_s, \delta'}$ where
$\scmap_c = \scmap_s = \emptyset$. By the definition of $\tgprogccs$
and $\tgkabccs$, we also have $s_{0\ctxb} \eqc s_{0s}$, and
$\delta' = \tgprogccs(\delta)$. Hence, by
\Cref{lem:cicgkab-to-sgkab-bisimilar-state}, we have
$s_{0\ctxb} \stbsim s_{0s}$. Therefore, by the definition of
ST-bisimulation, we have
$\ts{\csgkabsym}^{\csfilter_C} \stbsim
\ts{\tgkabccs(\csgkabsym)}^{\filter_S}$.  \ \
\end{proof}

Having \Cref{lem:cicgkab-to-sgkab-bisimilar-ts} in hand, in the
following, we show that the verification of \mulcs properties over
\cicgkabs can be recast as verification of \muladom properties over
S-GKABs as follows.

\begin{theorem}\label{thm:cicgkab-to-sgkab}
  Given a \cicgkab $\csgkabsym$ and a closed $\mulcs$ property
  $\Phi$, we have
\begin{center}
  $\ts{\csgkabsym}^{\csfilter_C} \models \Phi$ if and only if
  $ \ts{\tgkabccs(\csgkabsym)}^{\filter_S}\models \tfort(\Phi)$
\end{center}
\end{theorem}
\begin{proof}
  By \Cref{lem:cicgkab-to-sgkab-bisimilar-ts}, we have that
  $\ts{\csgkabsym}^{\csfilter_C} \stbsim
  \ts{\tgkabccs(\csgkabsym)}^{\filter_S}$.
  Hence, by \Cref{lem:stbisimilar-ts-satisfies-same-formula}, we have
  that the claim is proven.
\end{proof}



\subsection{From \eicgkabs into Standard GKABs}

Similar to our reductions in
\Cref{sec:from-bicgkabs-to-sgkabs,sec:from-cicgkabs-to-sgkabs}, to
mimic the evolution of \eicgkabs inside S-GKABs, we first adopt our
approach in \Cref{sec:transf-scsgkabs-to-sgkabs} to simulate the ABox
and context evolution of \eicgkabs within S-GKABs. Then we adopt our
approach in \Cref{EGKABToSGKAB} to simulate the computation of
bold-evolution. In the following we highlight some important aspects
on our transformation from \eicgkabs into S-GKABs:
\begin{compactitem}
\item We can not re-use the evolution action that we use to transform
  E-GKABs into S-GKABs in \Cref{EGKABToSGKAB}. The reason is simply
  because the TBox is changing based on the context. Thus, we require
  the evolution action to operate based on the current context that
  determines the TBox assertions that ``hold'' at a certain
  moment. Therefore, here we introduce the context-sensitive evolution
  action that simulate the bold-evolution computation while also aware
  of the context changing and adapts its computation based on the
  context.
\item Similar to our reduction from E-GKABs into S-GKABs in
  \Cref{EGKABToSGKAB}, we need a mechanism to keep track the newly
  added assertions in order to perform bold-evolution. As mentioned
  earlier, to do that, we use those additional concept/role names that
  has been introduced to keep track the temporary information about
  ABox assertions to be added/deleted. In particular, we use the ABox
  assertions that are made by the added fact marker concept names
  (i.e., those one with superscript $"^a"$).  Additionally, those kind
  of ABox assertions are also useful for checking the consistency of
  the newly added assertions.



\item As before, we also reserve
  a special concept assertion $\tmp$ in order to mark the intermediate
  states.

\end{compactitem}

\subsubsection{Translating \eicgkabs into S-GKABs}

This section is aimed to show how we transform \eicgkabs into
S-GKABs. To open this section, we start by introducing the notion of
duplicated action and duplicated action invocation that is obtained
from context-evolution rules. Basically they are similar to the one in
\Cref{def:action-and-action-invocation-obtained-from-context-evolution-rule},
except that for each newly added concept assertion $N(c)$, we do not
delete the corresponding concept assertion $N^a(c)$ where by $N^a$ is
an added fact marker concept name (similarly for roles).
%
%
%
The purpose of this modification are to keep the information about the
newly added ABox assertions and to enable the possibility to check the
consistency of the update.








\begin{definition}[Duplicated Action and Duplicated Action Invocation Obtained From
  Context-evolution Rule]
\label{def:dup-action-and-action-invocation-obtained-from-context-evolution-rule}
A \sidetext{Duplicated Action and Duplicated Action Invocation
  Obtained From Context-evolution Rule} \emph{duplicated action
  invocation obtained from a context-evolution rule}
$\tup{Q, \ctxe} \mapsto C_{new}$ in $\ctxprocset$, is an action
invocation $\gact{Q'}{\act^d_{\ctx}()}$ where
  \begin{compactenum}
  \item $Q' = Q_\ctxb \wedge q_{\ctxe}$
    where $Q_\ctxb$ is contextually compiled query of $Q$, and
    $q_{\ctxe}$ is the query obtained from the context expression
    $\ctxe$.

  \item $\act^d_\ctx$ is a 0-ary action obtained from
    $\tup{Q, \ctxe} \mapsto C_{new}$ as follows:

  \begin{compactenum}[(a)]

  \item For each $[d_i \mapsto v_j] \in C_{new}$, we have:
    \begin{enumerate}[i.]
    \item $\map{\true}{\add \set{\cdcc_i^{v_j}(\ctxconst), \cdcq_i^{v_j}(\ctxconst)  } }$ in
      $\eff{\act^d_\ctx}$, and 
    \item
      $\map{\true}{\del \set{ \cdcc_i^{v_k}(\ctxconst),
          \cdcq_i^{v_k}(\ctxconst) } }$
      in $\eff{\act^d_\ctx}$ \\ for every $v_k\in\cdom[d_i]$ such that
      $v_k \neq v_j$.
    \end{enumerate}

  \item For each concept name $N \in \voc(\ctbox)$, we have
    \begin{enumerate}[i.]
    \item $\map{N^a(x)}{\add \set{N(x)} }$ in $\eff{\act^d_\ctx}$,
    \item $\map{N^d(x)}{\del \set{N(x), N^d(x)} }$ in
      $\eff{\act^d_\ctx}$.
    \end{enumerate}
    Compare to
    \Cref{def:action-and-action-invocation-obtained-from-context-evolution-rule},
    the different is that in (i) for each newly added ABox assertion
    formed by concept $N$ we do not delete the ABox assertion that is formed
    by $N^a$. As mentioned before, essentially, the assertion formed
    by $N^a$ acts as a marker that marks the newly added assertion and
    we still need their information later.
  
  \item Similarly for the role names, we create the same effect as in
    the step (b) above.
 
  \end{compactenum}
  In this case we say that $\act^d_{\ctx}$ is \emph{a duplicated action obtained
    from the context-evolution rule $\tup{Q, \ctxe} \mapsto C_{new}$}.
\end{compactenum}
\ \
\end{definition}

We now proceed to lift the notion of evolution action in
\Cref{def:evolution-act} into the context-sensitive evolution action
as follows:

\begin{definition}[Context-sensitive Evolution
  Action]\label{def:cs-evol-act}
  Given \sidetext{Context-sensitive Evolution Action} a Contextualized
  TBox $\ctbox$, let $\ctxall{\cdimset}$ be the set of all possible
  context (see \Cref{def:set-all-possible-context}). We define a
  0-ary (i.e., has no action parameters) \emph{context-sensitive
    evolution action $\act^{\ctbox}_e$} over $\ctbox$, where
  $\eff{\act^{\ctbox}_e}$ is the smallest set containing the following
  effects: For each context $\ctx \in \ctxall{\cdimset}$, we have:
  \begin{compactitem}

  \item for each assertion $\funct{R} \in \ctbox^{\ctx}$, we have
    \[
    \map{q_\ctx \wedge \exists z.\qunsatf(\funct{R}, x, y, z) \land R^{a}(x, y)}
    {\del \set{R(x, z)}} \in \eff{\act^{\ctbox}_e},
    \]

  \item for each negative concept inclusion assertion
    $B_1 \ISA \neg B_2$ such that $\ctbox^{\ctx} \models B_1 \ISA \neg B_2$, we
    have
    \[
    \map{q_\ctx \wedge \qunsatn(B_1 \ISA \neg B_2, x) \land B_1^{a}(x)} \del
    \set{B_2(x)} \in \eff{\act^{\ctbox}_e},
    \]

  \item for each negative role inclusion assertion $R_1 \ISA \neg R_2$
    such that $\ctbox^{\ctx} \models R_1 \ISA \neg R_2$, we have:
    \[
    \map{q_\ctx \wedge \qunsatn(R_1 \ISA \neg R_2, x, y) \wedge R_1^{a}(x, y)} {\del
      \set{R_2(x, y)}} \in \eff{\act^{\ctbox}_e},
    \]

  \item for each concept name $N \in \voc(\ctbox)$, we have:
    \[
    \begin{array}{l}
      \map{N^{a}(x)}{\del \set{N^{a}(x)}} \in \eff{\act^{\ctbox}_e},
    \end{array}
    \]
    where $N^{a}$ is the reserved added fact marker concept name for
    $N$.

  \item for each role name $P \in \voc(\ctbox)$, we have:
    \[
    \begin{array}{l}
    \map{P^{a}(x, y)}{\del \set{P^{a}(x, y)}} \in \eff{\act^{\ctbox}_e},
    \end{array}
    \]
    where $P^{a}$ is the reserved added fact marker concept name for
    $P$.

  \item $\map{\true} {\del \set{\tmp} \in \eff{\act^{\ctbox}_e}}.$

  \end{compactitem}
\ \ 
\end{definition}

As the last preliminary before we formally define the translation of
\eicgkabs into S-GKABs, in the following we define the program
translation that will be used to translate the program in \eicgkabs.
Essentially we define a translation function $\tgprogecs$ that
concatenates each action invocation with a program that
non-deterministically choose an action that changes the context, and
then concatenates them with the update consistency checker action, and
also with the context-sensitive evolution action.  Additionally, the
translation function $\tgprogecs$ also serves as a one-to-one
correspondence (bijection) between the original and the translated
program (as well as between the sub-program).

\begin{definition}[Program Translation $\tgprogecs$]\label{def:prog-trans-eicgkab-sgkab}
  Given \sidetextb{Program Translation $\tgprogecs$} an \eicgkabs
  $\csgkabsym = \tup{\ctbox, \initabox, \actset, \ginitprog, \initctx,
    \ctxprocset}$
  we define a \emph{translation $\tgprogecs$} that translates a
  program $\delta$ into a program $\delta'$ inductively as follows:
  \[
\begin{array}{@{}l@{ }l@{ }l@{}}
  \tgprogecs(\gactc{Q(\vec{p})}{\ctxe}{\act(\vec{p})})  &= & 
                                                 \gact{Q'(\vec{p})}{\act'(\vec{p})}
                                                 ;\delta_{\ctxprocset};
                                                     \gact{\neg \qunsatecq{{T_a}}}{\act^{\ctbox}_e()}\\
  \tgprogecs(\gemptyprog) &=& \gemptyprog \\
  \tgprogecs(\delta_1|\delta_2) &=& \tgprogecs(\delta_1)|\tgprogecs(\delta_2) \\
  \tgprogecs(\delta_1;\delta_2) &=& \tgprogecs(\delta_1);\tgprogecs(\delta_2) \\
  \tgprogecs(\gif{\varphi}{\delta_1}{\delta_2}) &=& \gif{\varphi}{\tgprogecs(\delta_1)}{\tgprogecs(\delta_2)} \\
  \tgprogecs(\gwhile{\varphi}{\delta}) &=& \gwhile{\varphi}{\tgprogecs(\delta)}
\end{array}
\]
where 
\begin{compactitem}
\item $\gact{Q'(\vec{p})}{\act'(\vec{p})}$ is a action invocation
  obtained from $\gactc{Q(\vec{p})}{\ctxe}{\act(\vec{p})}$ as in
  \Cref{def:action-invoc-from-cs-action-invoc}.

\item $\delta_{\ctxprocset}$ is a context-change program obtained from
  $\ctxprocset$ as in \Cref{def:ctx-chg-prog}, except that it is
  formed by duplicated action invocation obtained from context
  evolution rule as in
  \Cref{def:dup-action-and-action-invocation-obtained-from-context-evolution-rule}.


\item $\act^{\ctbox}_e$ is a context-sensitive evolution action over
  $\ctbox$ as in \Cref{def:cs-evol-act}.

\item $\qunsatecq{{T_a}}$ is a context-sensitive Q-UNSAT-ECQ over
  $T_a$ (see \Cref{def:cs-q-unsat-ecq}), where $T_a$ is obtained from
  $\ctbox$ by renaming each concept name $N$ in $\ctbox$ into $N^a$
  (similarly for roles). Thus,
%
%
  with this mechanism, we can block any further execution when the
  newly added assertions are
  inconsistent. 

 \end{compactitem}
\ \ 
\end{definition}

Having all of the machinery in hand, we are ready to define a
translation $\tgkabecs$ that, given a \eicgkab, produces an S-GKAB
as follows:

\begin{definition}[Translation from \eicgkab to S-GKAB]
  We \sidetext{Translation from \eicgkab to S-GKAB} define a
  translation $\tgkabecs$ that, given an \eicgkab
  $\csgkabsym = \tup{\ctbox, \initabox, \actset, \ginitprog, \initctx,
    \ctxprocset}$,
  produces an S-GKAB $\tgkabecs(\csgkabsym) =
  \tup{T_\cdimset, 
    \initabox \cup A_{\initctx}, \actset', \ginitprog'}$, where
\begin{compactitem}

\item $T_\cdimset$ is a TBox obtained from a
  set of context dimensions $\cdimset$ (see
  \Cref{def:tbox-from-ctxdim}),


\item 
  $A_{\initctx}$ is an ABox obtained from $\initctx$ (see
  \Cref{def:abox-context}),

\item
  $\actset' = \actset_\act \cup \actset_\ctx \cup
  \set{\act^{\ctbox}_e}$ where:

\begin{compactitem}
\item $\actset_\act$ is obtained from $\actset$ such that for each
  action $\act \in \actset$, we have $\act' \in \actset_\act$ where
  $\act'$ is a delayed action obtained from $\act$ (see
  \Cref{def:delayed-action}),

\item $\actset_\ctx$ is obtained from $\ctxprocset$ such that for each
  context-evolution rule $\tup{Q, \ctxe} \mapsto C_{new}$ in
  $\ctxprocset$, we have $\act_\ctx \in \actset_\ctx$ where
  $\act_\ctx$ is \emph{a duplicated action obtained from the
    context-evolution rule $\tup{Q, \ctxe} \mapsto C_{new}$} (see
  \Cref{def:dup-action-and-action-invocation-obtained-from-context-evolution-rule}),

\item $\act^{\ctbox}_e$ is a context-sensitive evolution action over
  $\ctbox$ (see \Cref{def:cs-evol-act})

\end{compactitem}

\item $\ginitprog' = \tgprogecs(\ginitprog)$. 
\end{compactitem}
\ \ 
\end{definition}

A \mulcs property $\Phi$ over \eicgkabs $\csgkabsym$ can then be
recast as a corresponding property over $\tgkabecs(\csgkabsym)$ by
using the formula translation $\tfort$ (see \Cref{def:ttrip}).
Employing those two translations, later we show that
$\ts{\csgkabsym}^{\csfilter_E} \models \Phi$ if and only if
$\ts{\tgkabecs(\csgkabsym)}^{\filter_S} \models \tfort(\Phi)$. As a
consequence, we have that the verification of \mulcs over \eicgkabs
can be reduced to the corresponding verification of \muladom over
S-GKABs.

\subsubsection{Reducing the Verification of \eicgkabs into S-GKABs}

We now advanced further to show that the verification of \eicgkabs can
be recast into the verification of S-GKABs. Below, we first show that
the transition system of a \eicgkabs $\csgkabsym$ is ST-bisimilar with
the transition system of the corresponding S-GKAB
$\tgkabecs(\csgkabsym)$ that is obtained via translation $\tgkabecs$.

\begin{lemma}\label{lem:eicgkab-to-sgkab-bisimilar-state}
  Let $\csgkabsym$ be an \eicgkab with transition system
  $\ts{\csgkabsym}^{\csfilter_E}$, and let $\tgkabecs(\csgkabsym)$ be
  its corresponding S-GKAB $($with transition system
  $\ts{\tgkabecs(\csgkabsym)}^{\filter_S}$$)$
  obtained through
  $\tgkabecs$. 
  Consider
  \begin{inparaenum}[]
  \item a state $s_{cx} = \tup{A_{cx},\scmap_{cx}, \ctx, \delta_{cx}}$ of
    $\ts{\csgkabsym}^{\csfilter_E}$ and
  \item a state $s_s = \tup{A_s,\scmap_s, \delta_s}$ of
    $\ts{\tgkabecs(\csgkabsym)}^{\filter_S}$.
  \end{inparaenum}
  If
  \begin{inparaenum}[]
  \item $s_{cx} \eqc s_s$, $\scmap_{cx} = \scmap_s$ and
  \item $\delta_s = \tgprogecs(\delta_{cx})$,
  \end{inparaenum}
  then $\tup{A_{cx},\scmap_{cx},
    \ctx, \delta_{cx}} \stbsim \tup{A_s,\scmap_s, \delta_s}$.
\end{lemma}
\begin{proof}
  The proof is similar to the combination of the proof for
  \Cref{lem:scsgkab-to-sgkab-bisimilar-state,lem:egkab-to-sgkab-bisimilar-state},
  by also considering the following:
  \begin{compactenum}
    

  \item The different with the S-GKABs that capture \scsgkabs is that
    after changing the context and materializing the ABox changes,
    instead of executing an action that checks the inconsistency, the
    S-GKABs that capture \eicgkabs execute the evolution action which
    performs bold-evolution computation.

  \item The context-sensitive evolution action is executed after the
    the context has been updated. This is aligned with
    \Cref{def:evol-cs-filter} that \eicgkabs perform the bold
    evolution w.r.t.\ the TBox under the new context. Additionally, as
    it can be seen from the translation $\tgprogecs$ (see
    \Cref{def:prog-trans-eicgkab-sgkab}), before executing the
    evolution action, we also check the consistency of the updates
    w.r.t.\ the TBox under the new context. This guarantees that we
    fulfill the requirement in \Cref{def:evol-cs-filter} that the
    updates must be consistent w.r.t.\ the TBox under the new context.

  \item Similar to \Cref{lem:evol-prop,lem:eact-prop}, we can also
    easily show the correctness of context-sensitive evolution action
    that it performs the bold-evolution computation. The important
    observation is that the context-sensitive evolution action
    performs the bold-evolution based on the context, i.e., it only
    consider those assertions in the TBox that ``hold'' under the
    corresponding context.

  \end{compactenum}

\end{proof}

\begin{lemma}\label{lem:eicgkab-to-sgkab-bisimilar-ts}
  Given an \eicgkab $\csgkabsym$, we have
  $\ts{\csgkabsym}^{\csfilter_E} \stbsim
  \ts{\tgkabecs(\csgkabsym)}^{\filter_S}$
\end{lemma}
\begin{proof}
Let
\begin{compactenum}
\item
  $\csgkabsym = \tup{\ctbox, \initabox, \actset, \ginitprog, \initctx,
    \ctxprocset}$,
  and \\
  $\ts{\csgkabsym}^{\csfilter_E} = \tup{\const, \ctbox, \stateset_\ctxb, s_{0\ctxb},
    \abox_\ctxb, \cntx, \trans_\ctxb}$,
\item   $\tgkabecs(\csgkabsym) = \tup{T', \initabox', \actset',
    \ginitprog'}$ and 
  $\ts{\tgkabecs(\csgkabsym)}^{\filter_S} = \tup{\const, T',
    \stateset_s, s_{0s}, \abox_s, \trans_s}$.
\end{compactenum}
We have that $s_{0\ctxb} = \tup{\initabox, \scmap_c, \ctx_0, \delta}$
and $s_{0s} = \tup{\initabox', \scmap_s, \delta'}$ where
$\scmap_c = \scmap_s = \emptyset$. By the definition of $\tgprogecs$
and $\tgkabecs$, we also have $s_{0\ctxb} \eqc s_{0s}$, and
$\delta' = \tgprogecs(\delta)$. Hence, by
\Cref{lem:cicgkab-to-sgkab-bisimilar-state}, we have
$s_{0\ctxb} \stbsim s_{0s}$. Therefore, by the definition of
ST-bisimulation, we have
$\ts{\csgkabsym}^{\csfilter_E} \stbsim
\ts{\tgkabecs(\csgkabsym)}^{\filter_S}$.  \ \
\end{proof}

We now proceed to show that the verification of \mulcs properties over
\eicgkabs can be recast as verification of \muladom over S-GKABs as
follows.

\begin{theorem}\label{thm:eicgkab-to-sgkab}
  Given an \eicgkab $\csgkabsym$ and a closed $\mulcs$ property
  $\Phi$, we have
\begin{center}
  $\ts{\csgkabsym}^{\csfilter_E} \models \Phi$ if and only if
  $ \ts{\tgkabecs(\csgkabsym)}^{\filter_S}\models \tfort(\Phi)$
\end{center}
\end{theorem}
\begin{proof}
  By \Cref{lem:eicgkab-to-sgkab-bisimilar-ts}, we have that
  $\ts{\csgkabsym}^{\csfilter_E} \stbsim
  \ts{\tgkabecs(\csgkabsym)}^{\filter_S}$.
  Hence, by \Cref{lem:stbisimilar-ts-satisfies-same-formula}, we have
  that the claim is proven.
\end{proof}



\subsection{Bring It All Together: Verification of \icgkabs}

To sum up, we state the result of \icgkabs verification as follows:

\begin{theorem}
\label{thm:ictos}
Verification of \mulcs properties over \icgkabs can be recast as
verification over S-GKABs.
\end{theorem}
\begin{proof}
  As a consequence of
  \Cref{thm:bicgkab-to-sgkab,thm:cicgkab-to-sgkab,thm:eicgkab-to-sgkab},
  we essentially show that the verification of \mulcs properties over
  \icgkabs can be recast as verification over S-GKABs since we can
  recast the verification of \mulcs properties over \bicgkabs,
  \cicgkabs, and \eicgkabs as verification over S-GKABs.
\end{proof}

From \Cref{thm:ictos,thm:gtos}, we get our next result that
verification of inconsistency-aware variants of \csgkabs introduced in
\Cref{sec:icgkabs-execsem} can be compiled into verification of KABs,
by first reducing verification of \icgkabs into verification of
S-GKABs, and then reducing verification of S-GKABs into verification
of KABs.
\begin{theorem}
\label{thm:ictoverys}
Verification of \mulcs properties over \icgkabs can be recast as
verification over KABs.
\end{theorem}
\begin{proof}
  The proof is easily obtained from the \Cref{thm:ictos,thm:gtos},
  since by \Cref{thm:ictos} we can recast the verification of \mulcs
  over \icgkabs as verification over S-GKABs and then by
  \Cref{thm:gtos} we can recast the verification of \muladom over
  S-GKABs as verification over KABs. Thus, combining those two
  ingredients, we can reduce the verification of \mulcs over \icgkabs
  into the corresponding verification of \muladom over KABs.
\end{proof}

\bigskip
\subsection{Verification of Run-bounded \icgkabs}

We now aim to show that the reductions from \icgkabs to S-GKABs
preserve run-boundedness.

\begin{lemma}\label{lem:run-bounded-preservation-bicgkab}
  Let $\csgkabsym$ be a \bicgkab and $\tgkabbcs(\csgkabsym)$ be its
  corresponding S-GKAB. We have $\csgkabsym$ is run-bounded if and
  only if $\tgkabbcs(\csgkabsym)$ is run-bounded.
\end{lemma}
\begin{proof}
  Let
  \begin{compactenum}

  \item
    $\csgkabsym = \tup{\ctbox, \initabox, \actset, \ginitprog,
      \initctx, \ctxprocset}$,
    and 
    $\ts{\csgkabsym}^{\csfilter_B}$ be its transition system,


  \item $\ts{\tgkabbcs(\csgkabsym)}^{\filter_S}$ the transition system
    of $\tgkabbcs(\csgkabsym)$.

  \end{compactenum}
  The proof is easily obtained since
  \begin{itemize}

  \item the translation $\tgkabbcs$ essentially only appends each
    action invocation in $\delta$ with some additional programs to
    handle the context change and manage inconsistency.

  \item the program that manage inconsistency never inject new
    additional constants, but only remove facts causing
    inconsistency,


  \item the program that is used to simulate the context evolution
    does not inject unbounded number of new constants. In fact, we
    only reserve a constant $\ctxconst$ to simulate the context (i.e.,
    to construct the ABox assertions that represent the context
    dimension assignments).

  \item by \Cref{lem:bicgkab-to-sgkab-bisimilar-ts}, we have that
    $\ts{\csgkabsym}^{\csfilter_B} \cjbsim
    \tgkabbcs(\csgkabsym)$.
    Thus, basically they are ``equivalent'' modulo intermediate states
    (states containing $\tmp$) and also by considering that they
    represent context information in a different way (i.e., each two
    bisimilar states are equivalent modulo context ABox assertions).

  \end{itemize}
\end{proof}

\begin{lemma}\label{lem:run-bounded-preservation-cicgkab}
  Let $\csgkabsym$ be a \cicgkab and $\tgkabccs(\csgkabsym)$ be its
  corresponding S-GKAB. We have $\csgkabsym$ is run-bounded if and
  only if $\tgkabccs(\csgkabsym)$ is run-bounded.
\end{lemma}
\begin{proof}
  Similar to the proof of
  Lemma~\ref{lem:run-bounded-preservation-bicgkab}. 
\end{proof}

\begin{lemma}\label{lem:run-bounded-preservation-eicgkab}
  Let $\csgkabsym$ be an \eicgkab and $\tgkabecs(\csgkabsym)$ be its
  corresponding S-GKAB. We have $\csgkabsym$ is run-bounded if and
  only if $\tgkabecs(\csgkabsym)$ is run-bounded.
\end{lemma}
\begin{proof}
  Similar to the proof of
  Lemma~\ref{lem:run-bounded-preservation-bicgkab}. 
\end{proof}

Finally, we show the result on the verification of \mulcs properties
over run-bounded \icgkabs as follows.

\begin{theorem}
  Verification of \mulcs properties over run-bounded \icgkabs is
  decidable, and reducible to standard $\mu$-calculus finite-state
  model checking.
\end{theorem}
\begin{proof}
  By
  \Cref{lem:run-bounded-preservation-bicgkab,lem:run-bounded-preservation-cicgkab,lem:run-bounded-preservation-eicgkab},
  the translation from \icgkabs to S-GKABs preserves run-boundedness.
%
%
  Thus, the claim follows by combining \Cref{thm:ictos} and
  \Cref{thm:ver-run-bounded-sgkab}.
\end{proof}

\section[From Standard to Inconsistency-aware 
  Context-sensitive GKABs]{From Standard GKABs to Inconsistency-aware 
  Context-sensitive GKABs}\label{sec:cap-sgkabs-to-icsgkabs}

We have seen so far that we can transform \icgkabs into S-GKABs and
recast the verification of \icgkabs into S-GKABs.
Now, we show that we can also do the other direction. I.e., we show
that we can recast the verification of S-GKABs into the verification
of \icgkabs. As a consequence, we have that S-GKABs and \icgkabs are
reducible to each other in terms of the verification.

The general strategy to compile S-GKABs into \icgkabs is as follows:
\begin{compactitem}

\item Basically we combine the approach in
  \Cref{sec:cap-sgkab-to-scsgkabs,sec:cap-sgkabs-to-igkabs}.

\item We force the TBox to stay the same for all of the states along
  the system evolution by preventing the context change.

\item We prevent the repair when we encounter an inconsistent state
  and force to reject each action execution that leads to an
  inconsistent state. 

\end{compactitem}
Here we only show how we can reduce the verification of S-GKABs into
\bicgkabs. The reductions from S-GKABs into \cicgkabs and \eicgkabs
are similar.

To realize the strategy above, in the following we fix a set
$\cdimset$ of context dimension containing only a single context
dimension $d$ (i.e., $\cdimset = \set{d}$). Moreover, $d \in \cdimset$
has a tree shaped finite value domain $\tup{\cdom[d],\cover[d]}$ where
$\cdom[d]$ contains only a single value $\topv[d]$ (i.e.,
$\cdom[d] = \topv[d]$).

\clearpage

We now introduce the translation for program in S-GKABs. Particularly,
we define a translation function $\tgprogsic$ that basically 
\begin{compactenum}
\item replaces each action invocation with a context-sensitive action
  invocation in which its context expression always holds in any
  context, and then
\item concatenates it with an action invocation that does the
  inconsistency check.
\end{compactenum}
Additionally, the translation function $\tgprogsic$ also serves as a
one-to-one correspondence (bijection) between the original and the
translated program (as well as between the sub-program).

\begin{definition}[Program Translation $\tgprogsic$]\label{def:prog-trans-tgprogsic}
  Given \sidetextb{Program Translation $\tgprogsic$} a set of actions
  $\actset$, a program $\delta$ over $\actset$, and a TBox $T$, we
  define a \emph{translation $\tgprogsic$} which translates a program
  into a program inductively as follows:
\[
\begin{array}{@{}l@{}l@{}}
  \tgprogsic(\gact{Q(\vec{p})}{\act(\vec{p})}) &=  
                                                 \gactc{Q(\vec{p})}{\ctxe}{\act'(\vec{p})}
                                                 ; \gact{\neg\qunsatecq{T}}{\act_\bot()}\\
  \tgprogsic(\gemptyprog) &= \gemptyprog \\
  \tgprogsic(\delta_1|\delta_2) &= \tgprogsic(\delta_1)|\tgprogsic(\delta_2) \\
  \tgprogsic(\delta_1;\delta_2) &= \tgprogsic(\delta_1);\tgprogsic(\delta_2) \\
  \tgprogsic(\gif{\varphi}{\delta_1}{\delta_2}) &= \gif{\varphi}{\tgprogsic(\delta_1)}{\tgprogsic(\delta_2)} \\
  \tgprogsic(\gwhile{\varphi}{\delta}) &= \gwhile{\varphi}{\tgprogsic(\delta)}
\end{array}
\]
where 
\begin{compactitem}
\item $\qunsatecq{T}$ is a boolean Q-UNSAT-ECQ over $T$ (similar to
  Q-UNSAT-FOL in \Cref{def:qunsat-fol}) that is used to check the
  inconsistency. It will be evaluated to true if the ABox is
  $T$-inconsistent.

\item $\ctxe = \set{\cval{d}{\topv[d]}}$

\item action $\act'(\vec{p})$ is obtained from
  $\act(\vec{p}) \in \actset$, such that
  \[
  \eff{\act'} = \eff{\act} \cup \set{\map{\true}{\add \set{\tmp} } }
  \]

\item $\act_\bot$ is a 0-ary action of the form
  \[
  \act_{\bot}():\set{\map{\true}{\del \set{\tmp} }}
  \]
\end{compactitem}
\ \ 
\end{definition}

We then define the following translation that transform S-GKABs into
\bicgkabs as follows.

\begin{definition}[Translation from S-GKAB to \bicgkab]\label{def:trans-sgkab-bicgkab}
  \ \sidetext{Translation from S-GKAB to \bicgkab}
  We define a translation $\tgkabsic$ that, given an S-GKAB
  $\gkabsym = \tup{T, \initabox, \actset, \ginitprog}$, produces a
  \bicgkab
  $\tgkabsic(\gkabsym) = \tup{\ctbox, \initabox, \actset',
    \ginitprog', \initctx, \ctxprocset}$, where
\begin{compactitem}

\item $\ctbox$ is obtained from $T$ such that for each positive
  inclusion assertion $t \in T$, we have $\tup{t:\varphi}$ where
  $\varphi = \cval{d}{\topv[d]}$,

\item $\actset' = \actset_\act \cup \set{\act_\bot}$ where
  \begin{compactitem}
  \item $\actset_\act$ is obtained from $\actset$ such that for each
    $\act \in \actset$, we have $\act' \in \actset_\act$ where
    $ \eff{\act'} = \eff{\act} \cup \set{\map{\true}{\add \set{\tmp} }
    }$,
  \item $\act_\bot$ is a 0-ary action of the form
    $ \act_{\bot}():\set{\map{\true}{\del \set{\tmp} }} $.
  \end{compactitem}

\item $\ginitprog' = \tgprogsic(\ginitprog)$. 

\item $\initctx = \set{\cval{d}{\topv[d]}}$, 

\item $\ctxprocset = \set{\tup{\true, \cval{d}{\topv[d]}} \mapsto
    \set{\cval{d}{\topv[d]}} }$

\end{compactitem}
\ \ 
\end{definition}

Next, we show that given an S-GKAB $\gkabsym$, and \muladom formula
$\Phi$, we have that $\ts{\gkabsym}^{\filter_S} \models \Phi$ if and
only if $ \ts{\tgkabsic(\gkabsym)}^{\csfilter_B}\models \Phi$. The
strategy is as follows:
\begin{compactenum}

\item Recall the notion of S-Bisimulation in \Cref{sec:s-bsim}. Here
  we use a similar notion of bisimulation except that now the
  bisimulation relation is defined between a KB transition system and
  a context-sensitive transition system. However, the bisimulation
  condition are kept the same. Therefore, for compactness of
  presentation, here we do not redefine a new bisimulation
  relation. All notions related to S-Bisimulation that was introduced
  in \Cref{sec:s-bsim} can be seamlessly recast into this setting.

\item To utilize the S-Bisimulation relation and its properties, in
  the following we show that given an S-GKAB, its transition system is
  S-bisimilar to the transition system of its corresponding \bicgkab
  that is obtained through $\tgkabsic$.
  Then, by using \Cref{lem:sbisimilar-ts-satisfies-same-formula}
  (except that now we consider a KB transition system and a
  context-sensitive transition system) and also by considering that
  \mulcs without context expression is the same as \muladom, we can
  easily recast the verification of S-GKABs into \bicgkabs.

\end{compactenum}

In the following two lemmas we intend to show that given an S-GKAB,
its transition system is S-bisimilar to the transition system of its
corresponding \bicgkab that is obtained through $\tgkabsic$.

\begin{lemma}\label{lem:sgkab-to-bicgkab-bisimilar-state}
  Let $\gkabsym$ be an S-GKAB with transition system
  $\ts{\gkabsym}^{\filter_S}$, and let $\tgkabsic(\gkabsym)$ be its
  corresponding \bicgkab $($with transition system
  $\ts{\tgkabsic(\gkabsym)}^{\csfilter_B}$$)$
  obtained through
  $\tgkabsic$. 
  Consider
  \begin{inparaenum}[]
  \item a state $s_s = \tup{A_s,\scmap_s, \delta_s}$ of
    $\ts{\gkabsym}^{\filter_S}$, and  
%
  \item a state $s_{cx} = \tup{A_{cx},\scmap_{cx}, \ctx, \delta_{cx}}$ of
    $\ts{\tgkabsic(\gkabsym)}^{\csfilter_B}$.
  \end{inparaenum}
  If
  \begin{inparaenum}[]
  \item $A_{cx} = A_s$,
  \item $\scmap_{cx} = \scmap_s$, and
  \item $\delta_{cx} = \tgprogsic(\delta_{s})$,
  \end{inparaenum}
  then
  $\tup{A_{cx},\scmap_{cx}, \ctx, \delta_{cx}} \sbsim
  \tup{A_s,\scmap_s, \delta_s}$.
\end{lemma}
\begin{proof}
  Now, let
  \begin{compactenum}
  \item $\gkabsym = \tup{T, \initabox, \actset, \ginitprog}$ and
    $\ts{\gkabsym}^{\filter_S}
    = \tup{\const, T, \stateset_s, s_{0s}, \abox_s, \trans_s}$.
  \item $\tgkabsic(\gkabsym)
    = \tup{\ctbox, \initabox, \actset', \ginitprog', \initctx,
      \ctxprocset}$,
    and \\
    $\ts{\tgkabsic(\gkabsym)}^{\csfilter_S}
    = \tup{\const, \ctbox, \stateset_\ctxb, s_{0\ctxb}, \abox_\ctxb,
      \cntx, \trans_\ctxb}$,
\end{compactenum}
Now, we have to show the following: For every state $
\tup{A'_s,\scmap'_s, \delta'_s}$ such that $\tup{A_s,\scmap_s, \delta_s}
\trans_s \tup{A'_s,\scmap'_s, \delta'_s} $
there exists
$\tup{A'_{cx},\scmap'_{cx}, \ctx', \delta'_{cx}}$ such that
\begin{compactenum}[\bf (a)]
\item we have $
  \tup{A_{cx},\scmap_{cx}, \ctx, \delta_{cx}} \trans
  \tup{A'_{cx},\scmap'_{cx}, \ctx', \delta'_{cx}}, $
\item $A'_s = A'_{cx}$
\item $\scmap'_s = \scmap'_{cx}$;
\item $\delta'_{cx} = \tgprogsic(\delta'_{s})$.
\end{compactenum}
The proof can be easily obtained by considering that within \bicgkab,
by the definition of $\tgkabsic$
(see \Cref{def:trans-sgkab-bicgkab}), it is easy to see that the
following hold:
\begin{compactitem}

\item The initial context is $\initctx
  =
  \set{\cval{d}{\topv[d]}}$ and basically this is the only one
  possible context in the system. Furthermore, the context stays the
  same along the system evolution because we only have a single
  context evolution rule $\tup{\true,
    \cval{d}{\topv[d]}} \mapsto
  \set{\cval{d}{\topv[d]}}$ that never change the context.  Hence, the
  TBox stay the same for all states. Furthermore, since for each
  $\tup{t:\varphi}
  \in \ctbox$ we have $\varphi =
  \cval{d}{\topv[d]}$, it is easy to see that all of the TBox
  assertions hold in our only one possible context. As a consequence,
  essentially the situation of the TBox is the same as in the original
  S-GKAB.

\item We basically can ignore the context expression that guards each
  context-sensitive action invocation in $\delta'$ because it is alway
  be satisfied in any case. Hence, each context-sensitive action
  invocation in $\delta'$ is essentially the same as the usual action
  invocation in $\delta$.



\item Each state in the transition system of \bicgkab is always
  consistent, because we only keep the positive inclusion assertions
  when we translate an S-GKAB into a \bicgkab. Thus, the repair
  mechanism in B-GKAB will not change anything.

\item We transform each action invocation in the program in the given
  S-GKAB such that it will always be followed by the action invocation
  $\gact{\neg\qunsatecq{T}}{\act_\bot()}$
  where $\qunsatecq{T}$ will be evaluated to true when the
  corresponding ABox is $T$-inconsistent.
  Hence, the inconsistency check is basically delegated to the
  evaluation of the query that acts as the guard of the action
  $\act_\bot$ and it is triggered after each action execution.  The
  action $\act_\bot$ will not be executed if the previous action
  execution leads into an inconsistent state w.r.t.\ the TBox8
  $T$. 
  Thus, it is easy to see that when an action execution in S-GKAB is
  blocked because it leads into a $T$-inconsistent state, then the
  corresponding action execution in \bicgkab will not lead into a new
  state without $\tmp$ as well. However, when an execution in S-GKAB
  leads into a new $T$-consistent state, the corresponding action
  execution in \bicgkab will be followed by the execution of
  $\act_\bot$ and it leads into a new state without $\tmp$.

\end{compactitem}

\end{proof}

\begin{lemma}\label{lem:sgkab-to-bicgkab-bisimilar-ts}
  Given an S-GKAB $\gkabsym$, we have
  $\ts{\gkabsym}^{\filter_S} \sbsim
  \ts{\tgkabsic(\gkabsym)}^{\csfilter_B}$
\end{lemma}
\begin{proof}
Let
\begin{compactenum}
\item $\gkabsym = \tup{T, \initabox, \actset, \ginitprog}$ and
  $\ts{\gkabsym}^{\filter_S} = \tup{\const, T,
    \stateset_s, s_{0s}, \abox_s, \trans_s}$.
\item
  $\tgkabsic(\gkabsym) = \tup{\ctbox, \initabox, \actset', \ginitprog',
    \initctx, \ctxprocset}$,
  and \\
  $\ts{\tgkabsic(\gkabsym)}^{\csfilter_S} = \tup{\const, \ctbox,
    \stateset_\ctxb, s_{0\ctxb}, \abox_\ctxb, \cntx, \trans_\ctxb}$,
\end{compactenum}
We have that $s_{0s} = \tup{\initabox, \scmap_s, \delta}$ and
$s_{0\ctxb} = \tup{\initabox, \scmap_{cx}, \ctx_0, \delta'}$ where
$\scmap_s = \scmap_{cx} = \emptyset$. By the definition of
$\tgprogsic$ and $\tgkabsic$, we also have that their initial ABoxes
are the same, and $\delta' = \tgprogsic(\delta)$. Hence, by
\Cref{lem:sgkab-to-bicgkab-bisimilar-state}, we have
$s_{0s} \sbsim s_{0\ctxb}$. Therefore, by the definition of
S-bisimulation, we have
$\ts{\gkabsym}^{\filter_S} \sbsim
\ts{\tgkabsic(\gkabsym)}^{\csfilter_B}$.
\end{proof}

We close this section by showing that the verification of \muladom
properties over S-GKAB can be recast as verification over \bicgkab as
follows.

\begin{theorem}
  \label{thm:ver-sgkab-to-bicgkab}
Verification of closed \muladom properties over S-GKABs can be recast
as verification over \bicgkabs.
\end{theorem}
\begin{proof}
  By \Cref{lem:sgkab-to-bicgkab-bisimilar-ts}, we have that
  $\ts{\gkabsym}^{\filter_S} \sbsim
  \ts{\tgkabsic(\gkabsym)}^{\csfilter_B}$.
  Hence, by \Cref{lem:sbisimilar-ts-satisfies-same-formula} (but
  consider that now it is between a KB transition system and a
  context-sensitive transition system), for every $\muladom$ property
  $\Phi$, we have that
  \[
  \ts{\gkabsym}^{\filter_S} \models \Phi \textrm{ if and only if }
  \ts{\tgkabsic(\gkabsym)}^{\csfilter_B}\models \Phi
  \]
  Hence, by using the translation $\tgkabsic$ we can easily transform
  an S-GKAB into a \bicgkab and then the claim is easily follows due
  to the fact above. 
\end{proof}

%% file: 2.chapters/8-a-gkab.tex
\chapter[Alternating Golog-KAB\lowercase{s} (\agkabs)]{Alternating Golog-KAB\lowercase{s}}\label{ch:a-gkab}

\ifhidecontent
 
\fi

In \Cref{ch:cs-ia-gkab}, we have seen the \icgkabs which essentially
capture the manipulation of knowledge bases by actions while also
taking into account the contextual information and having a mechanism
to handle inconsistency. Additionally, we have also seen how the
problem of verifying sophisticated temporal properties over \icgkabs
can be tackled. Basically, we solve that problem by reducing it into
the problem of verifying temporal properties over S-GKABs which has
been discussed in \Cref{ch:gkab}. 

In \icgkabs, all computations of the successor states are encapsulated
in a single transition
that 
consists of:
\begin{compactenum}
\item the action execution which changes the ABox (add or delete
  facts) which might involve service calls,
\item the context changes, and
\item the inconsistency handling mechanism.
\end{compactenum}
As one might observed, there are several non-determinism sources while
computing the successor states of a state that cause several
non-deterministic transition from a state to another states. Those
sources of non-determinism are:
\begin{compactenum}
\item the choice of grounded actions (that is caused by different
  action parameters or when the actions are wrapped using
  non-deterministic choice program construct), 
\item The choice of service call results,
\item The choice among all possible new contexts, and
\item The choice of repaired ABoxes when there are several possible
  repairs (the case of b-repairs).
\end{compactenum}
By wrapping all of those non-determinism sources into a single
transition, we basically lose the capability to ``quantify'' over each
source of non-determinism above. Thus we lose the ability to do a
fine-grained analysis over the detail structure of \icgkabs transition
system. We can not check some properties such as ``\textit{no matter
  which action is executed, there exists a service call result in
  which no matter how the context is changing, there exists a repair
  that leads us into a certain state that satisfy a certain
  property}''.  To cope with this situation, here we introduce the so
called \emph{Alternating Golog-KABs} (\agkabs).

The important aspect of \agkabs is that they separate each source of
non-determinism. I.e., \agkabs explicitly present the alternation
among all of those sources of non-determinism where each source of
non-determinism is captured by a single transition. Thus, \agkabs give
us more fine-grained transition system. 
%
%
Furthermore, we can also ``quantify'' over each non-determinism
source. \agkabs can also be considered as a model of four player game
where the players responding to other players move.

In this chapter we also tackle the problem of verifying a
sophisticated temporal properties based on $\mu$-calculus over
\agkabs. We solved the problem by reducing it into the problem of
verifying \mula over S-GKABs.
  
In the following we use \dllitea for expressing KBs and we also do not
distinguish between objects and values (thus we drop attributes).
Moreover we make use of a countably infinite set $\const$ of
constants, which intuitively denotes all possible values in the
system.
Additionally, we consider a finite set of distinguished constants
$\iconst \subset \const$, and
a finite set $\servcall$ of \textit{function symbols} that represents
\textit{service calls}, which abstractly account for the injection of
fresh values (constants) from $\const$ into the system.
Additionally, for technical development of this chapter, we fix a set
$ \cdimset = \{d_1,\ldots,d_n\} $ of context dimensions. Each context
dimension $d_i \in \cdimset$ has its own tree-shaped finite value
domain $\tup{\cdom[d_i],\cover[d_i]}$, where $\cdom[d_i]$ represents
the finite set of domain values, and $\cover[d_i]$ represents the
predecessor relation forming the tree.

\section{\agkabs Formalism and Execution Semantics}

The formalism of \agkabs is similar to \csgkabs. We basically
formalize \agkabs as a tuple
$\agkabsym = \tup{\ctbox, \initabox, \actset, \ginitprog, \initctx,
  \ctxprocset}$
where $\ctbox, \initabox, \actset, \ginitprog, \initctx, \ctxprocset$
are the same as in \csgkabs (see \Cref{def:csgkabs-formalism}). In the
following, we proceed to define the execution semantics for \agkabs.


To define the execution semantics for \agkabs, we adopt the parametric
execution semantics of GKABs (see \Cref{ch:gkab}) in order to be able
to elegantly accommodate various inconsistency management mechanism.
Similar to \csgkabs, the execution semantics of a \agkabs
$\agkabsym = \tup{\ctbox, \initabox, \actset, \ginitprog, \initctx,
  \ctxprocset}$
is given in terms of a possibly infinite-state context-sensitive
transition system
$\ts{\agkabsym} = \tup{\const, \ctbox, \stateset, s_0, \abox, \cntx,
  \trans}$
(see \Cref{def:cs-trans-sys} for the detail of $\ts{\agkabsym}$
components).
However, differently from \csgkabs, the states we consider are tuples
of the form $\tup{id, A,\scmap, \ctx,\delta}$, where $id$ is a unique
identifier for the state, $A$ is an ABox, $\scmap$ is a service call map,
$\ctx$ is a context and $\delta$ is a program.
Furthermore, we distinguish the types of states in the transition
systems of \agkabs, namely \emph{stable} and \emph{intermediate}
states. Technically, we partition the set $\stateset$ of the states in
$\ts{\agkabsym}$ into the set $\ststateset$ of stable states and the
set $\imstateset$ of intermediate states (I.e.,
$\stateset = \ststateset \uplus \imstateset$).  \sidetext{Stable and
  Intermediate States} Those states in $\ststateset$ are called
\emph{stable states}, while those in $\imstateset$ are called \emph{
  intermediate states}.

Before we introduce the construction of \agkabs transition systems, 
%
%
in the following we define the notion of \agkabs program execution
relation by refining the notion of context-sensitive program execution
relation that has been defined in \Cref{def:cs-prog-exec-relation}.

\begin{definition}[\agkab Program Execution Relation]\label{def:agkab-prog-exec-relation}
  Given \sidetext{\agkab Program Execution Relation} an \agkab
  $\agkabsym = \tup{\ctbox, \initabox, \actset, \ginitprog, \initctx,
    \ctxprocset}$,
  and a context-sensitive filter relation $\csfilter$ (see
  \Cref{def:cs-filter-rel}), we define an \emph{\agkab program
    execution relation} $\gprogtrans{\act\sigma, \csfilter}$ as
  follows:
\begin{compactenum}
\item
  $\tup{A, \scmap, \ctx, \gactc{Q(\vec{p})}{\ctxe}{\act(\vec{p})} }
  \gprogtrans{\act\sigma, \csfilter} \tup{A, \scmap, \ctx,
    \gemptyprog}$, \\if the following hold:
  \begin{compactenum}


  \item $\sigma$ is a legal parameter assignment for $\act$ in $A$
    w.r.t.\ context $\ctx$ and action invocation
    $\gactc{Q(\vec{p})}{\ctxe}{\act(\vec{p})}$, 
  \item $\ctx \cup \ctxth \models \ctxe$.
  \end{compactenum}

%
\item $\tup{A, \scmap, \ctx, \delta_1|\delta_2} \gprogtrans{\act\sigma,
    \csfilter} \tup{A, \scmap, \ctx, \delta'}$, \\if $\tup{A, \scmap, \ctx,
    \delta_1} \!\gprogtrans{\act\sigma, \csfilter}\! \tup{A, \scmap, \ctx,
    \delta'}$ or $\tup{A, \scmap, \delta_2} \gprogtrans{\act\sigma,
    \csfilter} \tup{A, \scmap, \ctx, \delta'}$;
\item
  $\tup{A, \scmap, \ctx, \delta_1;\delta_2} \gprogtrans{\act\sigma,
    \csfilter} \tup{A, \scmap, \ctx, \delta_1';\delta_2}$,
  \\if
  $\tup{A, \scmap, \ctx, \delta_1} \gprogtrans{\act\sigma, \csfilter}
  \tup{A, \scmap, \ctx, \delta_1'}$;
\item
  $\tup{A, \scmap, \ctx, \delta_1;\delta_2} \gprogtrans{\act\sigma, \csfilter}
  \tup{A, \scmap, \ctx, \delta_2'}$, \\
  if $\final{\tup{A, \scmap, \ctx, \delta_1}}$, and
  $\tup{A, \scmap, \ctx, \delta_2} \gprogtrans{\act\sigma, \csfilter} \tup{A,
    \scmap, \ctx, \delta_2'}$;
\item
  $\tup{A, \scmap, \ctx, \gif{\varphi}{\delta_1}{\delta_2}}
  \gprogtrans{\act\sigma, \csfilter}
  \tup{A, \scmap, \ctx, \delta_1'}$, \\
  if $\ask(\varphi, T, A) = \true$, and
  $\tup{A, \scmap, \ctx, \delta_1} \gprogtrans{\act\sigma, \csfilter}
  \tup{A, \scmap, \ctx, \delta_1'}$;
\item $\tup{A, \scmap, \ctx, \gif{\varphi}{\delta_1}{\delta_2}} \gprogtrans{\act\sigma, \csfilter} \tup{A, \scmap, \ctx, \delta_2'}$,\\
  if $\ask(\varphi, T, A) = \false$, and
  $\tup{A, \scmap, \ctx, \delta_2} \gprogtrans{\act\sigma, \csfilter}
  \tup{A, \scmap, \ctx, \delta_2'}$;
\item $\tup{A, \scmap, \ctx, \gwhile{\varphi}{\delta}}
  \gprogtrans{\act\sigma, \csfilter} \tup{A, \scmap, \ctx, \delta';\gwhile{\varphi}{\delta}}$,\\
  if $\ask(\varphi, T, A) = \true$, and
  $\tup{A, \scmap, \ctx, \delta} \gprogtrans{\act\sigma, \csfilter}
  \tup{A, \scmap, \ctx, \delta'}$.
\end{compactenum}
\ \ 
\end{definition}

Now, we proceed to define the construction of \agkabs transition
systems that is parameterized with the filter relation
$\csfilter$ as follows. 
\begin{definition}[\agkabs Transition System]\label{def:agkabs-ts}
  Given \sidetext{\agkabs Transition System} an \agkab
  $\agkabsym = \tup{\ctbox, \initabox, \actset, \ginitprog, \initctx,
    \ctxprocset}$,
  and a context-sensitive filter relation $\csfilter$ (see
  \Cref{def:cs-filter-rel}), we define the \emph{transition system of
    $\agkabsym$ w.r.t.~$\csfilter$}, written
  $\ts{\agkabsym}^{\csfilter}$, as
  $\tup{\const, \ctbox, \stateset, s_0, \abox, \cntx, \trans}$, where
  \begin{compactitem}
  \item $s_0 = \tup{id_0, \initabox, \emptyset, \ctx_0, \ginitprog}$,
  \item $\stateset = \ststateset \uplus \imstateset$, and
  \item $\ststateset$, $\imstateset$ and $\trans$ are defined by
    simultaneous induction as the smallest sets such that
    \begin{compactitem}
    \item $s_0 \in \ststateset$, and
    \item if $\tup{id, A, \scmap, \ctx, \delta} \in \ststateset$  then
      we do the following
      \begin{compactenum}[(1)]
      \item for any action $\act$, and any substitution $\sigma$, let
        \[
        \begin{array}{@{}ll@{}} S^1 = \set{\tup{id^1, A, \scmap, \ctx,
              \delta'} ~\mid~& id^1 \mbox{ is a fresh
                               identifier, and } \\
                             &\tup{A, \scmap, \ctx, \delta}
                               \gprogtrans{\act\sigma, \csfilter}
                               \tup{A, \scmap, \ctx, \delta'} }
        \end{array}
        \]
      \item and then for each
        $s_1 = \tup{id^1, A, \scmap, \ctx, \delta'} \in S^1$ with\\
        $\tup{A, \scmap, \ctx, \delta} \gprogtrans{\act\sigma,
          \csfilter} \tup{A, \scmap, \ctx, \delta'}$,
        we do the following:
        \begin{compactenum}[(a)]
        \item let
        \[
        \begin{array}{@{}ll@{}} S^2 = \set{\tup{id^2, A, \scmap',
              \ctx, \delta'} ~\mid~&
                                     id^2 \mbox{ is a fresh identifier, } \\
                                   &\theta \in
                                     \eval{\addfacts{\ctbox^{\ctx}, A,
                                     \act\sigma}}, \\
                                   &\mbox{and }\scmap' = \scmap \cup
                                     \theta }
        \end{array}
        \]
      \item and then for each
        $s_2 = \tup{id^2, A, \scmap', \ctx, \delta'} \in S^2$ with\\
        $\scmap' = \scmap \cup \theta$, we do the following:
          \begin{compactenum}[(i)]
          \item let
            \[
            \begin{array}{@{}ll@{}} S^3 = \set{\tup{id^3, A, \scmap',
                  \ctx', \delta'} ~\mid~& id^3 \mbox{ is a
                                          fresh identifier, and } \\
                                        &\tup{A, \ctx, \ctx'} \in
                                          \ctxchg }
            \end{array}
            \]
          \item and then for each
            $s_3 = \tup{id^3, A, \scmap', \ctx', \delta'} \in S^3$
            with $\tup{A, \ctx, \ctx'} \in \ctxchg$, we do the
            following: \\ if there exists $A'$ such that
              \[ 
              \tup{A, \addfacts{\ctbox^{\ctx}, A, \act\sigma}\theta,
                \delfacts{\ctbox^{\ctx}, A, \act\sigma}, \ctx', A'}
              \in \csfilter,
              \]
              and $A'$ is $\ctbox^{\ctx'}$-consistent.  Then for each
              $A'$, we have
              \begin{compactitem}
              \item if there exists
                $\tup{id_s, A_s, \scmap_s, \ctx_s,
                  \delta_s}\in\ststateset$
                such that \\ $A_s = A'$, $\scmap_s = \scmap'$,
                $\ctx_s = \ctx'$, $\delta_s = \delta'$, then 
                \[
                s_3 \trans \tup{id_s, A_s, \scmap_s, \ctx_s,
                  \delta_s},
                \]
              \item otherwise we have
                $s_4 = \tup{id^4, A', \scmap', \ctx',
                  \delta'}\in\ststateset$,
                and $s_3 \trans s_4$, where $id^4$ is a fresh
                identifier.
            \end{compactitem}

            Additionally, we also have that $s_1\in\imstateset$,
            $s_2\in\imstateset$, $s_3\in\imstateset$, as well as
            $ \tup{id, A, \scmap, \ctx, \delta}\trans s_1$,
            $s_1\trans s_2$, and $s_2\trans s_3$.
 
          \end{compactenum}
        \end{compactenum}
      \end{compactenum}

    \end{compactitem}
  \end{compactitem}
 \ \ 
\end{definition}

\noindent
Notice that we never add the states and the transitions that do not
lead into a stable state. We use $id$ in each state in order to
enforce that the intermediate states between two stable states are
unique. 
This is important in order to prevent false reachability among two
stable states. This false reachability could be happened if we do not
distinguish two intermediate states that
could lead into two different stable states due to the fact that they
came from different stable states (Notice that the computation of
transition within intermediate states might require some information
from its predecessor states). 



Similar to \Cref{sec:icgkabs-execsem}, by exploiting the filter
relations $\csfilter_B$, $\csfilter_C$, and $\csfilter_E$ (see
\Cref{def:b-repair-cs-filter,def:c-repair-cs-filter,def:evol-cs-filter}),
we could obtain various execution semantics for \agkabs with different
mechanism to handle inconsistencies.
%
%
%
%
Now, employing the b-repair context-sensitive filter $\csfilter_B$
into \agkabs gives us \bagkabs that is an \agkabs with \emph{b-repair
  execution semantics}, i.e., where inconsistent ABoxes are repaired
by non-deterministically picking a b-repair. Formally, we define the
transition system which provide the b-repair execution semantics for
\agkabs as follows.

\begin{definition}[\agkab B-Transition System]
  Given \sidetext{\agkab B-Transition System} an \agkab $\agkabsym$
  and a b-repair filter $\csfilter_B$, the \emph{b-transition system
    of $\agkabsym$}, written $\ts{\agkabsym}^{\csfilter_B}$, is the
  transition system of $\agkabsym$ w.r.t.\ $\csfilter_B$ (see also
  \Cref{def:agkabs-ts}).
\end{definition}

\noindent
We call \emph{\bagkabs}the \agkabs adopting this semantics.

By utilizing the c-repair context-sensitive filter $\csfilter_C$, we obtain the
\emph{c-repair execution semantics} for \agkabs, where inconsistent
ABoxes are repaired by computing their unique c-repair.
The transition systems which provide the c-repair execution semantics
for \agkabs is then defined as follows.

\begin{definition}[\agkab C-Transition System]
  Given \sidetext{\agkab C-Transition System} an \agkab $\agkabsym$
  and a c-repair filter $\csfilter_C$, the \emph{c-transition system
    of $\agkabsym$}, written $\ts{\agkabsym}^{\csfilter_C}$, is the
  transition system of $\agkabsym$ w.r.t.\ $\csfilter_C$ (see also
  \Cref{def:agkabs-ts}).
\end{definition}

\noindent
We call \emph{\cagkabs}the \agkabs adopting this semantics. 

We now proceed to incorporate the b-evol context-sensitive filter
$\csfilter_E$ into \agkabs that leads us into the \emph{b-evol
  execution semantics} for \agkabs, where for updates leading to
inconsistent ABoxes, their unique bold-evolution is computed.
Basically, we define the transition systems which provide the b-evol
execution semantics for \agkabs as follows.

\begin{definition}[\agkab E-Transition System]
  Given \sidetext{\agkab E-Transition System} an \agkab $\agkabsym$
  and an e-repair filter $\csfilter_E$, the \emph{e-transition system
    of $\agkabsym$}, written $\ts{\agkabsym}^{\csfilter_E}$, is the
  transition system of $\agkabsym$ w.r.t.\ $\csfilter_E$.
\end{definition}
\noindent
We call \emph{\eagkabs}the \agkabs adopting this semantics.
%
%
Notice that by employing $\csfilter_E$ in the transition system of
\agkabs, we basically assume that the new ABox assertions are
consistent with the TBox under the new context (i.e., after the
context change). This means that we d etermine whether the updates are
applicable or not based on the new context. 
%
%
%


\begin{example}
  Recall our simple order processing scenario in
  \Cref{ex:context-formalization}. Consider the same context
  dimensions as in \Cref{ex:context-formalization}. To model such
  scenario we specify an \agkab
  $\agkabsym = \tup{\ctbox, \initabox, \actset, \ginitprog, \initctx,
    \ctxprocset}$
  where $\ctbox$, $\initabox$, $\actset$, $\ginitprog$, $\initctx$,
  and $\ctxprocset$ are the same as in \Cref{ex:csgkab}. 
  To give the intuition on how \agkabs are executed, here we provide
  an example of \bagkabs execution. Now, let $\agkabsym$ be a
  \bagkab. The initial state of $\ts{\agkabsym}^{\filter_B}$ is
  $s_0 = \tup{id_0, \initabox, \scmap_0, \initctx, \ginitprog}$, where
  \begin{compactitem}
  \item $id_0$ is the unique identifier for this state,
  \item
    $\initabox = \set{\exo{ReceivedOrder}(\excon{chair}),
      \exo{ApprovedOrder}(\excon{table})}$,
  \item $\scmap_0 = \emptyset$,
  \item
    $\initctx = \set{\cval{\exc{PP}}{\exv{N}},
      \cval{\exc{S}}{\exv{NS}}}$,
  \item
    $\ginitprog = \gwhile{ \exists \exvar{x}.[\exo{Order}(\exvar{x})]
      \wedge \neg[\exo{DeliveredOrder}(\exvar{x})] }{\delta_0}$ with
    \begin{compactenum}[1)]
    \item
      $\delta_0 = \delta_1 ; \delta_2 ; \delta_3 ; \delta_4 ; \delta_5
      $
    \item $\delta_1 = \gif{ \neg [\exists
        \exvar{x}.\exo{ApprovedOrder}(\exvar{x})] \\
        \hspace*{8mm}}{
        \gactc{\exo{ReceivedOrder}(\exvar{x})}{\cval{\exc{PP}}{\exv{AP}}
          \wedge
          \cval{\exc{S}}{\exv{AS}}}{\exa{approveOrder}(\exvar{x})} \\
        \hspace*{8mm}}{\gemptyprog}$,
    \item
      $\delta_2 = \gactc{\true}{\cval{\exc{PP}}{\exv{AP}} \wedge
        \cval{\exc{S}}{\exv{AS}}}{\exa{prepareOrders}()}$,
    \item
      $\delta_3 = \gactc{\true}{\cval{\exc{PP}}{\exv{AP}} \wedge
        \cval{\exc{S}}{\exv{AS}}}{\exa{assembleOrders}()}$,
    \item
      $\delta_4 = \gactc{\true}{ \neg ( \cval{\exc{PP}}{\exv{RE}} \vee
        \cval{\exc{S}}{\exv{PS}} ) }{\exa{checkAssembledOrders}()} \ | \\
      \hspace*{9mm}\gactc{\true}{ \cval{\exc{PP}}{\exv{RE}} \vee
        \cval{\exc{S}}{\exv{PS}} }{\exa{outsourceQualityCheck}()}$,
    \item
      $\delta_5 = \gactc{\true}{\cval{\exc{PP}}{\exv{AP}} \wedge
        \cval{\exc{S}}{\exv{AS}}}{\exa{deliverOrder}()}$.
    \end{compactenum}
  \end{compactitem}
  A snapshot of one possible run in $\ts{\agkabsym}^{\filter_B}$ is as follows:
  \[
  s_0 \trans s_1 \trans s_2 \trans s_3 \trans s_4 \trans \cdots
  \]
  where
  \begin{compactitem}

  \item
    $s_1 = \tup{id_1, \initabox, \scmap_0, \initctx, \ginitprog'}$,
    $s_2 = \tup{id_2, \initabox, \scmap', \initctx, \ginitprog'}$,
    $s_3 = \tup{id_3, \initabox, \scmap', \ctx', \ginitprog'}$, and 
    $s_4 = \tup{id_4, A', \scmap', \ctx', \ginitprog'}$,

  \item $id_1$, $id_2$, $id_3$ and $id_4$ are fresh unique
    identifiers,
   
  \item  
    $\ginitprog' = \delta_3 ; \delta_4 ; \delta_5 ; \gwhile{ \exists
      \exvar{x}.[\exo{Order}(\exvar{x})] \wedge
      \neg[\exo{DeliveredOrder}(\exvar{x})] }{\delta_0}$

  \item
    $\scmap' = \{ [\exs{getDesigner}(\excon{table}) \ra
    \excon{alice}], [\exs{getDesign}(\excon{table})
    \ra \excon{ecodesign}], $ \\
    \hspace*{11mm} $[\exs{assignAssemblingLoc}(\excon{table}) \ra \excon{bolzano}]$ \}.


  \item
    $\ctx' = \set{\cval{\exc{PP}}{\exv{WE}},
      \cval{\exc{S}}{\exv{PS}}}$ 
    (obtained from the application of context-evolution rule
    $ \carulex{\true}{\cval{\exc{PP}}{\exv{N}} \wedge
      \cval{\exc{S}}{\exv{NS}} }{ \cval{\exc{PP}}{\exv{WE}},
      \cval{\exc{S}}{\exv{PS}} }$).

  \item
    $A' = \{ \hspace*{1mm} \exo{ReceivedOrder}( \excon{chair} ),
    \exo{ApprovedOrder}( \excon{table} ),
    \exo{designedBy}( \excon{table}, \excon{alice} ), $\\
   \hspace*{11mm} $\exo{Designer}( \excon{alice} ), \exo{hasDesign}( \excon{table} ,
    \excon{ecodesign} ),$\\
   \hspace*{11mm} $ \exo{hasAssemblingLoc}( \excon{table}, \excon{bolzano} ) \}$

 \end{compactitem}
\end{example}

\section{Verification of \agkabs}

The interesting task on \agkabs is to verify whether the evolution of
\agkabs satisfy some temporal properties. In this section, we explain
the temporal properties formalisms, namely alternating
context-sensitive temporal logic \mulcsa, that will be used to specify
the temporal properties to be verified over \agkabs. Moreover, we also
exhibit the way how we solve the verification problem.
The problem definition of the \mulcsa formula verification over
\bagkabs, \cagkabs, and \eagkabs is defined similarly as in \csgkabs
(see \Cref{def:verification-csgkab}).
Here we solve this problem by compiling them into S-GKABs and show
that the verification of \mulcsa formulas over \bagkabs, \cagkabs, and
\eagkabs can be recast as verification of \muladom over S-GKABs. 




As the verification formalisms, we here we introduce the alternating
context-sensitive temporal logic \mulcsa, which is a fragment of
\mulcs. The syntax of \mulcsa is then defined as
follows: 
\sidetext{\ \\ Syntax of \mulcsa} \\
\hspace*{5mm}$
  \Phi ~:=~ Q 
              ~\mid~ \ctxe 
              ~\mid~ \lnot \Phi 
              ~\mid~ \Phi_1 \lor \Phi_2
              ~\mid~ \exists x.\Phi 
              ~\mid~ Z
              ~\mid~ \mu Z.\Phi 
              ~\mid~ 
$\\
\hspace*{16mm}$
        \DIAM{\DIAM{\DIAM{\DIAM{\Phi}}}} ~\mid~
          \DIAM{\DIAM{\DIAM{\BOX{\Phi}}}} ~\mid~
          \DIAM{\DIAM{\BOX{\DIAM{\Phi}}}} ~\mid~
          \DIAM{\DIAM{\BOX{\BOX{\Phi}}}} ~\mid~ 
$\\
\hspace*{16mm}$
        \DIAM{\BOX{\DIAM{\DIAM{\Phi}}}} ~\mid~
          \DIAM{\BOX{\DIAM{\BOX{\Phi}}}} ~\mid~
          \DIAM{\BOX{\BOX{\DIAM{\Phi}}}} ~\mid~
          \DIAM{\BOX{\BOX{\BOX{\Phi}}}} 
$


Essentially, \mulcsa is a fragment of \mulcs that is obtained by
quadruplicating the modal operator so that we can quantify over all
possible sources of non-determinism. Besides the standard
abbreviation, we also use the following abbreviation:
\[
\begin{array}{ll}
  \BOX{\BOX{\BOX{\BOX{\Phi}}}} = \neg\DIAM{\DIAM{\DIAM{\DIAM{\neg\Phi}}}},
  &\BOX{\BOX{\BOX{\DIAM{\Phi}}}} = \neg\DIAM{\DIAM{\DIAM{\BOX{\neg\Phi}}}}, \\
  \BOX{\BOX{\DIAM{\BOX{\Phi}}}} = \neg\DIAM{\DIAM{\BOX{\DIAM{\neg\Phi}}}},
  &\BOX{\BOX{\DIAM{\DIAM{\Phi}}}} = \neg\DIAM{\DIAM{\BOX{\BOX{\neg\Phi}}}}, \\
\BOX{\DIAM{\BOX{\BOX{\Phi}}}} = \neg\DIAM{\BOX{\DIAM{\DIAM{\neg\Phi}}}},
  &\BOX{\DIAM{\BOX{\DIAM{\Phi}}}} = \neg\DIAM{\BOX{\DIAM{\BOX{\neg\Phi}}}}, \\
\BOX{\DIAM{\DIAM{\BOX{\Phi}}}} = \neg\DIAM{\BOX{\BOX{\DIAM{\neg\Phi}}}},
  &\BOX{\DIAM{\DIAM{\DIAM{\Phi}}}} = \neg\DIAM{\BOX{\BOX{\BOX{\neg\Phi}}}}.
\end{array}
\]




\begin{example}
  As an example, the property 
\[
\begin{array}{r@{}l}
  \nu Z.(\forall x. \exo{Order}(x)
  \wedge& \cval{\exc{S}}{\exv{PS}} \ra \\
        &\mu Y.(\exo{DeliveredOrder}(x) \vee
          \DIAM{\BOX{\BOX{\BOX{Y}}}})) \wedge \BOX{\BOX{\BOX{\BOX{Z}}}}
\end{array}
\]
checks that along every path, it is always true that for every
customer order in the peak season, there exists a sequence of action
executions leading to the state where the order is delivered, no
matter how the service call is evaluated, no matter how the contexts
change, and no matter how is the repair behavior
%
%
%
%
%
%
%
\end{example}


We now proceed to refine the transition systems that provide the
execution semantics for \agkabs
in such a way that the refined transition systems
still provide the same structure except that
\begin{compactitem}
\item we distinguish the types of the three intermediate states
  between the two stable states, and
\item we provide a mechanism to extract more information from each of
  the intermediate states (e.g., the corresponding action that
  was executed before we reach a certain intermediate state).
\end{compactitem}
Hence, it is obvious that the refined version of the \agkab transition
system should satisfy the same \mulcsa properties.

Towards formalizing the refined transition systems for \agkabs, we
first introduce several notions as follows:
Given an \agkab $\agkabsym$, let $\ts{\agkabsym}$ be its
context-sensitive transition system. Consider the states $s_1$, $s_2$,
$s_3$, $s_4$, and $s_5$ of $\ts{\agkabsym}$ such that $s_1$ and $s_5$
are stable states, $s_2$, $s_3$, $s_4$ are intermediate states, and we
have the following transitions:
\[
s_1 \trans s_2 \trans s_ 3 \trans s_4 \trans s_5
\]
By the construction of $\ts{\agkabsym}$ (see \Cref{def:agkabs-ts}), it
is easy to see that the transitions $s_1 \trans s_2$ and
$ s_2 \trans s_ 3$ consecutively must be an action execution and an
service call evaluation, while the transition $s_ 3 \trans s_4$ and
$ s_4 \trans s_5$ consecutively must be the context change and the
filter application.
Now, \sidetext{source action and source action parameters} let
$s_1 = \tup{id, A, \scmap, \ctx, \delta}$,
$s'_2 = \tup{id, A, \scmap, \ctx, \delta'}$, $\act$ and $\sigma$
consecutively be the action and the legal parameter assignment that is
involved in the transition $s_1 \trans s_2$, and $\theta$ is the
corresponding substitution that is involved in the transition
$ s_2 \trans s_ 3$ (i.e., that replaces the ``corresponding service
calls in the execution of $\act\sigma$ over $A$''). In this case, we
call $\act$ (resp.\ $\sigma$) the \emph{source action} (resp.\
\emph{source action parameters}) of $s_2$, $s_3$ and $s_4$ (Note that
the source action and the source action parameters are only defined
for the intermediate states).
Additionally, we also say that the set
$\addfacts{\ctbox^{\ctx}, A, \act\sigma}\theta$, is the \emph{set of
  facts to be added} of $s_3$, and $s_4$, while the set
$\delfacts{\ctbox^{\ctx}, A, \act\sigma}$ is the \emph{set of facts to
  be deleted} of $s_3$, and $s_4$.
Furthermore, we also introduce three more types of states for the
refined transition system of \agkabs. Technically, we partition the
set of states of the \agkabs transition system into four different set
of states namely:
\begin{compactenum}[(i)]
\item the set $\ststateset$ of \emph{stable states}, 
\item the set $\scstateset$ of \emph{service call evaluation states}, 
\item the set $\cxstateset$ of \emph{context change states},
\item the set $\ftstateset$ of \emph{filter application states}.
\end{compactenum}
We call \emph{stable state} (resp.\ \emph{service call evaluation
  state}) a state that belong to $\ststateset$ (resp.\
$\scstateset$). Similarly, we say that a state is a \emph{context
  change state} (resp.\ \emph{filter application state}) if it belongs
to $\cxstateset$ (resp.\ $\ftstateset$).
Finally, having all necessary ingredients in hand, below we introduce
the notion of \emph{fine-grained context-sensitive transition system}
which is a refined transition system that provide the execution
semantics of an \agkab.

\begin{definition}[Fine-Grained Transition System]\label{def:fg-cs-trans-sys}
  A \sidetext{Fine-Grained Transition System} \emph{fine-grained
    transition system} is a tuple
  $\ts{\csgkabsym} = \tup{\const, \ctbox, \stateset, s_0, \abox,
    \cntx, \actsrc, \actpar, \fadd, \fdel, \trans}$, where:
\begin{compactenum}
\item $\ctbox$ is a contextualized TBox;
\item
  $\stateset = \ststateset \uplus \scstateset \uplus \cxstateset
  \uplus \ftstateset$ is a set of states that is partitioned into four
  different types of states;

\item $s_0 \in \ststateset$ is the initial state and belongs to
  $\ststateset$ (i.e., $s_0$ is a stable state);
\item $\abox$ is a function that, given a state $s\in\stateset$,
  returns the ABox associated to $s$;
\item $\cntx$ is a function that, given a state $s\in\stateset$,
  returns the context associated to $s$;

\item $\actsrc$ is a function that, given a state $s\in\stateset$,
  returns the source action name associated to $s$;

\item $\actpar$ is a function that, given a state $s\in\stateset$,
  returns the source action parameters associated to $s$;

\item $\fadd$ is a function that, given a state $s\in\stateset$,
  returns the corresponding set of facts to be added associated to
  $s$;

\item $\fdel$ is a function that, given a state $s\in\stateset$,
  returns the corresponding set of facts to be deleted associated to
  $s$;

\item $\trans \subseteq \stateset \times \stateset$ is a transition
  relation between pairs of states.
\end{compactenum}
Additionally, we require the following:
\begin{compactenum}
\item for each $s \in \ststateset$, if there exists $s' \in \stateset$ such
  that $s \trans s'$, then $s' \in \scstateset$.
\item for each $s \in \scstateset$, if there exists $s' \in \stateset$ such
  that $s \trans s'$, then $s' \in \cxstateset$.
\item for each $s \in \cxstateset$, if there exists $s' \in \stateset$ such
  that $s \trans s'$, then $s' \in \ftstateset$.
\item for each $s \in \ftstateset$, if there exists $s' \in \stateset$
  such that $s \trans s'$, then $s' \in \ststateset$.
\end{compactenum}
\ \ 
\end{definition}

\begin{figure}[tbp]
\centering
\includegraphics[width=0.9\textwidth]{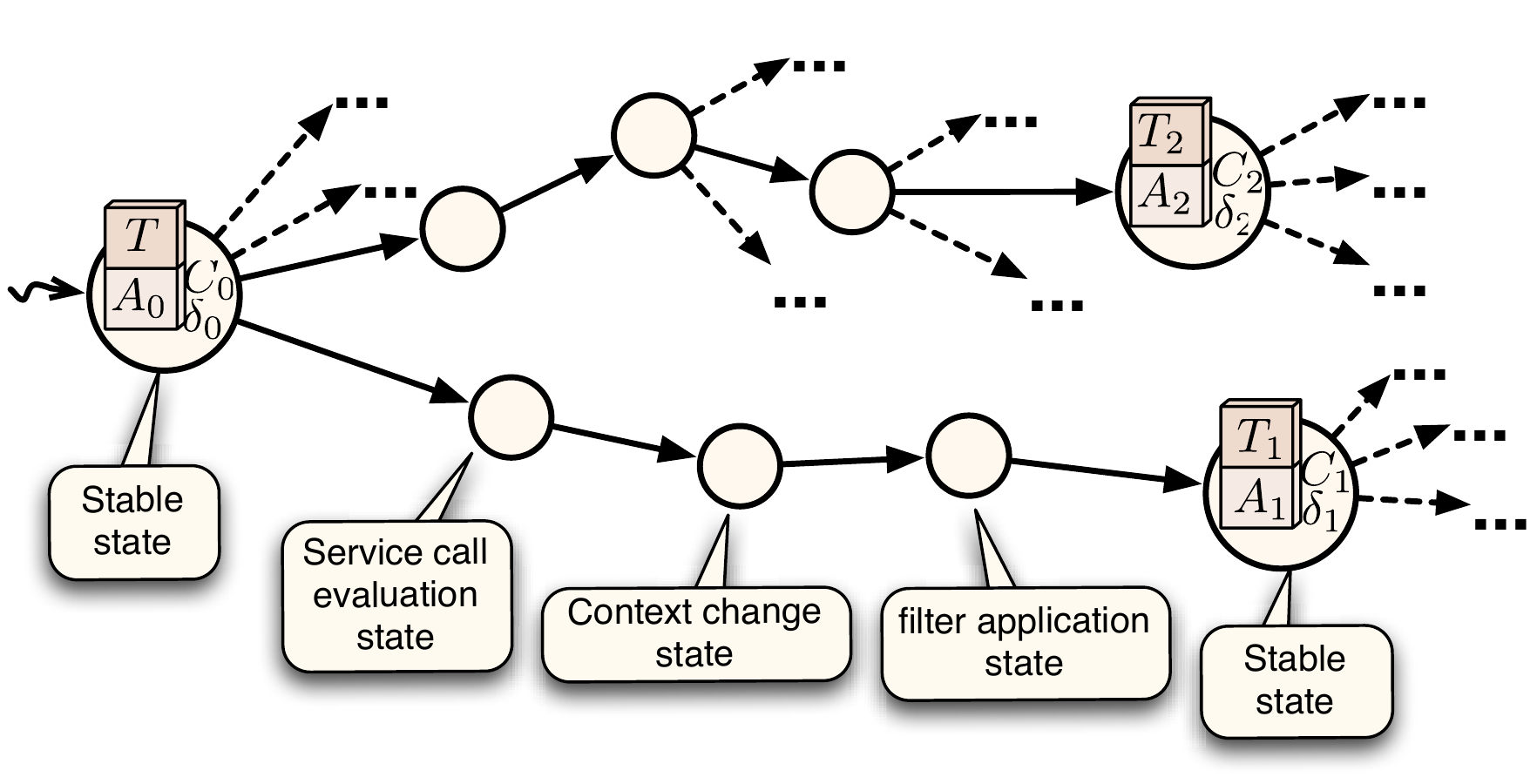}
\caption{Illustration of fine-grained transition system (Note: $T$ is
  a TBox, $A_i$ is an ABox, $C_i$ is a context, and $\delta_i$ is the
  remaining program to be
  executed).} \label{agkab-state-alternation-illustration}
\end{figure}

\Cref{agkab-state-alternation-illustration} illustrates the notion of
fine-grained transition system, in particular on the aspect of states
alternation.

The construction of a fine-grained transition system of an \agkab is
as follows:

\begin{definition}[\agkabs Fine-Grained Transition System]\label{def:agkabs-fg-ts}
  Given \sidetext{\agkabs Fine-Grained Transition System} an \agkab
  $\agkabsym = \tup{\ctbox, \initabox, \actset, \ginitprog, \initctx,
    \ctxprocset}$,
  and a context-sensitive filter relation $\csfilter$ (see
  \Cref{def:cs-filter-rel}), we define the \emph{transition system of
    $\agkabsym$ w.r.t.~$\csfilter$}, written
  $\ts{\agkabsym}^{\csfilter}$, as
  $\tup{\const, \ctbox, \stateset, s_0, \abox, \cntx, \actsrc,
    \actpar, \fadd, \fdel, \trans}$, where
  \begin{compactitem}
  \item $s_0 = \tup{id_0, \initabox, \emptyset, \ctx_0, \ginitprog}$,
    $\actsrc(s_0) = \fadd(s_0) = \fdel(s_0) = \emptyset$,
    $\actpar(s_0) = \sigma_\emptyset$ (where $\sigma_\emptyset$ is an
    empty substitution),

  \item $\stateset = \ststateset \uplus \scstateset \uplus \cxstateset \uplus \ftstateset$, and
  \item $\ststateset$, $\scstateset$, $\cxstateset$, $\ftstateset$ and
    $\trans$ are defined by simultaneous induction as the smallest
    sets such that
    \begin{compactitem}
    \item $s_0 \in \ststateset$, and
    \item if $\tup{id, A, \scmap, \ctx, \delta} \in \ststateset$  then
      we do the following
      \begin{compactenum}[(1)]
      \item for any action $\act$, and any substitution $\sigma$, let
        \[
        \begin{array}{@{}ll@{}} S^1 = \set{\tup{id^1, A, \scmap, \ctx,
              \delta'} ~\mid~& id^1 \mbox{ is a fresh
                               identifier, and } \\
                             &\tup{A, \scmap, \ctx, \delta}
                               \gprogtrans{\act\sigma, \csfilter}
                               \tup{A, \scmap, \ctx, \delta'} }
        \end{array}
        \]
      \item and then for each
        $s_1 = \tup{id^1, A, \scmap, \ctx, \delta'} \in S^1$ with\\
        $\tup{A, \scmap, \ctx, \delta} \gprogtrans{\act\sigma,
          \csfilter} \tup{A, \scmap, \ctx, \delta'}$,
        we do the following:
        \begin{compactenum}[(a)]
        \item let
        \[
        \begin{array}{@{}ll@{}} S^2 = \set{\tup{id^2, A, \scmap',
              \ctx, \delta'} ~\mid~&
                                     id^2 \mbox{ is a fresh identifier, } \\
                                   &\theta \in
                                     \eval{\addfacts{\ctbox^{\ctx}, A,
                                     \act\sigma}}, \\
                                   &\mbox{and }\scmap' = \scmap \cup
                                     \theta }
        \end{array}
        \]
      \item and then for each
        $s_2 = \tup{id^2, A, \scmap', \ctx, \delta'} \in S^2$ with\\
        $\scmap' = \scmap \cup \theta$, we do the following:
          \begin{compactenum}[(i)]
          \item let
            \[
            \begin{array}{@{}ll@{}} S^3 = \set{\tup{id^3, A, \scmap',
                  \ctx', \delta'} ~\mid~& id^3 \mbox{ is a
                                          fresh identifier, and } \\
                                        &\tup{A, \ctx, \ctx'} \in
                                          \ctxchg }
            \end{array}
            \]
          \item and then for each
            $s_3 = \tup{id^3, A, \scmap', \ctx', \delta'} \in S^3$
            with $\tup{A, \ctx, \ctx'} \in \ctxchg$, we do the
            following: \\if there exists $A'$ such that
              \[ 
              \tup{A, \addfacts{\ctbox^{\ctx}, A, \act\sigma}\theta,
                \delfacts{\ctbox^{\ctx}, A, \act\sigma}, \ctx', A'}
              \in \csfilter,
              \]
              and $A'$ is $\ctbox^{\ctx'}$-consistent.  Then for each
              of that $A'$, we have 
              \begin{compactitem}
              \item if there exists
                $\tup{id_s, A_s, \scmap_s, \ctx_s,
                  \delta_s}\in\ststateset$
                such that \\ $A_s = A'$, $\scmap_s = \scmap'$,
                $\ctx_s = \ctx'$, $\delta_s = \delta'$, then we have
                $s_3 \trans \tup{id_s, A_s, \scmap_s, \ctx_s,
                  \delta_s}$,

                
              \item otherwise we have
                $s_4 = \tup{id^4, A', \scmap', \ctx',
                  \delta'}\in\ststateset$, 
                where $id^4$ is a fresh identifier, and we have
                $s_3 \trans s_4$.
            \end{compactitem}

            Additionally, we also have that
            $s_1\in\scstateset$, $s_2\in\cxstateset$,
            $s_3\in\ftstateset$, 
            $\tup{id, A, \scmap, \ctx, \delta}\trans s_1$,
            $s_1 \trans s_2$, $s_2 \trans s_3$, and
            \begin{compactitem}
            \item $\actsrc(s_1) = \actsrc(s_2) = \actsrc(s_3) = \act$,
            \item $\actsrc(s_4) = \emptyset$
            \item $\actpar(s_1) = \actpar(s_2) =\actpar(s_3) = \sigma$,
            \item $\actpar(s_4) = \sigma_\emptyset$ (where
              $\sigma_\emptyset$ is an empty substitution),
            \item $\fadd(s_1) = $ the set of ABox
              assertions in $\addfacts{\ctbox^{\ctx}, A, \act\sigma}$
              (excluding the ground skolem terms).
            \item
              $\fadd(s_2) = \fadd(s_3) = \addfacts{\ctbox^{\ctx}, A,
                \act\sigma}\theta$,
            \item
              $\fdel(s_1) = \fdel(s_2) = \fdel(s_3) = \delfacts{\ctbox^{\ctx}, A,
                \act\sigma}$,
            \item $\fadd(s_4) = \fdel(s_4) = \emptyset$,
            \end{compactitem}
          \end{compactenum}
        \end{compactenum}
      \end{compactenum}

    \end{compactitem}
  \end{compactitem}
 \ \ 
\end{definition}

Notice that in the construction above we do not define the information
of $\actpar$, $\fadd$, and $\fdel$ for the stable states. The reason
is because a stable state might be reached from more than one
different run, hence it might be obtained through varios different way
of manipulating states (e.g., different executed action, different
fact to be added, and different facts to be deleted).

From this moment, unless explicitly stated differently, we consider
that the execution semantics of \agkabs is given by fine-grained
transition system.

\begin{lemma}
  Given an \agkab $\agkabsym$, a context-sensitive filter relation
  $\csfilter$, and a \mulcsa formula $\Phi$. Let $\ts{1}$ be the
  context-sensitive transition system of $\agkabsym$ w.r.t.\
  $\csfilter$ as in \Cref{def:agkabs-ts}, and $\ts{2}$ be the
  fine-grained transition system of $\agkabsym$ w.r.t.\ $\csfilter$ as
  in \Cref{def:agkabs-fg-ts}, we have that
\[
\ts{1} \models \Phi \mbox{  if and only if  } \ts{2} \models \Phi
\]
\end{lemma}
\begin{proof}
  The claim easily follows from \Cref{def:agkabs-ts,def:agkabs-fg-ts}
  since essentially both transition systems have the same structure
  except that the fine-grained transition systems provide more
  capabilities to access the information within a state.
\end{proof}

As the next preliminaries, in this chapter we reserve several fresh
concept/role names as follows:
\begin{compactitem}

\item In order to mimic the context evolution within S-GKABs, we
  simply adopt our approach in
  \Cref{sec:transf-scsgkabs-to-sgkabs}. Thus, similar to
  \Cref{sec:transf-scsgkabs-to-sgkabs}, for each context dimension
  assignment $\cval{d_i}{v_j}$ we reserve two fresh concept names
  $\cdcc_i^{v_j}$ and $\cdcq_i^{v_j}$ in order to represent it as an
  ABox assertion. Similarly, this kind of concept name is also called
  \emph{context dimension concept}.

\item Given any contextualized TBox $\ctbox$, for each concept name
  $N \in \voc(\ctbox)$, we reserve two fresh concept names $N^a$ and
  $N^d$ to keep track the temporary information about ABox assertions
  to be added/deleted before we materialize the update (similarly for
  role names). The concept names of the form $N^a$ (resp.\ $N^d$) is
  called \emph{added} (resp.\ \emph{deleted}) \emph{fact marker}
  concept names (similarly for role names, we call them \emph{added}
  (resp.\ \emph{deleted}) \emph{fact marker} role names).

\item To keep track the information of the source action, we reserve a
  fresh concept name $\actsrcconceptname$ and it will be populated
  only with special constants that is reserved to represent action
  names. Therefore, w.l.o.g., here we assume that each action name is
  unique. Note that we can easily enforce this assumption by renaming
  each action name that occurs more than once.

\item Given any action $\act$ with parameters $p_1, \ldots, p_m$, we
  reserve $m$ fresh concept names
  $\parconceptname{1}, \ldots, \parconceptname{m}$ (similarly for role
  names). This kind of concept names are called \emph{action parameter
    concept names}.  Therefore, w.l.o.g., we also assume that each
  action parameter has a unique name and each of them has a
  corresponding reserved concept name. Note that we can easily enforce
  this situation by renaming each action parameter name that occurs
  more than once.

\item To introduce several types of states for KB transition system,
  we reserve several ABox assertions namely $\sctmp$, $\cxtmp$,
  $\fttmp$, and $\ftitmp$.

\end{compactitem}
Here, we call \emph{special marker} the concept/role assertions made
by using the reserved concept/role names above.
Related to the reserved concept/role names above, we introduce
the notion of a set of added/deleted assertions as follows.

\begin{definition}[Added Assertions]\label{def:added-assertion}
  Given \sidetextb{Added Assertions} a contextualized TBox $\ctbox$,
  and an ABox $A$ over $\voc(\ctbox)$, we define the set of
  \emph{added assertions of $A$} as a set $\addass{A} \subseteq A$ of
  ABox assertions such that we have concept assertion
  $N^a(c) \in \addass{A}$ if and only if $N^a(c) \in A$ and $N^a$ is an added fact
  marker concept name (similarly for role assertions).
\end{definition}

\begin{definition}[Deleted Assertions]\label{def:deleted-assertion}
  Given \sidetextb{Deleted Assertions} a contextualized TBox $\ctbox$,
  and an ABox $A$ over $\voc(\ctbox)$, we define the set of
  \emph{deleted assertions of $A$} as a set $\delass{A} \subseteq A$
  of ABox assertions such that we have concept assertion
  $N^d(c) \in \delass{A}$ if and only if $N^d(c) \in A$ and $N^d$ is a
  deleted fact marker concept name (similarly for role assertions).
\end{definition}

\noindent
Furthermore, we also introduce the notion of a set of action parameter
assertions and a set of context assertions as follows.

\begin{definition}[Action Parameter Assertions]\label{def:actpar-assertion}
  Given \sidetextb{Action Parameter Assertions} a contextualized TBox
  $\ctbox$, and an ABox $A$ over $\voc(\ctbox)$, we define the set of
  \emph{action parameter assertions of $A$} as a set
  $\parass{A} \subseteq A$ of ABox assertions such that we have
  concept assertion $\parconceptname{}(c) \in \parass{A}$ if and only
  if $\parconceptname{}(c) \in A$ and $\parconceptname{}$ is an action
  parameter concept name.
\end{definition}

\begin{definition}[Context Assertions]\label{def:context-assertion}
  Given \sidetextb{Context Assertions} a contextualized TBox $\ctbox$,
  and an ABox $A$ over $\voc(\ctbox)$, we define the set of
  \emph{context assertions of $A$} as a set $\ctxass{A} \subseteq A$
  of ABox assertions such that we have concept assertion
  $\cdcc_i^{v_j}(c) \in \ctxass{A}$ if and only if
  $\cdcc_i^{v_j}(c) \in A$ and $\cdcc_i^{v_j}$ is a context dimension
  concept name.
\end{definition}

We now proceed to introduce specific fragment of KB transition systems
namely \emph{typed KB transition systems}. The idea is that they will
be the transition systems of S-GKABs that mimics the evolution of
\agkabs that is provided by the fine-grained transition systems.
As preliminaries, we now introduce several types of states namely
\emph{stable state}, \emph{service call evaluation state},
\emph{context change state}, \emph{filter application state}, and
\emph{filter application intermediate state}. All of them are formally
defined below:

\begin{definition}[Stable State]\label{def:stable-state-kb-ts}
  Let \sidetextb{Stable State} $\ts{}$ be a KB transition system, a state $s$ of $\ts{}$ is
  called a \emph{stable state} if $\sctmp\not\in\abox(s)$,
  $\cxtmp\not\in\abox(s)$, $\fttmp\not\in\abox(s)$, $\ftitmp\not\in\abox(s)$.
\end{definition}

\begin{definition}[Service Call Evaluation State]\label{def:serv-call-state-kb-ts}
  Let \sidetextb{Service Call Evaluation State} $\ts{}$ be a KB
  transition system, a state $s$ of $\ts{}$ is called a \emph{service
    call evaluation state} if $\sctmp\in\abox(s)$,
  $\cxtmp\not\in\abox(s)$, $\fttmp\not\in\abox(s)$,
  $\ftitmp\not\in\abox(s)$.
\end{definition}

\begin{definition}[Context Change State]\label{def:ctx-chg-state-kb-ts}
  Let \sidetextb{Context Change State} $\ts{}$ be a KB transition system, a state $s$ of $\ts{}$ is
  called a \emph{context change state} if $\sctmp\not\in\abox(s)$,
  $\cxtmp\in\abox(s)$, $\fttmp\not\in\abox(s)$, $\ftitmp\not\in\abox(s)$.
\end{definition}

\begin{definition}[Filter Application State]\label{def:filter-app-state-kb-ts}
  Let \sidetextb{Filter Application State} $\ts{}$ be a KB transition system, a state $s$ of $\ts{}$ is
  called a \emph{filter application state} if $\sctmp\not\in\abox(s)$,
  $\cxtmp\not\in\abox(s)$, $\fttmp\in\abox(s)$, $\ftitmp\not\in\abox(s)$.
\end{definition}

\begin{definition}[Filter Application Intermediate State]\label{def:inter-filter-app-state-kb-ts}
  Let \sidetextb{Filter Application Intermediate State} $\ts{}$ be a
  KB transition system, a state $s$ of $\ts{}$ is called a
  \emph{filter application state} if $\sctmp\not\in\abox(s)$,
  $\cxtmp\not\in\abox(s)$, $\fttmp\not\in\abox(s)$,
  $\ftitmp\in\abox(s)$.
\end{definition}


Notice that all of the types introduced above, except the filter
application intermediate states, are the same as the types of states
in the fine-grained transition system. The important different is that
for KB transition system, the type is determined by the special ABox
assertion that is contained by the state.

Having the necessary ingredients in hand, we are now ready to define
the notion of typed KB transition system as follows.

\begin{definition}[Typed KB Transition System]\label{def:typed-kb-ts}
  \ \sidetext{Typed KB Transition System} Given a KB transition system
  $\ts{} = \tup{\const,T,\stateset,s_0,\abox,\trans}$ we say that
  $\ts{}$ is a \emph{typed KB transition system} if the following
  hold:
  \begin{compactitem}

  \item $s_0$ is a stable state,

  \item for each state $s \in \stateset$, we have the following:

    \begin{compactitem}
    \item $s$ is either a stable state (see
      \Cref{def:stable-state-kb-ts}), a service call evaluation state
      (see \Cref{def:serv-call-state-kb-ts}), a context change state
      (see \Cref{def:ctx-chg-state-kb-ts}), a filter application state
      (see \Cref{def:filter-app-state-kb-ts}), or a filter application
      intermediate state (see
      \Cref{def:inter-filter-app-state-kb-ts}),

    \item if $s$ is a stable state, and there exists
      $s' \in \stateset$ such that $s \trans s'$, then $s'$ is a
      service call evaluation state.

    \item if $s$ is a service call evaluation state, and there exists
      $s' \in \stateset$ such that $s \trans s'$, then $s'$ is a
      context change state.

    \item if $s$ is a context change state, and there exists
      $s' \in \stateset$ such that $s \trans s'$, then $s'$ is a
      filter application state.

    \item if $s$ is a filter application state, and there exists
      $s' \in \stateset$ such that $s \trans s'$, then $s'$ is a
      filter application intermediate state.

    \item if $s$ is a filter application intermediate state, and there
      exists $s' \in \stateset$ such that $s \trans s'$, then $s'$ is
      either a filter application intermediate state or a stable
      state.
    \end{compactitem}

\end{compactitem}
\ \ 
\end{definition}

Intuitively, compare to the ordinary KB transition systems, the Typed
KB transition systems require that the states must be either a stable
state, service call evaluation state, context change state, filter
application state, and filter application intermediate
state. Additionally, the Typed KB transition systems require that
there is an alternation of state types among the transitions of the
states. In particular, the order of alternation should be as follows: 
\begin{inparaenum}[\it (i)]
\item a stable state should be followed by a service call evaluation
  state 
\item and then a service call evaluation state should be followed by
  a context change state,
\item after that the successor of a context change state should be
a filter application state
\item then a filter application state should be continued by filter
  application intermediate state.
\item last, the filter application intermediate state should be
  followed by either a filter application intermediate state or a
  stable state.
\end{inparaenum}

\subsection{Verification of 
\bagkabs}


This section is aimed to present the reduction of the \mulcsa
verification over \bagkabs into the \muladom verification over
S-GKABs. We first explain how we transform \bagkabs into S-GKABs, and
also how we translate \mulcsa formulas into \muladom formulas. Then,
we introduce a specific notion of bisimulation that will be used to
prove our reduction from \bagkabs into S-GKABs. Last, we show that we
can recast the verification of \bagkabs into S-GKABs.

\subsubsection{Translating \bagkabs to S-GKABs}

We devote this section to explain our translation that transforms
\bagkabs into S-GKABs. To do so, several preliminaries need to be
introduced first.

As it can be seen from the execution semantics of \agkabs, they
separate the service call evaluation from the action
execution. Moreover, the ABox is not directly updated with the changes
that is done by the action.
Thus, to simulate that situation inside S-GKABs, in the following we
introduce the notion of splitting an action as follows:

\begin{definition}[Split Action]\label{def:split-act}
  Given \sidetext{Split Action} an action
  $\act(p_1, \ldots, p_m):\set{e_1,\ldots,e_n}$ with
  $e_i = \map{[q_i^+]\land Q_i^-}{\add \facta_i, \del \factd_i}$. Let
  $\parconceptname{1}, \ldots, \parconceptname{m}$ be the fresh
  concept names for capturing the value of each action parameter, the
  \emph{split actions} obtained from $\act$ are two actions $\act_a$
  and $\act_b$ as follows:
\begin{compactenum}

\item $\act_a$ is a fresh action name and is of the form
  $\act_a(p_1,\ldots,p_m):\set{e'_1,\ldots,e'_n,e_{temp}, e_{par},
    e_{act}}$, where
  \begin{compactitem}
  \item $e_i'$ (for $i \in \set{1, \ldots, n}$) is obtained from $e_i$
    such that \\
    $e'_i = \map{\mbox{Q}^i_{\ctxb}}{\add {\facta_i}' \cup {\factd_i}'
    }$ where:
    \begin{itemize}

  \item $\mbox{Q}^i_{\ctxb}$ is a contextually compiled query of
    $[q_i^+]\land Q_i^-$ w.r.t.\ $\cdimset$.


  \item for each atom $N(t) \in \facta_i$, such that $t$ is not a
    skolem terms (representing a service call), we have
    $N^a(t) \in {\facta_i}'$.

  \item for each atom $P(t_1, t_2) \in \facta_i$, such that $t_1$ and
    $t_2$ are not skolem terms (representing a service call), we have
    $P^a(t_1, t_2) \in {\facta_i}'$.

  \item for each atom $N(t) \in \factd_i$ (resp.\
    $P(t_1, t_2) \in \factd_i$), \\ we have $N^d(t) \in {\factd_i}'$
    (resp.\ $P^d(t_1, t_2) \in {\factd_i}'$).

    \end{itemize}
  \item $e_{temp} = \set{\map{\true}{\add \set{\sctmp} } }$
  \item
    $e_{par} = \set{\map{\true}{\add \set{\parconceptname{1}(p_1),
          \ldots, \parconceptname{m}(p_m)} } }$
  \item
    $e_{act} = \set{\map{\true}{\add \set{\actsrcconceptname(\act) }}}$
  \end{compactitem}

\item $\act_b$ is a fresh action name and it is a 0-ary action of the
  form
  $\act_b():\set{e'_1,\ldots,e'_n, 
    e_{temp}}$, where
  \begin{compactitem}
  \item $e_i'$ (for $i \in \set{1, \ldots, n}$) is obtained from $e_i$
    such that
    \[
    e'_i = \map{\mbox{Q}^i_{\ctxb} \land \parconceptname{1}(p_1) \land \ldots
      \land \parconceptname{m}(p_m)}{\add {\facta_i}'}
    \]
    where
    \begin{compactitem}
    \item $\mbox{Q}^i_{\ctxb}$ is contextually compiled query of
      $[q_i^+]\land Q_i^-$ w.r.t.\ $\cdimset$.
    \item and ${\facta_i}'$ is obtained as follows: 
      \begin{compactitem}
      \item for each atom $N(t) \in \facta_i$, such that $t$ is a
        skolem terms (representing a service call), we have
        $N^a(t) \in {\facta_i}'$.
      \item for each atom $P(t_1, t_2) \in \facta_i$, such that $t_1$
        and $t_2$ are skolem terms (representing a service call), we
        have $P^a(t_1, t_2) \in {\facta_i}'$.
      \end{compactitem}
    \end{compactitem}
  \item $e_{temp} = \set{\map{\true}{\add \set{\cxtmp}, \del \set{\sctmp} } }$
  \end{compactitem}
\end{compactenum}
In this case we also say that $\act_a$ is the \emph{first split
  action} of $\act$ and $\act_b$ is the \emph{second split action} of
$\act$.
\end{definition}

The main purpose of introducing split actions is to separate the
non-determinism sources that come from the choice of actions and also
the choice of service call substitutions. Basically, the split actions
of an action are obtained by splitting an action into two actions
where one action do additions/deletions that do not involve any
service calls, while the other action do additions that involve
service calls. Additionally, the split actions of an action $\act$ do
not concretely add/delete the assertions but only give marks on the
assertions that need to be added/deleted by $\act$. This is done by
adding concept/role assertions made by the added/deleted fact marker
concept/role names. We will see later that the materialization of the
update (addition/deletion) is done by update action. As a further
intuition, in the second split action, we make use a query that is
made by action parameter concept names in order to enforce that it
uses the same action parameter as the corresponding first split
action.

Since inside a program an action is always appeared as an action
invocation, in the following we introduce the notion of split action
invocation that is based on the notion of split action.

\begin{definition}[Split Action
  Invocation]\label{def:split-action-invoc}
  A\sidetext{Split Action Invocation} \emph{split action invocation} obtained from a context-sensitive
  atomic action invocation $\gactc{Q(\vec{p})}{\ctxe}{\act(\vec{p})}$
  is a program 
  \[
  \gact{Q'(\vec{p})}{\act_a(\vec{p})};\gact{\true}{\act_b()}
  \]
  where
  \begin{compactitem}

  \item $Q' = Q_\ctxb \wedge q_{\ctxe}$, where $Q_\ctxb$ is a
    contextually compiled query of $Q$ (see
    \Cref{def:contextually-compiled-query}), and $q_{\ctxe}$ is the
    query that represents the context expression $\ctxe$ (see
    \Cref{def:query-rep-context-exp}).

  \item $\act_a$ and $\act_b$ are split actions obtained from $\act$ (see
    \Cref{def:split-act}).

  \end{compactitem}
\ \ 
\end{definition}


In \agkabs, after evaluating the service call, the next step is
changing the context. To mimic this step inside S-GKABs, we make use
the context-change program obtained from the set of context evolution
rule as introduced in \Cref{def:ctx-chg-prog}. However, we can not use
it directly since it materializes the updates
(additions/deletions). Thus, here we refine the notion of action and
action invocation obtained from context-evolution rule (introduced in
\Cref{def:action-and-action-invocation-obtained-from-context-evolution-rule})
and introduce the notion of sole action and sole action invocation
obtained from context-evolution rule. The core different is that in
the sole action obtained from context-evolution rule, the action only
changes the context and does not concretize the assertions to be
added/deleted.

\begin{definition}[Sole Action and Sole Action Invocation Obtained From
  Context-evolution Rule]
\label{def:sole-action-and-action-invocation-obtained-from-context-evolution-rule}
A \sidetext{Sole Action and Sole Action Invocation Obtained From
  Context-evolution Rule} \emph{sole action invocation obtained from a
  context-evolution rule} $\tup{Q, \ctxe} \mapsto C_{new}$ in
$\ctxprocset$, is an action invocation $\gact{Q'}{\act^s_{\ctx}()}$
where
  \begin{compactenum}
  \item $Q' = Q_\ctxb \wedge q_{\ctxe}$
    where $Q_\ctxb$ is contextually compiled query of $Q$, and
    $q_{\ctxe}$ is the query obtained from the context expression
    $\ctxe$.

  \item $\act^s_\ctx$ is a 0-ary action obtained from
    $\tup{Q, \ctxe} \mapsto C_{new}$ as follows: 

    \begin{compactitem}
    \item For each $[d_i \mapsto v_j] \in C_{new}$, we have:
      \begin{enumerate}[(i)]
      \item
        $\map{\true}{\add \set{\cdcc_i^{v_j}(\ctxconst),
            \cdcq_i^{v_j}(\ctxconst) } }$ in $\eff{\act^s_\ctx}$, and
      \item
        $\map{\true}{\del \set{ \cdcc_i^{v_k}(\ctxconst),
            \cdcq_i^{v_k}(\ctxconst) } }$
        in $\eff{\act^s_\ctx}$ \\ for every $v_k\in\cdom[d_i]$ such
        that $v_k \neq v_j$.
      \end{enumerate}
  
    \item Additionally, we have
      \begin{center}
        $\map{\true}{\add \set{\fttmp}, \del \set{\cxtmp} }$ in
      $\eff{\act^s_\ctx}$.
      \end{center}
    \end{compactitem}
  
    In this case we say that $\act^s_{\ctx}$ is \emph{a sole action
      obtained from the context-evolution rule
      $\tup{Q, \ctxe} \mapsto C_{new}$}.
\end{compactenum}
\ \
\end{definition}

To concretize the addition/deletion over the ABox, in the following we
introduce the notion of update action which essentially materializes
the updates.

\bigskip

\begin{definition}[Update Action]
\label{def:update-action}
Let \sidetext{Update Action}
$\csgkabsym = \tup{\ctbox, \initabox, \actset, \ginitprog, \initctx,
  \ctxprocset}$
be an \agkab.  An \emph{update action} $\act^u$ is 0-ary (i.e., has no
action parameters) action over $\ctbox$, where $\eff{\act^u}$ is the
smallest set containing the following effects:
\begin{compactitem}
\item For each concept name $N \in \voc(\ctbox)$, we have
  \begin{enumerate}[(i)]
  \item $\map{N^a(x)}{\add \set{N(x) }, \del \set{N^a(x) } }$ in
    $\eff{\act^u}$, and
  \item $\map{N^d(x)}{\del \set{N(x), N^d(x)} }$ in $\eff{\act^u}$.
  \end{enumerate}
\item Similarly for the role names, we create the same effect as
  above.


  \item
    $\map{\actsrcconceptname{}(x)}{\del
      \set{\actsrcconceptname{}(x)}}$ in $\eff{\act^{u}}$

  \item For each action parameter concept name $\parconceptname{}$, we
    have 
    \begin{center}
      $\map{\parconceptname{}(x)}{\del \set{\parconceptname{}(x)}}
      \mbox{ in } \eff{\act^{u}}$
    \end{center}

    \item Additionally, we have
      \begin{center}
        $\map{\true}{\add \set{\ftitmp}, \del \set{\fttmp} } $
        in $\eff{\act^u}$.
      \end{center}

  \end{compactitem}
\ \ 
\end{definition}

\noindent
Basically, the update action simply adds (resp.\ deletes) the ABox
assertions that has been marked to be added (resp.\ deleted).





As the last preliminary towards introducing our translation that
transforms \bagkabs into S-GKABs, in the following we introduce the
notion of program translation for \bagkabs.

\begin{definition}[Program Translation $\tgprogba$]\label{def:tran-prog-bagkab-sgkab}
  Given \sidetext{Program Translation $\tgprogba$} a \bagkabs
  $\agkabsym = \tup{\ctbox, \initabox, \actset, \ginitprog, \initctx,
    \ctxprocset}$
  we define a \emph{translation $\tgprogba$} which translates a
  program $\delta$ into a program $\delta'$ inductively as follows:
\[
\begin{array}{@{}lll@{}}
  \tgprogba(\gactc{Q(\vec{p})}{\ctxe}{\act(\vec{p})}) &= & 
                                                             \gact{Q'(\vec{p})}{\act_a(\vec{p})};\gact{\true}{\act_b()};\delta_{\ctxprocset};\\
                                                      &&
                                                         \gact{\true}{\act^u()};
                                                         \delta^{\ctbox}_b ; \gact{\true}{\act^-_{temp}()}\\
  \tgprogba(\gemptyprog) &=& \gemptyprog \\
  \tgprogba(\delta_1|\delta_2) &=& \tgprogba(\delta_1)|\tgprogba(\delta_2) \\
  \tgprogba(\delta_1;\delta_2) &=& \tgprogba(\delta_1);\tgprogba(\delta_2) \\
  \tgprogba(\gif{\varphi}{\delta_1}{\delta_2}) &= &\gif{\varphi}{\tgprogba(\delta_1)}{\tgprogba(\delta_2)} \\
  \tgprogba(\gwhile{\varphi}{\delta}) &=& \gwhile{\varphi}{\tgprogba(\delta)}
\end{array}
\]
where 
\begin{compactitem}
\item $\gact{Q'(\vec{p})}{\act_a(\vec{p})};\gact{\true}{\act_b()}$ is
  a split action invocation obtained from
  $\gactc{Q(\vec{p})}{\ctxe}{\act(\vec{p})}$ as in
  \Cref{def:split-action-invoc},

\item $\delta_{\ctxprocset}$ is a context-change program obtained from
  $\ctxprocset$ as in \Cref{def:ctx-chg-prog} except that it is formed
  by sole action invocation obtained from context evolution rule as in
  \Cref{def:sole-action-and-action-invocation-obtained-from-context-evolution-rule},

\item $\act^u$ is an update action (see \Cref{def:update-action}), 

\item $\delta^{\ctbox}_b$ is a context-sensitive b-repair program over
  $\ctbox$ as in \Cref{def:cs-brepair-prog},

\item $\act^-_{temp}() :\set{\map{\true}{\del \set{\ftitmp}}}$.
\end{compactitem}
\ \ 
\end{definition}

Essentially, the program translation $\tgprogba$ translates each
context-sensitive action invocation into a sequence of split action
invocation that is obtained from the given action invocation and then
concatenates it with the context-change program, the update action and
the program that simulates the b-repair computation. Thus,
intuitively, it can be seen that the obtained program separate the
source of non-determinism as in \bagkabs.

Having all ingredients in hand, we now proceed to define the
translation from \bagkabs into S-GKABs as follows.

\begin{definition}[Translation from \bagkab to S-GKAB]
  We \sidetext{Translation from \bagkab to S-GKAB} define a
  translation $\tgkabba$ that, given a \bagkab
  $\agkabsym = \tup{\ctbox, \initabox, \actset, \ginitprog, \initctx,
    \ctxprocset}$,
  produces an S-GKAB
  $\tgkabba(\agkabsym) = \tup{T_\cdimset, \initabox \cup A_{\initctx},
    \actset', \ginitprog'}$, where
\begin{compactitem}

\item $T_\cdimset$ is a TBox obtained from a
  set of context dimensions $\cdimset$ (see
  \Cref{def:tbox-from-ctxdim}),

\item 
  $A_{\initctx}$ is an ABox obtained from $\initctx$ (see
  \Cref{def:abox-context}),

\item
  $\actset' = \actset_\act \cup \actset_\ctx \cup \actset_b^{\ctbox}$
  $\cup \set{\act^u, \act^-_{temp}}$ where:

\begin{compactitem}
\item $\actset_\act$ is obtained from $\actset$ such that for each
  action $\act \in \actset$, we have $\act_1,\act_2 \in \actset_\act$
  where $\act_1$ and $\act_2$ are split action obtained from $\act$
  (see \Cref{def:split-act}),

\item $\actset_\ctx$ is obtained from $\ctxprocset$ such that for each
  context-evolution rule $\tup{Q, \ctxe} \mapsto C_{new}$ in
  $\ctxprocset$, we have $\act_\ctx \in \actset_\ctx$ where
  $\act_\ctx$ is \emph{a sole action obtained from the
    context-evolution rule $\tup{Q, \ctxe} \mapsto C_{new}$} (see
  \Cref{def:sole-action-and-action-invocation-obtained-from-context-evolution-rule}),

\item $\actset_b^{\ctbox}$ is the set of context-sensitive b-repair
  actions over $\ctbox$ (see \Cref{def:cs-brep-act-brep-actinv}),

\item $\act^u$ is an update action (see \Cref{def:update-action}), 

\item $\act^-_{temp}$ is an action of the form
  $\act^-_{temp}() :\set{\map{\true}{\del \set{\ftitmp}}}$.

\end{compactitem}

\item $\ginitprog' = \tgprogba(\ginitprog)$. 
\end{compactitem}
\ \ 
\end{definition}

In the following, we formally state that the transition system of the
S-GKAB that is obtained from a \bagkab is a typed KB transition
system. 

\begin{lemma}\label{lem:ts-sgkab-of-bagkab-is-typed}
  Given a \bagkab $\agkabsym$ with transition system
  $\ts{\agkabsym}^{\csfilter_B}$, let $\tgkabba(\agkabsym)$ be the
  S-GKAB obtained from $\agkabsym$ through $\tgkabba$ and
  $\ts{\tgkabba(\agkabsym)}^{\filter_S}$ be its transition system. We
  have that $\ts{\tgkabba(\agkabsym)}^{\filter_S}$ is typed KB
  transition system.
\end{lemma}
\begin{proof}
  The claim easily follows from the construction of
  $\tgkabba(\agkabsym)$ by also considering the following:
  \begin{compactitem}
  \item based on the program translation $\tgprogba$ (see
    \Cref{def:tran-prog-bagkab-sgkab}), especially on the part of
    atomic invocation translation, it is easy to see that we have an
    alternation between stable state, service call evaluation state,
    context-change state, and filter application state.
  \item also by the definition of $\tgprogba$ (see
    \Cref{def:tran-prog-bagkab-sgkab}) all states in
    $\tgkabba(\agkabsym)$ are either stable state, service call
    evaluation state, context-change state, or filter application
    state.
  \item the initial state of $\tgkabba(\agkabsym)$ is a stable state.
  \end{compactitem}
\ \ 
\end{proof}

The \mulcsa property $\Phi$ over a \bagkab $\agkabsym$ can then be
recast as a corresponding \muladom property over an S-GKAB
$\tgkabba(\agkabsym)$ using the following formula translation:

\begin{definition}[Translation $\tforja$]\label{def:tforja}
  We \sidetextb{Translation $\tforja$} define a \emph{translation
    $\tforja$} that transforms an arbitrary \mulcsa formula $\Phi$ (in
  NNF) into a \muladom formula $\Phi'$ inductively by recurring over
  the structure of $\Phi$ as follows:
  \begin{compactitem}
  \item $\tforja(Q) = Q_\ctxb $

  \item $\tforja(\ctxe) = q_{\ctxe}$

  \item $\tforja(\neg Q) = \neg Q_\ctxb$

  \item $\tforja(\Q x.\Phi) = \Q x. \tforja(\Phi)$

  \item $\tforja(\Phi_1 \circ \Phi_2) = \tforja(\Phi_1) \circ \tforja(\Phi_2) $

  \item $\tforja(\circledcirc Z.\Phi) = \circledcirc Z. \tforja(\Phi) $

  \item $\tforja(\bigodot \bigodot \bigodot\DIAM{\Phi}) =$ \\
    \hspace*{10mm}$\bigodot \bigodot \bigodot\DIAM{\mu Z.((\ftitmp \wedge
      \DIAM{Z}) \vee (\neg \ftitmp \wedge \tforja(\Phi)))}$

  \item $\tforja(\bigodot \bigodot \bigodot\BOX{\Phi}) =$\\
    \hspace*{10mm}$\bigodot \bigodot \bigodot\BOX{\mu Z.((\ftitmp \wedge
      \BOX{Z} \wedge \DIAM{\top}) \vee (\neg \ftitmp \wedge
      \tforja(\Phi)))}$
  \end{compactitem}








  \noindent
  where:
  \begin{compactitem}
  \item $\circ$ is a binary operator ($\vee, \wedge, \ra,$ or $\lra$),
  \item $\circledcirc$ is least ($\mu$) or greatest ($\nu$) fix-point
    operator,
  \item $\Q$ is forall ($\forall$) or existential ($\exists$)
    quantifier.
  \item $\bigodot$ is box ($\BOX$) or diamond ($\DIAM$) modal
    operator.
  \end{compactitem}
\ \ 
\end{definition}

\subsubsection{Alternating Jumping Bisimulation
  (\ajbsimabr-Bisimulation)}

As a vehicle to reduce the verification of \bagkabs into S-GKABs, in
this section we introduce the notion of alternating jumping
bisimulation (\ajbsimabr-Bisimulation). Additionally, here we also
show an important lemma about the situation where two
\ajbsimabr-bisimilar transition systems can not be distinguished by
certain temporal properties.

As a preliminary towards defining the notion of
\ajbsimabr-Bisimulation, below we introduce the notion of equality
between states that ignores the special markers and also consider
different representations of contextual information.
\begin{definition}[Contextually Equal Modulo Special
  Markers]\label{def:cs-equal-mod-markers} \ \\
  Let \sidetext{Contextually Equal Modulo Special Markers}
  $\ts{1}~=~\tup{\const, \ctbox, \stateset_1, s_{01}, \abox_1, \cntx,
    \actsrc, \actpar, \fadd, \fdel, \trans_1}$
  be a fine-grained transition system, and
  $\ts{2} = \tup{\const, T, \stateset_2, s_{02}, \abox_2, \trans_2}$
  be a KB transition system.
  Consider two states $s_1 \in \stateset_1$ and $s_2 \in \stateset_2$,
  we say $s_1$ is \emph{contextually equal modulo special markers} to
  $s_2$, written $s_1 \eqmc s_2$, if the following hold
  \begin{compactitem}
  \item $\voc(T) = \voc(\ctbox)$,
  \item For each concept name $N \in \voc(T)$ (i.e., $N$ is not a
    special marker concept name), we have a concept assertion
    $N(c) \in A_1$ if and only if a concept assertion $N(c) \in A_2$,
  \item For each role name $P \in \voc(T)$, we have a role assertion
    $P(c_1,c_2) \in A_1$ if and only if a role assertion
    $P(c_1,c_2) \in A_2$.
  \item We have context ABox assertion
    $\cdcq(\ctxconst) \in \abox_2(s_2)$ if and only if
    $\cdcq(\ctxconst) \in A_{\cntx(s_1)}$ (recall that
    $A_{\cntx(s_1)}$ is the set of ABox assertions that represents a
    context as defined in \Cref{def:abox-context}).
  \end{compactitem}
\ \ 
\end{definition}


\noindent
The \ajbsimabr-bisimulation is then defined as follows:

\begin{definition}[Alternating Jumping Bisimulation
  (\ajbsimabr-Bisimulation)] \
  \sidetext{Alternating Jumping Bisimulation (\ajbsimabr-Bisimulation)} \\
  Let
  $\ts{1} = \tup{\const, \ctbox, \stateset_1, s_{01}, \abox_1, \cntx,
    \actsrc, \actpar, \fadd, \fdel, \trans_1}$
  be a fine-grained transition system, and
  $\ts{2} = \tup{\const, T, \stateset_2, s_{02}, \abox_2, \trans_2}$
  be a typed KB transition system, with
  $\adom{\abox_1(s_{01})} \subseteq \const$,
  $\adom{\abox_2(s_{02})} \subseteq \const$, and $s_{01}$ as well as
  $s_{02}$ are stable states.
  An alternating jumping bisimulation (\ajbsimabr-Bisimulation)
  between $\ts{1}$ and $\ts{2}$ is a relation
  $\B \subseteq \Sigma_1 \times\Sigma_2$ such that
  $\tup{s_1, s_2} \in \B$ implies that one of the following condition
  must hold:
\begin{compactenum}

\item $s_{1}$ as well as $s_{2}$ both are either stable states,
  service call evaluation states, or context change states and we have
  the following:
  \begin{compactenum}
  \item $s_1 \eqmc s_2$
  \item for each $s_1'$, if $s_1 \Rightarrow_1 s_1'$ then there exists
    $s_2'$ with $ s_2 \Rightarrow_2 s_2'$ such that
    $\tup{s_1', s_2'}\in\B$.
  \item for each $s_2'$, if $ s_2 \Rightarrow_2 s_2'$, then there
    exists $s_1'$ with $s_1 \Rightarrow_1 s_1'$, such that
    $\tup{s_1', s_2'}\in\B$.
  \end{compactenum}
  or
\item $s_{1}$ and $s_{2}$ are both filter application states.
\begin{compactenum}
  \item $s_1 \eqmc s_2$
  \item for each $s_1'$, if $s_1 \Rightarrow_1 s_1'$ then there exists
    $t_1, \ldots, t_n$ (for $n\geq 0$) and $s_2'$ with
    \[
    s_2 \Rightarrow_2 t_1 \Rightarrow_2 \cdots \Rightarrow_2 t_n
    \Rightarrow_2 s_2'
    \]
    such that $\tup{s_1', s_2'}\in\B$, $t_1, \ldots, t_n$ are filter
    application intermediate states, $s_1'$ and $s_2'$ are stable
    states.
  \item for each $s_2'$, if 
    \[
    s_2 \Rightarrow_2 t_1 \Rightarrow_2 \cdots \Rightarrow_2 t_n
    \Rightarrow_2 s_2'
    \]
    (for $n\geq 0$) with $t_1, \ldots, t_n$ are filter application
    intermediate states and $s_2'$ is a stable state, then there
    exists $s_1'$ with $s_1 \Rightarrow_1 s_1'$, such that
    $\tup{s_1', s_2'}\in\B$, $s_1'$ and $s_2'$ are stable states.
  \end{compactenum}

\end{compactenum}
\ \ 
\end{definition}

\noindent
Let
$\ts{1} = \tup{\const, \ctbox, \stateset_1, s_{01}, \abox_1, \cntx,
  \actsrc, \actpar, \fadd, \fdel, \trans_1}$
be a fine-grained transition system, and
$\ts{2} = \tup{\const, T, \stateset_2, s_{02}, \abox_2, \trans_2}$ be
a KB transition system, 
a state $s_1 \in \stateset_1$ is \emph{\ajbsimabr-bisimilar} to
$s_2 \in \stateset_2$, written $s_1 \ajbsim s_2$, if there exists an
\ajbsimabr-bisimulation relation $\B$ between $\ts{1}$ and $\ts{2}$
such that $\tup{s_1, s_2}\in\B$.
A transition system $\ts{1}$ is \emph{\ajbsimabr-bisimilar} to
$\ts{2}$, written $\ts{1} \ajbsim \ts{2}$, if there exists an
\ajbsimabr-bisimulation relation $\B$ between $\ts{1}$ and $\ts{2}$
such that $\tup{s_{01}, s_{02}}\in\B$.

In the following two lemmas, we show some important properties of
\ajbsimabr-bisimilar states and transition systems that will be useful
later to show that we can recast the verification of \bagkabs into
S-GKABs. Essentially, we show that given a fine-grained transition
system $\ts{1}$ and a typed KB transition system $\ts{2}$ such that
they are \ajbsimabr-bisimilar, we have that $\ts{1}$ satisfies a
\mulcsa formula implies that $\ts{2}$ satisfies the same formula
modulo translation $\tforja$ and vice versa.

\begin{lemma}\label{lem:ajbisimilar-state-satisfies-same-formula}
  Let
  $\ts{1} = \tup{\const, \ctbox, \stateset_1, s_{01}, \abox_1, \cntx,
    \actsrc, \actpar, \fadd, \fdel, \trans_1}$
  be a fine-grained transition system, and
  $\ts{2} = \tup{\const, T, \stateset_2, s_{02}, \abox_2, \trans_2}$
  be a typed KB transition system,
  Consider two states $s_1 \in \stateset_1$ and $s_2 \in \stateset_2$
  such that $s_1 \ajbsim s_2$. Then for every formula $\Phi$ (in NNF)
  of $\mulcsa$, and every valuations $\vfo_1$ and $\vfo_2$ that assign
  to each of its free variables a constant
  $c_1 \in \adom{\abox_1(s_1)}$ and $c_2 \in \adom{\abox_2(s_2)}$,
  such that $c_1 = c_2$, we have that
  \[
  \ts{1},s_1 \models \Phi \vfo_1 \textrm{ if and only if } \ts{2},s_2
  \models \tforja(\Phi) \vfo_2.
  \]
\end{lemma}
\begin{proof}
  Similar to the combination of the proof for
  \Cref{lem:stbisimilar-state-satisfies-same-formula,lem:jumping-bisimilar-states-satisfies-same-formula}
  by also considering that by the definition of typed KB transition
  system and fine-grained transition system, both of them start from a
  stable state. Additionally, for the transition among the states, it
  is always be the case that there are three intermediate states
  (i.e., service call evaluation state, context change state, and
  filter application state) between the two stable states. On the
  other hand, the \mulcsa formula always have four consecutive modal
  operators. Hence, it is easy to see that we always verify query and
  context expression in the given \mulcsa formula over a stable state.
\end{proof}

\begin{lemma}\label{lem:ajbisimilar-ts-satisfies-same-formula}
  Consider a fine-grained transition system $\ts{1}$, and a typed KB
  transition system $\ts{2}$ such that $\ts{1} \ajbsim \ts{2}$.  For
  every closed \mulcsa formula $\Phi$ (in NNF), we have:
  \[
  \ts{1} \models \Phi \textrm{ if and only if } \ts{2} \models
  \tforja(\Phi)
  \]
\end{lemma}
\begin{proof} 
%
  Let
  $\ts{1} = \tup{\const, \ctbox, \stateset_1, s_{01}, \abox_1, \cntx,
    \actsrc, \actpar, \fadd, \fdel, \trans_1}$,
  and
  $\ts{2} = \tup{\const, T, \stateset_2, s_{02}, \abox_2, \trans_2}$.
  By the definition of \ajbsimabr-bisimilar transition system, we have
  that $s_{01} \ajbsim s_{02}$. Thus, we obtain the proof as a
  consequence of \Cref{lem:ajbisimilar-state-satisfies-same-formula},
  due to the fact that
  \[
  \ts{1}, s_{01} \models \Phi \textrm{ if and only if } \ts{2}, s_{02}
  \models \tforja(\Phi)
  \]
\ \ 
\end{proof}

\subsubsection{Reducing the Verification of \bagkabs into S-GKABs}

In this section we show that we can recast the verification of
\bagkabs into S-GKABs. In the following two lemmas we aim to show that
the transition system of a \bagkab $\agkabsym$ is \ajbsimabr-bisimilar
to the transition system of the corresponding S-GKAB
$\tgkabba(\agkabsym)$ that is obtained via translation $\tgkabba$.

\begin{lemma}\label{lem:bagkab-to-sgkab-bisimilar-state}
  Let $\agkabsym$ be a \bagkab with transition system
  $\ts{\agkabsym}^{\csfilter_B}$, and let $\tgkabba(\agkabsym)$ be
  its corresponding S-GKAB $($with transition system
  $\ts{\tgkabba(\agkabsym)}^{\filter_S}$$)$
  obtain through
  $\tgkabba$. 
  Consider
  \begin{inparaenum}[]
  \item a state $s_{cx}
    = \tup{id_{cx}, A_{cx},\scmap_{cx}, \ctx,
      \delta_{cx}}$ of $\ts{\agkabsym}^{\csfilter_B}$ and
  \item a state $s_s = \tup{A_s,\scmap_s, \delta_s}$ of
    $\ts{\tgkabba(\agkabsym)}^{\filter_S}$.
  \end{inparaenum}
  If the following hold:
  \begin{compactenum}
  \item $s_{cx}$ and $s_s$ are having the same state type,
    \item $s_{cx} \eqmc s_s$ (see \Cref{def:cs-equal-mod-markers}),
    \item $\scmap_{cx} = \scmap_s$,
    \item $\fadd(s_{cx}) = \addass{A_s}$ (see
      \Cref{def:added-assertion} for the definition of $\addass{A_s}$),
    \item $\fdel(s_{cx}) = \delass{A_s}$ (see
      \Cref{def:deleted-assertion} for the definition of $\delass{A_s}$),
    \item $\delta_s = \tgprogba(\delta_{cx})$ (if $s_{cx}$ and $s_s$ are stable states),
    \item $\actsrcconceptname(\act)
      \in A_s$ (if $s_{cx}$ and
      $s_s$ are not stable states, and $\actsrc(s_{cx}) = \act$),
    \item $\set{\parconceptname{1}(\sigma(p_1)),
        \ldots, \parconceptname{m}(\sigma(p_m))} = \parass{A_s}$ (if $s_{cx}$ and
      $s_s$ are not stable states, and $\actpar(s_{cx}) = \sigma$),
    \end{compactenum}
    then $s_{cx} \ajbsim s_s$,
\end{lemma}
\begin{proof}
  Let
\begin{compactitem}
\item
  $\agkabsym = \tup{\ctbox, \initabox, \actset, \ginitprog, \initctx,
    \ctxprocset}$, and \\
  $\ts{\agkabsym}^{\csfilter_B} = \tup{\const, \ctbox, \stateset_1,
    s_{01}, \abox_1, \cntx, \actsrc, \actpar, \fadd, \fdel,
    \trans_1}$,
\item
  $\tgkabba(\agkabsym) = \tup{T_\cdimset, \initabox \cup A_{\initctx},
    \actset', \ginitprog'}$, and \\
  $\ts{\tgkabba(\agkabsym)}^{\filter_S} = \tup{\const, T_\cdimset, \stateset_2,
    s_{02}, \abox_2, \trans_2}$.
\end{compactitem}
First, observe that by \Cref{lem:ts-sgkab-of-bagkab-is-typed}, we have
that $\ts{\tgkabba(\agkabsym)}^{\filter_S}$
is typed KB transition system. Hence, in
$\ts{\tgkabba(\agkabsym)}^{\filter_S}$
we have the alternation between state types.

To prove the lemmma, in the following we show that for every state
$s_{cx}'
= \tup{id_{cx}', A_{cx}',\scmap_{cx}', \ctx',
  \delta_{cx}'}$ such that $s_{cx} \trans_1
s_{cx}'$, there exists $s_s' = \tup{A_s',\scmap_s',
  \delta_s'}$ such that $s_{s} \trans_2 s_{s}'$ and the following hold
  \begin{compactenum}
  \item $s_{cx}'$ and $s_s'$ are having the same state type,
    \item $s_{cx}' \eqmc s_s'$,
    \item $\scmap_{cx}' = \scmap_s'$,
    \item $\fadd(s_{cx}') = \addass{A_s'}$,
    \item $\fdel(s_{cx}') = \delass{A_s'}$,
    \item $\delta_s' = \tgprogba(\delta_{cx}')$ (if $s_{cx}'$ and $s_s'$ are stable states),
    \item $\actsrcconceptname(\act)
      \in A_s'$ (if $s_{cx}'$ and
      $s_s'$ are not stable states, and $\actsrc(s_{cx}') = \act$),
    \item $\set{\parconceptname{1}(\sigma(p_1)),
        \ldots, \parconceptname{m}(\sigma(p_m))} = \parass{A_s'}$ (if $s_{cx}'$ and
      $s_s'$ are not stable states, and $\actpar(s_{cx}') = \sigma$),
    \end{compactenum}
%
%
    To show the claim, we have to separately discuss the case in
    which:
    \begin{compactenum}
    \item both $s_{cx}$ and $s_s$ are stable states,
    \item both $s_{cx}$ and $s_s$ are service call evaluation states,
    \item both $s_{cx}$ and $s_s$ are context change states, and
    \item both $s_{cx}$ and $s_s$ are filter application states.
    \end{compactenum}


\begin{itemize}

\item[\textbf{Base case.}]  trivially true from the shape of the
  initial states of $\ts{\agkabsym}^{\csfilter_B}$ and
  $\ts{\tgkabba(\agkabsym)}^{\filter_S}$.

\item[\textbf{Case 1} \textbf{- $s_{cx}$ and $s_s$ are stable
    states:}]  
  Now we have to show that for every state
  $s_{cx}' = \tup{id_{cx}', A_{cx}',\scmap_{cx}', \ctx',
    \delta_{cx}'}$
  such that $s_{cx} \trans_1 s_{cx}'$, there exists
  $s_s' = \tup{A_s',\scmap_s', \delta_s'}$ such that
  $s_{s} \trans_2 s_{s}'$ and the following hold
  \begin{compactenum}[(i)]
  \item $s_{cx}'$ and $s_s'$ are service call evaluation states,
  \item $s_{cx}' \eqmc s_s'$,
  \item $\scmap_{cx}' = \scmap_s'$,
    \item $\fadd(s_{cx}') = \addass{A_s'}$,
    \item $\fdel(s_{cx}') = \delass{A_s'}$,
    \item $\actsrc(s_{cx}') = \act$ and $\actsrcconceptname(\act) \in A_s'$,
    \item $\actpar(s_{cx}') = \sigma$ and $\set{\parconceptname{1}(\sigma(p_1)),
        \ldots, \parconceptname{m}(\sigma(p_m))} = \parass{A_s'}$.
    \end{compactenum}

Now, by definition of $\ts{\agkabsym}^{\csfilter_B}$,
since $s_{cx} \trans_1 s_{cx}'$, we have
\[
\tup{A_{cx},\scmap_{cx}, \ctx, \delta_{cx}}
\gprogtrans{\alpha\sigma, \csfilter_B}
\tup{A_{cx},\scmap_{cx}, \ctx, \delta'_{cx}}.
\]
Hence, by the definition of $\gprogtrans{\act\sigma, \csfilter_B}$
(\Cref{def:agkab-prog-exec-relation}), we have that:
\begin{compactitem}

\item $\sigma$ is a legal parameter assignment for $\act$ in $A_{cx}$
  w.r.t.\ context $\ctx$ and an action invocation
  $\gactc{Q(\vec{p})}{\ctxe}{\act(\vec{p})}$ (i.e.,
  $\ask(Q\sigma, \ctbox^\ctx, A_{cx}) = \true$). Notice that w.l.o.g.\
  $\gactc{Q(\vec{p})}{\ctxe}{\act(\vec{p})}$ is the next instruction
  that should be executed in $\delta_{cx}$.

\item $\ctx \cup \ctxth \models \ctxe$.

\end{compactitem}
Additionally, we have $\actsrc(s_{cx}') = \act$, and 
$\actpar(s_{cx}') = \sigma$.

Now, on the other hand, since $\delta_s = \tgprogba(\delta_{cx})$, by
the definition of $\tgprogba$ (see \Cref{def:tran-prog-bagkab-sgkab}),
we have
\[
\begin{array}{@{}lll@{}}
  \tgprogba(\gactc{Q(\vec{p})}{\ctxe}{\act(\vec{p})}) &= &
                                                             \gact{Q'(\vec{p})}{\act_a(\vec{p})};\gact{\true}{\act_b()};\delta_{\ctxprocset};\\
                                                      &&
                                                         \gact{\true}{\act^u()};
                                                         \delta^{\ctbox}_b ; \gact{\true}{\act^-_{temp}()}\\
\end{array}
\]
where $\gact{Q'(\vec{p})}{\act_a(\vec{p})};\gact{\true}{\act_b()}$ is
a split action invocation obtained from
$\gactc{Q(\vec{p})}{\ctxe}{\act(\vec{p})}$ as in
\Cref{def:split-action-invoc}. 
Thus, we have that $Q' = Q_\ctxb \wedge q_{\ctxe}$.
Since $\ctx \cup \ctxth \models \ctxe$ and $q_{\ctxe}$ only use
context dimension concept, by \Cref{lem:ctx-exp-and-query-about-it},
it is easy to see that $\Ans(q_{\ctxe}, T_\cdimset, A_s) = \true$.
Furthermore, by \Cref{lem:correctness-contextually-compiled-query}, we
have that
$\Ans(Q,\ctbox^{\ctx}, A_{cx}) = \Ans(Q_\ctxb, T_\cdimset, A_s)$.
Therefore, now we can construct $\sigma_s$ that maps parameters of
$\act'$ to constants in $\adom{A_s}$ such that $\sigma_c =
\sigma_s$.
Therefore, by definition of split action (see \Cref{def:split-act}),
we have $\actsrcconceptname(\act) \in A_s'$, and
$\set{\parconceptname{1}(\sigma(p_1)),
  \ldots, \parconceptname{m}(\sigma(p_m))} = \parass{A_s'}$.
Next, by the definition of split action it can be seen easily that 
\begin{compactitem}
\item $\scmap_{cx}' = \scmap_s'$, because $\act_a$ does not involve
  any service call and $\act$ does not update the service call map).
\item $\fadd(s_{cx}') = \addass{A_s'}$, because $\fadd(s_{cx}')$ is
  the set of ABox assertions in
  $\addfacts{\ctbox^{\ctx}, A_{cx}, \act\sigma}$ (excluding the ground
  skolem terms) and $\act_a$ only add information about ABox
  assertions to be added in which their construction do not involve
  any service call.
\item $\fdel(s_{cx}') = \delass{A_s'}$, because
  $\fdel(s_{cx}') = \delfacts{\ctbox^{\ctx}, A_{cx}, \act\sigma}$ and
  $\act_a$ add information about all ABox assertions to be deleted by
  action $\act$.
\item $s_{cx}' \eqmc s_s'$, because the context does not change and
  also the ABox stay the same. Additionally, $\act_a$ only add some
  assertions that is made by special reserved concept/role names.
\item $s_{cx}'$ and $s_s'$ are service call evaluation states. Because
  $\act_a$ adds the assertion $\sctmp$ and also by the construction of
  $\ts{\agkabsym}^{\csfilter_B}$.

\end{compactitem}

Since $s_{cx}'$ and $s_s'$ are service call evaluation states, then
the 
case 2 is applicable.

\item[\textbf{Case 2} \textbf{- $s_{cx}$ and $s_s$ are service call
    evaluation states.}]  
  Now we have to show that for every state
  $s_{cx}' = \tup{id_{cx}', A_{cx}',\scmap_{cx}', \ctx',
    \delta_{cx}'}$
  such that $s_{cx} \trans_1 s_{cx}'$, there exists
  $s_s' = \tup{A_s',\scmap_s', \delta_s'}$ such that
  $s_{s} \trans_2 s_{s}'$ and the following hold
  \begin{compactenum}[(i)]
  \item $s_{cx}'$ and $s_s'$ are context change states,
  \item $s_{cx}' \eqmc s_s'$,
  \item $\scmap_{cx}' = \scmap_s'$,
    \item $\fadd(s_{cx}') = \addass{A_s'}$,
    \item $\fdel(s_{cx}') = \delass{A_s'}$,
    \item $\actsrc(s_{cx}') = \act$ and $\actsrcconceptname(\act) \in A_s'$,
    \item $\actpar(s_{cx}') = \sigma$ and $\set{\parconceptname{1}(\sigma(p_1)),
        \ldots, \parconceptname{m}(\sigma(p_m))} = \parass{A_s'}$.
    \end{compactenum}

    Now, let $\actsrc(s_{cx}) = \act$ with parameters $p_1, \ldots, p_m$
    and $\actpar(s_{cx}) = \sigma$.
    By definition of $\ts{\agkabsym}^{\csfilter_B}$,
    since $s_{cx} \trans_1 s_{cx}'$, we have that there exists
    $ \theta \in \eval{\addfacts{\ctbox^{\ctx}, A, \act\sigma}}$, and
    $\scmap_{cx}' = \scmap_{cx} \cup \theta$.

    On the other hand, by the definition of $\tgprogba$ (see
    \Cref{def:tran-prog-bagkab-sgkab}) and the states alternation in
    the typed KB transition system, the next action invocation to be
    executed in the state $s_{s}$ is $\gact{\true}{\act_b()}$ where
    $\act_b$ is the second split action of $\act$. Thus, by the
    definition of execution semantics of S-GKABs, the transition
    $s_s \trans_2 s_s'$ involves the following:
    \begin{compactitem}
    \item the execution of action $\act_b$ with legal parameter
      assignments $\sigma_s$ where  $\sigma_s$  is an empty
      substitution because $\act_b$ is a 0-ary action. 
    \item the service call substitution $\theta_s$ which evaluates all
      ground skolem terms generated by the execution of $\act_b$.
    \end{compactitem}
    By the definition of split action, It is easy to see that
    $\theta_s = \theta$ because $m_{cx} = m_s$ and $\act_b$ is the
    second split action of $\act$ which add all of assertions that is
    added by $\act$ and involve service call ($\act_b$ and $\act$
    involve the same service call).

    Next, by the definition of split action, it can be seen easily
    that:
    \begin{compactitem}
    \item $m_{cx}' = m_s'$ because $\theta_s = \theta$,
      $m_{cx} = m_s$, $m_{cx}' = m_{cx} \cup \theta$ and
      $m_{s}' = m_{s} \cup \theta_s$.

    \item $\fadd(s_{cx}') = \addass{A_s'}$, because
      \begin{compactenum}[(1)]
      \item $\fadd(s_{cx}) = \addass{A_s}$ and they contain the set of
        ABox assertions in
        $\addfacts{\ctbox^{\ctx}, A_{cx}, \act\sigma}$ (excluding the
        ground skolem terms),
      \item
        $\fadd(s_{cx}') = \addfacts{\ctbox^{\ctx}, A_{cx},
          \act\sigma}\theta$,

      \item The set of ground skolem terms in
        $\addfacts{\ctbox^{\ctx}, A_{cx}, \act\sigma}$ and
        $\addfacts{T_{\cdimset}, A_{s}, \act_b\sigma_s}$ are the same
        because of the definition of split action and also the following:
        \begin{compactitem}
        \item
          $\set{\parconceptname{1}(\sigma(p_1)),
            \ldots, \parconceptname{m}(\sigma(p_m))} = \parass{A_s}$,
        \item $\act_b$ is the second split action of $\act$.
        \end{compactitem}
    
    \item $\act_b$ is the second split action of $\act$ which add all
      of assertions that is added by $\act$ and involve service call
      ($\act_b$ and $\act$ involve the same service call). Let
      $A_{\act_b}$ be the set of ABox assertions that are made by
      added fact marker concept/role names and added by $\act_b$, then
      it is easy to see that
      $\addass{A_s'} = \addass{A_s} \cup A_{\act_b}$.
      \item Now, since we also have that $\act$ and $\act_b$ involve
        the same service call, $m_{cx} = m_s$ and $\theta = \theta_s$,
        it is easy to see that
        $\addass{A_s'} = \addfacts{\ctbox^{\ctx}, A_{cx},
          \act\sigma}\theta$.

      \end{compactenum}

    \item $\fdel(s_{cx}') = \delass{A_s'}$, because
      $\fdel(s_{cx}) = \delass{A_s} = \delfacts{\ctbox^{\ctx}, A_{cx},
        \act\sigma}$,
      $\fdel(s_{cx}') = \fdel(s_{cx})$, and $\act_b$ does not add any
      ABox assertion made by the deleted fact marker concept/role names.

    \item $s_{cx}' \eqmc s_s'$, because $s_{cx} \eqmc s_s$, the
      context does not change and also the ABox (apart from the
      special markers) stay the same. Additionally, $\act_b$ only add/delete
      some assertions that is made by special reserved concept/role
      names.

    \item $\actsrcconceptname(\act) \in A_s'$ simply because
      $\actsrcconceptname(\act) \in A_s$ and $\act_b$ does nothing
      w.r.t.\ the concept assertion made by $\actsrcconceptname$,
    \item
      $\set{\parconceptname{1}(\sigma(p_1)),
        \ldots, \parconceptname{m}(\sigma(p_m))} = \parass{A_s'}$
      because
      $\set{\parconceptname{1}(\sigma(p_1)),
        \ldots, \parconceptname{m}(\sigma(p_m))} = \parass{A_s}$,
      and $\act_b$ does nothing w.r.t.\ the concept assertion made by
      action parameter concept names.
    \item $s_{cx}'$ and $s_s'$ are context change states. Because of
      the construction of $\ts{\agkabsym}^{\csfilter_B}$ as well as
      because $\act_b$ removes the assertion $\sctmp$ and adds the
      assertion $\cxtmp$.

    \end{compactitem}

    Since $s_{cx}'$ and $s_s'$ are context change states, then the
    case 3 is applicable.

  \item[\textbf{Case 3} \textbf{- $s_{cx}$ and $s_s$ are context
      change states.}]  
    Now we have to show that for every state
    $s_{cx}' = \tup{id_{cx}', A_{cx}',\scmap_{cx}', \ctx',
      \delta_{cx}'}$
    such that $s_{cx} \trans_1 s_{cx}'$, there exists
    $s_s' = \tup{A_s',\scmap_s', \delta_s'}$ such that
    $s_{s} \trans_2 s_{s}'$ and the following hold
    \begin{compactenum}[(i)]
    \item $s_{cx}'$ and $s_s'$ are filter application states,
    \item $s_{cx}' \eqmc s_s'$,
    \item $\scmap_{cx}' = \scmap_s'$,
    \item $\fadd(s_{cx}') = \addass{A_s'}$,
    \item $\fdel(s_{cx}') = \delass{A_s'}$,
    \item $\actsrc(s_{cx}') = \act$ and
      $\actsrcconceptname(\act) \in A_s'$,
    \item $\actpar(s_{cx}') = \sigma$ and
      $\set{\parconceptname{1}(\sigma(p_1)),
        \ldots, \parconceptname{m}(\sigma(p_m))} = \parass{A_s'}$.
    \end{compactenum}
    Now, let $\actsrc(s_{cx}) = \act$ with parameters $p_1, \ldots, p_m$
    and $\actpar(s_{cx}) = \sigma$.
    By definition of $\ts{\agkabsym}^{\csfilter_B}$,
    since $s_{cx} \trans_1 s_{cx}'$, we have that
    $\tup{A_{cx}, \ctx, \ctx'} \in \ctxchg$. 
    Therefore, by \Cref{def:ctx-chg-relation}, there exists a
    context-evolution rule $\tup{Q, \ctxe} \mapsto C_{new}$ in
    $\ctxprocset$ s.t.:
    \begin{compactenum}
    \item $\ask(Q, \ctbox^C, A_{cx})$ is $\true$;
    \item $C \cup \ctxth \models \ctxe$;
    \item for every context dimension $d\in\cdimset$ s.t.\
      $\cval{d}{v} \in C_{new}$, \\we have $\cval{d}{v} \in C'$;
    \item for every context dimension $d\in\cdimset$ s.t.\
      $\cval{d}{v} \in C$, and there does not exist any $v_2$ s.t.\
      $\cval{d}{v_2} \in C_{new}$, we have $\cval{d}{v} \in C'$.
    \end{compactenum}

    On the other hand, by the definition of $\tgprogba$ (see
    \Cref{def:tran-prog-bagkab-sgkab}) and the states alternation in
    the typed KB transition system, the part of the program to be
    executed in the state $s_{s}$ is $\delta_{\ctxprocset}$.
    By the definition of $\delta_{\ctxprocset}$, there exist an action
    invocation $\gact{Q'}{\act^s_{\ctx}()}$ that is obtained from
    $\tup{Q, \ctxe} \mapsto C_{new}$. Since $s_{cx} \eqmc s_s$ it is
    easy to see that there exists $A'_s$ such that
    $\ctxass{A'_s} = A_{\ctx'}$ and $A'_s$ is obtained by the
    execution of $\act^s_{\ctx}$.

    Next, it can be seen easily that:
    \begin{compactitem}
    \item $m_{cx}' = m_s'$ because $m_{cx} = m_s$ and both transitions
      $s_{cx} \trans_1 s_{cx}'$ and $s_{s} \trans_2 s_{s}'$ do not
      involve any service calls.

    \item $\fadd(s_{cx}') = \addass{A_s'}$, because
      $\fadd(s_{cx}) = \addass{A_s}$,
      $\fadd(s_{cx}') = \fadd(s_{cx})$, and
      $\addass{A_s'} = \addass{A_s}$ because $\act_\ctx^s$ does not
      add any ABox assertion made by the added fact marker
      concept/role names.

    \item $\fdel(s_{cx}') = \delass{A_s'}$, because
      $\fdel(s_{cx}) = \delass{A_s}$,
      $\fdel(s_{cx}') = \fdel(s_{cx})$, and
      $\delass{A_s'} = \delass{A_s}$ because $\act_\ctx^s$ does not
      add any ABox assertion made by the deleted fact marker
      concept/role names.

    \item $s_{cx}' \eqmc s_s'$, because $\ctxass{A'_s} = A_{\ctx'}$
      and because $\act_\ctx^s$ only changes the context related
      information.

    \item $\actsrcconceptname(\act) \in A_s'$ simply because
      $\actsrcconceptname(\act) \in A_s$ and $\act_\ctx^s$ does
      nothing w.r.t.\ the concept assertion made by
      $\actsrcconceptname$,

    \item
      $\set{\parconceptname{1}(\sigma(p_1)),
        \ldots, \parconceptname{m}(\sigma(p_m))} = \parass{A_s'}$
      because
      $\set{\parconceptname{1}(\sigma(p_1)),
        \ldots, \parconceptname{m}(\sigma(p_m))} = \parass{A_s}$,
      and $\act_\ctx^s$ does nothing w.r.t.\ the concept assertion
      made by action parameter concept names.

    \item $s_{cx}'$ and $s_s'$ are filter application states. Because
      $\act_\ctx^s$ removes the assertion $\cxtmp$ and adds the
      assertion $\fttmp$ and also because of the construction of
      $\ts{\agkabsym}^{\csfilter_B}$.

    \end{compactitem}

    Since $s_{cx}'$ and $s_s'$ are filter application states, then the
    case 4 is applicable.

\item[\textbf{Case 4} \textbf{- $s_{cx}$ and $s_s$ are filter application states.}]
%
  Now we have to show that for every state
  $s_{cx}' = \tup{id_{cx}', A_{cx}',\scmap_{cx}', \ctx',
    \delta_{cx}'}$
  such that $s_{cx} \trans_1 s_{cx}'$, there exists $t_1, \ldots, t_n$
  (for $n \geq 0$), and $s_s' = \tup{A_s',\scmap_s', \delta_s'}$ such
  that
  $s_{s} \trans_2 t_1 \trans_2 \ldots \trans_2 t_n \trans_2 s_{s}'$
  and the following hold
  \begin{compactenum}[(i)]
  \item $s_{cx}'$ and $s_s'$ are stable states,
  \item $s_{cx}' \eqmc s_s'$,
  \item $\scmap_{cx}' = \scmap_s'$,
  \item $\fadd(s_{cx}') = \addass{A_s'}$,
  \item $\fdel(s_{cx}') = \delass{A_s'}$,
  \item $\delta_s' = \tgprogba(\delta_{cx}')$.
  \end{compactenum}
  By the definition of $\ts{\agkabsym}^{\csfilter_B}$, since
  $s_{cx} \trans_1 s_{cx}'$, we have that
  \[
  \tup{A_{cx}, \fadd(s_{cx}), \fdel(s_{cx}), \ctx', A_{cx}'} \in
  \csfilter_B,
  \]
  On the other hand, by the definition of $\tgprogba$ (see
  \Cref{def:tran-prog-bagkab-sgkab}) and the states alternation in the
  typed KB transition system, the part of the program to be executed
  in the state $s_{s}$ is
  \[
  \gact{\true}{\act^u()}; \delta^{\ctbox}_b ;
  \gact{\true}{\act^-_{temp}()}.
  \]
  Hence, by considering those program above that need to be executed
%
%
%
%
%
  and also the correctness of b-repair program $\delta^{\ctbox}_b$ in
  simulating b-repair computation as in
  \Cref{sec:term-corr-cs-brep-prog},
  we can easily see that there exists states $t_1, \ldots, t_n$ (for
  $n \geq 0$), and $s_s' = \tup{A_s',\scmap_s', \delta_s'}$ with
  $s_{s} \trans_2 t_1 \trans_2 \ldots \trans_2 t_n \trans_2 s_{s}'$
  and the following hold:
  \begin{compactitem}
  \item $t_1, \ldots, t_n$ are filter application intermediate states,
    because $\act^u$ removes the assertion $\fttmp$ and add the
    assertion $\ftitmp$.

  \item $s_{cx}'$ and $s_s'$ are stable states, because
    $\act^-_{temp}$ removes the assertion $\ftitmp$ and also by the
    construction of $\ts{\agkabsym}^{\csfilter_B}$.

  \item $s_{cx}' \eqmc s_s'$, because of the following:
    \begin{compactitem}
    \item $s_{cx} \eqmc s_s$,
    \item the transition
      $s_s \trans_2 t_1 \trans_2 \ldots \trans_2 t_n \trans_2 s_s'$ do
      not change any context information,
    \item the correctness of b-repair program $\delta^{\ctbox}_b$ in
      simulating b-repair computation, and essentially $A_s'$ is the
      b-repair of $A_s$.
    \end{compactitem}
  
  \item $\scmap_{cx}' = \scmap_s'$, because both of the transitions
    $s_{cx} \trans_1 s_{cx}'$ and
    $s_s \trans_2 t_1 \trans_2 \ldots \trans_2 t_n \trans_2 s_s'$ do
    not involve any service calls.

  \item $\fadd(s_{cx}') = \addass{A_s'}$, because $\act^u$ removes
    all ABox assertions made by added fact marker concept/role names.

  \item $\fdel(s_{cx}') = \delass{A_s'}$, because $\act^u$ removes all
    ABox assertions made by deleted fact marker concept/role names.

  \item $\delta_s' = \tgprogba(\delta_{cx}')$, by the definition of
    $\tgprogba$ (see \Cref{def:tran-prog-bagkab-sgkab}) and also the
    construction of $\ts{\agkabsym}^{\csfilter_B}$.
  \end{compactitem}

\end{itemize}

The other direction can be shown similarly.
\end{proof}

\begin{lemma}\label{lem:bagkab-to-sgkab-bisimilar-ts}
  Given a \bagkab $\agkabsym$ with transition system
  $\ts{\agkabsym}^{\csfilter_B}$, let $\tgkabba(\agkabsym)$ be its
  corresponding S-GKAB $($with transition system
  $\ts{\tgkabba(\agkabsym)}^{\filter_S}$$)$
  obtained through $\tgkabba$.
  We have that $\ts{\agkabsym}^{\csfilter_B}
  \ajbsim \ts{\tgkabba(\agkabsym)}^{\filter_S}$
\end{lemma}
\begin{proof}
  Let
\begin{compactitem}
\item
  $\agkabsym = \tup{\ctbox, \initabox, \actset, \ginitprog, \initctx,
    \ctxprocset}$, and \\
  $\ts{\agkabsym}^{\csfilter_B} = \tup{\const, \ctbox, \stateset_1,
    s_{01}, \abox_1, \cntx, \actsrc, \actpar, \fadd, \fdel,
    \trans_1}$,
\item $\tgkabba(\agkabsym) = \tup{T_\cdimset, \initabox',
    \actset', \ginitprog'}$, and \\
  $\ts{\tgkabba(\agkabsym)}^{\filter_S} = \tup{\const, T_\cdimset,
    \stateset_2, s_{02}, \abox_2, \trans_2}$.
\end{compactitem}
We have that
$s_{01} = \tup{id, \initabox, \scmap_{cx}, \ctx_0, \delta}$ and
$s_{02} = \tup{\initabox', \scmap_s, \delta'}$ where
$\scmap_{cx} = \scmap_s = \emptyset$. By the definition of
$\tgprogba$ and $\tgkabba$, we also have:
\begin{inparaenum}[\it (i)]
\item $s_{01} \eqmc s_{02}$,
\item $s_{01}$ and $s_{02}$ are stable states,
\item $\fadd(s_{01}) = \addass{s_{02}} = \emptyset$
\item $\fdel(s_{01}) = \delass{s_{02}} = \emptyset$
\item $\delta' = \tgprogbcs(\delta)$.
\end{inparaenum}
Hence, by \Cref{lem:bagkab-to-sgkab-bisimilar-state}, we have
$s_{01} \ajbsim s_{02}$. Therefore, by the definition of
\ajbsimabr-bisimulation, we have
$\ts{\agkabsym}^{\csfilter_B} \ajbsim
\ts{\tgkabba(\agkabsym)}^{\filter_S}$. 
\end{proof}

Next, we show that the verification of \mulcsa properties over \bagkabs
can be recast as verification of \muladom over S-GKABs as follows.

\begin{theorem}\label{thm:ver-bagkab-to-sgkab}
  Given a \bagkab $\agkabsym$ and a closed $\mulcsa$ property $\Phi$
  (in NNF), we have
  \begin{center}
    $\ts{\agkabsym}^{\csfilter_B} \models \Phi$ if and only if
    $ \ts{\tgkabba(\agkabsym)}^{\filter_S}\models \tforja(\Phi)$
  \end{center}
\end{theorem}
\begin{proof}
  By \Cref{lem:bagkab-to-sgkab-bisimilar-ts}, we have that
  $\ts{\agkabsym}^{\csfilter_B} \ajbsim
  \ts{\tgkabba(\agkabsym)}^{\filter_S}$.
  Hence, by \Cref{lem:ajbisimilar-ts-satisfies-same-formula}, we have
  that for every $\mulcsa$ property $\Phi$
  \[
  \ts{\agkabsym}^{\csfilter_B} \models \Phi \textrm{ if and only if }
  \ts{\tgkabba(\agkabsym)}^{\filter_S}\models \tforja(\Phi)
  \]
\ \ \ 
\end{proof}



\subsection{Verification of 
\cagkabs}

This section is dedicated to show the reduction of \mulcsa
verification over \cagkabs into the \muladom verification over
S-GKABs. Basically, we start by presenting how we translate \cagkabs
into S-GKABs, and also how we translate \mulcsa formulas into \muladom
w.r.t.\ our purpose. Then, here we introduce a specific notion of
bisimulation that will be used to show that we can recast the
verification of \cagkabs into S-GKABs.

\subsubsection{Translating \cagkabs to S-GKABs}

To the aim of translating \cagkabs into S-GKABs, in the following we
first introduce the notion of program translation for the program in
\cagkabs.

\begin{definition}[Program Translation $\tgprogca$]
  Given \sidetextb{Program Translation $\tgprogca$} a \cagkabs
  $\agkabsym = \tup{\ctbox, \initabox, \actset, \ginitprog, \initctx,
    \ctxprocset}$,
  we define a \emph{translation $\tgprogca$} which translates a
  program $\delta$ into a program $\delta'$ inductively as follows:
  \[
\begin{array}{@{}l@{}l@{}l@{}}
  \tgprogca(\gactc{Q(\vec{p})}{\ctxe}{\act(\vec{p})})  &=&  
                                                           \gact{Q'(\vec{p})}{\act_a(\vec{p})};\gact{\true}{\act_b()};\\
                                                       &&\delta_{\ctxprocset}
                                                          ;\gact{\true}{\act^u()};\gact{\true}{\act^{\ctbox}_c()}\\
  \tgprogca(\gemptyprog) &=& \gemptyprog \\
  \tgprogca(\delta_1|\delta_2) &=& \tgprogca(\delta_1)|\tgprogca(\delta_2) \\
  \tgprogca(\delta_1;\delta_2) &=& \tgprogca(\delta_1);\tgprogca(\delta_2) \\
  \tgprogca(\gif{\varphi}{\delta_1}{\delta_2}) &=& \gif{\varphi}{\tgprogca(\delta_1)}{\tgprogca(\delta_2)} \\
  \tgprogca(\gwhile{\varphi}{\delta}) &=& \gwhile{\varphi}{\tgprogca(\delta)}
\end{array}
\]
where 
\begin{compactitem}
\item
  $\gact{Q'(\vec{p})}{\act_a(\vec{p})};\gact{\true}{\act_b(\vec{p})}$
  is a split action invocation obtained from
  $\gactc{Q(\vec{p})}{\ctxe}{\act(\vec{p})}$ as in
  \Cref{def:split-action-invoc},

\item $\delta_{\ctxprocset}$ is a context-change program obtained from
  $\ctxprocset$ as in \Cref{def:ctx-chg-prog} except that it is formed
  by sole action invocation obtained from context evolution rule as in
  \Cref{def:sole-action-and-action-invocation-obtained-from-context-evolution-rule}.

\item $\act^u$ is an update action (see \Cref{def:update-action}).

\item $\act^{\ctbox}_c$ is a context-sensitive c-repair action over
  $\ctbox$ as in \Cref{def:cs-c-repair-act}, except that we remove the
  effect $\map{\true} {\del \set{\tmp}}$ from $\eff{\act^{\ctbox}_c}$
  and add the effect $\map{\true} {\del \set{\ftitmp}}$ into
  $\eff{\act^{\ctbox}_c}$.

 \end{compactitem}
\ \ 
\end{definition}

We now step further to define the translation from \cagkabs into
S-GKABs as follows by also utilizing the program translation
$\tgprogca$ defined above.

\begin{definition}[Translation from \cagkab to S-GKAB]
  We \sidetext{Translation from \cagkab to S-GKAB} define a
  translation $\tgkabca$ that, given a \cagkab
  $\agkabsym = \tup{\ctbox, \initabox, \actset, \ginitprog, \initctx,
    \ctxprocset}$,
  produces an S-GKAB
  $\tgkabca(\agkabsym) = \tup{T_\cdimset, \initabox \cup A_{\initctx},
    \actset', \ginitprog'}$, where
\begin{compactitem}

\item $T_\cdimset$ is a TBox obtained from a
  set of context dimensions $\cdimset$ (see
  \Cref{def:tbox-from-ctxdim}),

\item 
  $A_{\initctx}$ is an ABox obtained from $\initctx$ (see
  \Cref{def:abox-context}),

\item
  $\actset' = \actset_\act \cup \actset_\ctx \cup \set{\act^u,
    \act^{\ctbox}_c}$ where:

\begin{compactitem}
\item $\actset_\act$ is obtained from $\actset$ such that for each
  action $\act \in \actset$, we have $\act_1,\act_2 \in \actset_\act$
  where $\act_1$ and $\act_2$ are split action obtained from $\act$
  (see \Cref{def:split-act}),

\item $\actset_\ctx$ is obtained from $\ctxprocset$ such that for each
  context-evolution rule $\tup{Q, \ctxe} \mapsto C_{new}$ in
  $\ctxprocset$, we have $\act_\ctx \in \actset_\ctx$ where
  $\act_\ctx$ is \emph{a sole action obtained from the
    context-evolution rule $\tup{Q, \ctxe} \mapsto C_{new}$} (see
  \Cref{def:sole-action-and-action-invocation-obtained-from-context-evolution-rule}),

\item $\act^u$ is an update action (see \Cref{def:update-action}).

\item $\act^{\ctbox}_c$ is a context-sensitive c-repair action over
  $\ctbox$ as in \Cref{def:cs-c-repair-act}, except that we remove the
  effect $\map{\true} {\del \set{\tmp}}$ from $\eff{\act^{\ctbox}_c}$
  and add the effect $\map{\true} {\del \set{\ftitmp}}$ into
  $\eff{\act^{\ctbox}_c}$.

\end{compactitem}

\item $\ginitprog' = \tgprogca(\ginitprog)$. 
\end{compactitem}
\ \ 
\end{definition}

The \mulcsa property $\Phi$ over a \cagkab $\agkabsym$ can then be
recast as a corresponding property over an S-GKAB
$\tgkabca(\csgkabsym)$ using the following formula translation:
\begin{definition}[Translation $\tforsa$]\label{def:tforsa}
  We \sidetextb{Translation $\tforsa$} define a \emph{translation
    $\tforsa$} that transforms an arbitrary \mulcsa formula
  $\Phi$ 
  into another \muladom formula $\Phi'$ inductively by recurring over
  the structure of $\Phi$ as follows:
  \[
  \begin{array}{lll}
    \bullet\ \tforsa(Q) &=& Q_\ctxb \\
    \bullet\ \tforsa(\ctxe) &=& q_{\ctxe} \\
    \bullet\ \tforsa(\neg \Phi) &=& \neg \tforsa(\Phi) \\
    \bullet\ \tforsa(\exists x.\Phi) &=& \exists x. \tforsa(\Phi) \\
    \bullet\ \tforsa(\Phi_1 \vee \Phi_2) &=& \tforsa(\Phi_1) \vee \tforsa(\Phi_2) \\
    \bullet\ \tforsa(\mu Z.\Phi) &=& \mu Z. \tforsa(\Phi) \\
    \bullet\ \tforsa(\bigodot \bigodot \bigodot\DIAM{\Phi}) &=& \bigodot \bigodot \bigodot \DIAM{\DIAM{\tforsa(\Phi)}}\\
    \bullet\ \tforsa(\bigodot \bigodot \bigodot\BOX{\Phi}) &=& \bigodot \bigodot \bigodot \BOX{\BOX{\tforsa(\Phi)}}
  \end{array}
  \]
  \noindent
  where $\bigodot$ is either box ($\BOX$) or diamond ($\DIAM$) modal
  operator.
\end{definition}

\subsubsection{Alternating Skip-one Bisimulation (\asbsimabr-Bisimulation)}

Here we introduce the notion of alternating skip-one bisimulation
(\asbsimabr-bisimulation) and show an important lemma about the
situation where two \asbsimabr-bisimilar transition systems can not be
distinguished by certain temporal properties. This bisimulation
relation is an important tool to show that we can recast the
verification of \mulcsa over \cagkabs (resp.\ \eagkabs) into the
verification of \muladom over S-GKABs.

\begin{definition}[Alternating Skip-one Bisimulation
  (\asbsimabr-Bisimulation)] \
  \sidetext{Alternating Skip-one Bisimulation (\asbsimabr-Bisimulation)} \\
  Let
  $\ts{1} = \tup{\const, \ctbox, \stateset_1, s_{01}, \abox_1, \cntx,
    \actsrc, \actpar, \fadd, \fdel, \trans_1}$
  be a fine-grained transition system, and
  $\ts{2} = \tup{\const, T, \stateset_2, s_{02}, \abox_2, \trans_2}$
  be a typed KB transition system, with
  $\adom{\abox_1(s_{01})} \subseteq \const$,
  $\adom{\abox_2(s_{02})} \subseteq \const$, and $s_{01}$ as well as
  $s_{02}$ are stable states.
  An alternating skip-one bisimulation (\asbsimabr-Bisimulation)
  between $\ts{1}$ and $\ts{2}$ is a relation
  $\B \subseteq \Sigma_1 \times\Sigma_2$ such that
  $\tup{s_1, s_2} \in \B$ implies that one of the following condition
  hold:
\begin{compactenum}

\item $s_{1}$ as well as $s_{2}$ are either stable states, service
  call evaluation states, or context change states and we have the
  following:
  \begin{compactenum}
  \item $s_1 \eqmc s_2$
  \item for each $s_1'$, if $s_1 \Rightarrow_1 s_1'$ then there exists
    $s_2'$ with $ s_2 \Rightarrow_2 s_2'$ such that
    $\tup{s_1', s_2'}\in\B$, $s_1'$ and $s_2'$ are service call
    evaluation states.
  \item for each $s_2'$, if $ s_2 \Rightarrow_2 s_2'$, then there
    exists $s_1'$ with $s_1 \Rightarrow_1 s_1'$, such that
    $\tup{s_1', s_2'}\in\B$, $s_1'$ and $s_2'$ are context change
    states.
  \end{compactenum}

\item $s_{1}$ and $s_{2}$ are filter application states.
\begin{compactenum}
  \item $s_1 \eqmc s_2$
  \item for each $s_1'$, if $s_1 \Rightarrow_1 s_1'$ then there exists
    $t$ and $s_2'$ with 
    \[
    s_2 \Rightarrow_2 t \Rightarrow_2 s_2'
    \]
    such that $\tup{s_1', s_2'}\in\B$, $t$ is a filter application intermediate state, $s_1'$ and $s_2'$ are stable
    states.
  \item for each $s_2'$, if 
    \[
    s_2 \Rightarrow_2 t \Rightarrow_2 s_2',
    \]
    with $t$ is a filter application intermediate state, then there exists $s_1'$
    with $s_1 \Rightarrow_1 s_1'$, such that $\tup{s_1', s_2'}\in\B$,
    $s_1'$ and $s_2'$ are stable states.
  \end{compactenum}

\end{compactenum}
\ \ 
\end{definition}

\noindent
Let
$\ts{1} = \tup{\const, \ctbox, \stateset_1, s_{01}, \abox_1, \cntx,
  \actsrc, \actpar, \fadd, \fdel, \trans_1}$
be a fine-grained transition system, and
$\ts{2} = \tup{\const, T, \stateset_2, s_{02}, \abox_2, \trans_2}$ be
a KB transition system, 
a state $s_1 \in \stateset_1$ is \emph{\asbsimabr-bisimilar} to
$s_2 \in \stateset_2$, written $s_1 \asbsim s_2$, if there exists an
\asbsimabr-bisimulation relation $\B$ between $\ts{1}$ and $\ts{2}$
such that $\tup{s_1, s_2}\in\B$.
A transition system $\ts{1}$ is \emph{\asbsimabr-bisimilar} to
$\ts{2}$, written $\ts{1} \asbsim \ts{2}$, if there exists an
\asbsimabr-bisimulation relation $\B$ between $\ts{1}$ and $\ts{2}$
such that $\tup{s_{01}, s_{02}}\in\B$.

In the following two lemmas we show some important properties of
\asbsimabr-bisimilar states and transition systems that will be useful
later to show that we can recast the verification of \cagkabs (as well
as \eagkabs) into S-GKABs.

\begin{lemma}\label{lem:asbisimilar-state-satisfies-same-formula}
  Let
  $\ts{1} = \tup{\const, \ctbox, \stateset_1, s_{01}, \abox_1, \cntx,
    \actsrc, \actpar, \fadd, \fdel, \trans_1}$
  be a fine-grained transition system, and
  $\ts{2} = \tup{\const, T, \stateset_2, s_{02}, \abox_2, \trans_2}$
  be a typed KB transition system,
  Consider two states $s_1 \in \stateset_1$ and $s_2 \in \stateset_2$
  such that $s_1 \asbsim s_2$. Then for every formula $\Phi$ 
  of $\mulcsa$, and every valuations $\vfo_1$ and $\vfo_2$ that assign
  to each of its free variables a constant
  $c_1 \in \adom{\abox_1(s_1)}$ and $c_2 \in \adom{\abox_2(s_2)}$,
  such that $c_1 = c_2$, we have that
  \[
  \ts{1},s_1 \models \Phi \vfo_1 \textrm{ if and only if } \ts{2},s_2
  \models \tforsa(\Phi) \vfo_2.
  \]
\end{lemma}
\begin{proof}
  Similar to the combination of the proof for
  \Cref{lem:stbisimilar-state-satisfies-same-formula,lem:sbisimilar-state-satisfies-same-formula}
  by also considering that by the definition of typed KB transition
  system and fine-grained transition system, both of them start from a
  stable state. Additionally, for the transition among the states in
  $\ts{1}$, it is always be the case that there are three intermediate
  states (i.e., service call evaluation state, context change state,
  and filter application state) between the two stable
  states. Moreover, for the case of $\ts{2}$, we have that there is an
  additional states namely a filter application intermediate state.
  On the other hand, the \mulcsa formula always have four consecutive
  modal operators and the translation $\tforsa$ always add an
  additional modal operator for the translation of modal operators in
  \mulcsa.
  Hence, it is easy to see that we always verify query and context
  expression in the given \mulcsa formula over a stable state.
\end{proof}

\begin{lemma}\label{lem:asbisimilar-ts-satisfies-same-formula}
  Consider a fine-grained transition system $\ts{1}$, and a typed KB
  transition system $\ts{2}$ such that $\ts{1} \asbsim \ts{2}$.  For
  every closed \mulcsa formula $\Phi$, we have:
  \[
  \ts{1} \models \Phi \textrm{ if and only if } \ts{2} \models
  \tforsa(\Phi)
  \]
\end{lemma}
\begin{proof} 
  Let
  $\ts{1} = \tup{\const, \ctbox, \stateset_1, s_{01}, \abox_1, \cntx,
    \actsrc, \actpar, \fadd, \fdel, \trans_1}$,
  and
  $\ts{2} = \tup{\const, T, \stateset_2, s_{02}, \abox_2, \trans_2}$.
  By the definition of \asbsimabr-bisimilar transition system, we have
  that $s_{01} \asbsim s_{02}$. Thus, we obtain the proof as a
  consequence of \Cref{lem:asbisimilar-state-satisfies-same-formula},
  due to the fact that
  \[
  \ts{1}, s_{01} \models \Phi \textrm{ if and only if } \ts{2}, s_{02}
  \models \tforsa(\Phi)
  \]
\ \ 
\end{proof}

\subsubsection{Reducing the Verification of \cagkabs into S-GKABs}

We now proceed to show that we can recast the verification of \cagkabs
into S-GKABs. In the following two lemmas we aim to show that the
transition system of a \cagkab $\agkabsym$ is \asbsimabr-bisimilar to
the transition system of the corresponding S-GKAB
$\tgkabca(\agkabsym)$ that is obtained via translation $\tgkabca$.

\begin{lemma}\label{lem:cagkab-to-sgkab-bisimilar-state}
  Let $\agkabsym$ be a \cagkab with transition system
  $\ts{\agkabsym}^{\csfilter_C}$, and let $\tgkabca(\agkabsym)$ be
  its corresponding S-GKAB $($with transition system
  $\ts{\tgkabca(\agkabsym)}^{\filter_S}$$)$
  obtain through
  $\tgkabca$. 
  Consider
  \begin{inparaenum}[]
  \item a state $s_{cx}
    = \tup{id_{cx}, A_{cx},\scmap_{cx}, \ctx,
      \delta_{cx}}$ of $\ts{\agkabsym}^{\csfilter_C}$ and
  \item a state $s_s = \tup{A_s,\scmap_s, \delta_s}$ of
    $\ts{\tgkabca(\agkabsym)}^{\filter_S}$.
  \end{inparaenum}
  If the following hold:
  \begin{compactenum}
  \item $s_{cx}$ and $s_s$ are having the same state type,
    \item $s_{cx} \eqmc s_s$ (see \Cref{def:cs-equal-mod-markers}),
    \item $\scmap_{cx} = \scmap_s$,
    \item $\fadd(s_{cx}) = \addass{A_s}$ (see
      \Cref{def:added-assertion} for the definition of $\addass{A_s}$),
    \item $\fdel(s_{cx}) = \delass{A_s}$ (see
      \Cref{def:deleted-assertion} for the definition of $\delass{A_s}$),
    \item $\delta_s = \tgprogca(\delta_{cx})$ (if $s_{cx}$ and $s_s$ are stable states),
    \item $\actsrcconceptname(\act)
      \in A_s$ (if $s_{cx}$ and
      $s_s$ are not stable states, and $\actsrc(s_{cx}) = \act$),
    \item $\set{\parconceptname{1}(\sigma(p_1)),
        \ldots, \parconceptname{m}(\sigma(p_m))} = \parass{A_s}$ (if $s_{cx}$ and
      $s_s$ are not stable states, and $\actpar(s_{cx}) = \sigma$),
    \end{compactenum}
    then $s_{cx} \asbsim s_s$,
\end{lemma}
\begin{proof}
  Similar to the proof of
  \Cref{lem:bagkab-to-sgkab-bisimilar-state}. The different is only in
  the case when $s_{cx}$
  and $s_s$
  are both filter application states. Here, instead of applying the
  b-repair program, we apply the c-repair action.  Similar to
  \Cref{thm:cact-equal-crep}, we can also easily show the correctness
  of context-sensitive c-repair action. The important observation is
  that the context-sensitive c-repair action do the repair based on
  the context, i.e., it only consider those assertion in the TBox that
  ``holds'' under the corresponding context. 
\end{proof}

\begin{lemma}\label{lem:cagkab-to-sgkab-bisimilar-ts}
  Given a \cagkab $\agkabsym$
  with transition system $\ts{\agkabsym}^{\csfilter_C}$,
  let $\tgkabca(\agkabsym)$
  be its corresponding S-GKAB $($with
  transition system $\ts{\tgkabca(\agkabsym)}^{\filter_S}$$)$
  obtained through $\tgkabca$.  We have that
  $\ts{\agkabsym}^{\csfilter_C} \ajbsim
  \ts{\tgkabca(\agkabsym)}^{\filter_S}$
\end{lemma}
\begin{proof}
  Let
\begin{compactitem}
\item
  $\agkabsym = \tup{\ctbox, \initabox, \actset, \ginitprog, \initctx,
    \ctxprocset}$, and \\
  $\ts{\agkabsym}^{\csfilter_C} = \tup{\const, \ctbox, \stateset_1,
    s_{01}, \abox_1, \cntx, \actsrc, \actpar, \fadd, \fdel,
    \trans_1}$,
\item $\tgkabca(\agkabsym) = \tup{T_\cdimset, \initabox',
    \actset', \ginitprog'}$, and \\
  $\ts{\tgkabca(\agkabsym)}^{\filter_S} = \tup{\const, T_\cdimset,
    \stateset_2, s_{02}, \abox_2, \trans_2}$.
\end{compactitem}
We have that
$s_{01} = \tup{id, \initabox, \scmap_{cx}, \ctx_0, \delta}$ and
$s_{02} = \tup{\initabox', \scmap_s, \delta'}$ where
$\scmap_{cx} = \scmap_s = \emptyset$. By the definition of
$\tgprogca$ and $\tgkabca$, we also have:
\begin{inparaenum}[\it (i)]
\item $s_{01} \eqmc s_{02}$,
\item $s_{01}$ and $s_{02}$ are stable states,
\item $\fadd(s_{01}) = \addass{s_{02}} = \emptyset$
\item $\fdel(s_{01}) = \delass{s_{02}} = \emptyset$
\item $\delta' = \tgprogca(\delta)$.
\end{inparaenum}
Hence, by \Cref{lem:cagkab-to-sgkab-bisimilar-state}, we have
$s_{01} \asbsim s_{02}$. Therefore, by the definition of
\asbsimabr-bisimulation, we have
$\ts{\agkabsym}^{\csfilter_C} \asbsim
\ts{\tgkabca(\agkabsym)}^{\filter_S}$. 
\end{proof}

Now, we show that the verification of \mulcsa properties over \cagkabs
can be recast as verification of \muladom over S-GKABs as follows.

\begin{theorem}\label{thm:ver-cagkab-to-sgkab}
  Given a \cagkab $\agkabsym$ and a closed $\mulcsa$ property
  $\Phi$, we have
  \begin{center}
    $\ts{\agkabsym}^{\csfilter_C} \models \Phi$ if and only if
    $ \ts{\tgkabca(\agkabsym)}^{\filter_S}\models \tforsa(\Phi)$
  \end{center}
\end{theorem}
\begin{proof}
  By \Cref{lem:cagkab-to-sgkab-bisimilar-ts}, we have that
  $\ts{\agkabsym}^{\csfilter_C} \asbsim
  \ts{\tgkabca(\agkabsym)}^{\filter_S}$.
  Hence, by \Cref{lem:asbisimilar-ts-satisfies-same-formula}, we have
  that for every $\mulcsa$ property $\Phi$
  \[
  \ts{\agkabsym}^{\csfilter_C} \models \Phi \textrm{ if and only if }
  \ts{\tgkabca(\agkabsym)}^{\filter_S}\models \tforsa(\Phi)
  \]
\ \ \ 
\end{proof}




\subsection{Verification of 
  \eagkabs}

This section is devoted to show that we can recast the verification of
\mulcsa over \eagkabs into the verification of \muladom over
S-GKABs. We open this section by explaining how we translate \eagkabs
into S-GKABs, and also how we transform \mulcsa formulas into \muladom
w.r.t.\ our purpose. By making use the \asbsimabr-bisimulation defined
before, later we show that we can reduce the verification of \mulcsa
over \eagkabs into the verification of \muladom over S-GKABs.

\subsubsection{Translating \eagkabs to S-GKABs}

In order to define our generic translation which transforms any
\eagkabs into S-GKABs, we first introduce the program translation for
the program inside \eagkabs as follows.


\begin{definition}[Program Translation
  $\tgprogea$]\label{def:prog-trans-eagkab-sgkab}
  Given \sidetextb{Program Translation $\tgprogea$} an \eagkabs
  $\agkabsym = \tup{\ctbox, \initabox, \actset, \ginitprog, \initctx,
    \ctxprocset}$,
  we define a \emph{translation $\tgprogea$} which translates a
  program $\delta$ into a program $\delta'$ inductively as follows:

  \[
\begin{array}{@{}l@{}l@{}l@{}}
  \tgprogea(\gactc{Q(\vec{p})}{\ctxe}{\act(\vec{p})})  &=&  
                                                           \gact{Q'(\vec{p})}{\act_a(\vec{p})};\gact{\true}{\act_b()};\\
                                                       &&\delta_{\ctxprocset}
                                                          ;\gact{\true}{\act^u()};\gact{\neg\qunsatecq{T_a}}{\act^{\ctbox}_e()}\\
  \tgprogea(\gemptyprog) &=& \gemptyprog \\
  \tgprogea(\delta_1|\delta_2) &=& \tgprogea(\delta_1)|\tgprogea(\delta_2) \\
  \tgprogea(\delta_1;\delta_2) &=& \tgprogea(\delta_1);\tgprogea(\delta_2) \\
  \tgprogea(\gif{\varphi}{\delta_1}{\delta_2}) &=& \gif{\varphi}{\tgprogea(\delta_1)}{\tgprogea(\delta_2)} \\
  \tgprogea(\gwhile{\varphi}{\delta}) &=& \gwhile{\varphi}{\tgprogea(\delta)}
\end{array}
\]
where 
\begin{compactitem}
\item
  $\gact{Q'(\vec{p})}{\act_a(\vec{p})};\gact{\true}{\act_b(\vec{p})}$
  is a split action invocation obtained from
  $\gactc{Q(\vec{p})}{\ctxe}{\act(\vec{p})}$ as in
  \Cref{def:split-action-invoc},

\item $\delta_{\ctxprocset}$ is a context-change program obtained from
  $\ctxprocset$ as in \Cref{def:ctx-chg-prog} except that it is formed
  by sole action invocation obtained from context evolution rule as in
  \Cref{def:sole-action-and-action-invocation-obtained-from-context-evolution-rule}.

\item $\act^u$ is an update action as in \Cref{def:update-action}
  except that we replace the effect
  $\map{N^a(x)}{\add \set{N(x) }, \del \set{N^a(x) } }$ in
  $\eff{\act^u}$ with $\map{N^a(x)}{\add \set{N(x) } }$. The different
  is only that we do not delete the assertions made by added fact
  marker concept/role names.

\item $\qunsatecq{{T_a}}$ is a context-sensitive Q-UNSAT-ECQ over
  $T_a$ (see \Cref{def:cs-q-unsat-ecq}), where $T_a$ is obtained from
  $\ctbox$ by renaming each concept name $N$ in $\ctbox$ into $N^a$
  (similarly for roles). Thus, with this mechanism, we can block any
  further execution when the newly added assertions are
  inconsistent. 

\item $\act^{\ctbox}_e$ is a context-sensitive evolution action over
  $\ctbox$ as in \Cref{def:cs-evol-act}, except that we replace the
  effect $\map{\true} {\del \set{\tmp}}$ in $\eff{\act^{\ctbox}_e}$
  with the effect $\map{\true} {\del \set{\ftitmp}}$.

 \end{compactitem}
\ \ 
\end{definition}

By utilizing the program translation $\tgprogea$ defined above, we
define the translation from \eagkabs into S-GKABs as follows:

\begin{definition}[Translation from \eagkab to S-GKAB]
  We \sidetext{Translation from \eagkab to S-GKAB} define a
  translation $\tgkabea$ that, given an \eagkab
  $\agkabsym = \tup{\ctbox, \initabox, \actset, \ginitprog, \initctx,
    \ctxprocset}$,
  produces an S-GKAB
  $\tgkabca(\agkabsym) = \tup{T_\cdimset, \initabox \cup A_{\initctx},
    \actset', \ginitprog'}$, where
\begin{compactitem}

\item $T_\cdimset$ is a TBox obtained from a
  set of context dimensions $\cdimset$ (see
  \Cref{def:tbox-from-ctxdim}),

\item 
  $A_{\initctx}$ is an ABox obtained from $\initctx$ (see
  \Cref{def:abox-context}),

\item
  $\actset' = \actset_\act \cup \actset_\ctx \cup \set{\act^u,
    \act^{\ctbox}_e}$ where:

\begin{compactitem}
\item $\actset_\act$ is obtained from $\actset$ such that for each
  action $\act \in \actset$, we have $\act_1,\act_2 \in \actset_\act$
  where $\act_1$ and $\act_2$ are split action obtained from $\act$
  (see \Cref{def:split-act}),

\item $\actset_\ctx$ is obtained from $\ctxprocset$ such that for each
  context-evolution rule $\tup{Q, \ctxe} \mapsto C_{new}$ in
  $\ctxprocset$, we have $\act_\ctx \in \actset_\ctx$ where
  $\act_\ctx$ is \emph{a sole action obtained from the
    context-evolution rule $\tup{Q, \ctxe} \mapsto C_{new}$} (see
  \Cref{def:sole-action-and-action-invocation-obtained-from-context-evolution-rule}),

\item $\act^u$ is an update action (see \Cref{def:update-action}).

\item $\act^{\ctbox}_e$ is a context-sensitive evolution action over
  $\ctbox$ as in \Cref{def:cs-evol-act}, except that we replace the
  effect $\map{\true} {\del \set{\tmp}}$ in $\eff{\act^{\ctbox}_e}$
  with the effect $\map{\true} {\del \set{\ftitmp}}$.

\end{compactitem}

\item $\ginitprog' = \tgprogea(\ginitprog)$. 
\end{compactitem}
\ \ 
\end{definition}

\subsubsection{Reducing the Verification of \eagkabs into S-GKABs}

We now step forward to show that the verification of \eagkabs can be
reduced into the verification S-GKABs. In the following two lemmas we
aim to show that the transition system of an \eagkabs $\agkabsym$ is
\asbsimabr-bisimilar to the transition system of the corresponding
S-GKAB $\tgkabea(\agkabsym)$ (that is obtained via translation
$\tgkabea$).

\begin{lemma}\label{lem:eagkab-to-sgkab-bisimilar-state}
  Let $\agkabsym$ be an \eagkab with transition system
  $\ts{\agkabsym}^{\csfilter_E}$, and let $\tgkabea(\agkabsym)$ be
  its corresponding S-GKAB $($with transition system
  $\ts{\tgkabea(\agkabsym)}^{\filter_S}$$)$
  obtain through
  $\tgkabea$. 
  Consider
  \begin{inparaenum}[]
  \item a state $s_{cx}
    = \tup{id_{cx}, A_{cx},\scmap_{cx}, \ctx,
      \delta_{cx}}$ of $\ts{\agkabsym}^{\csfilter_C}$ and
  \item a state $s_s = \tup{A_s,\scmap_s, \delta_s}$ of
    $\ts{\tgkabca(\agkabsym)}^{\filter_S}$.
  \end{inparaenum}
  If the following hold:
  \begin{compactenum}
  \item $s_{cx}$ and $s_s$ are having the same state type,
    \item $s_{cx} \eqmc s_s$ (see \Cref{def:cs-equal-mod-markers}),
    \item $\scmap_{cx} = \scmap_s$,
    \item $\fadd(s_{cx}) = \addass{A_s}$ (see
      \Cref{def:added-assertion} for the definition of $\addass{A_s}$),
    \item $\fdel(s_{cx}) = \delass{A_s}$ (see
      \Cref{def:deleted-assertion} for the definition of $\delass{A_s}$),
    \item $\delta_s = \tgprogea(\delta_{cx})$ (if $s_{cx}$ and $s_s$ are stable states),
    \item $\actsrcconceptname(\act)
      \in A_s$ (if $s_{cx}$ and
      $s_s$ are not stable states, and $\actsrc(s_{cx}) = \act$),
    \item $\set{\parconceptname{1}(\sigma(p_1)),
        \ldots, \parconceptname{m}(\sigma(p_m))} = \parass{A_s}$ (if $s_{cx}$ and
      $s_s$ are not stable states, and $\actpar(s_{cx}) = \sigma$),
    \end{compactenum}
    then $s_{cx} \asbsim s_s$,
\end{lemma}
\begin{proof}
  Similar to the proof of
  \Cref{lem:bagkab-to-sgkab-bisimilar-state}. The different is only in
  the case when $s_{cx}$
  and $s_s$
  are both filter application states, instead of applying the b-repair
  program, we apply the evolution action. Another aspect to observe in
  order to complete the proof is as follows:
  \begin{compactitem}

  \item Similar to \Cref{lem:evol-prop,lem:eact-prop}, we can also
    easily show the correctness of context-sensitive evolution action
    that it performs the bold-evolution computation. The important
    observation is that the context-sensitive evolution action
    performs the bold-evolution based on the context, i.e., it only
    consider those assertion in the TBox that ``holds'' under the
    corresponding context.

  \item As it can be seen from the translation $\tgprogea$ (see
    \Cref{def:prog-trans-eagkab-sgkab}), before executing the
    evolution action, we also check the consistency of the updates
    w.r.t.\ the TBox under the new context. This guarantees that we
    fulfill the requirement in \Cref{def:evol-cs-filter} that the
    updates must consistent w.r.t.\ the TBox under the new context.

  \end{compactitem}
\end{proof}

\begin{lemma}\label{lem:eagkab-to-sgkab-bisimilar-ts}
  Given an \eagkab $\agkabsym$ with transition system
  $\ts{\agkabsym}^{\csfilter_E}$, let $\tgkabea(\agkabsym)$ be its
  corresponding S-GKAB $($with transition system
  $\ts{\tgkabea(\agkabsym)}^{\filter_S}$$)$
  obtained through $\tgkabea$.
  We have that $\ts{\agkabsym}^{\csfilter_E}
  \asbsim \ts{\tgkabea(\agkabsym)}^{\filter_S}$
\end{lemma}
\begin{proof}
  Let
\begin{compactitem}
\item
  $\agkabsym = \tup{\ctbox, \initabox, \actset, \ginitprog, \initctx,
    \ctxprocset}$, and \\
  $\ts{\agkabsym}^{\csfilter_E} = \tup{\const, \ctbox, \stateset_1,
    s_{01}, \abox_1, \cntx, \actsrc, \actpar, \fadd, \fdel,
    \trans_1}$,
\item $\tgkabea(\agkabsym) = \tup{T_\cdimset, \initabox',
    \actset', \ginitprog'}$, and \\
  $\ts{\tgkabea(\agkabsym)}^{\filter_S} = \tup{\const, T_\cdimset,
    \stateset_2, s_{02}, \abox_2, \trans_2}$.
\end{compactitem}
We have that
$s_{01} = \tup{id, \initabox, \scmap_{cx}, \ctx_0, \delta}$ and
$s_{02} = \tup{\initabox', \scmap_s, \delta'}$ where
$\scmap_{cx} = \scmap_s = \emptyset$. By the definition of
$\tgprogea$ and $\tgkabea$, we also have:
\begin{inparaenum}[\it (i)]
\item $s_{01} \eqmc s_{02}$,
\item $s_{01}$ and $s_{02}$ are stable states,
\item $\fadd(s_{01}) = \addass{s_{02}} = \emptyset$
\item $\fdel(s_{01}) = \delass{s_{02}} = \emptyset$
\item $\delta' = \tgprogea(\delta)$.
\end{inparaenum}
Hence, by \Cref{lem:eagkab-to-sgkab-bisimilar-state}, we have
$s_{01} \asbsim s_{02}$. Therefore, by the definition of
\asbsimabr-bisimulation, we have
$\ts{\agkabsym}^{\csfilter_E} \asbsim
\ts{\tgkabea(\agkabsym)}^{\filter_S}$. 
\end{proof}

Having \Cref{lem:eagkab-to-sgkab-bisimilar-ts} in hand, we can now
easily show that the verification of \mulcsa properties over \eagkab
can be recast as verification of \muladom over S-GKAB as follows.

\begin{theorem}\label{thm:ver-eagkab-to-sgkab}
  Given an \eagkab $\agkabsym$ and a closed $\mulcsa$ property
  $\Phi$, we have
  \begin{center}
    $\ts{\agkabsym}^{\csfilter_E} \models \Phi$ if and only if
    $ \ts{\tgkabea(\agkabsym)}^{\filter_S}\models \tforsa(\Phi)$
  \end{center}
\end{theorem}
\begin{proof}
  By \Cref{lem:eagkab-to-sgkab-bisimilar-ts}, we have that
  $\ts{\agkabsym}^{\csfilter_E} \asbsim
  \ts{\tgkabea(\agkabsym)}^{\filter_S}$.
  Hence, by \Cref{lem:asbisimilar-ts-satisfies-same-formula}, we have
  that for every $\mulcsa$ property $\Phi$
  \[
  \ts{\agkabsym}^{\csfilter_E} \models \Phi \textrm{ if and only if }
  \ts{\tgkabea(\agkabsym)}^{\filter_S}\models \tforsa(\Phi)
  \]
\ \ \ 
\end{proof}




\subsection{Putting it all together: Verification of \agkabs}

Putting together all results above, from
\Cref{thm:gtos,thm:ver-bagkab-to-sgkab,thm:ver-cagkab-to-sgkab,thm:ver-eagkab-to-sgkab}
we get that verification of \bagkabs, \cagkabs, and \eagkabs can be
compiled into verification of KABs, by first translating them into
S-GKABs, and then into KABs.

\begin{theorem}
\label{thm:bce-agkabs-to-kab}
Verification of \mulcsa properties over \bagkabs, \cagkabs, and
\eagkabs can be reduced to verification over KABs.
\end{theorem}
\begin{proof}
  The proof can be easily obtained since from
  \Cref{thm:ver-bagkab-to-sgkab,thm:ver-cagkab-to-sgkab,thm:ver-eagkab-to-sgkab},
  we can reduce the verification of \mulcsa over \bagkabs, \cagkabs,
  and \eagkabs as verification of \muladom over S-GKABs and then by
  \Cref{thm:gtos} we can recast the verification of \muladom over
  S-GKABs as verification of \muladom over KABs. Thus, combining all
  of those ingredients, we have that the claim is proven. 
\end{proof}

\subsection{Verification of Run-bounded \agkabs}

Even more interesting, our reductions from \bagkabs, \cagkabs, and
\eagkabs into S-GKABs preserve run-boundedness.

\begin{lemma}\label{lem:run-bounded-preservation-bagkab}
  Let $\agkabsym$ be a \bagkabs and $\tgkabba(\agkabsym)$ be its
  corresponding S-GKAB. We have $\agkabsym$ is run-bounded if and only
  if $\tgkabba(\agkabsym)$ is run-bounded.
\end{lemma}
\begin{proof}
  Let
  \begin{compactitem}

  \item $\ts{\agkabsym}^{\csfilter_B}$ be the transition system of 
    $\agkabsym$, and

  \item $\ts{\tgkabba(\agkabsym)}^{\filter_S}$ be the transition
    system of $\tgkabba(\agkabsym)$.



  \end{compactitem}
  The claim can be easily shown by observing the following:
  \begin{compactitem}
  \item the translation $\tgkabba$ essentially do the following:
    \begin{compactenum}
    \item Splits each action $\act$ into two actions $\act_a$ and
      $\act_b$ where the former do the computation of $\act$ that do
      not involve any service calls while the latter do the computation
      of $\act$ that involve service calls.
    \item Appends the split actions with a context-change program that
      simulates context evolution.
    \item Appends context-change program with an additional program to
      simulate the b-repair computation.
    \end{compactenum}

  \item the split actions $\act_a$ and $\act_b$ of $\act$ essentially
    only split the computation that is done by $\act$ into two steps.

  \item the program that is used to simulate the context evolution
    does not inject unbounded number of new constants. In fact, we
    only reserve a constant $\ctxconst$ to simulate the context (i.e.,
    to construct the ABox assertions that represent the context
    dimension assignments).

  \item the program/actions that simulate the b-repair computation
    never inject new additional constants, but only remove facts
    causing inconsistency,

  \item by \Cref{lem:bagkab-to-sgkab-bisimilar-ts}, we have that
    $\ts{\agkabsym}^{\csfilter_B} \ajbsim \tgkabba(\agkabsym)$. Thus,
    basically they are ``equivalent'' modulo filter application
    intermediate states (states containing $\ftitmp$) and also by
    considering that they represent context information in a different
    way. 


  \end{compactitem}
\end{proof}

\begin{lemma}\label{lem:run-bounded-preservation-cagkab}
  Let $\agkabsym$ be a \cagkab and $\tgkabca(\agkabsym)$ be its
  corresponding S-GKAB. We have $\agkabsym$ is run-bounded if and only
  if $\tgkabca(\agkabsym)$ is run-bounded.
\end{lemma}
\begin{proof}
  Similar to the proof of
  Lemma~\ref{lem:run-bounded-preservation-bagkab}. 
\end{proof}

\begin{lemma}\label{lem:run-bounded-preservation-eagkab}
  Let $\agkabsym$ be a \eagkab and $\tgkabea(\agkabsym)$ be its
  corresponding S-GKAB. We have $\agkabsym$ is run-bounded if and only
  if $\tgkabea(\agkabsym)$ is run-bounded.
\end{lemma}
\begin{proof}
  Similar to the proof of
  Lemma~\ref{lem:run-bounded-preservation-bagkab}. 
\end{proof}

To close our tour on this chapter, we show the result on the
verification of \mulcsa properties over run-bounded \bagkabs,
\cagkabs, and \eagkabs as follows.

\begin{theorem}
  Verification of \mulcsa properties over run-bounded \bagkabs,
  \cagkabs, and \eagkabs are
  decidable, and reducible to standard $\mu$-calculus finite-state
  model checking.
\end{theorem}
\begin{proof}
  By
  \Cref{lem:run-bounded-preservation-bagkab,lem:run-bounded-preservation-cagkab,lem:run-bounded-preservation-eagkab},
  the translation from \bagkabs, \cagkabs, and \eagkabs to S-GKABs
  preserves run-boundedness.
%
%
  Thus, the claim follows by combining
  \Cref{thm:ver-bagkab-to-sgkab,thm:ver-cagkab-to-sgkab,thm:ver-eagkab-to-sgkab}
  and \Cref{thm:ver-run-bounded-sgkab}.
\end{proof}

\section{Emulating Standard GKABs in Alternating GKABs}

We have seen so far that we can recast the verification of \bagkabs,
\cagkabs, and \eagkabs into S-GKABs.
Now, we show that we can also do the other direction. In particular,
we show that we can recast the verification of S-GKABs into the
verification of \bagkabs. The reductions from \cagkabs and \eagkabs
into S-GKABs can be done similarly.


\subsection{Transforming S-GKABs into \bagkabs}

Similar to
\Cref{sec:cap-sgkabs-to-igkabs,sec:cap-sgkabs-to-icsgkabs,sec:cap-sgkab-to-scsgkabs},
some challenges in order to reduce \bagkabs into S-GKABs is to prevent
the repair and also the context change. Therefore, to cope with this
situation, here we simply adopt our previous approach.
%
%
Another important observation in order to reduce the verification of
S-GKABs into \bagkabs is that essentially S-GKABs ignore the various
quantification among all sources of non-determinisms. Thus, to mimics
this situation, we simply need to translate the given \muladom
formulas over S-GKABs into the corresponding \mulcsa formulas over
\bagkabs by duplicating the modal operator eight times (I.e., we
translate $\DIAM{\Phi}$ into
$\DIAM{\DIAM{\DIAM{\DIAM{\DIAM{\DIAM{\DIAM{\DIAM{\Phi}}}}}}}}$ and
similarly for $\BOX{}$). The reason why we need to duplicates the
modal operator eight times is that because a single action execution
in S-GKABs will correspond to two action execution in \bagkabs (One
for the same action execution, and one for the action that does the
inconsistency check). Additionally, notice that each action execution
in \bagkabs requires four transitions until it reaches another stable
state, we then need to quadruplicate the modal operator for each
action execution.

In the following we fix a set $\cdimset$ of context dimension
containing only a single context dimension $d$ (i.e.,
$\cdimset = \set{d}$). Moreover, $d \in \cdimset$ has a tree shaped
finite value domain $\tup{\cdom[d],\cover[d]}$ where $\cdom[d]$
contains only a single value $\topv[d]$ (i.e., $\cdom[d] = \topv[d]$).

To translate the program in the given S-GKABs, we basically can reuse
the program translation $\tgprogsic$ in
\Cref{def:prog-trans-tgprogsic}.  We then define the following
translation that transform S-GKABs into \bagkabs as follows.

\begin{definition}[Translation from S-GKAB to \bagkab]\label{def:trans-sgkab-bagkab}
  \ \sidetext{Translation from S-GKAB to \bagkab}
  We define a translation $\tgkabsba$ that, given an S-GKAB
  $\gkabsym = \tup{T, \initabox, \actset, \ginitprog}$, produces a
  \bagkab
  $\tgkabsba(\gkabsym) = \tup{\ctbox, \initabox, \actset',
    \ginitprog', \initctx, \ctxprocset}$, where
\begin{compactitem}

\item $\ctbox$ is obtained from $T$ such that for each positive
  inclusion assertion $t \in T$, we have $\tup{t:\varphi}$ where
  $\varphi = \cval{d}{\topv[d]}$,

\item $\actset' = \actset_\act \cup \set{\act_\bot}$ where
  \begin{compactitem}
  \item $\actset_\act$ is obtained from $\actset$ such that for each
    $\act \in \actset$, we have $\act' \in \actset_\act$ where
    $ \eff{\act'} = \eff{\act} \cup \set{\map{\true}{\add \set{\tmp} }
    }$,
  \item $\act_\bot$ is a 0-ary action of the form
    $ \act_{\bot}():\set{\map{\true}{\del \set{\tmp} }} $.
  \end{compactitem}

\item $\ginitprog' = \tgprogsic(\ginitprog)$. 

\item $\initctx = \set{\cval{d}{\topv[d]}}$, 

\item $\ctxprocset = \set{\tup{\true, \cval{d}{\topv[d]}} \mapsto
    \set{\cval{d}{\topv[d]}} }$

\end{compactitem}
\ \ 
\end{definition}

The \muladom property $\Phi$ over an S-GKAB $\gkabsym$ can then be
recast as a corresponding property over a \bagkab
$\tgkabsba(\gkabsym)$ using the following formula translation:
\begin{definition}[Translation $\tforsba$]\label{def:tforsba}
  We \sidetext{\muladom Formula Translation $\tforsba$} define a
  \emph{translation $\tforsba$} that takes a \muladom formula $\Phi$
  as an input and produces a \mulcsa formula $\tforsba(\Phi)$ by
  recurring over the structure of $\Phi$ as follows:
  \[
  \begin{array}{lll}
    \bullet\ \tforsba(Q) &=& Q \\
    \bullet\ \tforsba(\neg \Phi) &=& \neg \tforsba(\Phi) \\
    \bullet\ \tforsba(\exists x.\Phi) &=& \exists x. \tforsba(\Phi) \\
    \bullet\ \tforsba(\Phi_1 \vee \Phi_2) &=& \tforsba(\Phi_1) \vee \tforsba(\Phi_2) \\
    \bullet\ \tforsba(\mu Z.\Phi) &=& \mu Z. \tforsba(\Phi) \\
    \bullet\ \tforsba(\DIAM{\Phi}) &=& \DIAM{\DIAM{\DIAM{\DIAM{\DIAM{\DIAM{\DIAM{\DIAM{\tforsba(\Phi)}}}}}}}} 
  \end{array}
  \]
\ \ 
\end{definition}

\noindent
Having the translations $\tgkabsba$ and $\tforsba$ in hand, we show it
later that $\ts{\gkabsym}^{\filter_S} \models \Phi$ if and only if
$\ts{\tgkabsba(\gkabsym)}^{\csfilter_B} \models \tforsba(\Phi)$ which
consequently means that the verification of \muladom over S-GKABSs can
be reduced to the corresponding verification of \mulcsa over \bagkabs.
Towards this aim, in the next section we introduce a special
bisimulation relation that will ease to reduce the verification of
S-GKABs into \bagkabs.

\subsection{Skip-Seven Bisimulation (\ssevbsimabr-Bisimulation)}

In this section we the notion of \ssevbsimabr-Bisimulation and show that
two \ssevbsimabr-bisimilar transition system can not be distinguished by
\muladom formula modulo the translation $\tforsba$. Then, we will see
in the next section that the transition system of an S-GKAB is
actually \ssevbsimabr-bisimilar to the transition system of its
corresponding \bagkabs that is obtained via $\tgkabsba$.

\begin{definition}[Skip-Seven Bisimulation (\ssevbsimabr-Bisimulation)]
  \
  \sidetext{Skip-Seven Bisimulation (\ssevbsimabr-Bisimulation)} \\
  Let
  $\ts{1} = \tup{\const, T, \stateset_1, s_{01}, \abox_1, \trans_1}$
  be a KB transition system, and
  $\ts{2} = \tup{\const, \ctbox, \stateset_2, s_{02}, \abox_2, \cntx,
    \trans_2}$
  be context-sensitive transition system, with
  $\adom{\abox_1(s_{01})} \subseteq \const$ and
  $\adom{\abox_2(s_{02})} \subseteq \const$.  A \emph{skip-seven
    bisimulation} (\ssevbsimabr-Bisimulation) between $\ts{1}$ and
  $\ts{2}$ is a relation $\B \subseteq \Sigma_1 \times\Sigma_2$ such
  that $\tup{s_1, s_2} \in \B$ implies that:
  \begin{compactenum}
  \item $\abox_1(s_1) = \abox_2(s_2)$
  \item for each $s_1'$, if $s_1 \Rightarrow_1 s_1'$ then there exists
    $t_i$ (for $i \in \set{1,\ldots,7}$) and $s_2'$ with
    \[
    s_2 \Rightarrow_2 t_1 \Rightarrow_2 \cdots \Rightarrow_2 t_7
    \Rightarrow_2 s_2'
    \] 
    such that $\tup{s_1', s_2'}\in\B$, $\tmp \not\in \abox_2(s_2')$
    and $\tmp \in \abox_2(t_i)$ for $i \in \set{1, \ldots, 7}$.

  \item for each $s_2'$, if 
    \[
    s_2 \Rightarrow_2 t_1 \Rightarrow_2 \cdots \Rightarrow_2 t_7
    \Rightarrow_2 s_2'
    \] 
    with $\tmp \in \abox_2(t_i)$ for $i \in \set{1, \ldots, 7}$ and
    $\tmp \not\in \abox_2(s_2')$, then there exists $s_1'$ with
    $s_1 \Rightarrow_1 s_1'$, such that $\tup{s_1', s_2'}\in\B$.
 \end{compactenum}
\ \ 
\end{definition}

\noindent
Let
$\ts{1} = \tup{\const, T, \stateset_1, s_{01}, \abox_1, \trans_1}$ be
a KB transition system, and
$\ts{2} = \tup{\const, \ctbox, \stateset_2, s_{02}, \abox_2, \cntx,
  \trans_2}$
be a context-sensitive transition system, 
a state $s_1 \in \stateset_1$ is \emph{\ssevbsimabr-bisimilar} to
$s_2 \in \stateset_2$, written $s_1 \ssevbsim s_2$, if there exists an
\ssevbsimabr-bisimulation relation $\B$ between $\ts{1}$ and $\ts{2}$ such that
$\tup{s_1, s_2}\in\B$.
A transition system $\ts{1}$ is \emph{\ssevbsimabr-bisimilar} to
$\ts{2}$, written $\ts{1} \ssevbsim \ts{2}$, if there exists an
\ssevbsimabr-bisimulation relation $\B$ between $\ts{1}$ and $\ts{2}$
such that $\tup{s_{01}, s_{02}}\in\B$.

We now proceed to show that two \ssevbsimabr-bisimilar transition
system can not be distinguished by \muladom formula modulo the
translation $\tforsba$.

\begin{lemma}\label{lem:ssevbisimilar-ts-satisfies-same-formula}
  Consider a KB transition system $\ts{1}$ and a context-sensitive
  transition system $\ts{2}$ such that $\ts{1} \ssevbsim \ts{2}$.  For
  every closed \muladom formula $\Phi$, we have:
  \[
  \ts{1} \models \Phi \textrm{ if and only if } \ts{2} \models
  \tforsba(\Phi)
  \]
\end{lemma}
\begin{proof} 
  The proof is similar to the proof of
  \Cref{lem:stbisimilar-ts-satisfies-same-formula,lem:sbisimilar-ts-satisfies-same-formula}. The
  only different is that here we duplicate the modal operators eight
  times. All required supporting lemmas can be also easily recast into
  this case.
\end{proof}

\subsection{From Standard GKABs into \bagkabs}

To show that we can reduce the verification of S-GKABs into \bagkabs,
we first show that the transition system of an S-GKAB is
\ssevbsimabr-bisimilar to the transition system of its corresponding
\bagkabs that is obtained via $\tgkabsba$.

\begin{lemma}\label{lem:sgkab-to-bagkab-bisimilar-ts}
  Given an S-GKAB $\gkabsym$, we have
  $\ts{\gkabsym}^{\filter_S} \ssevbsim
  \ts{\tgkabsba(\gkabsym)}^{\csfilter_B}$
\end{lemma}
\begin{proof}
  The proof is similar to the proof of
  \Cref{lem:sgkab-to-bicgkab-bisimilar-ts}. The important difference
  is that, compare to \bicgkabs, \bagkabs basically elaborate each
  source of non-determinism therefore, there are seven intermediate states.
\end{proof}

Last, we close this tour by showing that the verification of \muladom
properties over S-GKAB can be recast as verification over \bagkab as
follows.

\begin{theorem}
  \label{thm:ver-sgkab-to-bagkab}
Verification of closed \muladom properties over S-GKABs can be recast
as verification over \bagkabs.
\end{theorem}
\begin{proof}
  By \Cref{lem:sgkab-to-bagkab-bisimilar-ts}, we have that
  $\ts{\gkabsym}^{\filter_S} \ssevbsim
  \ts{\tgkabsba(\gkabsym)}^{\csfilter_B}$.
  Hence, by \Cref{lem:ssevbisimilar-ts-satisfies-same-formula}, for
  every $\muladom$ property $\Phi$, we have that
  \[
  \ts{\gkabsym}^{\filter_S} \models \Phi \textrm{ if and only if }
  \ts{\tgkabsba(\gkabsym)}^{\csfilter_B}\models \tforsba(\Phi)
  \]
  Therefore, by using the translation $\tgkabsba$ we can easily
  transform an S-GKAB into a \bagkab and then the claim is easily
  follows due to the fact above.
\end{proof}

\section{Discussion: Connection between Inconsistency-aware Context-sensitive GKABs and \agkabs}

Notice that the crucial difference between \agkabs and \icgkabs (i.e.,
\bicgkabs, \cicgkabs, \eicgkabs) is that \icgkabs wrap several
non-determinism sources into a single transition while \agkabs
separate them into several transitions. Thus, it is easy to see that
we can easily reduce the verification of \bicgkabs, \cicgkabs, and
\eicgkabs into the corresponding verification of \bagkabs, \cagkabs,
and \eagkabs by simply quadruplicating the modal operator of \mulcs
properties to be verified over \icgkabs (I.e., we translate
$\DIAM{\Phi}$ into $\DIAM{\DIAM{\DIAM{\DIAM{\Phi}}}}$ and similarly
for $\BOX{}$) .



%% file: 2.chapters/9-sedap.tex
\chapter{Semantically-Enhanced Data-Aware Processes
  (\sgdss)}\label{ch:sedap}

\ifhidecontent
 
\fi

As we have seen, Data-centric Dynamic Systems (DCDSs) is built based
on relational database technology.
In a DCDS, processes operate over the data of the system and evolve it
by executing actions that may issue calls to external services.
%
%
On the other hand, the (Golog) Knowledge and Action Bases 
is a work in a form of DCDS but based on ontologies, i.e., the data
layer is represented in a rich ontology formalism, and actions perform
a form of instance level update of the ontology. 
The use of an ontology allows for a high-level conceptual view of the
data layer that is better suited for a business level treatment of the
manipulated information.

Here we introduce Semantically-Enhanced Data-Aware Processes (\sgdss),
in which we merge these two approaches by enhancing a \emph{relational
  layer} constituted by a DCDS-based system, with an ontology,
constituting a \emph{semantic layer}.  
This provides a mechanism to semantically enhance the existing
data-aware processes system that is built based on relational database
technology.
Essentially, in \sgdss, the ontology captures the domain of interest
in which a \sgds is executed. Additionally, it allows for seeing the
data and their manipulation at a conceptual level through an
ontology-based data access (OBDA) system~\cite{CDLL*09,RoCa12},
reflecting the relevant concepts and relations of the domain of
interest and abstracting away from how processes and data are
concretely realized and stored at the concrete implementation level.
It also provides us with a way of \emph{semantically governing} the
underlying DCDS-based system through the semantic layer by enabling us
to specify semantic constraints at the conceptual level. Those
constraints will prevent those actions that are executed at the
relational layer and would lead to new system states that violates
some constraints.
This setting, in turn, is the basis for different important reasoning
task such as verifying the evolving system through the conceptual
level.
Specifically, a \sgds is constituted by three main components:
\begin{compactenum}
\item an \emph{OBDA system}~\cite{CDLL*09} which includes (the
  intensional level of) an ontology, a relational database schema, and
  a mapping between the ontology and the database. Essentially, it
  keeps all the data of interest and provides a conceptual view over
  it.
\item a \emph{process component}, which characterizes the evolution of
  the system in 
  the relational layer.
\item a \emph{database instance}, which stores the initial data of the
  system that will be manipulated by the process component.
\end{compactenum}

In the following, we use \dllitea for expressing ontologies and we
also distinguish between objects and values.
We make use 
a countably infinite set $\valset$ to denote all possible values in
the system. Additionally, we consider a finite set of distinguished
values $\valset_0 \subset \valset$.
Note that the databases store values while in the ontological level,
the instance of concepts are objects. Thus, similar to
\cite{CDLL*09,PLCD*08}, to represent the objects, we make use a set
$\funcsym$ of function symbols, each with an associated arity and it
also contains a special function symbol $\val/1$ (that will be used to
wrap values).
The objects then are represented as terms of the form
$f(d_1,\ldots,d_n)$ where $f \in \funcsym$ and
$d_1,\ldots,d_n \in \valset$. Such kind of terms are called \emph{object
  terms}. We then also define the set $\const$ of constants as the
union of $\valset$ and the set
$\{f(d_1,\ldots,d_n) \mid f \in \funcsym \text{ and } d_1,\ldots,d_n
\in \valset \}$ of object terms.
%
%
Last but not least, we also
consider 
%
a finite set $\servcall$ of \textit{function symbols} that represents
\textit{service calls}, which abstractly account for the injection of
fresh values (constants) from $\const$ into the system.


The results in this chapter are published in
\cite{AS-KiBP-12,AS-RR-12a,AS-RR-12b,AS-ICSOC-13,AS-CORR-13b}

\section{Formalizing \sgdss}

Before we formally defined \sgdss in \Cref{sec:formalization-sedap},
we first briefly review the Ontology Based Data Access (OBDA) in
\Cref{sec:recap-obda} to give some necessary preliminaries. 

\subsection{Ontology Based Data Access (OBDA) at a glance}
\label{sec:recap-obda}

In an Ontology Based Data Access (OBDA) system, a relational database
is connected to an ontology that represents the domain of interest by
a mapping, which relates database values with values and (abstract)
objects in the ontology (c.f.\ \cite{CDLL*09,PLCD*08}).  The mapping
in OBDA is formally defined as follows:

\begin{definition}[OBDA Mapping]\label{def:obda-mapping}
  Given \sidetext{OBDA Mapping} a TBox $T$, and a database schema
  $\dbschema$, an \emph{OBDA mapping $\obdamap$ over $T$ and
    $\dbschema $} is a set of mapping assertions, each of the form
  \[
  \map{\Phi(\vec{x})}{\Psi(\vec{t} )},
  \]
  where:
  \begin{compactenum}

  \item $\vec{x}$ is a non-empty set of variables,

  \item $\vec{t}$ is a set of terms of the form $f(\vec{z})$, with $f
    \in \funcsym$ and $\vec{z} \subseteq \vec{x}$,

  \item $\Phi(\vec{x})$ is an SQL query over $\dbschema$, with
    $\vec{x}$ as output variables (Note that we only consider the core
    SQL fragment that corresponds to \difol), and

  \item $\Psi(\vec{t})$ is a CQ over $T$ without non-distinguished
    variables, whose atoms are the terms $\vec{t}$.

  \end{compactenum}
  Without loss of generality, we use a special function symbol
  $\val/1$ to map values from the relational layer to the range of
  attributes in the semantic layer.
\end{definition}

\noindent
Formally, an OBDA systems is then defined as follows.

\begin{definition}[OBDA System]\label{def:obda-system}
  Formally, \sidetext{OBDA System} an \emph{OBDA system} is a
  structure $\obdasys~=~\tup{T, \dbschema, \obdamap}$, where:
\begin{compactenum}

\item $T$ is a $\dllitea$ TBox;

\item $\dbschema = \set{R_1, \ldots, R_n}$ is a database schema,
  constituted by a finite set of relation schemas;


\item $\obdamap$ is an OBDA mapping over $T$ and $\dbschema$.

\end{compactenum}
\ \ 
\end{definition}

\begin{example}\label{ex:obda-sys}
  For the running example of this chapter, recall the simple order
  processing scenario that is used in \Cref{ex:dcds}.
%
%
  We now specify an OBDA system
  $\obdasys = \tup{T, \dbschema, \obdamap}$ where
  \begin{compactitem}
  \item $T$ is the same as the TBox specified in
    \Cref{ex:tbox-and-abox}. 
  \item $\dbschema$ is the same as the database schema specified as in
    \Cref{ex:dbschema-and-dbinst}.


  \item $\obdamap$ contains the following mapping assertions:\\
    \begin{tabular}{l@{ }l}
      $m_1:$ ~& \texttt{SELECT} \exra{id}, \exra{name}\ \texttt{FROM} \exr{ORDER}\\
              &\texttt{WHERE} \exra{processing\_status} \texttt{=} ``approved''\\
              & $\leadsto$ \exo{ApprovedOrder}(\exf{ord}(\exra{id}, \exra{name}))\\
    \end{tabular}\\
    \begin{tabular}{l@{ }l}
      $m_2:$ ~& \texttt{SELECT} \exra{id}, \exra{name}\ \texttt{FROM} \exr{ORDER}\\
                &\texttt{WHERE} \exra{processing\_status} \texttt{=} ``received''\\
              & $\leadsto$ \exo{ReceivedOrder}(\exf{ord}(\exra{id}, \exra{name}))\\
    \end{tabular}\\
    \begin{tabular}{l@{ }l}
      $m_3:$ ~& \texttt{SELECT} \exra{id}, \exra{name}\ \texttt{FROM} \exr{ORDER}\\
                &\texttt{WHERE} \exra{processing\_status} \texttt{=} ``assembled''\\
              & $\leadsto$ \exo{AssembledOrder}(\exf{ord}(\exra{id}, \exra{name}))\\
    \end{tabular}\\
    \begin{tabular}{l@{ }l}
      $m_4:$ ~& \texttt{SELECT} \exra{id}, \exra{name}\
                \texttt{FROM} \exr{ORDER} \exra{o}, \exr{DELIVERED\_ORDER} \exra{d},  \\
              &\texttt{WHERE} \exra{o.id} \texttt{=} \exra{d.id} \\
              & $\leadsto$ \exo{DeliveredOrder}(\exf{ord}(\exra{id}, \exra{name}))\\
    \end{tabular}\\
    \begin{tabular}{l@{ }l}
      $m_5:$ ~& \texttt{SELECT} \exra{assembler}\ \texttt{FROM} \exr{ORDER}\\
              &\texttt{WHERE} \exra{assembler} \texttt{IS NOT NULL}\\
              & $\leadsto$ \exo{Assembler}(\exf{emp}(\exra{assembler}))\\
    \end{tabular}\\
    \begin{tabular}{l@{ }l}
      $m_6:$ ~& \texttt{SELECT} \exra{designer}\ \texttt{FROM} \exr{ORDER}\\
              &\texttt{WHERE} \exra{designer} \texttt{IS NOT NULL}\\
              & $\leadsto$ \exo{Designer}(\exf{emp}(\exra{designer}))\\
    \end{tabular}\\
    \begin{tabular}{l@{ }l}
      $m_7:$ ~& \texttt{SELECT} \exra{quality\_controller}\ \texttt{FROM} \exr{ORDER}\\
              &\texttt{WHERE} \exra{quality\_controller} \texttt{IS NOT NULL}\\
              & $\leadsto$ \exo{QualityController}(\exf{emp}(\exra{quality\_controller}))\\
    \end{tabular}\\
    \begin{tabular}{l@{ }l}
      $m_8:$ ~& \texttt{SELECT} \exra{id}, \exra{name}, \exra{assembling\_loc}\ \texttt{FROM} \exr{ORDER}\\
              &\texttt{WHERE} \exra{assembling\_loc} \texttt{IS NOT NULL}\\
              & $\leadsto$ \exo{hasAssemblingLoc}(\exf{ord}(\exra{id}, \exra{name}), \exf{loc}(\exra{assembling\_loc}))\\
    \end{tabular}
    \begin{tabular}{l@{ }l}
      $m_9:$ ~& \texttt{SELECT} \exra{id}, \exra{name}, \exra{design}\ \texttt{FROM} \exr{ORDER}\\
              &\texttt{WHERE} \exra{design} \texttt{IS NOT NULL}\\
              & $\leadsto$ \exo{hasDesign}(\exf{ord}(\exra{id}, \exra{name}), \exf{des}(\exra{design}))\\
    \end{tabular}\\
    \begin{tabular}{l@{ }l}
      $m_{10}:$ ~& \texttt{SELECT} \exra{id}, \exra{name}, \exra{assembler}\ \texttt{FROM} \exr{ORDER}\\
              &\texttt{WHERE} \exra{assembler} \texttt{IS NOT NULL}\\
              & $\leadsto$ \exo{assembledBy}(\exf{ord}(\exra{id}, \exra{name}), \exf{emp}(\exra{assembler}))\\
    \end{tabular}\\
    
    \begin{tabular}{l@{ }l}
      $m_{11}:$ ~& \texttt{SELECT} \exra{id}, \exra{name}, \exra{designer}\ \texttt{FROM} \exr{ORDER}\\
              &\texttt{WHERE} \exra{designer} \texttt{IS NOT NULL}\\
              & $\leadsto$ \exo{designedBy}(\exf{ord}(\exra{id}, \exra{name}), \exf{emp}(\exra{designer}))\\
    \end{tabular}\\
    \begin{tabular}{l@{ }l}
      $m_{12}:$ ~& \texttt{SELECT} \exra{id}, \exra{name}, \exra{quality\_controller}\ \texttt{FROM} \exr{ORDER}\\
              &\texttt{WHERE} \exra{quality\_controller} \texttt{IS NOT NULL}\\
              & $\leadsto$ \exo{checkedBy}(\exf{ord}(\exra{id}, \exra{name}), \exf{emp}(\exra{quality\_controller}))\\
    \end{tabular}\\
  \end{compactitem}
  The intuition of the mappings above is as follows: 
  the mapping $m_1$ (resp.\ $m_2$ and $m_3$) maps every order in
  $\exr{ORDER}$ with $\exra{processing\_status}$ ``approved'' (resp.\
  ``received'' and ``assembled'') to an $\exo{ApprovedOrder}$ (resp.\
  \exo{ReceivedOrder} and \exo{AssembledOrder}). Such an order is
  constructed by ``objectifying'' the id and name using function
  $\exf{ord}/2$.
  The mapping $m_4$ generates delivered order by selecting only those
  orders in the $\exr{ORDER}$ table whose id is also contained in the
  $\exr{DELIVERED\_ORDER}$ table.
  The mapping $m_5$ (resp.\ $m_6$ and $m_7$) populates the concept
  $\exo{Assembler}$ (resp.\ $\exo{Designer}$ and
  $\exo{QualityController}$) with the $\exra{assembler}$ (resp.\
  $\exra{designer}$ and $\exra{quality\_controller}$) data in
  $\exr{ORDER}$.
  The role $\exo{hasAssemblingLoc}$ (resp.\ $\exo{hasDesign}$) is
  populated by the mapping $m_8$ (resp. $m_9$) using the data about
  orders in $\exr{ORDER}$ and their corresponding assembling location
  (resp.\ design). 
  Finally, the mapping $m_{10}$ (resp.\ $m_{11}$ and $m_{12}$)
  populate the role $\exo{assembledBy}$ (resp.\ $\exo{designedBy}$ and
  $\exo{checkedBy}$) with the orders in $\exr{ORDER}$ and their
  corresponding assembler (resp.\ designer and quality controller).
\end{example}

Given a database instance $\dbinst$ over a database schema $\dbschema$
with $\adom{\dbinst} \subseteq \valset$,
and given a mapping $\obdamap$, the \emph{virtual ABox} generated from
$\dbinst$ by a mapping assertion
$m=\map{\Phi(\vec{x})}{\Psi(\vec{t})}$ in $\obdamap$ is
$m(\dbinst) = \bigcup_{\sigma \in \ANS(\Phi,\dbinst)} \Psi\sigma$,
%
%
Then, the \emph{virtual ABox} generated from $\dbinst$ by the mapping
$\obdamap$ is
$\obdamap(\dbinst) = \bigcup_{m \in \obdamap} m(\dbinst)$.  
Notice that $\adom{\obdamap(\dbinst)} \subseteq \const$.  
%
%
Given an OBDA system $\obdasys = \tup{T, \dbschema, \obdamap}$ and a
database instance $\dbinst$ over a database schema $\dbschema$, a
\emph{model for $\obdasys$ wrt $\dbinst$} is an interpretation $\I$
such that $\I \models (T, \obdamap(\dbinst))$. We say that $\obdasys$
is satisfiable w.r.t.\ $\dbinst$ if it admits at least one model
w.r.t.\ $\dbinst$.

\begin{example}
  Continuing our running example (see \Cref{ex:obda-sys}), consider a
  database instance
  \[
  \dbinst = \set{\exr{ORDER}(\excon{123}, \excon{chair},
    \excon{approved}, \excon{456}, \excon{alice}, \excon{bob},
    \excon{john}, \excon{bolzano}, \excon{ecodesign}) }.
  \]
  The corresponding virtual ABox obtained from the application of the
  mapping $\obdamap$ is
\[
\begin{array}{r@{ \ }l}
  \obdamap(\dbinst) = \set{&\exo{ApprovedOrder}(\exf{ord}(123, chair)), \exo{Assembler}(\exf{emp}(bob)),\\
                     & \exo{Designer}(\exf{emp}(alice)), \exo{Quality\_Controller}(\exf{emp}(john)), \\
                     &\exo{hasAssemblingLoc}(\exf{ord}(123, chair), \exf{loc}(bolzano)), \\
                     &\exo{hasDesign}(\exf{ord}(123, chair), \exf{des}(ecodesign)), \\
                     &\exo{assembledBy}(\exf{ord}(123, chair), \exf{emp}(bob)), \\
                     &\exo{designedBy}(\exf{ord}(123, chair), \exf{emp}(alice)), \\
                     &\exo{checkedBy}(\exf{ord}(123, chair), \exf{emp}(john))
                       }
\end{array}
\]
\ \ 
\end{example}

A UCQ $q$ over an OBDA system
$\obdasys = \tup{T, \dbschema, \obdamap}$ is simply a UCQ over $T$. To
compute the certain answers of $q$ over $\obdasys$ wrt a database
instance $\dbinst$ for $\dbschema$, we follow a three-step approach:
\begin{compactenum}
\item $q$ is \emph{rewritten} to compile away $T$, obtaining
  $q_r = \rew(q, T)$;
\item the mapping $\obdamap$ is used to \emph{unfold} $q_r$ into a query over
  $\dbschema$, denoted by $\unfold{q_r}{\obdamap}$, which turns out to be an
  SQL query \cite{PLCD*08}; 
\item such a query is then evaluated over $\dbinst$, obtaining the
  certain answers.
\end{compactenum}
For an ECQ, we can proceed in a similar way, applying the rewriting and
unfolding steps to the embedded UCQs.  
It follows that computing certain answers to UCQs/ECQs in an OBDA
system is FO rewritable. Furthermore, applying the unfolding step to
$\qunsatfol{T}$, we obtain also that satisfiability in $\obdasys$
can be reduced into evaluating a query over $\dbinst$.




\subsection{Formalization of \sgdss} \label{sec:formalization-sedap}

Roughly speaking, a \sgds is constituted by:
\begin{inparaenum}[\it (i)]
\item A \emph{Relational Layer}, which captures the database evolution
  (manipulation) by actions.
\item A \emph{Semantic Layer}, which exploits the ontology for
  providing a conceptual view of the system evolution.
\item A set of \emph{mapping assertions} describing how to virtually
  project data concretely maintained at the Relational Layer into
  concepts and relations modeled in the Semantic Layer, thus providing
  a link between the data in the relational layer and the ontology.
\end{inparaenum}

To avoid unnecessary complication, here we only use a set of
condition-action rules (c.f.\ \Cref{def:dcds-ca-rules}) to formalize
the progression mechanism that evolves the data in the relational
layer. However, one can easily lift it into a more
complex/sophisticated formalism such as Golog program similar to
\Cref{ch:gkab}, but this is not our focus on this chapter.
Moreover, here we use the DCDS actions formalisms, and we will also
see later that we use the DCDS actions execution semantics that
rebuild the whole database instance at each action execution. However,
it is easy to see that actually we can change the setting such that it
uses the KAB actions formalisms (with a slight modification on the
queries) and KAB actions execution semantics at no cost. This is
possible due to the fact that we can simulate the KAB actions
execution semantics in DCDS actions execution semantics as we have
seen when we compile KABs into DCDSs in
\Cref{sec:trans-kab-to-dcds}. One reason why we use DCDS actions and
their execution semantics here is because we want to simplify the
proof and avoid unnecessary complication while focusing on presenting
the setting.

Formally, a \sgds is then defined as follows:

\begin{definition}[SEDAP]\label{def:sedap}
  A \sidetext{Semantically-Enhanced Data-Aware Processes (\sgds)}
  \emph{\sgds} $\sgdssym$ is a tuple
  $\tup{T, \dbschema, \obdamap, \idb, \sactset, \sprocset}$, where:
\begin{compactitem}
\item $T$ is a \dllitea TBox,
\item $\dbschema$ is a database schema,
\item $\obdamap$ is an OBDA mapping over $T$ and $\dbschema$ (see
  \Cref{def:obda-mapping}),
\item $\idb$ is a database instance over $\dbschema$,
\item $\sactset$ is a set of DCDS actions over $\dbschema$ and $\idb$
  (see \Cref{def:dcds-act}),
\item $\sprocset$ is a set of DCDS condition-action rules over
  $\dbschema$, $\idb$, and $\sactset$ (see \Cref{def:dcds-ca-rules}).
\end{compactitem}
 \ \ 
\end{definition}

\noindent
Together $T$, $\dbschema$, and $\obdamap$ constitute the OBDA system,
while $\sactset$ and $\sprocset$ form the \emph{process
  component}. 
%
%

\begin{example}\label{ex:sedap}
  Consider our running example, recall 
  the OBDA system $\obdasys = \tup{T, \dbschema, \obdamap}$ in
  \Cref{ex:obda-sys}.
%
%
  To model our simple order processing scenario, we specify a \sgds
  $\sgdssym = \tup{T, \dbschema, \obdamap, \idb, \sactset, \sprocset}$
  where 
\begin{compactitem}
\item $T, \dbschema$, and $\obdamap$ are the same as in
  \Cref{ex:obda-sys} (i.e., $T$ is specified in
  \Cref{ex:tbox-and-abox}, and $\dbschema$ is specified in
  \Cref{ex:dbschema-and-dbinst}).
\item $\sactset$ (resp.\ $\sprocset$) is the same as the set of
  actions (resp.\ condition-action rules) specified in \Cref{ex:dcds}.
\item The initial database $\idb$ 
  is specified as follows:
  \[
  \begin{array}{@{}r@{ \ }l@{}l@{}}
    \idb = \set{&\exr{ORDER}(\excon{123}, \excon{chair}, \excon{received},
                               \excon{456}, \excon{NULL}, &\excon{NULL},
                               \\
                &&\excon{NULL}, \excon{NULL},
                   \excon{ecodesign}), \\
                &\exr{ORDER}(\excon{321}, \excon{table}, \excon{approved},
                               \excon{654}, \excon{NULL}, &\excon{NULL},
                                \\
                &&   \excon{NULL}, \excon{NULL},
                   \excon{NULL})
                   }
  \end{array}
  \]
\end{compactitem}
\end{example}

\section{\sgdss Execution Semantics}

The semantics of \sgdss is provided in terms of possibly infinite
transition systems. More specifically, two transition systems are
constructed to describe the execution semantics of \sgdss:
\begin{compactenum}

\item A \emph{Relational Layer Transition System} (RTS), representing
  all allowed computations that, starting from the initial database $\idb$, the process
  component can do over the data in the relational layer, according to
  the constraints imposed at the semantic layer (semantic governance).

\item A \emph{Semantic Layer Transition System} (STS), representing
  the same computations at the semantic layer (abstracting the
  evolution in the relational layer).

\end{compactenum}
Both of them are formally defined as follows:

\begin{definition}[Relational Layer Transition System (RTS)]\label{def:rts}
  Given \sidetext{Relational Layer Transition System (RTS)} a \sgds
  $\sgdssym=\tup{T, \dbschema, \obdamap, \idb, \sactset, \sprocset}$,
  we define the \emph{Relational Layer Transition System} (RTS) of
  $\sgdssym$, written $\rts{\sgdssym}$, as a tuple
  $\tup{\const, \dbschema, \dstateset, s_0, \db, \trans}$, where:
  \begin{compactenum}
  \item $\stateset$ is a set of states where each state
    $s \in \stateset$ is defined as a tuple $\tup{\dbinst, \dscmap}$,
    where $\dbinst$ is a database instance and $\dscmap$ is a service
    call map,
  \item $s_0$ is an initial state,
  \item $\db$ is a function that, given a state in $\rts{\sgdssym}$,
    returns a corresponding database instance (conforming to
    $\dbschema$),
  \item $\trans \subseteq \dstateset \times \dstateset$ is the
    transition relation.
  \end{compactenum}
  The components $\dstateset$, $\trans$ and $\db$ of $\rts{\sgdssym}$
  are defined by simultaneous induction as the smallest sets
  satisfying the following conditions:

  \begin{compactitem}

  \item $s_0 = \tup{\idb, \emptyset} \in \dstateset$, $\db(s_0) = \idb$;
  \item if $s = \tup{\dbinst,\dscmap} \in \dstateset$, then for all
    actions $\dact \in \sactset$, for all legal parameter assignments
    $\sigma$ for $\dact$ in $\dbinst$ and
    for all $\tup{\dbinst',\dscmap'}$ such that
    \begin{compactenum}
    \item
      $\tup{\dbinst,\dscmap} \exect{\dact\sigma, \ \sgdssym\ }
      \tup{\dbinst',\dscmap'}$, and
    \item the OBDA system $\obdasys = \tup{T, \dbschema, \obdamap}$ is
      \emph{satisfiable} w.r.t.\ $\dbinst'$.
    \end{compactenum}
    we have $s' = \tup{\dbinst',\scmap'}\in \stateset$, and
    $s \trans s'$.
  \end{compactitem}
  (Note that the relation $ \exect{\dact\sigma, \ \sgdssym\ }$ is
  defined in \Cref{def:dcds-trans-rel}).
\end{definition}

\noindent
Observe that the satisfiability check that is done in the last step of
the RTS construction realizing the notion of \emph{semantic
  governance}. I.e., the system evolution in the relational layer
taking into account the constraints specified in the semantic layer.

The semantic layer transition system (STS) $\sts{\sgdssym}$ of a \sgds
$\sgdssym$ is basically a ``virtualization'' of the RTS
$\rts{\sgdssym}$ of $\sgdssym$ in the semantic layer.  It is basically
obtained from the RTS $\rts{\sgdssym}$.  Essentially,
STS maintains the structure of $\rts{\sgdssym}$ unaltered, reflecting
that the process component is executed over the relational layer, but
it associates each state $s$ to a virtual ABox obtained from the
application of the mapping $\obdamap$ to the database instance
associated by $\rts{\sgdssym}$ to the same state $s$. Formally, the
STS is then defined as follows:

\begin{definition}[Semantic Layer Transition System (STS)]\label{def:sts}
  Given \sidetext{Semantic Layer Transition System (STS)} a \sgds
  $\sgdssym=\tup{T, \dbschema, \obdamap, \idb, \sactset, \sprocset}$,
  let
  $\rts{\sgdssym} = \tup{\const, \dbschema, \dstateset, s_0, \db,
    \trans}$
  be its RTS. We define the \emph{Semantic Layer Transition System
    (STS)} of $\sgdssym$, written $\sts{\sgdssym}$, as a tuple
  $\tup{\const, T, \dstateset, s_0, \abox, \trans}$ such that for each
  $s \in \dstateset$, $\abox(s) = \obdamap(\db(s))$. (Note that
  $\dstateset$, $s_0$, and $\trans$ are the same as in
  $\rts{\sgdssym}$).
\end{definition}

\noindent
The intuition of the \sgds setting (the semantic layer transition
system and the relational layer transition system) is depicted in
\Cref{sedap-intuition}.

\begin{example}\label{ex:sedap-exec}
  Continuing our running example in \Cref{ex:sedap} and reconsider the
  \sgds $\sgdssym$ specified in that example. The construction of
  RTS $\rts{\sgdssym}$ is started from the
  initial state $s_0 = \tup{\idb,\emptyset}$, where:
 \[
  \begin{array}{@{}r@{ \ }l@{}l@{}}
    \idb = \set{&\exr{ORDER}(\excon{123}, \excon{chair}, \excon{received},
                  \excon{456}, \excon{NULL}, &\excon{NULL},
                                               \excon{NULL}, \excon{NULL},
                                               \excon{ecodesign}), \\
                &\exr{ORDER}(\excon{321}, \excon{table}, \excon{approved},
                  \excon{654}, \excon{NULL}, &\excon{NULL},
                                               \excon{NULL}, \excon{NULL},
                                               \excon{NULL})
                                               }
  \end{array}
  \]
  An example of a sucessor of state $s_0$ is a state
  $s_1 = \tup{\dbinst_1, \scmap_1}$, where
  \[
  \begin{array}{@{}r@{}l@{}}
    \dbinst_1 = \set{&\exr{ORDER}(\excon{123}, \excon{chair}, \excon{approved},
                       \excon{456}, \excon{NULL}, \excon{NULL}, \excon{NULL}, \excon{NULL},
                       \excon{ecodesign}), \\
                     &\exr{ORDER}(\excon{321}, \excon{table}, \excon{approved},
                       \excon{654}, \excon{NULL}, \excon{NULL}, \excon{NULL}, \excon{NULL},
                       \excon{NULL})
                       },
  \end{array}
  \]
  and $s_1$ is obtained similarly as in \Cref{ex:dcds-exec}. Note that
  we use the same set of actions and condition-action rules as in
  \Cref{ex:dcds-exec}. The construction of $\rts{\sgdssym}$ is
  continued further by applying all possible actions and so on.

  On the other hand, the STS $\sts{\sgdssym}$ is obtained by
  virtualizing $\rts{\sgdssym}$ into the semantic layer by utilizing
  the mapping $\obdamap$. The corresponding ABox of each state is
  obtained by applying mapping $\obdamap$ to the database instance of
  the corresponding states. For instance, we have that
\begin{center}
$ 
 \begin{array}{r@{ \ }l}
    \abox(s_0) = \obdamap(\db(s_0)) = \set{
    &\exo{ReceivedOrder}(\exf{ord}(123, chair)), \\
    &\exo{hasDesign}(\exf{ord}(123, chair), \exf{des}(ecodesign)), \\
    &\exo{ApprovedOrder}(\exf{ord}(321, table)) 
      }
\end{array}
$
\end{center}
and also a successor of $s_0$ is $s_1$ where
\begin{center}
$ 
 \begin{array}{r@{ \ }l}
    \abox(s_1) = \obdamap(\db(s_1)) = \set{
    &\exo{ApprovedOrder}(\exf{ord}(123, chair)), \\
    &\exo{hasDesign}(\exf{ord}(123, chair), \exf{des}(ecodesign)), \\
    &\exo{ApprovedOrder}(\exf{ord}(321, table)) 
      }
\end{array}
$
\end{center}
\end{example}

\section{Correspondence Between \sgdss and DCDSs}

The interesting task in \sgdss is to verify the compliance of \sgdss
evolution against the conceptual temporal properties specified over
the semantic layer. To tackle this issue, later we will see that the
verification of \sgdss can be reduced to the verification of DSDSs and
hence we can take the advantages from the well-established results in
DCDSs \cite{BCDDM13}. To this aim, in this section we establish an
interesting correspondence between \sgdss and DCDSs. In particular,
here we present the mechanism of compiling \sgdss into DCDSs.

We now define a translation $\todcds$ that takes a \sgds $\sgdssym$ as
input and produce a DCDS $\todcds(\sgdssym)$ such that the transition
system $\ts{\todcds(\sgdssym)}$ of $\todcds(\sgdssym)$ is equivalent
to the relational transition system $\rts{\sgdssym}$ of $\sgdssym$.

\begin{definition}[Translation From \sgds to DCDS]\label{def:trans-sedap-dcds}
  We \sidetext{Translation From \sgds to DCDS} define a translation
  $\todcds$ that, given a \sgds
  $\sgdssym=\tup{T, \dbschema, \obdamap, \idb, \sactset, \sprocset}$,
  produces a DCDS $\todcds(\sgdssym) = \tup{\dcomp, \pcomp}$ such that
  \begin{compactitem}
  \item $\dcomp = \tup{\dbschema,\idb, \ecset}$ is a DCDS data
    component with
    $\ecset = \set{\unfold{\qunsatfol{T}}{\obdamap} \ra \false}$,
    where $\qunsatfol{T}$ is an FOL query defined in
    \Cref{def:qunsat-fol}.
    Intuitively, we encode the constraints in the TBox $T$ into the
    equality constraints $\ecset$ in DCDS.
  \item $\pcomp = \tup{\dactset,\dprocset}$ is a DCDS process
    component over $\dcomp$.
\end{compactitem}
\ \ 
\end{definition}

\noindent
Essentially, to obtain a DCDS from a \sgds, we compile the negative
inclusion and functionality assertions in the TBox into equality
constraints. Thus, roughly speaking we delegate the consistency check
into the relational layer. Having the translation $\todcds$ in hand,
we can easily show the following theorem.

\begin{theorem}\label{thm:sedap-to-dcds-reduction}
  Given a \sgds $\sgdssym$ with RTS $\rts{\sgdssym}$, let
  $\todcds(\sgdssym)$ be its corresponding DCDS obtained via $\todcds$
  and $\ts{\todcds(\sgdssym)}$ be the corresponding TS of
  $\todcds(\sgdssym)$.  We have that $\rts{\sgdssym}$ is equivalent to
  $\ts{\todcds(\sgdssym)}$
\end{theorem}
\begin{proof}
  Let
  $\sgdssym=\tup{T, \dbschema, \obdamap, \idb, \sactset, \sprocset}$
  and $\todcds(\sgdssym) = \tup{\dcomp, \pcomp}$ where
  $\dcomp = \tup{\dbschema,\idb, \ecset}$,
  $\ecset = \set{\unfold{\qunsatfol{T}}{\obdamap} \ra \false}$, and
  $\pcomp = \tup{\dactset,\dprocset}$.  The proof can be easily
  obtained by considering the following:
\begin{compactitem}

\item Both $\sgdssym$ and $\todcds(\sgdssym)$ start from the same
  initial database instance $\idb$.

\item The satisfiability check of an OBDA system
  $\obdasys = \tup{T, \dbschema, \obdamap}$ w.r.t.\ a database
  instance $\dbinst$ can be delegated to checking whether $\dbinst$
  satisfy $\ecset$ due to the correctness of rewriting and unfolding
  procedure in answering queries over an OBDA system \cite{PLCD*08}
  and the fact that the satisfiability check in \dllitea can be
  delegated into query answering \cite{CDLLR07}. Thus, it follows that
  for each state $s_s$ in $\rts{\sgdssym}$ and state $s_d$ in
  $\ts{\todcds(\sgdssym)}$ such that $\db(s_s) = \db(s_d)$, $s_s$ is a
  consistent state if and only if $s_d$ is a consistent state.

\item both $\rts{\sgdssym}$ and $\ts{\todcds(\sgdssym)}$ has the same
  structure since their process component are the same and each
  corresponding states contain the same database instance.

\end{compactitem}
\end{proof}

\section{Verifying \sgdss}\label{sec:ver-sedap}

Given a \sgds $\sgdssym$, we are interested in studying the
verification of semantic temporal properties specified over the
Semantic Layer. Technically, this means that properties are verified
against the \sgds's STS $\sts{\sgdssym}$. Moreover, the temporal
properties to be verified combines temporal operators with queries
posed over the ontologies obtained by combining the TBox $T$ with the
ABoxes associated to the states of $\sts{\sgdssym}$. As verification
formalisms, here we consider $\muladom$. The verification problem of
\muladom over \sgdss is then formally defined as follows:

\begin{definition}[Verification of $\muladom$ over \sgdss]
  Given \sidetext{Translation From \sgds to DCDS} a \sgds $\sgdssym$
  and a $\muladom$ formula $\Phi$, the \emph{verification of $\Phi$
    over $\sgdssym$} is the problem of checking whether
  $\sts{\sgdssym} \models \Phi$, where $\sts{\sgdssym}$ is the
  semantic layer transition system of $\sgdssym$.
\end{definition}

The problem is that the temporal properties are specified over the
semantic layer but the system actually evolves at the relational
layer. So, to reconcile these pieces, we bring the verification down
to the relational layer. We show that verification of $\muladom$
properties over the STS $\sts{\sgdssym}$ can be reduced to
verification of $\mula$ properties over the corresponding RTS
$\rts{\sgdssym}$.

The reduction is realized by providing a translation mechanism from
$\Phi$ into a corresponding $\mula$ property $\Phi'$ specified over
$\dbschema$, and then showing that $\sts{\sgdssym} \models \Phi$ if
and only if $\rts{\sgdssym} \models \Phi'$. This translation is based
on the notion of rewriting and unfolding as the procedure to compute
the certain answers over an OBDA system. Given a \sgds
$\sgdssym = \tup{T, \dbschema, \obdamap, \idb, \sactset, \sprocset}$
and a \muladom formula $\Phi$, the general strategy about the
translation for $\Phi$ is as follows:
\begin{compactenum}

\item We keep the temporal part unaltered.

\item We rewrite each query in $\Phi$ w.r.t.\ the TBox $T$ in order to
  compile away the TBox and incorporate the knowledge encoded in
  $T$. Formally this step is defined in
  \Cref{def:perf-ref-muladom}. Essentially this step transforms each
  query in $\Phi$ into a DI-FOL query.

\item We unfold the rewritten temporal formula $\rew(\Phi, T)$ based
  on the given mapping $\obdamap$ in order to transform each
  rewritten query in $\Phi$ into a query over $\dbschema$. 

\end{compactenum}


An important observation while unfolding the query is that in the
semantic layer the elements of the active domain are objects while in
the relational layer the elements of the active domain are
values. Hence, we also need to transform the quantification over the
objects into the quantification of the corresponding values that form
the objects.
The unfolding mechanism for the rewritten temporal formula is formally
defined as follows: 


\clearpage
\begin{definition}[Unfolding Mechanism for $\muladom$]
  Given \sidetext{Unfolding Mechanism for $\muladom$} a TBox $T$, a
  database schema $\dbschema$, a $\muladom$ formula $\Phi$ over $T$, a
  mapping $\obdamap$ over $T$ and $\dbschema$. Let
  $\Phi' = \rew(\Phi, T)$ be the rewritten formula of $\Phi$ w.r.t.\
  $T$, we define the unfolding of $\Phi'$ w.r.t.\ $\obdamap$, written
  $\unfold{\Phi'}{\obdamap}$, recursively as follows:

  \[
  \unfold{\Phi'}{\obdamap} = \left\{
    \begin{array}{@{}r@{}ll}
      & \unfold{Q}{\obdamap}   &\mbox{if}\ {\Phi' = Q} \\
      & \unfold{\Psi_1}{\obdamap} \vee \unfold{\Psi_2}{\obdamap}   & \mbox{if}\ {\Phi' = \Psi_1 \vee \Psi_2} \\
      & \bigvee_{(f/n) \in \fsset{\obdamap}}\exists x_1,\ldots,x_n. & \mbox{if}\ {\Phi' = \exists x. \Psi} \\
      & \hspace*{10mm}\unfold{\Psi[x/f(x_1,\ldots,x_n)]}{\obdamap} &  \\
      & \DIAM{\unfold{\Psi}{\obdamap}}   & \mbox{if}\ {\Phi' = \DIAM{\Psi}} \\
      & \mu Z.\unfold{\Psi}{\obdamap}   & \mbox{if}\ {\Phi' = \mu Z.\Psi} \\
    \end{array}
  \right.
  \]

  \noindent
  where:
  \begin{compactenum}

  \item $\unfold{Q}{\obdamap}$ is as in the usual unfolding in OBDA
    (c.f.\ \cite{PLCD*08,CDLL*09}).

  \item $\fsset{\obdamap}$ is the set of function symbols that are
    used to form the object terms and occur in $\obdamap$ (including
    the special function symbol $\val/1$).

  \end{compactenum}
  \ \ 
\end{definition}


\noindent
For unfolding the query, we unfold the query of the form
$\exists x. Q'$ as follows:\\
\hspace*{3mm}$\unfold{\exists x. Q'}{\obdamap} = $
\begin{center}
  $\bigvee_{(f/n) \in \fsset{\obdamap}}\exists
  x_1,\ldots,x_n.\unfold{Q'[x/f(x_1,\ldots,x_n)]}{\obdamap}$.
\end{center}
I.e., we unfold $\exists x. Q'$ into a disjunction of formulas, where
each formula is obtained from $Q$ by replacing $x$ with one of the
possible terms constructed from function symbols in $\obdamap$, and
then existentially quantify each variable that form the corresponding term.
%
%
The reason of this unfolding is to rephrase the quantification over
object terms into the corresponding quantification over values in the
relational layer that could lead to produce such object terms and
values through the application of $\obdamap$.
%
%
This is done by unfolding $\exists x.Q$ into a disjunction of
formulas, where each of the formula is obtained from $Q$ by replacing
$x$ with one of the possible variable terms constructed from function
symbols in $\obdamap$, and quantifying over the existence of values
that could form a corresponding object term.
%

For unfolding the UCQ, the atoms in the UCQ are unified with the heads
of the mapping assertions in $\obdamap$. For each successful
unification, each atom is replaced with the body of the corresponding
mapping. The unfolding of the UCQ is then obtained as the union of all
queries obtained in this way.
%
%
Other cases of query are simply managed by pushing the unfolding down
to the sub-formulas.

\begin{theorem}\label{thm:corr-unf-rew-muladom}
  Let
  $\sgdssym = \tup{T, \dbschema, \obdamap, \idb, \sactset, \sprocset}$
  be a \sgds,
  $\rts{\sgdssym} = \tup{\const, \dbschema, \dstateset, s_0, \db,
    \trans}$
  and
  $\sts{\sgdssym} = \tup{\const, T, \dstateset, s_0, \abox, \trans}$
  consecutively be the RTS and STS of $\sgdssym$.  Consider 
  a $\muladom$ formula $\Phi$ over
  $T$. Then:
  \[
  \sts{\sgdssym} \models \Phi \quad\text{if and only if}\quad
  \rts{\sgdssym} \models \unfold{\rew(\Phi, T)}{\obdamap}
  \]
\end{theorem}
\begin{proof}
The proof can be simply obtained by observing the following:
\begin{compactenum}
\item Both $\sts{\sgdssym}$ and $\rts{\sgdssym}$ have the same
  structure. In fact, essentially they can be seen as a single
  transition system such that each state $s$ in the transition system
  has its own associated database instance and ABox and
  $\abox(s) = \obdamap(\db(s))$.
\item The unfolding and the rewriting process do not alter the
  temporal part of the formula. 
\item The correctness of local queries is obtained by the correctness
  of the unfolding and rewriting in OBDA (see \cite{PLCD*08}). 
\item The obtained formula $\unfold{\rew(\Phi, T)}{\obdamap}$ is a
  $\mula$ property that can be verified over RTS (note that RTS is the
  same as database transition systems that define the semantics of
  $\mula$).
\end{compactenum}
\end{proof}

Due to the injection of new, fresh data into the system due to call to
external services, $\rts{\sgdssym}$ (as well as $\sts{\sgdssym}$) is
in general infinite-state. This causes verification to be undecidable
in general, even for the very simple case of a \sgds in which the
TBox contains no assertions and directly reflects the database schema
via simple one-to-one mappings. This boils down to the undecidability
result of DCDS \cite{BCDDM13}.


An extensive study concerning some decidability boundaries for the
verification of Data-Centric Dynamic Systems (DCDSs) with
non-deterministic external services has been provided in
\cite{BCDDM13}.  One of the interesting conditions for decidability
that have been studied is \emph{run-boundedness}. Thus, to gain
decidability, we adopt such restriction and we show that the
verification of run-bounded \sgdss can be reduced to the verification
of run-bounded DCDSs.

\begin{definition}[Run-bounded \sgds]\label{def:run-bounded-sedap}
  Given \sidetext{Run-bounded \sgds} a \sgds $\sgdssym$ with RTS
  $\rts{\sgdssym} = \tup{\const, T, \stateset, s_{0}, \db, \trans}$,
  we say \emph{$\sgdssym$ is run-bounded} if there exists an integer
  bound $b$ such that for every run $\pi = s_0s_1\cdots$ of
  $\rts{\sgdssym}$, we have that
  $\card{\bigcup_{s \textrm{ state of } \pi}\adom{\db(s)}} < b$.
\end{definition}

Utilizing \Cref{thm:corr-unf-rew-muladom} and the well-established
result for DCDS, in the following we show the decidability of the
\muladom verification over run-bounded \sgdss.

\begin{theorem}
Verification of $\muladom$ properties over run-bounded \sgdss is
decidable, and can be reduced to conventional finite-state model
checking.
\end{theorem}
\begin{proof}
  Let
  $\sgdssym$ 
  be a \sgds with RTS $\rts{\sgdssym}$ and STS $\sts{\sgdssym}$,
  $\todcds(\sgdssym)$ 
  be its corresponding DCDS obtained through translation $\todcds$ and
  has transition system $\ts{\todcds(\sgdssym)}$. By
  \Cref{thm:sedap-to-dcds-reduction}, we have that $\rts{\sgdssym}$
  and $\ts{\todcds(\sgdssym)}$ are equivalent. Thus, $\sgdssym$ is
  run-bounded if and only if $\todcds(\sgdssym)$ is run-bounded.
  Furthermore, by using \Cref{thm:corr-unf-rew-muladom}, since
  $ \sts{\sgdssym} \models \Phi \quad\text{if and only if}\quad
  \rts{\sgdssym} \models \unfold{\rew(\Phi, T)}{\obdamap}$,
  it follows that
  $ \sts{\sgdssym} \models \Phi \quad\text{if and only if}\quad
  \ts{\todcds(\sgdssym)} \models \unfold{\rew(\Phi, T)}{\obdamap}$
  (consider also that $\unfold{\rew(\Phi, T)}{\obdamap}$ is a $\mula$
  formula).
  The proof is then completed since according to
  \Cref{thm:verification-dcds}, the verification of $\mula$ over
  run-bounded DCDSs is decidable and can be reduced to conventional
  finite-state model checking.
\end{proof}

\section{From Theory to Practice: \sgdss Instantiation}

So far we have introduced \sgdss as a formal framework for
representing data-aware processes system equipped with a Semantic
Layer. In particular, we have focused our attention to the usage of
lightweight Description Logics, belonging to the DL-Lite family, to
conceptually capture the relevant domain entities and relationships at
the Semantic Layer. At the same time, we have shown that, thanks to
the FO rewritability of DL-Lite, verification of temporal properties
over the evolution of a data-aware processes system understood through
the lens of the Semantic Layer can be faithfully reduced to
verification of properties directly carried out at the Relational
Layer.

In this section, we show how the formal framework \sgdss can be
concretely instantiated. This work is part of the deliverables of EU
FP7 Project namely ACSI (``Artifact-Centric Service Interoperation'',
see \url{http://www.acsi-project.eu/}). Specifically, the results
presented here is part of the ACSI deliverable in
\cite{ACSI-D2.4.2}. To concretize the idea of \sgdss, here we consider
a specific setting where the transition relation at the Relational
Layer is obtained from \emph{artifacts} (\emph{artifact-centric
  systems}) specified using the \emph{Guard-Stage-Milestone} (GSM)
\cite{HDDF*11,DaHV13} approach. In particular, we exhibit how existing
techniques and tools can be 
suitably combined into a tool, called \emph{\obgsm,} which enables the
verification of GSM-based data-aware processes system equipped with a
Semantic Layer. To this aim, we leverage on the following tools:
\begin{compactenum}

\item \emph{\ontop}\footnote{\url{http://ontop.inf.unibz.it}}, a JAVA-based
  framework for OBDA, and in particular the Quest reasoner, which is the
  component dedicated to handle query rewriting and unfolding;

\item the \emph{GSMC model checker}, developed within ACSI project to
  verify GSM-based artifact-centric systems against temporal
  properties \cite{GGL12,ACSI-D2.2.3,BeLP12b}. 

\end{compactenum}

The main purpose of \obgsm is: given a temporal property specified
over the Semantic Layer of the system, together with mapping
assertions whose language is suitably shaped to work with GSMC,
automatically rewrite and unfold the property by producing a
corresponding translation that can be directly processed by the GSMC
model checker.  This cannot be done by solely relying on the
functionalities provided by \ontop, for two reasons:
\begin{compactenum}
\item \obgsm deals with temporal properties specified in a fragment of
  $\muladom$, and not just (local) ECQs;
\item the mapping assertions are shaped so as to reflect the specific query
  language supported by GSMC, guaranteeing that the rewriting and unfolding
  process produces a temporal property expressed in the input language of GSMC.
\end{compactenum}
In particular, we note that GSMC is not able to process the entire
\muladom logic, but only its CTL fragment. Here we denote such
fragment by \ctla. This requires also to restrict the \muladom
verification formalism accordingly, in particular focusing on its CTL
fragment, denoted by \ctladl.

An important observation related to semantic governance in this
setting is that since the construction of the RTS for GSM is handled
internally by GSMC, it is not possible (at least for the time being)
to prune it so as to remove inconsistent states. Therefore, in the
following we assume that all the states in the RTS are consistent with
the constraints of the Semantic Layer. This can be trivially achieved
by, e.g., avoiding to use negative inclusion in the
TBox. 

Before we proceed with the system specification of \obgsm in
\Cref{sec:obgsm-sys-spec}, we first briefly review the
Guard-Stage-Milestone (GSM) in \Cref{sec:gsm-short}. In the following
we might use the terms artifact layer and relational layer
interchangeably in order to refer to the notion of relational layer as
introduced in \sgdss.

\subsection{Guard-Stage-Milestone (GSM) at a Glance}\label{sec:gsm-short}


Guard-Stage-Milestone (GSM) \cite{HDDF*11} has been proposed as a
framework for modeling/specifying artifact-centric systems
\cite{NiCa03,Hull08} which combine both static and dynamic aspects of
the systems. In artifact-centric systems,
%
%
an \emph{artifact} is characterized by an \emph{information model},
which maintains the artifact data, and by a \emph{lifecycle} that
specifies the allowed ways to progress the information model (i.e.,
characterize the evolution of the system).  Among the different
proposals for artifact-centric process modeling, the GSM approach has
been proposed to model artifacts and their lifecycle in a declarative,
flexible way. 
GSM is equipped with a formal execution
semantics \cite{DaHV13}, which unambiguously characterizes the
artifact progression in response to external events. Notably, several
key constructs of the emerging OMG standard on Case Management and
Model Notation\footnote{\url{http://www.omg.org/spec/CMMN/}} have
been borrowed from GSM.

Here we only provide a general overview of the Guard-Stage-Milestone
(GSM) methodology and we refer to \cite{HDDF*11,DaHV13} for more
detailed and formal definitions.
%
%
%
Technically, a GSM model consists of a set of artifact types where
each artifact type has its own information model as well as lifecycle
and when an artifact type is instantiated, its instance has a
corresponding identifier. During the execution, each artifact instance
is interacting to each other forming the evolution of the system.
The GSM information model uses (possibly nested) attribute/value
tuples 
to capture the domain of interest.  The key elements of a lifecycle
model are \emph{stages}, \emph{milestones} and
\emph{guards}. 
Stages 
are possibly hierarchical clusters of activities, intended to update
and extend the data of the information model. They are associated to
milestones, 
business operational objectives which can be achieved while the stage
is under execution. Each stage has one or more guards, 
which control the activation of stages and, like milestones, are
described in terms of data-aware expressions, involving conditions
over the artifact information model.

\subsection{\obgsm System Specification}\label{sec:obgsm-sys-spec}

The \obgsm tool takes a \textit{conceptual temporal property}
specified over the Semantic Layer of a GSM-based artifact system, and
then producing a corresponding temporal property that can be directly
verified by GSMC over the GSM specification, without involving the
Semantic Layer anymore.
Specifically, \obgsm has three inputs:
\begin{compactenum}
\item a conceptual temporal property $\Phi$; 
\item an OWL 2
  QL\footnote{\url{http://www.w3.org/TR/2008/WD-owl2-profiles-20081008/\#OWL\_2\_QL}. OWL
    2 QL is the OWL2 profile that closely corresponds to the DL-Lite
    family of Description Logics.}  TBox; 
\item a mapping specification $\obgsmmap$. 
\end{compactenum}
In the following, we detail the languages used to specify $\Phi$ and
$\obgsmmap$.

\subsubsection{Specification of Conceptual Temporal Properties}
%
For the temporal component of the conceptual temporal properties, we
rely on CTL, in accordance to the input verification language of GSMC.
%
Remember that CTL is subsumed by $\mu$-calculus \cite{Dam92,Emer96b}.
As far as the local queries over the ontology are concerned, the
language relies on a fragment of SPARQL, in accordance to the query
language supported by Ontop. More specifically, the syntax of the
conceptual temporal properties is as follows:
%
{\small
\[
\begin{array}{rl}
  \formula ~::=&~ [~\query~] \\
~\mid&~ ( \formula ) \\
~\mid&~ \andTemp{\formula}{\formula} \\
~\mid&~ \orTemp{\formula}{\formula} \\
~\mid&~ \implTemp{\formula}{\formula} \\
~\mid&~ \notTemp~\formula \\
~\mid&~ \AG~\formula \\ 
~\mid&~ \EG~\formula \\
~\mid&~ \AF~\formula \\
~\mid&~ \EF~\formula \\
~\mid&~ \AX~\formula \\
~\mid&~ \EX~\formula \\
~\mid&~ \Auntil{\formula}{\formula} \\ 
~\mid&~ \Euntil{\formula}{\formula}  \\
~\mid&~ \forallTemp\ \variables\ .\ \forallQuantification \\
~\mid&~ \existsTemp\ \variables\ .\ \existsQuantification\\ \\
\forallQuantification ~::=&~ \implTemp{[~\query~]}{\formula} \\
~\mid&~ \forallTemp\ \variables\ .\ \forallQuantification \\
~\mid&~ [~\query~] \\ \\
\existsQuantification ~::=&~ \andTemp{[~\query~]}{\formula} \\
~\mid&~ \existsTemp\ \variables\ .\ \existsQuantification\\ 
~\mid&~ [~\query~] \\ \\
\end{array}
\]
}
%
where $[\query]$ is a 
\sparql
\footnote{\url{http://www.w3.org/TR/sparql11-query/}} \sparqlSELECT
query.
\sparqlSELECT
queries, in turn, obey to the following grammar:
{\small
\begin{equation*}
\begin{split}
  \query ~::=&~ \prefixDeclaration\ \sparqlSELECT \ \variables\ \sparqlWHERE\  \textsc{\{}
  \triples\ \mbox{\textsc{Filter}} (\sparqlFILTER) \textsc{\}}  \\ 
\end{split}
\end{equation*}
\begin{equation*}
\begin{split}
  \sparqlFILTER ~::=&~ \sparqlFILTEREXPRESSION\  \\
~\mid&~ \sparqlFILTEREXPRESSION\ \sparqlFILTERAND\ \sparqlFILTER \\
\end{split}
\end{equation*}
\begin{equation*}
\begin{split}
  \sparqlFILTEREXPRESSION ~::=&~ \sparqlVarConst\ \sparqlLT\ \sparqlVarConst \\
~\mid&~ \sparqlVarConst\ \sparqlLTE\ \sparqlVarConst \\
~\mid&~ \sparqlVarConst\ \sparqlGT\ \sparqlVarConst \\
~\mid&~ \sparqlVarConst\ \sparqlGTE\ \sparqlVarConst \\
~\mid&~ \sparqlVarConst\ \sparqlEQ\ \sparqlVarConst \\
~\mid&~ \sparqlVarConst\ \sparqlNEQ\ \sparqlVarConst \\
\end{split}
\end{equation*}
\begin{equation*}
\begin{split}
  \sparqlVarConst &~::=~ \variables \\
~\mid&~ \sparqlInteger \\
~\mid&~ "\sparqlString" \\
~\mid&~ "\sparqlString" \doubleCarret \mbox{\tt http://www.w3.org/2001/XMLSchema\#string} \\
~\mid&~ "\sparqlString" \doubleCarret \mbox{\tt http://www.w3.org/2001/XMLSchema\#integer} \\
~\mid&~ "\sparqlString" \doubleCarret \mbox{\tt http://www.w3.org/2001/XMLSchema\#decimal} \\
~\mid&~ "\sparqlString" \doubleCarret \mbox{\tt http://www.w3.org/2001/XMLSchema\#double} \\
~\mid&~ "\sparqlString" \doubleCarret \mbox{\tt http://www.w3.org/2001/XMLSchema\#dateTime} \\
~\mid&~ "\sparqlString" \doubleCarret \mbox{\tt http://www.w3.org/2001/XMLSchema\#boolean} \\
~\mid&~ "\sparqlString" \doubleCarret \mbox{\tt http://www.w3.org/1999/02/22-rdf-syntax-ns\#Literal} \\
\end{split}
\end{equation*}
}
where:
\begin{compactitem}
\item $\variables$ is a variable that obeys to the pattern
\verb!?([a-z]|[A-Z])+!;
\item $\triples$ and $\prefixDeclaration$ follow the usual triple
  patterns and prefix declarations of \sparql;
\item $\sparqlInteger$ and $\sparqlString$ are the standard integer
  and string built-in domains.
\end{compactitem}
Additionally, we require that all variables present in the \sparqlSELECT clause
of the query also appear in the \sparqlWHERE clause, and vice-versa;
in other words, all variables in the query must be answer variables.

The semantics of the temporal operators is as in CTL
\cite{BaKa08}. 
For the first order quantification, we impose the following
restrictions:
\begin{compactitem}
\item Only closed temporal formulas are supported for verification.
\item Each first-order quantifier must be ``guarded'', in such a way
  that it ranges over constants present in the current active
  domain. This active domain quantification is in line with GSMC, and
  also with the $\muladom$ 
  logics. 
  As attested by the grammar above, this is syntactically guaranteed
  by requiring quantified variables to appear in a $[\query]$
  according to the following guidelines:
 \[
\begin{array}{l}
\forall \vec{x}. \query(\vec{x}) \ra \phi\\
\exists \vec{x}. \query(\vec{x}) \wedge \phi\\
\end{array}
\]

\item Quantified variables must obey to specific restrictions,
  depending on whether they quantify over object terms or values. This can
  be syntactically recognized by checking whether the variable appears
  in the second component of an attribute (in this case, it ranges
  over values) or not. The restriction is as follows: for each
  variable $y$ ranging over values, there must be at least one
  variable $x$ that ranges over object terms and that appears in the
  first component of the corresponding attribute (i.e., $Attr(x,y)$ is
  present in the query, with $Attr$ being an attribute of the TBox), such
  that $x$ is quantified ``before'' $y$. For example, $\forall x. C(x)
  \implies \exists y. Attr(x.y)$ satisfies this condition, whereas
  $\exists y \exists x. Attr(x,y)$ does not.
\end{compactitem}
%
%
%
%
These restrictions have been introduced so as to guarantee that the
conceptual temporal property can be translated into a corresponding
GSMC temporal property. In fact, GSMC poses several restrictions on
the way values 
can be accessed.

\begin{example}\label{ex:students}
 We consider a simple university information system in order to
 provide an example while explaining the system specification of
 \obgsm. The following TBox is used to capture the relevant concepts
 and relations of the university domain at the Semantic Layer:
\[
\begin{array}{rl@{~~~~~~}rl@{~~~~~~}l}
\exo{Bachelor} \sqsubseteq &\exo{Student}&\delta(\exo{MNum}) \sqsubseteq &\exo{Student}  & \\
%
\exo{Master} \sqsubseteq & \exo{Student} & \delta(\exo{HasAge}) \sqsubseteq &\exo{Student} &\\
\exo{Graduated} \sqsubseteq & \exo{Student}&\exists \exo{Attend} \sqsubseteq &\exo{Student} &\\
& &\exists \exo{Attend}^- \sqsubseteq &\exo{Course} &\\
\end{array}
\]
The Artifact Layer contains the following artifact types:
\begin{enumerate}
\item {\tt ENROLLEDSTUDENT}, whose instances represent the enrolled
  students. For each enrolled student, these data attributes are maintained: {\tt ID, MNum, Name,
    Age, Type}, where {\tt ID, Name} and {\tt Type} are of type
  String, while
  {\tt Age} and {\tt MNum} are of type Integer.
\item {\tt GRAD}, whose instances represent those students who have been
  graduated. The following data attributes are maintained: {\tt ID, MNum}.
\item {\tt COURSE}, whose instances represent the courses offered by the
  university. They have the following data attributes: {\tt ID, CourseName}.
\end{enumerate}
An example of temporal property specified over the Semantic Layer is:
{\small
\begin{verbatim}
EF FORALL ?x.([PREFIX : <http://acsi/example/student/student.owl#> 
               PREFIX rdf:<http://www.w3.org/1999/02/22-rdf-syntax-ns#> 
               SELECT ?x WHERE {  ?x rdf:type :Bachelor }] ->
              [PREFIX : <http://acsi/example/student/student.owl#> 
               SELECT ?x WHERE {  ?x rdf:type :Graduated }]
          );
\end{verbatim}
}
  which says that ``eventually there is a state in the future where
  all bachelor students are graduated''. 
  Another example is: 
{\small
\begin{verbatim}
EF FORALL ?x. ([PREFIX : <http://acsi/example/student/student.owl#> 
                PREFIX rdf:<http://www.w3.org/1999/02/22-rdf-syntax-ns#> 
                PREFIX xsd: <http://www.w3.org/2001/XMLSchema#> 
                SELECT ?x WHERE 
                     {?x rdf:type :Master; :HasAge "26"^^xsd:integer}] 
               -> [PREFIX : <http://acsi/example/student/student.owl#> 
                  SELECT ?x WHERE {  ?x rdf:type :Graduated }]
          );
\end{verbatim}
}
  It states that ``eventually in the future there is a state where
  all bachelor students who are at least 26 years old are graduated''.
\end{example}

Notice that in the second temporal property of \Cref{ex:students}, the
typed value \verb!"26"^^xsd:integer! is used to denote the age of
students. More in general, according to the current implementation of
\ontop, there is support for the following type of values:
\begin{compactitem}
\item  xsd:string
\item  xsd:integer
\item  xsd:decimal
\item  xsd:double
\item  xsd:dateTime
\item  xsd:boolean
\item  rdf:Literal
\end{compactitem}
where ``{\tt xsd:}'' and ``{\tt rdf:}'' are predefined prefixes,
respectively defined as ``{\tt xsd:
  http://www.w3.org/2001/XMLSchema\#}'' and ``{\tt rdf:
  http://www.w3.org/1999/02/22-rdf-syntax-ns\#}''. Whenever an input
value is not typed, we consider it to be, by default, of type {\tt
  rdf:Literal}.

\subsubsection{Specification of the Input Mapping}
%
The structure of our mapping language is borrowed from the one of
\ontop. More specifically, the expected file format is:
\begin{center}
\fbox{
\parbox{12 cm}{
{\tt
\prefixDeclarationMapping \\
... \\

\classDeclarationMapping @collection [[\\
...\\
]]\\

\objectPropertyDeclarationMapping @collection [[\\
...\\
]]\\

\dataPropertyDeclarationMapping @collection [[\\
...\\
]]\\

\mappingDeclaration  @collection [[\\
...\\
]]
}
}
}
\end{center}
In the following, we detail the different parts of this format.




The \prefixDeclarationMapping part contains the definition of the
URI (Uniform Resource Identifier) prefixes that will be used in the
remainder of the file. 

\begin{example}\label{ex:prefix-declaration}
We provide a simple prefix declaration that could be contained in a mapping
specification file:

\begin{center}
\fbox{\parbox{12 cm}{
{\tt[PrefixDeclaration] \\
xsd:		http://www.w3.org/2001/XMLSchema\#\\
rdf:		http://www.w3.org/1999/02/22-rdf-syntax-ns\#\\
:			http://acsi/example/student/student.owl\#

}}}
\end{center}
\end{example}
  
Parts \classDeclarationMapping, \objectPropertyDeclarationMapping, and
\dataPropertyDeclarationMapping, respectively contain the declaration
of \emph{concepts}, \emph{roles}, and \emph{attributes} name that will
be mentioned in the mapping declarations. They are also specified in
terms of URIs, where each entry is separated by comma.

\begin{example}\label{ex:class-object-property-and-data-property-declaration}
We provide three sample declarations for a class, an object property,
and a data property, respectively:

\begin{center}
\fbox{\parbox{12 cm}{
{\tt

[ClassDeclaration] @collection [[\\
:Student, :Bachelor, :Graduated, :Master, :Course\\
]]\\

[ObjectPropertyDeclaration] @collection [[\\
:Attend\\
]]\\

[DataPropertyDeclaration] @collection [[\\
:MNum, :HasAge\\
]]\\

}}}
\end{center}
\end{example}


The \mappingDeclaration contains the declaration of mapping assertions
(cf. \Cref{def:obda-mapping}).
When constructing object terms starting from 
Relational Layer, we require that only unary function symbols are
used. As in \ontop, such unary function symbols are in turn
represented by \emph{URI templates} (i.e., a preset format for URIs).
For example, the object term $stud(x)$ is represented as {\tt
  <http://www.acsi-project.eu/example/\#stud\{x\}>}.
%

Each mapping assertion is then described by three components:
\begin{compactenum}
\item {\tt mappingId}, which provides a unique identifier for the
  mapping assertion.
\item {\tt target}, which contains the \emph{target query} (i.e., the
  head of the mapping). Technically, a target query is a CQ over the
  vocabulary of the ontology. For the specification of such target
  query, we adopt the \ontop syntax, which is 
  based on the Turtle\footnote{\url{http://www.w3.org/TR/turtle/}}
  syntax to represent RDF triples. Each atom in the CQ
  is in fact represented as an RDF-like triple
  template.
  There are three kinds of possible atoms in the target query:
   \begin{enumerate}
  \item Concepts, expressed as
    \begin{center} {\tt [URI\_Template] rdf:type [ConceptName] }
    \end{center}
    where {\tt [ConceptName]} is an URI, and {\tt rdf:} is the prefix
    {\tt rdf: http://www.w3.org/1999/02/22-rdf-syntax-ns\#}. For
    example, to represent the atom
    \[ {ConceptName(c(x))}
    \]
    (where $ConceptName$ is a concept name in the ontology), the
    following notation is used:
    \begin{center} {\tt <"\&:;c\{\$x\}"> rdf:type :ConceptName }
    \end{center}
    where ``{\tt :}'' is a predefined prefix.

  \item Roles, again expressed as triples:
    \begin{center} {\tt [URI\_Template] [RoleName] [URI\_Template] }
    \end{center}
    where {\tt [RoleName]} is an URI. For example, the atom
    \[ {RoleName(r1(x), r2(y))}
    \]
    (where $RoleName$ is a role name in the ontology) is represented
    as:
    \begin{center} {\tt <"\&:;r1\{\$x\}"> :RoleName <"\&:;r2\{\$y\}">}
    \end{center}
    where ``{\tt :}'' is a predefined prefix.
  \item Attributes, whose definition resembles the one of roles:
    \begin{center} {\tt [URI\_Template] [AttributeName]
        [TypedOrUntypedVariable]}
    \end{center}
    where {\tt [AttributeName]} is an URI. For example,
    \[ {AttributeName (att(x), integer(y))}
    \]
    is represented as:
    \begin{center} {\tt <"\&:;att\{\$x\}"> :AttributeName
        \$y\^{}\^{}xsd:integer .}
    \end{center}
    where ``{\tt :}'' and ``{\tt xsd:}'' are predefined prefixes, and
    ``{\tt xsd:}'' is defined as ``{\tt xsd:
      http://www.w3.org/2001/XMLSchema\#}''. It is worth noting that
    the second component of an attribute is a value. We assume that it
    originates from a value attribute contained inside the information
    model of an artifact, we use dedicated function symbols to wrap
    the value into an object term, ensuring that this choice does not
    overlap with any function symbol chosen for ``real'' object
    terms. In the example above, we use ``{\tt
      http://www.w3.org/2001/XMLSchema\#integer}'', but in general,
    \ontop supports all the following special data types:
    \begin{itemize}
    \item \url{http://www.w3.org/2001/XMLSchema\#string}
    \item \url{http://www.w3.org/2001/XMLSchema\#integer}
    \item \url{http://www.w3.org/2001/XMLSchema\#decimal}
    \item \url{http://www.w3.org/2001/XMLSchema\#double}
    \item \url{http://www.w3.org/2001/XMLSchema\#dateTime}
    \item \url{http://www.w3.org/2001/XMLSchema\#boolean}
    \item \url{http://www.w3.org/1999/02/22-rdf-syntax-ns\#Literal}
    \end{itemize}

  \end{enumerate}

\item {\tt source}, which describes the \emph{source query}, i.e., the
  body of the mapping. The grammar of the source query is borrowed
  from the grammar of the GSMC input language \cite{ACSI-D2.2.3}, with
  extensions that allow to ``extract'' artifact identifiers and their
  value attributes, so as to link them to the ontology. The extended
  syntax is:
  \begin{equation*}
    \begin{split}
      \expression ~::=&~ \constantGSM \\
      ~\mid&~ \expression\ == \ ?\variableGSM \\
      ~\mid&~ \expression\ \aopGSM \ \expression \\
      ~\mid&~ \expression\ \lopGSM \ \expression \\
      ~\mid&~ \{\variableGSM./path/attributeID\} \\
      ~\mid&~ \textbf{GSM.isStageActive}('\variableGSM','stageID') \\
      ~\mid&~ \textbf{GSM.isMilestoneAchieved}('\variableGSM','milestoneID') \\
      ~\mid&~ \variableGSM.attributeID1 \mbox{ -$>$ } \textbf{exists}(attributeID2 = \expression) \\
    \end{split}
  \end{equation*}
  \begin{equation*}
    \begin{split}
      \formula ~::=&~  \expression \\
      ~\mid&~(\formula)\\
      ~\mid&~\formula\ \andGSM\ \formula\\
      ~\mid&~\formula\ \orGSM\ \formula\\
      ~\mid&~\notGSM\ \formula\\
      ~\mid&~\existsGSM ('variable',\ 'artifactID')(\formula)\\
      ~\mid&~\forallGSM ('variable',\ 'artifactID')(\formula)\\
      ~\mid&~\getGSM('variable',\ 'artifactID')(\formula)\\
    \end{split}
  \end{equation*}
%
\end{compactenum}

The two key additional features rely in the possibility of introducing
a variable assigning it to an expression (see the second line in the
grammar definition), and the possibility of ``getting'' a variable
 representing an instance of the specified artifact (see the last line
in the grammar definition). These variables are considered to be free
in the specified query, and can be consequently used to ``transport''
values and artifact identifiers into the Semantic Layer, respectively
as attributes and object terms.

\begin{example}\label{ex:obgsm-mapping-declaration}
  Consider again the Artifact and Semantic Layer introduced in
  \Cref{ex:students}. We specify the following mapping assertions to
  link the three artifacts and their information models to the
  ontology present in the Semantic Layer:
\begin{center}
\fbox{\parbox{12.9 cm}{
{\tt \small [MappingDeclaration] @collection [[ \\

mappingId \hspace*{1 mm}	BachelorStudent\\
target \hspace*{6 mm}		<"\&:;stud/\{\$x\}/"> rdf:type :Bachelor .\\
source \hspace*{6 mm}     get('x','ENROLLEDSTUDENT')\\
\hspace*{18 mm} (\{x./ENROLLEDSTUDENT/Type\} == "Bachelor")\\

mappingId \hspace*{1 mm}	MasterStudent\\
target \hspace*{6 mm}		<"\&:;stud/\{\$x\}/"> rdf:type :Master . \\
source \hspace*{6 mm}      get('x','ENROLLEDSTUDENT')\\
\hspace*{18 mm} (\{x./ENROLLEDSTUDENT/Type\} == "Master")\\

mappingId \hspace*{1 mm}	MatriculationNumber\\
target \hspace*{6 mm}		<"\&:;stud/\{\$x\}/"> :MNum \$y\^{}\^{}xsd:integer . \\
source \hspace*{6 mm}      get('x','ENROLLEDSTUDENT')(\{x./ENROLLEDSTUDENT/MNum\} == ?y) \\

mappingId \hspace*{1 mm}	GraduatedStudent \\
target \hspace*{6 mm}		<"\&:;stud/\{\$x\}/"> rdf:type :Graduated . \\
source \hspace*{6 mm} get('x','ENROLLEDSTUDENT')(exists('y', 'GRAD')( \\
\hspace*{18 mm} \{x./ENROLLEDSTUDENT/MNum\} == \{y./GRAD/MNum\})) \\

mappingId \hspace*{1 mm}	Age \\
target \hspace*{6 mm}		<"\&:;stud/\{\$x\}/"> :HasAge \$y\^{}\^{}xsd:integer . \\
source \hspace*{6 mm}      get('x','ENROLLEDSTUDENT')({x./ENROLLEDSTUDENT/Age} == ?y) \\

mappingId \hspace*{1 mm}	AttendingCourse \\
target \hspace*{6 mm}		<"\&:;stud/\{\$x\}/"> :Attend <"\&:;course/\{\$y\}/"> . \\
source \hspace*{6 mm}		get('x','ENROLLEDSTUDENT')(get('y','COURSE')\\
\hspace*{18 mm} (x./ENROLLEDSTUDENT/AttendedCourses->exists(\\
\hspace*{18 mm} ID == \{y./COURSE/ID\}))) \\

]]

}}}
\end{center}









The first two mapping assertions are used to populate bachelor and
master students in the Semantic Layer, by extracting information from
artifact instances of type \texttt{ENROLLEDSTUDENT}, respectively
selecting those instances whose \texttt{Type} field corresponds to the
string ``Bachelor'' or ``Master''. Notice that the artifact instance identifier
$x$ is used to create the corresponding student object term $stud(x)$
in the ontology.

The following three mapping assertions are used to populate attributes
in the Semantic Layer, starting from specific artifacts and fields in
their information models. According to the previously discussed
restrictions, the first component of attributes is always associated
to an object term constructed starting from an artifact instance identifier,
and the second from a value in its information model.

The last mapping assertions is used to populate a relation in the
Semantic Layer, starting from pairs of artifact identifiers in the
Artifact Layer. In particular, the source query is used to extract all
pairs of artifact instances of type \texttt{ENROLLEDSTUDENT} and
\texttt{COURSE}, such that the course artifact instance is among the
attended courses by the student artifact instance (notice the
navigation \texttt{x./ENROLLEDSTUDENT/AttendedCourses} to select all attended
courses, and the consequent join used to check whether the considered
course instance is among the attended ones). In this case, both the
first and the second component of the association are object terms
constructed from artifact instance identifiers.
\end{example}

\subsubsection{\obgsm Workflow and Components}

As depicted in the \Cref{obgsm-system-architecture}, the workflow of
\obgsm is as follows.
\begin{compactenum}

\item The tool reads and parses the input conceptual temporal property
  $\Phi$, the input ontology (TBox) $T$, and the input mapping declaration
  $\obgsmmap$.

\item The tool \emph{rewrites} the input conceptual temporal property
  $\Phi$ based on the input ontology (TBox) $T$, in order to compile
  away the TBox.  This step produces \emph{rewritten temporal
    property} $\rew(\Phi,T)$.

\item The rewritten property $rew(\Phi, T)$ is \emph{unfolded} by
  exploiting $\obgsmmap$.  The final temporal property
  $\Phi_{GSM} = \unfold{\rew(\Phi,T)}{\obgsmmap}$ obeys to the syntax
  expected by GSMC, and is such that verifying $\Phi$ over the
  transition system of the GSM model under study after projecting its
  states into the Semantic Layer through $\obgsmmap$, is equivalent to
  verifying $\Phi_{GSM}$ directly over the GSM model (without
  considering the Semantic Layer).  

\item GSMC is invoked by passing $\Phi_{GSM}$ together with the
  specification file of the GSM model under study.
\end{compactenum}
Notice that the correctness of the translation is guaranteed by the
fact \obgsm manipulates the local components of the query $\Phi$
according to the standard rewriting and unfolding algorithms, while
maintaining untouched the temporal structure of the property. This has
been proven to be the correct way of manipulating the temporal
property as in \sgdss (cf. \Cref{sec:ver-sedap}). The proof has been
done for $\muladom$, and since first-order CTL with active domain
quantification is a fragment of $\muladom$, the result directly
applies also in our setting. This result also shows that \obgsm can
largely rely on state-of-the-art existing rewriting and unfolding
techniques to manipulate the temporal properties. Indeed, \obgsm
exploits \ontop to accomplish this task, by also adding a last step to
deal with the constructs that have been introduced for the mapping
assertions, but that are not directly supported by GSMC. In
particular, \obgsm turn ``get'' statements into corresponding the
quantifications.



\obgsm consists of the following main components (see also
\Cref{obgsm-system-architecture}):

\begin{figure}[bp]
\centering
\includegraphics[width=1.0\textwidth]{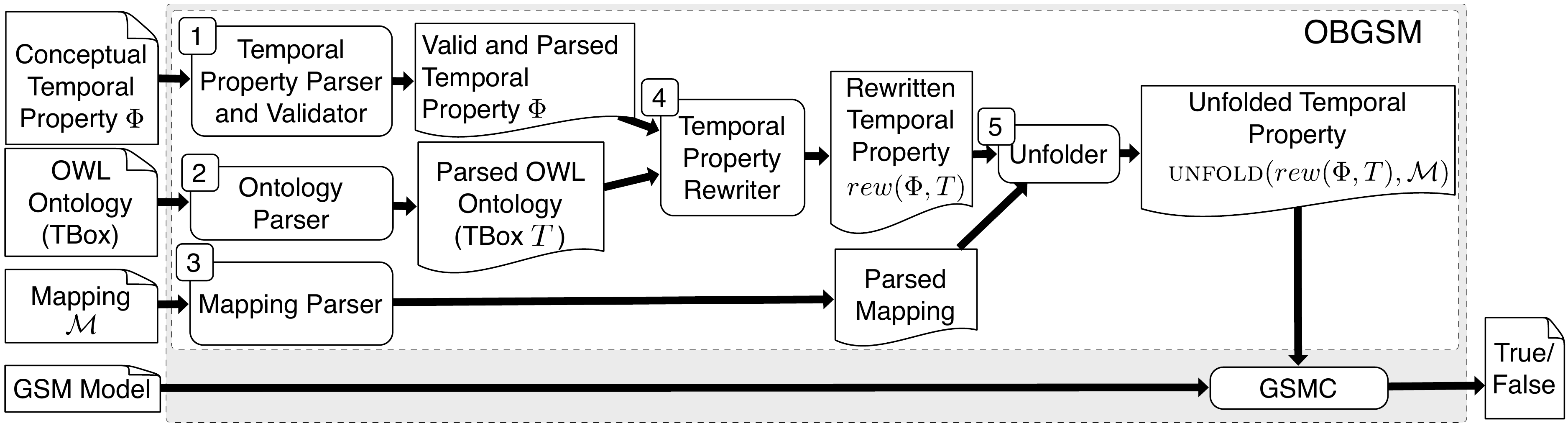}
\caption{\obgsm System Architecture \label{obgsm-system-architecture}}
\end{figure}

\begin{compactenum}
\item \emph{Temporal Property Parser and Validator}. This component
  parses and validates a conceptual temporal property, checking its
  well-formedness and whether it guarantees the required restrictions
  or not. The parser for the temporal part of the property is
  implemented using Antlr
  4.0\footnote{\url{http://www.antlr.org}}. For parsing the local
  queries in the temporal property, the \sparql parser component from
  \ontop is extensively used, together with the \Jena
  library\footnote{\url{http://jena.apache.org}}.

\item \emph{Ontology Parser}. This component, entirely provided by
  \ontop, reads and parse an OWL 2 QL ontology.

\item \emph{Mapping Parser}. This component reads and parses the file
  containing mapping assertions. As for the implementation, the parser
  already present in \ontop is reused and complemented with the
  additional features by using Antlr 4.0.

\item \emph{Temporal Property Rewriter}. When the input temporal
  property and the input ontology have been parsed, \obgsm
  \emph{rewrites} the parsed temporal property based on the given
  ontology, using this component. The implementation of the query
  rewriting functionality is fully inherited from \ontop.

\item \emph{Unfolder}. This component takes the specification of
  mapping assertions and the property produced by the rewriter,
  producing the final unfolded property. To do so, it extends the base
  unfolding functionality already present in \ontop.

\end{compactenum}


\subsection{A Use Case Example}


As a case study to demonstrate our approach we refer to the fragment
of the Energy Use Case Scenario developed within the ACSI Project
\cite{ACSI-D5.1,ACSI-D5.2,ACSI-D5.3,ACSI-D5.5}.  We show how the
Semantic Layer can be exploited in order to facilitate the
specification of temporal properties of interest, and discuss how
these are automatically translated into properties that can be
directly verified by GSMC over the Energy GSM model.

\subsubsection{ACSI Energy Use Case at a Glance}
We sketch here the main aspects of the Energy use case, and referring
the interested reader to
\cite{ACSI-D5.1,ACSI-D5.2,ACSI-D5.3,ACSI-D5.5} for further details.

The ACSI Energy use case focuses on the electricity supply exchange
process between electric companies inside a distribution network.  The
electricity exchange between companies occurs at \emph{control
  points}. Within a control point, a measurement of electricity supply
exchange takes place in order to calculate the fair remuneration that
the participating companies in the control point should receive. The
measurement is done by a \emph{meter reader company}, which
corresponds to one of the companies pertaining to that particular
certain control point. The measurement results from the control points
are then submitted to the \emph{system operator}, who is in charge of
processing the results and publishing a \emph{control point monthly
  report}. A participating company can raise an objection concerning
the published measurement.  Once all the risen objections are
resolved, the report is closed. The collection of CP monthly reports
is then represented as a \emph{CP monthly report accumulator}.




\subsubsection{The GSM Model for ACSI Energy Use Case}\label{sec:energy-usecase-gsm-model}

\begin{figure}[tbp]
\centering
\hspace*{-10mm}
\includegraphics[width=1.2\textwidth]{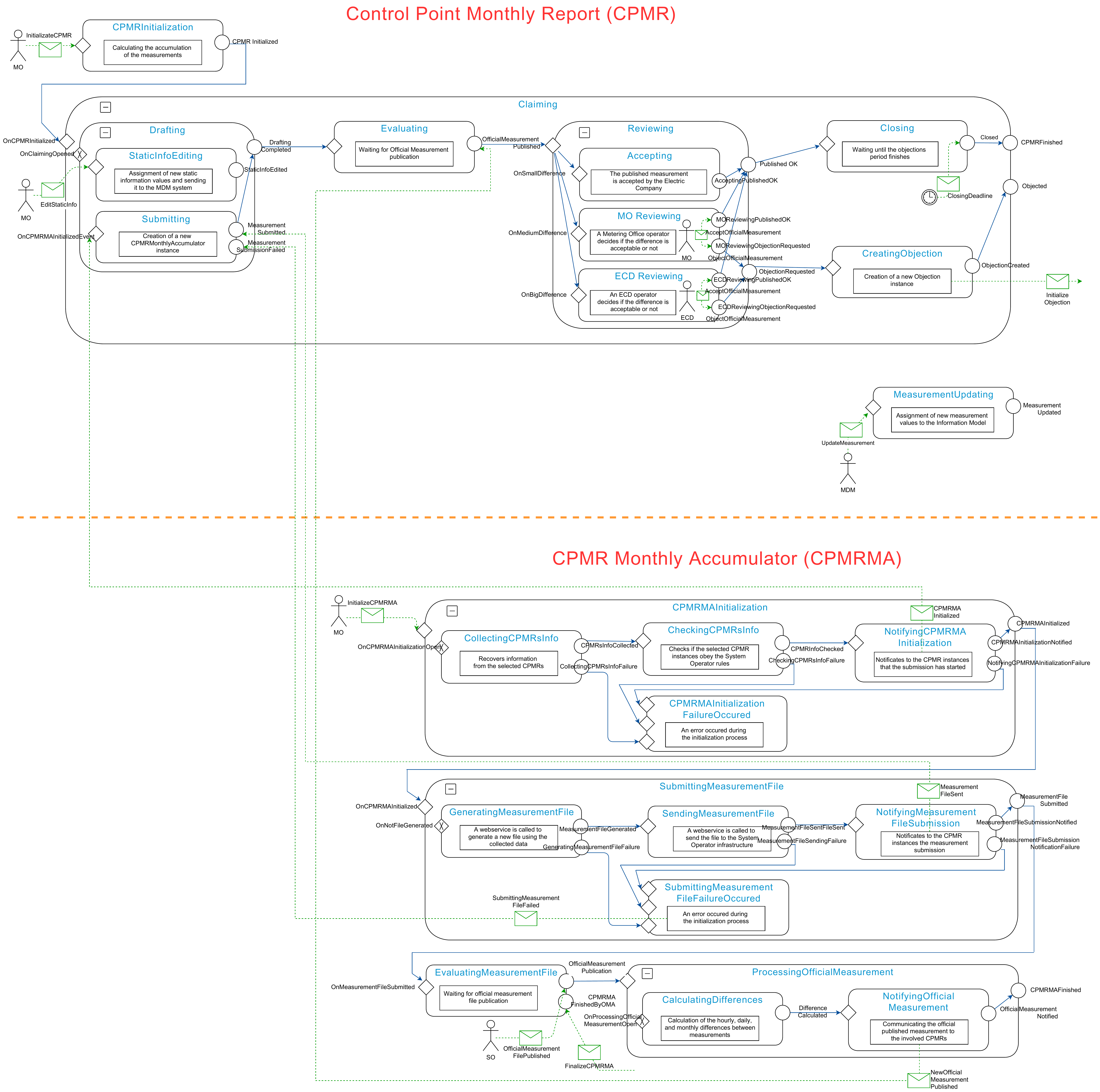}
\caption{GSM Model for ACSI Energy Use
  Case (The picture is from \cite{ACSI-D2.4.2}) \label{fig:gsm-model} }
\end{figure}

The sketched ACSI energy use case scenario is implemented by
considering two artifacts:
\begin{itemize}
\item \emph{Control Point Monthly Report $($CPMR$)$}. This artifact
  contains the information about hourly measurements done in a control
  point within a certain month. The lifecycle of an instance of this
  artifact is started when the Metering Data Management (MDM) system
  provides the hourly measurements, and runs until the liquidation for
  the CP measurements is started. This artifact consists of three root
  stages:
  \begin{compactitem}
  \item \emph{CPMRInitialization}, activated when a new instance of
    the CPMR artifact is created.
  \item \emph{Claiming}, This stage handles the submission of
    measurements, and the consequent reviewing stage, where objections
    may be raised. Five sub-stages are used to deal with this
    lifecycle in a fine-grained way: \emph{Drafting},
    \emph{Evaluating}, \emph{Reviewing}, \emph{Closing}, and
    \emph{CreateObjection}.
  \item \emph{MeasurementUpdating}, activated when there is an event
    requesting for updating the measurement results.
  \end{compactitem}
\item \emph{CPMR Monthly Accumulator $($CPMRMA$)$}, responsible for
  the measurement files. It submits measurements to the system
  operator, and receives back from the operator the value for the
  corresponding official measurements.  It consists of four root
  stages:
  \begin{compactitem}
  \item \emph{CPMRMAInitialization}, which handles the generation of
    measurement files.
  \item \emph{SubmittingMeasurementFile}, which submits the
    measurement files to the system operator.
  \item \emph{EvaluatingMeasurementFile}, which waits the official
    measurements from the system operator.
  \item \emph{ProcessingOfficialMeasurement}, which calculates the
    differences between the official measurements and the submitted
    measurements, and consequently notify the CPMR instances about the
    differences. The \emph{CalculatingDifferences} and
    \emph{NotifyingOfficialMeasurement} sub-stages are used to handle
    this portion of the lifecycle.
  \end{compactitem}
\end{itemize}
Figure \ref{fig:gsm-model} shows the GSM model for the two artifacts
described above.

\subsubsection{The Semantic Layer Specification}\label{sec:energy-use-case-sem-layer}

We provide a Semantic Layer on top of the GSM model for the ACSI
Energy use case, restricting our attention to the Published Control
Point Measurement Report (CPMR).

\subsubsection{The Ontology}\label{sec:energy-use-case-ontology}
\begin{figure}[tbp]
\centering
\includegraphics[width=0.9\textwidth]{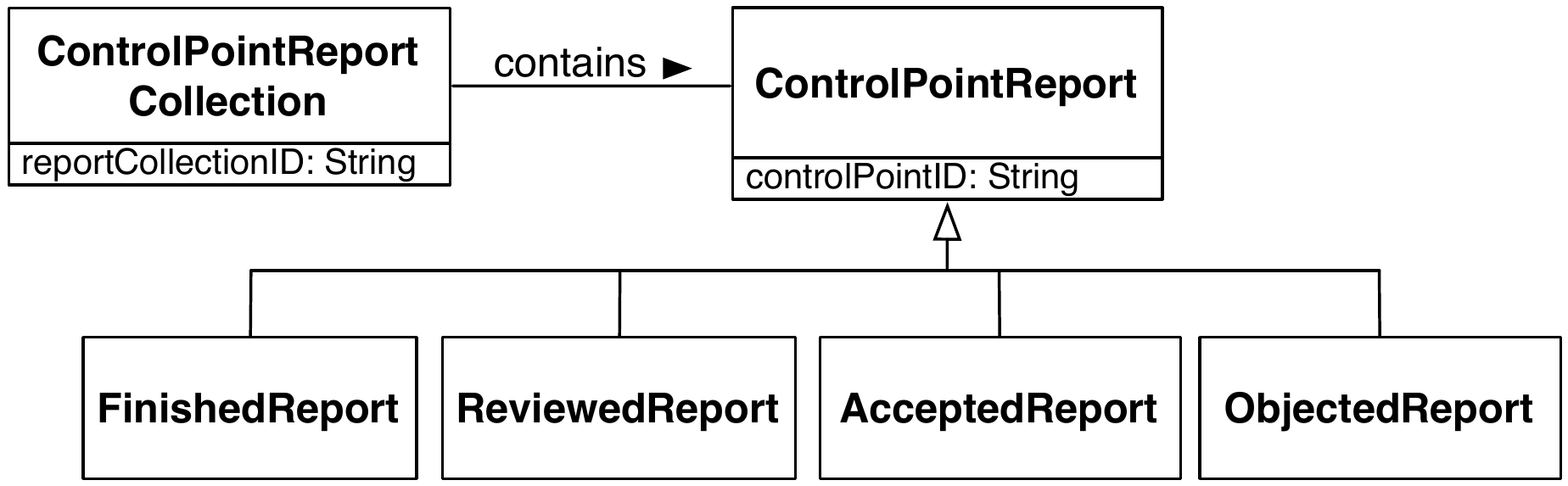}
\caption{Ontology for the CPMR reviewing process in ACSI Energy Use
  Case \label{fig:energy-ontology} }
\end{figure}

A UML model for the ontology of the Semantic Layer is depicted in
\Cref{fig:energy-ontology}.  A control point measurement report can be
either:
\begin{compactitem}
\item a finished CPMR (when the milestone \emph{CPMRFinished} is achieved),
\item a reviewed CPMR (after finishing the review inside the \emph{Reviewing} stage),
\item an accepted CPMR (when the milestone \emph{PublishedOK} is achieved),
\item an objected CPMR. 
\end{compactitem}
We formalize the UML model in \dllitea:
\[
\begin{array}{rl@{~~~~~~}rl@{~~~~~~}l}
\exo{FinishedReport} &\sqsubseteq \exo{ControlPointReport}  \\
\exo{ReviewedReport} &\sqsubseteq \exo{ControlPointReport} \\
\exo{AcceptedReport} &\sqsubseteq \exo{ControlPointReport} \\
\exo{ObjectedReport} &\sqsubseteq \exo{ControlPointReport} \\
%
%
\exists \exo{contains} &\sqsubseteq \exo{ControlPointReportCollection} \\
\exists \exo{contains}^- &\sqsubseteq \exo{ControlPointReport} \\
\delta(\exo{controlPointID}) &\sqsubseteq \exo{ControlPointReport} \\
\rho(\exo{controlPointID}) &\sqsubseteq \exo{String} \\
\end{array}
\]
This ontology is shown visually in Figure \ref{fig:energy-ontology}.

\subsubsection{The Mapping Assertions}\label{sec:energy-usecase-the-mappings}

We use the following mapping assertions in order to link the
information model in the GSM model to the Semantic Layer. The
assertions are written in the mapping specification language of
\obgsm as follows.
%
\begin{center}
\line(1,0){403}\\
\vspace*{-4.3mm}
\line(1,0){403}\\
\vspace*{-4.3mm}
\end{center}
{\small

\begin{verbatim}
[PrefixDeclaration]
  xsd:		http://www.w3.org/2001/XMLSchema#
  owl:		http://www.w3.org/2002/07/owl#
  :			http://acsi/example/ACSIEnergy/ACSIEnergy.owl#
  rdf:		http://www.w3.org/1999/02/22-rdf-syntax-ns#

[ClassDeclaration] @collection [[
  owl:Thing, :ControlPointReportCollection, :ControlPointReport, 
  :ObjectedReport, :AcceptedReport, :ReviewedReport, :FinishedReport
]]

[ObjectPropertyDeclaration] @collection [[
  :contains
]]

[DataPropertyDeclaration] @collection [[
  :hasControlPointID
]]

[MappingDeclaration] @collection [[
 mappingId  ControlPointReportCollectionMapping
 target     <"&:;cpmrma/{$x}/"> rdf:type :ControlPointReportCollection . 
 source     get('x', 'CPMRMA')(TRUE)

 mappingId  ControlPointIDMapping
 target     <"&:;cpmr/{$x}/"> :hasControlPointID $y^^xsd:string . 
 source     get('x', 'CPMR')({x./CPMR/CPID} == ?y)

 mappingId  CPMRMAContainsCPRMMapping
 target     <"&:;cpmrma/{$x}/"> :contains <"&:;cpmr/{$y}/"> . 
 source     get('x', 'CPMRMA')(get('y', 'CPMR')(
                   x./CPMRA/CPMRDATA->exists(CPMRID == {y./CPMR/ID})))

 mappingId  ReviewedReportMapping
 target     <"&:;cpmr/{$x}/"> rdf:type :ReviewedReport . 
 source     get('x','CPMR')(
                GSM.isMilestoneAchieved('x','AcceptingPublishedOK') OR
                GSM.isMilestoneAchieved('x','MOReviewingPublishedOK') OR
                GSM.isMilestoneAchieved('x','ECDReviewingPublishedOK'))

 mappingId  AcceptedReportMapping
 target     <"&:;cpmr/{$x}/"> rdf:type :AcceptedReport . 
 source     get('x','CPMR')(GSM.isMilestoneAchieved('x','PublishedOK'))

 mappingId  ObjectedReportMapping
 target     <"&:;cpmr/{$x}/"> rdf:type :ObjectedReport . 
 source     get('x','CPMR')(GSM.isMilestoneAchieved('x','Objected') OR
               GSM.isMilestoneAchieved('x','ObjectionRequested') OR
               GSM.isMilestoneAchieved('x','ObjectionCreated') OR
               GSM.isMilestoneAchieved('x','MOReviewingObjectionRequested') OR
               GSM.isMilestoneAchieved('x','ECDReviewingObjectionRequested'))

 mappingId  FinishedReportMapping
 target     <"&:;cpmr/{$x}/"> rdf:type :FinishedReport . 
 source     get('x','CPMR')(GSM.isMilestoneAchieved('x','CPMRFinished'))
]]
\end{verbatim}
}
\begin{center}
\vspace*{-4.3mm}
\line(1,0){403}\\
\vspace*{-4.3mm}
\line(1,0){403}
\end{center}
%
%
Where ``:'' is a prefix declared as {\tt
  http://acsi/example/energy/energy.owl\#}. The intuition of some 
mapping assertions above is as follows:
\begin{compactitem}
\item {\tt ControlPointReportMapping} and {\tt
    ControlPointReportCollectionMapping} populate the concepts
  $ControlPointReport$ and $ControlPointReportCollection$ with the
CPMR and CPMRMA artifact instances respectively.
\item {\tt CPMRMAContainsCPMRMapping} populates the 
  $contains$ role, which relates CPMRMA with the CPMRs it contains.
\item {\tt ControlPointIDMapping} populates the attribute
  $hasControlPointID$ by relating $ControlPointReport$ with its control point ID.
\item {\tt ReviewedReportMapping} populates the
  $ReviewedReport$ concept with the CPMR artifact instance, given the
  achievement of  one of the three milestones \emph{AcceptingPublishedOK} 
  \emph{MOReviewingPublishedOK} or
  \emph{ECDReviewingPublishedOK}. In the Semantic Layer,
  $ReviewedReport$ intuitively represents a CPMR that has been
  reviewed. In the Artifact Layer, this corresponds to the situation
  in which the
  CPMR has been reviewed and accepted either by the electric company,
  or by the metering office (MO), or by the Electric Control Department (ECD). This
  example show how such details can be hidden from the Semantic Layer,
  which does not show the fact that a $ReviewedReport$ is obtained by
  a (possibly complex) chaining of achieved milestones in the
  underlying GSM model.
\item {\tt AcceptedReportMapping} populates the
  $AcceptedReport$ concept with the CMPR artifact instance, when
  milestone \emph{PublishedOK} is achieved. This intuitively
  means that the $AcceptedReport$ is a published CPMR that has been
  approved.
\item {\tt ObjectedReportMapping} populates the
  $ObjectedReport$ concept with an objected CMPR artifact instance.
  This situation is recognized, at the Artifact Layer, by combining different related milestones.
\item {\tt FinishedReportMapping} populates the
  $FinishedReport$ concept with the finished CMPRs, i.e., those that
  have achieved milestone \emph{CPMRFinished}.
\end{compactitem}

\subsubsection{Verification}\label{sec:energy-usecase-verification}

In this section, we demonstrate how the presence of the Semantic Layer
can help in the specification of temporal properties of interests. We
consider in particular the following properties:
\begin{compactenum}
\item All control point reports will eventually will be finished.
\item All control point reports that are accepted must have been
  reviewed. This property is used to ensure that there is no way to
  achieve a state in which a certain CPMR is accepted, without going
  through the review for that CPMR. Notice that ``having being
  reviewed'' is considered to be a permanent property of CPMRs.
\item All control point reports that are finished must not be objected
  control point reports. This property ensures that control point
  reports cannot be classified as finished as long as they are still
  objected.
\item All objected control point reports must not be finished control
  point reports.
\end{compactenum}
Such four properties can be expressed as conceptual temporal
properties over the Semantic Layer. In particular, we encode them
using the language provided by \obgsm as follows (Note: for
compactness of the presentation, in the following we do not write the
queries in SPARQL):
\begin{compactenum}
\item $\AG (\forallTemp\ x\  .\ (\implTemp{[\exo{ControlPointReport}(x)]}{ \EF [\exo{FinishedReport}(x)]}) )$
\item $\AG (\forallTemp\ x\  .\ (\implTemp{[\exo{AcceptedReport}(x)]}{[\exo{ReviewedReport}(x)]}) )$
\item $\AG (\forallTemp\ x\  .\ (\implTemp{[\exo{FinishedReport}(x)]}{\notTemp [\exo{ObjectedReport}(x)]}) )$
\item $\AG (\forallTemp\ x\  .\ (\implTemp{[\exo{ObjectedReport}(x)]}{\notTemp [\exo{FinishedReport}(x)]}) )$
\end{compactenum}
We now show how this high-level property are compiled by \obgsm into
underlying temporal properties that can be fed into the GSMC model
checker.  We stress that, without the presence of the Semantic Layer,
the user would be forced to write this low-level properties manually.

\begin{enumerate}
\item The rewriting step for the first conceptual temporal property
 produces the following formula, which ``embeds'' the constraints of
 the ontology present at the Semantic Layer:
{\small
\begin{equation*}
\begin{split}
\AG (\forallTemp\ x\  .\ (
(&\existsTemp\ y\  .\ [\exo{hasControlPointID}(x,y)]) \orTempB \\
&[\exo{ObjectedReport}(x)] \orTempB \\
&[\exo{AcceptedReport}(x)] \orTempB \\
&[\exo{ControlPointReport}(x)] \orTempB \\
&[\exo{ReviewedReport}(x)] \orTempB \\
&(\existsTemp\ z\  .\ [\exo{contains}(z,x)]) \orTempB \\
&[\exo{FinishedReport}(x)] \orTempB \\
&\implTempB \EF [\exo{FinishedReport}(x)] ) )
\end{split}
\end{equation*}
}
%
The expansion of the queries contained in the property is done by
embedding the following cases, reflected by the ontology constraints:
\begin{compactitem}
\item Those objects that have a control point ID (i.e., are in the domain of the
  attribute $\exo{hasControlPointID}$), are instances of $\exo{ControlPointReport}$.
\item $\exo{ObjectedReport}$ is a $\exo{ControlPointReport}$.
\item $\exo{AcceptedReport}$ is a $\exo{ControlPointReport}$.
\item $\exo{ReviewedReport}$ is a $\exo{ControlPointReport}$.
\item Those objects that are in the range of the role $\exo{contains}$
  are instances of $\exo{ControlPointReport}$.
\item $\exo{FinishedReport}$ is a $\exo{ControlPointReport}$.
\end{compactitem}
By exploiting the mapping assertions, \obgsm unfolds the rewritten
property into this final result:
{\small
\begin{equation*}
\begin{split}
\AG\ ( &\forallGSM ('x',\ 'CPMR')( ! ( GSM.isMilestoneAchieved('x','Objected') \orTempB  \\
& GSM.isMilestoneAchieved('x','ObjectionRequested')  \orTempB  \\
& GSM.isMilestoneAchieved('x','ObjectionCreated') \orTempB  \\
& GSM.isMilestoneAchieved('x','MOReviewingObjectionRequested')  \orTempB  \\
& GSM.isMilestoneAchieved('x','ECDReviewingObjectionRequested')   \orTempB  \\
& GSM.isMilestoneAchieved('x','PublishedOK') \orTempB  \\
& ( false ) \orTempB\\
& GSM.isMilestoneAchieved('x','AcceptingPublishedOK') \orTempB  \\
& GSM.isMilestoneAchieved('x','MOReviewingPublishedOK') \orTempB  \\
& GSM.isMilestoneAchieved('x','ECDReviewingPublishedOK') \orTempB  \\
& \existsGSM( 'y', 'CPMRMA' )  ( y./CPMRA/CPMRDATA \implTempB \\
& \ \ \ \ \ \ \ \ \ \ \ \ \ \ \ \ \textbf{exists}(CPMRID == {x./CPMR/ID}) )  \orTempB  \\
& GSM.isMilestoneAchieved('x','CPMRFinished')  )   \orTempB  \\
& EF  ( GSM.isMilestoneAchieved('x','CPMRFinished') )  ) )  \\
\end{split}
\end{equation*}
}
Notice that the absence of a mapping assertion for the concept
$ControlPointReport$ results into a $false$ disjunct in the unfolding.
\item The translation of the second conceptual temporal property
  produces this final result:
{\small
\begin{equation*}
\begin{split}
\AG\ &\forallGSM ('x',\ 'CPMR')(
\orTemp{\notTemp (GSM.isMilestoneAchieved('x','Published OK'))}{(\\
&\orTemp{\orTemp{ GSM.isMilestoneAchieved('x','AcceptingPublishedOK')} \\
&{GSM.isMilestoneAchieved('x','MOReviewingPublishedOK')}} \\
&{GSM.isMilestoneAchieved('x','ECDReviewingPublishedOK')})}\\
&)
\end{split}
\end{equation*}
}
\item The translation of the third conceptual temporal property
  produces this final result:
{\small
\begin{equation*}
\begin{split}
\AG\ &\forallGSM ('x',\ 'CPMR')(
\orTemp{\notTemp GSM.isMilestoneAchieved('x',CPMRFinished')}{\\
&\notTemp
(\orTemp{\orTemp{\orTemp{\orTemp{GSM.isMilestoneAchieved('x','Objected')} \\
&{GSM.isMilestoneAchieved('x','ObjectionRequested')}} \\
&{GSM.isMilestoneAchieved('x','ObjectionCreated')}} \\
&{GSM.isMilestoneAchieved('x','MOReviewingObjectionRequested')}} \\
&{GSM.isMilestoneAchieved('x','ECDReviewingObjectionRequested')})}\\
&)
\end{split}
\end{equation*}
}
\item The translation of the fourth conceptual temporal property
  produces this final result:
{\small
\begin{equation*}
\begin{split}
\AG\ &\forallGSM ('x',\ 'CPMR')(
\orTemp{\notTemp (\orTemp{\orTemp{\orTemp{\orTemp{GSM.isMilestoneAchieved('x','Objected')} \\
&{GSM.isMilestoneAchieved('x','ObjectionRequested')}} \\
&{GSM.isMilestoneAchieved('x','ObjectionCreated')}} \\
&{GSM.isMilestoneAchieved('x','MOReviewingObjectionRequested')}} \\
&{GSM.isMilestoneAchieved('x','ECDReviewingObjectionRequested')}) }{\\
&\notTemp GSM.isMilestoneAchieved('x',CPMRFinished')
}\\
&)
\end{split}
\end{equation*}
}
\end{enumerate}

\noindent 
By comparing the properties specified over the Semantic Layer and
their corresponding translations, it is apparent that, even in this
simple case study, the presence of the Semantic Layer hides low-level
details, helps the modeler in focusing on the domain under study, and
allows for using the vocabulary he/she is familiar with (i.e., the
vocabulary of the ontologies).

%% file: 2.chapters/10-conclusion.tex
\chapter{Conclusion}\label{ch:conclusion}

\ifhideconclusion
 
\fi


We close this thesis by recapping our journey so far, and also
discussing several plausible future directions.

\section{Summary}\label{sec:summary}

Within this thesis, we have proposed several frameworks for specifying
semantically-rich data-aware business processes systems that also
take into account various aspects. For each setting, we have addressed
the problem of verifying temporal properties over the evolution of the
system.

Specifically, in \Cref{ch:gkab}, we introduced a framework for
specifying semantically-rich data-aware processes systems, namely
Golog-KABs (GKABs), by leveraging on Knowledge and Action Bases (KABs)
\cite{BCMD*13}.
Fundamentally, GKABs capture the manipulation of Knowledge Bases (KBs)
by the Golog program \cite{LRLLS97}. We have also introduced standard
execution semantics for GKABs that do nothing regarding inconsistency
(i.e., updates that lead to an inconsistent state are simply
rejected). 
A GKAB with standard execution semantics is called S-GKAB.
%
%
Furthermore, we have shown that verification of rich temporal
properties over S-GKABs can be reduced to corresponding verification
of KABs and vice versa. 

In \Cref{ch:ia-gkab}, we extended GKABs towards Inconsistency-aware
GKABs (I-GKABs) by incorporating several inconsistency management
approaches (based on the notion of repairs). 
Concerning about verification of I-GKABs, we have proven that they can
be reduced to corresponding verification of S-GKABs and vice
versa. 

Next, in \Cref{ch:cs-gkab}, we proposed Context-Sensitive GKABs
(\csgkabs), which are an extension of GKABs that takes into account
contextual information. In \Cref{ch:cs-gkab}, we only focused on
\scsgkabs that is \csgkabs with standard execution semantics (i.e., do
nothing w.r.t.\ inconsistency).
%
%
We have proved that verification of sophisticated context-sensitive
temporal properties over \scsgkabs is reducible to corresponding
verification of S-GKABs and vice versa. 

In \Cref{ch:cs-ia-gkab}, we introduced an extension of
GKABs 
that takes into account contextual information as well as employs
a sophisticated inconsistency handling
mechanism. 
Additionally, we have shown that verification of such an extension can
be reduced to verification of S-GKABs and vice
versa. 
As a deeper investigation, in \Cref{ch:a-gkab} we proposed another
extension, namely Alternating GKABs (\agkabs), that does not only
consider contextual information and employ a sophisticated
inconsistency handling mechanism, but also exposes each source of
non-determinisms (e.g., the choice of action) within a single
evolution step. Essentially, \agkabs allow us to have a more
fine-grained understanding over the system evolution. We studied
verification of temporal properties over \agkabs and showed that it
can be reduced to verification of S-GKABs. In addition, the other
direction also has been shown.

Prominently, all of our reductions from various GKABs into KABs
preserve run-boundedness, which is a restriction that guarantees the
decidability of KABs verification.

As an orthogonal approach for specifying semantically-rich data-aware
processes systems, in \Cref{ch:sedap} we proposed a novel framework
called Semantically-Enhanced Data-Aware Processes (\sgdss) that
provides a high-level conceptual view over the evolution of a
data-aware processes system, by making use of ontologies. We have
successfully addressed the problem of verification of temporal
properties expressed over the conceptual level in \sgdss, by showing
that the verification can be reduced to the corresponding verification
over DCDSs. Not only theoretical results, we have also concretized the
concept of \sgdss into an implementation of a tool.

\iffinal
\else
\clearpage
\fi
\section{Discussion and Related Works}


%
%

Concerning the restriction to get the decidability of verification,
apart from the run-boundedness condition that we have borrowed from
\cite{BCDDM13}, the work in \cite{BCDDM13} also introduces another
restriction called \emph{state boundedness}. Essentially, state
boundedness constraints the system by requiring that the number of
constants in each state of the system is bounded by a generic bound
$b$.
Clearly, run boundedness is more restrictive than state
boundedness. I.e., run-boundedness implies state boundedness (consider
that the bound on the number of constants in each run also bounds
the number of constants in each state of the system).

In \cite{DeLP12}, the authors propose so called \emph{bounded
  action theories}, which are situation calculus basic action theories
\cite{Reit01} in which in each situation, the number of domain objects 
that belong to any fluent
is bounded (i.e., the number of ground fluent atoms in each situation
is bounded).
Such restriction on the action theories is similar to the
notion of state boundedness, but in the setting of situation
calculus. The work by \cite{DeLP12} studies the verification of
expressive temporal properties based on $\mu$-calculus over dynamic
systems formalized in bounded action theories, and shows
that such problem is decidable.
%
%
Furthermore \cite{DeLP12} also explores several ways to obtain bounded
action theories such as
\begin{inparaenum}[\it (i)]
\item blocking the  execution of actions that
  would destroy the bound;
\item requiring that for each action and situation, the number of
  tuples that are added to the fluent is less than or equal to those
  that are deleted;
\item axiomatizing the notion of \emph{fading fluents} that basically
  enforce each fluent to eventually become false if it is not
  (re)-added for some period of time.
\end{inparaenum}
The work in \cite{DLPV14} studies a framework for modeling and
verifying agents expressed in situation calculus bounded action
theories \cite{DeLP12}. 
Moreover, \cite{DLPV14} shows the decidability of temporal properties
verification over online executions of the agent, i.e., those
executions resulting from the actions that can be performed by the
agent. For specifying the temporal properties, such work considers
expressive temporal properties based on $\mu$-calculus.
%

The works in \cite{BeLP12,BeLP14} propose data-aware multi-agent
systems called \emph{Artifact-Centric Multi Agent Systems} (AC-MASs)
and study the verification of a first order variant of CTL in such a
setting. In brief, AC-MASs capture the combined behavior of agents in
which each agent has a database to maintain its internal data and can
perform actions that manipulate the data. To get decidability of
verification, those works assume, in addition to
\emph{state-boundedness}, also \emph{uniformity}. The notion of
uniformity is actually borrowed from the notion of \emph{genericity}
in databases \cite{AbHV95}, which basically says that \emph{a query is
  generic} if it is insensitive w.r.t.\ renaming of constants. 
In the setting of ``relational transition systems'' (i.e., transition
systems where each state contains relational data and each transition
represents an action execution), roughly speaking, the notion of
uniformity says that, if we can go from state $s$ to state $t$ by
executing an action $\act(\vec{x})$ with parameters $\vec{p}$ (i.e.,
$\vec{x}$ is instantiated with $\vec{p}$), then if we have a bijection
$h$ that renames the constants $\vec{c}$ into $\vec{c'}$, we have that
we can go from state $s'$ to state $t'$ by executing the action
$\act(\vec{x})$ with parameters $\vec{p'}$, where $s'$, $t'$ and
$\vec{p'}$ are obtained by applying $h$ respectively to $s$, $t$ and
$\vec{p}$ (i.e., renaming according to $h$, all constants therein).
Concerning uniformity, DCDSs, KABs and our various GKABs variants
satisfy uniformity as a consequence of the definition of their
execution semantics (see \cite{CDMP13,CDMP15}).
Still related to the boundedness condition, the work by \cite{BGL15}
introduce so called Open Multi-Agent Systems (OMAS), i.e., a framework
for modeling multi-agent systems in which the agents may enter and
leave the system at run time. Moreover, \cite{BGL15} studies the
verification of temporal properties over such setting, in which the
temporal properties are specified in first order variant of CTL. The
authors obtained decidability of verification by restricting that in
each state of the system, the number of constants and the agents are
bounded (which is similar to state-boundedness).
In addition to the works above, also the following works rely on the
state-boundedness assumption to obtain decidability result:
\cite{DLPV14b,CDMP13,CDMP15,MoCD14}.

As a remark on state boundedness, although our work in this thesis
assumes run-boundedness to get decidability, it can be shown that our
results carry over even if we adopt state boundedness. The core
intuition is that our translations from the various kinds of GKABs
into KABs do not introduce an unbounded number of constants (i.e., we
only make use of finitely many additional constants).


Concerning our translation that basically ``unfold'' the given Golog
program into a set of condition-action rules and a set of actions in
\Cref{sec:transform-sgkab-to-kab},
it is worth to mention that there are also some related works on
``unfolding'' Golog programs into another formalisms with a particular
purpose.
In \cite{ClLa08}, the authors provide a systematic mechanism to unfold
Golog program into the so called \emph{characteristic graphs} that aim
to capture all possible evolution of program (i.e., encode all
reachable program configurations). Technically, each vertex in the
graph captures a reachable program configurations that denote the
remaining program to be executed, and each edge in the graph capture
some informations that is related to the changes of remaining program
from one vertex to another vertex (e.g., the action that is executed
and causes the changes).
The work in \cite{FBM08} presents an algorithm to compile arbitrary
Golog programs into basic action theories in situation calculus. Such
compilation also involves a transformation from Golog program into
petri nets. Furthermore, the authors of \cite{BFM07} ``unfold'' Golog
programs into finite state automata.


The works by \cite{BLMSW05,BLMSW05b} introduce a DL-based action
formalism.  Moreover, building on~\cite{BLMSW05,BLMSW05b}, the
works~\cite{BZ13,ZC15} study Golog programs~\cite{LRLLS97} in which
the atomic actions are formalized as DL-based actions. Within this
setting,~\cite{BZ13,ZC15} tackle the problem of verifying LTL-based
temporal properties over the execution of Golog program with respect
to the given DL KB. 
In contrast to our work and also the works
in~\cite{BCDD*12,BCMD*13,MoCD14}, the works of
\cite{BLMSW05,BLMSW05b,BZ13,ZC15} consider the semantics of DL-based
actions to be specified in terms of manipulation of DL
interpretations.
Whereas, in our setting, we have that the actions manipulate the data
within the DL KB. Therefore, in our transition systems, we have that
each state is labeled by KB instead of DL-models as in
\cite{BLMSW05,BLMSW05b,BZ13,ZC15}.



The combination of DLs with temporal logics has been studied
extensively within the line of research called Temporal Description
Logics (TDLs)
\cite{LuWZ08,ArFr00,WoZa98,BaGL08,AKLWZ07,ArLT06}. Although they do
not have actions that progress the knowledge base over time as in our
setting, the augmented temporal operators in TDL describe the dynamic
aspects of the knowledge base. Such combinations between DL and
temporal logic are based on a two-dimensional semantics, where one
dimension is for time and the other dimension is for the DL domain.




\section{Future Works}

We elaborate several plausible future directions of this research as
follows:

\begin{itemize}



\item \textbf{Inconsistency-aware GKABs based on Consistent Query
    Answering.} \xspace  
  Within this thesis, the \ask operation in GKABs corresponds to
  certain answers computation, and
  we have approached the problem of inconsistency handling in GKABs,
  by resorting to an approach based on ABox repairs. This is achieved
  through our GKABs parametric execution semantics that intuitively
  allows us to parameterize the \tell operation via defining various
  filters.
  An orthogonal approach to the one taken is to maintain ABoxes that
  are inconsistent with the TBox as states of the transition system,
  and rely, both for the progression mechanism and for answering
  queries used in verification, on consistent query answering
  \cite{Bert06,LLRRS10}. Concerning progression mechanism in GKABs, we
  can accommodate this changing easily by modifying the \ask operation
  such that it corresponds to consistent query answering
  computation. Then, we need to redefine the semantics for query using
  the consistent query answering semantics. The next question is how
  should we deal with the verification?
%
%
  A plausible approach to answer this question is exploring whether we
  can emulate the computation of consistent query answer as a Golog
  program. Then we can try to reduce the verification problem into the
  verification of S-GKABs that mimics this GKABs extension.
  Furthermore, it is also challenging to investigate the
  correspondence between the consistent query answering based approach
  and the repair-based approach in dealing with inconsistency in
  GKABs.



\item \textbf{Adopting Repair-based Semantics to \sgdss.}\xspace 
%
  In a \sgds, an action execution that leads to inconsistency in the
  semantic layer is rejected. Hence, we reject inconsistent states.
  However, the inconsistency in a state might be caused by only a
  small portion of the ABox in the semantic layer and the other
  consistent portion might be still useful for some
  reasoning. Therefore, keeping the state by removing the small
  portion of ABox that made the state inconsistent might be useful. A
  possible solution to this situation is to ``repair'' the ABox (in
  the semantic layer) and allow the action that was rejected to be
  executed together with some repair in order to maintain the
  consistency. Notice that the repair happens in the semantic layer
  and it might lead to several possible repaired states.
  One question is how we will do the repair.  In addition, one needs
  to understand how the repair in the semantic layer will affect the
  relational layer, i.e., the question is how we propagate down the
  repaired ABoxes in the semantic layer into the corresponding
  database instances in the relational layer. 
  This problem might has some connections with the problem of view
  updates \cite{BaSp81,DaBe82,GPZ88}, because we can consider the
  semantic layer as a view of the relational layer. To formalize the
  relation between the two layers, one might also look into
  bidirectional transformations (c.f. \cite{Pier12}).
  Also an interesting task to investigate is the verification of
  conceptual temporal properties over this
  setting. 



\item \textbf{Embracing Context into \sgdss.}\xspace
A further challenging research direction that can be pursued
immediately is about adding the notion of context to \sgds. The idea
is as follows: In \sgds, the intensional knowledge about the domain
(which is expressed as a \dllitea TBox) is fixed along the evolution
of the system.  However, this situation is in general too restrictive,
since specific knowledge might hold or be applicable only in specific
\emph{context-dependent} circumstances.  Ideally, one should be able
to express the knowledge that is known to be true in certain cases,
but not necessarily in all.
Having gained the understanding of how to integrate contextual
information into GKABs, we are now investigating how we could extend
this result to the \sgds setting. Within \sgds, we need to understand
how to introduce the contextual information over the setting as well
as how the contextual information affects the system
evolution. Moreover, notice that the verification of \sgds relies on
the notion of ``rewriting'' and ``unfolding'', which takes into
account the mapping and relies on the assumption that the TBox is
fixed.  Given that in the presence of contextual information the TBox
is changing over the time with the context, the question is how we
should ``rewrite'' the given conceptual properties. Also, which TBox
should we use? A promising route towards tackling this issue is by:
\begin{inparaenum}[\it (i)]
\item Compiling the given query into a disjunction of all possible
  rewritten queries based on all possible contexts.
\item Additionally, we conjunct each rewritten query with a query that
  represents the corresponding context.
\end{inparaenum}
Essentially, this approach is similar to the way of how we introduce
contextually compiled query in \Cref{def:contextually-compiled-query}.



\item \textbf{Inconsistency-Aware Context-Sensitive SEDAP.} \xspace
  Following the last two future works, i.e., adopting repair-based
  semantics to \sgdss as well as incorporating context into \sgdss, it
  is natural to proceed further by combining those directions towards
  Inconsistency-Aware Context-Sensitive SEDAP. The presence of context
  that indirectly changes the TBox require us to adapt the repair
  mechanism based on the context changes since we need to do the
  repair based on the TBox.  In addition, the setting of \sgds, that
  separates the semantic and relational layer, also complicates the
  problem.


\item \textbf{Accommodating Update on Semantic Layer.} \xspace Within
  the setting of \sgdss, another interesting direction is to
  investigate the situation where there is a change explicitly over
  the ontologies in the semantic layer (in the conceptual level). For
  example when there is a new process introduced in the semantic
  layer. In this case, the new introduced process might change a
  certain ABox in the semantic layer. The question here is how such a
  change will/can affect the relational layer.

\item \textbf{Process Synthesis.}\xspace In the \sgdss, the processes
  are specified over the relational layer, and we have a conceptual
  view provided by the semantic layer which is obtained by projecting
  the evolution happening in the relational layer to the semantic
  layer by using the mapping.
  One interesting alternative setting is to consider the situation
  where the processes are specified over the semantic layer. In other
  words, in this setting we have a relational database in the
  relational layer, and the processes are specified in a high-level
  way through the semantic layer. Then, we are interested on how this
  high-level process specification can be ``brought down'' into the
  relational layer and executed in the relational layer as well as
  evolve the existing database. This leads to the problem of
  synthesizing the process in the relational layer from the given high
  level specification in the semantic layer, which might also be
  expected to satisfy some temporal properties specified over the
  semantic layer. Still about synthesis, it is also interesting to
  investigate a setting where the process is partly specified and then
  we synthesize additional process components such that the system
  satisfies some set of specified temporal properties.


\item \textbf{Semantic Compliance Checking.} \xspace Another
  interesting setting related to \sgdss is to consider the situation
  where we have the process specification on both semantic and
  relational layer. Then it is interesting to see how the process
  specifications in these two layers are matched and how they evolve
  the data as well as satisfying the specified temporal properties. In
  this scenario, the interesting task is to check if the evolution of
  our actual/concrete system complies with a certain ``evolution
  requirement''. In this setting, the processes specified in the
  relational layer can be considered as our actual/concrete system and
  the processes specified in the semantic layer can be considered as
  the ``evolution requirement'' to be checked. Hence in this case we
  are interested to check if the evolution in the relational layer
  matches the evolution in the semantic layer. We call the problem
  ``Semantic Compliance Checking'' because intuitively we can think
  that the system evolution happened in the semantic layer is the
  semantic of the system evolution.



\item \textbf{Bringing Theory to Practice.} \xspace an obvious future
  direction of this work is of course to implement the theoretical
  results that has been obtained. Since it has been shown that all
  variants of our proposed frameworks can be reduced to DCDSs, one
  possible direction is to take advantage from the works that attempt
  to implement DCDSs (e.g., \cite{RMPM13,CMPR15,CMPR15a}). I.e., we
  could simply implement a translation to transform our systems into
  DCDSs.
  Furthermore, there is also a recent work provided by \cite{CMPS16}
  that studies the planning problem in the setting of
  KABs. Specifically, \cite{CMPS16} studies the problem of plan
  existence over KABs as well as the problem of plan synthesis. Not
  only providing theoretical results, \cite{CMPS16} also implements
  the results (by also employing off-the-shelf planner
  system). Essentially, the planning over KABs addresses the problem
  of checking whether starting from the initial state of a KAB, we can
  have a sequence of actions (i.e., plan) that brought us into a KAB
  state in which the goal (that is expressed in ECQs query) is
  satisfied. I.e., it can be seen as a particular instance of
  verification problem.  Hence, since verification of our proposed
  frameworks (i.e., GKABs and their variants) has been shown to be
  reducible to verification of KABs, to implement (some of) our
  results, we could try to make use the available results in
  \cite{CMPS16}.
  Concerning the specification language, to bring these theoretical
  results into the people in business processes management area, it
  might also be desirable to study the correspondence between typical
  business processes specification language (e.g., BPMN \cite{BPMN})
  with our specification language. Once the correspondence is
  established, we can implement a translation between two
  formalisms. Some works that might be related to this direction can
  be found in \cite{MR14,MR16}, which involve translations from Petri
  Nets to DCDSs.



\item \textbf{Syntactic Restriction Based on Golog Structure.}
  \xspace The work by \cite{BCDDM13} proposes two
  syntactic restrictions for obtaining decidability in
  DCDSs. 
%
%
%
  Basically, those restrictions are properties over a data flow graph
  that is constructed from the actions specification. Moreover, such
  data flow graph essentially captures all possible data flow that can
  be induced by the specified actions. However, since it is only
  constructed from the actions specification, it over estimates the
  possible data flow.  This is the case because some actions might not
  be executable, or some actions might never be executed after a
  particular action.
%
%
%
  On the other hand, a Golog program basically captures some
  information about the possible data flow. Essentially, it has some
  information about some possible sequence of actions. Hence, it might
  be interesting to investigate whether we can exploit such
  information, get a better data flow information, and hence devise a
  better syntactic restriction (i.e., obtain a better decidable
  class).





\item \textbf{Embracing Quantitative Aspects.}
  Another further interesting research direction is to consider the
  quantitative aspects of the system. One possibility is to deal with
  the problem of verifying whether the system under study satisfies
  some quantitative properties specified in a certain language. For
  instance we can have that each transition in the transition system
  can be decorated with some values which possibly representing the
  cost of execution the corresponding action (similar to the
  transition systems in the work by \cite{AS-ZEUS-12}). Then, one
  might be interested to check whether we can reach a particular state
  in such a way that the total cost of actions that are executed is
  less than a particular value.




\end{itemize}

%% file: 3.bibliography/bibliography.tex
\manualmark
\markboth{\spacedlowsmallcaps{\bibname}}{\spacedlowsmallcaps{\bibname}} 
\refstepcounter{dummy}
\addtocontents{toc}{\protect\vspace{\beforebibskip}} 
\addcontentsline{toc}{chapter}{\tocEntry{\bibname}}
\bibliographystyle{plainnat}
\label{app:bibliography} 
\bibliography{3.Bibliography/string-medium,3.Bibliography/string-medium-plus,3.Bibliography/bib}


%% file: 1.front-back-matter/closing.tex
\pagestyle{empty}

\pdfbookmark[0]{Closing}{closing}

\bigskip
\bigskip
\bigskip
\begin{flushleft}
  ``\textit{There comes a time \\ when you have to choose \\ between turning
    the page \\ and closing the book.}'' \\--Josh Jameson
\end{flushleft}
\hfill

\vfill


\vfill

\begin{flushright}
 Thanks for reading.\\
``\textit{A great article might be meaningless \\ without the
  readers.}'' \\
\end{flushright}

 
\bigskip

%% file: main.bbl
\begin{thebibliography}{186}
\providecommand{\natexlab}[1]{#1}
\providecommand{\url}[1]{\texttt{#1}}
\expandafter\ifx\csname urlstyle\endcsname\relax
  \providecommand{\doi}[1]{doi: #1}\else
  \providecommand{\doi}{doi: \begingroup \urlstyle{rm}\Url}\fi

\bibitem[Abiteboul et~al.(1995)Abiteboul, Hull, and Vianu]{AbHV95}
Serge Abiteboul, Richard Hull, and Victor Vianu.
\newblock \emph{Foundations of Databases}.
\newblock Addison Wesley Publ.\ Co., 1995.

\bibitem[Abiteboul et~al.(2004)Abiteboul, Benjelloun, and Milo]{ABM04}
Serge Abiteboul, Omar Benjelloun, and Tova Milo.
\newblock Positive {A}ctive {XML}.
\newblock In \emph{Proc.\ of the 23rd ACM SIGACT SIGMOD SIGART Symp.\ on
  Principles of Database Systems (PODS)}, pages 35--45. {ACM} Press, 2004.

\bibitem[Abiteboul et~al.(2008{\natexlab{a}})Abiteboul, Benjelloun, and
  Milo]{ABM08}
Serge Abiteboul, Omar Benjelloun, and Tova Milo.
\newblock The {A}ctive {XML} project: an overview.
\newblock \emph{Very Large Database J.}, 17\penalty0 (5):\penalty0 1019--1040,
  2008{\natexlab{a}}.

\bibitem[Abiteboul et~al.(2008{\natexlab{b}})Abiteboul, Segoufin, and
  Vianu]{AbSV08}
Serge Abiteboul, Luc Segoufin, and Victor Vianu.
\newblock Static analysis of {A}ctive {XML} systems.
\newblock In \emph{Proc.\ of the 27th ACM SIGACT SIGMOD SIGART Symp.\ on
  Principles of Database Systems (PODS)}, pages 221--230, 2008{\natexlab{b}}.
\newblock \doi{10.1145/1376916.1376948}.

\bibitem[Abiteboul et~al.(2009{\natexlab{a}})Abiteboul, Bourhis, Galland, and
  Marinoiu]{ABGM09b}
Serge Abiteboul, Pierre Bourhis, Alban Galland, and Bogdan Marinoiu.
\newblock The {AXML} artifact model.
\newblock In \emph{Proc.\ of the 16th Int.\ Symp.\ on Temporal Representation
  and Reasoning (TIME)}, pages 11--17, 2009{\natexlab{a}}.

\bibitem[Abiteboul et~al.(2009{\natexlab{b}})Abiteboul, Segoufin, and
  Vianu]{AbSV09b}
Serge Abiteboul, Luc Segoufin, and Victor Vianu.
\newblock Static analysis of {A}ctive {XML} systems.
\newblock \emph{ACM Trans.\ on Database Systems}, 34\penalty0 (4):\penalty0
  23:1--23:44, 2009{\natexlab{b}}.
\newblock \doi{10.1145/1620585.1620590}.

\bibitem[Arsac and Kodratoff(1982)]{AK82}
J.~Arsac and Y.~Kodratoff.
\newblock Some techniques for recursion removal from recursive functions.
\newblock \emph{TOPLAS}, 4\penalty0 (2):\penalty0 295--322, 1982.

\bibitem[Artale and Franconi(2000)]{ArFr00}
Alessandro Artale and Enrico Franconi.
\newblock A survey of temporal extensions of description logics.
\newblock \emph{Ann.\ of Mathematics and Artificial Intelligence},
  1--4:\penalty0 171--210, 2000.

\bibitem[Artale et~al.(2006)Artale, Lutz, and Toman]{ArLT06}
Alessandro Artale, Carsten Lutz, and David Toman.
\newblock A description logic of change.
\newblock In \emph{Proc.\ of the 19th Int.\ Workshop on Description Logic
  (DL)}, volume 189 of \emph{CEUR Electronic Workshop Proceedings,
  {\upshape\protect\url{http://ceur-ws.org/}}}, 2006.

\bibitem[Artale et~al.(2007)Artale, Kontchakov, Lutz, Wolter, and
  Zakharyaschev]{AKLWZ07}
Alessandro Artale, Roman Kontchakov, Carsten Lutz, Frank Wolter, and Michael
  Zakharyaschev.
\newblock Temporalising tractable description logics.
\newblock In \emph{Proc.\ of the 14th Int.\ Symp.\ on Temporal Representation
  and Reasoning (TIME)}, pages 11--22, 2007.

\bibitem[Artale et~al.(2009)Artale, Calvanese, Kontchakov, and
  Zakharyaschev]{ACKZ09}
Alessandro Artale, Diego Calvanese, Roman Kontchakov, and Michael
  Zakharyaschev.
\newblock The \textit{DL-Lite} family and relations.
\newblock \emph{J.\ of Artificial Intelligence Research}, 36:\penalty0 1--69,
  2009.

\bibitem[Baader(1999)]{Bad99}
Franz Baader.
\newblock Logic-based knowledge representation.
\newblock In \emph{Artificial Intelligence Today}, volume 1600 of \emph{Lecture
  Notes in Computer Science}, pages 13--41. Springer, 1999.

\bibitem[Baader and Zarrie{\ss}(2013)]{BZ13}
Franz Baader and Benjamin Zarrie{\ss}.
\newblock Verification of {G}olog programs over description logic actions.
\newblock In \emph{Proc.\ of the 9th Int.\ Symp.\ on Frontiers of Combining
  Systems (FroCoS)}, volume 8152 of \emph{Lecture Notes in Computer Science},
  pages 181--196. Springer, September 2013.

\bibitem[Baader et~al.(2005{\natexlab{a}})Baader, Lutz, Milicic, Sattler, and
  Wolter]{BLMSW05}
Franz Baader, Carsten Lutz, Maja Milicic, Ulrike Sattler, and Frank Wolter.
\newblock Integrating description logics and action formalisms: First results.
\newblock In \emph{Proc.\ of the 20th Nat.\ Conf.\ on Artificial Intelligence
  (AAAI)}, 2005{\natexlab{a}}.

\bibitem[Baader et~al.(2005{\natexlab{b}})Baader, Lutz, Milicic, Sattler, and
  Wolter]{BLMSW05b}
Franz Baader, Carsten Lutz, Maja Milicic, Ulrike Sattler, and Frank Wolter.
\newblock Integrating description logics and action formalisms: First results.
\newblock In \emph{Proc.\ of the 18th Int.\ Workshop on Description Logic
  (DL)}, number 147 in CEUR Electronic Workshop Proceedings,
  {\upshape\protect\url{http://ceur-ws.org/}}, 2005{\natexlab{b}}.

\bibitem[Baader et~al.(2007)Baader, Calvanese, McGuinness, Nardi, and
  Patel-Schneider]{BCMNP07}
Franz Baader, Diego Calvanese, Deborah McGuinness, Daniele Nardi, and Peter~F.
  Patel-Schneider, editors.
\newblock \emph{The Description Logic Handbook: {T}heory, Implementation and
  Applications}.
\newblock Cambridge University Press, 2nd edition, 2007.

\bibitem[Baader et~al.(2008)Baader, Ghilardi, and Lutz]{BaGL08}
Franz Baader, Silvio Ghilardi, and Carsten Lutz.
\newblock {LTL} over description logic axioms.
\newblock In \emph{Proc.\ of the 11th Int.\ Conf.\ on the Principles of
  Knowledge Representation and Reasoning (KR)}, pages 684--694, 2008.

\bibitem[Baader et~al.(2012{\natexlab{a}})Baader, Ghilardi, and Lutz]{BaGL12}
Franz Baader, Silvio Ghilardi, and Carsten Lutz.
\newblock {LTL} over description logic axioms.
\newblock \emph{ACM Trans.\ on Computational Logic}, 13\penalty0 (3):\penalty0
  21:1--21:32, 2012{\natexlab{a}}.

\bibitem[Baader et~al.(2012{\natexlab{b}})Baader, Knechtel, and
  Pe{\~n}aloza]{BaKP12}
Franz Baader, Martin Knechtel, and Rafael Pe{\~n}aloza.
\newblock Context-dependent views to axioms and consequences of semantic web
  ontologies.
\newblock \emph{John Wiley \& Sons}, 12--13:\penalty0 22--40,
  2012{\natexlab{b}}.

\bibitem[Bagheri~Hariri et~al.(2011{\natexlab{a}})Bagheri~Hariri, Calvanese,
  De~Giacomo, and De~Masellis]{BCDD11}
Babak Bagheri~Hariri, Diego Calvanese, Giuseppe De~Giacomo, and Riccardo
  De~Masellis.
\newblock Verification of conjunctive-query based semantic artifacts.
\newblock In \emph{Proc.\ of the 24th Int.\ Workshop on Description Logic
  (DL)}, volume 745 of \emph{CEUR Electronic Workshop Proceedings,
  {\upshape\protect\url{http://ceur-ws.org/}}}, 2011{\natexlab{a}}.

\bibitem[Bagheri~Hariri et~al.(2011{\natexlab{b}})Bagheri~Hariri, Calvanese,
  De~Giacomo, De~Masellis, and Felli]{BCDDF11}
Babak Bagheri~Hariri, Diego Calvanese, Giuseppe De~Giacomo, Riccardo
  De~Masellis, and Paolo Felli.
\newblock Foundations of relational artifacts verification.
\newblock In \emph{Proc.\ of the 9th Int.\ Conf.\ on Business Process
  Management (BPM)}, volume 6896 of \emph{Lecture Notes in Computer Science},
  pages 379--395. Springer, 2011{\natexlab{b}}.

\bibitem[Bagheri~Hariri et~al.(2012{\natexlab{a}})Bagheri~Hariri, Calvanese,
  De~Giacomo, De~Masellis, Montali, and Felli]{BCDD*12}
Babak Bagheri~Hariri, Diego Calvanese, Giuseppe De~Giacomo, Riccardo
  De~Masellis, Marco Montali, and Paolo Felli.
\newblock Verification of description logic {K}nowledge and {A}ction {B}ases.
\newblock In \emph{Proc.\ of the 20th Eur.\ Conf.\ on Artificial Intelligence
  (ECAI)}, pages 103--108, 2012{\natexlab{a}}.

\bibitem[Bagheri~Hariri et~al.(2012{\natexlab{b}})Bagheri~Hariri, Calvanese,
  De~Giacomo, Deutsch, and Montali]{BCDDM12}
Babak Bagheri~Hariri, Diego Calvanese, Giuseppe De~Giacomo, Alin Deutsch, and
  Marco Montali.
\newblock Verification of relational data-centric dynamic systems with external
  services.
\newblock CoRR Technical Report arXiv:1203.0024, arXiv.org e-Print archive,
  2012{\natexlab{b}}.
\newblock Available at \protect\url{http://arxiv.org/abs/1203.0024}.

\bibitem[Bagheri~Hariri et~al.(2013{\natexlab{a}})Bagheri~Hariri, Calvanese,
  De~Giacomo, Deutsch, and Montali]{BCDDM13}
Babak Bagheri~Hariri, Diego Calvanese, Giuseppe De~Giacomo, Alin Deutsch, and
  Marco Montali.
\newblock Verification of relational data-centric dynamic systems with external
  services.
\newblock In \emph{Proc.\ of the 32nd ACM SIGACT SIGMOD SIGAI Symp.\ on
  Principles of Database Systems (PODS)}, pages 163--174, 2013{\natexlab{a}}.

\bibitem[Bagheri~Hariri et~al.(2013{\natexlab{b}})Bagheri~Hariri, Calvanese,
  Montali, Santoso, and Solomakhin]{AS-CORR-13b}
Babak Bagheri~Hariri, Diego Calvanese, Marco Montali, Ario Santoso, and Dmitry
  Solomakhin.
\newblock Verification of semantically-enhanced artifact systems (extended
  version).
\newblock CoRR Technical Report arXiv:1308.6292, arXiv.org e-Print archive,
  2013{\natexlab{b}}.
\newblock Available at \protect\url{http://arxiv.org/abs/1308.6292}.

\bibitem[Bagheri~Hariri et~al.(2013{\natexlab{c}})Bagheri~Hariri, Calvanese,
  Montali, Santoso, and Solomakhin]{AS-ICSOC-13}
Babak Bagheri~Hariri, Diego Calvanese, Marco Montali, Ario Santoso, and Dmitry
  Solomakhin.
\newblock Verification of semantically-enhanced artifact systems.
\newblock In \emph{Proc.\ of the 11th Int.\ Joint Conf.\ on Service Oriented
  Computing (ICSOC)}, volume 8274 of \emph{Lecture Notes in Computer Science},
  pages 600--607. Springer, 2013{\natexlab{c}}.

\bibitem[Baier and Katoen(2008)]{BaKa08}
Christel Baier and Joost-Pieter Katoen.
\newblock \emph{Principles of Model Checking}.
\newblock The MIT Press, 2008.

\bibitem[Baier et~al.(2007)Baier, Fritz, and McIlraith]{BFM07}
Jorge~A. Baier, Christian Fritz, and Sheila~A. McIlraith.
\newblock Exploiting procedural {Domain Control Knowledge} in state-of-the-art
  planners.
\newblock In \emph{Proc.\ of the 17th Int.\ Conf.\ on Automated Planning and
  Scheduling (ICAPS)}, pages 26--33, 2007.

\bibitem[Bancilhon and Spyratos(1981)]{BaSp81}
Fran\c{c}ois Bancilhon and Nicolas Spyratos.
\newblock Update semantics of relational views.
\newblock \emph{ACM Trans.\ on Database Systems}, 6\penalty0 (4):\penalty0
  557--575, 1981.

\bibitem[Belardinelli et~al.(2012{\natexlab{a}})Belardinelli, Lomuscio, and
  Patrizi]{BeLP12}
Francesco Belardinelli, Alessio Lomuscio, and Fabio Patrizi.
\newblock An abstraction technique for the verification of artifact-centric
  systems.
\newblock In \emph{Proc.\ of the 13th Int.\ Conf.\ on the Principles of
  Knowledge Representation and Reasoning (KR)}, pages 319--328,
  2012{\natexlab{a}}.

\bibitem[Belardinelli et~al.(2012{\natexlab{b}})Belardinelli, Lomuscio, and
  Patrizi]{BeLP12b}
Francesco Belardinelli, Alessio Lomuscio, and Fabio Patrizi.
\newblock Verification of {GSM}-based artifact-centric systems through finite
  abstraction.
\newblock In \emph{Proc.\ of the 10th Int.\ Joint Conf.\ on Service Oriented
  Computing (ICSOC)}, volume 7636 of \emph{Lecture Notes in Computer Science},
  pages 17--31. Springer, 2012{\natexlab{b}}.
\newblock \doi{10.1007/978-3-642-34321-6_2}.

\bibitem[Belardinelli et~al.(2014)Belardinelli, Lomuscio, and Patrizi]{BeLP14}
Francesco Belardinelli, Alessio Lomuscio, and Fabio Patrizi.
\newblock Verification of agent-based artifact systems.
\newblock \emph{J.\ of Artificial Intelligence Research}, 51:\penalty0
  333--376, 2014.
\newblock \doi{10.1613/jair.4424}.

\bibitem[Belardinelli et~al.(2015)Belardinelli, Grossi, and Lomuscio]{BGL15}
Francesco Belardinelli, Davide Grossi, and Alessio Lomuscio.
\newblock Finite abstractions for the verification of epistemic properties in
  open multi-agent systems.
\newblock In \emph{Proc.\ of the 24th Int.\ Joint Conf.\ on Artificial
  Intelligence (IJCAI)}, pages 854--860, 2015.

\bibitem[Berardi et~al.(2005)Berardi, Calvanese, and De~Giacomo]{BeCD05}
Daniela Berardi, Diego Calvanese, and Giuseppe De~Giacomo.
\newblock Reasoning on {UML} class diagrams.
\newblock \emph{Artificial Intelligence}, 168\penalty0 (1--2):\penalty0
  70--118, 2005.

\bibitem[Bertossi(2006)]{Bert06}
Leopoldo~E. Bertossi.
\newblock Consistent query answering in databases.
\newblock \emph{SIGMOD Record}, 35\penalty0 (2):\penalty0 68--76, 2006.

\bibitem[Bhattacharya et~al.(2007{\natexlab{a}})Bhattacharya, Caswell, Kumaran,
  Nigam, and Wu]{BCKNW07}
K.~Bhattacharya, N.~S. Caswell, S.~Kumaran, A.~Nigam, and F.~Y. Wu.
\newblock Artifact-centered operational modeling: {L}essons from customer
  engagements.
\newblock \emph{IBM Systems J.}, 46\penalty0 (4):\penalty0 703--721,
  2007{\natexlab{a}}.

\bibitem[Bhattacharya et~al.(2007{\natexlab{b}})Bhattacharya, Gerede, Hull,
  Liu, and Su]{BGHLS07}
K.~Bhattacharya, C.~Gerede, R.~Hull, R.~Liu, and J.~Su.
\newblock Towards formal analysis of artifact-centric business process models.
\newblock In \emph{Proc.\ of the 5th Int.\ Conf.\ on Business Process
  Management (BPM)}, volume 4714 of \emph{Lecture Notes in Computer Science},
  pages 288--234. Springer, 2007{\natexlab{b}}.

\bibitem[Bhattacharya et~al.(2005)Bhattacharya, Guttman, Lyman, Heath, Kumaran,
  Nandi, Wu, Athma, Freiberg, Johannsen, and Staudt]{BGLH*05}
Kamal Bhattacharya, Robert Guttman, Kelly Lyman, Fenno~F. Heath, Santhosh
  Kumaran, Prabir Nandi, Frederick~Y. Wu, Prasanna Athma, Christoph Freiberg,
  Lars Johannsen, and Andreas Staudt.
\newblock A model-driven approach to industrializing discovery processes in
  pharmaceutical research.
\newblock \emph{IBM Systems J.}, 44\penalty0 (1):\penalty0 145--162, 2005.

\bibitem[Bienvenu(2012)]{Bien12}
Meghyn Bienvenu.
\newblock On the complexity of consistent query answering in the presence of
  simple ontologies.
\newblock In \emph{Proc.\ of the 26th AAAI Conf.\ on Artificial Intelligence
  (AAAI)}, pages 705--711, 2012.

\bibitem[Bienvenu and Rosati(2015)]{BiRo15}
Meghyn Bienvenu and Riccardo Rosati.
\newblock Query-based comparison of {OBDA} specifications.
\newblock In \emph{Proc.\ of the 28th Int.\ Workshop on Description Logic
  (DL)}, volume 1350 of \emph{CEUR Electronic Workshop Proceedings,
  {\upshape\protect\url{http://ceur-ws.org/}}}, 2015.
\newblock URL \url{http://ceur-ws.org/Vol-1350/paper-11.pdf}.

\bibitem[B\"orger(2012)]{B12}
Egon B\"orger.
\newblock Approaches to modeling business processes: a critical analysis of
  {BPMN}, workflow patterns and {YAWL}.
\newblock \emph{J.\ of Software and Systems Modeling}, 11:\penalty0 305--318,
  2012.

\bibitem[Bozzato et~al.(2013)Bozzato, Ghidini, and Serafini]{BoGS13}
Loris Bozzato, Chiara Ghidini, and Luciano Serafini.
\newblock Comparing contextual and flat representations of knowledge: a
  concrete case about football data.
\newblock In \emph{Proc.\ of the 7th Int.\ Conf.\ on Knowledge Capture
  (K-CAP)}, pages 9--16. {ACM} Press, 2013.

\bibitem[Bradfield and Stirling(2007)]{BrSt07}
Julian Bradfield and Colin Stirling.
\newblock Modal mu-calculi.
\newblock In \emph{Handbook of Modal Logic}, volume~3, pages 721--756.
  Elsevier, 2007.

\bibitem[Cadoli et~al.(1996)Cadoli, Palopoli, and Lenzerini]{CaLP96}
Marco Cadoli, Luigi Palopoli, and Maurizio Lenzerini.
\newblock Datalog and description logics: {E}xpressive power. {P}reliminary
  report.
\newblock In \emph{Proc.\ of the 9th Int.\ Workshop on Description Logic (DL)},
  number WS-96-05 in AAAI Technical Report. AAAI Press, 1996.

\bibitem[Cal\`{\i} et~al.(2010)Cal\`{\i}, Gottlob, Lukasiewicz, Marnette, and
  Pieris]{CGLMP10}
Andrea Cal\`{\i}, Georg Gottlob, Thomas Lukasiewicz, Bruno Marnette, and
  Andreas Pieris.
\newblock {Datalog+/-}: A family of logical knowledge representation and query
  languages for new applications.
\newblock In \emph{Proc.\ of the 25th IEEE Symp.\ on Logic in Computer Science
  (LICS)}, pages 228--242, 2010.

\bibitem[Calvanese et~al.(2013{\natexlab{a}})Calvanese, Giese, Haase, Horrocks,
  Hubauer, Ioannidis, Jim{\'e}nez-Ruiz, Kharlamov, Kllapi, Kl{\"u}wer,
  Koubarakis, Lamparter, M{\"o}ller, Neuenstadt, Nordtveit, {\"O}zcep,
  Rodriguez-Muro, Roshchin, Savo, Schmidt, Soylu, Waaler, and
  Zheleznyakov]{CGHH*13}
D.~Calvanese, M.~Giese, P.~Haase, I.~Horrocks, T.~Hubauer, Y.~Ioannidis,
  E.~Jim{\'e}nez-Ruiz, E.~Kharlamov, H.~Kllapi, J.~Kl{\"u}wer, M.~Koubarakis,
  S.~Lamparter, R.~M{\"o}ller, C.~Neuenstadt, T.~Nordtveit, {\"O}.~{\"O}zcep,
  M.~Rodriguez-Muro, M.~Roshchin, F.~Savo, M.~Schmidt, A.~Soylu, A.~Waaler, and
  D.~Zheleznyakov.
\newblock {O}ptique: {OBDA} solution for big data.
\newblock In \emph{Revised Selected Papers of ESWC~2013 Satellite Events},
  volume 7955 of \emph{Lecture Notes in Computer Science}, pages 293--295.
  Springer, 2013{\natexlab{a}}.
\newblock ISBN 978-3-642-41241-7.
\newblock \doi{10.1007/978-3-642-41242-4_48}.

\bibitem[Calvanese et~al.(2013{\natexlab{b}})Calvanese, Giese, Haase, Horrocks,
  Hubauer, Ioannidis, Jim{\'e}nez-Ruiz, Kharlamov, Kllapi, Kl{\"u}wer,
  Koubarakis, Lamparter, M{\"o}ller, Neuenstadt, Nordtveit, {\"O}zcep,
  Rodriguez-Muro, Roshchin, Ruzzi, Savo, Schmidt, Soylu, Waaler, and
  Zheleznyakov]{CGHH*13b}
D.~Calvanese, M.~Giese, P.~Haase, I.~Horrocks, T.~Hubauer, Y.~Ioannidis,
  E.~Jim{\'e}nez-Ruiz, E.~Kharlamov, H.~Kllapi, J.~Kl{\"u}wer, M.~Koubarakis,
  S.~Lamparter, R.~M{\"o}ller, C.~Neuenstadt, T.~Nordtveit, {\"O}.~{\"O}zcep,
  M.~Rodriguez-Muro, M.~Roshchin, M.~Ruzzi, F.~Savo, M.~Schmidt, A.~Soylu,
  A.~Waaler, and D.~Zheleznyakov.
\newblock The {O}ptique project: {T}owards {OBDA} systems for industry ({S}hort
  paper).
\newblock In \emph{Proc.\ of the 10th Int.\ Workshop on OWL: Experiences and
  Directions (OWLED)}, volume 1080 of \emph{CEUR Electronic Workshop
  Proceedings, {\upshape\protect\url{http://ceur-ws.org/}}},
  2013{\natexlab{b}}.

\bibitem[Calvanese and Santoso(2012)]{AS-ZEUS-12}
Diego Calvanese and Ario Santoso.
\newblock Best service synthesis in the weighted roman model.
\newblock In \emph{Proc. of the 4th Central-European Workshop on Services and
  their Composition (ZEUS 2012)}, volume 847 of \emph{CEUR Electronic Workshop
  Proceedings, {\upshape\protect\url{http://ceur-ws.org/}}}, pages 42--49,
  2012.

\bibitem[Calvanese et~al.(2007{\natexlab{a}})Calvanese, De~Giacomo, Lembo,
  Lenzerini, Poggi, and Rosati]{CDLL*07b}
Diego Calvanese, Giuseppe De~Giacomo, Domenico Lembo, Maurizio Lenzerini,
  Antonella Poggi, and Riccardo Rosati.
\newblock Ontology-based database access.
\newblock In \emph{Proc.\ of the 15th Ital.\ Conf.\ on Database Systems
  (SEBD)}, pages 324--331, 2007{\natexlab{a}}.

\bibitem[Calvanese et~al.(2007{\natexlab{b}})Calvanese, De~Giacomo, Lembo,
  Lenzerini, and Rosati]{CDLLR07}
Diego Calvanese, Giuseppe De~Giacomo, Domenico Lembo, Maurizio Lenzerini, and
  Riccardo Rosati.
\newblock Tractable reasoning and efficient query answering in description
  logics: The \textit{DL-Lite} family.
\newblock \emph{J.\ of Automated Reasoning}, 39\penalty0 (3):\penalty0
  385--429, 2007{\natexlab{b}}.

\bibitem[Calvanese et~al.(2007{\natexlab{c}})Calvanese, De~Giacomo, Lembo,
  Lenzerini, and Rosati]{CDLLR07b}
Diego Calvanese, Giuseppe De~Giacomo, Domenico Lembo, Maurizio Lenzerini, and
  Riccardo Rosati.
\newblock {EQL-Lite}: {E}ffective first-order query processing in description
  logics.
\newblock In \emph{Proc.\ of the 20th Int.\ Joint Conf.\ on Artificial
  Intelligence (IJCAI)}, pages 274--279, 2007{\natexlab{c}}.

\bibitem[Calvanese et~al.(2008)Calvanese, De~Giacomo, Lembo, Lenzerini, and
  Rosati]{CDLLR08b}
Diego Calvanese, Giuseppe De~Giacomo, Domenico Lembo, Maurizio Lenzerini, and
  Riccardo Rosati.
\newblock Path-based identification constraints in description logics.
\newblock In \emph{Proc.\ of the 11th Int.\ Conf.\ on the Principles of
  Knowledge Representation and Reasoning (KR)}, pages 231--241, 2008.

\bibitem[Calvanese et~al.(2009)Calvanese, De~Giacomo, Lembo, Lenzerini, Poggi,
  Rodr{\'\i}guez-Muro, and Rosati]{CDLL*09}
Diego Calvanese, Giuseppe De~Giacomo, Domenico Lembo, Maurizio Lenzerini,
  Antonella Poggi, Mariano Rodr{\'\i}guez-Muro, and Riccardo Rosati.
\newblock Ontologies and databases: The \textit{DL-Lite} approach.
\newblock In Sergio Tessaris and Enrico Franconi, editors, \emph{Reasoning Web.
  Semantic Technologies for Informations Systems -- 5th Int.\ Summer School
  Tutorial Lectures (RW)}, volume 5689 of \emph{Lecture Notes in Computer
  Science}, pages 255--356. Springer, 2009.

\bibitem[Calvanese et~al.(2010)Calvanese, Kharlamov, Nutt, and
  Zheleznyakov]{CKNZ10b}
Diego Calvanese, Evgeny Kharlamov, Werner Nutt, and Dmitriy Zheleznyakov.
\newblock Evolution of \textit{DL-Lite} knowledge bases.
\newblock In \emph{Proc.\ of the 9th Int.\ Semantic Web Conf.\ (ISWC)}, volume
  6496 of \emph{Lecture Notes in Computer Science}, pages 112--128. Springer,
  2010.

\bibitem[Calvanese et~al.(2011{\natexlab{a}})Calvanese, De~Giacomo, Lenzerini,
  and Rosati]{CDLR11}
Diego Calvanese, Giuseppe De~Giacomo, Maurizio Lenzerini, and Riccardo Rosati.
\newblock Actions and programs over description logic knowledge bases: {A}
  functional approach.
\newblock In Gerhard Lakemeyer and Sheila~A. McIlraith, editors, \emph{Knowing,
  Reasoning, and Acting: {E}ssays in Honour of {H}ector {L}evesque}. College
  Publications, 2011{\natexlab{a}}.

\bibitem[Calvanese et~al.(2011{\natexlab{b}})Calvanese, Giacomo, Lembo,
  Lenzerini, Poggi, Rodriguez-Muro, Rosati, Ruzzi, and Savo]{CDLLPRRRS11}
Diego Calvanese, Giuseppe~De Giacomo, Domenico Lembo, Maurizio Lenzerini,
  Antonella Poggi, Mariano Rodriguez-Muro, Riccardo Rosati, Marco Ruzzi, and
  Domenico~Fabio Savo.
\newblock The {Mastro} system for ontology-based data access.
\newblock \emph{Semantic Web J.}, 2\penalty0 (1):\penalty0 43--53,
  2011{\natexlab{b}}.
\newblock Listed among the 5 most cited papers in the first five years of the
  Semantic Web Journal.

\bibitem[Calvanese et~al.(2012{\natexlab{a}})Calvanese, De~Giacomo, Lembo,
  Montali, and Santoso]{AS-KiBP-12}
Diego Calvanese, Giuseppe De~Giacomo, Domenico Lembo, Marco Montali, and Ario
  Santoso.
\newblock Semantically-governed data-aware processes.
\newblock In \emph{Proc. of the 1st Int. Workshop on Knowledge-intensive
  Business Processes (KiBP 2012)}, volume 861 of \emph{CEUR Electronic Workshop
  Proceedings, {\upshape\protect\url{http://ceur-ws.org/}}}, pages 21--32,
  2012{\natexlab{a}}.

\bibitem[Calvanese et~al.(2012{\natexlab{b}})Calvanese, De~Giacomo, Lembo,
  Montali, and Santoso]{AS-RR-12a}
Diego Calvanese, Giuseppe De~Giacomo, Domenico Lembo, Marco Montali, and Ario
  Santoso.
\newblock Ontology-based governance of data-aware processes.
\newblock In \emph{Proc.\ of the 6th Int.\ Conf.\ on Web Reasoning and Rule
  Systems (RR)}, volume 7497 of \emph{Lecture Notes in Computer Science}, pages
  25--41. Springer, 2012{\natexlab{b}}.

\bibitem[Calvanese et~al.(2012{\natexlab{c}})Calvanese, Kharlamov, Montali, and
  Zheleznyakov]{CKMZ12}
Diego Calvanese, Evgeny Kharlamov, Marco Montali, and Dmitriy Zheleznyakov.
\newblock Inconsistency tolerance in {OWL 2 QL} knowledge and action bases -
  statement of interest.
\newblock In \emph{Proc.\ of the 9th Int.\ Workshop on OWL: Experiences and
  Directions (OWLED)}, volume 849 of \emph{CEUR Electronic Workshop
  Proceedings, {\upshape\protect\url{http://ceur-ws.org/}}},
  2012{\natexlab{c}}.

\bibitem[Calvanese et~al.(2013{\natexlab{c}})Calvanese, Bagheri~Hariri,
  De~Masellis, Lembo, Montali, Santoso, Solomakhin, and Tessaris]{ACSI-D2.4.2}
Diego Calvanese, Babak Bagheri~Hariri, Riccardo De~Masellis, Domenico Lembo,
  Marco Montali, Ario Santoso, Dimitry Solomakhin, and Sergio Tessaris.
\newblock Techniques and tools for {KAB}, to manage action linkage with the
  {A}rtifact {L}ayer -- {I}teration~2.
\newblock Deliverable ACSI-D2.4.2, ACSI Consortium, May 2013{\natexlab{c}}.

\bibitem[Calvanese et~al.(2013{\natexlab{d}})Calvanese, De~Giacomo, Lembo,
  Lenzerini, and Rosati]{CDLLR13}
Diego Calvanese, Giuseppe De~Giacomo, Domenico Lembo, Maurizio Lenzerini, and
  Riccardo Rosati.
\newblock Data complexity of query answering in description logics.
\newblock \emph{Artificial Intelligence}, 195:\penalty0 335--360,
  2013{\natexlab{d}}.
\newblock \doi{10.1016/j.artint.2012.10.003}.

\bibitem[Calvanese et~al.(2013{\natexlab{e}})Calvanese, Giacomo, and
  Montali]{CDM13}
Diego Calvanese, Giuseppe~De Giacomo, and Marco Montali.
\newblock Foundations of data-aware process analysis: A database theory
  perspective.
\newblock In \emph{Proc.\ of the 32nd ACM SIGACT SIGMOD SIGAI Symp.\ on
  Principles of Database Systems (PODS)}, pages 1--12, 2013{\natexlab{e}}.

\bibitem[Calvanese et~al.(2013{\natexlab{f}})Calvanese, Giacomo, Montali, and
  Patrizi]{CDMP13}
Diego Calvanese, Giuseppe~De Giacomo, Marco Montali, and Fabio Patrizi.
\newblock Verification and synthesis in description logic based dynamic
  systems.
\newblock In \emph{Proc.\ of the 7th Int.\ Conf.\ on Web Reasoning and Rule
  Systems (RR)}, volume 7994 of \emph{Lecture Notes in Computer Science}, pages
  50--64. Springer, 2013{\natexlab{f}}.

\bibitem[Calvanese et~al.(2013{\natexlab{g}})Calvanese, Kharlamov, Montali,
  Santoso, and Zheleznyakov]{AS-CORR-13a}
Diego Calvanese, Evgeny Kharlamov, Marco Montali, Ario Santoso, and Dmitriy
  Zheleznyakov.
\newblock Verification of inconsistency-aware knowledge and action bases
  (extended version).
\newblock CoRR Technical Report arXiv:1304.6442, arXiv.org e-Print archive,
  2013{\natexlab{g}}.
\newblock Available at \protect\url{http://arxiv.org/abs/1304.6442}.

\bibitem[Calvanese et~al.(2013{\natexlab{h}})Calvanese, Kharlamov, Montali,
  Santoso, and Zheleznyakov]{AS-DL-13}
Diego Calvanese, Evgeny Kharlamov, Marco Montali, Ario Santoso, and Dmitriy
  Zheleznyakov.
\newblock Verification of inconsistency-aware knowledge and action bases.
\newblock In \emph{Proc.\ of the 26th Int.\ Workshop on Description Logic
  (DL)}, volume 1014 of \emph{CEUR Electronic Workshop Proceedings,
  {\upshape\protect\url{http://ceur-ws.org/}}}, pages 107--119,
  2013{\natexlab{h}}.

\bibitem[Calvanese et~al.(2013{\natexlab{i}})Calvanese, Kharlamov, Montali,
  Santoso, and Zheleznyakov]{AS-IJCAI-13}
Diego Calvanese, Evgeny Kharlamov, Marco Montali, Ario Santoso, and Dmitriy
  Zheleznyakov.
\newblock Verification of inconsistency-aware knowledge and action bases.
\newblock In \emph{Proc.\ of the 23rd Int.\ Joint Conf.\ on Artificial
  Intelligence (IJCAI)}, pages 810--816. AAAI Press, 2013{\natexlab{i}}.

\bibitem[Calvanese et~al.(2014{\natexlab{a}})Calvanese, Ceylan, Montali, and
  Santoso]{AS-ARCOE-14}
Diego Calvanese, {\.I}smail~{\.I}lkan Ceylan, Marco Montali, and Ario Santoso.
\newblock Adding context to knowledge and action bases.
\newblock In \emph{Workshop Notes of the 6th Int. Workshop on Acquisition,
  Representation and Reasoning about Context with Logic (ARCOE-Logic 2014)},
  volume arXiv:1412.7965 of \emph{CoRR Technical Report}, pages 25--36.
  arXiv.org e-Print archive, 2014{\natexlab{a}}.
\newblock Available at \protect\url{http://arxiv.org/abs/1412.7965}.

\bibitem[Calvanese et~al.(2014{\natexlab{b}})Calvanese, Ceylan, Montali, and
  Santoso]{AS-JELIA-14}
Diego Calvanese, {\.I}smail~{\.I}lkan Ceylan, Marco Montali, and Ario Santoso.
\newblock Verification of context-sensitive knowledge and action bases.
\newblock In \emph{Proc.\ of the 14th Eur.\ Conf.\ on Logics in Artificial
  Intelligence (JELIA)}, volume 8761 of \emph{Lecture Notes in Computer
  Science}, pages 514--528. Springer, 2014{\natexlab{b}}.

\bibitem[Calvanese et~al.(2014{\natexlab{c}})Calvanese, Lanti, Rezk, Slusnys,
  and Xiao]{CLRSX14}
Diego Calvanese, Davide Lanti, Martin Rezk, Mindaugas Slusnys, and Guohui Xiao.
\newblock A scalable benchmark for {OBDA} systems: {P}reliminary report.
\newblock In \emph{Proc.\ of the 3rd Int.\ Workshop on OWL Reasoner Evaluation
  (ORE)}, volume 1207 of \emph{CEUR Electronic Workshop Proceedings,
  {\upshape\protect\url{http://ceur-ws.org/}}}, 2014{\natexlab{c}}.

\bibitem[Calvanese et~al.(2015{\natexlab{a}})Calvanese, Cogrel, Kalayci, Ebri,
  Kontchakov, Lanti, Rezk, Rodriguez-Muro, and Xiao]{CCKEKLRRX15}
Diego Calvanese, Benjamin Cogrel, Elem~Guzel Kalayci, Sarah~Komla Ebri, Roman
  Kontchakov, Davide Lanti, Martin Rezk, Mariano Rodriguez-Muro, and Guohui
  Xiao.
\newblock {OBDA} with the ontop framework.
\newblock In \emph{Proc.\ of the 23th Ital.\ Symp.\ on Advanced Database
  Systems (SEBD)}, 2015{\natexlab{a}}.

\bibitem[Calvanese et~al.(2015{\natexlab{b}})Calvanese, Giacomo, Montali, and
  Patrizi]{CDMP15}
Diego Calvanese, Giuseppe~De Giacomo, Marco Montali, and Fabio Patrizi.
\newblock Description logic based dynamic systems: Modeling, verification, and
  synthesis.
\newblock In \emph{Proc.\ of the 24th Int.\ Joint Conf.\ on Artificial
  Intelligence (IJCAI)}, pages 4247--4253, 2015{\natexlab{b}}.

\bibitem[Calvanese et~al.(2015{\natexlab{c}})Calvanese, Giacomo, and
  Soutchanski]{CDS15}
Diego Calvanese, Giuseppe~De Giacomo, and Mikhail Soutchanski.
\newblock On the undecidability of the situation calculus extended with
  description logic ontologies.
\newblock In \emph{Proc.\ of the 24th Int.\ Joint Conf.\ on Artificial
  Intelligence (IJCAI)}, pages 2840--2846, 2015{\natexlab{c}}.

\bibitem[Calvanese et~al.(2015{\natexlab{d}})Calvanese, Montali, Patrizi, and
  Rivkin]{CMPR15}
Diego Calvanese, Marco Montali, Fabio Patrizi, and Andrey Rivkin.
\newblock Leveraging relational technology for data-centric dynamic systems.
\newblock In \emph{Proc.\ of the 23th Ital.\ Symp.\ on Advanced Database
  Systems (SEBD)}, 2015{\natexlab{d}}.

\bibitem[Calvanese et~al.(2015{\natexlab{e}})Calvanese, Montali, Patrizi, and
  Rivkin]{CMPR15a}
Diego Calvanese, Marco Montali, Fabio Patrizi, and Andrey Rivkin.
\newblock Implementing data-centric dynamic systems over a relational dbms.
\newblock In \emph{Proc.\ of the 9th Alberto Mendelzon Int.\ Workshop on
  Foundations of Data Management (AMW)}, volume 1378 of \emph{CEUR Electronic
  Workshop Proceedings, {\upshape\protect\url{http://ceur-ws.org/}}},
  2015{\natexlab{e}}.

\bibitem[Calvanese et~al.(2015{\natexlab{f}})Calvanese, Montali, and
  Santoso]{AS-CORR-15}
Diego Calvanese, Marco Montali, and Ario Santoso.
\newblock Verification of generalized inconsistency-aware knowledge and action
  bases (extended version).
\newblock CoRR Technical Report arXiv:1504.08108, arXiv.org e-Print archive,
  2015{\natexlab{f}}.
\newblock Available at \protect\url{http://arxiv.org/abs/1504.08108}.

\bibitem[Calvanese et~al.(2015{\natexlab{g}})Calvanese, Montali, and
  Santoso]{AS-DL-15}
Diego Calvanese, Marco Montali, and Ario Santoso.
\newblock Inconsistency management in generalized knowledge and action bases.
\newblock In \emph{Proc.\ of the 28th Int.\ Workshop on Description Logic
  (DL)}, volume 1350, 2015{\natexlab{g}}.

\bibitem[Calvanese et~al.(2015{\natexlab{h}})Calvanese, Montali, and
  Santoso]{AS-IJCAI-15}
Diego Calvanese, Marco Montali, and Ario Santoso.
\newblock Verification of generalized inconsistency-aware knowledge and action
  bases.
\newblock In \emph{Proc.\ of the 24th Int.\ Joint Conf.\ on Artificial
  Intelligence (IJCAI)}, pages 2847--2853. AAAI Press, 2015{\natexlab{h}}.

\bibitem[Calvanese et~al.(2016)Calvanese, Montali, Patrizi, and
  Stawowy]{CMPS16}
Diego Calvanese, Marco Montali, Fabio Patrizi, and Michele Stawowy.
\newblock Plan synthesis for knowledge and action bases.
\newblock In \emph{Proc.\ of the 25th Int.\ Joint Conf.\ on Artificial
  Intelligence (IJCAI)}. AAAI Press, 2016.
\newblock To appear.

\bibitem[Cangialosi et~al.(2010)Cangialosi, De~Giacomo, De~Masellis, and
  Rosati]{CDDR10}
Piero Cangialosi, Giuseppe De~Giacomo, Riccardo De~Masellis, and Riccardo
  Rosati.
\newblock Conjunctive artifact-centric services.
\newblock In \emph{Proc.\ of the 8th Int.\ Joint Conf.\ on Service Oriented
  Computing (ICSOC)}, volume 6470 of \emph{Lecture Notes in Computer Science},
  pages 318--333. Springer, 2010.

\bibitem[Ceylan and Pe{\~n}aloza(2014)]{CePe14}
{\.I}smail~{\.I}lkan Ceylan and Rafael Pe{\~n}aloza.
\newblock The {B}ayesian description logic $\mathcal{BEL}$.
\newblock In \emph{Proc.\ of the 7th Int.\ Joint Conf.\ on Automated Reasoning
  (IJCAR)}, volume 8562 of \emph{Lecture Notes in Computer Science}, pages
  480--494. Springer, 2014.

\bibitem[Chao et~al.(2009)Chao, Cohn, Flatgard, Hahn, Linehan, Nandi, Nigam,
  Pinel, Vergo, and y~Wu]{CCFHLNNPVW09}
Tian Chao, David Cohn, Adrian Flatgard, Sandy Hahn, Mark Linehan, Prabir Nandi,
  Anil Nigam, Florian Pinel, John Vergo, and Frederick y~Wu.
\newblock Artifact-based transformation of {IBM} {G}lobal {F}inancing, a case
  study.
\newblock In \emph{Proc.\ of the 7th Int.\ Conf.\ on Business Process
  Management (BPM)}, Lecture Notes in Computer Science. Springer, 2009.

\bibitem[Clarke and Wing(1996)]{CW96}
Edmund~M. Clarke and Jeannette~M. Wing.
\newblock Formal methods: State of the art and future directions.
\newblock \emph{ACM Computing Surveys}, 28\penalty0 (4):\penalty0 626--643,
  1996.

\bibitem[Clarke et~al.(1999)Clarke, Grumberg, and Peled]{ClGP99}
Edmund~M. Clarke, Orna Grumberg, and Doron~A. Peled.
\newblock \emph{Model checking}.
\newblock The MIT Press, Cambridge, MA, USA, 1999.

\bibitem[Cla{\ss}en and Lakemeyer(2008)]{ClLa08}
Jens Cla{\ss}en and Gerhard Lakemeyer.
\newblock A logic for non-terminating {Golog} programs.
\newblock In \emph{Proc.\ of the 11th Int.\ Conf.\ on the Principles of
  Knowledge Representation and Reasoning (KR)}, pages 589--599, 2008.

\bibitem[Cohn and Hull(2009)]{CoHu09}
David Cohn and Richard Hull.
\newblock Business artifacts: A data-centric approach to modeling business
  operations and processes.
\newblock \emph{Bull.\ of the IEEE Computer Society Technical Committee on Data
  Engineering}, 32\penalty0 (3):\penalty0 3--9, 2009.

\bibitem[Dam(1992)]{Dam92}
Mads Dam.
\newblock {CTL*} and {ECTL*} as fragments of the modal $\mu$-calculus.
\newblock In \emph{Proc.\ of the 17th Colloquium on Trees in Algebra and
  Programming (CAAP'92)}, volume 581 of \emph{Lecture Notes in Computer
  Science}, pages 145--164. Springer, 1992.

\bibitem[Damaggio et~al.(2013)Damaggio, Hull, and Vacul\'{\i}n]{DaHV13}
Elio Damaggio, Richard Hull, and Roman Vacul\'{\i}n.
\newblock On the equivalence of incremental and fixpoint semantics for business
  artifacts with {G}uard-{S}tage-{M}ilestone lifecycles.
\newblock \emph{Information Systems}, 38\penalty0 (4):\penalty0 561--584, 2013.

\bibitem[Dayal and Bernstein(1982)]{DaBe82}
Umeshwar Dayal and Philip~A. Bernstein.
\newblock On the correct translation of update operations on relational views.
\newblock \emph{ACM Trans.\ on Database Systems}, 7\penalty0 (3):\penalty0
  381--416, September 1982.
\newblock ISSN 0362-5915.

\bibitem[De~Giacomo et~al.(1997)De~Giacomo, Lesp{\'e}rance, and
  Levesque]{DeLL97}
Giuseppe De~Giacomo, Yves Lesp{\'e}rance, and Hector~J. Levesque.
\newblock Reasoning about concurrent execution, prioritized interrupts, and
  exogenous actions in the situation calculus.
\newblock In \emph{Proc.\ of the 15th Int.\ Joint Conf.\ on Artificial
  Intelligence (IJCAI)}, pages 1221--1226, 1997.

\bibitem[De~Giacomo et~al.(2000)De~Giacomo, Lesp{\'e}rance, and
  Levesque]{DeLL00}
Giuseppe De~Giacomo, Yves Lesp{\'e}rance, and Hector~J. Levesque.
\newblock {ConGolog}, a concurrent programming language based on the situation
  calculus.
\newblock \emph{Artificial Intelligence}, 121\penalty0 (1--2):\penalty0
  109--169, 2000.

\bibitem[De~Giacomo et~al.(2012)De~Giacomo, Lesperance, and Patrizi]{DeLP12}
Giuseppe De~Giacomo, Yves Lesperance, and Fabio Patrizi.
\newblock Bounded {S}ituation {C}alculus action theories and decidable
  verification.
\newblock In \emph{Proc.\ of the 13th Int.\ Conf.\ on the Principles of
  Knowledge Representation and Reasoning (KR)}, pages 467--477, 2012.

\bibitem[de~Medeiros et~al.(2007)de~Medeiros, Pedrinaci, van~der Aalst,
  Domingue, Song, Rozinat, Norton, and Cabral]{DPVDSRNC07}
A.~de~Medeiros, C.~Pedrinaci, W.~van~der Aalst, J.~Domingue, M.~Song,
  A.~Rozinat, B.~Norton, and L.~Cabral.
\newblock An outlook on semantic business process mining and monitoring.
\newblock In \emph{Proc.\ of On the Move to Meaningful Internet Systems 2007:
  OTM 2007 Workshops}, volume 4806 of \emph{Lecture Notes in Computer Science},
  pages 1244--1255. Springer, 2007.

\bibitem[Deutsch et~al.(2009)Deutsch, Hull, Patrizi, and Vianu]{DHPV09}
Alin Deutsch, Richard Hull, Fabio Patrizi, and Victor Vianu.
\newblock Automatic verification of data-centric business processes.
\newblock In \emph{Proc.\ of the 12th Int.\ Conf.\ on Database Theory (ICDT)},
  pages 252--267, 2009.

\bibitem[Dumas(2012)]{Dum12}
Marlon Dumas.
\newblock Integrated data and process management: Finally?
\newblock In \emph{Proc.\ of the 1st Int.\ Workshop on Knowledge-intensive
  Business Processes (KiBP)}, volume 861 of \emph{CEUR Electronic Workshop
  Proceedings, {\upshape\protect\url{http://ceur-ws.org/}}}, pages 21--32,
  2012.

\bibitem[Ehrig et~al.(2007)Ehrig, Koschmider, and Oberweis]{EKO07}
Marc Ehrig, Agnes Koschmider, and Andreas Oberweis.
\newblock Measuring similarity between semantic business process models.
\newblock In \emph{Proc.\ of the fourth Asia-Pacific conference on Conceptual
  modelling - Volume 67}, pages 71--80. Australian Computer Society, Inc.,
  2007.

\bibitem[Eiter and Gottlob(1992)]{EiGo92}
Thomas Eiter and Georg Gottlob.
\newblock On the complexity of propositional knowledge base revision, updates
  and counterfactuals.
\newblock \emph{Artificial Intelligence}, 57:\penalty0 227--270, 1992.

\bibitem[ElMasri and Navathe(2007)]{ElNa07}
Ramez~A. ElMasri and Shamkant~B. Navathe.
\newblock \emph{Fundamentals of Database Systems}.
\newblock Addison Wesley Publ.\ Co., 5th edition, 2007.

\bibitem[Emerson(1996)]{Emer96b}
E.~Allen Emerson.
\newblock Model checking and the {M}u-calculus.
\newblock In N.~Immerman and P.~Kolaitis, editors, \emph{Proceedings of the
  DIMACS Symposium on Descriptive Complexity and Finite Models}, DIMACS Series
  in Discrete Mathematics and Theoretical Computer Science, pages 185--214.
  American Mathematical Society Press, 1996.
\newblock ISBN~0-8218-0517-7.

\bibitem[Fagin et~al.(2003)Fagin, Kolaitis, Miller, and Popa]{FKMP03}
Ronald Fagin, Phokion~G. Kolaitis, Ren{\'e}e~J. Miller, and Lucian Popa.
\newblock Data exchange: Semantics and query answering.
\newblock In \emph{Proc.\ of the 9th Int.\ Conf.\ on Database Theory (ICDT)},
  pages 207--224, 2003.

\bibitem[Fagin et~al.(2005)Fagin, Kolaitis, Miller, and Popa]{FKMP05}
Ronald Fagin, Phokion~G. Kolaitis, Ren{\'e}e~J. Miller, and Lucian Popa.
\newblock Data exchange: {S}emantics and query answering.
\newblock \emph{Theoretical Computer Science}, 336\penalty0 (1):\penalty0
  89--124, 2005.

\bibitem[Fikes and Kehler(1985)]{FiKe85}
Richard Fikes and Tom Kehler.
\newblock The role of frame-based representation in reasoning.
\newblock \emph{Communications of the {ACM}}, 28\penalty0 (9):\penalty0
  904--920, 1985.

\bibitem[Fikes and Nilsson(1971)]{FN71}
Richard~E. Fikes and Nils~J. Nilsson.
\newblock {STRIPS}: A new approach to the application of theorem proving to
  problem solving.
\newblock \emph{Artificial Intelligence}, 2\penalty0 (3):\penalty0 189 -- 208,
  1971.
\newblock ISSN 0004-3702.

\bibitem[Flouris et~al.(2008)Flouris, Manakanatas, Kondylakis, Plexousakis, and
  Antoniou]{FMKPA08}
Giorgos Flouris, Dimitris Manakanatas, Haridimos Kondylakis, Dimitris
  Plexousakis, and Grigoris Antoniou.
\newblock Ontology change: Classification and survey.
\newblock \emph{Knowledge Engineering Review}, 23\penalty0 (2):\penalty0
  117--152, 2008.

\bibitem[Fowler and Scott(1997)]{FoSc97}
Martin Fowler and Kendall Scott.
\newblock \emph{{UML} Distilled -- {A}pplying the Standard {O}bject {M}odeling
  {L}aguage}.
\newblock Addison Wesley Publ.\ Co., 1997.

\bibitem[Franconi et~al.(2012)Franconi, Mosca, and Solomakhin]{FMS12}
Enrico Franconi, Alessandro Mosca, and Dmitry Solomakhin.
\newblock {ORM2} encoding into description logics.
\newblock In \emph{Proc.\ of the 25th Int.\ Workshop on Description Logic
  (DL)}, volume 846 of \emph{CEUR Electronic Workshop Proceedings,
  {\upshape\protect\url{http://ceur-ws.org/}}}, 2012.

\bibitem[Fritz et~al.(2008)Fritz, Baier, and McIlraith]{FBM08}
Christian Fritz, Jorge~A. Baier, and Sheila~A. McIlraith.
\newblock {ConGolog}, {Sin Trans}: Compiling {ConGolog} into {Basic Action
  Theories} for planning and beyond.
\newblock In \emph{Proc.\ of the 11th Int.\ Conf.\ on the Principles of
  Knowledge Representation and Reasoning (KR)}, pages 600--610, 2008.

\bibitem[Gerede and Su(2007)]{GeSu07}
C.~E. Gerede and J.~Su.
\newblock Specification and verification of artifact behaviors in business
  process models.
\newblock In \emph{Proc.\ of the 5th Int.\ Conf.\ on Service Oriented Computing
  (ICSOC)}, 2007.

\bibitem[Gerede et~al.(2007)Gerede, Bhattacharya, and Su]{GeBS07}
C.~E. Gerede, K.~Bhattacharya, and J.~Su.
\newblock Static analysis of business artifact-centric operational models.
\newblock In \emph{Proc.\ of the 5th Int.\ Conf.\ on Service Oriented Computing
  (ICSOC)}, 2007.

\bibitem[Gerogiannis et~al.(1998)Gerogiannis, Kameas, and Pintelas]{GKP98}
Vasilis~C Gerogiannis, Achilles~D Kameas, and Panayotis~E Pintelas.
\newblock Comparative study and categorization of high-level petri nets.
\newblock \emph{J.\ of Systems and Software}, 43\penalty0 (2):\penalty0
  133--160, 1998.

\bibitem[Giacomo et~al.(2014{\natexlab{a}})Giacomo, Lesp{\'{e}}rance, Patrizi,
  and Vassos]{DLPV14}
Giuseppe~De Giacomo, Yves Lesp{\'{e}}rance, Fabio Patrizi, and Stavros Vassos.
\newblock Progression and verification of situation calculus agents with
  bounded beliefs.
\newblock In \emph{Proc.\ of the 13th Int.\ Conf.\ on Autonomous Agents and
  Multiagent Systems (AAMAS)}, pages 141--148, 2014{\natexlab{a}}.

\bibitem[Giacomo et~al.(2014{\natexlab{b}})Giacomo, Lesp{\'{e}}rance, Patrizi,
  and Vassos]{DLPV14b}
Giuseppe~De Giacomo, Yves Lesp{\'{e}}rance, Fabio Patrizi, and Stavros Vassos.
\newblock {LTL} verification of online executions with sensing in bounded
  situation calculus.
\newblock In \emph{Proc.\ of the 21st Eur.\ Conf.\ on Artificial Intelligence
  (ECAI)}, pages 369--374, 2014{\natexlab{b}}.

\bibitem[Giunchiglia and Bouquet(1997)]{GiBo97}
Fausto Giunchiglia and Paolo Bouquet.
\newblock Introduction to contextual reasoning. an artificial intelligence
  perspective.
\newblock In \emph{Perspectives on Cognitive Science}, pages 138--159. NBU
  Press, 1997.

\bibitem[Gonzales et~al.(2013)Gonzales, Griesmayer, and Lomuscio]{ACSI-D2.2.3}
Pavel Gonzales, Andreas Griesmayer, and Alessio Lomuscio.
\newblock Model checking tool for artifact interoperations {(MOCAI)} --
  {I}teration~3.
\newblock Deliverable ACSI-D2.2.3, ACSI Consortium, May 2013.

\bibitem[Gonzalez et~al.(2012)Gonzalez, Griesmayer, and Lomuscio]{GGL12}
Pavel Gonzalez, Andreas Griesmayer, and Alessio Lomuscio.
\newblock Verifying {GSM}-based business artifacts.
\newblock In \emph{Proc.\ of the 19th IEEE Int.\ Conf.\ on Web Services
  (ICWS)}, pages 25--32, 2012.

\bibitem[Gonzalez et~al.(2014)Gonzalez, Griesmayer, and Lomuscio]{GGL14}
Pavel Gonzalez, Andreas Griesmayer, and Alessio Lomuscio.
\newblock Model checking {GSM}-based multi-agent systems.
\newblock In \emph{Proc.\ of Service-Oriented Computing - ICSOC 2013
  Workshops}, volume 8377 of \emph{Lecture Notes in Computer Science}, pages 54
  -- 68. Springer, 2014.

\bibitem[Gottlob et~al.(1988)Gottlob, Paolini, and Zicari]{GPZ88}
Georg Gottlob, Paolo Paolini, and Roberto Zicari.
\newblock Properties and update semantics of consistent views.
\newblock \emph{ACM Trans.\ on Database Systems}, 13\penalty0 (4):\penalty0
  486--524, October 1988.
\newblock ISSN 0362-5915.

\bibitem[Haase et~al.(2013)Haase, Horrocks, Hovland, Hubauer, Jim{\'e}nez-Ruiz,
  Kharlamov, Kl{\"u}wer, Pinkel, Rosati, Santarelli, Soylu, and
  Zheleznyakov]{HHHH*13}
Peter Haase, Ian Horrocks, Dag Hovland, Thomas Hubauer, Ernesto
  Jim{\'e}nez-Ruiz, Evgeny Kharlamov, Johan~W. Kl{\"u}wer, Christoph Pinkel,
  Riccardo Rosati, Valerio Santarelli, Ahmet Soylu, and Dmitriy Zheleznyakov.
\newblock {O}ptique system: Towards ontology and mapping management in {OBDA}
  solutions.
\newblock In \emph{Proc.\ of the 2nd Int.\ Workshop on Debugging Ontologies and
  Ontology Mappings (WoDOOM)}, pages 21--32, 2013.

\bibitem[Halpin(2010)]{TH10}
Terry Halpin.
\newblock Object-role modeling: Principles and benefits.
\newblock \emph{Int.\ J.\ of Information System Modeling and Design},
  1\penalty0 (1):\penalty0 33--57, January 2010.
\newblock ISSN 1947-8186.

\bibitem[Halpin(2015)]{TH15}
Terry Halpin.
\newblock \emph{Object-Role Modeling Fundamentals}.
\newblock Technics Publications, 2015.

\bibitem[Halpin and Bloesch(1999)]{HaBl99}
Terry~A. Halpin and Anthony~C. Bloesch.
\newblock Data modeling in {UML} and {ORM}: A comparison.
\newblock \emph{J.\ of Database Management}, 10\penalty0 (4):\penalty0 4--13,
  1999.

\bibitem[Hariri et~al.(2013)Hariri, Calvanese, Montali, Giacomo, Masellis, and
  Felli]{BCMD*13}
Babak~Bagheri Hariri, Diego Calvanese, Marco Montali, Giuseppe~De Giacomo,
  Riccardo~De Masellis, and Paolo Felli.
\newblock Description logic {K}nowledge and {A}ction {B}ases.
\newblock \emph{J.\ of Artificial Intelligence Research}, 46:\penalty0
  651--686, 2013.
\newblock ISSN 1076-9757.
\newblock \doi{10.1613/jair.3826}.

\bibitem[Harrison and Khoshnevisan(1992)]{HK92}
Peter~G. Harrison and Hessam Khoshnevisan.
\newblock A new approach to recursion removal.
\newblock \emph{Theoretical Computer Science}, 93\penalty0 (1):\penalty0 91 --
  113, 1992.

\bibitem[Hepp et~al.(2005)Hepp, Leymann, Domingue, Wahler, and Fensel]{HLDWF05}
Martin Hepp, Frank Leymann, John Domingue, Alexander Wahler, and Dieter Fensel.
\newblock Semantic business process management: a vision towards using semantic
  web services for business process management.
\newblock In \emph{Proc.\ of IEEE International Conference on e-Business
  Engineering (ICEBE 2005)}, pages 535--540, Oct 2005.

\bibitem[Heymans et~al.(2008)Heymans, Ma, Anicic, Ma, Steinmetz, Pan, Mei,
  Fokoue, Kalyanpur, Kershenbaum, Schonberg, Srinivas, Feier, Hench, Wetzstein,
  and Keller]{HMAM*08}
Stijn Heymans, Li~Ma, Darko Anicic, Zhilei Ma, Nathalie Steinmetz, Yue Pan,
  Jing Mei, Achille Fokoue, Aditya Kalyanpur, Aaron Kershenbaum, Edith
  Schonberg, Kavitha Srinivas, Cristina Feier, Graham Hench, Branimir
  Wetzstein, and Uwe Keller.
\newblock Ontology reasoning with large data repositories.
\newblock In Martin Hepp, Pieter De~Leenheer, Aldo de~Moor, and York Sure,
  editors, \emph{Ontology Management, Semantic Web, Semantic Web Services, and
  Business Applications}, volume~7 of \emph{Semantic Web And Beyond Computing
  for Human Experience}, pages 89--128. Springer, 2008.

\bibitem[Hull(2008)]{Hull08}
Richard Hull.
\newblock Artifact-centric business process models: {B}rief survey of research
  results and challenges.
\newblock In \emph{Proc.\ of the 7th Int.\ Conf.\ on Ontologies, DataBases, and
  Applications of Semantics (ODBASE)}, volume 5332 of \emph{Lecture Notes in
  Computer Science}, pages 1152--1163. Springer, 2008.
\newblock \doi{10.1007/978-3-540-88873-4_17}.

\bibitem[Hull et~al.(2011)Hull, Damaggio, De~Masellis, Fournier, Gupta, Heath,
  Hobson, Linehan, Maradugu, Nigam, Sukaviriya, and Vaculin]{HDDF*11}
Richard Hull, Elio Damaggio, Riccardo De~Masellis, Fabiana Fournier, Manmohan
  Gupta, Fenno~Terry Heath, III, Stacy Hobson, Mark Linehan, Sridhar Maradugu,
  Anil Nigam, Piwadee~Noi Sukaviriya, and Roman Vaculin.
\newblock Business artifacts with {G}uard-{S}tage-{M}ilestone lifecycles:
  {M}anaging artifact interactions with conditions and events.
\newblock In \emph{Proc.\ of the 5th ACM Int.\ Conf.\ on Distributed
  Event-Based Systems (DEBS~2011)}, pages 51--62, 2011.

\bibitem[Kharbili et~al.(2008)Kharbili, Stein, Markovic, and
  Pulverm{\"u}ller]{KSMP08}
Marwane~El Kharbili, Sebastian Stein, Ivan Markovic, and Elke Pulverm{\"u}ller.
\newblock Towards a framework for semantic business process compliance
  management.
\newblock In \emph{Proc.\ of the 1st Int.\ Workshop on Governance, Risk and
  Compliance - Applications in Information Systems (GRCIS)}, volume 339 of
  \emph{CEUR Electronic Workshop Proceedings,
  {\upshape\protect\url{http://ceur-ws.org/}}}, 2008.

\bibitem[Lanti et~al.(2015)Lanti, Rezk, Xiao, and Calvanese]{LRXC15}
Davide Lanti, Martin Rezk, Guohui Xiao, and Diego Calvanese.
\newblock The {NPD} benchmark: Reality check for {OBDA} systems.
\newblock In \emph{Proc.\ of the 18th Int.\ Conf.\ on Extending Database
  Technology (EDBT)}, 2015.

\bibitem[L\'ecu\'e and L\'eger(2006)]{LL06}
Freddy L\'ecu\'e and Alain L\'eger.
\newblock A formal model for semantic web service composition.
\newblock In \emph{Proc.\ of the 5th Int.\ Semantic Web Conf.\ (ISWC)}, volume
  4273 of \emph{Lecture Notes in Computer Science}, pages 385--398. Springer,
  2006.

\bibitem[Lehmann(1992)]{Lehm92}
Fritz Lehmann, editor.
\newblock \emph{Semantic Networks in {A}rtificial {I}ntelligence}.
\newblock Pergamon Press, Oxford (United Kingdom), 1992.

\bibitem[Lembo and Ruzzi(2007)]{LeRu07}
Domenico Lembo and Marco Ruzzi.
\newblock Consistent query answering over description logic ontologies.
\newblock In \emph{Proc.\ of the 1st Int.\ Conf.\ on Web Reasoning and Rule
  Systems (RR)}, 2007.

\bibitem[Lembo et~al.(2010)Lembo, Lenzerini, Rosati, Ruzzi, and Savo]{LLRRS10}
Domenico Lembo, Maurizio Lenzerini, Riccardo Rosati, Marco Ruzzi, and
  Domenico~Fabio Savo.
\newblock Inconsistency-tolerant semantics for description logics.
\newblock In \emph{Proc.\ of the 4th Int.\ Conf.\ on Web Reasoning and Rule
  Systems (RR)}, pages 103--117, 2010.

\bibitem[Lembo et~al.(2011)Lembo, Lenzerini, Rosati, Ruzzi, and Savo]{LLRRS11}
Domenico Lembo, Maurizio Lenzerini, Riccardo Rosati, Marco Ruzzi, and
  Domenico~Fabio Savo.
\newblock Query rewriting for inconsistent \textit{DL-Lite} ontologies.
\newblock In \emph{Proc.\ of the 5th Int.\ Conf.\ on Web Reasoning and Rule
  Systems (RR)}, 2011.

\bibitem[Levesque et~al.(1997)Levesque, Reiter, Lesperance, Lin, and
  Scherl]{LRLLS97}
H.~J. Levesque, R.~Reiter, Y.~Lesperance, F.~Lin, and R.~Scherl.
\newblock {GOLOG}: {A} logic programming language for dynamic domains.
\newblock \emph{J.\ of Logic Programming}, 31:\penalty0 59--84, 1997.

\bibitem[Levesque(1984)]{Leve84}
Hector~J. Levesque.
\newblock Foundations of a functional approach to knowledge representation.
\newblock \emph{Artificial Intelligence}, 23:\penalty0 155--212, 1984.

\bibitem[Liu and Stoller(1999)]{LS99}
Yanhong~A. Liu and Scott~D. Stoller.
\newblock From recursion to iteration: What are the optimizations?
\newblock In \emph{Proc.\ of the 2000 ACM SIGPLAN Workshop on Partial
  Evaluation and Semantics-based Program Manipulation}, pages 73--82. {ACM}
  Press, 1999.

\bibitem[Lu and Sadiq(2007)]{LS07}
Ruopeng Lu and Shazia Sadiq.
\newblock A survey of comparative business process modeling approaches.
\newblock In \emph{Proc.\ of 10th Int.\ Conf.\ on Business Information Systems
  (BIS)}, volume 4439 of \emph{Lecture Notes in Computer Science}, pages
  82--94. Springer, 2007.

\bibitem[Lutz et~al.(2008)Lutz, Wolter, and Zakharyaschev]{LuWZ08}
Carsten Lutz, Frank Wolter, and Michael Zakharyaschev.
\newblock Temporal description logics: {A} survey.
\newblock In \emph{Proc.\ of the 15th Int.\ Symp.\ on Temporal Representation
  and Reasoning (TIME)}, pages 3--14, 2008.

\bibitem[Lutz et~al.(2009)Lutz, Toman, and Wolter]{LuTW09}
Carsten Lutz, David Toman, and Frank Wolter.
\newblock Conjunctive query answering in the description logic $\mathcal{EL}$
  using a relational database system.
\newblock In \emph{Proc.\ of the 21st Int.\ Joint Conf.\ on Artificial
  Intelligence (IJCAI)}, pages 2070--2075, 2009.

\bibitem[McCarthy(1993)]{McCa93}
John McCarthy.
\newblock Notes on formalizing context.
\newblock In \emph{Proc.\ of the 13th Int.\ Joint Conf.\ on Artificial
  Intelligence (IJCAI)}, pages 555--560, 1993.

\bibitem[McIlraith et~al.(2001)McIlraith, Son, and Zeng]{MSZ01}
Sheila~A. McIlraith, Tran~Cao Son, and Honglei Zeng.
\newblock Semantic web services.
\newblock \emph{IEEE Intelligent Systems}, 16\penalty0 (2):\penalty0 46 -- 53,
  Mar/Apr 2001.

\bibitem[Medjahed et~al.(2003)Medjahed, Bouguettaya, and Elmagarmid]{MBE03}
Brahim Medjahed, Athman Bouguettaya, and Ahmed~K. Elmagarmid.
\newblock Composing web services on the semantic web.
\newblock \emph{Very Large Database J.}, 2003.

\bibitem[Meyer et~al.(2011)Meyer, Smirnov, and Weske]{MSW11}
Andreas Meyer, Sergey Smirnov, and Mathias Weske.
\newblock \emph{Data in business processes}.
\newblock Universit{\"a}tsverlag Potsdam, 2011.

\bibitem[Montali and Rivkin(2014)]{MR14}
Marco Montali and Andrey Rivkin.
\newblock Formal verification of petri nets with names.
\newblock In \emph{Proc.\ of the 11th Int.\ Workshop on Web Services and Formal
  Methods (WS-FM)}, Lecture Notes in Computer Science. Springer, 2014.

\bibitem[Montali and Rivkin(2016)]{MR16}
Marco Montali and Andrey Rivkin.
\newblock Model checking petri nets with names using data-centric dynamic
  systems.
\newblock \emph{Formal Aspects of Computing}, 2016.
\newblock To appear.

\bibitem[Montali et~al.(2014)Montali, Calvanese, and Giacomo]{MoCD14}
Marco Montali, Diego Calvanese, and Giuseppe~De Giacomo.
\newblock Verification of data-aware commitment-based multiagent systems.
\newblock In \emph{Proc.\ of the 13th Int.\ Conf.\ on Autonomous Agents and
  Multiagent Systems (AAMAS)}, pages 157--164, 2014.
\newblock ISBN 978-1-4503-2738-1.

\bibitem[Nigam and Caswell(2003)]{NiCa03}
A.~Nigam and N.~S. Caswell.
\newblock Business artifacts: {A}n approach to operational specification.
\newblock \emph{IBM Systems J.}, 42\penalty0 (3):\penalty0 428--445, 2003.

\bibitem[(OMG)(2014)]{BPMN}
Object Management~Group (OMG).
\newblock Business process model and notation ({BPMN}).
\newblock Technical report, Object Management Group (OMG), January 2014.
\newblock Available at \protect\url{http://www.omg.org/spec/BPMN/2.0.2/}.

\bibitem[Park(1976)]{Park76}
David Michael~Ritchie Park.
\newblock Finiteness is {M}u-ineffable.
\newblock \emph{Theoretical Computer Science}, 3\penalty0 (2):\penalty0
  173--181, 1976.

\bibitem[Pierce(2012)]{Pier12}
Benjamin~C. Pierce.
\newblock Linguistic foundations for bidirectional transformations: invited
  tutorial.
\newblock In \emph{Proc.\ of the 31st ACM SIGACT SIGMOD SIGART Symp.\ on
  Principles of Database Systems (PODS)}, pages 61--64. {ACM} Press, 2012.

\bibitem[Poggi et~al.(2008{\natexlab{a}})Poggi, Lembo, Calvanese, De~Giacomo,
  Lenzerini, and Rosati]{PLCD*08}
Antonella Poggi, Domenico Lembo, Diego Calvanese, Giuseppe De~Giacomo, Maurizio
  Lenzerini, and Riccardo Rosati.
\newblock Linking data to ontologies.
\newblock \emph{J.\ on Data Semantics}, X:\penalty0 133--173,
  2008{\natexlab{a}}.
\newblock \doi{10.1007/978-3-540-77688-8_5}.

\bibitem[Poggi et~al.(2008{\natexlab{b}})Poggi, Rodriguez-Muro, and
  Ruzzi]{PoRR08}
Antonella Poggi, Mariano Rodriguez-Muro, and Marco Ruzzi.
\newblock Ontology-based database access with {DIG-Mastro} and the {OBDA}
  {P}lugin for {Prot\'eg\'e}.
\newblock In Kendall Clark and Peter~F. Patel-Schneider, editors, \emph{Proc.\
  of the 4th Int.\ Workshop on OWL: Experiences and Directions (OWLED~DC)},
  2008{\natexlab{b}}.

\bibitem[Recker et~al.(2009)Recker, Rosemann, Indulska, and Green]{RRIG09}
Jan Recker, Michael Rosemann, Marta Indulska, and Peter Green.
\newblock Business process modeling - a comparative analysis.
\newblock \emph{J.\ of the Association of Information Systems}, 10\penalty0
  (4), 2009.

\bibitem[Reichert(2012)]{Reic12}
Manfred Reichert.
\newblock Process and data: {T}wo sides of the same coin?
\newblock In \emph{Proc.\ of the On the Move Confederated Int.\ Conf.\
  (OTM~2012)}, volume 7565 of \emph{Lecture Notes in Computer Science}, pages
  2--19. Springer, 2012.
\newblock \doi{10.1007/978-3-642-33606-5_2}.

\bibitem[Reiter(2001)]{Reit01}
Raymond Reiter.
\newblock \emph{Knowledge in Action: {L}ogical Foundations for Specifying and
  Implementing Dynamical Systems}.
\newblock The MIT Press, 2001.

\bibitem[Richardson(2010)]{Ric10}
Clay Richardson.
\newblock Warning: Don't assume your business processes use master data.
\newblock In \emph{Proc.\ of the 8th Int.\ Conf.\ on Business Process
  Management (BPM)}, volume 6336 of \emph{Lecture Notes in Computer Science},
  pages 11--12. Springer, 2010.

\bibitem[Rodriguez-Muro and Calvanese(2008)]{RoCa08}
Mariano Rodriguez-Muro and Diego Calvanese.
\newblock Towards an open framework for ontology based data access with
  {Prot\'eg\'e} and {DIG~1.1}.
\newblock In \emph{Proc.\ of the 5th Int.\ Workshop on OWL: Experiences and
  Directions (OWLED)}, volume 432 of \emph{CEUR Electronic Workshop
  Proceedings, {\upshape\protect\url{http://ceur-ws.org/}}}, 2008.

\bibitem[Rodriguez-Muro and Calvanese(2011{\natexlab{a}})]{RoCa11}
Mariano Rodriguez-Muro and Diego Calvanese.
\newblock Dependencies: {M}aking ontology based data access work in practice.
\newblock In \emph{Proc.\ of the 5th Alberto Mendelzon Int.\ Workshop on
  Foundations of Data Management (AMW)}, volume 749 of \emph{CEUR Electronic
  Workshop Proceedings, {\upshape\protect\url{http://ceur-ws.org/}}},
  2011{\natexlab{a}}.

\bibitem[Rodriguez-Muro and Calvanese(2011{\natexlab{b}})]{RoCa11b}
Mariano Rodriguez-Muro and Diego Calvanese.
\newblock Dependencies to optimize ontology based data access.
\newblock In \emph{Proc.\ of the 24th Int.\ Workshop on Description Logic
  (DL)}, volume 745 of \emph{CEUR Electronic Workshop Proceedings,
  {\upshape\protect\url{http://ceur-ws.org/}}}, 2011{\natexlab{b}}.

\bibitem[Rodriguez-Muro and Calvanese(2012)]{RoCa12}
Mariano Rodriguez-Muro and Diego Calvanese.
\newblock High performance query answering over \textit{DL-Lite} ontologies.
\newblock In \emph{Proc.\ of the 13th Int.\ Conf.\ on the Principles of
  Knowledge Representation and Reasoning (KR)}, pages 308--318, 2012.

\bibitem[Rodriguez-Muro et~al.(2008)Rodriguez-Muro, Lubyte, and
  Calvanese]{RoLC08}
Mariano Rodriguez-Muro, Lina Lubyte, and Diego Calvanese.
\newblock Realizing ontology based data access: {A} plug-in for {Prot\'eg\'e}.
\newblock In \emph{Proc.\ of the ICDE Workshop on Information Integration
  Methods, Architectures, and Systems (IIMAS~2008)}, pages 286--289. IEEE
  Computer Society Press, 2008.

\bibitem[Rosemann et~al.(2008)Rosemann, Recker, and Flender]{RRF08}
Michael Rosemann, Jan Recker, and Christian Flender.
\newblock Contextualisation of business processes.
\newblock \emph{Int. J. of Business Process Integration and Management},
  3\penalty0 (1):\penalty0 47--60, 2008.

\bibitem[Russo et~al.(2013)Russo, Mecella, Patrizi, and Montali]{RMPM13}
Alessandro Russo, Massimo Mecella, Fabio Patrizi, and Marco Montali.
\newblock Implementing and running data-centric dynamic systems.
\newblock In \emph{Proc.\ of the 7th IEEE Int.\ Conf.\ on Service Oriented
  Computing and Applications (SOCA)}, pages 225--232. IEEE, 2013.

\bibitem[Santoso(2012)]{AS-RR-12b}
Ario Santoso.
\newblock When data, knowledge and processes meet together.
\newblock In \emph{Proc.\ of the 6th Int.\ Conf.\ on Web Reasoning and Rule
  Systems (RR)}, volume 7497 of \emph{Lecture Notes in Computer Science}, pages
  291--296. Springer, 2012.

\bibitem[Sequeda et~al.(2014)Sequeda, Arenas, and Miranker]{SeAM14}
Juan~F. Sequeda, Marcelo Arenas, and Daniel~P. Miranker.
\newblock {OBDA}: {Q}uery rewriting or materialization? {I}n practice, both!
\newblock In \emph{Proc.\ of the 13th Int.\ Semantic Web Conf.\ (ISWC)}, volume
  8796 of \emph{Lecture Notes in Computer Science}, pages 535--551. Springer,
  2014.

\bibitem[Serafini and Homola(2012)]{SeHo12}
Luciano Serafini and Martin Homola.
\newblock Contextualized knowledge repositories for the semantic web.
\newblock \emph{J.\ of Web Semantics}, 12:\penalty0 64--87, 2012.

\bibitem[Smullyan(1968)]{Smul68}
R.~M. Smullyan.
\newblock \emph{First Order Logic}.
\newblock Springer, Berlin (Germany), 1968.

\bibitem[Sowa(1991)]{Sowa91}
John~F. Sowa, editor.
\newblock \emph{Principles of Semantic Networks: Explorations in the
  Representation of Knowledge}.
\newblock Morgan Kaufmann, 1991.

\bibitem[Soylu et~al.(2014)Soylu, Kharlamov, Zheleznyakov, Jim{\'e}nez-Ruiz,
  Giese, and Horrocks]{SKZJ*14}
Ahmet Soylu, Evgeny Kharlamov, Dmitriy Zheleznyakov, Ernesto Jim{\'e}nez-Ruiz,
  Martin Giese, and Ian Horrocks.
\newblock {OptiqueVQS}: Visual query formulation for {OBDA}.
\newblock In \emph{Proc.\ of the 27th Int.\ Workshop on Description Logic
  (DL)}, volume 1193 of \emph{CEUR Electronic Workshop Proceedings,
  {\upshape\protect\url{http://ceur-ws.org/}}}, pages 725--728, 2014.
\newblock URL \url{http://ceur-ws.org/Vol-1193/paper_88.pdf}.

\bibitem[Stirling(2001)]{Stir01}
Colin Stirling.
\newblock \emph{Modal and Temporal Properties of Processes}.
\newblock Springer, 2001.

\bibitem[Swartz(2002)]{Swa02}
A.~Swartz.
\newblock {M}usic{B}rainz: a semantic web service.
\newblock \emph{IEEE Intelligent Systems}, 17\penalty0 (1):\penalty0 76 -- 77,
  Jan/Feb 2002.

\bibitem[TC(2007)]{WSBPEL}
OASIS Web Services Business Process Execution Language~(WSBPEL) TC.
\newblock Web services business process execution language version 2.0.
\newblock Technical report, OASIS, April 2007.
\newblock Available at
  \protect\url{http://docs.oasis-open.org/wsbpel/2.0/OS/wsbpel-v2.0-OS.html}.

\bibitem[Toribio~Gomez et~al.(2010)Toribio~Gomez, Murphy O~Connor, De~Leenheer,
  Malarme, De~Vos, Fournier, Boaz, van Dongen, Fahland, de~Leoni, and
  Dumas]{ACSI-D5.1}
David Toribio~Gomez, Catherine Murphy O~Connor, Pieter De~Leenheer, Pierre
  Malarme, Jordi De~Vos, Fabiana Fournier, David Boaz, Boudewijn van Dongen,
  Dirk Fahland, Massimiliano de~Leoni, and Marlon Dumas.
\newblock Energy and {FRIS} use case definition and requirements.
\newblock Deliverable ACSI-D5.1, ACSI Consortium, March 2010.

\bibitem[Toribio~Gomez et~al.(2011)Toribio~Gomez, Murphy O~Connor, De~Leenheer,
  Malarme, De~Vos, Christiaens, Fournier, and Limonad]{ACSI-D5.2}
David Toribio~Gomez, Catherine Murphy O~Connor, Pieter De~Leenheer, Pierre
  Malarme, Jordi De~Vos, Stijn Christiaens, Fabiana Fournier, and Lior Limonad.
\newblock Energy and {FRIS} ``as-is'' assessment.
\newblock Deliverable ACSI-D5.2, ACSI Consortium, May 2011.

\bibitem[Toribio~G{\'o}mez et~al.(2012)Toribio~G{\'o}mez, Murphy-O~Connor,
  Leenheer, and Malarme]{ACSI-D5.3}
David Toribio~G{\'o}mez, Catherine Murphy-O~Connor, Pieter~De Leenheer, and
  Pierre Malarme.
\newblock Deployment and evaluation of pilots using the {ACSI} {H}ub {S}ystem
  -- {I}teration~1.
\newblock Deliverable ACSI-D5.3, ACSI Consortium, June 2012.

\bibitem[Toribio~G{\'o}mez et~al.(2013)Toribio~G{\'o}mez, Murphy-O~Connor,
  Leenheer, and Malarme]{ACSI-D5.5}
David Toribio~G{\'o}mez, Catherine Murphy-O~Connor, Pieter~De Leenheer, and
  Pierre Malarme.
\newblock Deployment and evaluation of pilots using the {ACSI} {H}ub {S}ystem
  -- {R}esults and evaluation.
\newblock Deliverable ACSI-D5.5, ACSI Consortium, May 2013.

\bibitem[UML()]{UML05}
UML.
\newblock {U}nified {M}odeling {L}anguage ({UML}) superstructure, version 2.0.
\newblock Available at \protect\url{http://www.uml.org/}, August 2005.

\bibitem[van~der Aalst(2013)]{aalst13}
Wil~MP van~der Aalst.
\newblock Business process management: A comprehensive survey.
\newblock \emph{ISRN Software Engineering}, 2013, 2013.

\bibitem[van~der Aalst and ter Hofstede(2005)]{YAWL}
W.M.P. van~der Aalst and A.H.M. ter Hofstede.
\newblock {YAWL}: yet another workflow language.
\newblock \emph{Information Systems}, 30\penalty0 (4):\penalty0 245 -- 275,
  2005.
\newblock ISSN 0306-4379.

\bibitem[Weske(2007)]{Wesk07}
Mathias Weske.
\newblock \emph{Business Process Management: Concepts, Languages,
  Architectures}.
\newblock Springer, 2007.

\bibitem[Wetzstein et~al.(2007)Wetzstein, Ma, Filipowska, Kaczmarek, Bhiri,
  Losada, Lopez-Cobo, and Cicurel]{WMFKBLLC07}
Branimir Wetzstein, Zhilei Ma, Agata Filipowska, Monika Kaczmarek, Sami Bhiri,
  Silvestre Losada, Jose-Manuel Lopez-Cobo, and Laurent Cicurel.
\newblock Semantic business process management: A lifecycle based requirements
  analysis.
\newblock In \emph{Proc.\ of the Workshop on Semantic Business Process and
  Product Lifecycle Management (SBPM)}, volume 251 of \emph{CEUR Electronic
  Workshop Proceedings, {\upshape\protect\url{http://ceur-ws.org/}}}, 2007.

\bibitem[Winslett(1990)]{Wins90}
Marianne Winslett.
\newblock \emph{{U}pdating {L}ogical {D}atabases}.
\newblock Cambridge University Press, 1990.

\bibitem[Wolter and Zakharyaschev(1999)]{WoZa98}
Frank Wolter and Michael Zakharyaschev.
\newblock Temporalizing description logics.
\newblock In M.~de~Rijke and D.~Gabbay, editors, \emph{Proc.\ of the 2th Int.\
  Workshop on Frontiers of Combining Systems (FroCoS)}, Amsterdam, 1999. Wiley.

\bibitem[Woods(1991)]{Wood91}
William~A. Woods.
\newblock Understanding subsumption and taxomony: {A} framework for progress.
\newblock In J.~F. Sowa, editor, \emph{Principles of Semantic Networks}, pages
  45--94. Morgan Kaufmann, 1991.

\bibitem[Zarrie{\ss} and Cla{\ss}en(2014)]{ZC14}
Benjamin Zarrie{\ss} and Jens Cla{\ss}en.
\newblock Verifying {CTL} properties of {GOLOG} programs over local-effect
  actions.
\newblock In \emph{Proc.\ of the 21st Eur.\ Conf.\ on Artificial Intelligence
  (ECAI)}, 2014.

\bibitem[Zarrie{\ss} and Cla{\ss}en(2015)]{ZC15}
Benjamin Zarrie{\ss} and Jens Cla{\ss}en.
\newblock Verification of knowledge-based programs over description logic
  actions.
\newblock In \emph{Proc.\ of the 24th Int.\ Joint Conf.\ on Artificial
  Intelligence (IJCAI)}, pages 3278--3284, 2015.

\end{thebibliography}
